\documentclass[11pt]{elsarticle} 
\pdfoutput=1
\usepackage[letterpaper,dvips,hmargin={1in,1in},vmargin={1.2in,0.8in}]{geometry}
\usepackage[utf8]{inputenc}
\usepackage{datetime2}
\usepackage{lineno}
\modulolinenumbers[5]
\usepackage[monochrome]{color}
\usepackage[usenames,dvipsnames]{xcolor}
\usepackage{amsmath,amssymb,amsthm,amsfonts}
\usepackage{hyperref}
\usepackage{cleveref}
\usepackage{blindtext} 
\usepackage{graphicx,tabularx}
\usepackage{wrapfig,caption}
\usepackage{subfigure}
\usepackage{multicol}
\usepackage{xspace}
\usepackage{rviewport}
\usepackage{ragged2e}
\usepackage{array}
\usepackage{setspace}
\usepackage{comment}
\usepackage{bm}
\usepackage{braket}
\usepackage{enumerate}
\usepackage[shortlabels]{enumitem}

\setlist[description]{leftmargin=0.3cm}
\setlist[itemize]{leftmargin=0.5cm}
\usepackage{mathtools}
\usepackage{mathrsfs} 
\usepackage{sidecap}
\sidecaptionvpos{figure}{c}
\usepackage{url}

\usepackage[printonlyused,withpage]{acronym}
\usepackage{changepage}

\usepackage{glossaries}

\usepackage[nottoc,notlot,notlof]{tocbibind}

\usepackage{titlesec}
\titleformat{\subsection}
  {\normalfont\large\bfseries}{\thesubsection}{1em}{}
\titlespacing{\subsection}{0pt}{5mm}{2mm}

\titleformat{\subsubsection}
  {\normalfont\normalsize\bfseries}{\thesubsubsection}{1em}{}
\titlespacing{\subsubsection}{0pt}{2mm}{2mm}

\titleformat{\paragraph}
  {\normalfont\normalsize\bfseries}{\theparagraph}{.5em}{}
\titlespacing{\paragraph}{0pt}{1mm}{0mm}

\titleformat{\subparagraph}[runin]{\normalfont\normalsize\bfseries}{\thesubparagraph}{0em}{}
\titlespacing{\subparagraph}{0pt}{5mm}{2mm}

\makeatletter
 \def\@textbottom{\vskip \z@ \@plus 1pt}
 \let\@texttop\relax
\makeatother

\graphicspath{{./}{}}

\def\figureautorefname~#1\null{Fig.\,#1\null}
\def\tableautorefname~#1\null{Tab.\,#1\null}

\def\equationautorefname~#1\null{Eq.\,(#1)\null}

\crefname{chapter}{Ch.\negthinspace}{Chs.\negthinspace}
\crefname{section}{Ch.\negthinspace}{Chs.\negthinspace}
\crefname{subsection}{Sec.\negthinspace}{Secs.\negthinspace}
\crefname{subsubsection}{Sec.\negthinspace}{Secs.\negthinspace}
\crefname{paragraph}{Sec.\negthinspace}{Secs.\negthinspace}
\crefname{figure}{Fig.\negthinspace}{Figs.\negthinspace}
\crefname{equation}{Eq.\negthinspace}{Eqs.\negthinspace}
\crefname{table}{Table}{Tables}
\crefname{appendix}{Appendix}{Appendices}

\definecolor{darkBlue}{rgb}{0, 0, 0.8}
\hypersetup{
    bookmarksopen=true,     
    bookmarksopenlevel=1,
    unicode=false,          
    pdftoolbar=true,        
    pdfmenubar=true,        
    colorlinks=true,        
    linkcolor=darkBlue,     
    citecolor=darkBlue,      
    filecolor=darkBlue,      
    urlcolor=darkBlue        
}

\makeatletter
\AtBeginDocument{%
  \renewcommand*{\AC@hyperlink}[2]{%
    \begingroup
      \hypersetup{hidelinks}%
      \hyperlink{#1}{#2}%
    \endgroup
  }%
}
\makeatother

\definecolor{coat1}{RGB}{0, 110, 120 }
\definecolor{hookgreen}{rgb}{0.0,0.44,0.0}

\newcommand{\tmv}[1]{\color{black}{#1}}
\newcommand{\fsarazin}[1]{\color{black}{#1}}
\newcommand{\ewm}[1]{\color{black}{#1}}
\newcommand{\fgs}[1]{\color{black}{#1}}
\newcommand{\alan}[1]{\color{black}{#1}}
\newcommand{\ds}[1]{\color{black}{#1}}
\newcommand{\je}[1]{\color{black}{#1}}

\let\oldbibliography\thebibliography
\renewcommand{\thebibliography}[1]{%
  \oldbibliography{#1}%
  \setlength{\itemsep}{-1pt}%
}

\newcommand\snowmass{
\begin{center}
  \rule[-0.2in]{\hsize}{0.01in}\\
  \rule{\hsize}{0.01in}\\
  \vskip 0.1in
  Submitted to the US Community Study\\ 
  on the Future of Particle Physics (Snowmass 2021)\\
  \rule{\hsize}{0.01in}\\
  \rule[+0.2in]{\hsize}{0.01in}\\[-2em]
\end{center}
}

\newcommand{\COT}{CO$_2$}

\newcommand{\eV}{\mathrm{eV}}
\newcommand{\EeV}{\mathrm{EeV}}
\newcommand{\km}{\mathrm{km}}

\newcommand{\Hinf}{H_\mathbf{inf}}

\newcommand{\Geff}{\Gamma_\mathbf{eff}}

\newcommand{\sqrts}{\ensuremath{\sqrt{s}}\xspace}
\newcommand{\sqrtsnn}{\ensuremath{\sqrt{s_\mathrm{NN}}}\xspace}

\newcommand{\tev}{\,TeV\xspace}

\newcommand{\pp}{p-p\xspace}
\newcommand{\pPb}{p-Pb\xspace}
\newcommand{\pO}{p-O\xspace}
\newcommand{\OO}{O-O\xspace}
\newcommand{\pion}{p-ion\xspace}
\newcommand{\ionion}{ion-ion\xspace}
\newcommand{\PbPb}{Pb-Pb\xspace}
\newcommand{\XeXe}{Xe-Xe\xspace}
\newcommand{\NASixtyOne}{NA61\slash SHINE\xspace}
\newcommand{\rhoz}{\ensuremath{\rho^0}\xspace}
\newcommand{\piz}{\ensuremath{\pi^0}\xspace}

\newcommand{\meanlnA}{\ensuremath{\langle\ln A\rangle}\xspace}

\newcommand{\xmax}{\ensuremath{X_\mathrm{max}}\xspace}
\newcommand{\xmumax}{\ensuremath{\xmax^\mu}\xspace}
\newcommand{\nmu}{\ensuremath{N_\mu}\xspace}
\newcommand{\nelec}{\ensuremath{N_e}\xspace}

\newcommand{\gcm}{\ensuremath{\mathrm{g\,cm}^{-2}}\xspace}

\newcommand{\meanXmax}{\ensuremath{\langle X_{\mathrm{max}}\rangle}\xspace}
\newcommand{\sigmaXmax}{\ensuremath{\sigma(X_\mathrm{max})}\xspace}
\newcommand{\epos}{\textsc{EPOS}\xspace}
\newcommand{\eposlhc}{\textsc{EPOS-LHC}\xspace}
\newcommand{\sibyll}[1]{\textsc{Sibyll#1}\xspace}
\newcommand{\qgsii}{\textsc{QGSJet-II.04}\xspace}
\newcommand{\qgs}{\textsc{QGSJet}\xspace}

\newcommand{\lsim}{\mathrel{\hbox{\rlap{\lower.75ex \hbox{$\sim$}} \kern-.3em \raise.4ex \hbox{$<$}}}}
\newcommand{\gsim}{\mathrel{\hbox{\rlap{\lower.75ex \hbox{$\sim$}} \kern-.3em \raise.4ex \hbox{$>$}}}}

\newcommand{\TAxFour}{TA\ensuremath{\times}4\xspace}

\def\eV{\ifmmode {\mathrm{e\kern -0.07em V}}
        \else \textrm{e\kern -0.07em V}\fi\xspace}

\def\ccite#1{}

\newcommand{\unitspace}[2]{%
  #1%
  \texorpdfstring{\,}{\space}%
  #2%
}

\newcommand{\fakesection}[1]{%
  \par\refstepcounter{section}
  \sectionmark{#1}
  \addcontentsline{toc}{section}{\protect\numberline{\thesection}#1}
}

\begin{document}

\clearpage
\begin{titlepage}
\thispagestyle{empty}	


\begin{center}
{\Large }
\end{center}

\vspace*{-2cm}
\snowmass
\vspace*{-1.35cm}
\makebox[0pt][l]{%
  \raisebox{-\totalheight}[0pt][0pt]{%
    \includegraphics[width=16.2cm]{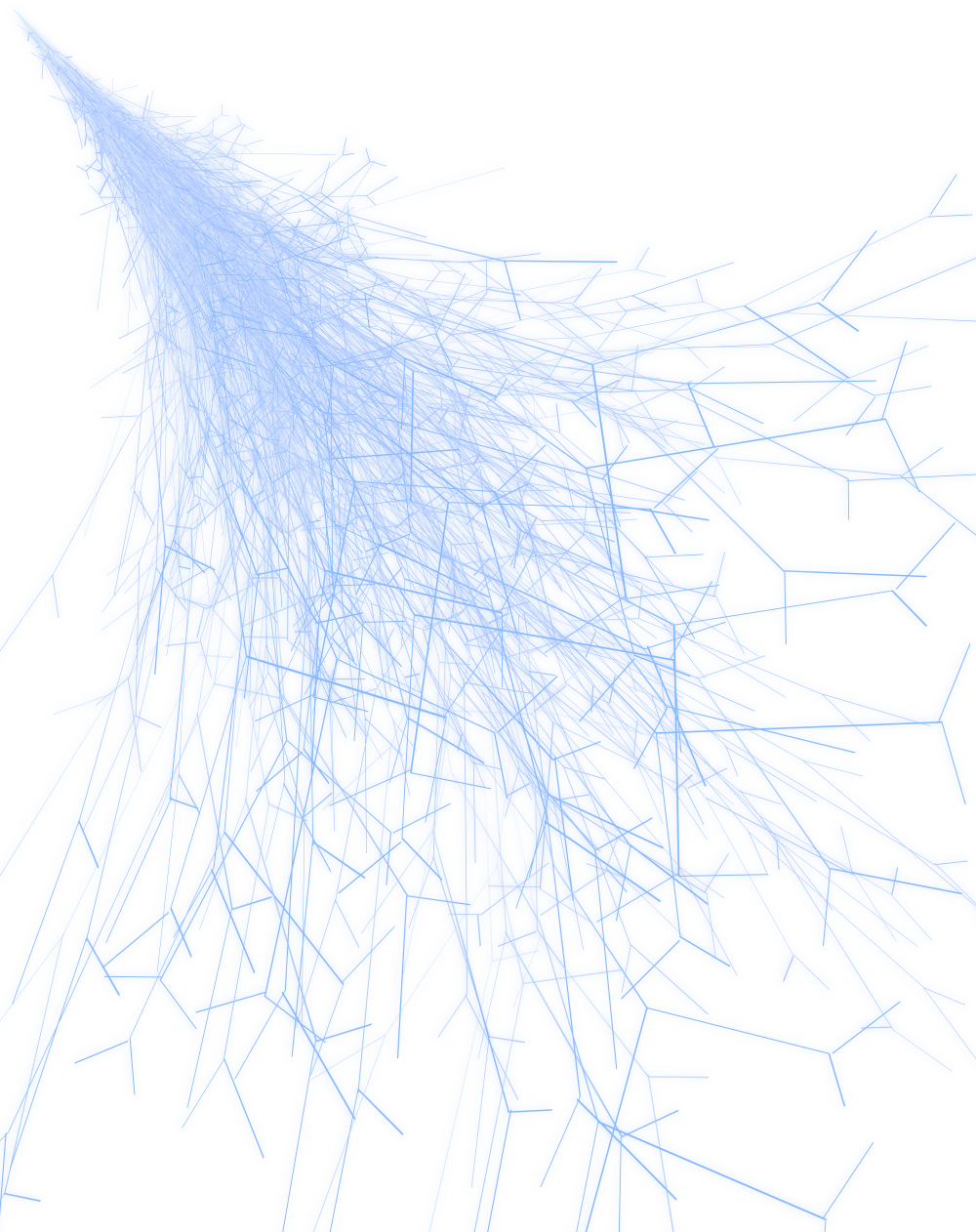}}}%

\vspace*{1.5cm}

\vspace*{0.8cm}
\begin{center}{\textbf{\huge Ultra-High-Energy Cosmic Rays}\ \\ \vspace*{0.4cm}{\LARGE
\bf The Intersection of the Cosmic and Energy Frontiers}}\end{center}	

\vspace*{4.8cm}

{\bf Abstract:} The present white paper is submitted as part of the ``Snowmass'' process to help inform the long-term plans of the United States Department of Energy and the National Science Foundation for high-energy physics. It summarizes the science questions driving the Ultra-High-Energy Cosmic-Ray (UHECR) community and provides recommendations on the strategy to answer them in the next two decades.

\vspace*{0.2in}
\end{titlepage}

\clearpage
\pagenumbering{roman}
\thispagestyle{empty}	
\vspace*{0.2cm}
\begin{center}{{\large \textsc{Conveners}}\\
\vspace*{0.2cm}

{\alan A.~Coleman}$^{1}$, 
{\je J.~Eser}$^{2}$, 
{\ewm E.~Mayotte}$^{3}$, 
{\fsarazin F.~Sarazin}$^{\dag\,3}$, 
{\fgs F.\,G.~Schr\"oder}$^{\dag\,1,4}$, 
{\ds D.~Soldin}$^{1,5}$,
{\tmv T.\,M.~Venters}$^{\dag\,6}$
}\end{center}

\vspace*{-0.2cm}
\begin{center}{{\large \textsc{Topical Conveners}}\\ 
\vspace*{0.2cm}

{R.~Aloisio}$^{7}$,
{J. Alvarez-Mu\~niz}$^{8}$,
{R.~{Alves~Batista}}$^{9}$,
{D.~Bergman}$^{10}$,
{M.~Bertaina}$^{11}$,
{L.~Caccianiga}$^{12}$,
{O.~Deligny}$^{13}$,
{H.\,P.~Dembinski}$^{14}$,
{P.\,B.~Denton}$^{15}$,
{A.~di~Matteo}$^{16}$,
{N.~Globus}$^{17,18}$,
{J.~Glombitza}$^{19}$,
{G.~Golup}$^{20}$,
{A.~Haungs}$^{4}$,
{J.\,R.~H\"{o}randel}$^{21}$,
{T.\,R.~Jaffe}$^{22}$,
{J.\,L.~Kelley}$^{23}$,
{J.\,F.~Krizmanic}$^{6}$,
{L.~Lu}$^{23}$,
{J.\,N.~Matthews}$^{10}$,
{I.~Mari\c{s}}$^{24}$,
{R.~Mussa}$^{16}$,
{F.~Oikonomou}$^{25}$,
{T.~Pierog}$^{4}$,
{E.~Santos}$^{26}$,
{P.~Tinyakov}$^{24}$,
{Y.~Tsunesada}$^{27,28}$,
{M.~Unger}$^{4}$,
{A.~Yushkov}$^{26}$
}\end{center}
\vspace*{-0.2cm}
\begin{center}{{\large \textsc{Contributors}}\\
\vspace*{0.2cm}
{M.\,G.~Albrow}$^{29}$,
{L.\,A.~Anchordoqui}$^{30}$,
{K.~Andeen}$^{31}$,
{E.~Arnone}$^{11,16}$,
{D.~Barghini}$^{11,16}$,
{E.~Bechtol}$^{23}$,
{J.\,A.~Bellido}$^{32}$,
{M.~Casolino}$^{33,34}$,
{A.~Castellina}$^{16,35}$,
{L.~Cazon}$^{8}$,
{R.~Concei\c c\~ ao}$^{36}$,
{R.~Cremonini}$^{37}$,
{H.~Dujmovic}$^{4}$,
{R.~Engel}$^{4,5}$,
{G.~Farrar}$^{38}$,
{F.~Fenu}$^{11,16}$,
{S.~Ferrarese}$^{11}$,
{T.~Fujii}$^{39}$,
{D.~Gardiol}$^{16,35}$,
{M.~Gritsevich}$^{40,41}$,
{P.~Homola}$^{42}$,
{T.~Huege}$^{4,43}$,
{K.\,-H.~Kampert}$^{44}$,
{D.~Kang}$^{4}$,
{E.~Kido}$^{45}$,
{P.~Klimov}$^{46}$,
{K.~Kotera}$^{43,47}$,
{B.~Kozelov}$^{48}$,
{A.~Leszczy\'{n}ska}$^{1,5}$,
{J.~Madsen}$^{23}$,
{L.~Marcelli}$^{34}$,
{M.~Marisaldi}$^{49}$,
{O.~Martineau-Huynh}$^{50}$,
{S.~Mayotte}$^{3}$,
{K.~Mulrey}$^{21}$,
{K.~Murase}$^{51,52}$,
{M.~S.~Muzio}$^{51}$,
{S.~Ogio}$^{28}$,
{A.~V.~Olinto}$^{2}$,
{Y.~Onel}$^{53}$,
{T.~Paul}$^{30}$,
{L.~Piotrowski}$^{54}$,
{M.~Plum}$^{55}$,
{B.~Pont}$^{21}$,
{M.~Reininghaus}$^{4}$,
{B.~Riedel}$^{23}$,
{F.~Riehn}$^{36}$,
{M.~Roth}$^{4}$,
{T.~Sako}$^{56}$,
{F.~Schl\"{u}ter}$^{4,57}$,
{D.\,H.~Shoemaker}$^{58}$,
{J.~Sidhu}$^{59}$,
{I.~Sidelnik}$^{20}$,
{C. Timmermans}$^{21,60}$,
{O.~Tkachenko}$^{4}$,
{D.~Veberic}$^{4}$,
{S.~Verpoest}$^{61}$,
{V.~Verzi}$^{34}$,
{J.~V\'icha}$^{26}$,
{D.~Winn}$^{53}$,
{E.~Zas}$^{8}$,
{M.~Zotov}$^{46}$

}\end{center}
\begin{center}
$^\dag$Correspondence\label{sec:correspondence}: \href{mailto:fsarazin@mines.edu}{fsarazin@mines.edu}, \href{mailto:fgs@udel.edu}{fgs@udel.edu}, \href{mailto:tonia.m.venters@nasa.gov}{tonia.m.venters@nasa.gov}
\end{center}

\vspace*{-0.2cm}
\begin{adjustwidth}{-1cm}{-1cm}
\begin{center}{\it\footnotesize
$^{1}${Bartol Research Institute, Department of Physics and Astronomy, University of Delaware, Newark DE, USA}\\
$^{2}${Department of Astronomy and Astrophysics, University of Chicago, Chicago IL, USA}\\
$^{3}${Department of Physics, Colorado School of Mines, Golden CO, USA}\\
$^{4}${Institute for Astroparticle Physics, Karlsruhe Institute of Technology (KIT), Karlsruhe, Germany}\\
$^{5}${Institute of Experimental Particle Physics, Karlsruhe Institute of Technology (KIT), Karlsruhe, Germany}\\
$^{6}${Astroparticle Physics Laboratory, NASA Goddard Space Flight Center, Greenbelt, MD, USA}\\
$^{7}${Gran Sasso Science Institute (GSSI), L’Aquila, Italy}\\
$^{8}${Instituto Galego de F\' \i sica de Altas Enerx\' \i as (IGFAE), University of Santiago de Compostela, Santiago, Spain}\\
$^{9}${Instituto de Física Teórica UAM/CSIC, U. Autónoma de Madrid, Madrid, Spain}\\
$^{10}${Department of Physics and Astronomy, University of Utah, Salt Lake UT, USA}\\
$^{11}${Dipartimento di Fisica, Università degli studi di Torino, Torino, Italy}\\
$^{12}${Istituto Nazionale di Fisica Nucleare - Sezione di Milano, Italy}\\
$^{13}${Laboratoire de Physique des 2 Infinis Ir\`ene Joliot-Curie (IJCLab), CNRS/IN2P3, Universit\'{e} Paris-Saclay, Orsay, France}\\
$^{14}${Faculty of Physics, TU Dortmund University, Germany}\\
$^{15}${High Energy Theory Group, Physics Department, Brookhaven National Laboratory, Upton NY, USA}\\
$^{16}${Istituto Nazionale di Fisica Nucleare (INFN), sezione di Torino, Turin, Italy}\\
$^{17}${Department of Astronomy and Astrophysics, University of California, Santa Cruz CA, USA}\\
$^{18}${Center for Computational Astrophysics, Flatiron Institute, Simons Foundation, New York NY, USA}\\
$^{19}${III. Physics Institute A, RWTH Aachen University, Aachen, Germany}\\
$^{20}${Centro At\'omico Bariloche, CNEA and CONICET, Bariloche, Argentina}\\
$^{21}${Department of Astrophysics/IMAPP, Radboud University, Nijmegen, The Netherlands}\\
$^{22}${HEASARC Office, NASA Goddard Space Flight Center, Greenbelt, MD, USA}\\
$^{23}${WIPAC / University of Wisconsin, Madison WI, USA}\\
$^{24}${Universit\'e Libre de Bruxelles (ULB), Brussels, Belgium}\\
$^{25}${Institutt for fysikk, Norwegian University of Science and Technology (NTNU), Trondheim, Norway}\\
$^{26}${Institute of Physics of the Czech Academy of Sciences, Prague, Czech Republic}\\
$^{27}${Graduate School of Science, Osaka Metropolitan University, Osaka, Japan}\\
$^{28}${Nambu Yoichiro Institute for Theoretical and Experimental Physics, Osaka City University, Osaka, Japan}\\
$^{29}${Fermi National Accelerator Laboratory, USA}\\
$^{30}${Lehman College, City University of New York, Bronx NY, USA}\\
$^{31}${Department of Physics, Marquette University, Milwaukee WI, USA}\\
$^{32}${The University of Adelaide, Adelaide, Australia}\\
$^{33}${RIKEN Cluster for Pioneering Research, Advanced Science Institute (ASI), Wako Saitama, Japan}\\
$^{34}${INFN, sezione di Roma ``Tor Vergata", Roma, Italy}\\
$^{35}${INAF, Osservatorio Astrofisico di Torino, Pino Torinese, Italy}\\
$^{36}${Laborat\'orio de Instrumenta\c{c}\~{a}o e F\'isica Experimental de Part\'iculas, Instituto Superior T{\'e}cnico, Lisbon, Portugal}\\
$^{37}${ARPA Piemonte, Turin, Italy}\\
$^{38}${Center for Cosmology and Particle Physics, New York University, New York NY, USA}\\
$^{39}${Hakubi Center for Advanced Research, Kyoto University, Kyoto, Japan}\\
$^{40}${Finnish Geospatial Research Institute (FGI), Espoo, Finland}\\
$^{41}${Department of Physics, University of Helsinki, Helsinki, Finland}\\
$^{42}${Institute of Nuclear Physics Polish Academy of Sciences, Krak\'ow, Poland}\\
$^{43}${Astrophysical Institute, Vrije Universiteit Brussel, Brussels, Belgium}\\
$^{44}${Department of Physics, University of Wuppertal, Wuppertal, Germany}\\
$^{45}${RIKEN Cluster for Pioneering Research, Astrophysical Big Bang Laboratory (ABBL), Saitama, Japan}\\
$^{46}${Skobeltsyn Institute of Nuclear Physics (SINP), Lomonosov Moscow State University (SINP MSU), Moscow, Russia}\\
$^{47}${Institut d'Astrophysique de Paris (IAP), Paris, France}\\
$^{48}${Polar Geophysical Institute, Apatity, Russia}\\
$^{49}${Birkeland Centre for Space Science, Department of Physics, University of Bergen, Bergen, Norway}\\
$^{50}${Sorbonne Universit\'e, CNRS/IN2P3, LPNHE, CNRS/INSU, IAP, Paris, France}\\
$^{51}${Pennsylvania State University, University Park PA, USA}\\
$^{52}${Yukawa Institute for Theoretical Physics (YITP), Kyoto University, Kyoto, Japan}\\
$^{53}${Department of Physics and Astronomy, University of Iowa, Iowa City IA, USA}\\
$^{54}${University of Warsaw, Warsaw, Poland}\\
$^{55}${South Dakota School of Mines \& Technology, Rapid City SD, USA}\\
$^{56}${ICRR, University of Tokyo, Kashiwa, Chiba, Japan}\\
$^{57}${Instituto de Tecnolog{\'i}as en Detecci{\'o}n y Astropart{\'i}culas, Universidad Nacional de San Mart{\'i}n, Buenos Aires, Argentina}\\
$^{58}${Kavli Institute, Massachusetts Institute of Technology, Cambridge, Massachusetts, USA}\\
$^{59}${Cleveland Clinic, Cleveland OH, USA}\\
$^{60}${National Institute for Subatomic Physics (NIKHEF), Amsterdam, The Netherlands}\\[-1mm]
$^{61}${Dept.~of Physics and Astronomy, University of Gent, Gent, Belgium}
}
  \rule[-0.2in]{\hsize}{0.01in}\\[.15mm]
  \rule{\hsize}{0.01in}\\
\end{center}
\end{adjustwidth}
\thispagestyle{empty}

\vspace*{-0.2cm}
\pagebreak
\begin{center}{
{\large \textsc{In addition, 200$+$ scientists from 32 countries endorse this white paper}
}
\\
\vspace*{0.5cm}
{M.~Ackermann},
{G.~Alex~Anastasi},
{K.~Almeida~Cheminant},
{S.~Andringa},
{J.~Antonio~Aguilar~S\'anchez},
{C.~Aramo},
{P.~Assis},
{R.~Attallah},
{J.~Augusto~Chinellato},
{M.~Bagheri
% ...
},

}\end{center}
\pagebreak
\pagestyle{plain}

\tableofcontents

\section*{List of Acronyms}
\addcontentsline{toc}{section}{\protect\textbf{List of Acronyms}}
\begin{acronym}[TDMA]
    \acro{AAI}{Applied Artificial Intelligence}
    \acro{ADEME}{French Environment and Energy Management Agency}
	\acro{AERA}{Auger Engineering Radio Array}
	\acro{AGASA}{Akeno Giant Air Shower Array}
	\acro{AGN}{active galactic nuclei}
	\acro{AMON}{Astrophysical Multi-messenger Observatory Network}
	\acro{ANITA}{Antarctic Impulsive Transient Antenna}
	\acro{ARIANNA}{Antarctic Ross Ice-Shelf Antenna Neutrino Array}
	\acro{ASIM}{Atmosphere-Space Interactions Monitor}
	\acro{BDT}{boosted decision tree}
	\acro{BSM}{beyond the Standard Model}
	\acro{CSCCE}{Center for Scientific Collaboration and Community Engagement}
	\acro{CIC}{constant-intensity cut}
	\acro{CLF}{Central Laser Facility}
	\acro{CMB}{cosmic microwave background}
	\acro{CRAFFT}{Cosmic Ray Air Fluorescence Fresnel-lens Telescope}
	\acro{CRE}{Cosmic Ray Ensemble}
	\acro{CREDO}{The Cosmic Ray Extremely Distributed Observatory}
	\acro{CNN}{convolutional neural network}
	\acro{CPU}{central processing unit}
	\acro{CR}{cosmic ray}
	\acro{CTA}{Cherenkov Telescope Array}
	\acro{CTH}{cloud-top height}
	\acro{CVMFS}{CernVM File System}
	\acro{DEC}{declination}
	\acro{DM}{dark matter}
	\acro{DNN}{deep neural network}
	\acro{DOE}{Department of Energy}
	\acro{DOM}{digital optical module}
	\acro{DPU}{data processing unit}
	\acro{DSA}{diffusive shock acceleration}
	\acro{EAS}{extended air shower}
	\acro{EBL}{extragalactic background light}
	\acro{EDI}{equity diversity and inclusion}
	\acro{EMP}{electromagnetic pulse}
	\acro{EUSO}{(Joint Experiment Missions) Extreme Universe Space Observatory}
	\acro{FAIR}{findability, accessibility, interoperability, and reusability}
	\acro{FAST}{Fluorescence detector Array of Single-pixel Telescopes}
	\acro{FCC}{Future Circular Collider}
	\acro{FD}{fluorescence detector}
	\acro{FoV}{field of view}
	\acro{FPF}{Forward Physics Facility}
	\acro{FPGA}{field-programmable gate array}
	\acro{FRB}{fast radio burst}
	\acro{GCN}{Gamma-Ray Coordinates Network}
	\acro{GCOS}{Global Cosmic Ray Observatory\acroextra{, see \autoref{sec:GCOS}}}
	\acro{GDM}{Galactic dark matter}
	\acro{GMF}{Galactic magnetic field}
	\acro{GNN}{graph neural network}
	\acro{GPU}{graphics processing unit}
	\acro{GRAND}{Giant Radio Array for Neutrino Detection\acroextra{, see \autoref{sec:GRAND}}}
	\acro{GRB}{gamma-ray burst}
	\acro{GSF}{Global Spline Fit}
	\acro{GUT}{grand unified theory}
	\acrodefplural{GUT}[GUTs]{grand unified theories}
	\acro{GW}{graviational wave}
	\acro{GZK}{Greisen-Zatsepin-Kuzmin}
	\acro{HAWC}{High-Altitude Water Cherenkov}
	\acro{HEAT}{High Elevation Auger Telescope}
	\acro{HEP}{high-energy physics}
	\acro{HESS}{High Energy Stereoscopic System}
	\acro{HiRes}{High Resolution Fly's Eye}
	\acro{HL-LHC}{high-luminosity LHC}
	\acro{HLGRB}{high-luminosity GRB}
	\acro{HPC}{high-performance computing}
	\acro{IACT}{Imaging air Cherenkov telescope}
	\acro{IR}{infrared}
	\acro{IGM}{intergalactic medium}
	\acrodefplural{IGM}{intergalactic media}
	\acro{IGMF}{intergalactic magnetic field}
	\acro{ISM}{interstellar medium}
	\acro{ISS}{International Space Station}
	\acro{ISUAL}{Imager of Sprites and Upper Atmospheric Lightning}
	\acro{iRODS}{Integrated Rule Oriented Data System}
	\acro{KCDC}{KASCADE Cosmic-ray Data Centre}
	\acro{KG}{KASCADE-Grande}
	\acro{KM3NeT}{Cubic Kilometre Neutrino Telescope}
	\acro{QGSJET}{Quark Gluon String Model with JETs}
	\acro{LAGO}{The Latin American Giant Observatory}
	\acro{LDF}{lateral distribution function}
	\acro{LEO}{low Earth orbit}
	\acro{LHC}{Large Hadron Collider}
	\acro{LIV}{Lorentz-invariance violation}
	\acro{LOFAR}{Low-Frequency Array}
	\acro{LLGRB}{Low-luminosity GRB}
	\acro{LMA}{Lightning Mapping Array}
	\acro{LOPES}{LOFAR prototype station}
	\acro{MAPMT}{Multi-Anode Photomultiplier}
	\acro{MC}{Monte Carlo}
	\acro{MD}{Middle Drum}
	\acro{MDN}{Multi-messenger Diversity Network}
	\acro{MDR}{modified dispersion relation}
	\acro{MHD}{magnetohydrodynamics}
	\acro{mini-EUSO}[Mini-EUSO]{Multiwavelength Imaging New Instrument for the Extreme Universe Space Observatory}
	\acro{MLT}{Magnetic Local Time}
	\acro{MMA}{multi-messenger astrophysics}
	\acro{MPIA}{the Max Planck Institute for Astronomy}
	\acro{NFS}{Network File System}
	\acro{NIAC}{Non-imaging air Cherenkov}
	\acro{NICHE}{non-imaging Cherenkov array}
	\acro{NSF}{National Science Foundation}
	\acro{NWP}{Numerical Weather Prediction}
	\acro{PCC}{POEMMA Cherenkov Camera}
	\acro{PDM}{Photo Detector Module}
	\acro{PFC}{POEMMA Fluorescence Camera}
	\acro{PMT}{photomultiplier tube}
	\acro{POEMMA}{Probe of MultiMessenger Astrophysics\acroextra{, see \autoref{sec:POEMMA}}}
	\acro{PPSC}{Perseus-Pieces Super Cluster}
	\acro{PUEO}{Payload for Ultrahigh Energy Observations}
	\acro{PsA}{Pulsating Aurora}
	\acro{QCD}{quantum chromodynamics}
	\acro{RA}{right ascension}
	\acro{RD}{radio detector}
	\acro{RDG}{relativistic dust grain}
	\acro{RHIC}{Relativistic Heavy-Ion Collider}
	\acro{RM}{rotation measure}
	\acro{RNN}{recurrent neural network}
	\acro{SBG}{starburst galaxy}
	\acrodefplural{SBG}{starburst galaxies}
	\acro{SCIMMA}{Scalable CyberInfrastructure for Multi-Messenger Astrophysics}
	\acro{SD}{surface detector}
	\acro{SGP}{Super-Galactic Plane}
	\acro{SHDM}{super-heavy dark matter}
	\acro{SiPM}{silicon photo-multiplier}
	\acro{SKA}{Square Kilometer Array}
	\acro{SKA-low}{The Square Kilometer Array low-frequency array}
	\acro{SPS}{Super Proton Synchrotron}
	\acro{SSD}{surface scintillator detector}
	\acro{SM}{Standard Model}
	\acro{SNR}{supernova remnant}
	\acro{STEM}{science, technology, engineering and math}
	\acro{TA}{Telescope Array\acroextra{, see \autoref{sec:TA_DesignAndTimeline}}}
	\acro{TALE}{Telescope Array Low Energy}
	\acro{TDE}{tidal disruption event}
	\acro{TGF}{terrestrial gamma-ray flashes}
	\acro{TLE}{transient luminous event}
	\acro{TUS}{Tracking Ultraviolet Setup}
	\acro{TREND}{Tianshan Radio Experiment for Neutrino Detection}
	\acro{UHE}{ultra-high-energy}
	\acrodefplural{UHE}{ultra-high-energies}
	\acro{UHECR}{ultra-high-energy cosmic ray}
	\acro{UMD}{underground muon detector}
	\acro{UrQMD}{Ultra-relativistic Quantum Molecular Dynamics}
	\acro{UV}{ultraviolet}
	\acro{VFHS}{Very Forward Hadron Spectrometer}
	\acro{VHE}{very-high-energy}
	\acrodefplural{VHE}{very-high-energies}
	\acro{VHECR}{very-high-energy cosmic ray}
	\acro{WCD}{water Cherenkov detector}
	\acro{WHISP}{Working group for Hadronic Interactions and Shower Physics}
	\acro{WIMP}{weakly-interacting massive particle}
	\acro{WMAP}{Wilkinson Microwave Anisotropy Probe}
	\acro{WRF}{weather research and forecasting}
\end{acronym}

\vspace{-10mm}
\pagenumbering{arabic}
\fakesection{The exciting future ahead}
{\noindent \LARGE \textbf{Chapter 1}}\\[.8cm]
\textbf{\noindent \huge The exciting future ahead:}\\[3mm]
\textbf{\LARGE Probing the fundamental physics of the nature\\[1.5mm] and origin of UHECRs}
\label{sec:Intro}
\vspace{1cm}

\noindent\Acp{UHECR} ($E>100$\,PeV for the purpose of this white paper) sit in a unique position at the intersection of the Cosmic and Energy Frontiers. They have the potential to simultaneously inform our knowledge of the most extreme processes in the Universe and of particle physics well beyond the energies reachable by terrestrial (i.e., human-made) accelerators.

While there has been very significant progress in astroparticle physics over the past twenty years, the nature and origin(s) of \acp{UHECR}, and in particular, the identity of their sources and acceleration mechanisms, largely remain open questions~\cite{Sarazin:2019fjz, AlvesBatista:2019tlv, Anchordoqui:2018qom}. The complex picture that has emerged from recent advances in the field also poses the question: to what degree will charged-particle astronomy, the ability to study individual (classes of) sources with cosmic rays, be possible? This question has serious consequences for multi-messenger astrophysics because it has implications for the extent to which cosmic rays can be used as a \emph{messenger} and because \acp{UHECR} themselves are fundamental to the production of \ac{UHE} photons and neutrinos and to the interpretation of their measurement~\cite{halzen2003multi, Barwick:1998xq, Kotera:2010yn}. Additionally, \acp{UHECR} represent a unique laboratory to both probe particle physics~\cite{Ulrich:2010rg,Albrecht:2021cxw} and discover physics \ac{BSM}~\cite{PierreAuger:2021tog,Saveliev:2011vw,Diaz:2016dpk,Anchordoqui:2017pmf,Klinkhamer:2017puj,GuedesLang:2017sfl,Torri:2020fao,Duenkel:2021gkq,PierreAuger:2021mve, ThePierreAuger:2022} at the extreme end of the Energy Frontier. However, fully leveraging these capabilities will require accurate measurement and characterization of \ac{UHECR} interaction processes in order to provide a higher-energy complement to traditional accelerator data. This endeavor represents a promising avenue for a strong test of the Standard Model as it requires the extrapolation of existing hadronic interaction models to energies well past the constraints provided by terrestrial accelerators, where there are already hints of tensions with data~\cite{EAS-MSU:2019kmv, Cazon:2020zhx, Soldin:2021wyv}. Hence, through \acp{UHECR}, there is a high potential for discoveries at both the Energy and Cosmic Frontiers.



This white paper has been primarily written to help inform the long term plans of the United States \ac{DOE} and the \ac{NSF} for high-energy physics as part of the ``Snowmass'' process. It is however also an opportunity to outline the international \ac{UHECR} community's road map for addressing the above open questions over the next two decades. In summary, we are approaching a golden age in astroparticle physics and its ability to finally address these questions. The largest \ac{UHECR} observatories are currently undergoing upgrades~\cite{PierreAuger:2016qzd, Kido:2019enj, IceCube:2014gqr} that will provide higher-resolution experimental data for the next decade. These upgrades have been specifically designed to address the new realities of the evolving scientific case that has emerged since the construction of the giant arrays in the early 2000s. Due to these upgrades, the next decade also promises to be rich in further technical advances that will be folded into the design of the next-generation \ac{UHECR} experiments that will be built beyond 2030~\cite{POEMMA:2020ykm, GRAND:2018iaj,Horandel:2021prj}. 
To make this plan a reality, a comprehensive approach needs to be established that extends beyond the field of \acp{UHECR} itself and into other areas of both particle physics and astrophysics. The objectives of this white paper are therefore to outline this strategy and then to provide clear recommendations 
on how to implement it through the upgraded and next-generation instruments.

To set the stage for the road map, it is necessary to understand why, after more than 100 years of study, answers to the central questions of the origin(s) of \acp{UHECR} are still elusive. Though \acp{UHECR} have routinely been detected for decades with energies up to several $10^{20}$\,eV~\cite{Bird:1994uy}, their study is notoriously challenging for several reasons: 

\begin{itemize}
    \item The cosmic-ray spectrum measured at Earth can be described by a series of power laws spanning many orders of magnitude that eventually lead to a vanishingly small flux (less than 1 \ac{UHECR} per square kilometer per century) at the highest energies.
    
    \item Propagation effects change the energy and composition of \acp{UHECR} as they travel. Therefore, the properties of the UHECR beam measured at Earth can not be easily related to its properties at the sources.  
    
    \item The properties of \ac{UHECR} primaries (arrival direction, energy, composition) at Earth can only be inferred from indirect measurements through the \acp{EAS} they induce in the Earth’s atmosphere. Thus, a direct energy calibration is not possible, and an event-by-event determination of cosmic-ray primary composition is complicated by the statistical nature of the UHECR interactions in the upper layers of the atmosphere.
    
   \item The physics needed to describe \ac{EAS} development relies on extrapolations of particle physics processes constrained at much lower energies by terrestrial accelerators.
    
    \item Unlike photons and neutrinos, cosmic rays are charged subatomic particles and are therefore deflected by the \acp{GMF} and the \acp{IGMF}. Hence, their arrival directions, as measured at Earth, may only approximately point back to their actual sources. 

\end{itemize}
Given these measurement challenges, progress in the field has been arduous. Yet, the long lasting heritage of the pioneering arrays of the 20\textsuperscript{th} century lives on through
the critical technical developments and methods that are
now in use at the giant modern experiments, such as the Pierre Auger Observatory (Auger) in Argentina~\cite{PierreAuger:2015eyc}, \acf{TA} in Utah~\cite{TelescopeArray:2008toq}, and the IceCube Neutrino Observatory (IceCube) in Antarctica~\cite{IceCube:2016zyt}.

As discussed in \cref{sec:CurrentStatus}, in the last two decades, a steady stream of fundamental discoveries has come out of the most recent generation of experiments, leading to a transformation of our understanding of \acp{UHECR}, their underlying physics and their potential source class(es). As a result, the entire field has undergone a paradigm shift. Through ever more precise measurements~\cite{Kampert:2012mx}, the old and simplistic picture of \acp{UHECR} as protons at the highest energies has been replaced by a much richer and more nuanced one (see \cref{sec:MassCurrentStatus}). Long-held beliefs about \acp{UHECR} are being called into question. Chief among them is the interpretation of the now firmly established~\cite{PierreAuger:2020qqz, TelescopeArray:2021zox} flux suppression as the telltale sign of the \ac{GZK} process~\cite{Greisen:1966jv, Zatsepin:1966jv} (see \cref{sec:EnergySpectrum}). Despite the tremendous progress of the field in the past two decades, critical questions remain to be answered. While there is conclusive evidence that \acp{UHECR} above 8\,EeV originate from outside our galaxy~\cite{PierreAuger:2017pzq}, there is as yet no consensus on how to interpret the cosmic-ray spectrum as it transitions from galactic to extragalactic origins. This particular point partly motivates the extension of the scope of this white paper down to 100\,PeV.
The quest for the identification of extragalactic sources has so far yielded regional hot spots in the northern~\cite{TelescopeArray:2014tsd} and southern skies~\cite{PierreAuger:2018qvk} with only hints of potential source classes; hence, the nature and origin(s) of \acp{UHECR} largely remains an open question (see \cref{sec:Anisotropy}). Similarly, as outlined in \cref{sec:AccelSyn}, the use of \acp{UHECR} as a probe to particle physics beyond the reach of terrestrial accelerators has made great strides, but also revealed some challenges. In the first decade of operation of Auger, the proton-air and proton-proton cross sections at energies well-beyond the reach of the \ac{LHC} were measured for the first time~\cite{PierreAuger:2012egl, Ulrich:2015yoo}, and most top-down scenarios arising from \ac{BSM} physics were strongly constrained through strict limits on the \ac{UHE} photon flux \cite{Rautenberg:2021vvt}. However, systematic studies have confirmed earlier observations of a muon excess in the data (or a muon deficit in the \ac{EAS} physics models)~\cite{EAS-MSU:2019kmv, Cazon:2020zhx, Soldin:2021wyv, PierreAuger:2014ucz, PierreAuger:2016nfk}, hinting at some processes in the accelerator-based hadronic interaction models that have not been taken into account~\cite{Ulrich:2010rg, Albrecht:2021cxw}. The quality of measurements obtained by current \ac{UHECR} experiments enables narrowing down the potential root causes of the muon problem, thereby informing new investigations to be performed at accelerators.

This revolution of understanding, based on increasingly precise measurements and progress in detection technologies and computational techniques, is ushering in a new and very exciting era of \ac{UHECR} studies. The enormous advances made possible by giant arrays demonstrate that \ac{UHECR} physics has achieved a level of maturity that make it possible to not only probe but discover new fundamental physics in a unique phase space far from the reach of current and future terrestrial accelerators. Addressing the major goals outlined earlier appears to be within reach in the next two decades through a combination of advances in \ac{UHECR} physics, astrophysics, and particle physics. The close synergy between \acp{UHECR} and particle physics outlined earlier is explored in \cref{sec:AccelSyn}, while the astrophysics background as related to the highest energy processes in the universe is discussed in \cref{sec:Astrophysics}. The new \ac{UHECR} paradigm and the evolving science case have prompted the experimental collaborations to consider upgrades of their respective instruments, such as AugerPrime, the upgrade of the Pierre Auger Observatory~\cite{PierreAuger:2016qzd}, \TAxFour, the extension of \ac{TA}~\cite{Kido:2019enj}, and IceCube-Gen2, the extension of the IceCube Neutrino Observatory~\cite{IceCube:2014gqr}. Combined with advances in detectors, refinements in data analysis, and the emergence of new computing methods, the next decade promises an exciting set of new results. This is discussed in \cref{sec:EvolvingScienceCase}.

\begin{figure}[b]
     \centering
     \includegraphics[width=0.8\textwidth]{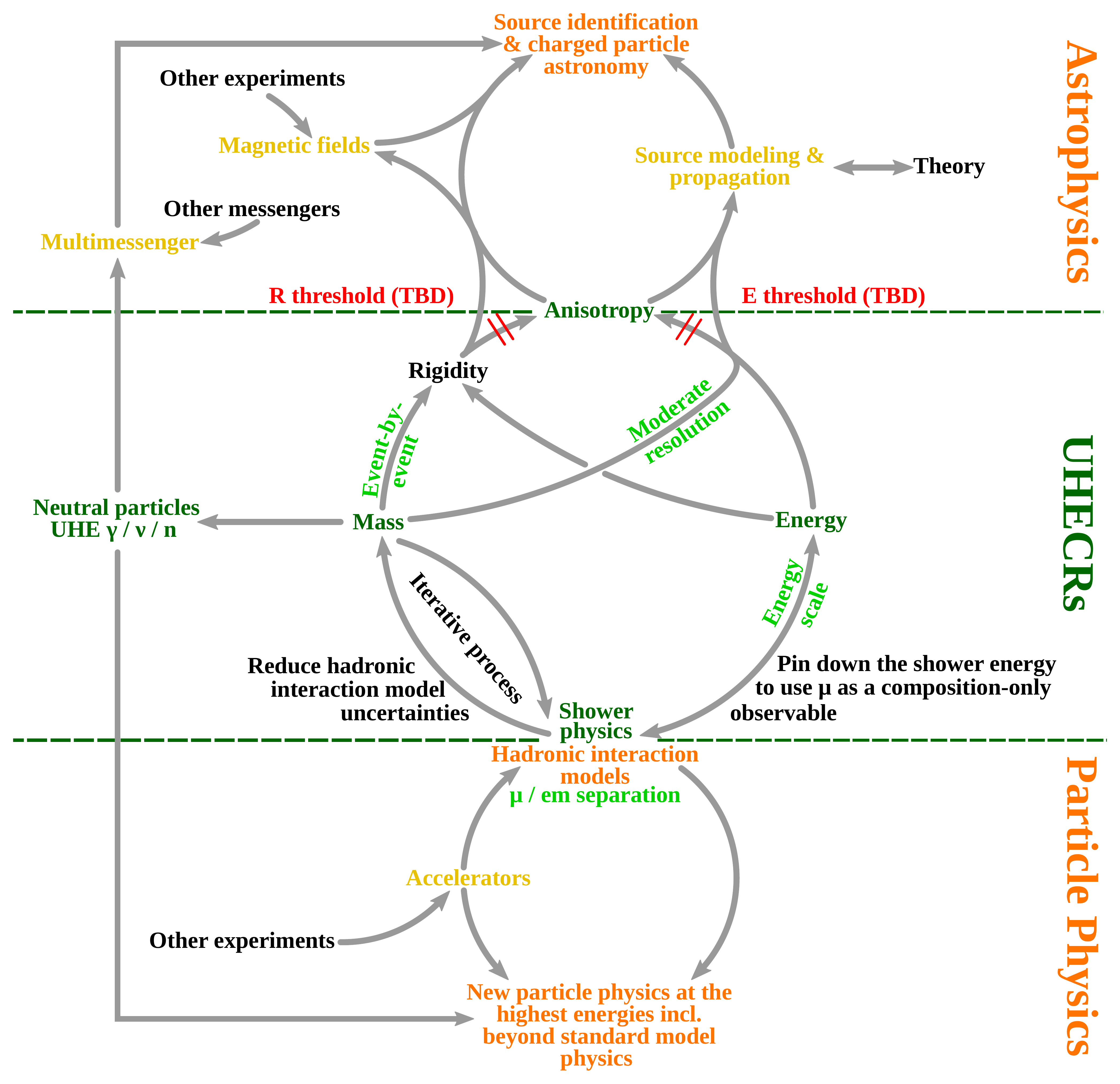}
     \caption{Diagram summarizing the strong connections of \acp{UHECR} with particle physics and astrophysics, and the strategies to attain the fundamental objectives (in orange) in the next two decades (see text for details).}
     \label{fig:the_diagram}
 \end{figure}

The major change in our understanding of \acp{UHECR} comes primarily from the observation that the average mass composition of the primaries becomes heavier with increasing energy. Understanding this evolution is critical to our quest to identify the class(es) of sources responsible for the emission of \acp{UHECR}. As highlighted in \cref{sec:CurrentStatus} and \cref{sec:AccelSyn}, accurately identifying the primary mass groups depends strongly on pinning down the underlying hadronic interaction models used to describe shower physics. Doing so will close the loop between particle physics and astrophysics. In this context, 
the diagram shown in \cref{fig:the_diagram} summarizes how \acp{UHECR} can inform both the Cosmic and Energy Frontiers. A more detailed version of this diagram, including how existing and future experiments complement each other and collectively contribute to the fundamental goals (shown in orange), can be found in \cref{sec:TNG}, and in particular, \cref{fig:roadmap_with_experiments}.

With the primary mass composition playing a pivotal role, there is a need to improve mass resolution, preferably on an \emph{event-by-event} basis. The concept of ``event-by-event'' mass resolution can be understood in two ways:
\begin{enumerate}
    \item Event-by-event composition sensitivity, where there is an available observable for each event which can be statistically related to the primary's mass range, (e.g., heavy/light);
    \item Event-by-event composition reconstruction, where the specific mass group (p, He, C, Si, Fe) of a well-measured primary can be inferred with a confidence interval approaching 50\%.
\end{enumerate}
To date, the term has often been used without differentiation or definition. However, in this work event-by-event mass resolution is defined solely by the second definition as it represents a significant improvement over current capabilities and therefore represents a major goal for the field.
Precise mass determination is currently limited by the systematic uncertainties between hadronic model predictions and the known issues with the modeling of the \ac{EAS} muon component for example~\cite{Albrecht:2021cxw,Kampert:2012mx}. Over the last few years, some hadronic models, such as \eposlhc~\cite{Pierog:2013ria} or the latest version of \sibyll{}~\cite{Riehn:2019jet}, have integrated new accelerator data, especially from the \ac{LHC}, but more heavy ion data need to be collected. There appears to be a path to partially address the muon problem in the next decade using hybrid data from AugerPrime and IceCube-Gen2. In both cases, the principle relies on using multiple, independent detectors to simultaneously measure the \ac{EAS} energy (whose estimators are dominated by the electromagnetic component of the shower) on the one hand, and the muon content on the other. However, it is anticipated that at least one of the next-generation ground arrays will need to tackle this issue by achieving higher energy resolution and better separation of the electromagnetic and muonic parts of the shower. In the lower sector of the diagram, pinning down the parameters of the hadronic interaction models through a comprehensive strategy that includes new accelerator measurements will surely yield new results, which will directly inform new particle physics at the highest energies, including possible hints of new \ac{BSM} physics. 

In the upper sector of the diagram, the traditional approach to anisotropy studies has been to perform model-dependent and model-independent scans as a function of energy to find significant excesses in the arrival directions of \acp{UHECR}. More recent approaches have included limited mass composition information afforded by statistical considerations. This approach will benefit from a better determination of the mass groups resulting from improved hadronic interaction modeling. Space instruments with enormous apertures and relying on the precise determination of \xmax are bound to directly benefit from these advances. A more sophisticated approach combines precise energy and mass composition measurements to estimate the \ac{UHECR} rigidity on an event-by-event basis. Scans in rigidity will be more powerful to reveal anisotropy signals as they naturally relate to the predicted deflections in galactic and extragalactic magnetic fields. Based on our current knowledge, only a future large ground array will be able to explore this avenue beyond what will be achievable by AugerPrime and IceCube-Gen2 (at lower energies). Ultimately, determining the \ac{UHECR} sources and their characteristics will also necessitate inputs from astrophysicists in the areas of source modeling and \ac{UHECR} propagation. Whether charged-particle astronomy will ever be possible may depend on progress in magnetic field modeling, in particular. A wide variety of experiments is expected to contribute to this.

Finally, on the left side of the diagram, \ac{UHE} neutral particles, especially photons and neutrinos, are highlighted as critically important to the field. \ac{UHECR} observatories are naturally sensitive to \ac{UHE} photons and neutrinos. As mentioned earlier, limits on \ac{UHE} photons have already strongly constrained most top-down models for the origin(s) of \acp{UHECR}. In principle, the observation of a single \ac{UHE} (cosmogenic) neutrino or photon would be a game-changer in our understanding of the flux suppression, as well as indicate the existence of a proton component at the highest energies. As such, they have the potential to contribute both to astrophysics and particle physics.


Stepping up to these scientific challenges will require a new generation of air-shower experiments beyond the  upgraded existing instruments. 
These experiments are enabled by recent and future progress in detector and computational technologies, such as the rise of digital radio detection of air showers or the application of machine-learning techniques for data analysis. 
The various open questions of the particle and astrophysics of \acp{UHECR} call for experiments capable of achieving higher accuracy in measuring the properties of the primary particle, as well as huge exposures at the highest energies. 
The highest exposures will be provided by observations from space with the \acf{POEMMA}~\cite{POEMMA:2020ykm} and from the ground with the cosmic-ray measurements of the \acf{GRAND}~\cite{GRAND:2018iaj}. Such instruments are perhaps the only ones capable of looking for ZeV particles and a recovery in the flux beyond the suppression. 
The \acf{GCOS}~\cite{Horandel:2021prj} on the other hand will combine an order of magnitude higher exposure than current ground arrays with the high measurement accuracy provided by combining several detection techniques.
These technology developments and next-generation experiments, as well as their expected contributions to solving the big science questions of the field are described in \cref{sec:TNG}.

The opportunities for broader impacts and advances in interdisciplinary sciences while studying \acp{UHECR} are discussed in \cref{sec:SuppSci}. Applications range broadly from astrobiology to earth sciences. In particular, all \ac{UHECR} instruments use the atmosphere as detector material. As a result, the atmospheric conditions above or below the instruments need to be well characterized. This naturally provides opportunities for advances in atmospheric sciences, especially in the area of transient luminous events that occur during thunderstorms, due to the sensitivity and timing of the fluorescence detectors used by current experiments such as Auger and \ac{TA} at ground level, and Mini-EUSO (part of the \ac{EUSO}
program) \cite{Bacholle:2020emk} on board the \ac{ISS}. The need to observe large volumes of atmosphere with sensitive detectors also opens the opportunity to detect other transient events produced in the atmosphere by anything from macroscopic dark matter and nuclearites to relativistic dust grains to space debris.

Finally, the continuation of highly-collaborative research activities and future construction and operation of even larger observatories call for a fully integrated effort, requiring the examination of the societal and environmental impacts of carrying out such projects. This is discussed in \cref{sec:NewChallenges}. First of all, the scientific community needs to become a model for diversity, equity, inclusion, and accessibility, in which underrepresented groups not only feel welcomed and supported, but are actively provided with opportunities to succeed. 
While there have been some positive trends developing over the past decade or so, physics in particular largely remains a white male dominated field at every level, from (under)graduate students to senior faculty and researchers.
Big science has always been at the forefront of open science for reasons ranging from scientific considerations, such as having the data available on a global scale to facilitate data analysis and archiving at multiple locations, to more practical ones, such as fulfilling pledges to release data in exchange for public funding. With only rich countries able to afford contributions to big science, open access to the data helps close the wealth gap between scientists around the world.
Finally, the scientific community needs to lead the way in assessing and minimizing its own environmental impact. This not only applies to the operation of the experiments themselves, but also to the environmental cost of developing and building such experiments, using ever increasing computing resources, and attending meetings and conferences all over the world.

\fakesection{UHECR physics comes of age}
\vspace{3cm}
{\noindent \LARGE \textbf{Chapter 2}}\\[.8cm]
\textbf{\noindent \huge UHECR physics comes of age:}\\[3mm]
\textbf{\LARGE Two decades of fundamental discoveries}
\label{sec:CurrentStatus}
\vspace{1cm}

\noindent Our current understanding of \ac{UHECR} physics has been built upon almost a century of observations of air showers. The steeply-falling flux in this energy region has required the construction of increasingly larger expansive arrays of detectors. 
The results of this effort have allowed us to refine our interpretation of the highest energy particles which arrive at Earth, probe sources and  related processes which impart up to tens of joules in energy per particle, and make measurements of particle physics at beyond-\ac{LHC} energy scales.

To start this chapter the design of three \ac{UHECR} experiments is highlighted: the Pierre Auger Observatory (\cref{sec:Auger_DesignAndTimeline}), the Telescope Array experiment (\cref{sec:TA_DesignAndTimeline}), and the IceCube Neutrino Observatory (\cref{sec:IceCube_DesignAndTimeline}), chosen for their impact to our understanding of \ac{UHECR} science. Additionally, their impending upgrades during the upcoming decade are briefly described (also see \cref{sec:FutureDetectors} for more extensive information). Results from this current generation of experiments, which have dispelled the pre-existing simple \ac{UHECR} picture, are then reviewed.
These findings, which have informed this new interpretation of the nature of \acp{UHECR}, are described in several sections, the energy spectrum in \cref{sec:EnergySpectrum}, primary mass composition in \cref{sec:MassCurrentStatus}, arrival directions in \cref{sec:Anisotropy}, and other neutral messengers that are studied using air shower arrays in \cref{sec:NeutralParticles}. 
From these results, a new paradigm is emerging which still needs to be clarified and understood.
Therefore, while this section primarily describes the measurements, their particle physics implications are covered in \cref{sec:UHECRPartSyn} possible astrophysical interpretations of these measurements can be found in \cref{sec:Astrophysics}.
Additionally, the outlook for the future of the field over the next decade(s) can be found in \cref{sec:EvolvingScienceCase,sec:TNG}.

\subsection[Entering the 21st century]{Entering the 21\textsuperscript{st} century: Go big or go home}\label{sec:GoBigOrHome}

\subsubsection{The Pierre Auger Observatory}
\label{sec:Auger_DesignAndTimeline}
The Pierre Auger Observatory~\cite{PierreAuger:2015eyc} is currently the largest cosmic-ray observatory in the world.  It is located on a semi-arid plateau in the province of Mendoza, western Argentina ($35.2^\circ$~S\@, $69.2^\circ$~W\@, 1400\,m~a.s.l.)\@.
Its main array for detecting the highest-energy cosmic rays consists of 1,600 water-Cherenkov \ac{SD} stations on a 1500\,m-spacing triangular grid (hereafter ``SD-1500'') covering an area of $3000\,\km^2$, plus four \ac{FD} buildings at the periphery each containing six telescopes overlooking the atmosphere above the array.
Each \ac{SD} station consists of a cylindrical plastic tank with $10\,\mathrm{m}^2$~base area and 1.2\,m~height, filled with 12\,000 liters of ultra-pure water, and surmounted by three $9''$-diameter \acp{PMT} detecting the Cherenkov light emitted by relativistic charged particles in air showers when they pass through the water.  Each \ac{FD} telescope consists of a $13\,\mathrm{m}^2$-area curved mirror focusing the fluorescence light emitted in air showers onto a camera composed of 440 hexagonal \acp{PMT}, and has a $30^\circ\times30^\circ$ \ac{FoV} with a minimum elevation of $1.5^\circ$~above the horizon. In order to extend the sensitivity to lower-energy showers, in a $23.5\,\km^2$~region of the array, 61 \ac{SD} stations have been deployed with a 750\,m~spacing (``SD-750'')~\cite{PierreAuger:2021hun} and 19 stations with a 433\,m~spacing (``SD-433'')~\cite{PierreAuger:2021tmd}, overlooked by three extra \ac{FD}~telescopes looking at elevations of~$30^\circ$ to~$58^\circ$ above the horizon (\ac{HEAT}).  The Observatory also contains various other facilities for calibration, atmosphere monitoring, R\&D, and interdisciplinary purposes, such as the \ac{AERA}.

\begin{wrapfigure}{r}{0.5\columnwidth}
    \vspace{-5mm}
    \includegraphics[width=0.5\columnwidth]{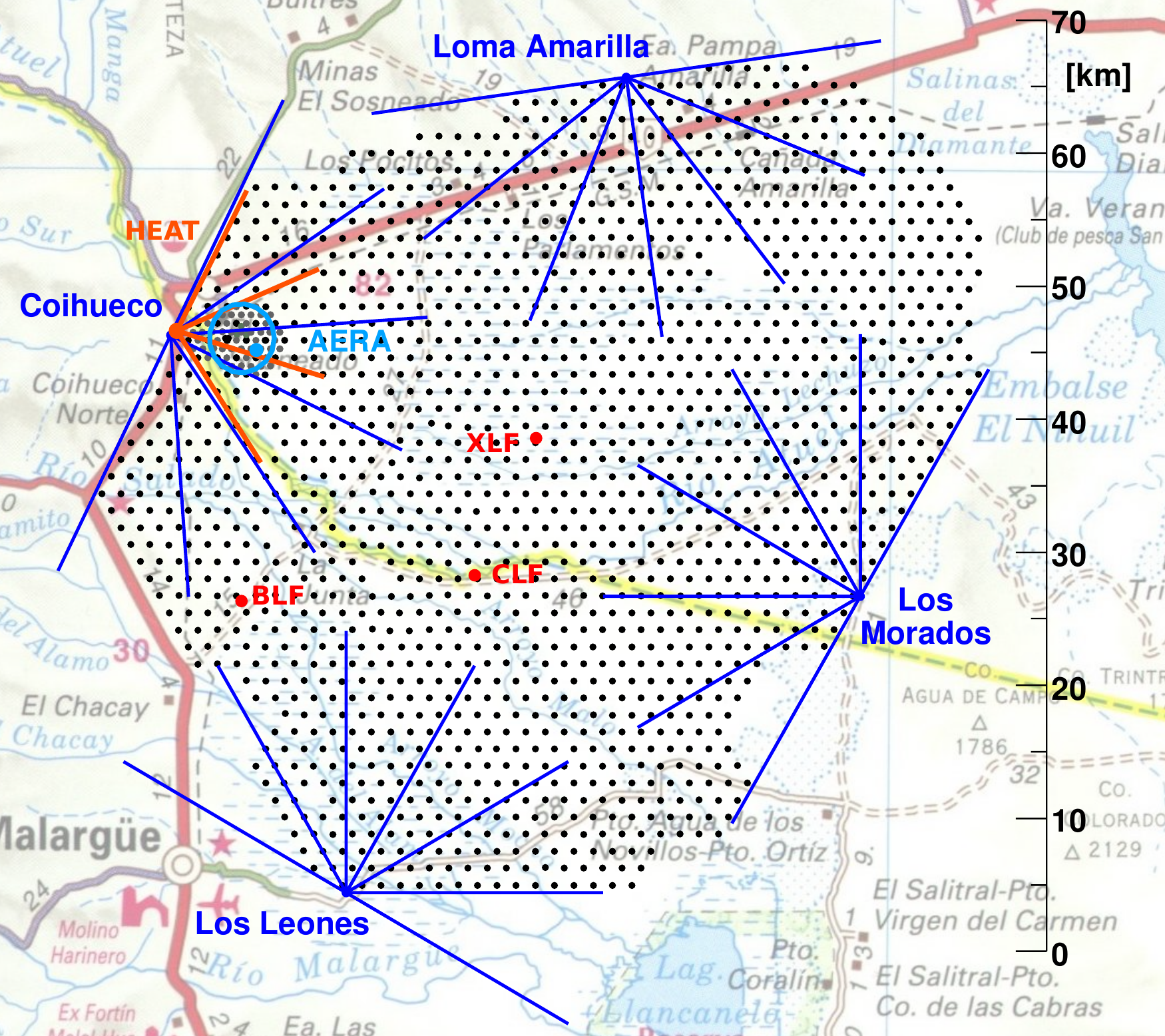}
    \vspace{-4mm}
    \caption{Map of of the Pierre Auger Observatory and its various components. Black dots: the detector stations of the \ac{SD}. Blue lines: the \ac{FoV} of each of the 24 fluorescence telescopes in the \ac{FD}. Red lines: the \ac{FoV} of the 3 fluorescence that make up the low energy extension to the \ac{FD}, \acs{HEAT}. The extent of the \acs{AERA} radio array and the locations of various atmospheric monitoring stations are also shown (taken from~\cite{augerWebsite}).}
    \label{fig:PAO-layout}
    \vspace{-4mm}
\end{wrapfigure}

The deployment of the array lasted from 2002 to 2008, and data taking started in January 2004.
Applying the broadest selection cuts (used for arrival direction studies at energies above~$32\,\EeV$), the exposure of the Observatory exceeded~$120\,000\,\mathrm{km^2\,yr\,sr}$ in 2020~\cite{PierreAuger:2021rfz}, which no other experiment is expected to achieve until at least the late 2020s (see \cref{sec:nextgen_spectrum}). 

The Observatory is also currently undergoing an upgrade named AugerPrime (see \cref{sec:AugerPrime}), which aims to significantly increase its sensitivity to the characteristics of an \ac{EAS}. The main components of the upgrade consist of the addition of \acp{SSD} and \acp{RD} to each of majority of the surface detector array. This will allow for multi-hybrid observations resulting in a high resolution separation of the electromagnetic and muonic components of measured air showers. This in turn will provide the full duty cycle \ac{SD} with enhanced composition sensitivity and provide better constraints to be made for shower physics studies.

\paragraph{Scientific Capabilities}
\label{sec:ScientificCapabilities}

\subparagraph*{Studies at the highest energies}

The main goal of the Observatory is the detection of cosmic rays at the highest energies. The SD-1500 array has a detection efficiency of approximately $100\%$ for vertical showers (zenith angles~$\theta < 60^\circ$) with energies~$E \ge 10^{18.4}\,\eV$ and inclined showers ($60^\circ \le \theta < 80^\circ$) with~$E \ge 10^{18.6}\,\eV$.  Counting only the vertical events passing the most stringent quality cuts, it has registered 215\,030 events allowing us to reconstruct the \ac{UHECR} energy spectrum with unprecedented precision~\cite{PierreAuger:2020qqz}, confirming the previously observed \emph{ankle} and \emph{cutoff} features at approximately~$5\,\EeV$ and~$50\,\EeV$ respectively, and finding a new \emph{instep} feature at~$(13 \pm 1_\mathrm{stat} \pm 2_\mathrm{syst})\,\EeV$.  The energy resolution of these events decreases from around~$20\%$ at~$2\,\EeV$ to~$7\%$ above~$20\,\EeV$, and the systematic uncertainty is~$14\%$, dominated by the uncertainty in the \ac{FD} calibration.

Using relaxed selection criteria, the angular distribution of \ac{UHECR} arrival directions has been studied with unprecedented statistics at the Pierre Auger Observatory.  A modulation in the right ascension distribution of events with~$E \ge 8\,\EeV$ first discovered in 2017~\cite{PierreAuger:2017pzq} has now reached a statistical significance of~$6.6\sigma$~\cite{PierreAuger:2021dqp}.  It can be interpreted as a dipole moment of amplitude~$d = (5.0\pm0.7)\times(E/10\,\EeV)^{0.98\pm0.15}\%$ towards celestial coordinates~$(\alpha_d, \delta_d)=(95^\circ \pm 8^\circ, -36^\circ \pm 9^\circ)$, with no statistically significant evidence for a quadrupole moment.  The strength of the dipole is much weaker than expectations assuming Galactic sources, and its direction is about $115^\circ$~away from the Galactic Center, suggesting an extragalactic origin for these particles.  At higher energies and smaller angular scales, there have been several indications of excesses towards certain regions of the sky or classes of objects~\cite{PierreAuger:2021rfz}, none of which reaching the discovery level so far.  The most significant is a correlation between events with~$E \ge 38\,\EeV$ and nearby starburst galaxies, with a best-fit equivalent top-hat radius of~$\Psi=\left(25^{+11}_{-7}\right)^\circ$ and signal fraction~$\alpha=\left(9^{+6}_{-4}\right)\%$, with a $4.0\sigma$~post-trial significance.  This signal strengthens to $4.2\sigma$ post-trial significance when Auger and \ac{TA} data are combined and analyzed together~\cite{TelescopeArray:2021gxg}.  In the future, continued data taking may strengthen this finding to the discovery level: assuming the excess continues growing linearly with time, the Auger-only significance is expected to reach $5\sigma$ by the end of~$2026 \pm 2$~years.

As for \ac{UHECR} mass composition, it is currently mainly estimated is via \xmax, as measured by \ac{FD} telescopes~\cite{Yushkov:2020nhr}.  This method is affected by major systematic uncertainties and model dependence, as it relies on simulations of the hadronic interactions in air showers in kinematic regimes where they are poorly known, but it shows that the composition is lightest around~$2\,\EeV$ (where the geometric mean mass is most likely between hydrogen and helium) and gradually becomes heavier at lower and higher energies (being most likely between helium and carbon at~$10^{17.2}\,\eV$ and between carbon and calcium at~$10^{19.7}\,\eV$, the precise values depending on the hadronic interaction model assumed), and that it gradually becomes less mixed with increasing energies. The \xmax{} resolution of the \ac{FD} decreases from around~25\,\gcm{} at~$10^{17.8}\,\eV$ to~15\,\gcm{} above~$10^{19}\,\eV$ and the systematic uncertainties range from around~$7$ to~10\,\gcm{}, whereas the predictions of various hadronic models differ by up to~26\,\gcm{}; for comparison, all other things being equal a 17\,\gcm{}~difference in the average~\xmax{} approximately corresponds to a factor of~2 in the mass number.  Simultaneously using \ac{FD} and SD observables allows us to estimate certain features of the mass composition in a much more model-independent way, for example that near the ``ankle'' energy it is a mix of both light (H, He) and heavier nuclei, with any pure element excluded at, $6\sigma$ and any H+He-only mixture at $>$\,$5\sigma$ with any of the hadronic models considered~\cite{Yushkov:2020nhr}.  The composition also appears to be heavier at low than at high Galactic latitudes~\cite{PierreAuger:2021jlg}.

In principle, another way to estimate the mass composition is from the muon content of showers, but it has been seen that all currently available hadronic models are inadequate for the task as they all predict many fewer muons in average for any realistic composition than actually observed by any experiment~\cite{Soldin:2021wyv}.  Conversely, the size of shower-to-shower fluctuations in the muon number as measured by the Observatory does agree with model predictions, indicating that the mismatch in the average cannot be due only to a major mis-modeling of extreme-energy interactions at the top of the shower, but must be due to a small effect compounding throughout the shower development, including in lower-energy interactions close to the ground~\cite{PierreAuger:2021qsd}.
The \xmax{} and muon content of showers can also be estimated from \ac{SD} data using machine learning techniques~\cite{PierreAuger:2021fkf,PierreAuger:2021nsq}, and the new AugerPrime detectors are going to further reduce statistical and systematic uncertainties on the \ac{UHECR} mass composition, shed more light on hadronic interactions at extreme energies, and allow us to compile proton-enhanced samples of events for anisotropy studies.

The Observatory is also sensitive to $\EeV$-energy gamma rays and neutrinos, making it suitable for multi-messenger observations and searches for new physics~\cite{PierreAuger:2019fdm}.  The limits on the diffuse neutrino fluxes~\cite{Pedreira:2021gcl} are competitive with IceCube ones above~$1\,\EeV$ and those on gamma-ray fluxes~\cite{Rautenberg:2021vvt} are the most stringent available above a few hundred PeV\@; such limits have been used to set constraints to properties of \ac{UHECR} sources~\cite{PierreAuger:2019ens}. Limits on neutrino~\cite{PierreAuger:2021asv} and gamma-ray~\cite{PierreAuger:2021oks} emission by black-hole mergers have also been set, as well as on \ac{UHE} neutrinos from the blazar TXS\,0506+056~\cite{PierreAuger:2020llu} and from the neutron star merger GW170817~\cite{LIGOScientific:2017ync,ANTARES:2017bia} (which by fortunate coincidence occurred around~$2^\circ$ below the horizon at the Auger site, close to the maximum of the neutrino sensitivity).
Machine-learning techniques and the new AugerPrime detectors are going to improve the discrimination between photon candidates and the hadronic background, improving the limits on EeV gamma-ray fluxes.

\subparagraph*{The low-energy extension}
The low-energy extensions of the Observatory allows studies to be extended into the energy range where Galactic cosmic rays are expected to dominate.
The SD-750 has a detection efficiency of approximately $100\%$ for events with~$\theta < 40^\circ$ and~$E \ge 10^{17}\,\eV$, and has been used to measure the energy spectrum of cosmic rays down to the so-called \emph{second-knee}~\cite{PierreAuger:2021hun}. The SD-433 will extend the full efficiency further to~$10^{16.6}\,\eV$~\cite{PierreAuger:2021tmd}, while preliminary studies using the \ac{HEAT} \ac{FD} to detect the air Cherenkov emissions from showers reach down to~$10^{15.8}\,\eV$, below the so-called \emph{low-energy ankle}~\cite{PierreAuger:2021ibw}.  
As for arrival directions, the SD-750 has been used to extend the measurements of the \ac{RA} modulation down to~$1/32^{\rm nd}\,\EeV$~\cite{PierreAuger:2020fbi}. Though not yet statistically significant below 8\,EeV, the dipole direction is consistent with the direction of Galactic Center from ${1/16}^{\rm th}\,\EeV$ to $2\,\EeV$, after which it gradually approaches that of the $E \ge 8\,\EeV$~dipole.

The Observatory is also sensitive to a variety of atmospheric, solar, and geophysical phenomena, such as elves~\cite{PierreAuger:2020lri} with the \ac{FD}, and terrestrial gamma-ray flashes~\cite{PierreAuger:2021int}, Forbursh decreases, and even earthquakes with the \ac{SD}~\cite{Wiencke:2012dj}.

\subsubsection{The Telescope Array Project}
\label{sec:TA_DesignAndTimeline}
\begin{wrapfigure}{r}{0.5\columnwidth}
    \vspace{-13mm}
    \includegraphics[width=0.5\columnwidth]{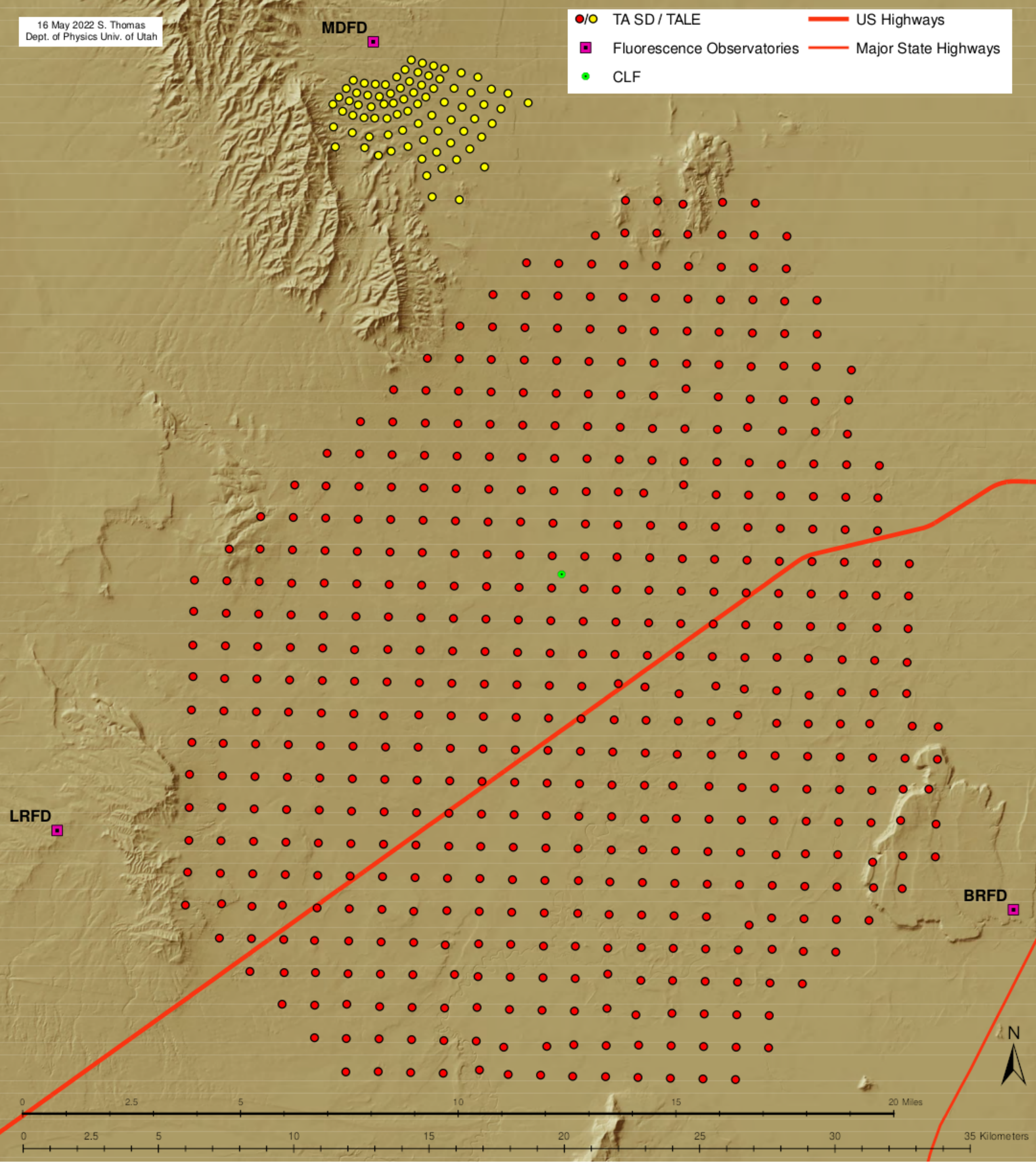}
    \vspace{-6mm}
    \caption{Map of \ac{TA}.}
    \label{fig:ta-layout}
\end{wrapfigure}

The \acf{TA} (\cref{fig:ta-layout}) is located 274\,km south of Salt Lake City in central Utah, USA. It is the largest cosmic ray detector in the northern hemisphere.  It measures the properties of cosmic rays over more than five orders of magnitude in energy with a series of overlapping detector components.  

The original \ac{TA} construction consists of 507 scintillator detectors (which comprise the \ac{SD}) deployed on a 1.2\,km square grid deployed over approximately 700\,km$^{\rm 2}$. The array samples the charge particle density of cosmic ray induced extensive air showers when they reach the Earth's surface.  The active portion of each detector consists of two layers of 1.2-cm-thick scintillator, each 3\,m$^{\rm 2}$ in area. Wavelength shifting optical fibers are installed into grooves in the extruded scintillators. The fibers gather the signal light generated when the shower particles pass through the each scintillator layer and guide the light to the \acp{PMT} for that layer.  Three telescope stations (which comprise the \ac{FD}), at the vertices of a $\sim$\,30\,km triangle, are instrumented with 38 telescopes and view the skies 3-31$^{\circ}$ in elevation above the array of scintillator detectors.  The telescope's segmented mirrors focus the light generated when the extensive air shower passes through the atmosphere onto cameras which are composed of a 16$\times$16 array of hexagonal \acp{PMT} each viewing about 1$^{\circ}$ of sky.

Showers from lower energy events reach maximal development higher in the atmosphere and have smaller footprints at the Earth. The \ac{TALE} extension added ten additional telescopes at the \ac{MD} station viewing 31-59$^{\circ}$ in elevation above the main telescopes to study these events and the transition from galactic to extra-Galactic sources.
By utilizing the shower's Cerenkov light in addition to its fluorescence light, events are well reconstructed down to $\sim$\,$10^{\rm 15.3}$\,eV.
In addition, new scintillator detectors were deployed in a graded; 400\,m, 600\,m, and 1200\,m, spacing near the station. 

To better understand the excess in events seen just off the \ac{SGP} in the vicinity of Ursa Major reported in 2014~\cite{TelescopeArray:2014tsd} (see below), the Telescope Array collaboration set about to expand the area of the \ac{SD} by a factor of 4 to $\sim$\,3000\,km$^{2}$ by adding 500 new scintillator detectors with a spacing of 2.08\,km. 
In this upgrade, called \TAxFour, the spacing was optimized to maximize aperture for detecting showers with $E>57$\,EeV (provides a better than 95\,\% reconstruction efficiency at these energies), while reducing the overall costs~\cite{Kido:2019enj}. 
The first 257 of the new \TAxFour \acp{SD} have been deployed in sites to maximize the hybrid aperture (see \cref{sec:TAx4}). The remaining counters have been delayed due to COVID-19.  
Plans are presently being explored on how best to quickly complete the array.  
Twelve new telescopes have already been added viewing 3-17$^{\circ}$ above the \TAxFour expansion detectors both to calibrate the scintillator array, with its new spacing, as well as to measure composition via hybrid measurement of events at the highest energies.

\paragraph{Scientific Capabilities}
\label{sec:TA_ScientificCapabilities}

\vspace{-0mm}
\begin{figure}
    \begin{minipage}[t]{0.65\columnwidth}
        \centering
        \includegraphics[width=\columnwidth]{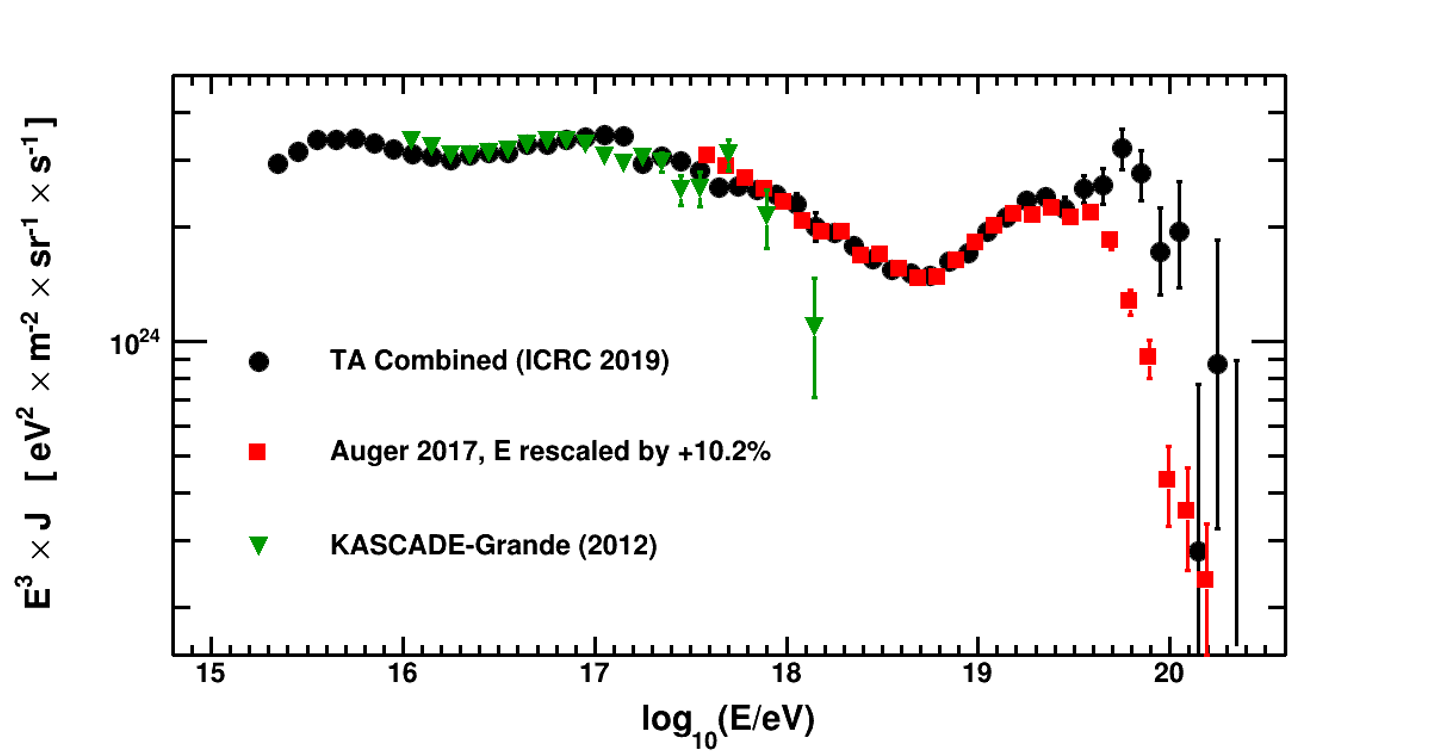}
    \end{minipage}%
    \hfill
    \begin{minipage}[t]{0.34\columnwidth}
        \centering
        \includegraphics[width=\columnwidth]{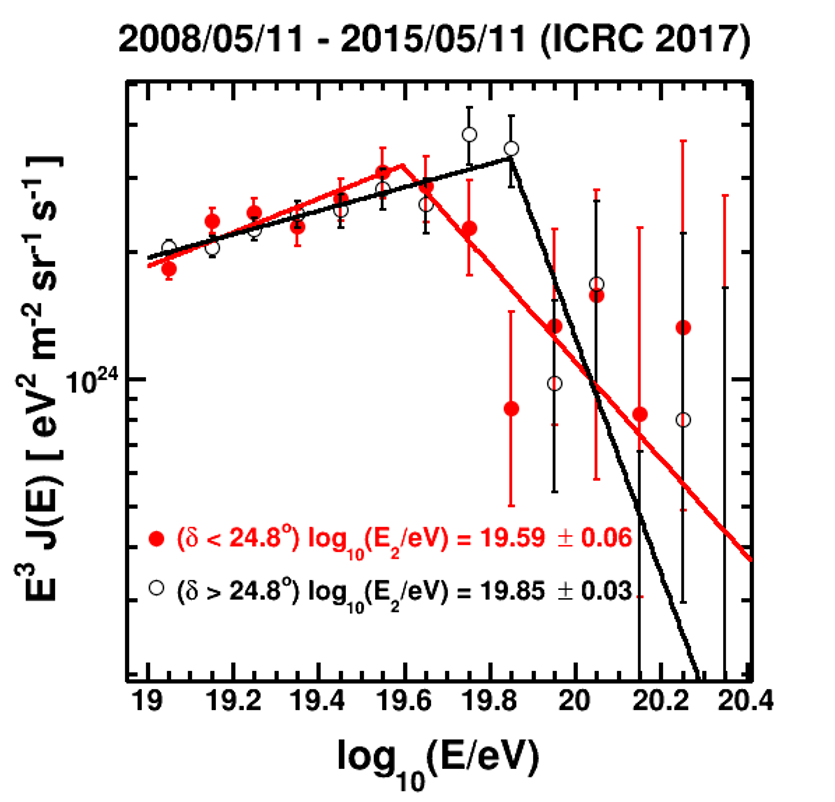}
        \label{fig:spectrum-declination}
    \end{minipage}
     \vspace{-8mm}
     \caption{\small {\bf Left:} {The 11-year \ac{TA} \ac{SD} flux spectrum and the 2-year \ac{TALE} monocular \ac{FD} spectrum (black points) compared with that of the Auger (red points, energy rescaled in this plot by +10.2\%) and KASCADE-Grande (green points). Five features are clearly seen: (1) Suppression above about $10^{19.8}$\,eV, (2) a newly observed break at $10^{19.2}$ \,eV, tentatively called the \emph{instep}, (3) the ankle at $10^{18.7}$\,eV, (4) the \emph{second-knee} near $10^{17.1}$\,eV, and (5) another ankle-like break at about $10^{16.2}$\,eV. The rescaled Auger spectrum begins to diverge from that of \ac{TA} above $10^{19.3}$ eV.} {\bf Right:} {\ac{TA} \ac{SD} spectrum fits in two declination bands. There is a $3.9\sigma$ difference in the break: $\lg(E/{\rm eV}) = 19.59(6)$ vs. $\lg(E/{\rm eV}) =19.85(3)$
        }}
    \label{fig:sd-spectrum}
    \vspace{-5mm}
\end{figure}

The Telescope Array measures the cosmic ray spectrum from $\sim$\,10$^{\rm 15.3}$\,eV to the highest energies and observes multiple structures in the cosmic ray spectrum from the knee and what looks like a Peter's cycle thru the \ac{GZK} suppression. 
The Telescope Array spectrum is shown in \cref{fig:sd-spectrum}, overlaid with the spectra measured by Auger and KASCADE-Grande. The
cutoff appears in the Telescope Array data with $\sim$\,6$\sigma$ significance. The Telescope Array \ac{SD} spectrum is in good agreement with that of Auger, the latter with a $+10$\% adjustment in energy scale (within the combined systematic uncertainties of both measurements). However, above $10^{19.3}$\,eV, the two diverge significantly; the high-energy cutoff appears at a lower energy in Auger than is observed with Telescope Array~\cite{PACollab-2017-ICRC-35}. 

The Telescope Array collaboration investigated the high energy region where the spectra diverge (see \cref{fig:sd-spectrum}). In the high declination band, 24.8$^{\circ}$--90$^{\circ}$, the cut-off occurs at a higher energy.  In the lower declination band, $-16^{\circ}$--24.8$^{\circ}$, where the sky is viewed in common by both experiments, the cut-off occurred significantly lower in energy. The significance of the difference was  $\sim$\,4$\sigma$. Recently, a flattening in the cosmic ray spectrum was observed in the Auger data between 1.3 and 4.6$\times$10$^{19}$\,eV. The same flattening can also be observed with more than $5\sigma$ significance if one combines the data of the Telescope Array with that of \ac{HiRes}. 

At the lowest energy ranges, monocular \ac{FD} data has been collected using the \ac{TALE} telescopes at the \ac{MD} \ac{FD} site since 2014 \cite{TelescopeArray:2018bya}. 
From this data, two additional features are clearly seen: a \emph{second-knee} like softening of the spectrum at $\sim$\,$10^{17.1}$\,eV, and a second \emph{ankle} like hardening of the spectrum at $\sim$\,$10^{16.2}$\,eV. At the lowest reach of \ac{TALE} appears to be the cosmic ray \emph{knee} at about $10^{15.7}$\,eV. The ratio of energies between the two \emph{knees} is $10^{17.1-15.7}\simeq25$, tantalizingly close to the charge ratio of 26 between iron nuclei and protons.


The \ac{TA} data is consistent with a light, unchanging composition from $10^{18.2}$\,eV up to $10^{19.1}$\,eV, within statistical uncertainties. Within systematics the results are also in agreement between the telescope stations~\cite{Abbasi:2014sfa}. The interpretation of the absolute \meanXmax values is limited by varying predictions for different high-energy interaction packages, and it is not possible to distinguish whether \ac{TA} \meanXmax data represent protons or helium from these results. On the other hand, the width of \xmax{} distributions are far less model dependent. Because reliable measurement of widths requires about $5\times$ more data than reliable measurements of averages, the energy range was restricted to $10^{18.2}$--$10^{19.1}$\,eV~\cite{TelescopeArray:2018xyi}. 
More data is needed to extend the \sigmaXmax measurement to the \ac{GZK} cutoff. The \TAxFour expansion will provide extra hybrid aperture for this effort.

\begin{figure}
    \begin{minipage}{0.3\columnwidth}
    \begin{center}
    \includegraphics[width=\columnwidth]{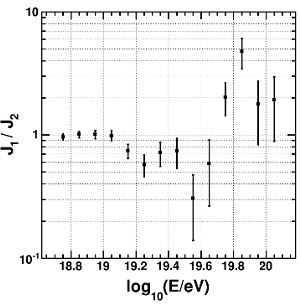}
    \end{center}
    \end{minipage}%
    \hfill
    \begin{minipage}{0.65\columnwidth}
    \begin{center}
    \includegraphics[width=\columnwidth]{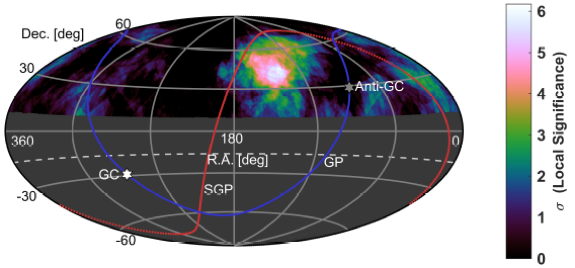}
    \end{center}
    \end{minipage}%
    \caption{\small {\bf Left:} {The ratio of the \ac{TA} \ac{SD} flux (7 years) inside the \ac{TA} hotspot circle divided by that outside, plotted against energy.} {\bf Right:} {The local pre-trial energy spectrum anisotropy one-sided significance, for each spherical cap bin of (average) radius $30^\circ$ and $\log_{10}(E/{\rm eV})>19.2$. The maximum significance is $6.17\sigma$ at $139^\circ$ \ac{RA}, $45^\circ$ \ac{DEC} \cite{Lundquist:2017fjo}. This is $7^\circ$ from the  published \ac{TA} hotspot location~\cite{TelescopeArray:2014tsd}.}}
    \label{fig:spectrum-anisotropy}
\end{figure}
In 2014, the \ac{TA} Collaboration reported an indication of an excess in the arrival directions of \acp{UHECR} just off the \ac{SGP} in the vicinity of Ursa Major \cite{TelescopeArray:2014tsd}. A total of 19 of the 72 \ac{TA} events above $5.7\times10^{19}$\,eV were found within a $20^\circ$-radius circle, corresponding to a $5.1\sigma$ excess. The chance probability of seeing the \ac{TA} \emph{hotspot} is $3.7\times10^{-4}$, or $\sim$\,$3.4\sigma$. With six additional years of data, another 19 events events have been observed within a $25^\circ$
radius \cite{Kawata-2019-ICRC-36-310}. 
The overall signal significance has dropped slightly to $2.9\sigma$. No corresponding excess is seen when the event selection threshold is lowered to $4.0\times10^{19}$\,eV or $1.0\times10^{19}$\,eV. The cut-off energy of $5.7\times10^{19}$\,eV is very close to the \ac{GZK} threshold for photo-pion ($\Delta^+$ resonance) production from cosmic protons propagating though the \ac{CMB}. Hence most of the events likely originated from within 50\,Mpc of the Earth. The magnetic deflection of protons over this distance in the \ac{IGMF} and \ac{GMF} should be limited to at most $\sim$\,$5^\circ$, so that arrival directions retain some memory of their origin. Events below the \ac{GZK} threshold come from much further, and their arrival directions would be smeared out. The \ac{TA} hotspot may represent a local source of \acp{UHECR}. A confirmation of this discovery with additional data would represent a transformative advance in \ac{UHECR} physics.

The spectra reported by \ac{TA} is significantly higher than that of Auger at energies greater than $10^{19.3}$\,eV. 
This raises the tantalizing possibility that the sources of the highest energy cosmic rays, and hence their energy spectra, may differ between disparate parts of the sky. This hypothesis was tested by splitting the data set into two equal sets by arrival declination \cite{Ivanov:2017rwl}.  
There is a $3.9\sigma$ difference in the location of the spectral break between the two: $\log_{10}E=19.85\pm0.03$ for higher declination band, and at $\log_{10}E=19.59\pm0.06$ for lower declination band. The events in the hotspot clearly contribute to the harder spectrum in the higher declinations. However, a clear difference remains when the $20^\circ$ circle of the hotspot is excluded \cite{Abbasi:2018ygn}.

Another possible spectral anisotropy is illustrated on the left side of \cref{fig:spectrum-anisotropy} shows the ratio of the \ac{TA} \ac{SD} flux inside the hotspot to that outside. The hotspot itself is the excess of events above $\log_{10}E=19.75$. Surprisingly, a deficit is seen (a \emph{coldspot}) in the range $19.1<\log_{10}E<19.75$. A scan was carried out on the full sky to look for other possible coldspots using a binned maximum likelihood test comparing the spectra inside and outside circles of radius 15, 20, 25, and 30 degrees. The right side of \cref{fig:spectrum-anisotropy} shows a sky map in equatorial coordinates of local significances indicating that a spectral anisotropy occurs only in the hotspot region. A numerical study found the global significance of this effect to be $3.4\sigma$ \cite{TelescopeArray:2018rtg}.

\subsubsection{The IceCube Neutrino Observatory}
\label{sec:IceCube_DesignAndTimeline}
\begin{wrapfigure}{r}{0.5\columnwidth}
    \vspace{-5mm}
\centering
\includegraphics[width=0.4\textwidth]{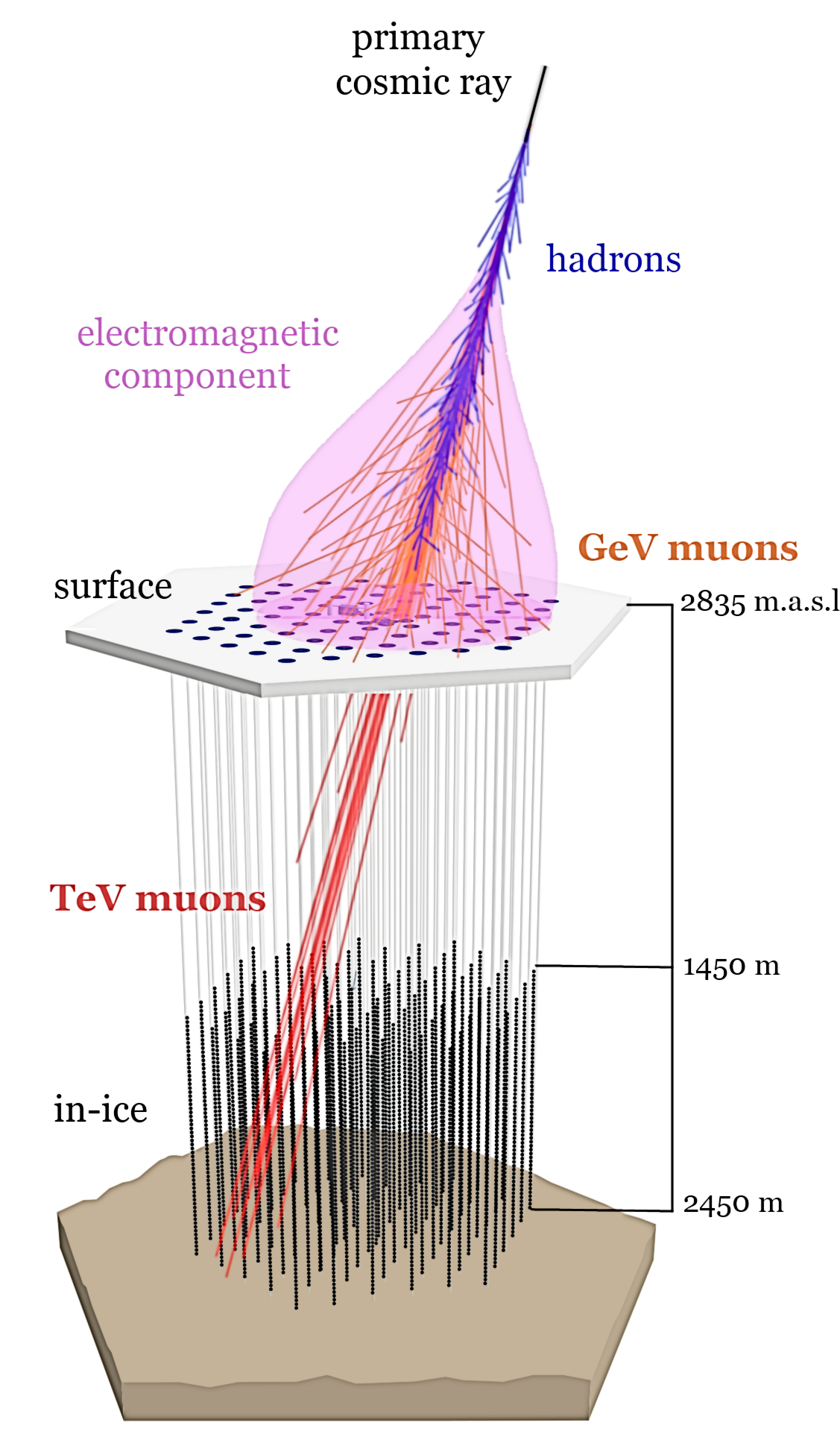}
\caption{\label{fig:ic_array} Layout of IceCube with its surface array and deep detector. Built primarily for neutrino detection, IceCube also constitutes a unique detector for cosmic-ray air showers: TeV and PeV muons are measured in the deep detector and electromagnetic particle and low-energy muons of the same showers are measured at the surface~\cite{IceCube:2019hmk,DeRidder:2017alk, IceCube:2021ixw}.}
\vspace{-5mm}
\end{wrapfigure}

The IceCube Neutrino Observatory is a cubic-kilometer-scale particle detector at the South Pole operating in its completed configuration since 2011. IceCube employs over 5000 \acp{DOM} to detect Cherenkov light produced by secondary particles from neutrino and \ac{CR} interactions \cite{IceCube:2016zyt}. The detector consists of both a deep in-ice array of \acp{DOM} and a square-kilometer surface component (\cref{fig:ic_array}), IceTop~\cite{IceCube:2012nn}, consisting of 81 surface stations, each with two ice-Cherenkov tanks containing two \acp{DOM}. The combination of surface air shower and deep in-ice muon measurements provides unique capabilities for various analyses of \acp{CR}, probing the Galactic to extragalactic transition region.

IceCube makes important contributions to Galactic cosmic-ray physics from below the TeV to the EeV energy range, i.e., it covers the highest energies of Galactic cosmic rays and the transition to extragalactic cosmic rays.
IceTop has measured the all-particle \ac{CR} spectrum at PeV to EeV energies~\cite{Aartsen:2013wda,IceCube:2019hmk}; this has been recently extended below the knee to measure the energy spectrum down to $250\,\mathrm{TeV}$~\cite{Aartsen_2020_IceTopLE}. 
Using its unique combination of the surface and in-ice array, IceCube provides a high mass-separation power that has facilitated the first measurement of individual spectra from four elemental mass groups between $2.5\,\mathrm{PeV}$ and $1\,\mathrm{EeV}$~\cite{IceCube:2019hmk}. 
Moreover, such hybrid observations have also allowed for searches for PeV gamma-ray emission from the southern hemisphere~\cite{IceCube:2012vku,IceCube:2019scr}.
Although the sensitivity achieved with the data and analysis methods available now has not lead to a discovery, by improving analysis techniques and continued operation, IceCube may eventually discover PeV photon sources, in particular, since PeV photon sources are meanwhile known to exist and be observable with a square-kilometer size array \cite{Huang:2021hjc, LHAASO:2021cbz}.

With its surface and deep detectors, IceCube is well suited to study the particle physics in air showers, especially, the production of atmospheric leptons.
IceCube has recently reported preliminary results of the measurement of GeV muons in air showers~\cite{Gonzalez:2019epd, IceCube:2021tuv, IceCube:2022yap} and simultaneous measurements of GeV muons measured with IceTop in coincidence with TeV muons in the deep ice~\cite{DeRidder:2017alk, IceCube:2021ixw}. 
IceCube has measured the lateral separation of TeV muons in the deep ice~\cite{Abbasi:2012kza, Soldin:2015iaa, Soldin:2018vak} and the spectrum of muons with energies above $10\,\mathrm{TeV}$~\cite{Soldin:2018vak,IceCube:2015wro, Fuchs:2017nuo}, with evidence for a prompt muon flux above $\sim$\,$1$~PeV in muon energy at the $\sim$\,$3\sigma$ level. IceCube also has measured the $\sim$\,$20\%$ seasonal variations in the muon intensity over the individual years with a statistical significance that is sensitive to daily stratospheric temperature variations of a few degrees~\cite{Desiati:2011hea, Tilav:2010hj, Gaisser:2013lrk, Tilav:2019xmf}. 

Finally, IceCube is the only ground-based experiment that has measured the CR anisotropy in the TeV--PeV energy range in the southern hemisphere. It was the first experiment to detail the anisotropy’s energy dependence in this energy range~\cite{IceCube:2016biq}, and the first to show the angular power spectrum of the spherical harmonic expansion as a means to quantify how the medium/small angular scales of the anisotropy are  distributed~\cite{IceCube:2011fxx}. In collaboration with the HAWC gamma-ray observatory, IceCube also produced the first full-sky view of the 10~TeV cosmic-ray anisotropy~\cite{HAWC:2018wju}, demonstrating how the increased field of view affects the observation at large angular scales. 

A surface enhancement of IceTop comprised of scintillation and radio antennas has mainly been planned to mitigate and calibrate the effect of snow accumulation, and will also increase the measurement accuracy for cosmic-ray air showers. 
A prototype station of that enhancement is successfully operating at the South Pole \cite{IceCube:2021epf}, and the deployment of further stations over the full IceTop array is foreseen ahead of IceCube-Gen2 (see \cref{sec:IceCubeGen2}). 
Consequently, IceCube will continue to make leading contributions to the field of Galactic cosmic-ray physics and hadronic interactions in the ongoing decade, before its capabilities will be magnified by the planned IceCube-Gen2 extension (see \cref{sec:TNG}).

\subsection[Energy spectrum]{Energy spectrum: Well established but not well explained} 
\label{sec:EnergySpectrum}
The flux of cosmic rays as a function of energy, i.e., the energy spectrum, is one of the most fundamental observables to infer on the nature of \acp{UHECR}.
The production mechanisms, the source type and distribution and the propagation environment, shape the spectrum in a non-trivial way, imprinting on the spectrum several features deviating from a pure power law.
The shape is thus an object of detailed scrutiny for studying the combined effects of the evolution of the arrival directions and mass composition with primary energy.
The precise measurements of the spectrum have been used to put strong constraints on astrophysical models of the sources, particularly when combined with other measurements like \xmax~\cite{PierreAuger:2021mmt,Bergman:2021djm} (see \cref{sec:Astrophysics}).

 \begin{figure}[htb]
       \centering
        \includegraphics[width=0.99\textwidth]{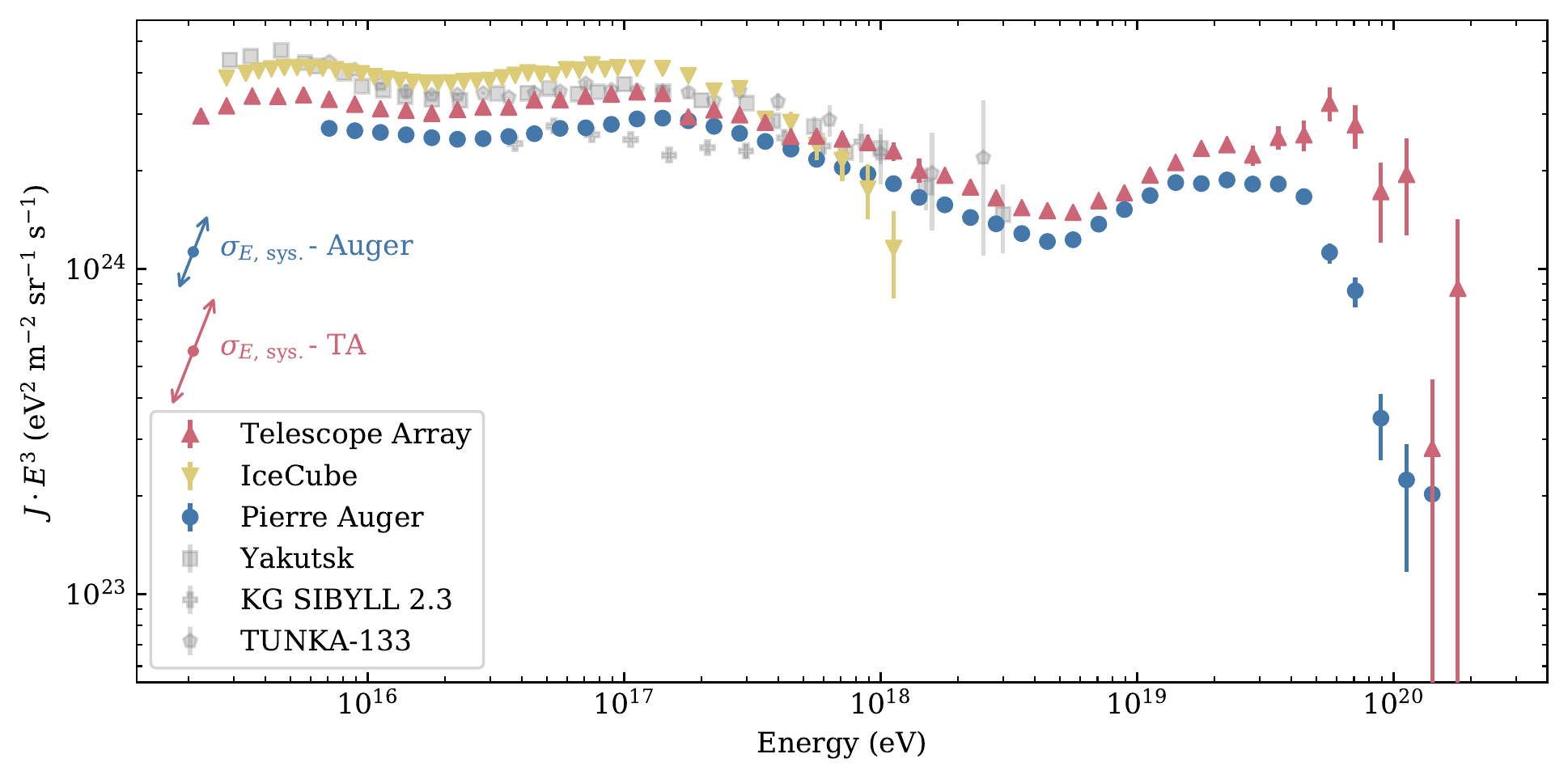}
        \caption{Recent measurements of the all-particle flux from the \ac{TA}~\cite{Ivanov:2020rqn}, IceCube~\cite{IceCube:2019hmk}, Pierre Auger~\cite{PierreAuger:2020qqz,PierreAuger:2021hun,PierreAuger:2021ibw}, Yakutsk~\cite{Knurenko:2013dia}, KASCADE-Grande~\cite{KASCADEGrande:2017gtn}, and TUNKA~\cite{Budnev:2020oad} experiments, which define the spectral features in the \ac{UHE} region, are shown.
        Those with upgrades specifically described in this white paper are shown in color. The direction and magnitude of the systematic uncertainty in the energy scale for Auger and \ac{TA} is indicated by the corresponding arrows.
        }
        \label{fig:all_spectra}
\end{figure}

The spectra measured by the Auger (\cref{sec:Auger_DesignAndTimeline}) and \ac{TA} (see \cref{sec:TA_DesignAndTimeline}) collaborations are shown in \cref{fig:all_spectra}, scaled by $E^{3}$ to highlight the deviation from a pure power law. Despite being conceived as \ac{UHECR} detectors, the two observatories achieve an impressive 5 orders of magnitude spectrum in energy. This feature, other than being visually extremely powerful, allows to construct a single overview of the spectrum from the low energy up to the highest. This allows to give a single description of the transition from the galactic to extragalactic cosmic rays, reducing the systematic uncertainties that would result from different measurements. Modelling efforts can now rely on data from single experiments, both in the northern and southern hemispheres, over an impressively wide ranges of energy. Several features are now well established, the 
\emph{knee} at $\simeq$\,$5\times 10^{15}$\,eV, the so-called \emph{low energy ankle} just above 
$10^{16}$\,eV, the \emph{second-knee} at $\simeq$\,$10^{17}$\,eV, the \emph{ankle} at $\simeq$\,$5\times 10^{18}$\,eV, the \emph{instep} at $\simeq$\,$10^{19}$\,eV, and the \emph{suppression} beginning at $\simeq$\,$5\times 10^{19}$\,eV.
In the following, measurements which cover the final two decades in energy, in the \ac{UHECR} range, where Auger and \ac{TA} are the only experiments available are mainly covered. The developments needed for a better understanding  of the transition from galactic to extragalactic component will be also briefly discussed.

There are two techniques to measure the energies of primary cosmic rays at ultra-high energies. The first is to use the lateral distribution of charged particles in an air shower, observed on ground with particle detectors. This is a traditional method employed in early experiments, e.g., the Volcano Ranch experiment in New Mexico in the US~\cite{Linsley:1963km}, the Haverah Park Experiment in the UK, and AGASA in Japan~\cite{Chiba:1991nf} (see Ref.~\cite{Nagano:2000ve} for a review). The second consists in measuring the fluorescence photons emitted from air molecules excited by the charged particles in an air shower (for the new additional method of using radio measurements see \cref{sec:RD_tech_development}).
The fluorescence technique was proposed in 1960's~\cite{Greisen:1960wc,delvaille1962spectrum,Suga1962} and firstly implemented in the Fly's Eye experiment in Utah in the US~\cite{Baltrusaitis:1985mx}, and followed by the \ac{HiRes} experiment~\cite{Abu-Zayyad:2000vin}.
This is a calorimetric measurement of the cosmic ray energy, that is therefore less dependent on the details of hadronic interactions beyond accelerator energies (the \ac{LHC} energy corresponds to a cosmic ray proton of $\sim$\,$10^{17}$\,eV interacting with a nitrogen nucleus at rest). There were two differences in the energy spectra of \acp{UHECR} in the results of these $20^{\rm th}$ century experiments~\cite{Takeda:2002at,HiRes:2007lra}. The first is in the energy scale of the two techniques (apparent in the difference of the position of the ankle), and the shape of the spectra at the highest energies (the AGASA spectrum extended beyond $10^{20}$\,eV, whereas the \ac{HiRes} result exhibited a steepening at $10^{19.75}$\,eV). It was difficult to identify the origin of the difference in the spectrum measurements with different techniques.

The discrepancy observed by these early experiments led to the construction of the Pierre Auger Observatory (see \cref{sec:Auger_DesignAndTimeline}) in the southern hemisphere and \ac{TA} (see \cref{sec:TA_DesignAndTimeline}) in the north. They are the largest cosmic ray observatories ever built, covering 3000\,km$^2$ and 700\,km$^2$ respectively, and have an hybrid design, employing both a \ac{SD} and a \ac{FD}. Using a sub-sample of high quality events recorded by both detectors, the \ac{SD} signals are calibrated against the energies measured with the \ac{FD}. In this way, the tiny flux of cosmic rays at \ac{UHE} can be measured with the largest possible exposure achievable, using direct particle detection on ground (the \ac{FD} can operate only during moonless nights) and with a calorimetric, almost model-independent, measurement of the shower energy.

Both Auger and \ac{TA} represent an enormous increase in exposure with respect to AGASA and \ac{HiRes}. The Auger and \ac{TA} collaborations have indeed achieved a cumulative exposure of about 70\,000\,km$^2$\,sr\,yr on the full sky, to be compared with the total exposure of about 5000\,km$^2$\,sr\,yr achieved by the previous generation of experiments, AGASA and \ac{HiRes}.

\subsubsection{Current measurements of the energy spectrum at UHE}
\begin{figure}[tb]
    \centering
    \includegraphics[width=0.56\textwidth]{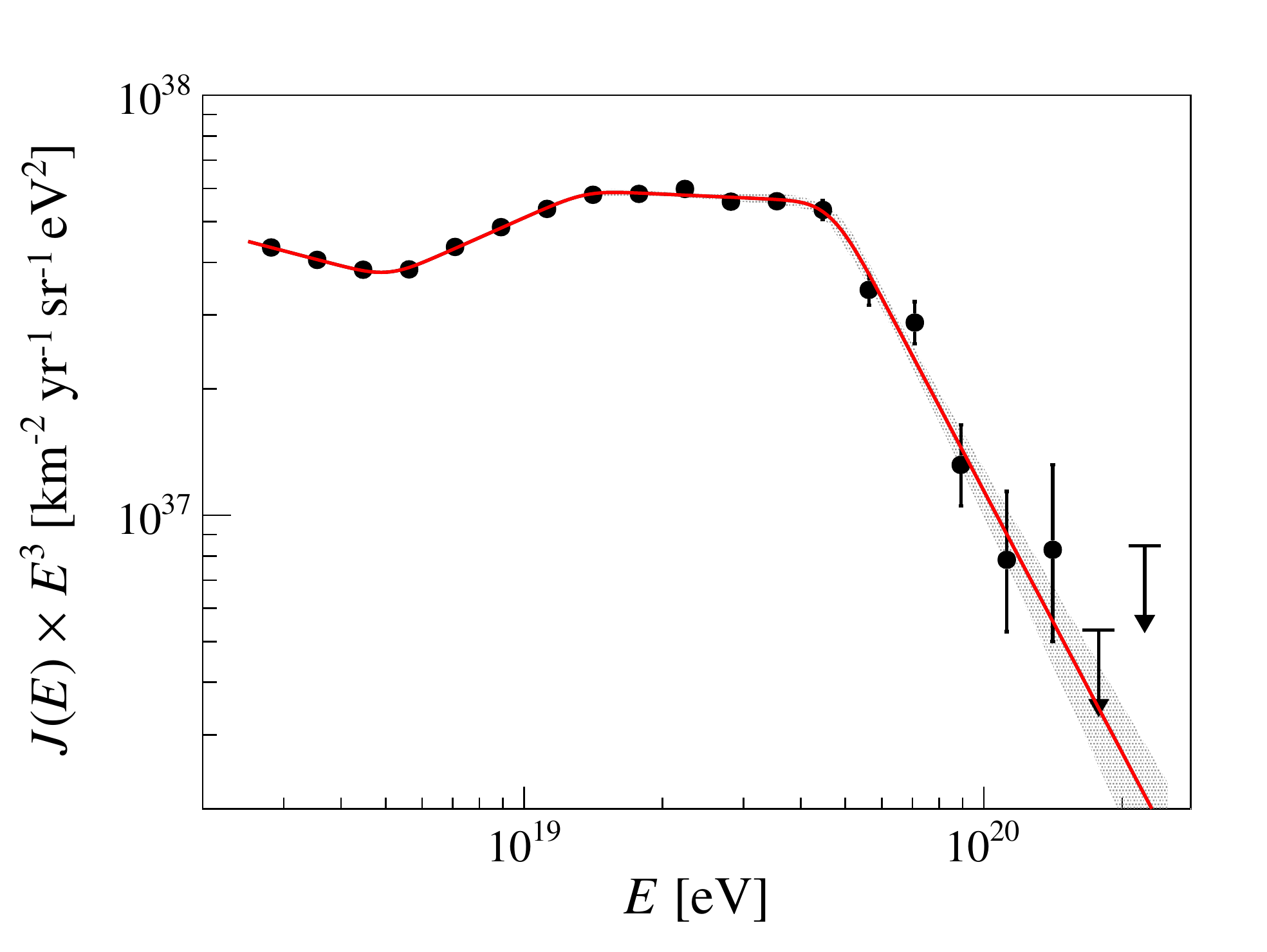}
    \includegraphics[width=0.4\textwidth]{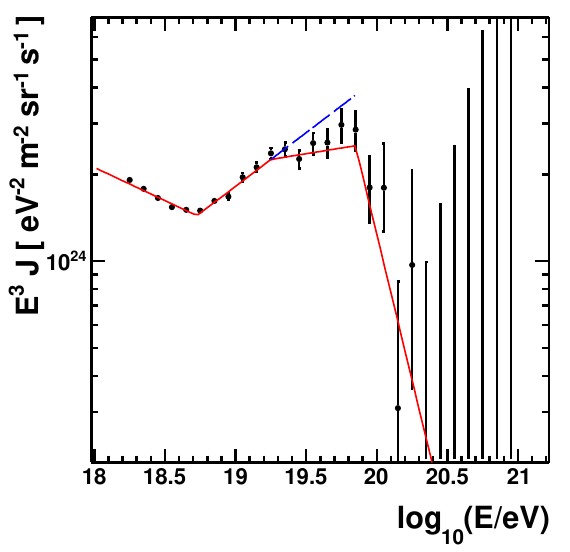}
    
    \caption{Left: the Auger energy spectrum above $2.5 \times 10^{18}$\,eV~\cite{PierreAuger:2020qqz}. The red line is a fit to the data using a smoothed broken power-law. Right: The \ac{TA} spectrum above $10^{18.2}$\,eV measured with SD~\cite{Ivanov:2021mkn}. The red line is a fit to the combination of the shown data with the FD measurements of \ac{TA} and the results from \ac{HiRes}.}
    \label{fig:newFeature}
\end{figure}
The energy spectrum measured by Auger~\cite{PierreAuger:2020qqz} and \ac{TA}~\cite{Ivanov:2021mkn} at and above the ankle are shown in \cref{fig:newFeature}. 
The spectra are measured with the high statistics obtained with the surface detectors of both observatories. In Auger, the \ac{SD} units are water-Cherenkov detectors and the energy estimator is corrected 
for the attenuation in the atmosphere with the so-called Constant Intensity Cut method~\cite{Hersil:1961zz}. The corrected energy estimator is then calibrated against the \ac{FD} energies using a power-law relationship. The measurements are performed above the energies at which the \ac{SD} array is fully efficient and the entire analysis to derive the energy spectrum is data-driven and does not make assumptions about the hadronic physics and mass composition~\cite{PierreAuger:2020qqz}. In \ac{TA}, the \ac{SD} units are scintillator detectors and the signal at ground is converted into shower energy using a Monte Carlo lookup table that accounts also for the attenuation effects. The hybrid events are then used to rescale the reconstructed energies to the values estimated with the \ac{FD}~\cite{TelescopeArray:2012qqu}.  Auger measures the spectrum above $2.5 \times 10^{18}$\,eV with an energy resolution of 10$\%$ at $10^{19}$\,eV and a systematic uncertainty on the energy scale of 14$\%$~\cite{Dawson:2020bkp}. The \ac{TA} measurements starts at $1.6 \times 10^{18}$\,eV. The energy resolution is 19$\%$ and the uncertainty in the energy scale is 21$\%$~\cite{Abu-Zayyad:2011ugz}.

The spectral features obtained from a fit to the data using a sequence of 4 power-laws, shown in red in the plots, are given in \cref{tab:spectral_features}.  Generally, good agreement is found between the two experiments, the \emph{ankle} is determined with high precision and the measurements confirm with higher statistical significance previous reports of the suppression at highest energies~\cite{HiRes:2007lra,TelescopeArray:2012qqu,PierreAuger:2008rol}.
A new feature has been recently discovered by both collaborations: the \emph{instep}.
It was observed for the first time by Auger with a significance of $3.9\,\sigma$ \cite{PierreAuger:2020qqz}. The significance has been calculated with a 
likelihood ratio procedure estimating the improvement of the fit with the additional break at $10^{19}$\,eV with respect to an old model with a single smooth suppression. 
This finding was later confirmed by the \ac{TA} collaboration, by using a combination of the observations of the \ac{SD} and \ac{FD} of \ac{TA} along with the measurements from \ac{HiRes}.
With this combination a single power-law model between the ankle and the suppression is rejected with a $5.3\,\sigma$ significance~\cite{Ivanov:2021mkn}. 
The \emph{instep} feature is an observation of fundamental importance to constrain astrophysical models and, as shown in Ref.~\cite{PierreAuger:2020kuy}, it can be reproduced by a model with an energy-dependent mass composition (see also \cref{sec:Astrophysics}).
\begin{table}[tb]
    \centering
    \begin{tabular}{cll}
        Parameter & Pierre Auger Obs. & Telescope Array\\
        \hline
        $E_{\rm ankle}$ / EeV & $\phantom{0}5.0 \pm 0.1$ & $\phantom{0}5.4 \pm 0.1$ \\
        $E_{\rm instep}$ / EeV & $13\phantom{.0} \pm 1\phantom{.0}$ & $18\phantom{.0} \pm 1\phantom{.0}$ \\
        $E_{\rm cut}$ / EeV & $46\phantom{.0} \pm 3\phantom{.0}$ & $71\phantom{.0} \pm 5\phantom{.0}$ \\
        $\gamma_{1}$ & $3.29 \pm 0.02 $ & $3.23 \pm 0.01$ \\
        $\gamma_{2}$ & $2.51 \pm 0.03 $ & $2.63 \pm 0.02$ \\
        $\gamma_{3}$ & $3.05 \pm 0.05 $ & $2.92 \pm 0.06$ \\
        $\gamma_{4}$ & $5.1\phantom{0} \pm 0.3\phantom{0} $ & $5.0\phantom{0} \pm 0.4\phantom{0}$ \\
    \end{tabular}
    \caption{The values of the shape of the spectrum in the \ac{UHE} region, as measured by Auger~\cite{PierreAuger:2020qqz} and \ac{TA}~\cite{TelescopeArray:2021zox}, are given above. The spectral indices describe the average power-law slope, $E^{-\gamma}$, between the spectral-break energies. Only statistical uncertainties are given.}
    \label{tab:spectral_features}
\end{table}

\begin{figure}
    \centering
    \includegraphics[width=0.5\textwidth]{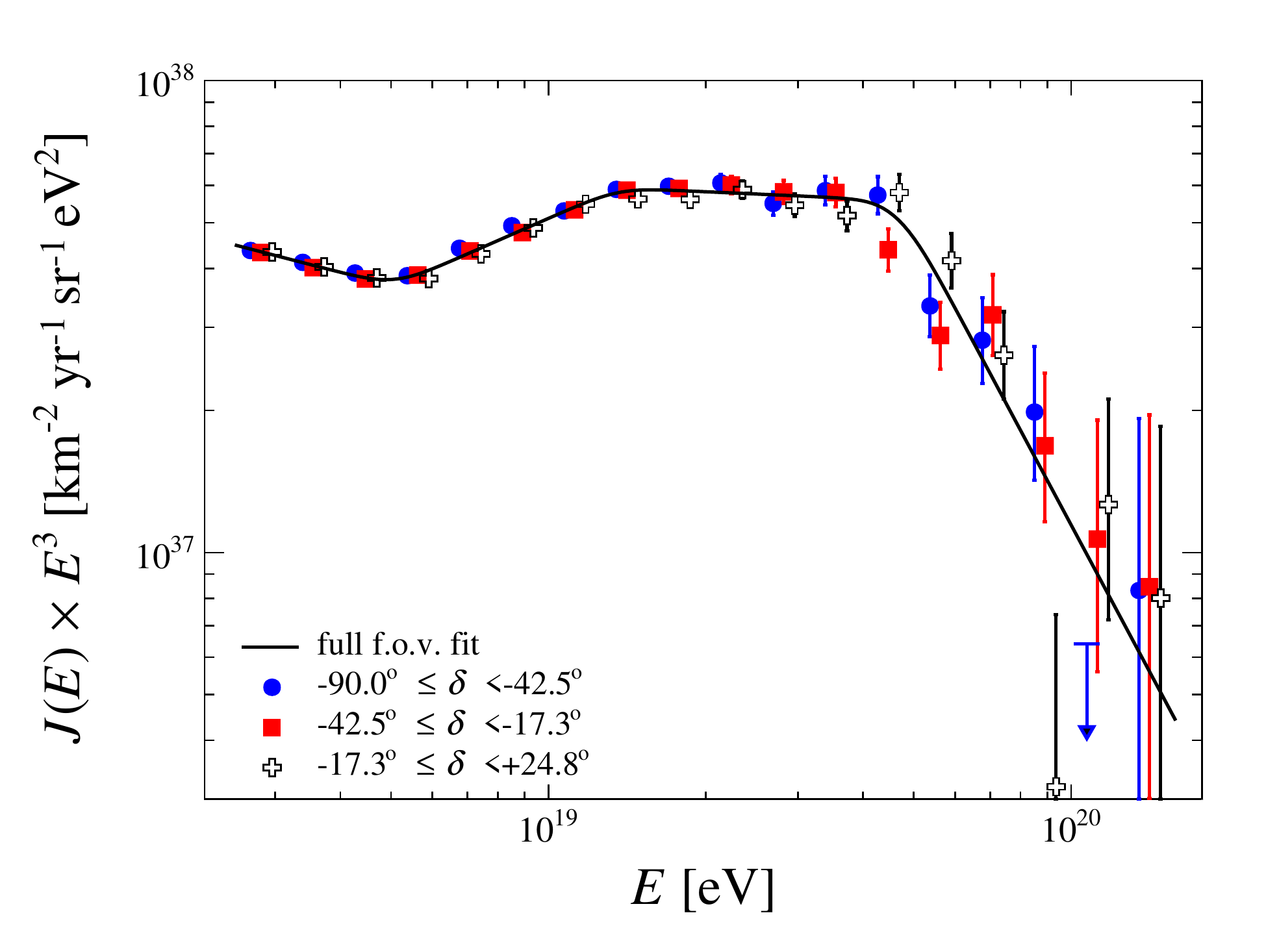}
    \includegraphics[width=0.4\textwidth]{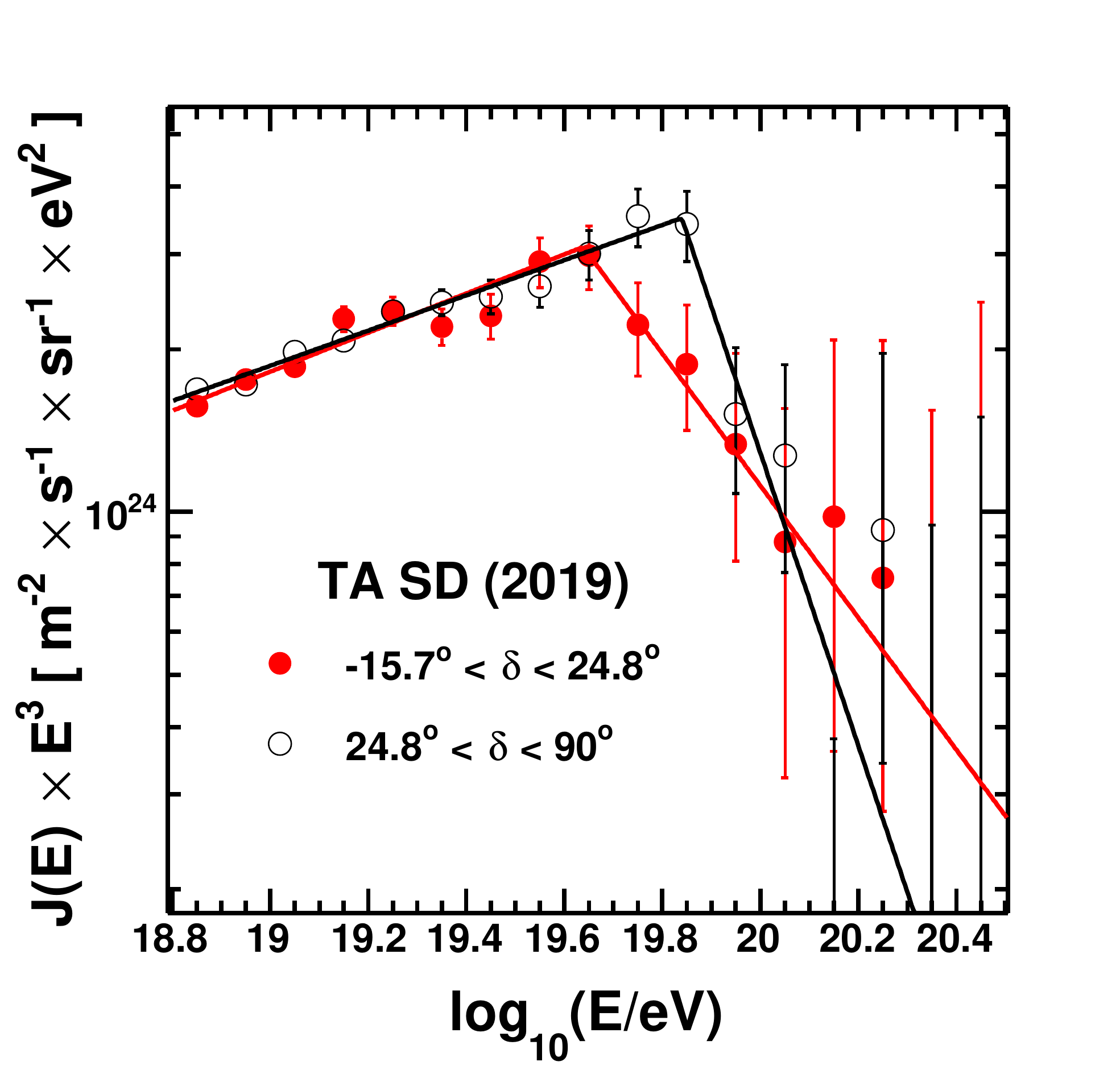}
    \caption{The energy spectrum in different declination bands measured by the Pierre Auger  Observatory~\cite{PierreAuger:2020qqz} (left) and the Telescope Array~\cite{Abbasi:2018ygn} (right).}
    \label{fig:decbands}
\end{figure}

The enormous statistical power achieved by both collaborations has allowed for the production of spectra in different declination bands. The goal of such studies is to investigate the spectrum in different parts of the sky.
The measurements from Auger~\cite{PierreAuger:2020qqz,PierreAuger:2020kuy} are obtained using the showers with zenith angle below $60^\circ$ and cover the declination range between $-90^\circ$ and $24.8^\circ$. They are shown in the left panel of \cref{fig:decbands} in three declination bands of equal exposure, and they do not give any evidence of a declination dependence of the spectrum other than the mild excess from Southern Hemisphere, consistent with the directional anisotropy above $8 \times  10^{18}$\,eV \cite{PierreAuger:2017pzq}. The \ac{TA} measurements, shown in the right panel of \cref{fig:decbands}, are in the declination bands ($-15.7^\circ$,~24.8$^\circ$) and (24.8$^\circ$,~90$^\circ$) and suggest different positions of the steepening at highest energies~\cite{Ivanov:2020rqn}.
It is worth noting that the southernmost declination band of \ac{TA} overlaps with the \ac{FoV} of Auger, and in this band the steepening position is at about $3.9 \times 10^{19}$\,eV, significantly below to what is observed in the full sky and therefore in better agreement with the Auger measurement (see \cref{tab:spectral_features}).  

\subsubsection{Detailed studies at the highest energies from the joint working groups}
\label{subsec:AugerTAWG}

\begin{figure}[tb]
        \centering
        \includegraphics[width=0.48\textwidth]{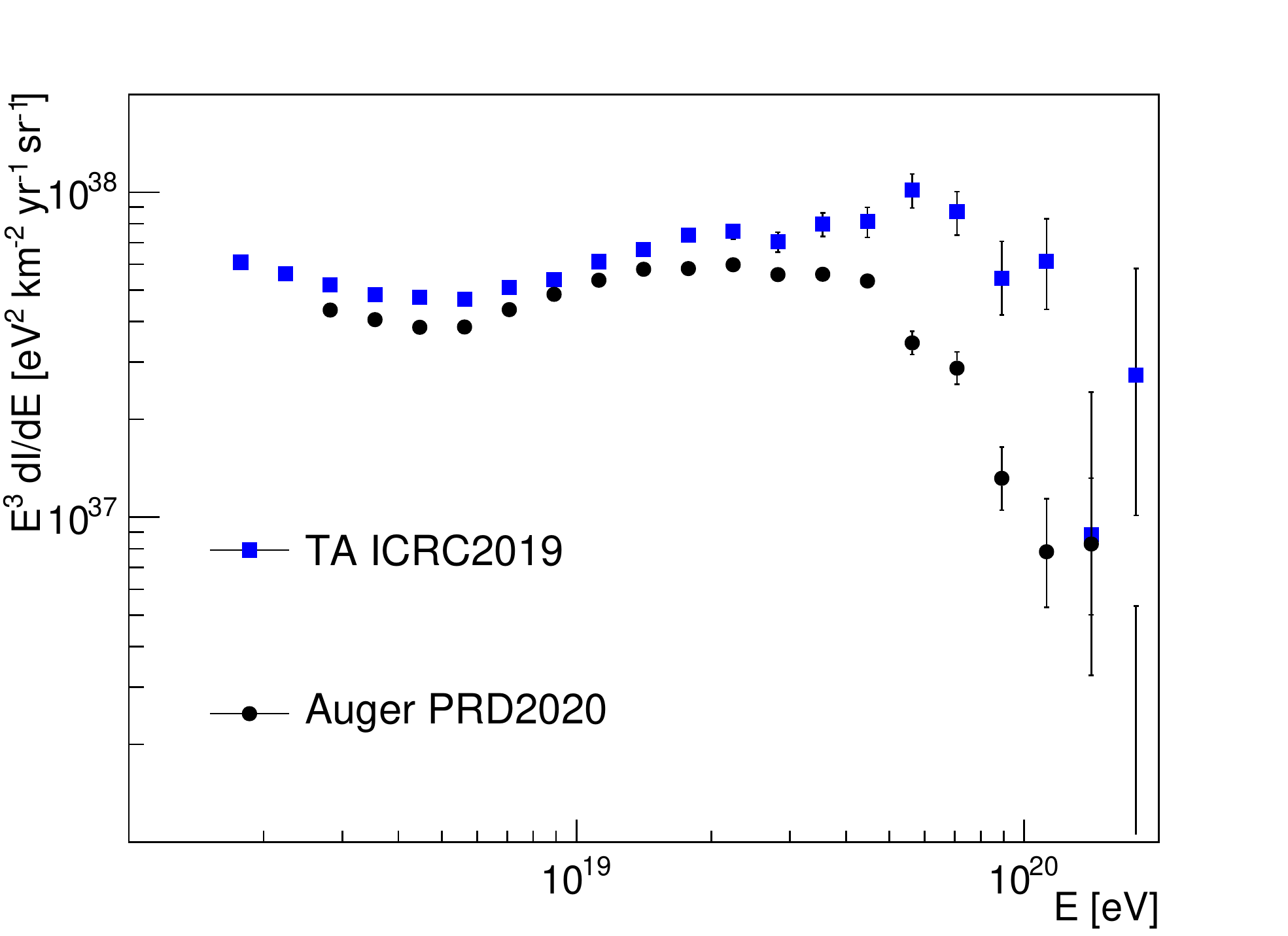}
        \includegraphics[width=0.48\textwidth]{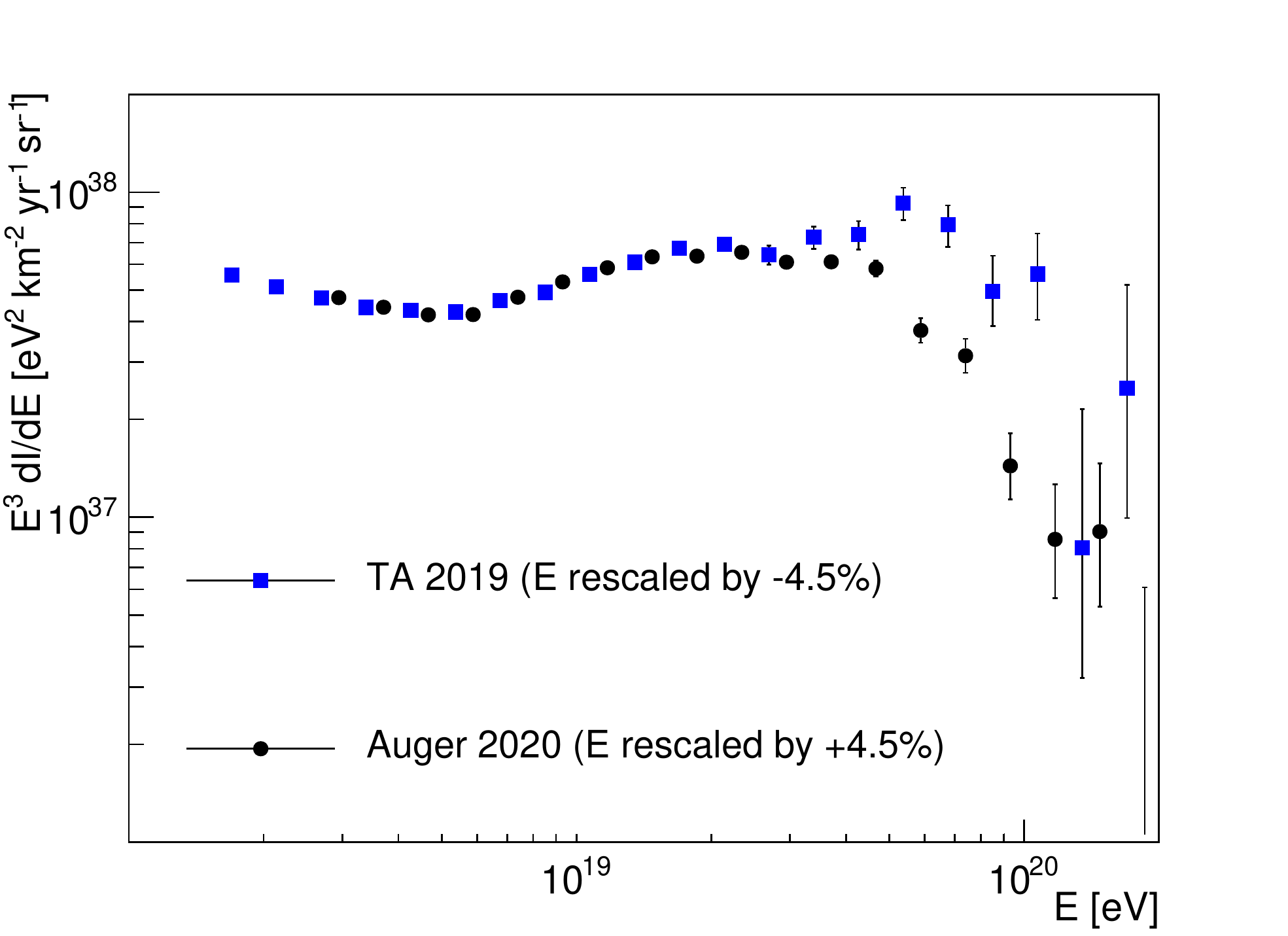}
        \includegraphics[width=0.48\textwidth]{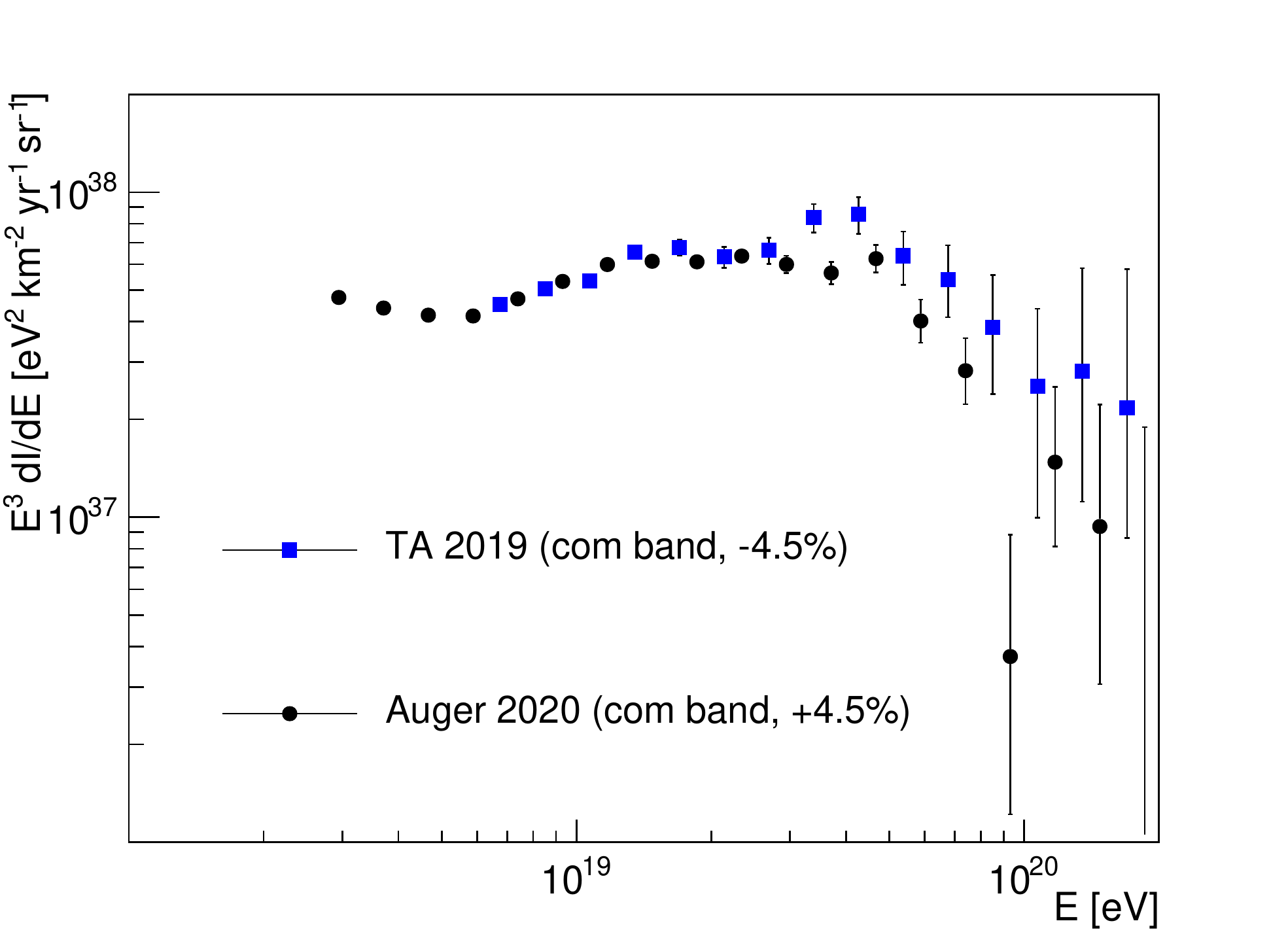}
         \includegraphics[width=0.48\textwidth]{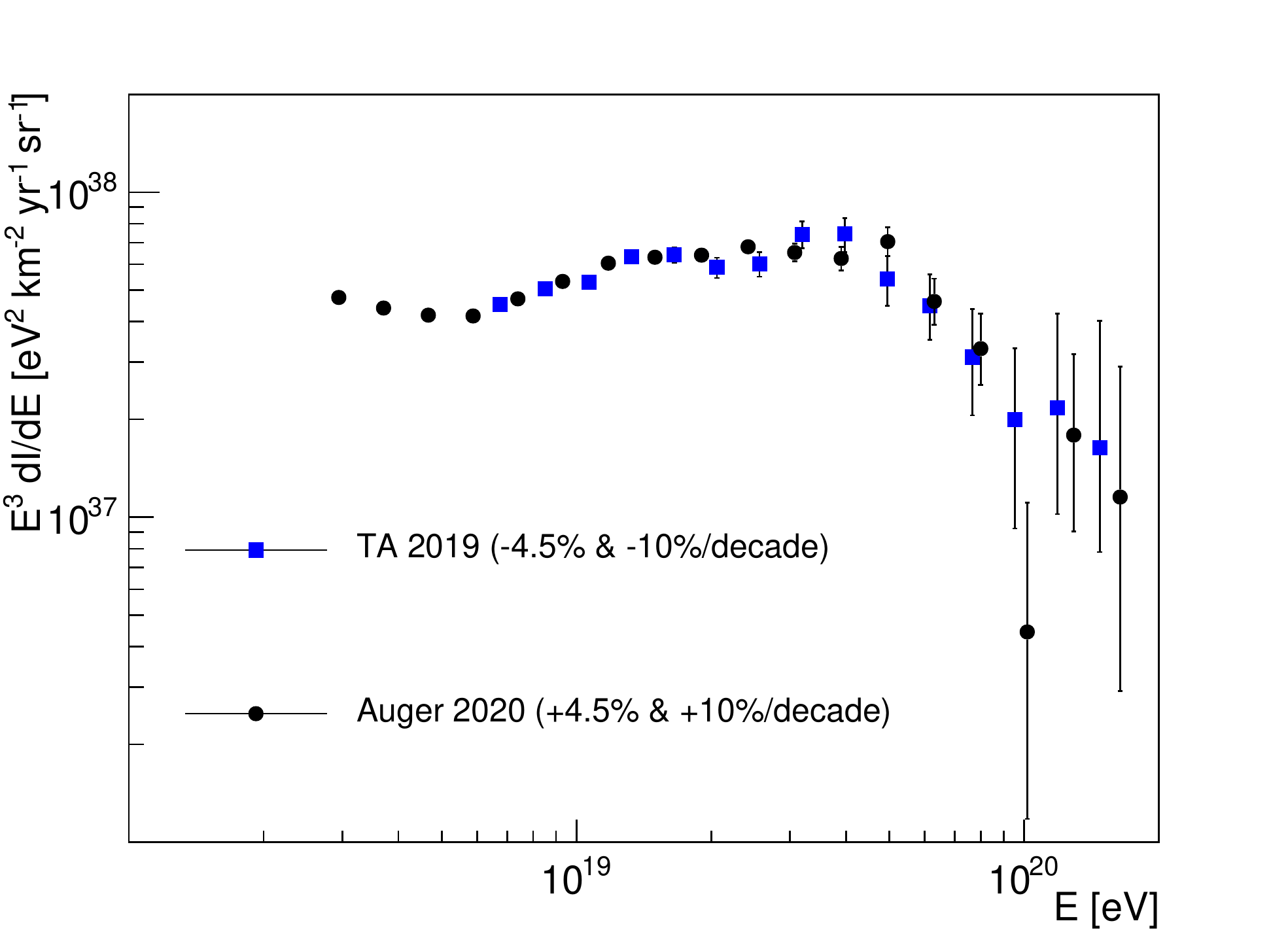}
        \caption{Upper panel: Comparison of the Auger and \ac{TA} spectra in the full \acp{FoV} (left) and same comparison after a rescaling of the energy scale by the same amount ($\pm 4.5\%$) and in opposite directions (right). Lower panels: Auger and \ac{TA} spectra in the common declination band after the above rescaling of $\pm 4.5\%$ (left) 
        and after an additional energy-dependent shift of $\pm 10\% \times \log_{10}(E/10^{19}~\mathrm{eV})$ 
        for $E > 10^{19}$\,eV. Original figures from Ref.~\cite{TelescopeArray:2021zox}.
       The full \acp{FoV} of Auger and \ac{TA} are $-90^{\circ} < \delta < +24.8^{\circ}$ and $-15.7^{\circ} < \delta < +90^{\circ}$, respectively, while the common declination band is $-15.7^\circ < \delta < 24.8^\circ$.}
        \label{fig:PAO_TA_spec_comp}
    \end{figure}

A fruitful collaboration between the Auger and \ac{TA} observatories is underway to give a unique and consistent interpretation of the cosmic ray flux. The results of such activities were reported
in the UHECR and ICRC conference series~\cite{PierreAuger:2013dyy,Verzi:2017hro,Ivanov:2017juh,AbuZayyad:2018aua,PierreAuger:2019vpk,Deligny:2020gzq,TelescopeArray:2021zox}.
The most updated results are presented in \cref{fig:PAO_TA_spec_comp}~\cite{TelescopeArray:2021zox}. In the upper panels of this figure, the measurements are compared in the full \acp{FoV} of the two observatories. As shown in the right panel, the two spectra are in agreement up to few $10^{19}$\,eV once a $\pm$4.5$\%$ shift in the energy scale of each experiment is applied. 
Such differences would be further reduced  once one accounts
for the different models used by the two collaborations for the fluorescence yield and the so-called invisible energy, 
the energy of the primary carried to ground by muons and neutrinos that has to be added to the calorimetric energy measured 
by the \ac{FD} in order to obtain the total shower energy. 
The Auger collaboration uses the high precision measurement of the fluorescence yield performed by the Airfly experiment~\cite{AIRFLY:2012msg,AIRFLY:2007msg} while \ac{TA} the measurements from Kakimoto et al.~\cite{Kakimoto:1995pr} and the FLASH experiment~\cite{Abbasi:2007am}.
For the invisible energy, Auger uses a data driven estimation exploiting the muon sensitivity of the \ac{SD}~\cite{PierreAuger:2019dhr}
while \ac{TA} obtains it from Monte Carlo simulations~\cite{Tsunesada:2011mp}. Using the Airfly fluorescence yield in the \ac{TA} reconstruction would reduce the shower energy by 14\% while using the Auger invisible energy, the \ac{TA} energies would be increased by 7\%. 
Therefore, the net effect of synchronizing both the fluorescence yield and invisible energy would reduce the overall offset of 9\% to well below 5\% which is an indication that the systematic uncertainties in the energy scales of the two experiments, like the one on the absolute calibration of the FD telescopes, are well under control.

As can be seen in the upper-right panel of \cref{fig:PAO_TA_spec_comp}, the $\pm$4.5$\%$ shift is not enough to put the 
spectra in agreement at the highest energies. In order to understand if this disagreement is due to astrophysical or experimental effects, the two collaborations have compared the measurements in the declination band accessible by both observatories, namely between $-15.7^\circ$ and 24.8$^\circ$. 
The results of such studies are presented in lower panels of \cref{fig:PAO_TA_spec_comp}. As a consequence of the indication of the declination dependence of the \ac{TA} spectrum addressed in the previous section, after the $\pm$4.5$\%$ shift the two spectra are in better in agreement in comparison to what is observed in the full band. However, this is still not enough and a further energy-dependent shift of $\pm 10\% \times \log_{10}(E/10^{19}~\mathrm{eV})$ for $E > 10^{19}$\,eV is needed to get the full consistency up to the highest energies.
The same conclusion is attained once the different directional exposures of the two observatories are accounted for~\cite{Deligny:2020gzq}. Such an energy-dependent shift cannot 
be explained by the systematic uncertainties since their energy dependence is expected to be small~\cite{PierreAuger:2019vpk}, and a Monte Carlo study is underway to disentangle systematic from statistical effects.

\subsubsection{Understanding the transition to extragalactic sources}
While there is strong evidence that the sources of cosmic rays above 8\,EeV are extragalactic~\cite{PierreAuger:2017pzq}, the transition between the source population(s) between 100\,PeV and 10\,EeV are not well understood.
In this energy region, the shape of the all-particle spectrum has been measured with high statistical precision.
Measurements of the flux spectra for primary mass groups at-and-above the second knee have been performed by several collaborations~\cite{IceCube:2019hmk,KASCADEGrande:2017gtn,Epimakhov:2013cnw,TelescopeArray:2020bfv}.

The second-knee corresponds to a point in the spectrum where the average nuclear mass is relatively heavy, mostly CNO-like or heavier, but where this flux of heavy elements exhibits a softening.
The flux that makes up the second-knee has been postulated to be the high-rigidity counterpart of the knee, based on a maximum-rigidity acceleration scenario from, for example, \acp{SNR}~\cite{Blasi:2013rva}.
Between $10^{17.0}$\,eV and $10^{17.5}$\,eV, the proton flux hardens which may indicate the beginning of a new source class which produces the flux between the second-knee and the extragalactic sources.
However, it is uncertain if the power-law flux leading up to the ankle is galactic~\cite{Hillas:2005cs} or extragalactic~\cite{Aloisio:2013hya}.

The separation of the all-particle flux into different mass groups has large systematic uncertainties coming from the interpretation of air-shower measurements using simulations.
A better understanding of the transition to extragalactic sources is expected from improvements in our understanding of hadronic interaction models (see \cref{sec:AccelSyn}) as well as improved cosmic rays observatories and techniques (see \cref{sec:comp10yr-MLM}).

\subsection[Primary mass composition]{Primary mass composition: When nature throws curve-balls}
\label{sec:MassCurrentStatus}
The understanding of the composition of \acp{UHECR} and the role composition plays in the wider study of \acp{UHECR}, has undergone a dramatic change in the last 20\,years. The field has moved from a picture of relative simplicity to one with deep nuances and critical questions. With this change the over all view has become richer, providing powerful tools for understanding the sources of \acp{UHECR}.
This new and more complex reality means that a more precise determination of the cosmic-ray composition is crucial to expanding our knowledge about \acp{UHECR}. In particular, a finer-grained identification of primary cosmic-ray composition, through increases in available statistics and/or mass resolution, will enable:
\begin{itemize}[topsep=2pt]\setlength\itemsep{-0.2em}
    \item[a)] the measurement of primary composition at post-suppression energies, in turn providing
    \vspace{-3mm}
    \begin{itemize}
        \item stronger constraints of the properties of Z\eV{\kern -0.07em}atron accelerators of \acp{UHECR} and
        \item strict constraints on \ac{UHECR} propagation and cutoff scenarios;
    \end{itemize}
    \item[b)] event-by-event charged-particle astronomy and the study of magnetic fields;
    \item[c)] more precise predictions of cosmogenic fluxes of high-energy photons and neutrinos;
    \item[d)] more precise predictions of atmospheric neutrino backgrounds for neutrino telescopes;
    \item[e)] precision studies of hadronic interactions at energies way beyond human-made accelerators;
    \item[f)] higher-efficiency direct searches for \ac{UHE} neutrinos and photons;
    \item[g)] expanded searches of new physics, e.g., signatures of \ac{LIV} or \ac{SHDM}.
\end{itemize}
\vspace{2mm}

Two particularly important recent examples of the synergy between \ac{UHE} hadronic interactions and cosmic-ray experiments were the measurement of the proton-air cross section around a center-of-mass energy of 60~T\eV~\cite{PierreAuger:2012egl, Abbasi:2020chd}, and the identification of a deficit of muons in most current hadronic interaction models used in air shower simulations~\cite{PierreAuger:2014ucz, PierreAuger:2016nfk, TelescopeArray:2018eph}. A prerequisite for the proton-air cross section studies is the establishment and identification of protons in the cosmic-ray particle beam. Likewise, the precise quantification of the muon deficit depends crucially on the primary cosmic-ray composition. Furthermore, any search for proposed new physics at \ac{UHE}~\cite{Farrar:2019cid, Anchordoqui:2016oxy, Anchordoqui:2019laz, Alcantara:2019sco, Supanitsky:2019ayx, Aloisio:2015lva, Ishiwata:2019aet} relies on having a good handle on the nature of primary cosmic rays. Lastly, in the search for cosmogenic neutrinos, above PeV energies the uncertainties in atmospheric neutrino production due to primary composition are comparable to those due to \ac{QCD}, greatly complicating background estimation and removal in neutrino telescopes~\cite{Garzelli:2016xmx, Zenaiev:2019ktw}. Each of these studies clearly stands to benefit with an increase in precision and statistics in \ac{UHECR} cosmic ray composition information.

The astrophysical importance of the cosmic-ray composition is multifaceted. Through the charge number $Z$, primary composition determines the rigidity of cosmic rays ($R\propto E/Z$), which directly governs the acceleration of \acp{UHECR} in sources and their propagation in magnetic fields (see \cref{sec:Astrophysics}). This means that knowledge of it is critical to charged particle astronomy and the differentiation of acceleration scenarios~\cite{PierreAuger:2020kuy, Sommers:2008ji, Erdmann:2016vle, PierreAuger:2016use, Unger:2015laa}. Additionally, the primary mass number $A$ is required to extract the Lorentz factor ($\gamma \propto E/A$) of primaries, which in turn determines interactions with photon fields during both acceleration in sources and propagation. Therefore, measuring \ac{UHECR} mass is pivotal to modeling the production of secondaries and therefore multi-messenger astronomy~\cite{Heinze:2019jou, AlvesBatista:2019rhs, Fang:2016hop}. This also makes it fundamental to any tomographic analyses of source distributions \cite{Globus:2007bi, Ding:2021emg}.

Substantial efforts of the \ac{UHECR} community are thus focused on the extension of measurements of the mass composition to extreme energies and reduction of the uncertainties in the description of hadronic interactions. Due to the $10\%-15\%$ duty cycle of \acp{FD}, data on the depth of shower maximum (\xmax{}) at $E>40$~EeV are scarce. There are plans to directly address this in the future with large-aperture \ac{FD}-based experiments such as \ac{POEMMA} (see \cref{sec:POEMMA})~\cite{Olinto:2019euf, Anchordoqui:2019omw}, and new technologies like those developed for the \ac{CRAFFT} project~\cite{Tameda:2019wmj} and the \ac{FAST}~\cite{Fujii:2015dra, Malacari:2019uqw}, all of which have concrete designs and have deployed prototype detectors. However, an immediate possibility to address this is to use the data collected by \ac{SD} and \ac{RD} arrays for mass composition studies as their nearly 100\,\% duty cycle would increase the available statistics by around an order of magnitude with respect to those currently available from \ac{FD} measurements. The main obstacle here is that due to deficiencies in the modern hadronic generators~\cite{PierreAuger:2014zay,  PierreAuger:2014ucz,  PierreAuger:2016nfk, PierreAuger:2017tlx, PierreAuger:2020gxz, PierreAuger:2021qsd} that have to be fixed at energies and phase spaces not accessible to particle accelerators, the \ac{SD} data are not even always bracketed by the predictions of these generators for protons and iron nuclei. The solution of this problem can be achieved using data on \acp{UHECR} and analyses with a reduced sensitivity to uncertainties in the description of hadronic interactions. To address this need for fine grained shower component information with sufficient statistics, as well as to extend the studies of cosmic rays at the highest energies $>40$~EeV, upgrades to current generation observatories are under way (AugerPrime~\cite{PierreAuger:2016qzd, Taboada:2020spx, Pont:2021pwd} and \TAxFour~\cite{Kido:2019enj}) and future large-scale observatories such as GCOS~\cite{gcos_uhecr2018} are being designed.

\subsubsection{Primary composition: 100\unitspace{}PeV--1\unitspace{}EeV}
\begin{wrapfigure}{r}{0.5\columnwidth}
  \vspace{-3mm}
  \centering
  \includegraphics[width=.5\textwidth]{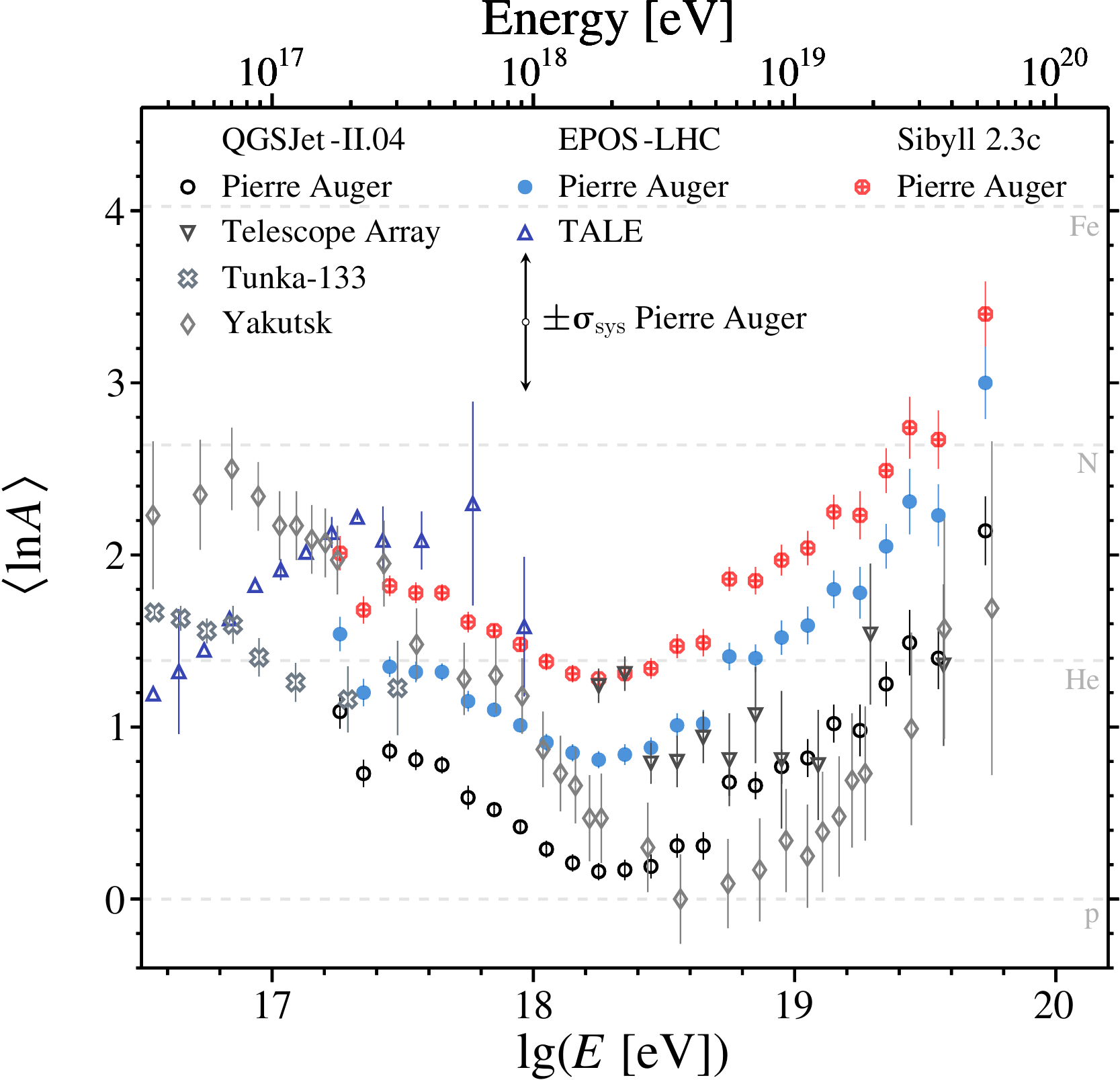}
  \caption{Mean logarithmic mass obtained from the experimental data on \meanXmax using various hadronic models. Pierre Auger Observatory~\cite{Yushkov:2020nhr}, Telescope Array~\cite{TelescopeArray:2018xyi}, \ac{TALE}~\cite{TelescopeArray:2020bfv}, Tunka-133~\cite{TAIGA:2021fgx}, Yakutsk~\cite{Knurenko:2019oil}. }
  \label{fig:LnA}
\end{wrapfigure}
\begin{figure}[!htb]
  \centering
    \includegraphics[width=.8\textwidth]{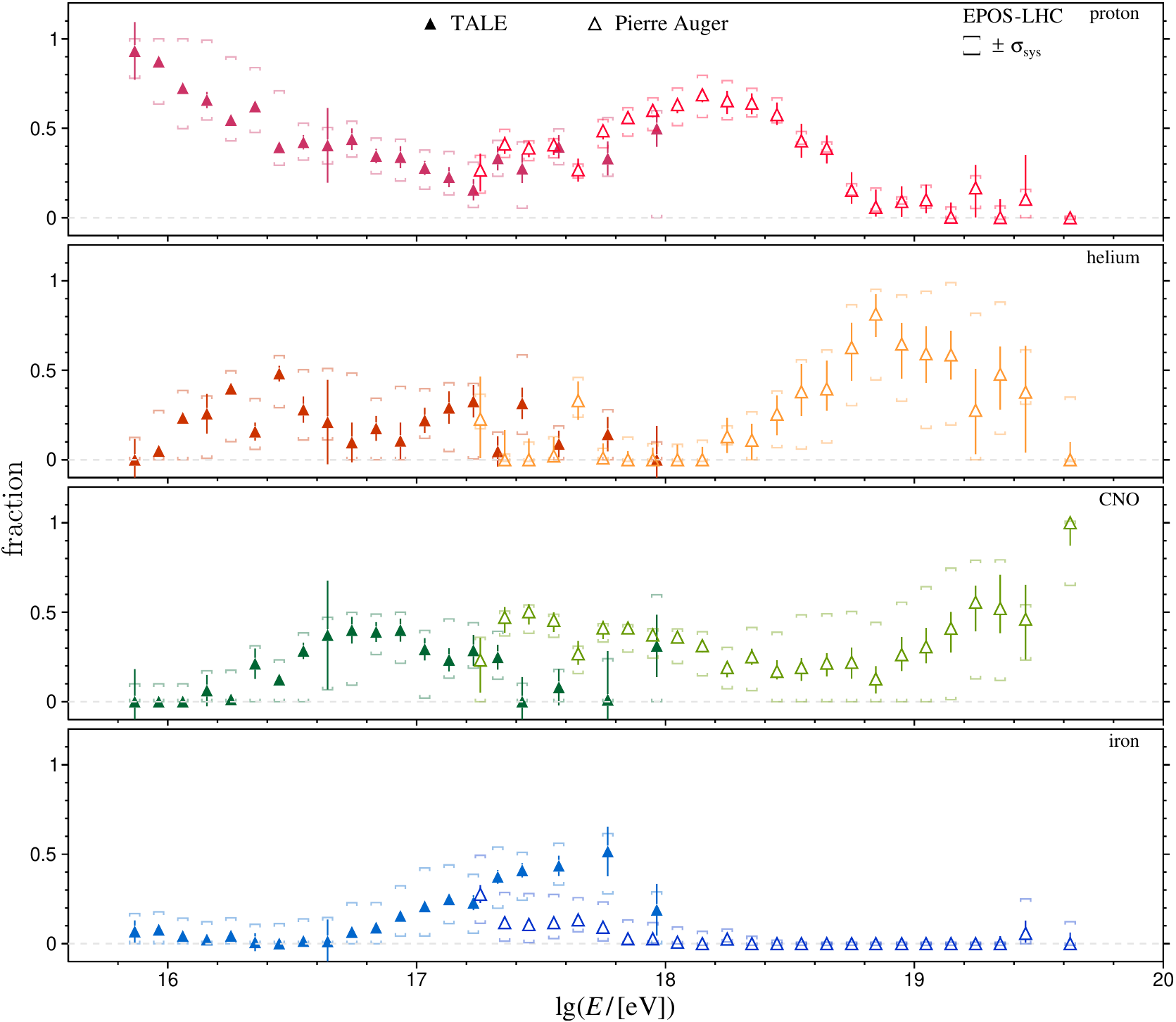}

  \caption{Fractions of primary nuclei from the mass composition fits of \xmax distributions measured at the Telescope Array (\ac{TALE})~\cite{TelescopeArray:2020bfv} and the Pierre Auger Observatory~\cite{Bellido:2017cgf} inferred with \eposlhc.}
  \label{fig:XmaxFractions}
\end{figure}

The transition from galactic to extragalactic \acp{UHECR} is believed to take place in the energy range from 100\,PeV to a few EeV.
Several \ac{CR} detectors measure the primary mass composition in this region using different techniques.
The low-energy enhancements \ac{TALE} \cite{TelescopeArray:2020bfv} of \ac{TA} and \ac{HEAT} \cite{Yushkov:2020nhr} of Auger measure the \xmax{} of \ac{FD} events which reach their points of maximum shower development high in the atmosphere.
The radio arrays \ac{AERA}~\cite{PierreAuger:2021rio}, LOFAR~\cite{Corstanje:2021kik}, Tunka-Rex~\cite{Bezyazeekov:2018yjw}, and Yakutsk-Radio~\cite{Petrov:2020edv} measure \xmax{} using the shape of the radio emission footprint emanating from the electromagnetic part of the shower.
KASCADE-Grande~\cite{KASCADEGrande:2017gtn} used a surface scintillator array measurement and focused on the disentanglement of the muonic and the electromagnetic signal in the detectors from air showers to measure the composition.
IceCube/IceTop~\cite{IceCube:2019hmk} utilize the combination of an ice-Cherenkov tank surface detector and a deep in-ice detector to measure the primary energy and mass composition simultaneously using the electromagnetic/low-energy muonic air shower component from the surface and the high-energy ($\geq 500$ TeV) muon bundles in the deep ice.
KASCADE-Grande and IceCube/IceTop results rely on the comparison to hadronic interaction model data sets to reconstruct the data, which causes a larger systematic uncertainty on the final results. \ac{TALE} and \ac{HEAT}, on the other hand, directly measure two orthogonal air shower properties (calorimetric energy and shower maximum depth). In all cases, however, the interpretation of the mass-sensitive observables relies on hadronic interaction models. This means that for most astrophysical or particle physics analyses, there is still considerable uncertainty in this energy range due to the models.
Due to conflicting observations, the general behaviour of \meanlnA around 100\,PeV is not yet established well (\cref{fig:LnA}). The data from all \ac{FD} measurements show a change in the composition with energy but there are significant differences in the absolute interpretation and even slopes of \meanlnA. There is a clear difference between the \meanlnA behaviour of \ac{TALE}, Tunka-133, and Yakutsk which might indicate the presence of unaccounted systematic measurement uncertainties in some of these experiments.
If the region of the knee at 3\,PeV is dominated by protons (for discussion of experimental results see~\cite{ARGO-YBJ:2015isx}), the iron-knee is expected to appear at around 80\,PeV as a signature of the end of the galactic component. Above this energy more rigid and lighter extragalactic component is expected to take over. The increase of \meanlnA until around 200\,PeV observed by \ac{TALE} might be already in tension with these expectations. This tension looks even stronger from the detailed information on the energy evolution of individual primary components obtained from the composition fits of \xmax distribution measured with \ac{TALE}. From \cref{fig:XmaxFractions} one can see, that there is no evidence for the iron knee below 100\,PeV in the \ac{TALE} data and that the observed spectrum of iron nuclei is harder compared to the spectrum of protons.

The situation with the \ac{SD} measurements is no better. Features compatible with the transition from a softer heavy galactic component to a harder light extragalactic component were observed at KASCADE-Grande, where it was found that a steepening in the spectrum of the heavy component at around 80\,PeV is followed by the hardening of the light component at around 120\,PeV~\cite{KASCADEGrande:2011kpw,Apel:2013ura}. Nevertheless, the observations at IceCube/IceTop~\cite{IceCube:2019hmk} are in tension with these findings, which makes the picture of the tail end of the galactic flux inconclusive at present. The onset of the muon deficit in the \ac{MC} simulations (Muon Puzzle) at around the same energies can cause larger systematic uncertainties in the interpretation of \ac{SD} data compared to the data from the \ac{FD}, and can be thus one possible reason for this discrepancy.

Therefore, improvements both in the detection techniques and in the description of hadronic interactions are required for understanding of characteristics of the galactic and extragalactic components, and acceleration mechanisms in sources in this energy range.

\subsubsection{Primary composition above 1\unitspace{}EeV}\smallskip\label{sec:CompAbove1EeV}

\begin{figure}[!htb]
  \includegraphics[height=0.475\textwidth]{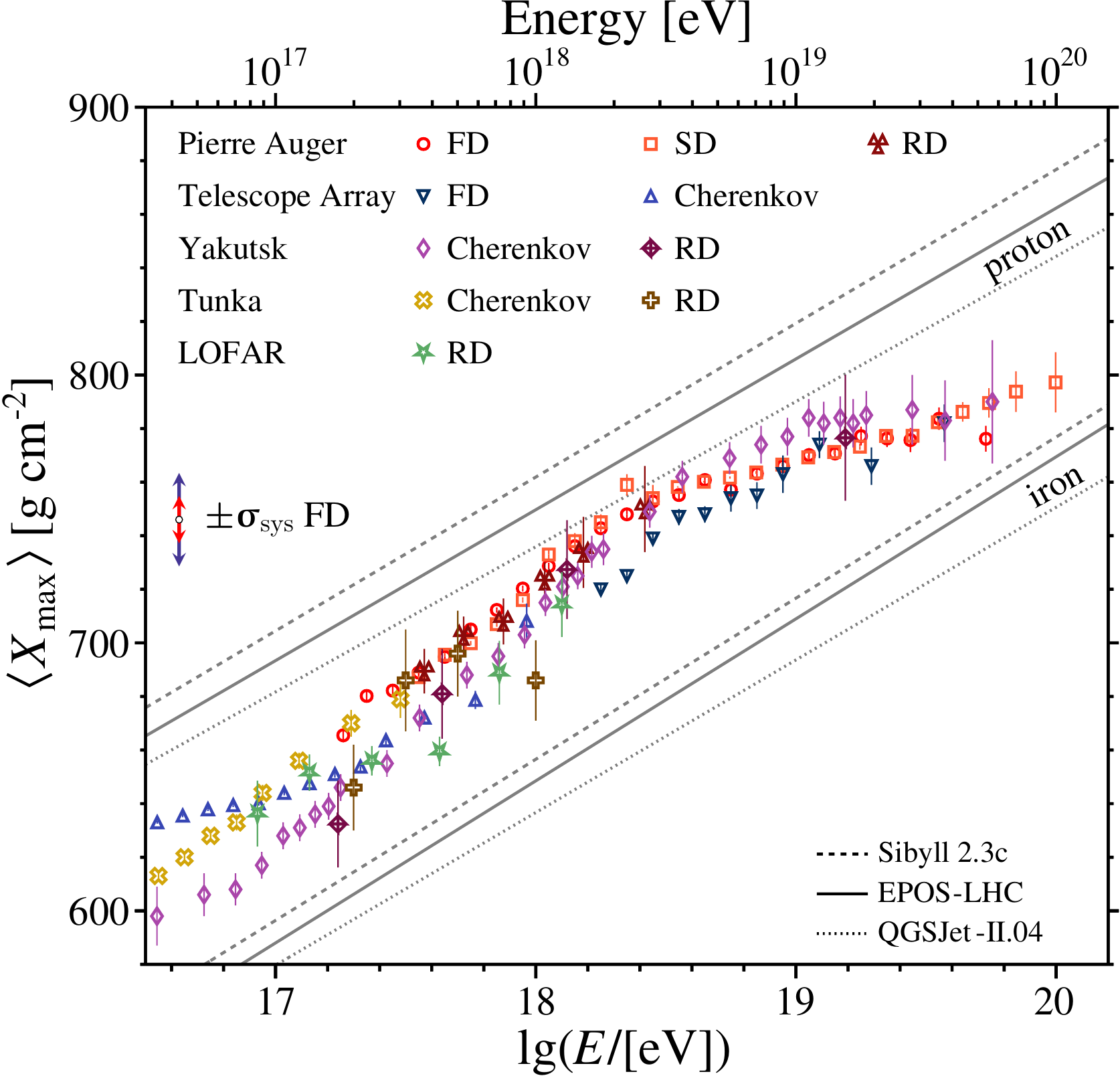}~~\includegraphics[height=0.475\textwidth]{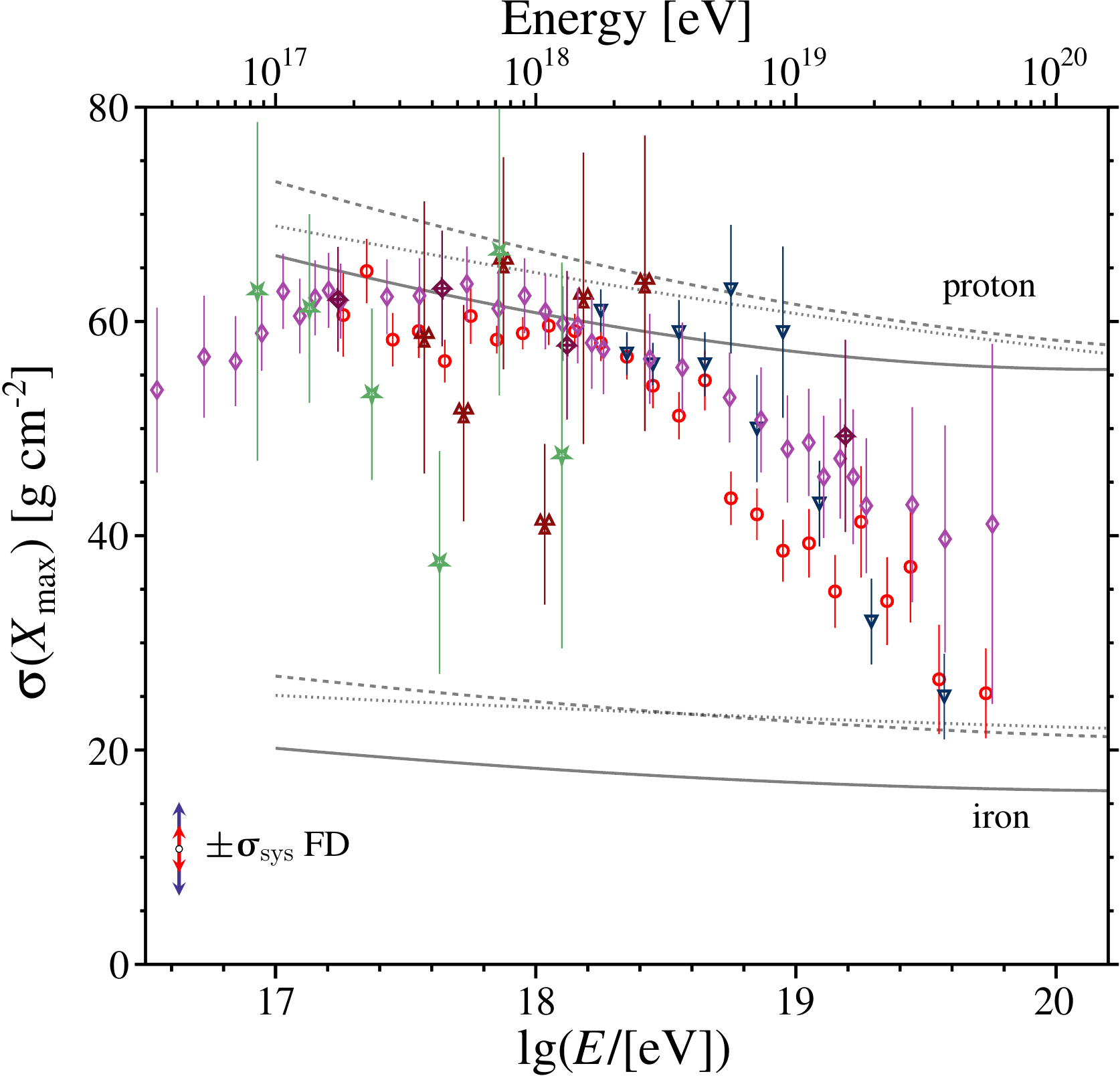}

  \caption{Measurements of \meanXmax (left) and \sigmaXmax (right) compared to the predictions for proton and iron nuclei of the hadronic models \sibyll{2.3c}, \eposlhc and \qgsii. Detection techniques: fluorescence (\ac{FD}), Cherenkov, using time traces in the \ac{SD}, and \ac{RD}. \\
  Pierre Auger Observatory: \ac{FD}~\cite{Yushkov:2020nhr}, \ac{SD}~\cite{ToderoPeixoto:2020rta}, RD (\ac{AERA})~\cite{PierreAuger:2021rio}; Telescope Array: \ac{FD}~\cite{TelescopeArray:2018xyi} (\meanXmax and \sigmaXmax are corrected for reconstruction and detector biases same as was done in Ref.~\cite{AlvesBatista:2019tlv} except here there is no correction of the energy scale), Cherenkov (\ac{TALE})~\cite{TelescopeArray:2020bfv}; Yakutsk: Cherenkov~\cite{Knurenko:2019oil}, \ac{RD}~\cite{Petrov:2020edv}; Tunka: Cherenkov~\cite{TAIGA:2021fgx}, \ac{RD}~\cite{Bezyazeekov:2018yjw}; LOFAR~\cite{Corstanje:2021kik}.
  Systematic uncertainties of the \ac{FD} measurements at $10^{18.5}$~eV are indicated for the Pierre Auger (red arrows) and Telescope Array (blue arrows) data.
}
  \label{fig:CombinedXmax}
\end{figure}

Our current knowledge of the cosmic-ray composition at moderately high energies ($\gtrsim 10^{18}$\,\eV{}) is dominantly inferred from the observation of the development of air showers using the fluorescence and Cherenkov techniques. The corresponding data for the two first \xmax{} moments are shown in \cref{fig:CombinedXmax}. 
At EeV energies, the two state-of-the-art experiments for \acp{UHECR}, Auger and \ac{TA}, report air shower observations\,\cite{PierreAuger:2010ymv, PierreAuger:2014sui, PierreAuger:2021rio, Abbasi:2014sfa,TelescopeArray:2018xyi} that point consistently to a predominantly light composition with a large fraction of primary protons~\cite{Hanlon:2018dhz, deSouza:2017wgx, Yushkov:2019hoh}, as clearly seen in \cref{fig:LnA} which shows the average logarithm of the primary masses of observed \acp{UHECR}. 

Above an energy of $E>2\times10^{18}$~eV the data from both the Pierre Auger Observatory and Yakutsk indicate that the composition of primary cosmic rays is mixed with the mean mass steadily growing due to a gradual depletion of protons and helium nuclei from the primary beam~\cite{PierreAuger:2016qzj, PierreAuger:2014gko, Bellido:2017cgf} as shown in \cref{fig:XmaxFractions}.
Though the published measurements of \xmax{}~\cite{Abbasi:2014sfa, TelescopeArray:2018xyi, Hanlon:2019onl} at \ac{TA}~\cite{TelescopeArray:2008toq} seem to be in tension with this picture, they are compatible with the results of Auger within the current statistical and systematic uncertainties~\cite{Hanlon:2018dhz, deSouza:2017wgx, Yushkov:2019hoh}. 

The above picture is strengthened by an analysis of the collection of apparent elongation rates of northern and southern observatories. An analysis of \xmax{} measurements taken from peer-reviewed publications of the Fly's Eye, \ac{HiRes}, Telescope Array, Yakutsk, and Pierre Auger Observatories, shows that statistically there is generally good agreement in trends of the elongation rate above 1\,EeV between the northern and southern skies. 
Nearly all published data are consistent with the description of having a steep rate up to an apparent change to a flatter rate in the vicinity of 3\,EeV. This transition supports the growing evidence of a transition from a lighter proton dominated composition to a heavier composition as energy climbs \cite{Sokolsky:2021xto, Watson:2021rfb} in both hemispheres.

At energies above the suppression ($E>10^{19.6}$\,\eV{}), the total number of detected events with a high-precision measurement with \acp{FD} is less than a hundred~\cite{Hanlon:2019onl, Yushkov:2020nhr} and therefore the composition at these energies is still an open question. However, with a reliable identification of the nature of the \acp{UHECR} at these energies a more precise determination of the parameters of astrophysical models, composition enhanced anisotropy studies, tests of the hadronic interactions at the energies way beyond human-made accelerators, searches of signatures of \ac{LIV}, and improved estimations of the photon and neutrino fluxes will become possible.

These statistical limitations will be overcome by observing \acp{UHECR} with the larger exposure of the upgraded current and next generation detectors. The first step in this direction was made at the Pierre Auger Observatory where the information on arrival times of particles in the \ac{SD} stations was calibrated with \ac{FD} \xmax~\cite{PierreAuger:2017tlx, ToderoPeixoto:2020rta}. This way the measurements of \meanXmax could be extended up to 100~EeV with 237 events available for $E>50$~EeV~(see \cref{fig:CombinedXmax}); still, larger statistics are required to confirm whether the trend towards heavier composition and increasing beam purity continues for these energies. Detailed information on the distribution of nuclear masses using the SD-FD \xmax{} calibration can be obtained using other novel methods like those based on the deep learning~\cite{PierreAuger:2021xnt}. 
The determination of the cosmic-ray composition directly from the \ac{SD} variables suffers from relatively large theoretical uncertainties in the hadronic interaction models used to interpret air shower data, see e.g., Refs.~\cite{Ulrich:2010rg, Engel:2011zzb, Abbasi:2018bzb} and \cref{sec:tensions}. These systematic effects
can be reduced by further laboratory measurement of multiparticle production in hadronic interactions as proposed in Refs.~\cite{LoI_EFmeetsCF21,LoI_WHISP21}. At the same time, a high-statistics observation of cosmic rays at UHE can help significantly to resolve the mass vs.\ interaction ambiguity.

Moreover, both the Pierre Auger and Telescope Array Collaborations observe that simulation-based composition analyses using the surface detector indicate a heavier composition than determined by fluorescence observations (see \cref{fig:lnA_SD_FD})~\cite{PierreAuger:2017tlx, TelescopeArray:2018bep}. This was thought to mainly derive from uncertainties in muon production; however, recent studies\footnote{Machine learning methods cross calibrated with \acp{FD} \cite{PierreAuger:2021fkf} and mass/energy/arrival direction combined fit results~\cite{PierreAuger:2021mmt, PierreAuger:2021oxo} both suggest an offset between the \xmax{} scale predicted models and that seen in \ac{UHECR} observations.} indicate that an energy independent shift of the \xmax\ scale, on the order of 20--30\,\gcm, could also be needed. This is in good agreement with studies that estimate the influence of hadronic models on the shower maximum and the signals in the surface detector~\cite{PierreAuger:2016nfk}.
The extent to which these observations are related to the muon deficit, and the \xmax{} scale, of simulations must be determined in further studies.

%

\subsubsection{The self-consistency of hadronic interaction models}\smallskip
\label{sec:tensions}

\begin{figure}[!htb]
  \centering
  \includegraphics[width=0.7\textwidth]{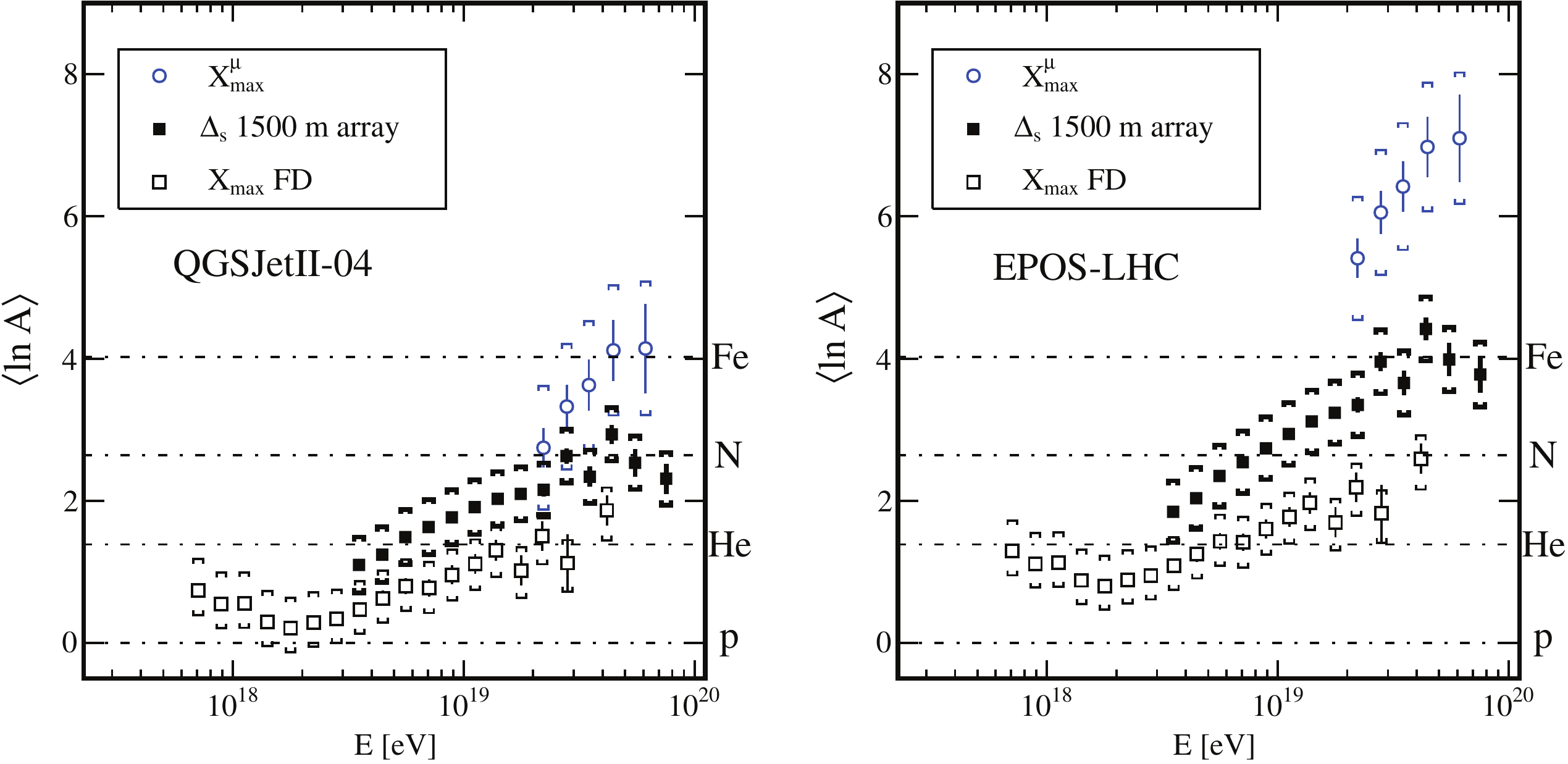}\\[0.2cm]
  \includegraphics[width=0.59\textwidth]{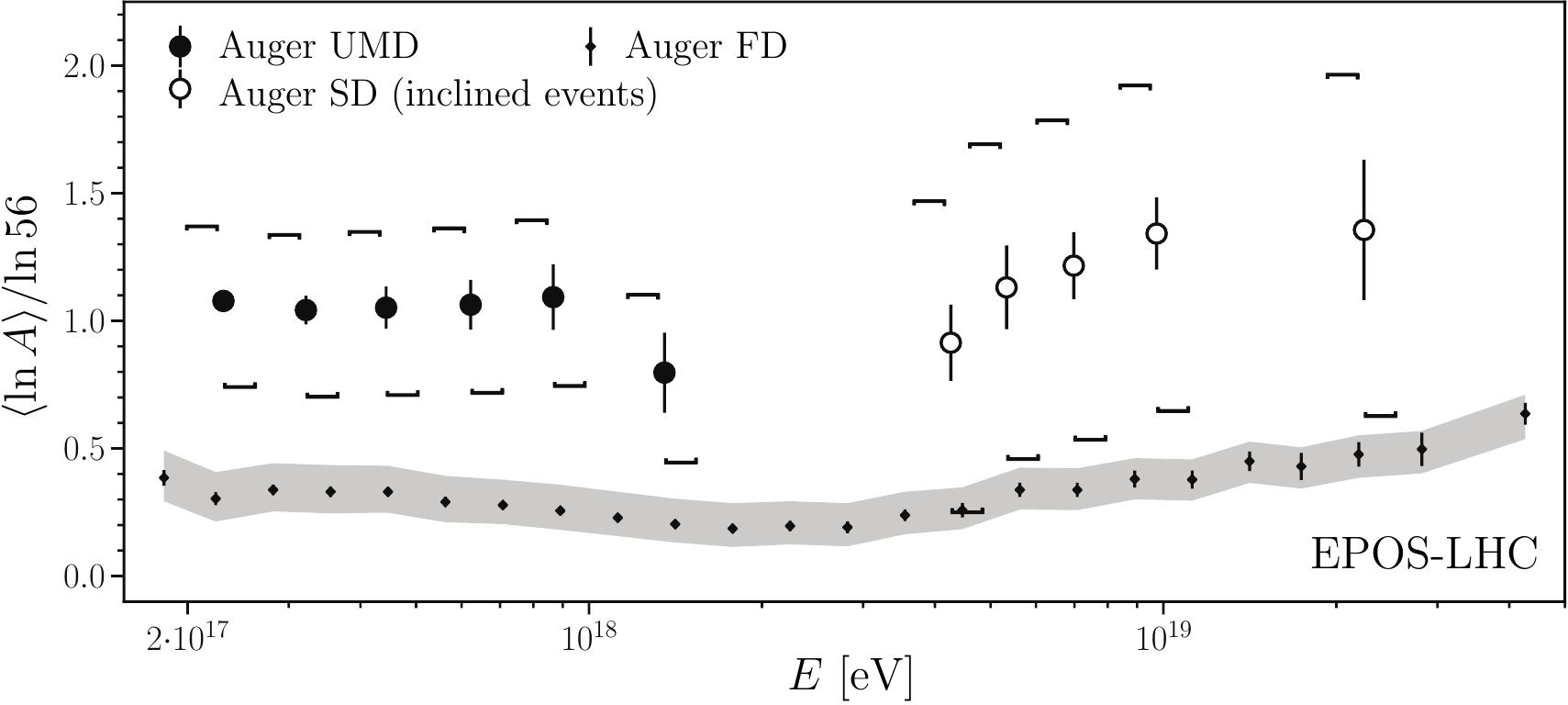}
  \includegraphics[width=0.39\textwidth]{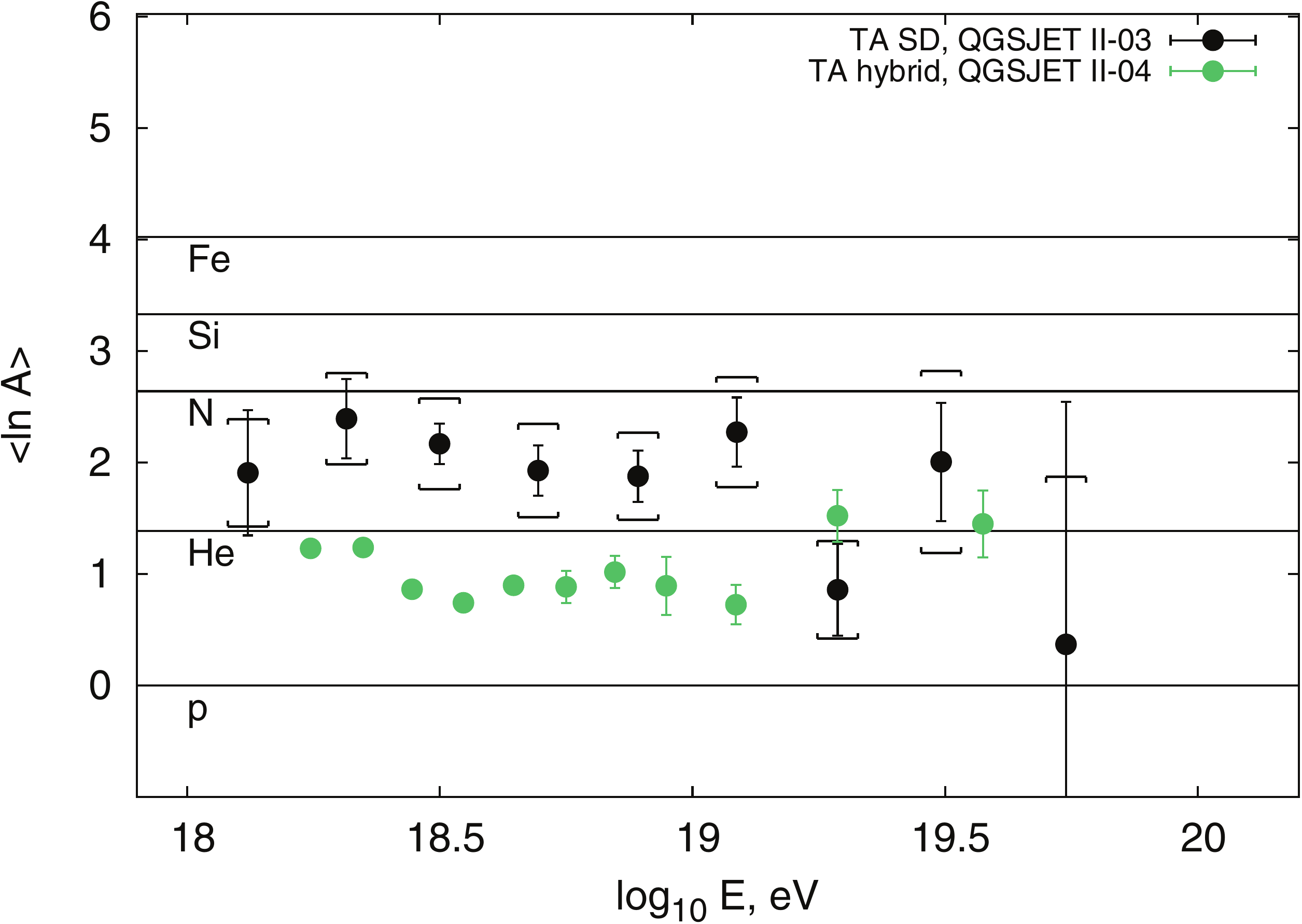}
  \caption{$\langle \text{ln}A \rangle$ inferred from \ac{FD} \meanXmax and pseudo-$\langle \text{ln}A \rangle$ taken from \nmu{}, $X_{\mu \text{max}}$ or Delta.\\
    \ac{SD} \meanlnA: top panels~---~Auger measurements
    of the muon production depth ($\xmax^\mu$)~\cite{PierreAuger:2014zay} and results from Delta
    method ($\Delta_{\rm s}$)~\cite{PierreAuger:2017tlx}; bottom left~---~Auger muon
    density from \ac{UMD}~\cite{PierreAuger:2020gxz} and muon number in inclined \ac{SD}
    events~\cite{PierreAuger:2014ucz}; bottom right~---~Telescope Array analysis of
    complex of \ac{SD} data~\cite{TelescopeArray:2018bep}.\\
    \ac{FD} \meanlnA: Auger~\cite{Porcelli:2015jli} (top panels) and \cite{Bellido:2017cgf}
  (bottom left); \ac{TA}~\cite{TelescopeArray:2018xyi}.}
    \label{fig:lnA_SD_FD}
    \vspace{-3mm}
\end{figure}

\ac{SD} measurements run nearly 100\% of the time and require rather simple event selection criteria, meaning they can offer around an order of magnitude more data than measurements from \acp{FD}. 
However, due to the lack of the accelerator data relevant for the description of \ac{UHECR} interactions, current inaccuracies in the modeling of high-energy nuclear collisions remain relatively large.
As a result the mass compositions inferred from \ac{SD} measurements with the current hadronic models often turn out to be outside the expectations of any realistic astrophysical scenarios. 
Being inconsistent as well with \ac{FD} results (see \cref{fig:lnA_SD_FD}), the absolute values of \meanlnA from the \ac{SD} data can currently be only used for describing the trends in the changes of the mass compositions with energy which are found to be very similar to those from the \ac{FD} data.

The discrepancy between \ac{SD} and \ac{FD} results is larger for \ac{SD} measurements of the characteristics of the muon component of showers. 
This indicates that the observed differences likely arise due to an inadequate description of the muon production mechanisms in air showers. 
\cref{fig:lnA_SD_FD} shows two such examples, the measurements of the atmospheric depth (\xmumax) at which the production of muons reaches its maximum~\cite{PierreAuger:2014zay} and muon density~\cite{PierreAuger:2020gxz}/muon number~\cite{PierreAuger:2014ucz} at ground (Auger \ac{UMD} and \ac{SD} (inclined events)) (see \cref{sec:AccelSyn} for more detailed discussion of the `Muon Puzzle').
For Auger \ac{SD}~\cite{PierreAuger:2016tar, PierreAuger:2017tlx, ToderoPeixoto:2020rta} and \ac{TA} \ac{SD} measurements~\cite{TelescopeArray:2018bep} where a comparison of the EM and muon signals is used, the observed discrepancy with the FD data is smaller.

The calibration of SD data with FD \xmax{} is possible in some cases, but is not a fully satisfactory solution of the problem since the uncertainties in the predictions of \xmax{} (\cref{fig:LnA} and \cref{fig:CombinedXmax}) are non-negligible, amounting to approximately 30\,\gcm. 
Still, this value might not be representative of the full range of possible \xmax{} uncertainties. 
This was indicated by a recent Auger analysis~\cite{PierreAuger:2021xah} of distributions of \xmax and signals in \ac{SD} stations, which suggests that the \xmax scale of the current interaction models could be underestimated and thus may also be partly responsible for the FD-SD \meanlnA discrepancies.

Multi-hybrid observations, which include data from water-Cherenkov detectors, scintillator surface detectors, underground muon detectors, radio detectors, and \acp{FD}, will provide us with crucial information necessary for reduction of the uncertainties in the description of hadronic interactions. The AugerPrime upgrade will allow for simultaneous observations of showers using all of these detector types, potentially making it possible to consistently determine primary mass composition with each of these detectors independently.

\subsection[Arrival directions]{Arrival directions: The slow emergence of source class candidates}
\label{sec:Anisotropy}
The discovery of the production mechanisms of the highest-energy particles in the Universe and the identification of the astrophysical hosts of the remarkable engines responsible for their acceleration, are the most important and challenging ambitions of multi-messenger Astrophysics. The two essential messengers for this task are \acp{UHECR} and \ac{VHE} astrophysical neutrinos, with energies of PeV and above. \Ac{VHE} neutrinos are likely to be progeny of \acp{UHECR} (see \cref{sec:photonAndNeutrinoProduction}) but whether they are produced in the original \ac{UHECR} source or its environment, and whether only a subset of \ac{UHECR} sources are copious producers of neutrinos, is part of the long list of unknowns. Since there may be multiple mechanisms and sites, a well-balanced observational program in upcoming decades is needed to tease apart the physics and astrophysics of \ac{UHECR} and \ac{VHE} neutrino production.

Fortunately, the virtues and limitations of these two messengers, \acp{UHECR} and \ac{VHE} neutrinos, are highly complementary so that together -- but likely only together -- the mystery of the origin of the highest energy particles in the Universe can be tackled and potentially cracked in the upcoming decade. The \ac{VHE} neutrinos have the virtue of traveling directly to us without deflection or energy loss, apart from red-shifting. However, the directional resolution of \ac{VHE} neutrinos can be relatively poor ($\sim$\,$0.5^\circ$ for track-like events but $\sim$\,$10^\circ$ for cascade-like ones), and at best a few hundreds of astrophysical-candidate events of 0.1\,PeV and higher energy can be expected in the next decade (e.g., the IceCube high-energy starting event sample has 60 events with deposited energy above 60\,TeV from 7.5 years of data \cite{IceCube:2020wum}). Lowering the energy threshold to have more events is not a solution because then the flux is strongly contaminated by atmospheric neutrinos polluting the signal in correlation studies.  Another challenge to finding the sources of the astrophysical neutrinos is essentially Olbers' paradox:  unless individual sources are rare and extremely powerful or transient, individual sources may only contribute zero or one events each and integrating over radial distance averages out the structure, leading to a nearly isotropic arrival direction distribution.   

\acp{UHECR}, by contrast, are blessed by the \ac{GZK} effect: which imposes a horizon to possible sources thanks to energy losses on the cosmic background light. This horizon is longest ($\sim250$\.Mpc) for protons and heavy (i.e., iron) nuclei, but much shorter in between (e.g., $\sim5$\,Mpc for helium and $\sim100$\,Mpc for silicon). The sharp cutoff at highest energies induced by photo-pion production off \ac{CMB} photons, gives way to a more gentle but real decrease with distance for lower energy \acp{UHECR}.  Having only a limited range of source distances contributing to the signal, with a known energy dependence given the mass composition, makes it potentially feasible to identify sources or infer their properties statistically.  On the other hand, magnetic deflections produce time delays which make temporal correlations futile with optical or other emissions in candidate sources.  Discovering sources via spatial correlations of \acp{UHECR} with source candidates is in principle feasible if the charges of individual \acp{UHECR} and the magnetic field of Galaxy are well-enough determined. However, correlation between arrival directions and individual sources is not the only tool.

\subsubsection{Large-scale anisotropies}
\begin{figure} [!ht]
\centering
\begin{tabular}{cc}
\includegraphics[width=0.45\textwidth]{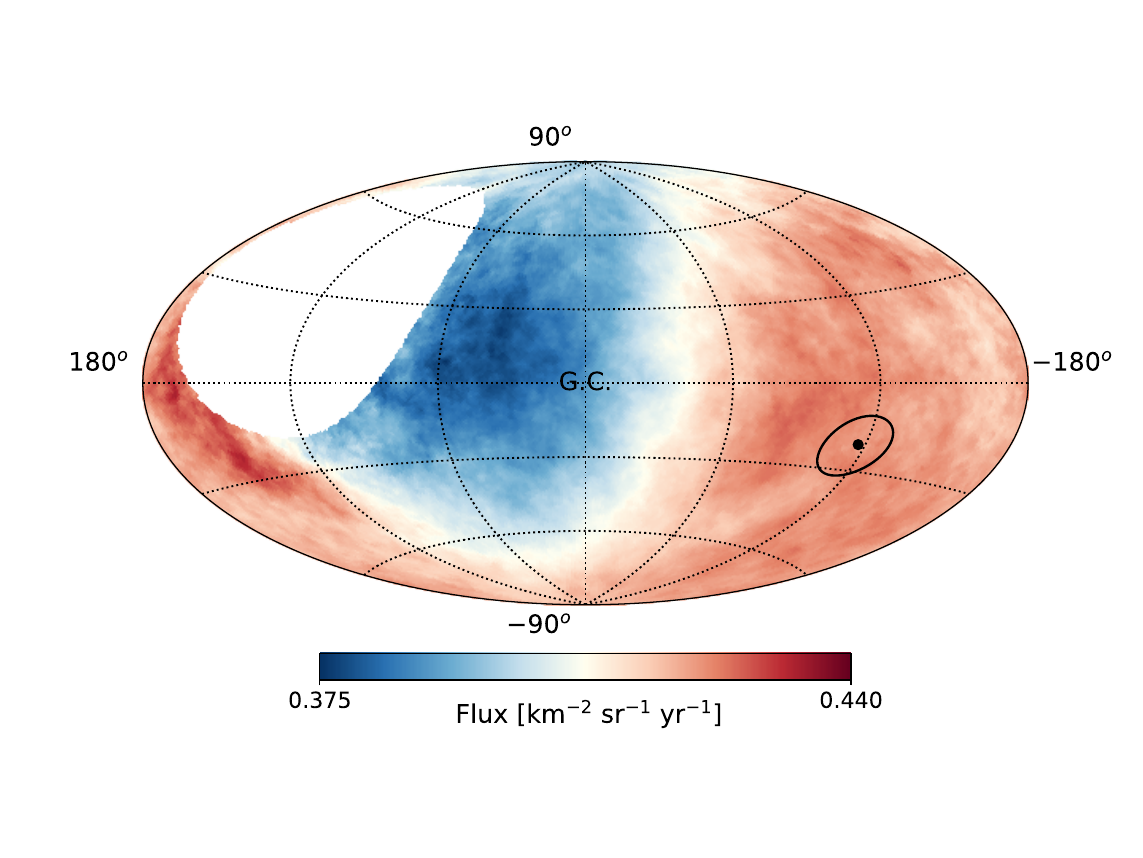} &
\includegraphics[width=0.45\textwidth]{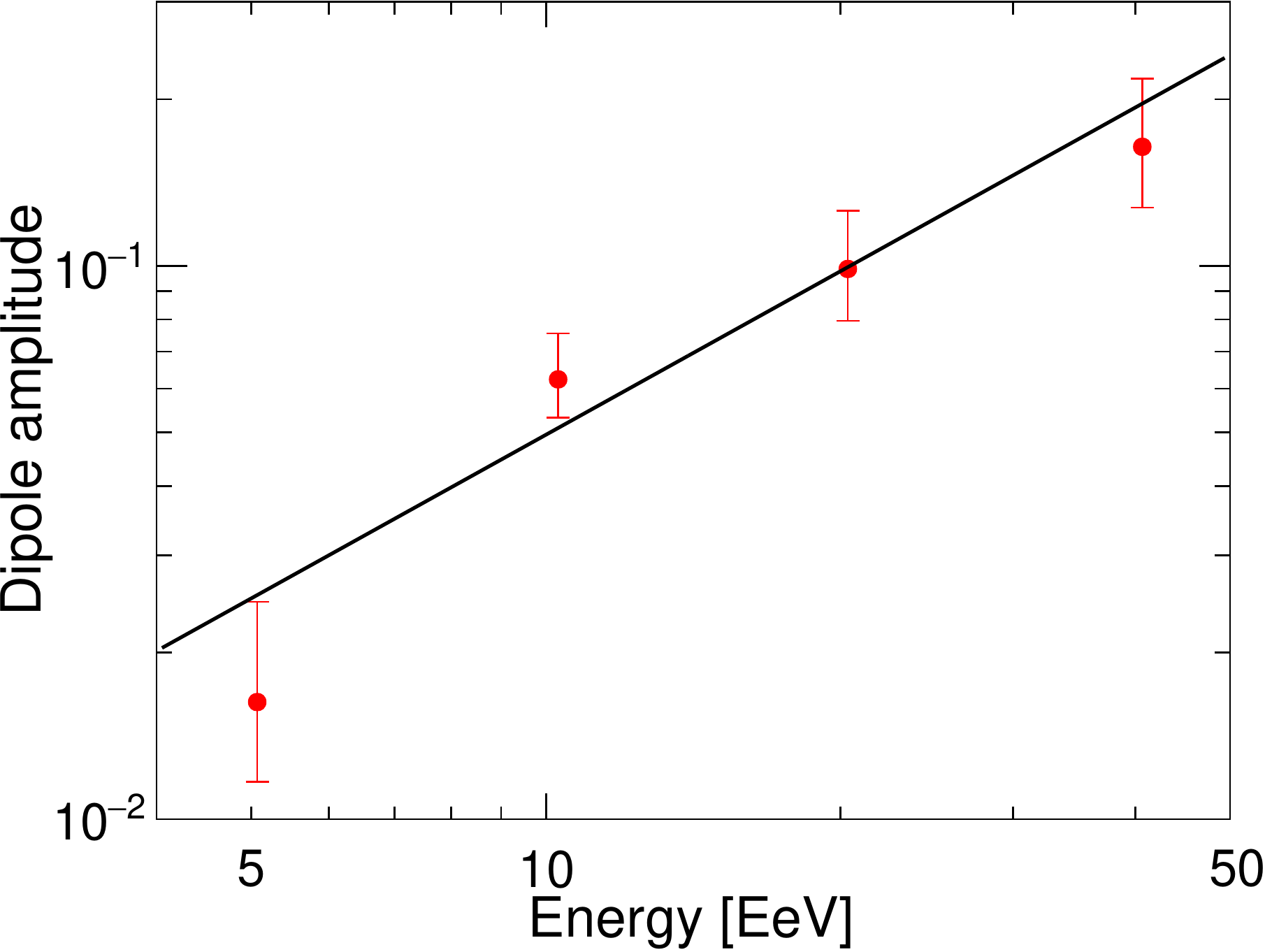}\\
\includegraphics[width=0.45\textwidth]{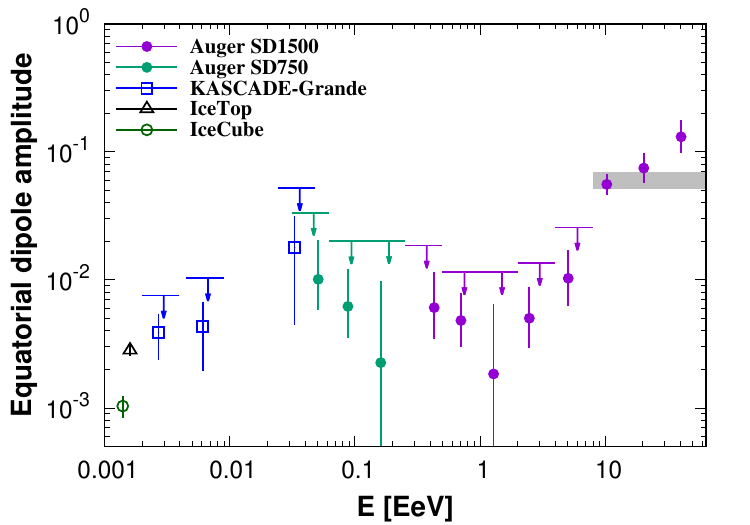} &
\includegraphics[width=0.45\textwidth]{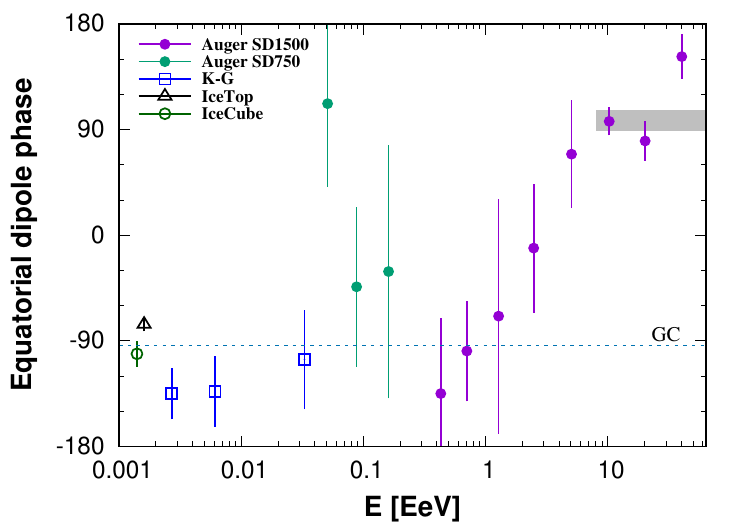}
\end{tabular}
\vspace{-2mm}
\caption {Upper left panel: map showing the cosmic-ray flux detected by the Pierre Auger Observatory above $8$\,EeV, in Galactic coordinates, smoothed with a 45$^\circ$ top-hat function (the Galactic Center, GC, is at the origin). The dot indicates the measured dipole direction and the contour denotes the 68\% confidence level region, from Ref.~\cite{PierreAuger:2021dqp}. Upper right panel: amplitude of the 3D dipole determined in four energy bins above $4\,$EeV with the Auger data set, from Ref.~\cite{PierreAuger:2021dqp}. Lower panels: reconstructed equatorial dipole amplitude (left) and phase (right), published in Ref.~\cite{PierreAuger:2020fbi} by the Pierre Auger Collaboration. The gray bands indicate the amplitude and phase for the energy bin above $8\,$EeV. Results from other experiments are shown for comparison.}
\label{fig:LS}
\end{figure}

Anisotropy in arrival directions is a key ingredient for discovering the sources. To draw robust conclusions, however, a large number of events is needed;  this limited the analyses prior to the advent of the current detectors in the last decade. The Auger and \ac{TA} Collaborations have made significant progress both with their individual data sets and in joint efforts. An extremely important milestone achieved by the Pierre Auger Collaboration is the $> 6 \sigma$ measurement  of a large scale dipole anisotropy above $8\,$EeV~\cite{PierreAuger:2017pzq}, with an amplitude of $d=0.073 ^{+0.011}_{-0.009}$ obtained with the latest published data set \cite{PierreAuger:2021dqp}. The map showing the cosmic-ray flux, smoothed with a 45$^\circ$ top-hat function, is illustrated on the upper left panel of \cref{fig:LS}. Given that the dipole direction is $\sim$\,$115^\circ$ away from the Galactic Center, this is evidence of the extragalactic origin of cosmic rays above this energy threshold. Intriguingly, the dipole direction is not aligned with the \ac{CMB} dipole, or the local matter over-density, or any obvious individual source.

A compelling feature, first published in Ref.~\cite{PierreAuger:2018zqu} and shown in the upper right panel of \cref{fig:LS}, is the growth of the amplitude of the dipole with energy (although the $p$-values for the higher energy bins are not at the $5\sigma$ level due to lower statistics at the highest energies).  This growth is in agreement with the prediction of various models, with particles of higher rigidity being less deflected by the magnetic fields they transverse and with nearby, non-homogeneously-located sources making a larger relative contribution to the flux. 
At lower energies the amplitude of the dipole (\cref{fig:LS}, lower left panel) is smaller and not so significantly established.  However the phases of the equatorial dipole  -- always quicker to produce a robust determination than the amplitude -- line up close to the right ascension of the Galactic center (lower right panel of \cref{fig:LS}).  This suggests that the transition between Galactic and extragalactic origin occurs at energies in-between \cite{PierreAuger:2020fbi}.

Motivated by these results, the Telescope Array Collaboration has searched for a large-scale anisotropy in the northern hemisphere \cite{TelescopeArray:2020cbq}. The events collected during 11 years of operation have been projected onto the equatorial plane and fitted with the dipole distribution. The fit yielded the amplitude of $3.3\pm 1.9\%$ and a phase of $131^\circ \pm 33^\circ$, albeit still with low significance. The \ac{TA} data are compatible with isotropy with a probability of 14\%, and with the dipole found by the Pierre Auger Observatory with a probability of 20\%, small statistics being the main limiting factor of this analysis. 

\begin{figure} [ht]
\centering
\begin{tabular}{cc}
\includegraphics[width=0.45\textwidth]{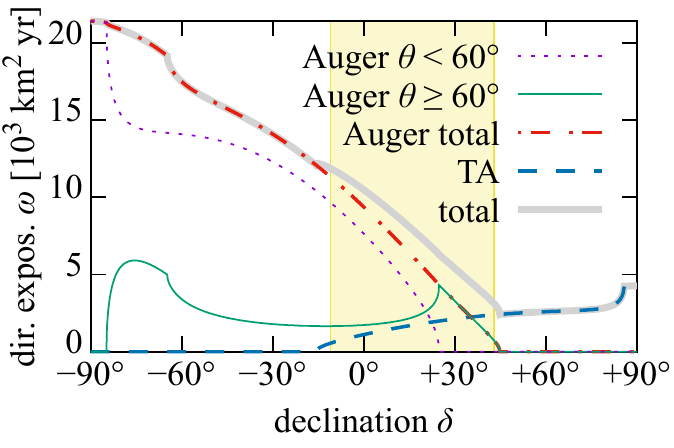} &
\includegraphics[width=0.45\textwidth]{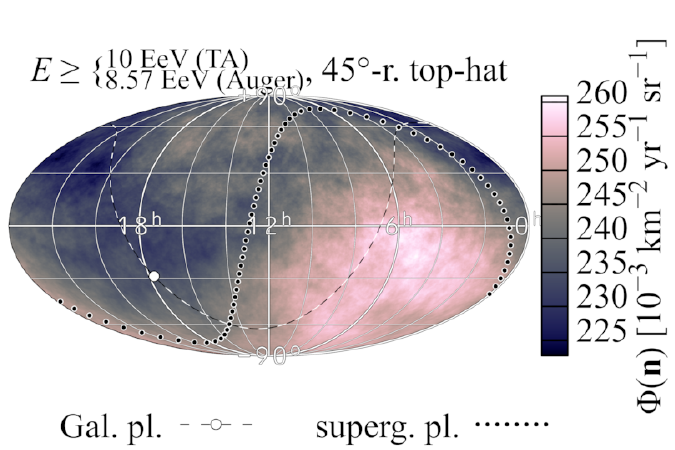}\\
 \end{tabular}
 \vspace{-2mm}
\caption {Left panel: Auger and TA effective exposure; the yellow band shows the common declination band used for the cross-calibration of energy scales. Right panel: map showing the cosmic-ray flux detected by the Pierre Auger Observatory and Telescope Array above $8.57\,$EeV and $10\,$EeV, respectively, in equatorial coordinates, smoothed with a $45^\circ$ top-hat function. From Ref.~\cite{TelescopeArray:2021ygq}.  }
\label{fig:LSjoint}
\end{figure}

The Pierre Auger and Telescope Array Collaborations have joined forces and have worked together on several analyses making use of the fact that the two data sets together have full-sky coverage. The combination of both data sets was done by cross-calibrating the energy scales using the equatorial band where the exposures of both observatories overlap (see \cref{fig:LSjoint}, left panel). The latest results for the search of large scale anisotropies with the combined data sets was presented in Ref.~\cite{TelescopeArray:2021ygq}. Thanks to the full-sky coverage, the dipole and quadrupole moments could be computed without any assumptions about higher order multipoles and with smaller statistical uncertainty. The results are compatible with the Auger-only results.  The combined sky map, smoothed with a $45^\circ$ top-hat function, is shown in the right panel of \cref{fig:LSjoint}. 

An interpretation of the large scale dipolar anisotropy could be the following~\cite{Ding:2021emg}:
the sources of \acp{UHECR} above $8\,$EeV are numerous, such that individual nearby sources do not stand out; rather, the sources form a relatively continuous distribution following the matter density of the Universe.  This inhomogeneous source distribution, in combination with the relatively short \ac{UHECR} horizon due to energy losses, results in the \ac{UHECR} illumination of the Milky Way being anisotropic.  Finally, in the last stage of their journey, the \acp{UHECR} are deflected by the \ac{GMF}.  The matter distribution is known to reasonable fidelity out to hundreds of Mpc -- the relevant distance given the \ac{UHECR} horizon -- and the \ac{GMF} is approximately known based on more than 40,000 Faraday rotation measures of extragalactic sources and Planck synchrotron emission data; the distribution of \ac{UHECR} charges is approximately known from the composition.  The resultant model~\cite{Ding:2021emg} gives a good fit to the observed anisotropy and its evolution with energy shown in \cref{fig:LS}.  Other models such as discussed in Ref.~\cite{Allard:2021ioh} are also able to explain the observed large scale anisotropy and its energy dependence
-- the point being made here is that high quality data with complete sky coverage yields valuable information about the nature of the sources even if the sources are transient or too numerous to allow for individual correlation.  In time, as the knowledge of the \ac{GMF} and the ability to infer composition and hence \ac{UHECR} charge assignments improve, the model can be more and more accurately tested and refined. As confidence in this or another picture builds, it will serve as a complementary constraint on the \ac{GMF} model. {\color{red} Section \ref{subsec:magneticfields} discusses our knowledge of \ac{GMF} and \ac{IGMF} in more details.}

\subsubsection{Small- and intermediate-scale anisotropies}
Searches for small (i.e., few degrees) and intermediate-scale (i.e., few tens of degrees) anisotropies with data from the Pierre Auger Observatory and the Telescope Array have been performed since the very beginning of operation of these detectors, as the technical demands are much less\footnote{Detecting a small-amplitude but large-scale anisotropy demands exquisite control over systematics like seasonal and daily variations; high statistics is not only invaluable directly but also enables detailed studies of systematics.}
than for revealing a large-scale anisotropy, even with a detector of $3000\,$km$^2$ such as the Auger Observatory.  

\begin{figure} [ht]
\centering
\begin{tabular}{cc}
\includegraphics[width=0.45\textwidth]{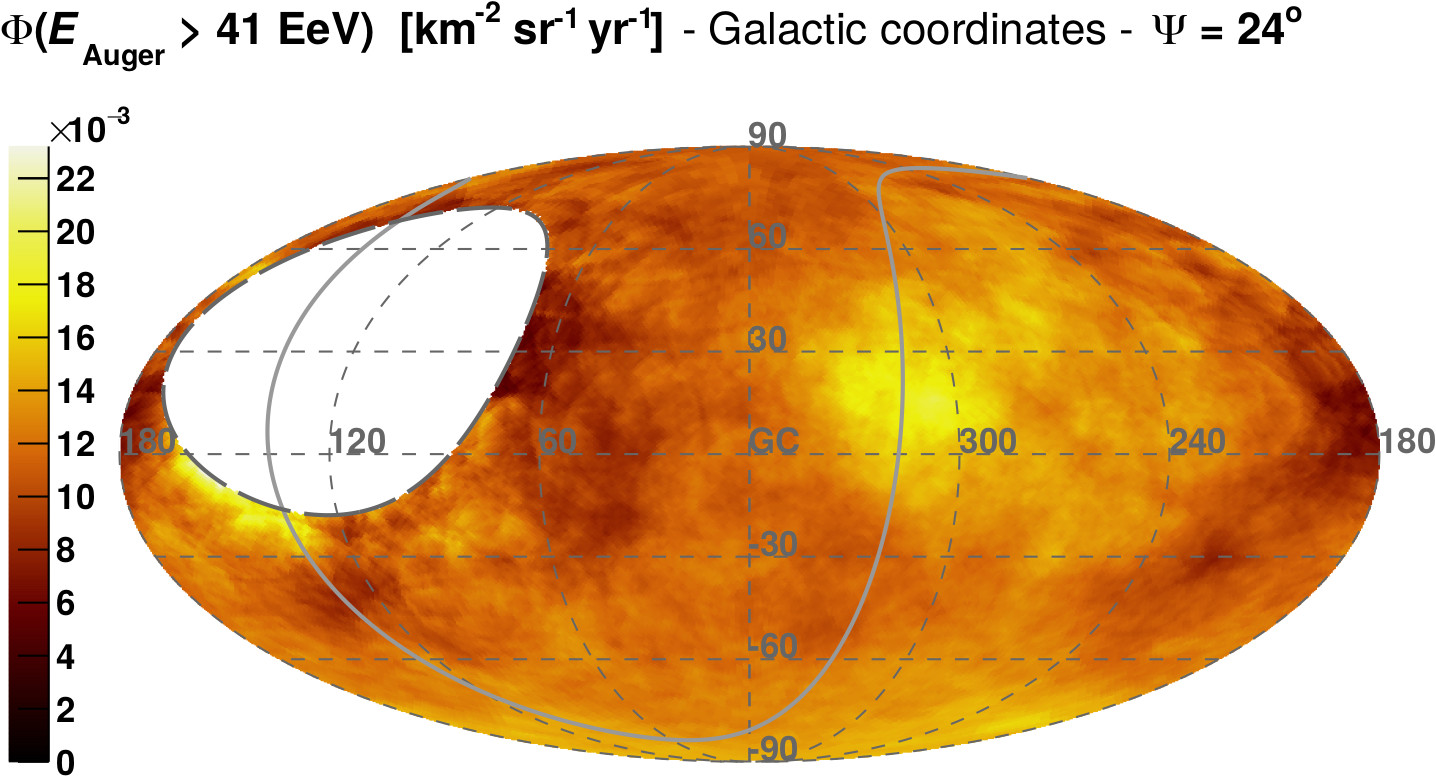} &
\includegraphics[width=0.45\textwidth]{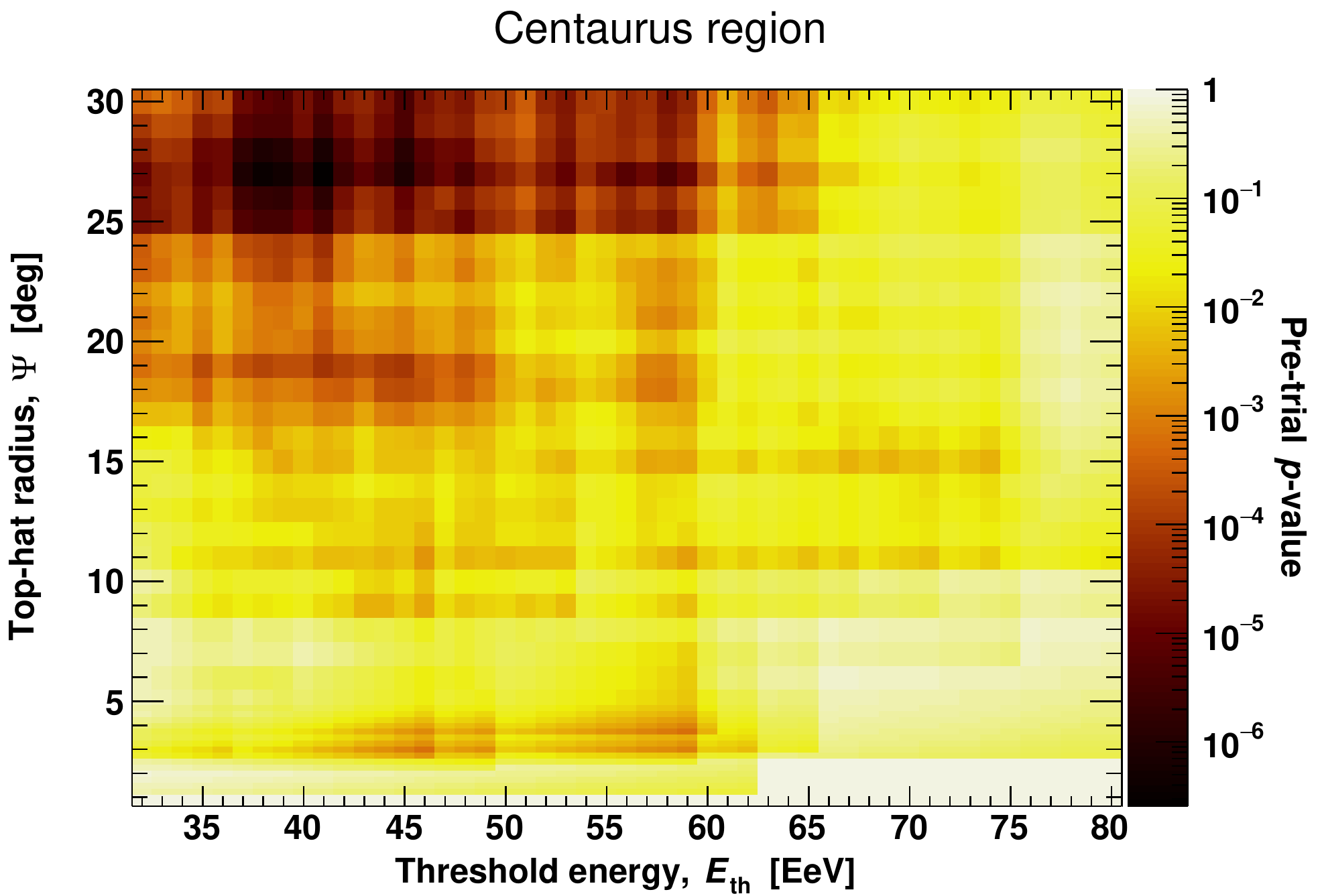}\\
\includegraphics[width=0.45\textwidth]{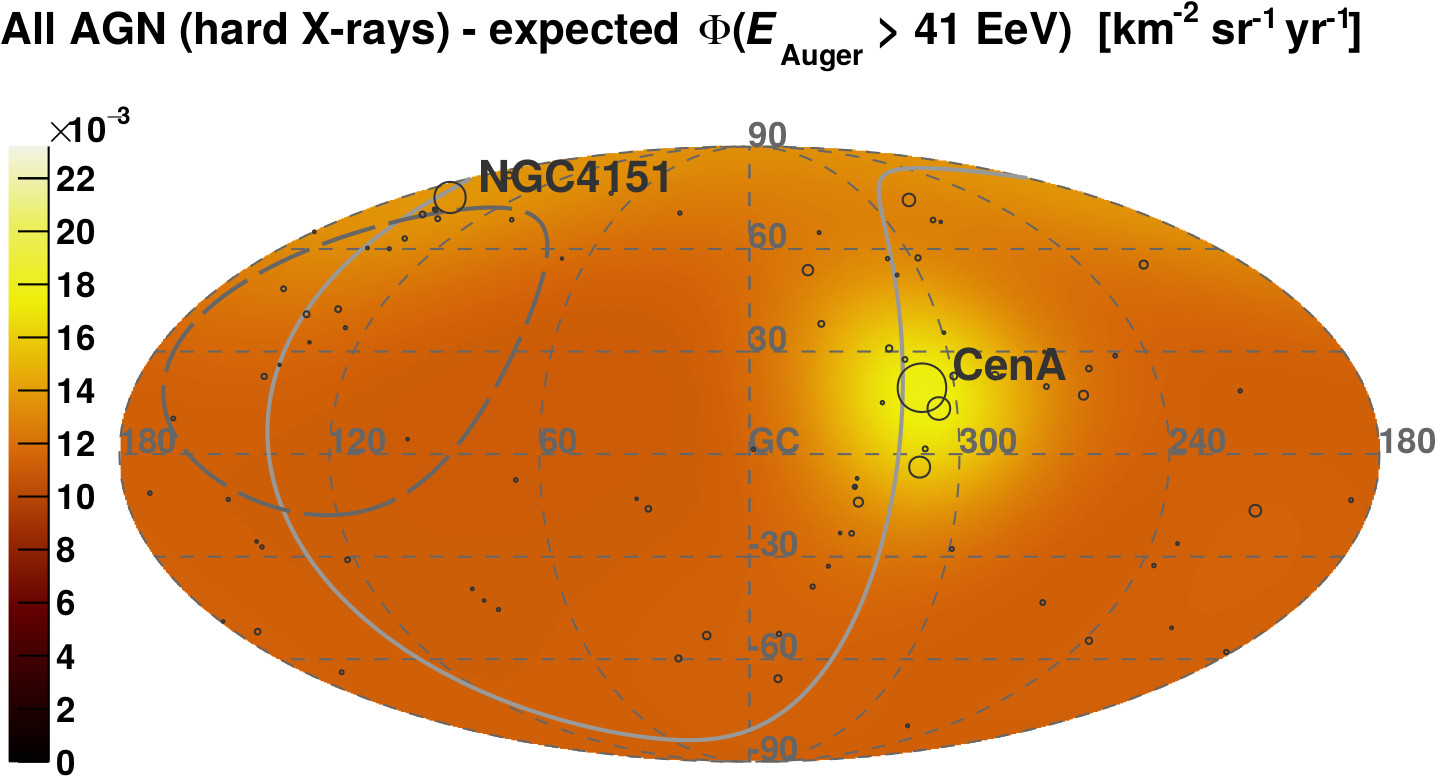} &
\includegraphics[width=0.45\textwidth]{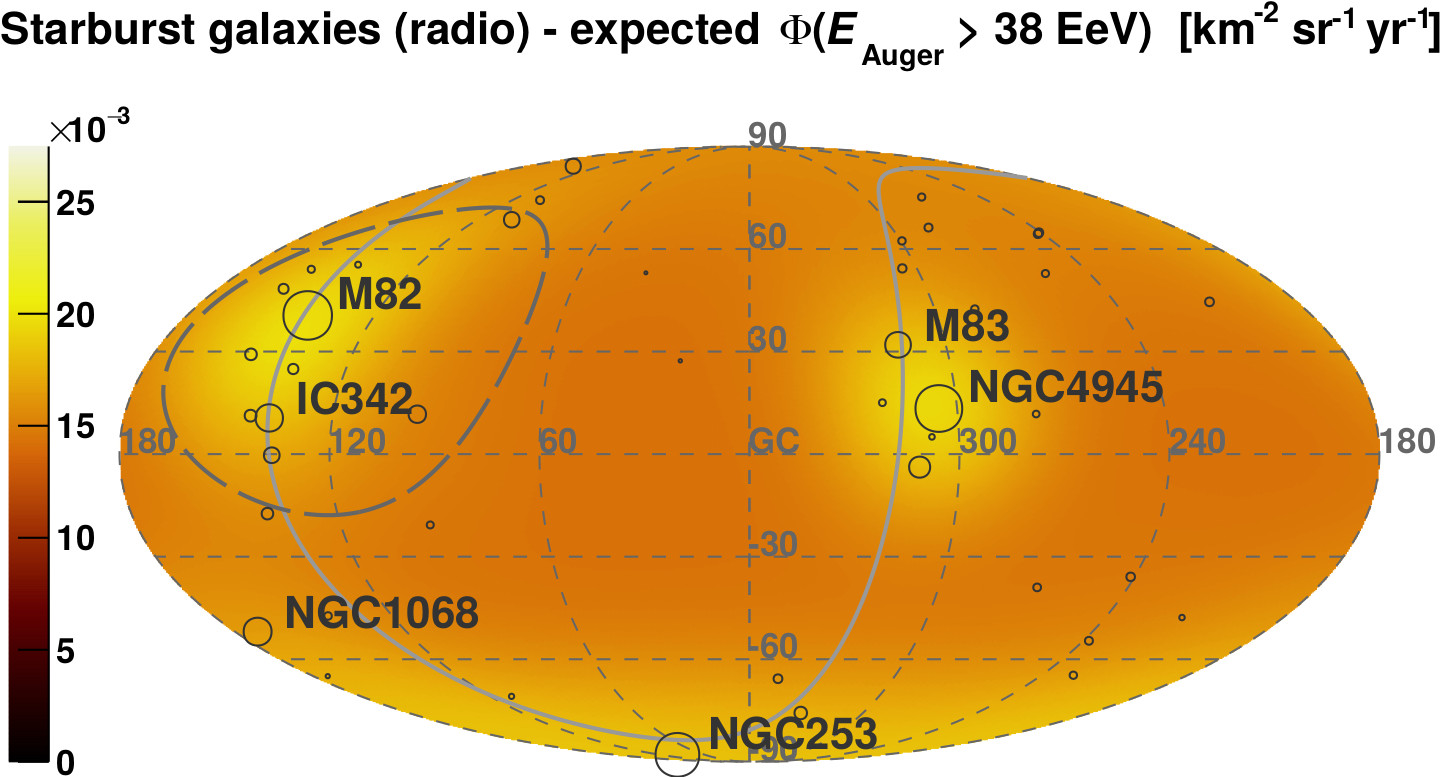}
\end{tabular}
\vspace{-2mm}
\caption {Upper left panel: map showing the \ac{CR} flux detected by the Pierre Auger Observatory above $41\,$EeV, in Galactic coordinates, smoothed with a 24$^\circ$ top-hat function. Upper right panel: Pre-trial $p$-value as a function of the energy threshold and top-hat radius for an overdensity search centered in the Centaurus region. Lower panels: best-fit models of the All \ac{AGN} (left) and starburst galaxies (right) catalogs used in Galactic coordinates. From Ref.~\cite{PierreAuger:2021rfz}.}
\label{fig:SS}
\end{figure}

Even so, demonstrating a statistically-ironclad intermediate scale anisotropy is very challenging. Even after 17 years of operation and $120{,}000\; \textrm{km}^2\;\textrm{sr}\;\textrm{yr}$ of accumulated exposure, the small and intermediate angular scale signals in the current Auger data set do not reach the $5\sigma$ level. However, exciting hints of correlation have been confirmed in multiple analyses, e.g., Refs.~\cite{PierreAuger:2014yba,PierreAuger:2018qvk, PierreAuger:2021rfz}. With more than 1200 events above $41\,$EeV recorded between 2004 and the end of 2020, a $3.9\sigma$ excess was found in the region centered on Centaurus A, at an angular scale of $27^\circ$. The CR flux map above this energy threshold is shown in the upper left panel of \cref{fig:SS}, while the results of the scan in energy threshold and angular size of the search window are illustrated in the upper right panel. 

The Centaurus region excess also drives the hint of correlations found by the Auger Collaboration with a more elaborate likelihood analysis searching for correlations with catalogs of potential sources.  Several catalogs were tested: 2MRS to map nearby matter, Swift-BAT to test all \ac{AGN}, Fermi $\gamma$-AGN to test jetted active galaxies and a catalog of starburst galaxies selected using radio emission \cite{PierreAuger:2018qvk}. 
The starburst galaxies catalog shows the highest significance compared to an isotropic flux of cosmic rays,  with a post-trial significance at $4\sigma$ \cite{PierreAuger:2021rfz}, for an energy threshold of $38\,$EeV and a best-fit equivalent top-hat radius of $\approx 25^\circ$. Other catalogs show significances of $\sim$\,$3.2\sigma$ under the same analysis. The most prominent source in the two \ac{AGN} catalogs is indeed Cen-A, while the starburst galaxies NGC4945 and M83 are within few degrees of Cen-A itself.  The starburst model also benefits from one prominent source candidate, NGC253, being close to the southern Galactic pole where a \textit{warm spot} of Auger events is found. For more details, see the lower panels of \cref{fig:SS} and Ref.~\cite{PierreAuger:2018qvk, PierreAuger:2021rfz}. 

\begin{figure}[ht]
\centering
\includegraphics[width=1.0\textwidth]{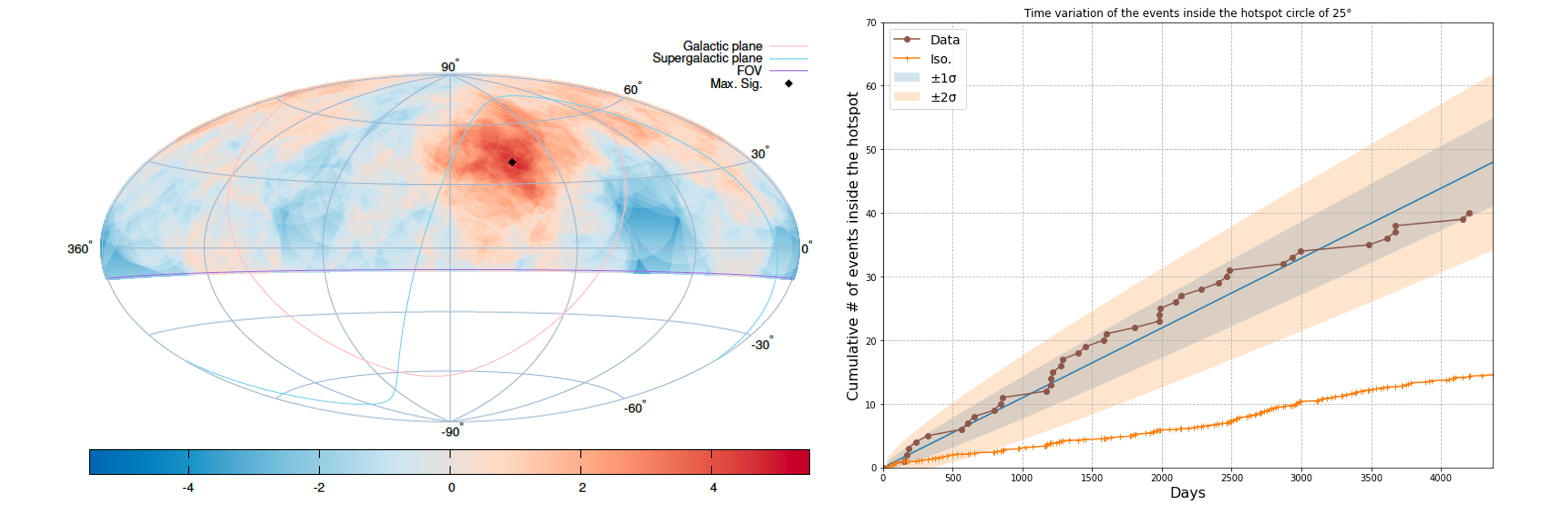} 
\caption { Left panel: The sky map of the Li-Ma significance of the cosmic ray excess observed by the Telescope Array in the circle of $25^\circ$ radius. Black dot shows the position of the most significant excess (Hammer projection, Equatorial coordinates). Right panel: brown dots show evolution of the cumulative number of observed events inside the hot spot region with time. Orange crosses indicate that of isotropic background events. The bands show $\pm 1\sigma$ and $\pm 2\sigma$ deviations from a linear increase rate.
From Ref.~\cite{Kim:2021Aj}.}
\label{fig:hotspot}
\end{figure}

The Telescope Array Collaboration has reported a similar intermediate-scale excess in the northern sky \cite{TelescopeArray:2014tsd}. Their blind search for a cosmic ray excess in a moving window of $20^\circ$ revealed an excess of events (the hot spot) in the 5 year data set with energies above 57~EeV in the direction of 
${\rm R.A.} = 146.7^\circ$, ${\rm Dec.} = 43.2^\circ$, with a significance of $3.4\sigma$ when penalized for the search trials. 
Since the initial study the data set at these energies has more than doubled, but the statistical significance has remained about the same ($3.2\sigma$)~\cite{Kim:2021Aj}. The results of the updated analysis are shown in \cref{fig:hotspot} together with the time evolution of the excess. Interestingly, a slightly less significant excess has also been found in the most recent TA data set at $E \geq 25$~EeV \cite{TelescopeArray:2021dfb} which coincides in position with the Perseus--Pisces supercluster, whose center in equatorial coordinates is ($20.9^\circ, 27.9^\circ$).
 
Joining efforts, the Pierre Auger and Telescope Array Collaborations, have performed a search for intermediate scale anisotropies, testing correlations with two galaxy catalogs, 2MASS and starburst galaxies \cite{TelescopeArray:2021gxg}. The result for the catalog of starburst galaxies is mildly stronger than the Auger-only result, with a post-trial significance of $4.2\sigma$.

The Pierre Auger and Telescope Array Collaborations have also performed searches for correlations between the arrival directions of \acp{UHECR} and neutrino candidates detected by the IceCube and ANTARES Collaborations. The potentially interesting result with astrophysical neutrino candidates first reported in Refs.~\cite{IceCube:2015xib,IceCube:2015afa}, was not confirmed with more statistics \cite{IceCube:2017qeh,ANTARES:2019ufk,Barbano:2020scg}.

The Auger Collaboration has also performed searches for neutron excesses. Similarly to neutrinos, neutrons are expected to be produced by charged \ac{UHECR} interactions near the sources.  Due to the huge Lorentz-boost factor at \ac{UHECR} energies, they can reach Earth as neutrons if produced within the Galaxy. Neutron-induced showers are indistinguishable from proton-induced ones, but since neutrons travel undeflected they should produce, if present, excesses on very small angular scales (of the order of the angular resolution of the Observatory). Both blind searches \cite{PierreAuger:2012dqz} and ones targeted towards interesting Galactic candidate sources~\cite{PierreAuger:2014tey} have been performed, but no significant excesses have appeared so far. Upper limits on the fluxes of neutrons from different classes of sources thereby improve limits on the \ac{UHECR} emissions from Galactic sources.

\subsection[The search for neutral particles]{The search for neutral particles: Bringing the bigger picture into focus}
\label{sec:NeutralParticles}
The spectacular discovery of the coalescence of two neutron stars with gravitational waves and with practically all bands of the electromagnetic spectrum, from radio waves to very high energy gamma rays, has brought multi-messenger astronomy to the forefront of Physical Science~\cite{LIGOScientific:2017ync}. The results from this transient event have shown that combining the detection of particles or radiation of completely different nature from the same objects, great leaps in understanding of the Universe can be made~\cite{LIGOScientific:2017vwq, LIGOScientific:2017adf, Drout:2017ijr, Metzger:2017wot}. 

Gravitational waves and electromagnetic radiation travel in straight lines at the speed of light and can be combined to search for correlations with given objects in space or with given events in time (transients). Gravitational and electromagnetic radiation in the conventional astronomy bands up to the TeV scale can reach the Earth from cosmological distances. Gamma rays above ($\sim$\,$100$~TeV) are limited to distances below the Mpc scale. Neutrinos keep directional information and are also candidates to be combined with other messengers produced in the same sources or events. They also reach us from cosmological distances and upper limits to the neutrino masses imply negligible time delays with respect to light travel even if they come from the confines of the Universe. Neutrinos are practically the only messengers that sample energies above the PeV which can arrive from arbitrarily large distances. Finally, \ac{UHE} neutrons are also neutral but they can only reach us if produced in our vicinity, within a few times the decay distance, $l_d \sim 6 \times (E/\text{EeV)}$\,kpc, which is proportional to their energy $E$. Such limited ranges implies that neutrons above 100 PeV are only delayed with respect to light by less that one millisecond. The potential of each one of these messengers is large and they all are complementary in multi-messenger astronomy since they sample different distance ranges. These messengers are searched for with dedicated telescopes and observatories. Gravitational waves, \ac{UHE} photons and neutrinos have already played a prominent role in multi-messenger astronomy due to their time correlation with transients. 

Understanding how and where \acp{UHECR} are produced remains one of the oldest and most important questions of particle astrophysics and the study of \acp{UHECR} plays a multiple role in multi-messenger astronomy. 
Firstly, fundamental questions about \ac{UHECR} such as the spectrum, their composition and the sources which are not completely settled, are highly relevant for multi messenger searches because there is a close connection between \acp{UHECR} and the production of neutrinos and photons, messengers that have a high potential and play a crucial role in multi messenger astronomy. These high energy photons and neutrinos are believed to be a direct product of the interactions of \acp{UHECR}. Secondly, the study of the \ac{UHECR} arrival directions, including their interaction with the intervening magnetic fields and their angular correlation with potential sources, converts them in messengers too. Depending on distance these charged particles may keep directional information provided their energy is sufficiently large. Such studies constrain the sources of these \ac{UHE} messengers and the still poorly known galactic and extragalactic magnetic fields. There is a third connection from the detection point of view because UHECR observatories can also be used to search for \ac{UHE} neutral messengers, such as \ac{UHE} photons, neutrinos, and also neutrons, complementing the multi messenger capabilities of other dedicated neutrino and photon observatories.  

\subsubsection{The connection between UHE cosmic-rays, neutrinos, and photons}

The interactions of \acp{UHECR} with matter and/or radiation produces secondary hadrons, mostly pions, which decay to produce photons and neutrinos as secondary particles. If the target population is in the source or its vicinity it leads to {\it astrophysical} photons and neutrinos that point back to their sources. Those neutrinos and photons that make it to Earth without being attenuated are potential messengers that can be correlated with the sources of UHECR providing most valuable information. Inevitably, as the accelerated \acp{UHECR} propagate to Earth, they will interact with the diffuse photon backgrounds that permeate the Universe, mostly the \ac{CMB}, but also radio, infrared and optical backgrounds, leading to a diffuse flux of {\it cosmogenic} photons and neutrinos. The neutrino energy, $E_\nu$, traces the primary \ac{UHECR} energy $E_\nu\simeq 0.05 \cdot E_{\mathrm CR}/A$, where $A$ is the mass number of the cosmic-ray nucleus, and the diffuse neutrino spectrum directly reflects baryon acceleration in the sources. In the case of \ac{UHE} photons there is a further complication because they can have further interactions with the electromagnetic background fields as they propagate to Earth to produce electron-positron pairs that also interact with the magnetic fields dumping their energy in a diffuse flux of lower energy photons in the MeV to GeV range providing a sort of calorimetric measurement of the energy deposit. 

The production of UHE photons and neutrinos is directly related to the acceleration of cosmic rays and the establishment of the UHECR spectrum and its composition is thus crucial for multi-messenger astrophysics. 
Composition measurements at the highest energies have a crucial role in determining the expected astrophysical or cosmogenic photon and neutrino fluxes that should result from interactions of the cosmic rays in the sources or during transport to Earth. This is because the energies of neutral secondaries (neutrinos, photons, and neutrons)
produced by CR interactions with matter will be proportional to $E_{\rm{CR}}/A$ with $E_{\rm CR}$ the primary cosmic-ray energy. As a result, the peak flux of neutral secondaries will shift towards lower energies as the mass of the CR increases. In the case of cosmogenic photons and neutrinos resulting from the \ac{GZK}-effect, i.e., from photo-pion or photo-disintegration interactions of UHECR in the \ac{CMB}, the effect of the \ac{UHECR} mass will be even more dramatic: UHECR nuclei disintegrating at the \ac{GZK}-threshold ($E_{\rm{GZK}} \simeq 5\cdot 10^{19}$\,eV) will produce protons, neutrons, and lighter nuclei of energies $\simeq (E/A)_{\rm{CR}}$, which generally will be below the photo-pion production threshold. As a result, the cosmogenic photon and neutrino fluxes arising from UHECR nuclei will be dramatically reduced as compared to those expected from proton primaries. This is demonstrated e.g., in Ref.~\cite{AlvesBatista:2018zui} where 
composition models resulting from combined fits of the energy spectrum and depth of maximum measurements of the Pierre Auger Observatory~\cite{PierreAuger:2016use} were used as input to CRPropa~\cite{AlvesBatista:2016vpy} simulations. 
The connection works both ways. On the one hand, the mere existence of \acp{UHECR} guarantees fluxes of \ac{UHE} photons and neutrinos and the \ac{UHECR} spectrum and composition will determine the fluxes. On the other, while observed energetic photons from given astrophysical sources can have leptonic or hadronic origin, the detection of a neutrino flux from them provides direct evidence for hadronic acceleration giving information about the \ac{UHECR} sources. 

\paragraph{Astrophysical neutrinos and photons} 

Astrophysical neutrinos, photons and also neutrons (if produced near enough the Earth to reach it) are valuable messengers that can be combined with gravitational wave detection and more conventional astronomy to greatly improve our knowledge about their sources and their dynamics in the case of transients, as happened with the discovery of the neutron star merger event in 2017~\cite{LIGOScientific:2017ync}. 
One of the greatest discoveries of the past ten years is the flux of astrophysical neutrinos discovered by IceCube~\cite{IceCube:2013low, IceCube:2014stg, IceCube:2015qii,  Kowalski:2021oda}. It is most likely of extragalactic origin because of the lack of directional correlation with the Galactic plane. However, individual sources remain unidentified with the possible exception of the BL-Lac blazar TXS 0506+056, from which a neutrino with most probable energy of 290 TeV was detected on 22 September 2017 while the source was in a period of flaring activity in gamma-rays \cite{IceCube:2018dnn}. An excess of neutrino events was also found in archival data in 2014/15 \cite{IceCube:2018cha} although the source was not flaring in gamma-rays during that period. How the astrophysical neutrino flux is connected to the \ac{UHECR} flux and to the diffuse gamma-ray background from unresolved extragalactic sources detected by the {\it Fermi} satellite, are two questions still under investigation, motivated by the fact that approximately the same amount of energy is contained in the three types of particles when integrating their spectra \cite{Ahlers:2018fkn} (see~\cref{fig:UHECR-nu}). 

Neutrino observatories and very high-energy gamma-ray detectors have also contributed to constrain the sources of \ac{UHECR} by providing measurements or limits of the neutrino and photon fluxes that, combined with \ac{UHECR} measurements, constrain \ac{UHECR} sources in what is a genuinely multi-messenger observation. 
Early versions of these approaches are the Waxman-Bahcall bound to the diffuse neutrino flux~\cite{Waxman:1998yy} which was obtained by calculating the maximum neutrino flux that could be produced by accelerated protons interacting and producing pions without overproducing the \ac{UHECR} spectrum. While this calculation was limited because it ignored the possibility of multiple proton interactions in the sources and other technical details~\cite{Mannheim:1998wp}, it represented significant progress. Other similar examples of multi-messenger approaches performed with \ac{UHECR} detectors are limits to diffuse fluxes of \ac{UHE} photons~\cite{Ave:2000nd,PierreAuger:2007hjd,TelescopeArray:2018rbt,PierreAuger:2021mjh} and neutrinos~\cite{PierreAuger:2007vvh,PierreAuger:2015ihf}, that ruled out a family of ``exotic mechanisms" for the production \acp{UHECR}, the \emph{top-down} scenarios. In these models fragmentation of quarks from decays of massive particles produced by topological defects~\cite{Bhattacharjee:1991zm} was the source of the \acp{UHECR} and of neutrino and photon fluxes that exceeded the experimental limits. In some of these examples the double connection with multi-messenger observations is apparent because both the UHECR measurements and the limits to \ac{UHE} neutrinos and photons were obtained with \ac{UHECR} observatories. 

\paragraph{Cosmogenic neutrinos and photons}\label{sec:photonAndNeutrinoProduction}

Because of the strong link between \ac{UHECR}, photons, and neutrinos (see for instance Ref.~\cite{PierreAuger:2019fdm}), a flux of \ac{UHE} cosmogenic neutrinos \cite{Kotera:2010yn,AlvesBatista:2018zui,Heinze:2019jou} and photons \cite{Gelmini:2005wu,Kampert:2011hkm,Muzio:2019leu,Bobrikova:2021kuj} is guaranteed by the detection of \ac{UHECR} beyond the \ac{GZK}-threshold and the existence of the \ac{CMB} and other diffuse extragalactic radiation fields which act as background targets. The processes involved are:\begin{align*}
p+\gamma_{\rm CMB} &\to p+\pi^0 \to p+\gamma \gamma\text{, and}\\ p+\gamma_{\rm CMB} &\to n+\pi^+ \to p+\nu_{e,\mu}.
\end{align*}

The shape and magnitude of the cosmogenic neutrino and photon fluxes are very uncertain, being strongly dependent on the maximum energy and composition of the \ac{UHECR} at the sources and, in case of neutrinos, on the redshift evolution of the potential \ac{UHECR} sources in the Universe. Recent descriptions of the observed \ac{UHECR} spectrum \cite{PierreAuger:2020kuy, PierreAuger:2020qqz} and composition \cite{PierreAuger:2014sui}, indicate that the maximum CR energy observed at Earth could be limited by the accelerators themselves, depending on rigidity as $E^{\mathrm  {max}}_{\mathrm CR}\propto Z$ with $Z$ the charge of the accelerated nuclei \cite{PierreAuger:2016use}. This framework leads to very low cosmogenic neutrino and photon fluxes \cite{AlvesBatista:2018zui, Heinze:2019jou} at \ac{UHE}, unless there is a subdominant proton component emerging at the \ac{GZK}-threshold \cite{PierreAuger:2019ens,vanVliet:2019nse}. Reversing the argument, the non-observation of \ac{UHE} neutrinos and photons either in UHECR observatories such as Auger \cite{PierreAuger:2019ens,PierreAuger:2019azx} or with dedicated \ac{UHE} neutrino observatories such as IceCube \cite{IceCube:2018fhm}, constrains the fraction of protons that can be accelerated in them \cite{PierreAuger:2019ens,vanVliet:2019nse, Heinze:2015hhp}, disfavoring sources that would produce a proton fraction larger than $\sim$\,$30\%$ in the \ac{GZK} energy range, provided the density of sources follows a strong evolution with redshift~\cite{PierreAuger:2019ens,vanVliet:2019nse}. 
It has to be emphasized that an independent measurement of the \ac{UHECR} composition via air shower experiments provides a reliable prediction of the cosmogenic UHE neutrino flux for the nearby universe. A comparison with model-dependent composition constrains obtained by neutrino measurements will determine if the origin of UHECR is consistent between the nearby and distant universe.

The constraints on \ac{UHECR} sources from observations of the \ac{UHECR} spectrum and mass composition are also complemented by multi-messenger observations of GeV-TeV energy photons, 100\,TeV-PeV neutrinos, and the lack of detection of neutrinos at EeV energies, see e.g., Refs.~\cite{Muzio:2019leu, Muzio:2021zud}. 
It is also important to keep in mind that neutrinos and (in case of nearby sources) photons that are produced directly in the \ac{UHECR} sources may outshine cosmogenic fluxes in some scenarios \cite{Muzio:2021zud, Rodrigues:2020pli}. If the neutrino flux found cannot be correlated with the sources then it may be that the sources are too numerous and cannot be resolved or it may be that the detected neutrinos are cosmogenic. Cosmogenic neutrino fluxes constitute a background in the search of source correlations and it is important to quantify their fluxes. In fact, the low fluxes of cosmogenic neutrinos and photons that are expected from the limited maximum rigidity of the sources, may provide favorable conditions for identifying neutrinos from point sources.

\begin{figure}[t!]\centering
\includegraphics[width=0.99\linewidth]{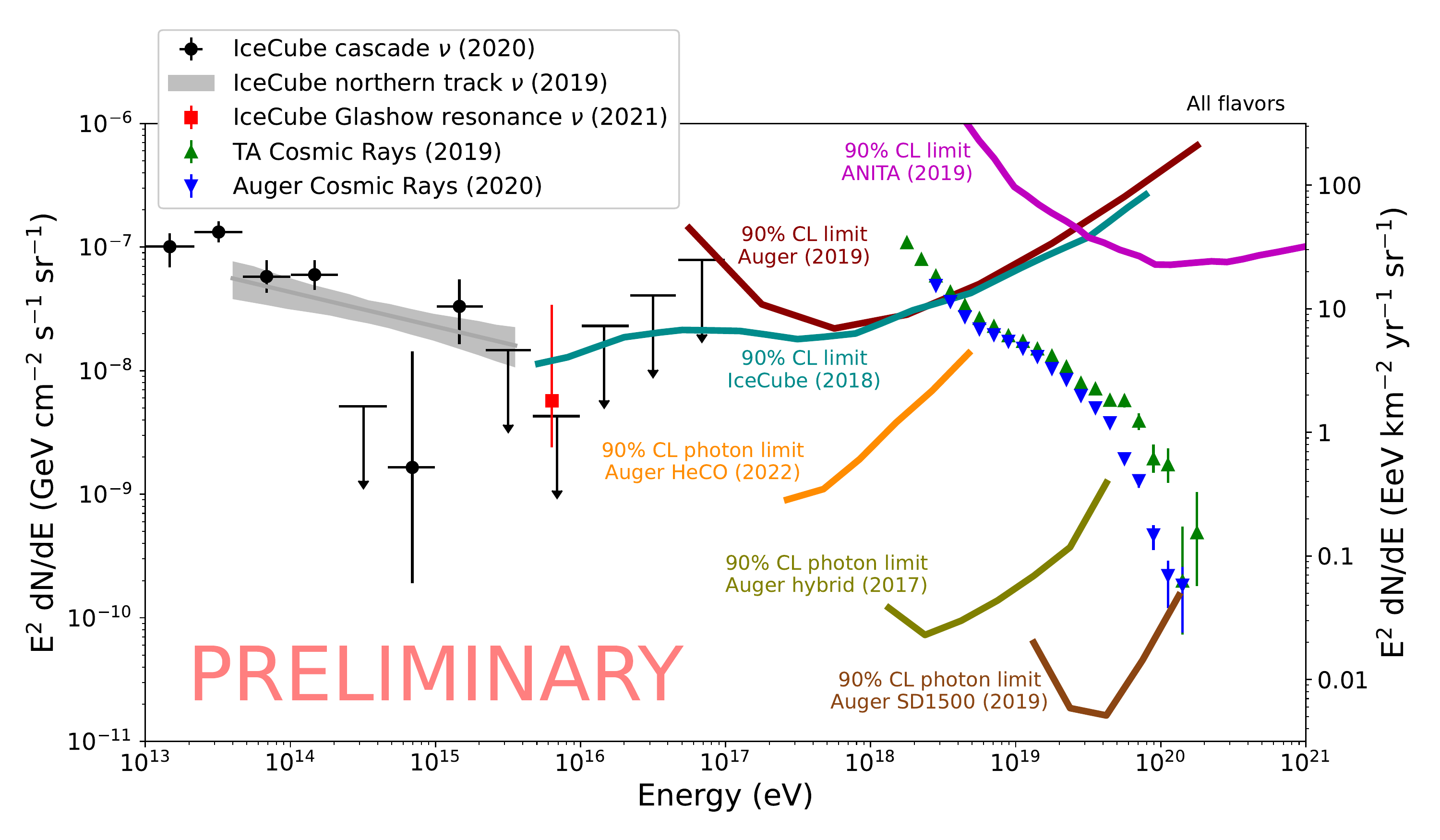}\vspace{-2mm}
\caption{Panorama of VHE astrophysical neutrino measurements from the IceCube Neutrino Observatory \cite{IceCube:2020acn, Stettner:2019tok, IceCube:2021rpz}, and UHE and constraints from IceCube \cite{IceCube:2018fhm}, the Pierre Auger Observatory \cite{PierreAuger:2019ens}, and the ANITA detector \cite{ANITA:2019wyx}, in terms of energy flux (all flavors). Also plotted are the UHECR spectrum as measured at the Pierre Auger Observatory \cite{PierreAuger:2020qqz, PierreAuger:2020kuy} and the Telescope Array \cite{Ivanov:2020rqn}.}
\label{fig:nu_panorama}
\vspace{-2mm}
\end{figure}

\subsubsection{Correlations with the arrival directions: UHECRs as messengers}

\acp{UHECR} are a class of messengers of their own that are however deviated in the magnetic fields that permeate the Universe in proportion to their rigidity. As a result they lose both time and directional correlation with other messengers emitted at the same time and/or from the same sources where they are produced. As deviation depends on the particle nature and magnetic fields are generally poorly known, it is particularly hard to use arrival directions to constrain \ac{UHECR} sources. However, at the highest energies of order $100$~EeV, deflections can be reduced to a few degrees because of the high rigidities, $R=p/Z$, ($\Delta \vartheta \propto 1/R$) and because at these energies the \ac{UHECR} interactions with the cosmic microwave and other electromagnetic backgrounds prevent them from traveling freely over distances exceeding $\sim$\,$200$~Mpc. The latter is known as the \ac{GZK} effect. \acp{UHECR} uniquely complement neutrinos from sources in the interesting distance range between order 1 to 200\,Mpc. By studying their arrival directions it is then possible to constrain the nature and the sources of the UHECR. Further progress in this area will be closely related to advances in the study of UHECR composition and also in the mapping of the magnetic fields. 

The possibility of \ac{UHE} proton astronomy was one of the motivations to search for the highest energy cosmic rays at the turn of the century, and in particular to build the Pierre Auger Observatory \cite{PierreAuger:2015eyc}. As angular deviations are inversely proportional to rigidity, few tens of degrees can be expected for protons of 3 to 10\,EeV or for oxygen of 30 to 100\,EeV, which can be small enough for the arrival directions to still carry relevant directional information that could constrain the sources of the highest energy particles known to mankind. 
Two of the main contributions that have been revealed by the largest \ac{UHECR} detectors~\cite{HiRes:2007lra,PierreAuger:2008rol} turned out to reduce early expectations in this respect. 
Firstly, at the highest energies the energy spectrum is now established with great precision  \cite{PierreAuger:2020kuy,PierreAuger:2020qqz}, and this puts a limit to the maximal energy observed in UHECR not far beyond the $100$\,EeV benchmark and reduces the flux quite dramatically beyond an exponential-like cut-off around 40\,EeV. 
The second finding is related to the mean mass of the particles which has been shown to change gradually from being light-dominated at the EeV region to being dominated by a mixture of intermediate mass nuclei in the 20 to 50\,EeV region. This conclusion has been reached by measuring for each shower the depth of shower maximum with the fluorescence technique~\cite{PierreAuger:2014sui,PierreAuger:2014gko}. However, this conclusion is not free of uncertainties due to the lack of knowledge about the interactions and subsequent shower development at the highest energies. The duty cycle of the fluorescence technique and the suppressed flux at the highest energies prevents measurements of mass beyond 50\,EeV with current statistics. While the determination of the average mass of the primary nuclei is uncertain and events detected with Telescope Array have not confirmed an intermediate mass in the $30 - 80$\,EeV range, they do confirm the observed change in the elongation rate~\cite{Sokolsky:2021xto} which is considered a quite robust evidence of a composition change~\cite{Watson:2021rfb}, unless the cross sections involved have dramatic deviations from Standard Model predictions. 

Despite the statistical obstacles and trend towards intermediate mass primaries as the \ac{UHECR} energy increases, the study of the arrival directions of \acp{UHECR} has revealed very important deviations from an isotropic distribution \cite{PierreAuger:2017pzq,PierreAuger:2018zqu,TelescopeArray:2018rtg}. Spatial correlations have been searched for between the arrival directions of the highest energy cosmic rays and classes of objects that are known to emit in the very high-energy regime such as \ac{AGN} and \acp{SBG} and with the overall matter distribution~\cite{PierreAuger:2018qvk}. There are indications of correlations that slightly favor \acp{SBG}, but the significance is not yet strong enough to claim a causal correlation with a given class of sources. 

Since the change from a diffusive to a ballistic regime within the Galaxy is at a rigidity of a few EV \cite{Kaeaepae-icrc21}, an inference of composition on an event by event basis would allow the selection of high rigidity events to enhance the anisotropies that seem to be washed out by the fact that there is a mix of masses at given energies and that the average mass is intermediate at the highest measured energies. Rigidity selection requires independent measurements of the energy and an observable which is sensitive to composition such as the depth of shower maximum or the number of muons, which can at the moment only be achieved with limited statistics using events which are simultaneously measured with \acp{SD} and fluorescence detectors \acp{FD}. 
The study of these {\sl hybrid} events has already been exploited by the Pierre Auger collaboration in two most important directions. By taking full advantage of the greater exposure of the \ac{SD} and inferring new observables from the \ac{SD} that are sensitive to composition, composition measurements could be extended up to 80\,EeV~\cite{PierreAuger:2017tlx}. Hybrid events have already been used to explore possible anisotropies that are related to composition. Indeed, a study has been made that suggests that the subsample of events whose arrival directions are within $30^\circ$ of galactic declination from the galactic plane, may be of higher average mass than the rest of the events~\cite{PierreAuger:2021jlg}. While no discovery has been claimed yet, it is apparent that a more definitive statement should be made in the years to come. 

\paragraph{Arrival directions of UHECR and neutrinos} Neutrinos are produced in hadronic interactions and are good tracers to point back to \ac{UHECR} sources. To search for such correlations in arrival directions, neutrino data from the IceCube Neutrino Observatory, ANTARES Neutrino Telescope, \ac{UHECR} data with energies greater than 50\,EeV from the Pierre Auger Observatory and Telescope Array have been analysed. Three independent methods were tested~\cite{ANTARES:2022pdr, Schumacher:2019qdx, Barbano:2020scg}. The first analysis looks for a clustering of neutrino events along the approximated direction of \acp{UHECR}. The second analysis reverses the logic considering high-energy neutrinos as sources and searches for an excess of cosmic-ray clustering in the direction of neutrinos. The last analysis counts correlating pairs of neutrinos and \acp{UHECR} with a dynamic angular distance. No significant correlations were found from 7-years of neutrino and 11-years of \ac{UHECR} data.   

A non-correlation is somewhat expected knowing the mass composition of the highest energy cosmic rays is mixed. The deflection during propagation in the galactic magnetic field for a 100\,EeV proton is expected to be $\sim$\,${3}^\circ$. For heavier mass the deflection is greater since it scales with charge. Thus, without event-by-event information on the mass composition it is challenging to make correct assumptions on uncertainties of \ac{UHECR} source positions. Another difficulty is that \acp{UHECR} are from the local Universe with a horizon of $\sim$\,${200}$\,Mpc while neutrinos can travel from as far as the entire visible Universe $\sim$\,${4}$\,Gpc. For example, the first source candidate reported by IceCube \cite{IceCube:2018dnn,IceCube:2018cha} - TXS0506+056 - is located at $\sim$\,${1.3}$\,Gpc far beyond the \ac{UHECR} zone.

\subsubsection{UHECR detectors as neutrino, photon, and neutron telescopes}

All \ac{UHECR} observatories that have been constructed or are being designed to detect extensive air showers, are naturally also potential detectors of any other particle that induces a similar shower in the same medium, in particular neutral particles of equivalent energies such as neutrons, neutrinos, and photons. Such searches can be made provided that methods are devised to separate these showers from those produced by the more abundant cosmic rays. These particles carry directional information from the sources and allow the exploitation of both directional and time correlations with all bands of astronomy and gravitational waves. Inevitably, \ac{UHECR} observatories are thus also multi-messenger observatories that could observe \ac{UHE} neutrons, photons, and neutrinos and exploit in these cases both directional and time correlation with all bands of astronomy and gravitational waves, fully contributing into the multi-messenger endeavour. 

\paragraph{Neutrinos} The detection of photons, neutrinos, and neutrons has to be performed in a background of the more abundant cosmic rays and requires special techniques that allow the separation of potential signals from the background. By looking at the depth development of the extensive air showers 
it is relatively easy to identify neutrinos in \ac{UHECR} observatories because, contrary to cosmic rays that typically interact within the first hundred $\mathrm{g\,cm}^{-2}$, they can induce showers at any depth because their interaction length exceeds that of the atmosphere. The search for down-going neutrinos with these observatories relies on detecting showers that start their development in the lower layers of the atmosphere \cite{Berezinsky:1969erk,Zas:2005zz} and with this technique the Pierre Auger Observatory has achieved large effective areas~\cite{PierreAuger:2019azx}. Moreover, neutrino oscillations lead us to expect approximately equivalent fluxes of all flavor neutrinos \cite{Learned:1994wg,Athar:2000yw}, and tau neutrinos offer a unique and most interesting window that involves the detection of air showers and outweights in acceptance other search strategies \cite{Fargion:2000iz,Bertou:2001vm}.
Tau neutrinos that reach the Earth surface slightly below the horizon can interact below the surface producing a tau lepton that exits into the atmosphere where it decays producing a slightly up-going shower~\cite{Fargion:2000iz, Bertou:2001vm}. No \ac{UHECR} is expected to cross the Earth and even if nearly horizontal \ac{UHECR} showers are misidentified as upcoming showers, they would start developing very high in the atmosphere in contrast with those from tau decay that tend to develop closer to ground. Since the effective area of \ac{UHECR} detectors must have a scale of hundreds or even thousands of square kilometers and the density of the target for the interactions is that of the Earth crust, about $2-3~\mathrm{g\,cm}^{-3}$, the target mass for neutrinos becomes huge compensating the small solid angle acceptance for the shower directions to be nearly horizontal. Thus, it is not surprising that earth-skimming showers can be extremely effective to search for neutrinos in specific directions of the sky. This is precisely the reason why the Pierre Auger Observatory was able to set the most stringent limits to the neutrino flux at \ac{UHE} from the spectacular neutron star coalescence that marked the onset of multi-messenger astronomy in 2017~\cite{LIGOScientific:2017ync,ANTARES:2017bia}.

\paragraph{Photons} Photons can be also discerned from the background of cosmic rays because the produced showers have a reduced number of muons and they also develop on average deeper into the than CR induced showers. The difference, however, is more subtle than for neutrinos and thus the requirements on the detector performance are more demanding to be able to separate them from the more abundant \acp{CR}. Photon searches have been accomplished both with surface detector arrays that provide high statistical power \cite{Ave:2000nd,PierreAuger:2007hjd,TelescopeArray:2018rbt,Rautenberg:2021vvt}, as well as with fluorescence telescopes that provide high separation power on event-by-event basis \cite{PierreAuger:2016kuz,PierreAuger:2016ppv,PierreAuger:2021mjh}.

Using such techniques, important results about photon searches could be derived from a number of air shower observatories. They comprise bounds both on a diffuse flux of high energy photons as well as on point sources.

Diffuse flux bounds have served to constrain \ac{SHDM} models (see e.g., Refs.~\cite{Kalashev:2016cre,Anchordoqui:2021crl}), topological defects, and cosmogenic photon fluxes from the \ac{GZK}-effect \cite{TelescopeArray:2013yze,PierreAuger:2016kuz,PierreAuger:2021mjh}. Targeted searches for point sources, on the other hand, allowed to (\textsl{i}) constrain the continuation of measured TeV photon fluxes to EeV energies, (\textsl{ii}) predictions of EeV proton emission models from non-transient Galactic sources including the galactic center region and from nearby extragalactic sources \cite{PierreAuger:2016ppv, TelescopeArray:2020hey}, as well as (\textsl{iii}) the lifetime of \ac{SHDM} particles branching to the $q\bar{q}$ channel in the mass range $10^{19} - 10^{25}$\,eV \cite{Kalashev:2020hqc}.
Moreover, gravitationally-produced \ac{SHDM} particles that may arise e.g., from coupling between the dark sector and gravitons (motivated by Standard Model of particle physics) or from the inflaton field in the early Universe can be constrained. This provides an interesting link to fundamental cosmological aspects, such as the Hubble rate and the curvature of space-time \cite{ThePierreAuger:2022}. The search for photons has also served to constrain physics beyond the Standard Model, for instance models that violate Lorentz invariance~\cite{Maccione:2010sv, Galaverni:2007tq, Galaverni:2008yj, Stecker:2017gdy, PierreAuger:2021tog}.

\paragraph{Neutrons} Finally, although there is no known possibility to separate neutron-induced showers from those induced by charged cosmic rays on the basis of the shower development, it is in principle possible to identify nearby sources of neutrons just by looking at an excess of \ac{EAS} from given directions \cite{PierreAuger:2012dqz, PierreAuger:2014tey}, or by exploiting potential time and directional correlations to other messengers. In fact, this latter procedure can be in principle applied to any type of neutral particle that induces a shower in the atmosphere.

\paragraph{Follow-Up Observations} \ac{UHECR} observatories also contribute to campaigns of follow-up observations \cite{PierreAuger:2016efk, Schimp:2020gxx, PierreAuger:2021oks}, with the search for neutrinos from the binary neutron star merger GW170817 \cite{LIGOScientific:2017ync} being the most prominent example. 
At the time of the \ac{GW} detection, the source was located at a zenith angle of $91.9^\circ$ at the site of the Auger Observatory, just below the horizon and extremely close to the sweet-spot for Earth-skimming neutrinos. When considered in a time interval of $\pm 500$\,s about the detection ($93.3^\circ < \theta < 90.4^\circ$), the EeV exposure has been larger than that of dedicated neutrino telescopes with allowed stringent upper limit to the neutrino fluence \cite{ANTARES:2017bia}. Another example is the search for \ac{UHE} neutrinos from TXS\,0506+056 using the Pierre Auger Observatory. Despite the fact that the source is located at an unfavorable position, relevant upper bounds on the \ac{UHE} $\nu$-flux complementing the detection by IceCube at lower energies, could be provided by the Auger observatory \cite{PierreAuger:2020llu}.

The Pierre Auger Observatory is both a triggering and a follow-up partner in the \ac{AMON}~\cite{AyalaSolares:2019iiy} which establishes and distributes alerts for immediate follow-up by subscribed observatories with private or \ac{GCN} notices. It initiates automatized follow-up observations on gravitational wave events and sends back alerts to \ac{GCN} in case of a positive detection.

\fakesection{Particle physics at the Cosmic Frontier}
\vspace{3cm}
{\noindent \LARGE \textbf{Chapter 3}}\\[.8cm]
\textbf{\noindent \huge Particle physics at the Cosmic Frontier:}\\[3mm]
\textbf{\LARGE Bridging terrestrial and natural accelerators}
\label{sec:AccelSyn}
\vspace{1cm}

Throughout the history of elementary particle physics, discoveries have been made through the observation of cosmic rays and neutrinos. This includes, for example, the discovery of new elementary particles, the confirmation of neutrino oscillations, as well as measurements of particle interactions far beyond current collider energies. In this chapter, the synergies between modern \ac{UHECR} measurements and high-energy particle physics will be discussed and described how \ac{UHECR} experiments and particle physics can inform each other to improve the understanding of fundamental particle interactions at the highest energies. How to leverage \ac{UHECR} experiments in order to inform particle physics will be described in \cref{sec:UHECRPartSyn} and relevant collider measurements will be discussed in turn in \cref{sec:PartUHECRSyn}. In \cref{sec:PartBSMDM}, unique opportunities for searches for beyond Standard Model physics and dark matter with \ac{UHECR} observatories will be presented. Finally, an outlook for the next decade and perspectives for future synergies between modern astroparticle and high-energy particle physics will be discussed in  \cref{sec:PartUHECROutlook}.

\subsection[Particle physics with UHECRs]{Leveraging UHECR experiments to inform particle physics}\label{sec:UHECRPartSyn}

When a cosmic ray enters the atmosphere and collides with an air nucleus, it produces hadronic secondaries, mostly pions. This initiates an extensive air shower (\ac{EAS}) in the atmosphere where the decay of neutral pions feeds an accompanying electromagnetic cascade, while the charged pions, baryons and kaons interact again with air nuclei deeper down on their way to the ground. The process is self-sustained until most energy is dissipated through the electromagnetic cascade, and charged pions reach an energy where the decay into muons becomes more likely than interactions with air nuclei. Muons are thus tracers of the hadronic activity of the air shower. Integrated over time, this gives rise to an atmospheric lepton flux, which is also the background for the observation of astrophysical neutrinos in modern large-scale neutrino telescopes. 

As previously described in \cref{sec:GoBigOrHome}, there are generally two methods to observe air showers:
\begin{itemize}
    \item[(i)] The detection of radiation emitted by the interaction of the charged particles, mostly electrons from the electromagnetic cascade, with the atmosphere. Such radiation can be measured in the UV band (i.e., Cherenkov and fluorescence light), or in the MHz band (i.e., radio emission). 
    \item[(ii)] Direct sampling of the secondary air shower particles at ground (or underground) by means of large particle detector arrays. 
\end{itemize}
 
The detected air showers are reconstructed and a set of observables can be retrieved: typically the arrival direction, the electron and muon content, \nelec and \nmu, the atmospheric depth at which the longitudinal development of the electromagnetic shower reached its maximum, \xmax, and the depth where the production rate of muons reached its maximum, \xmumax. Also, other shower observables which are related to the lateral spread of the particles can be determined.

The reconstruction of fundamental properties of the primary \ac{UHECR}, such as its energy and mass, requires the use of accurate air shower simulations. The cosmic ray community has developed sophisticated simulation packages that integrate state-of-the-art models of electromagnetic and hadronic interactions (see e.g., Ref.~\cite{Albrecht:2021cxw} for a recent review). Commonly used hadronic interaction models are \sibyll{}~\cite{Riehn:2019jet,Ahn:2009wx, Riehn:2015oba,Riehn:2017mfm}, \qgs \cite{Ostapchenko:2005nj, Ostapchenko:2006vr, Ostapchenko:2010vb, Ostapchenko:2019few}, \epos \cite{Pierog:2013ria, Werner:2005jf, Pierog:2009zt, Pierog:2017awp}, and \textsc{DPMJet} \cite{Ranft:1994fd, Ranft:1999fy, Ranft:2002rj, Roesler:2000he, Fedynitch:2015kcn}. All these models are based on different realizations of perturbative \ac{QCD} associated with Gribov-Regge effective quantum field theory and rely on fundamental principles like conservation laws. However, the particle production is dominated by non-perturbative \ac{QCD} which is treated by more phenomenological approaches. As a consequence, the necessary parameters are tuned to a large data set covering many orders of magnitude in energy (from few $10\,\mathrm{GeV}$ to $\,\mathrm{TeV}$ with current colliders) but limited by what the accelerator experiments can measure, thus leading to extrapolations both in energy and phase space. Indeed the \ac{EAS} development is driven by the particles carrying most of the energy while the latter are the most challenging to measure in collider experiments (i.e., forward particle production). As previously described in \cref{sec:tensions}, large uncertainties remain both due to theoretical limitations and the lack of data from existing collider experiments. Nevertheless, \acp{EAS} simulated with these packages generally describe real air showers quite successfully, and are also used to predict the propagation and interaction of \acp{UHECR} in space and around the sources~\cite{AlvesBatista:2016vpy}.

In the following, current limitations in our understanding of air shower physics will be discussed and it will be highlighted how \ac{UHECR} measurements can inform particle physics beyond the phase space of existing collider experiments to improve current hadronic interaction models.

\subsubsection[Measurements of the proton-air cross section]{Measurements of the proton-air cross section}

As described in \cref{sec:MassCurrentStatus}, various \ac{EAS} observables are sensitive to the average mass composition of the primary cosmic ray by direct comparison of observations to predictions from hadronic interaction models. The electromagnetic component does not suffer much from theoretical uncertainties and when experimentally accessible, it is typically used to assess the energy and mass (number of nucleons). However, measurements of the electromagnetic shower maximum, \xmax, can also be used to determine the proton-air cross section. This is done by selecting the most proton-like \acp{UHECR} to determine the attenuation length, $\Lambda_\eta$, of proton showers in the atmosphere. The attenuation length is then converted into the proton-air cross section, $\sigma_\mathrm{p-Air}$, based on \ac{EAS} simulations.

\begin{figure}[tb]
    \centering
    \includegraphics[width=0.75\textwidth]{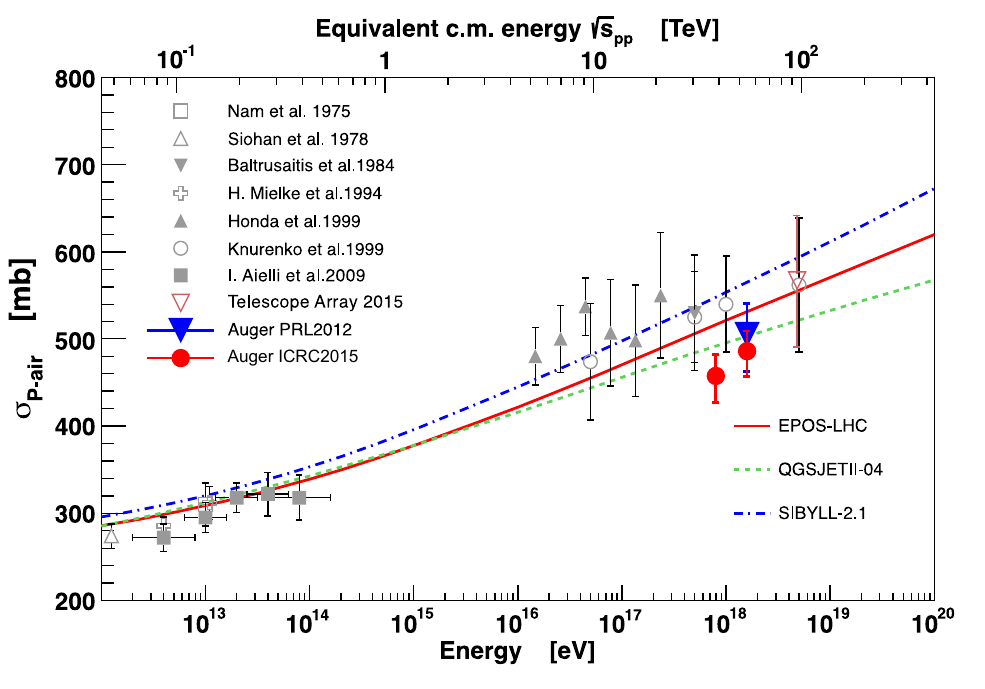}
    \caption{Proton-air cross section obtained from different experiments compared to predictions obtained from \eposlhc, \qgsii, and \sibyll{2.1}. Figure adapted from Ref.~\cite{Ulrich:2015yoo}.}
    \label{fig:AugerCrossSection}%
\end{figure}

\cref{fig:AugerCrossSection} shows the proton-air cross section recently measured by the Pierre Auger Collaboration~\cite{PierreAuger:2012egl, Ulrich:2015yoo} in the two energy intervals from $10^{17.8}\,\mathrm{eV}$ to $10^{18.0}\,\mathrm{eV}$ and from $10^{18.0}\,\mathrm{eV}$ to $10^{18.5}\,\mathrm{eV}$, and by the Telescope Array Collaboration~\cite{Abbasi:2015fdr, Abbasi:2020chd} in the interval between $10^{18.3}\,\mathrm{eV}$ and $10^{19.3}\,\mathrm{eV}$. Also shown are previous results from other experiments (see Ref.~\cite{Ulrich:2015yoo} for details) and predictions from the hadronic interaction models \eposlhc~\cite{Pierog:2013ria} and \qgsii~\cite{Ostapchenko:2019few}, which already have been tuned to \ac{LHC} data (post-\ac{LHC}), as well as \sibyll{2.1}~\cite{Ahn:2009wx}, which was developed before the \ac{LHC} era (pre-\ac{LHC}). Indeed, the cross sections obtained from the post-\ac{LHC} models appear to be in better agreement with current \ac{EAS} data than the pre-\ac{LHC} model \sibyll{2.1}. 

These \ac{EAS} measurements provide complementary particle physics data far beyond the energies reachable by any current collider experiment. Thereby, they very clearly demonstrate the large potential for synergies between astroparticle physics and high-energy particle physics.

\subsubsection[Hadronic interactions and the Muon Puzzle in EASs]{Hadronic interactions and the Muon Puzzle in EASs}

The muonic component in the air shower is generally used as a probe of the hadronic interactions during the shower development. Various measurements of atmospheric muons with energies around $1\,\mathrm{GeV}-10\,\mathrm{GeV}$ have revealed a discrepancy between simulated and observed muon production in air showers (see also \cref{sec:tensions}). A muon deficit in simulations was directly observed for the first time more than 20 years ago by the HiRes-MIA collaboration~\cite{HiRes:1999ioa}. Further indirect evidence for a muon discrepancy was found by several other air shower experiments, but the situation remained inconclusive until the Pierre Auger Observatory also reported a muon deficit in simulations in a direct measurement at even higher cosmic ray energies~\cite{PierreAuger:2014ucz,PierreAuger:2016nfk}. 

\begin{figure}[tb]
    \vspace{-1em}
    
    \includegraphics[width=0.51\textwidth]{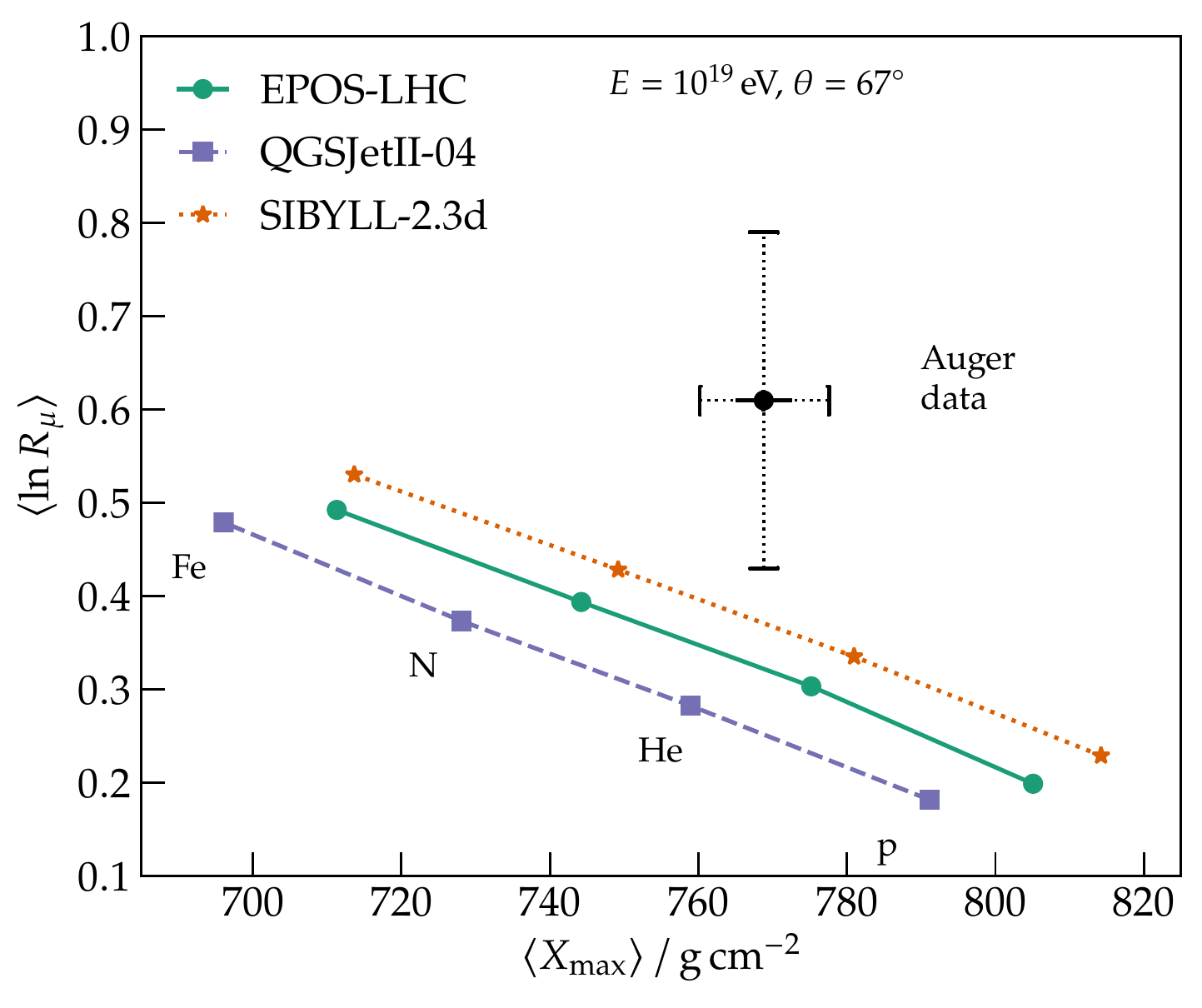}\;\;
    \includegraphics[width=0.485\textwidth]{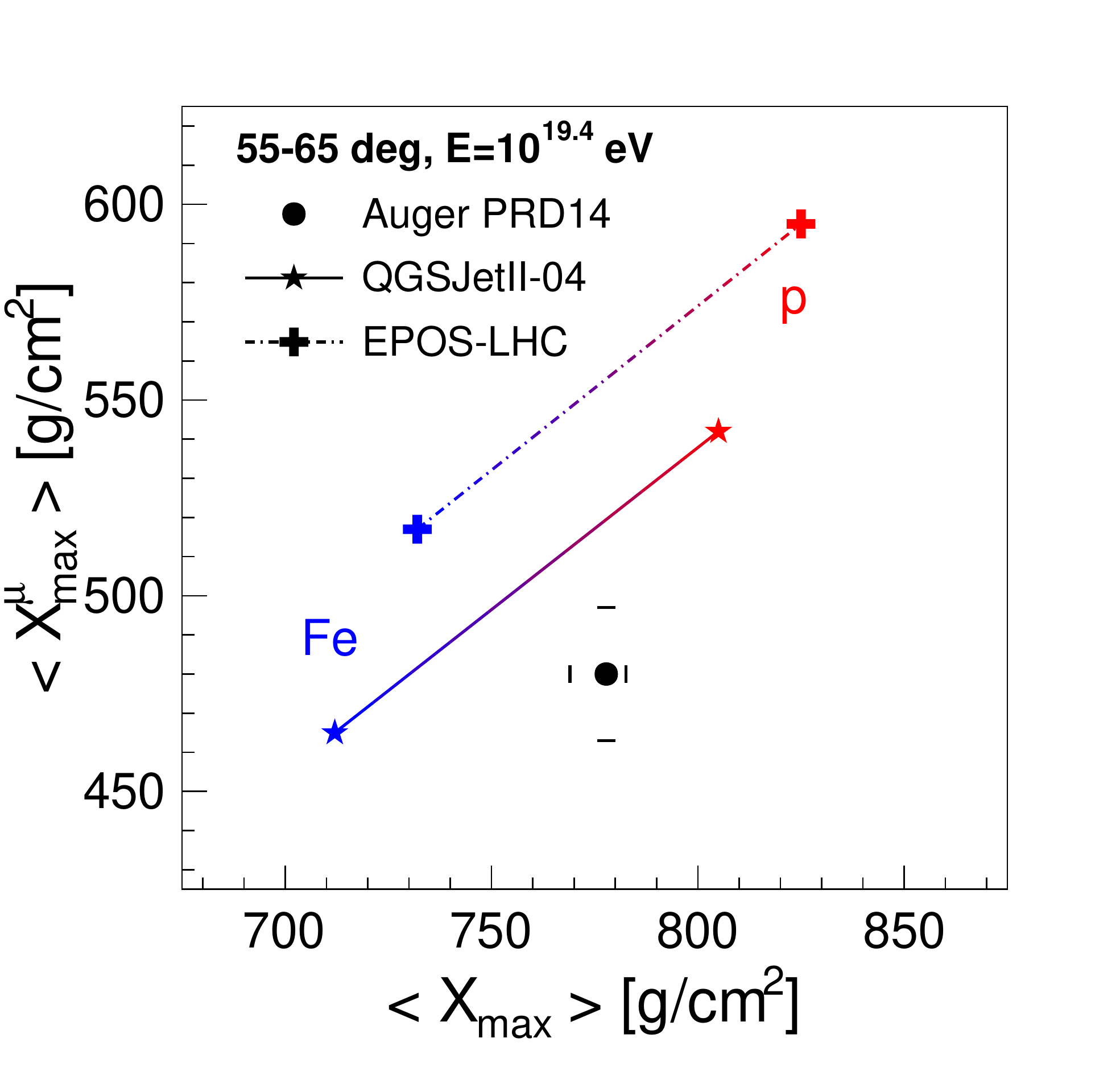}
    \vspace{-1.5em}
    
    \caption{Measurement of the muon content (left)~\cite{PierreAuger:2021qsd}, where $R_\mu=1$ corresponds to $2.148\times 10^{7}$ muons, and depth of maximum muon production , \xmumax, (right)~\cite{Cazon:2019mtd}, compared to measurements of \xmax, for air showers at $10^{19}$\,eV. Also shown is the phase space occupied by all possible primary mass combinations between proton and iron according simulations with several hadronic models.}
    \label{fig:XmaxXmumax}%
\end{figure}

The simultaneous comparison of independent air shower observables, such as \xmax and \nmu, constrain the phase space of hadronic models. \cref{fig:XmaxXmumax} (left) shows the mean logarithmic muon number compared to the average shower maximum measured by Auger in air showers at $10^{19}\,\mathrm{eV}$~\cite{PierreAuger:2014zay,Cazon:2019mtd}. Also shown are predictions from recent hadronic interaction models for different cosmic ray masses, as well as interpolations (lines), which are clearly inconsistent with the experimental data. A similar picture can be obtained by comparing the maximum depth of muon production, \xmumax, with the electromagnetic shower maximum, \xmax, as shown in \cref{fig:XmaxXmumax} (right). Here, the experimental data is also inconsistent with model predictions for proton and iron showers, indicating a \ac{UHECR} mass composition heavier than iron. 

These discrepancies are referred to as the Muon Puzzle in astroparticle physics and their observation led to the formation of the \ac{WHISP} from members of eight (now nine) air shower experiments, to systematically combine the existing data on muons for the first time~\cite{EAS-MSU:2019kmv, Cazon:2020zhx, Soldin:2021wyv}. The most recent meta-analysis includes data from \mbox{HiRes-MIA}~\cite{HiRes:1999ioa}, the Pierre Auger Observatory~\cite{PierreAuger:2020gxz, PierreAuger:2021qsd}, Telescope Array~\cite{TelescopeArray:2018eph}, the IceCube Neutrino Observatory~\cite{IceCube:2021tuv, IceCube:2022yap}, KASCADE-Grande~\cite{Apel:2017thr}, NEVOD-DECOR~\cite{Bogdanov:2018sfw}, the Yakutsk EAS array~\cite{Glushkov}, EAS-MSU~\cite{Fomin:2016kul}, SUGAR~\cite{Bellido:2018toz}, and AGASA~\cite{Gesualdi:2021yay}. 

Since the raw data are not directly comparable due to variations in the conditions between these experiments, the \ac{WHISP} introduced $z$-values, defined as
\begin{equation}
    z = \frac{\ln N_\mu - \ln N_{\mu,\mathrm{p}}}{\ln N_{\mu,\mathrm{Fe}} - \ln N_{\mu,\mathrm{p}}},
    \label{eq:z-values}
\end{equation}
where \nmu is the measured muon number or a proxy thereof, while $N_{\mu,\mathrm{p}}$ and $N_{\mu,\mathrm{Fe}}$ are the corresponding numbers for proton and iron cosmic rays with the same energy obtained from fully simulated events that are analysed like the data, where the simulation covers the air shower and the detector response~\cite{Dembinski:2017kpa,Gesualdi:2021yay}. Since the $z$-values depend on air shower simulations, one obtains different $z$-values for each hadronic interaction model. 

\begin{figure}[tb]
    \centering
    \includegraphics[width=\textwidth]{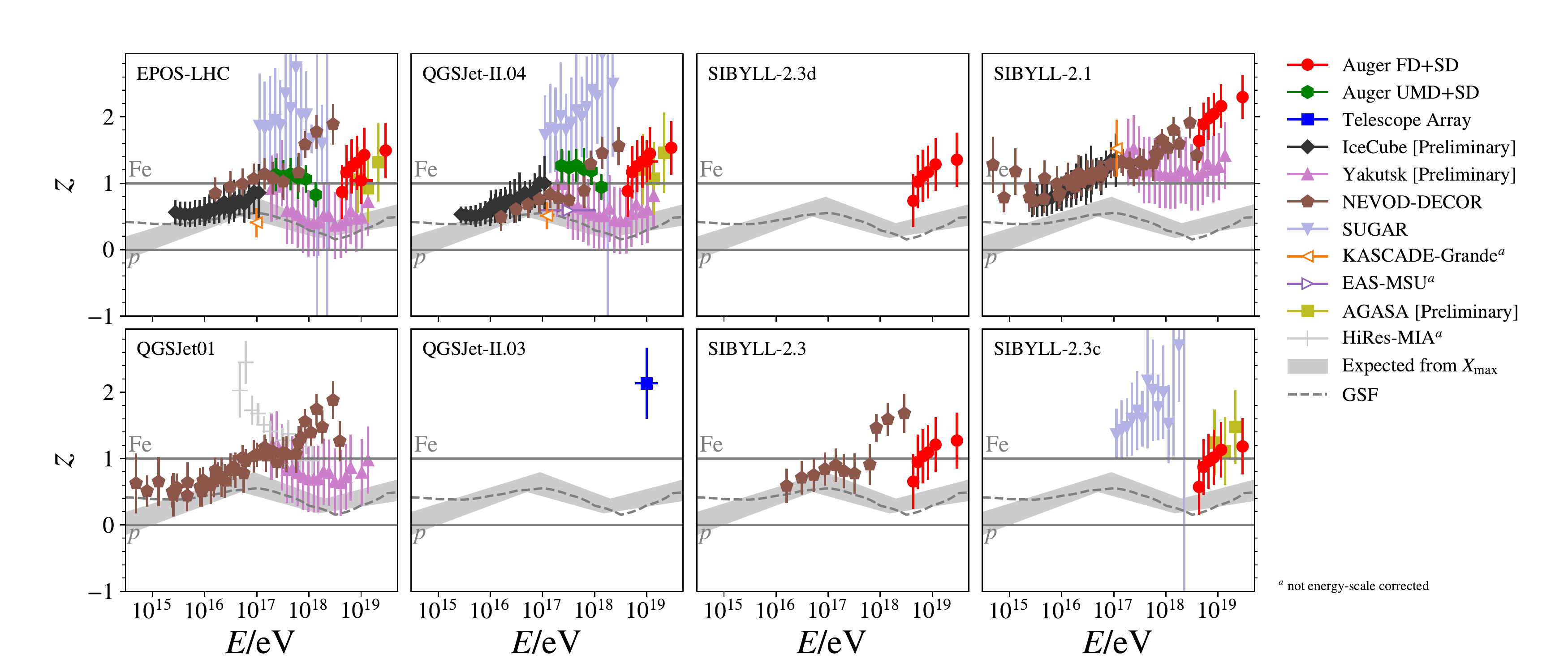}
    \caption{Muon density measurements converted to the z-scale, as defined in \cref{eq:z-values}, for different hadronic interaction models, after applying energy-scale cross-calibration, as described in the text. The data of KASCADE-Grande and EAS-MSU cannot be cross-calibrated and are only included for comparison. Also shown for comparison are $z$-values expected for a mixed composition from measurements of the electromagnetic \ac{EAS} component (\xmax), based on an update of Ref.~\cite{Kampert:2012mx}, and from the \acs{GSF} flux model~\cite{Dembinski:2015xtn}. Figure adapted from Ref.~\cite{Soldin:2021wyv}.}
    \label{fig:rho_mu_rescaled}%
\end{figure}

Air shower experiments usually have independently calibrated energy scales with systematic uncertainties at the $10\%-20\%$ level (see also \cref{sec:EnergySpectrum}). However, two experiments with an energy-scale offset of $20\%$ would find a $18\%$ offset in the measured muons numbers because equivalent measurements are compared to air showers simulated at different apparent energies. Thus, the \ac{WHISP} also introduced a cross-calibration of the energy scales of the experiments, an important correction to account for the known systematic offsets between experiments. Assuming that the cosmic ray flux is a universal reference and that all deviations in measured fluxes between different experiments arise from energy scale offsets, a relative scale can be determined for each experiment such that the all-particle fluxes match~\cite{Dembinski:2015xtn}. The resulting $z$-values from nine air shower experiments after energy-scale cross-calibration are shown in \cref{fig:rho_mu_rescaled} for eight different hadronic interaction models~\cite{Soldin:2021wyv}. The $z$-values depend only on the mass composition of the cosmic rays at a given shower energy, which can be nearly independently obtained from the electromagnetic component of the air shower, as described in \cref{sec:MassCurrentStatus}. Hence, also shown are the expected $z$-values from measurements of the electromagnetic shower depth, \xmax, and the \ac{GSF} flux model~\cite{Dembinski:2015xtn}, which is mostly consistent with these measurements. At energies above around $100\,\mathrm{PeV}$, for most experiments, inconsistencies between \xmax measurements and muon data can be observed, with the latter indicating a \ac{UHECR} mass composition heavier than iron.

\begin{figure}[tb]
    \centering
    \mbox{\hspace{-0.5em}%
    \includegraphics[width=0.49\textwidth]{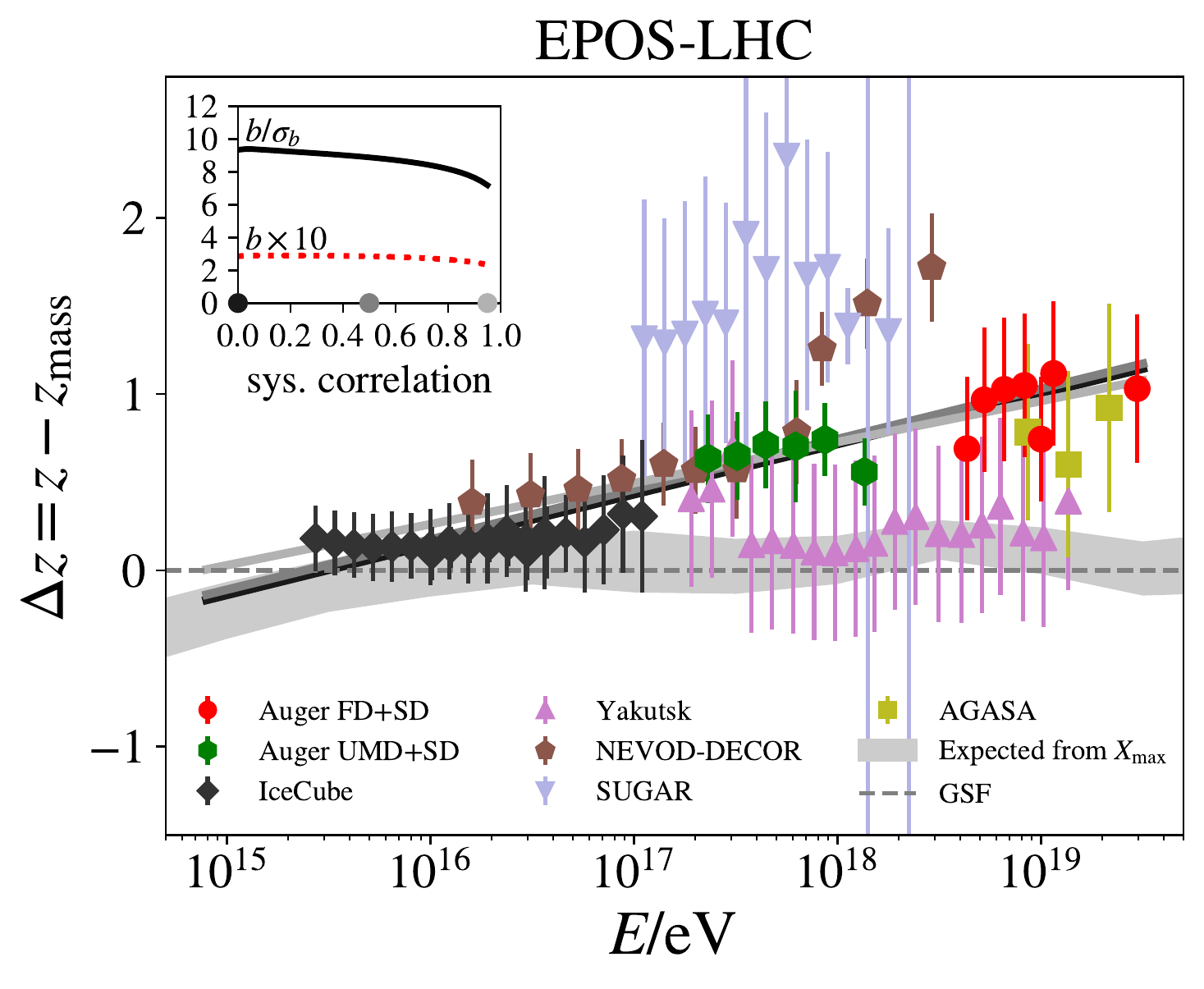}\;
    \includegraphics[width=0.49\textwidth]{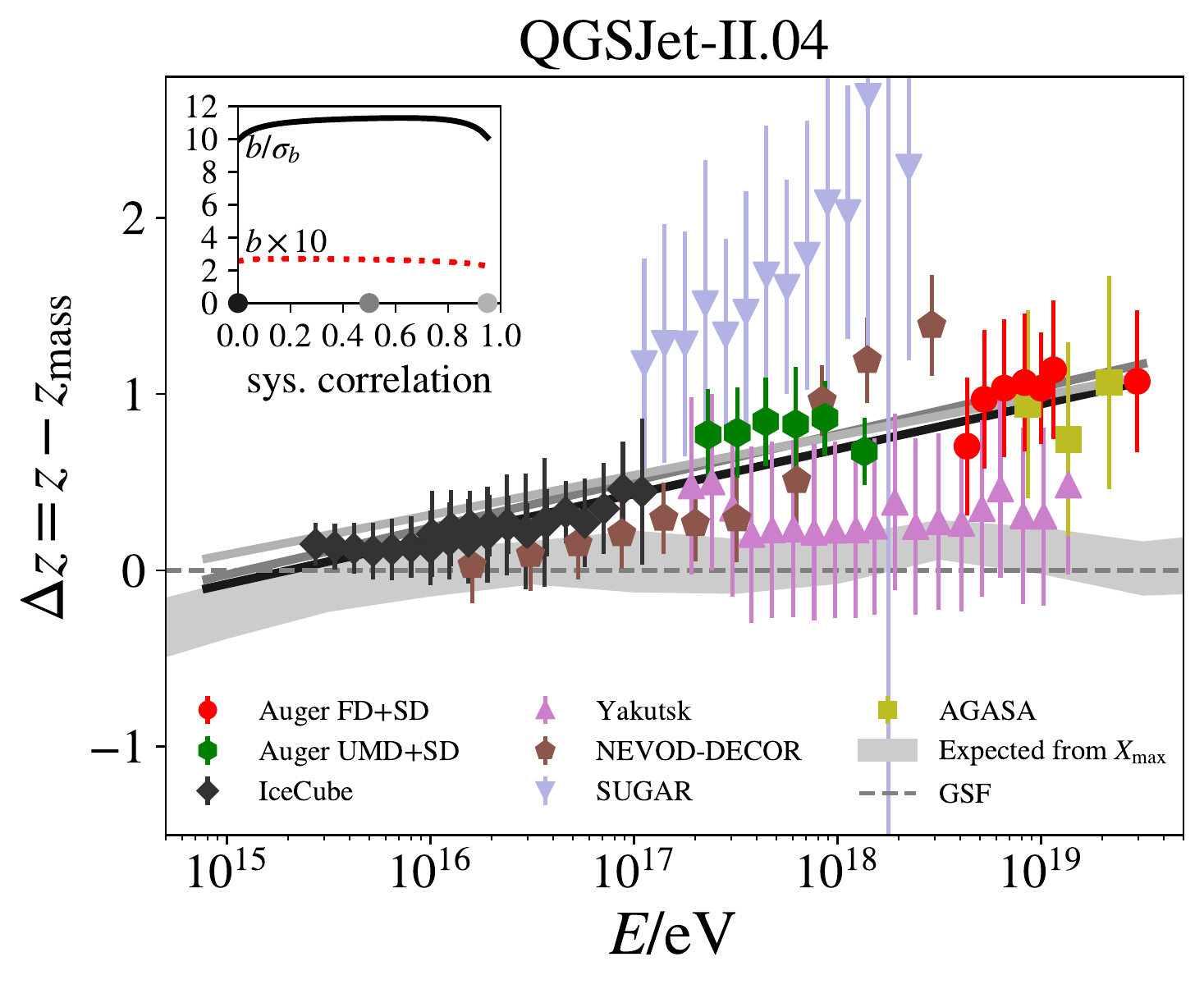}%
    }
    \caption{Linear fits of the form $\Delta z=a+b\cdot \log_{10}(E/10^{16}\,\mathrm{eV})$ to the $\Delta z=z-z_\mathrm{mass}$ distributions, as described in Ref.~\cite{Dembinski:2017kpa}. Shown in the inset are the slope, $b$, and its deviation from zero in standard deviations for an assumed correlation of the point-wise uncertainties within each experiment (for details see Ref.~\cite{Soldin:2021wyv}). Examples of the fits are shown for a correlation of $0.0$, $0.5$, and $0.95$ in varying shades of gray. Figure taken from Ref.~\cite{Soldin:2021wyv}.}
    \label{fig:main_fits}%
\end{figure}

By subtracting the expected evolution of  the $z$-values based on the \acs{GSF} flux model, $z_\mathrm{mass}$, \cref{fig:main_fits} is obtained (for \eposlhc and \qgsii). The remaining trend appears to be approximately linear with the logarithm of the energy, indicating significant discrepancies between hadronic model predictions and data. In fact, the slope of a line model fitted to this data differs from zero (agreement between simulations and data) at the level of $8\sigma$ or higher~\cite{EAS-MSU:2019kmv, Cazon:2020zhx, Soldin:2021wyv}. These discrepancies are currently not understood and indicate significant shortcomings in the description of multi-particle production in the far-forward region. It is also important to keep in mind that the absolute scale depends on the mass used as reference and this mass depends on \xmax. However, the latter suffers from other uncertainties, either experimental or from the model, which could be larger than usually foreseen. Indeed, in order to resolve the observed discrepancies, it is probably necessary not only to increase the muon production in the models but also to change \xmax predictions~\cite{PierreAuger:2021xah}. To reduce uncertainties on the shower maximum which are dominated by the first interaction starting the air shower development, new and precise \ac{LHC} data is required, as discussed in \cref{sec:PartUHECROutlook}.

Another important air shower measurement is that of muon number fluctuations in air showers, $\sigma_{\nmu}$, recently published by the Pierre Auger Collaboration~\cite{PierreAuger:2021qsd}. Because the fluctuations in the muon number are mainly driven by the early interactions in the \ac{EAS}, this measurement is particularly sensitive to the first hadronic interactions of the shower development. It has been observed that despite the well-known deficit in the simulated mean number of muons, the simulated fluctuations are in good agreement with the measurement assuming current hadronic interaction models. This indicates that the observed discrepancies in the number of muons accumulate throughout the shower development rather than being driven by the first few interactions of the \ac{EAS}. This observation constrains exotic explanations of the muon discrepancy. In fact, various approaches have been made to explain the Muon Puzzle within more exotic scenarios, such as Lorentz-invariance violation, for example, and current upper bounds for these models determined from \ac{UHECR} observations will be discussed in more detail in \cref{sec:LIV}.

In addition to the recent measurements of \acp{UHECR} by Auger, IceCube has reported various measurements of the high-energy ($\sim\mathrm{TeV}$) muon content in air showers at lower cosmic ray energies (i.e., in the PeV to EeV region) \cite{Abbasi:2012kza, Soldin:2015iaa, Soldin:2018vak,IceCube:2015wro, Fuchs:2017nuo, Desiati:2011hea, Tilav:2010hj, Gaisser:2013lrk, Tilav:2019xmf}. Recent preliminary results \cite{DeRidder:2017alk, IceCube:2021ixw} indicate inconsistencies between different components in the models, in particular in the GeV muon content in \acp{EAS}. However, the predicted TeV muon flux seems to agree with current experimental data in all analyses within the rather large systematic uncertainties. This favors explanations of the muon discrepancies which enhance the GeV muon number in hadronic models while the TeV muon flux remains at the same level~\cite{Riehn:2019jet}. Despite the large uncertainties, this observation thereby also indicates that the discrepancies in the GeV muon content accumulate in soft-\ac{QCD} processes during the \ac{EAS} development and dedicated studies will further constrain exotic explanations as they typically have an impact on the first few interactions of the air shower development. 

Further measurements at lower cosmic ray energies of the seasonal variations of the high-energy muon flux measured in IceCube can be also be used to infer the kaon to pion ratio from the size of the flux variation for a given temperature variation~\cite{Desiati:2011hea, Tilav:2010hj, Gaisser:2013lrk, Tilav:2019xmf}, for example. This measurement thereby constraints the $K/\pi$ ratio in hadronic interaction models. 

In addition, the measurement of the muon attenuation length in the atmosphere, as reported by the KASCADE-Grande Collaboration~\cite{KASCADE-Grande:2017wfe}, provides further constraints on the muon production in \acp{EAS}. Simulations based on various hadronic interaction models predict smaller attenuation lengths than observed, with smaller deviations for the two post-\acs{LHC} models. This shows that the muon number in simulations decreases more rapidly with zenith angle than observed in experimental data, which indicates that the muon energy spectra are harder than expected from current model predictions. These measurements also show a dependence on the lateral distribution of muons, which is also not reproduced correctly by current \ac{EAS} simulations.

Although measurements over the entire cosmic ray energy range need to be considered to understand the observed discrepancies, the contribution from experiments at lower energies, i.e., below $100\,\mathrm{PeV}$, is beyond the scope of this report and a comprehensive review can be found in Ref.~\cite{Albrecht:2021cxw}.

\subsection[Collider measurements]{Leveraging colliders to inform hadronic interaction models}\label{sec:PartUHECRSyn}

To describe the interactions of \acp{UHECR} with matter (atmospheric, around the source, and interstellar), the hadron production cross sections for \pp, \pion, $\pi$-p, $\pi$-ion, $K$-ion, and \ionion collisions must be known over a wide energy range of center-of-mass energies \sqrts from GeV to hundreds of TeV. Collisions with center-of-mass energy up to $\sqrts = 13$\tev have been studied at the \ac{LHC}~\cite{Evans:2008zzb}. Collisions between protons, lead ions, and xenon ions have been recorded so far. Collisions of oxygen ions are planned in the next years~\cite{Citron:2018lsq,Bruce:2021hii}, which will be an ideal reference for atmospheric interactions. Important data is also collected at the \ac{SPS}, the pre-accelerator of the \ac{LHC} and the \ac{RHIC}. Data on $\pi$ and $K$ collisions is only available at energies up to tens of GeV in \sqrts~\cite{French-Soviet:1975yfh,Brick:1981sy,DeWolf:1986mb,EHSNA22:1990otw,Prado:2017hub} because interactions can only be studied with secondary beams. Since \ac{QCD} is flavour-blind and the hadron multiplicity increases fast with \sqrts, the flavour and number of valence quarks becomes less important at high energies~\cite{Brick:1981sy}, however, it would be desirable to confirm this experimentally.

\begin{figure}[tb]
    \centering
    \hspace{-1.5em}\includegraphics[width=1.\textwidth]{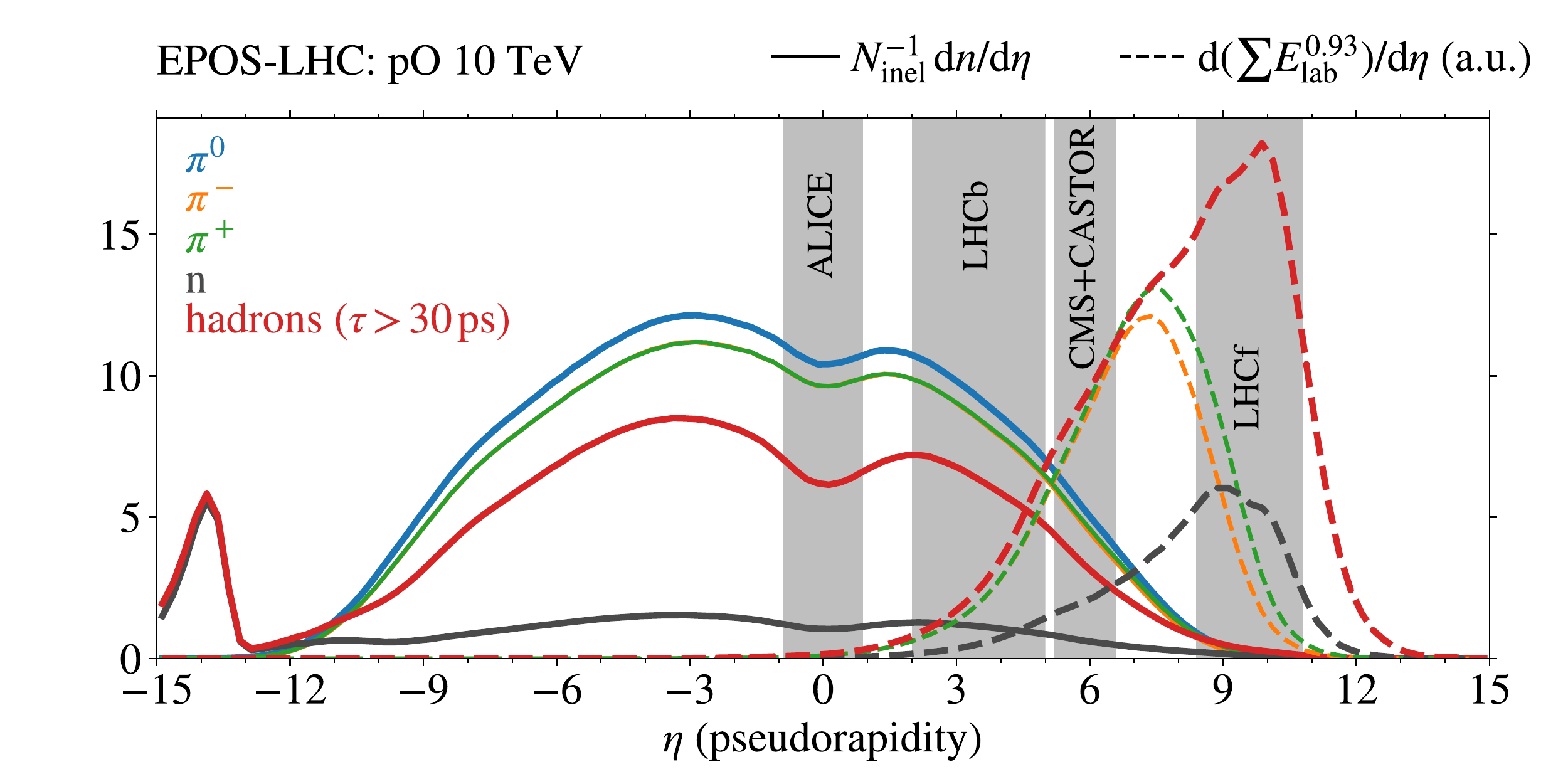}
    \caption{Simulated densities of prompt hadrons (solid lines) in proton-oxygen collisions at $10\,\mathrm{TeV}$ as a function of pseudorapidity. The estimated number of muons that would be produced by the secondaries in an air shower is also shown (dashed lines). Figure taken from Ref.~\cite{Albrecht:2021cxw}.}
    \label{fig:eta-spectra}
\end{figure}

Most experiments do not observe forward production with pseudorapidity of $\eta > 5$ in detail, since this region is not attractive for discovery and study of new particles. However, this particular region is the focus for astroparticle physics. This is illustrated in \cref{fig:eta-spectra}, which shows how many muons would be produced by the secondary particles in a p-O collision at 10\tev, if the secondaries would proceed to form an air shower. Forward produced particles have the largest energies in the fixed-target frame and generate the largest number of particles in the following interactions. However, collider measurements nevertheless provide crucial information for the understanding of hadronic interactions during the \ac{EAS} development.

In the following, it will be discussed how collider experiments inform air shower physics and how they can help to reduce uncertainties of current hadronic interaction models in order to understand the discrepancies observed in \ac{EAS} measurements.

\subsubsection[Constraining hadronic interaction models at the LHC]{Constraining hadronic interaction models at the LHC}

\begin{figure}[tb]
    \centering
    \vspace{1em}
    \includegraphics[width=1.\textwidth]{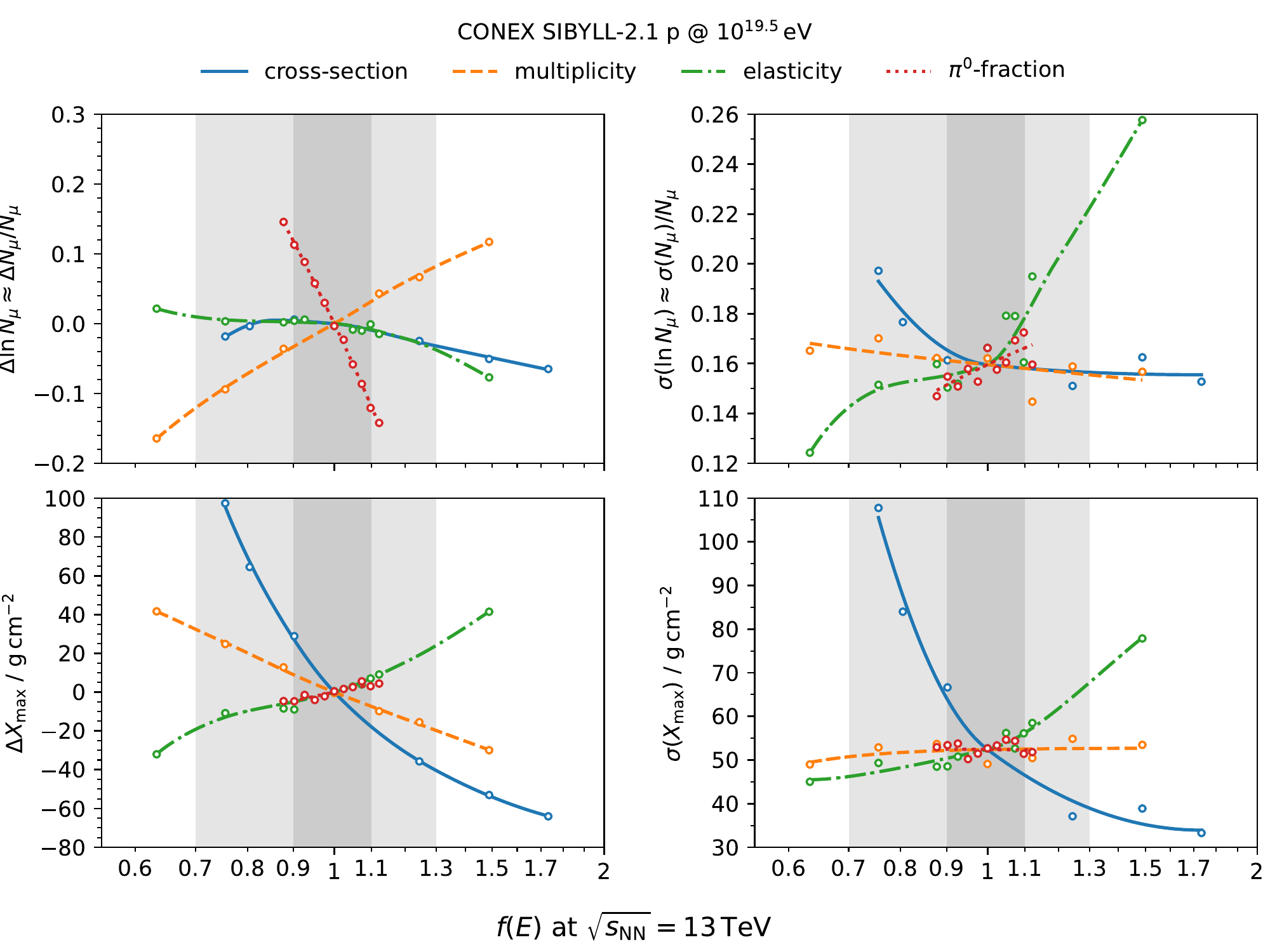}
    \caption{Impact of changing basic parameters of hadronic interactions at $\sqrts = 13$\tev and extrapolating logarithmically (see text for details) on the means and standard deviations of the logarithm of the muon number $N_\mu$ (top row) and the depth \xmax of the shower maximum (bottom row) for a $10^{19.5}\,\mathrm{eV}$ proton shower simulated with \sibyll{2.1} as the baseline model~\cite{Albrecht:2021cxw}. Relative shifts to the mean values are shown on the left-hand side. Fluctuations are shown on the right-hand side. The shaded bands highlight a $\pm 10\,\%$ and $\pm 30\,\%$ modification, respectively. The figure is an update of the original data from Ref.~\cite{Ulrich:2010rg}, taken from Ref.~\cite{Albrecht:2021cxw}.}
    \label{fig:adhoc}
\end{figure}

Four basic aspects of hadronic interactions are relevant for astroparticle physics: the inelastic cross section for hadrons in air, the hadron multiplicity, the elasticity (the energy fraction carried by the most energetic particle), and finally the ratio of electromagnetic to hadronic energy flow -- or more generally, the hadron composition. The impact of modifications of these aspects on air shower observables has been investigated with full air shower simulations \cite{Ulrich:2010rg}. The key results are shown in \cref{fig:adhoc}. The baseline prediction of the \sibyll{2.1} model \cite{Ahn:2009wx} was modified with a factor that is $1$ for a beam energy of $10^{15}\,\mathrm{eV}$ and then increases or decreases logarithmically with the beam energy, depending on the value $f(E)$ at some intermediate scale (here $\sqrtsnn= 13\,\mathrm{TeV}$, which corresponds to a beam energy of about $10^{17}\,\mathrm{eV}$). For a thorough description of $f(E)$ and a detailed discussion of the modifications in the modeling see Refs.~\cite{Albrecht:2021cxw, Ulrich:2010rg}.

The inelastic cross section has a high impact on the first two moments of the depth of shower maximum \xmax. It has been very accurately measured to a level of a few percent in proton-proton collisions at the \ac{LHC}, in particular by TOTEM and ATLAS, which resolved the $1.9\sigma$ ambiguity in earlier Tevatron data~\cite{CDF:1993wpv,E710:1991bcl}. This had a noted impact on the systematic uncertainty of \xmax predictions. A measurement of the \pPb inelastic cross section with CMS \cite{CMS:2015nfb} at 5.02\tev recently validated the standard Glauber model to better than about $10\,\%$, which is used to extrapolate from \pp to \pion and \ionion. There is still a remaining uncertainty in the extrapolation of the inelastic cross section from \pp to p-air, which can be reduced with future data from \pO collisions, but the inelastic cross section is now comparably safe to extrapolate to higher energies.

The experimental proxy for the hadron multiplicity is the charged particle multiplicity, which has been measured in \pp, \pPb, \PbPb, and \XeXe collisions at the \ac{LHC}. Very accurate data is available up to $\eta = 5$ by ALICE \cite{ALICE:2017pcy}, LHCb \cite{LHCb:2021abm, LHCb:2021vww}, and TOTEM \cite{CMS:2014kix}. Another experimental proxy is the energy flow, which has been measured in the forward direction by LHCb \cite{LHCb:2012gpm} and CMS with CASTOR \cite{CMS:2011xjg, CMS:2017dou, CMS:2018lqt, CMS:2019kap} up to $\eta = 6.4$. These measurements strongly constrain the shape of the $\eta$ distribution, which is important, since models deviate by less than 5\,\% at $|\eta| < 1$ in \pp collisions, but up to 20\,\% in the forward region.

As shown in \cref{fig:adhoc}, the elasticity has an impact on all observables, but is particularly important for the fluctuations of \xmax and \nmu. A measurement of elasticity can be performed with zero-degree calorimeters like LHCf, which has measured the neutron-elasticity \cite{LHCf:2020hjf}.

The neutral pion fraction in \cref{fig:adhoc} is a proxy for the ratio of electromagnetic to hadronic energy flow, which has a strong impact on the mean of \nmu. High-precision data on the relative yields of pions, kaons, and protons in \pp, \pPb, and \PbPb collisions is available at mid-rapidity $|\eta| < 1$ from ALICE and CMS. These data are important for model tuning and validation, but do not directly constrain the hadron composition in the forward region. In the very forward region, the hadron composition was measured with the CASTOR calorimeter of CMS in \pp  collisions, which provide a direct measurement of the ratio of electromagnetic to hadronic energy flow \cite{CMS:2019kap}. In the very forward region, the ratio is constrained by LHCf with measurements of photon-production, $\pi^0$ production, and neutron production in \pp  and \pPb  collisions \cite{LHCf:2020hjf,LHCf:2011hln,LHCf:2012stt,LHCf:2012mtr,LHCf:2014gqm,LHCf:2015nel,LHCf:2015rcj,LHCf:2018gbv}.

ALICE studied the production cross sections of strange hadrons at mid-rapidity, $|\eta| < 1$, and discovered an universal rise in the production ratios of strange hadrons to pions as a function of the charged particle density at mid-rapidity, which is independent of the collision system or \sqrts (within a few TeV) \cite{ALICE:2016fzo,Vasileiou:2020rov}. This behavior was previously known only from heavy-ion collisions and not expected in \pPb and \pp collisions. The universality is remarkable, since the hadron density in the central region rises rapidly with \sqrts and thus the relative yield of strange hadrons rises as well. This is accompanied by a reduction of the neutral pion yield, which could potentially solve the muon discrepancy in air showers \cite{Baur:2019cpv}. It is important to study this effect also in the forward region, where two hadron production mechanisms contribute, string fragmentation and remnant fragmentation. Studying strange decays requires a full tracking system with vertex resolution and magnetic field, which currently only LHCb offers in the forward region. CMS with CASTOR has studied the ratio of electromagnetic to hadronic energy flow as a function of the charged particle density in \pp collisions, but found no significant reduction \cite{CMS:2019kap}. 

\subsubsection[Fixed-target experiments]{Fixed-target experiments}

\begin{figure}[tb]
    \mbox{\hspace{-0.5em}\includegraphics[width=0.51\linewidth]{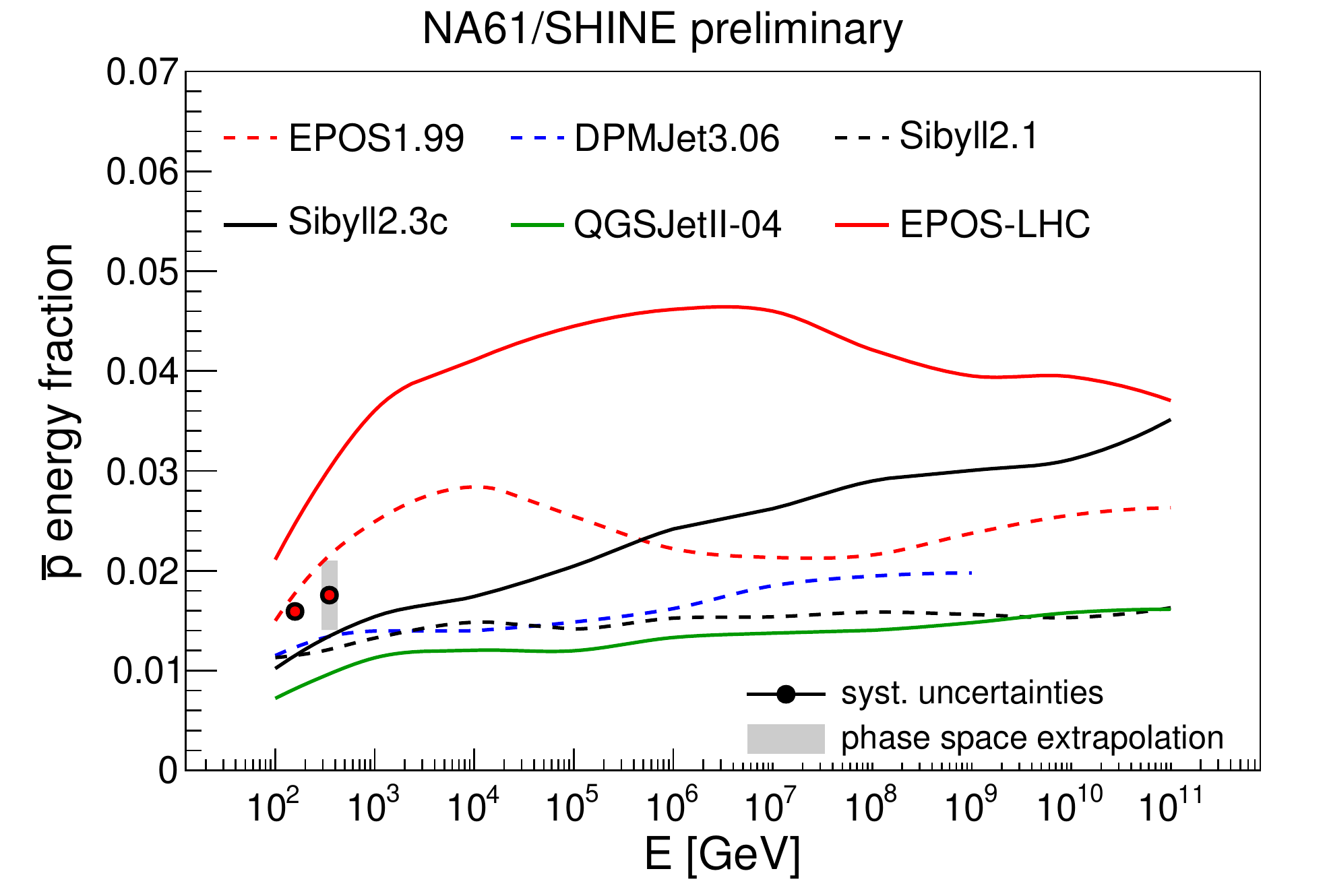}
    \includegraphics[width=0.51\linewidth]{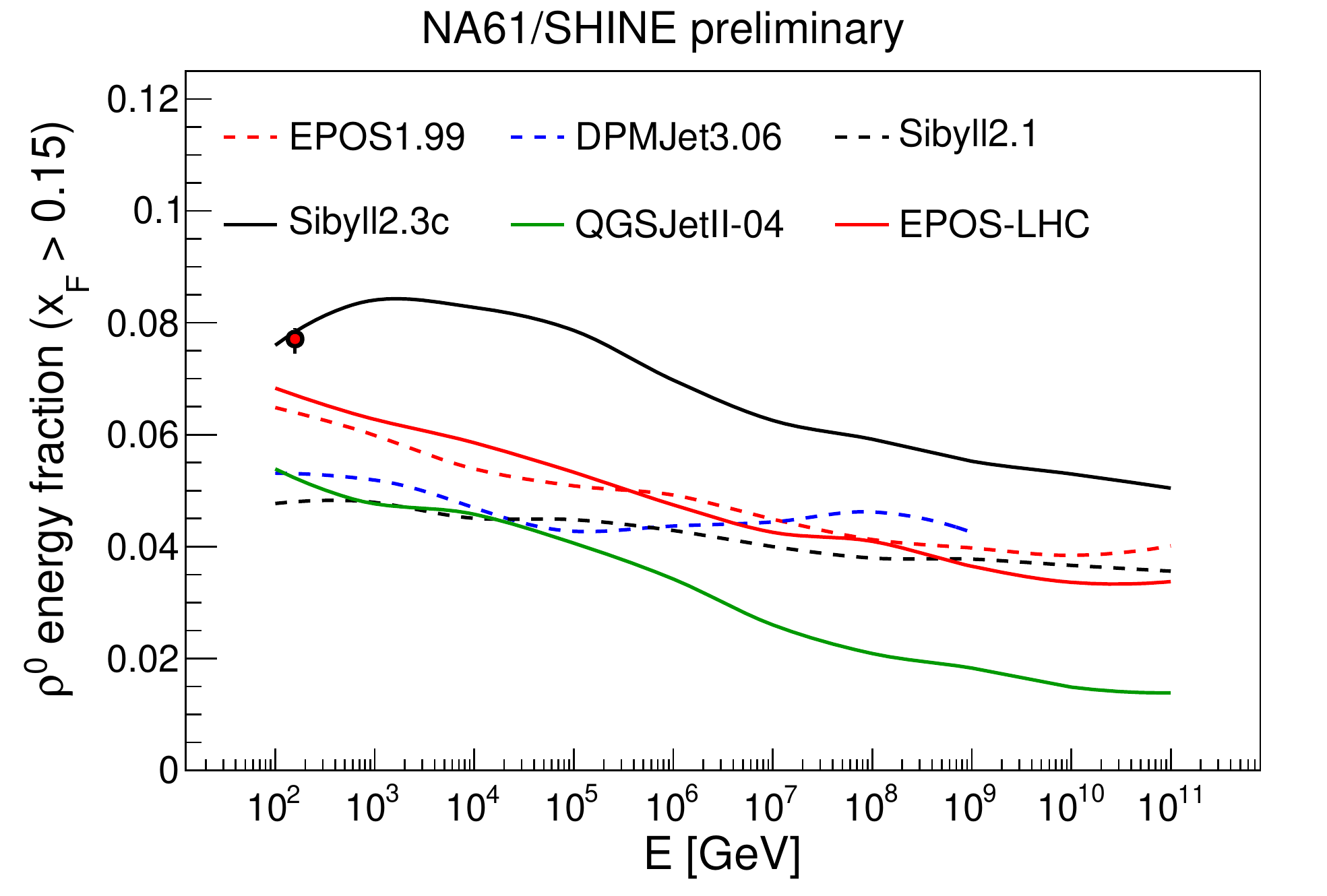}
    }
    \caption{Energy fraction transferred to anti-protons (left) and
    \rhoz-mesons (right) in $\pi$-C collisions as measured by \NASixtyOne (data points)
    and as predicted by hadronic interaction models over the whole
    range of beam energies relevant for air showers. Figures taken from Refs.~\cite{Prado:2018wsv,Unger:2019nus}.}
    \label{fig:efrac}
\end{figure}

Since hadronic interactions in an air shower span over many orders of magnitude in energy, there are also opportunities to improve our knowledge at lower values of $\sqrtsnn$ reached by fixed-target experiments~\cite{Meurer:2005dt}. The \NASixtyOne experiment~\cite{NA61:2014lfx} at the \ac{SPS}, the pre-accelerator of the \ac{LHC}, has measured hadron production in p-p, $\pi$-C, and p-C collisions, where carbon is used as a proxy for air. The corresponding measurements of the forward $\rhoz$ and anti-proton production are shown in \cref{fig:efrac}, where the differential cross section was integrated over the energy of the secondary particles~\cite{NA61SHINE:2017vqs,Prado:2017hub}. The $\rhoz$ production is important since it is an alternative to producing a $\piz$-meson in the charge-exchange reaction $\pi^- + p \to \piz + n + X$. An increase of the $\rhoz/\piz$ ratio subsequently also enhances the muon number in air shower simulations. In addition, anti-protons are a measure of baryogenesis in the air shower, which also increases the muon number. The compounding effect over many interactions leads to an increase at a level of 60\,\% in the muon number produced in \acp{EAS} run with the recent version of \sibyll{2.3d}~\cite{Riehn:2019jet} over \sibyll{2.1} without these effects. This has not resolved the muon discrepancy observed in air showers, however, it demonstrates the impact and importance of dedicated studies of the hadron composition also at lower center-of-mass energies, $\sqrtsnn$.

At the \ac{LHC}, fixed-target experiments are performed by LHCb with the SMOG device \cite{LHCb:2014vhh} which injects small amounts of gas into the detector. The fixed-target data has been used to place limits on the intrinsic charm inside the proton \cite{LHCb:2018jry} and to measure the anti-proton production cross section in p-He collisions \cite{LHCb:2018ygc}, which is also an important ingredient to compute a background in searches for primordial anti-matter.

\subsection[Beyond Standard Model physics with UHECRs]{Beyond Standard Model physics with UHECRs}\label{sec:PartBSMDM}

\ac{UHECR} observatories also offer unique opportunities to probe physics beyond the Standard Model and provide complementary constraints for various dark matter scenarios. Examples are searches for \acf{LIV} and dark matter in \ac{UHECR} observations. While \ac{LIV} would have an effect on the propagation of \acp{UHECR} in the Universe and the development of air showers on Earth, the origin of \acf{SHDM} particles can be connected to inflationary cosmologies and their decay to instanton-induced processes. These decays would produce primarily a cosmic flux of extreme energy neutrinos and photons and their non-observation sets restrictive constraints on the gauge couplings of the dark sector, for example. In the following, these exotic scenarios will be discussed in further detail.

\subsubsection[Lorentz invariance violation in EASs]{Lorentz invariance violation in EASs}\label{sec:LIV}

The variety of air shower observations described in the previous sections strongly constrains exotic explanations of the Muon Puzzle, such as Lorentz invariance violation (\ac{LIV}), for example. The effects of \ac{LIV} can be written as a Taylor expansion of the generic \ac{MDR}, which relates the energy~$E_i$ of a particle~$i$ (with mass $m_i$) to its momentum~$p_i$, as 
\begin{align}
    E_i^2 &= m_i^2 +  
    (1 + \delta_i^{(0)}) p_i^2 
    + \delta_i^{(1)} p_i^3
    + \delta_i^{(2)} p_i^4
    + \cdots \label{eq:liv_delta} \\
    &= m_i^2 + 
    (1 + \eta_i^{(0)}) p_i^2 
    + \frac{\eta_i^{(1)}}{M_\text{P}} p_i^3
    + \frac{\eta_i^{(2)}}{M_\text{P}^2} p_i^4
    + \cdots \; ,
    \label{eq:mdr}
\end{align} 
where $M_\text{P} \approx 1.22\times 10^{28}\,\mathrm{eV}$~is the Planck mass and $\eta_i^{(n)}=\delta_i^{(n)}M_\text{P}^n\ll 1$ gives the scaling of the deviation from the standard model. Searches for non-zero values of $\eta_i^{(n)}$ can be used to constrain \ac{MDR} coefficients for various particle types (see e.g., Refs.~\cite{PierreAuger:2021tog,Saveliev:2011vw,Diaz:2016dpk,Anchordoqui:2017pmf,Klinkhamer:2017puj,GuedesLang:2017sfl,Torri:2020fao,Duenkel:2021gkq,PierreAuger:2021mve}). \ac{LIV} can cause certain photo-nuclear processes which are allowed for non-relativistic nuclei to be forbidden for an ultra-relativistic ones, or vice versa (and some processes can be allowed in both cases but with different rates). This can affect the propagation of \acp{UHECR}, as described in detail in \cref{sec:Astrophysics}, and it can cause deviations of the \ac{EAS} development from standard predictions assuming special relativity. 

For instance, the decay of a photon into an electron-positron pair is kinematically forbidden in special relativity, but in the case of an isotropic nonbirefringent \ac{LIV}, characterized by one dimensionless parameter~$\kappa$, it can be allowed if~$\kappa < 0$ for photons with energies greater than~$2 m_e \sqrt{(1-\kappa)/(-2\kappa)}$, where $m_e$ is the electron mass. This would speed up the development of air showers, which results in a \xmax higher in the atmosphere than in the Lorentz-invariant case. Hence, comparing air shower simulations with experimental data from the Pierre Auger Observatory results in an upper bound of $-\kappa < 6 \times 10^{-21}$ at the $98\%$~C.L.\ \cite{Duenkel:2021gkq}. 
Each neutral pion in an \ac{EAS} normally decays into two photons which initiate electromagnetic sub-showers, thereby transferring energy from the hadronic to the electromagnetic component of the \ac{EAS}. However, in the presence of a negative $\eta_\pi^{(n)}$, the decay becomes kinematically forbidden above a certain pion energy, so that such pions continue the hadronic cascade instead.  The final result of this is an \ac{EAS} EAS with larger muon content and reduced muon shower-to-shower fluctuations than in the Lorentz-invariant case. By comparing \ac{EAS} simulations with Auger data a preliminary upper bound for $-\eta_\pi^{(1)}$ of~$6 \times 10^{-6}$ is obtained at the $90.5\%$~C.L.~\cite{PierreAuger:2021mve}.

Although many more attempts have been made to describe muon production in extensive air showers correctly, including various exotic scenarios (see e.g., Refs. \cite{ Ulrich:2010rg, Albrecht:2021cxw, Farrar:2013sfa} for a more comprehensive discussion), large discrepancies in the description of muons remain. In order to understand these discrepancies and to discover their origin, complementary measurements from collider experiments are required, as described in \cref{sec:PartUHECROutlook}. In addition, \ac{LIV} can have an impact on the propagation of cosmic rays that can potentially be observed in \ac{UHECR} observatories. The effects of \ac{LIV} on \ac{UHECR} propagation will be further discussed in \cref{sec:UHECR_transport}.

\subsubsection{Super-heavy dark matter searches and constraint-based modeling of Grand Unified Theories}
\label{sec:PartSHDM}

Currently, the concordance model used in Big-Bang cosmology is the $\Lambda$CDM model, which states that the Universe is $\simeq$\,$13.8{\times}10^9$ years old and made up of $\simeq$\,$5$\% baryonic matter, $\simeq$\,$26$\% dark matter, and $\simeq$\,$69$\% dark energy~\cite{Planck:2018nkj}. Although multiple hypotheses have been proposed to describe dark matter, the leading scenario is one in which dark matter consists of particles that only engage in gravitational interactions or interactions that are as weak or weaker than the scale of the weak nuclear force, i.e., \acp{WIMP}. This arises from the observation that the present-day \ac{WIMP} relic abundance determined by the freeze-out condition in the early Universe, combined with the expected annihilation cross section for a new particle with weak-scale interactions, is surprisingly close to the present-day abundance of dark matter (the so-called ``\ac{WIMP} miracle,'' see e.g., Ref.~\cite{Jungman:1995df}). However, \acp{WIMP} have thus far escaped detection, whether by underground direct detection experiments~\cite{MarrodanUndagoitia:2015veg} or through indirect astrophysical searches~\cite{Gaskins:2016cha}. Furthermore, \acs{LHC} experiments have yet to observe new physics at the TeV scale~\cite{Rappoccio:2018qxp}. Overall, the various null results push the originally expected masses towards larger values and the couplings towards weaker ones. This gives increasingly strong constraints for the \ac{WIMP} scenario and motivates searching elsewhere for an explanation for dark matter. 

Models of \ac{SHDM} particles, first put forward in the  1990s~\cite{Ellis:1983ew, Nanopoulos:1982bv, Khlopov:1984pf, Olive:1984bi, Ellis:1990iu, Ellis:1990nb, Berezinsky:1997hy, Chung:1998zb, Kuzmin:1997jua, Birkel:1998nx, Berezinsky:1998ft}, were recently revived as an alternative to \ac{WIMP} scenarios~\cite{Garny:2015sjg,Ellis:2015jpg,Dudas:2017rpa,Kaneta:2019zgw,Mambrini:2021zpp}. In fact, if the assumption of naturalness is relaxed, precision measurements carried out at the \acs{LHC} may even suggest the existence of \ac{SHDM}. For instance, \ac{LHC} measurements of the masses of the Higgs boson and the top quark signify the energy scale, $\Lambda_\mathrm{I}$, above which new physics is necessary to stabilize the meta-stable Standard Model vacuum state as indicated by analysis of the running of the Higgs quartic self-coupling parameter, $\lambda$, with energy~\cite{Buttazzo:2013uya,Alekhin:2012py,Bednyakov:2015sca}. Once propagating all uncertainties stemming from the input values of the observables, this energy is found to be $\sim 10^{10}\mathrm{-}10^{12}~\mathrm{GeV}.$ Furthermore, in order to guarantee the survival of the current meta-stable state throughout the history of the Universe, the rate of decay of the meta-stable vacuum into a lower-energy vacuum state must be slow, both today and in the past~\cite{Markkanen:2018pdo}. Hence, the running of $\lambda$ must slow down above the energy scale $\Lambda_\mathrm{I}$, possibly even up to the scale of the Planck mass, $M_{\mathrm{P}}$. As such, the search for new physics in the intermediate scale between $\Lambda_\mathrm{I}$ and $M_{\mathrm{P}}$ is well motivated in the context of current \ac{LHC} measurements, and it is in this energy range that the lightest particle of \ac{SHDM} models in the spectrum of the hidden sector can be found.

If \ac{SHDM} particles do exist, they may decay into Standard Model particles, secondary products in the form of \ac{UHE} photons, neutrinos, and nucleons that can be detected by \ac{UHECR} observatories. Furthermore, if the \ac{UHECR} flux does include a component arising from \ac{SHDM}, there should be a signature anisotropy signal reflecting the dark matter distribution in the Galaxy. Thus far, searches for \ac{UHE} photons and neutrinos have produced only upper limits ~\cite{Rautenberg:2021vvt}, and the strongest anisotropy signals show no signs of a galactic component at \acp{UHE} (see also \cref{sec:Anisotropy}). 
These null results translate into stringent constraints on \ac{SHDM} parameters. 

\begin{figure}[tb]
    \centering
    \mbox{\hspace{-1em}\includegraphics[width=0.53\textwidth]{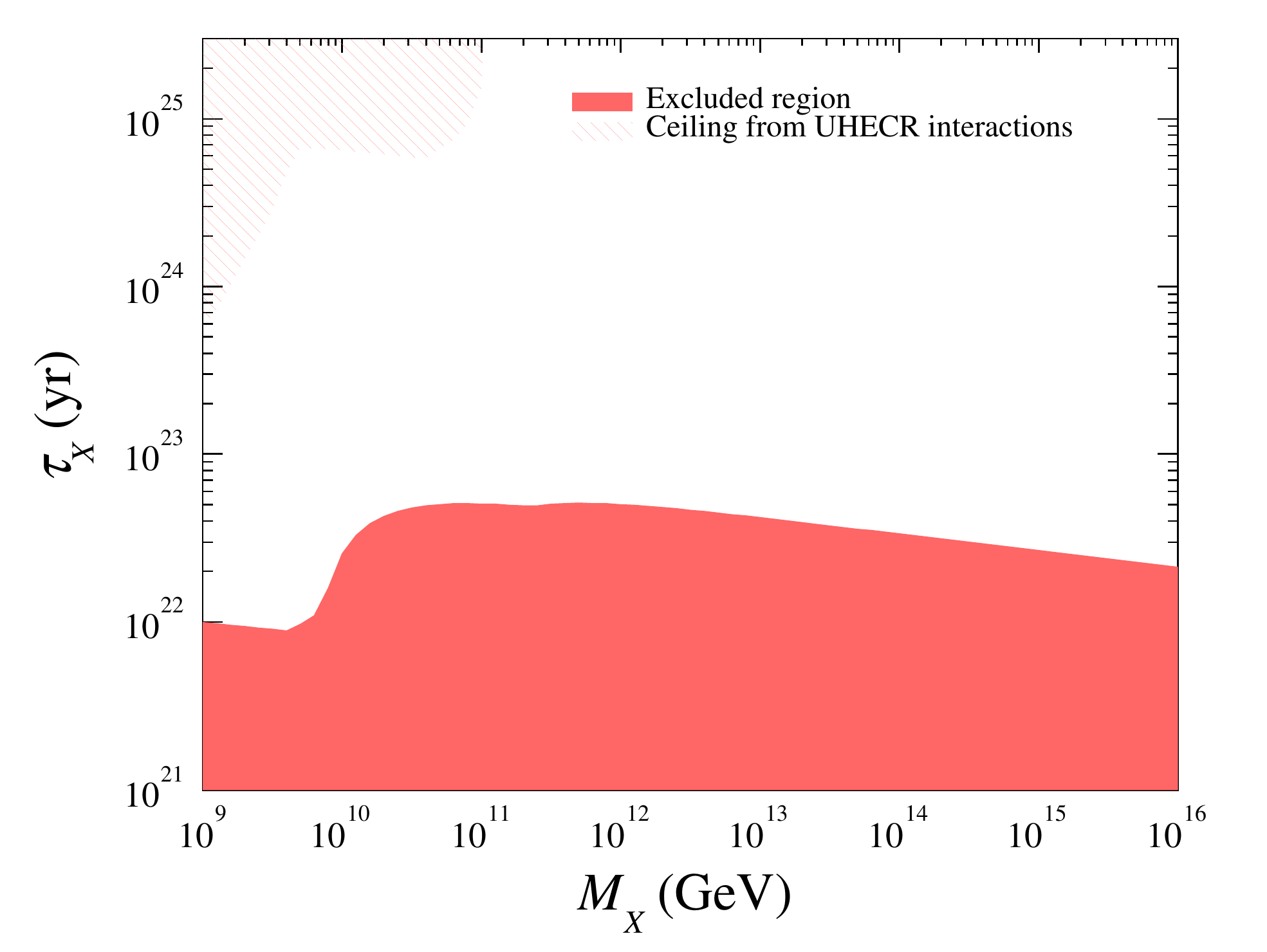}\hspace{-1em}
    \centering\includegraphics[width=0.53\textwidth]{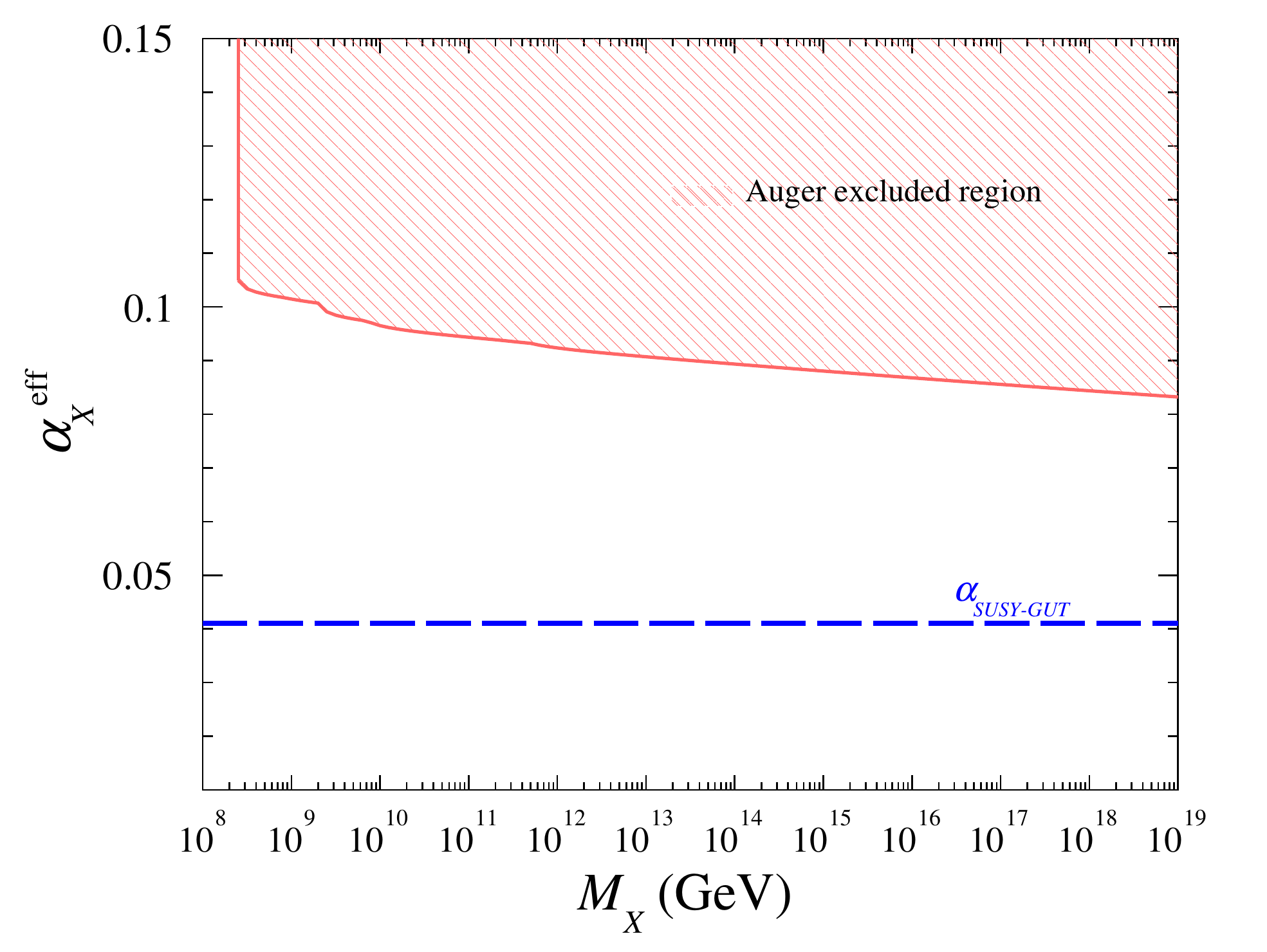}}
    \caption{Constraints on the mass and lifetime of \ac{SHDM} particles as obtained from the upper limits on photons~\cite{Berat:2022iea} (left) and upper limits at 95\% C.L. on the effective coupling constant of a hidden gauge interaction as a function of the mass for a dark matter particle decaying into $q \bar q$~\cite{ThePierreAuger:2022} (right). For reference, the unification of the three \acs{SM} gauge couplings is shown as the blue dashed line in the framework of supersymmetric \acp{GUT}~\cite{ParticleDataGroup:2020ssz}. Figures taken from Refs.~\cite{Berat:2022iea, ThePierreAuger:2022}.}
    \label{fig:mtau0}
\end{figure}

\cref{fig:mtau0} (left) shows current limits obtained from searches for \ac{UHE} photons by the Pierre Auger Observatory~\cite{ThePierreAuger:2022}, on the lifetime, $\tau_X$, and effective coupling constant, $\alpha_X$, of \ac{SHDM} particles, as a function of their mass, $M_X$. It is seen that particles are required to be stable more than a few $10^{22}$~yr for a wide range of masses. The hatched region corresponds to a constraint induced by cosmogenic photon fluxes expected from the interactions of \acp{UHECR} with the matter in the Galactic disk~\cite{Berat:2022iea} or with the background photon fields in the Universe~\cite{Bobrikova:2021kuj}. 

Using a generic form of the decay rate of the $X$ particle, constraints on the coupling constant and the dimensions of the interaction operators can also be obtained. For a given energy scale $\mathcal{E}$, the upper limit on the coupling constant $\alpha_X$ can be calculated as a function of the mass $M_X$ by fixing a specific value of the dimension $n$ of the operator responsible for the perturbative decay. It results that the mass of the particles can, in principle, approach $\mathcal{E}$ for very large dimension values of $n>100$ and/or for allowing for masses that approach $\mathcal{E}$. It is difficult to find fundamental motivations to justify such a fine-tuning. By contrast, instanton-induced decays are not that strongly constrained by current data and are an interesting possibility to further explore. Constraints on $\alpha_X$ of less than around $0.09$--$0.10$ can be obtained for a wide range of masses $M_X$ from data taken by the Pierre Auger Observatory~\cite{ThePierreAuger:2022}, as shown in \cref{fig:mtau0} (right).

\subsection[Outlook and perspectives]{Outlook and perspectives: The future of particle physics measurements at \ac{UHECR} observatories}\label{sec:PartUHECROutlook}

In order to understand the discrepancies observed in current air shower simulations, both precise air shower data, as well as dedicated measurements at colliders are required. In the following, the future prospects for \ac{EAS} and collider measurements in the next decade will be discussed that will help to understand multi-particle production in the forward region in order to discover the origin of the Muon Puzzle and enable detailed studies of elementary particle physics processes with \acp{EAS}. Moreover, the perspectives for future searches of macroscopic dark matter and nuclearites with \ac{UHECR} observatories will also be discussed.

\subsubsection[Air shower physics and hadronic interactions]{Air shower physics and hadronic interactions}

Previous studies of GeV muons in \acp{EAS} have been focused on measurements of the average muon number and very recently the muon number fluctuation (see \cref{sec:UHECRPartSyn}). Higher moments of the muon number distribution have not yet been measured. Similarly to the relation of the \xmax with the p-Air inelastic cross section, the slope of the tail of the muon number distribution in p-Air showers is a direct reflection of the high-energy $\pi^0$ production cross section \cite{Cazon:2020jla}.

\begin{figure}[tb]
    \centering
    \includegraphics[width=0.7\textwidth]{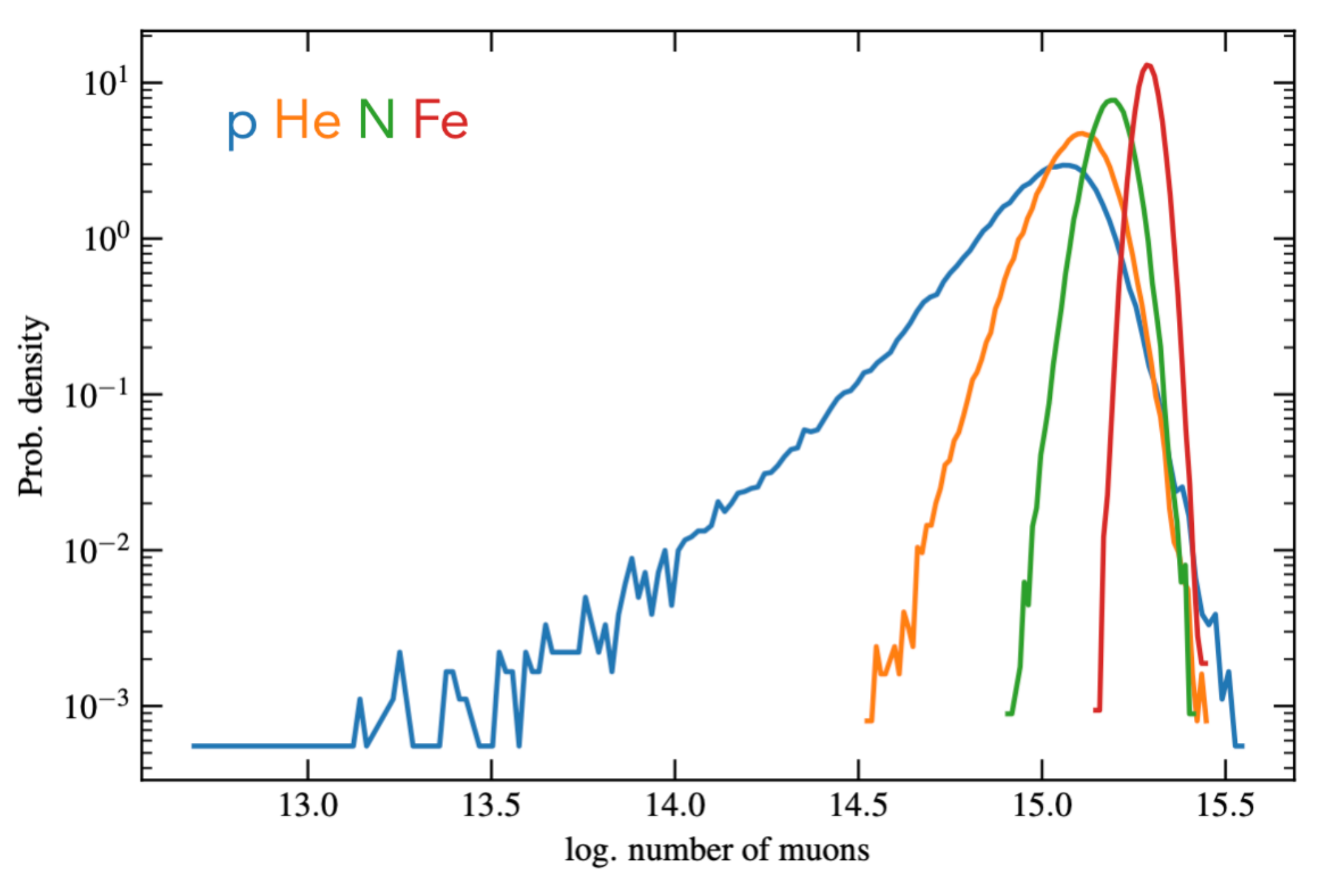}
    \caption{Shower-to-shower distribution of $N_\mu$ for different primaries (adapted from Ref.~\cite{Cazon:2018gww}). The shape of the distributions carries information on the multi-particle production of the first interactions of the \ac{EAS}. The left tail of the $N_\mu$ distribution for proton primaries has been demonstrated to be a direct transformation of the high-energy tail of the $\pi^{0}$ production cross section.}
    \label{fig:Nmu}%
\end{figure}

In general, the full event-to-event muon distributions encode important information about different aspects of the hadronic interaction of \acp{EAS} which will be studied throughout the next decade. \cref{fig:Nmu}, for instance, shows the shower-to-shower distribution of $N_\mu$ for different primary masses, which could potentially be probed in future \ac{EAS} observatories. A fit of the hadronic model predictions to the observed $N_\mu$ distributions must be consistent with the \xmax fits which have been used to produce the different primary abundance. These studies will provide important tests for current hadronic interaction models and are crucial to further constrain possible explanations for the Muon Puzzle in \acp{EAS}~\cite{Cazon:2018gww}.

\clearpage

The typical resolution of muon measurements is around $15\%-20\%$, given by the experimental variance with respect to its true physical value. This variance is of the same order as the variance of physical fluctuations of proton showers, but much larger than the variance of iron showers, which is $3-4\%$. If the experimental resolution is larger than the size of the physical muon fluctuations, it is difficult to measure higher orders of the physical muon fluctuations. Therefore, future experiments should aim for achieving resolutions better than $10\%$. Larger muon detectors and detectors closer to the shower axis, which become operational within the next decade, will presumably already improve the experimental resolution to around $15\%$ or better (see also \cref{sec:FutureDetectors}). To precisely measure the physical muon distributions, the experimental resolution has to be unfolded from the raw experimental distributions. Hence, stable experimental event resolutions are crucial. Dedicated studies of the uncertainties of existing detectors within the next decade will improve the stability of the measurements and minimize, or at least account for their fluctuations with time. Both efforts will significantly improve the unfolding of the muon distributions and contribute to the understanding of muon production in \acp{EAS}.

In addition, ongoing multi-hybrid air shower measurements will allow a better understanding of the origin of the Muon Puzzle. For example, the simultaneous measurement of the shower energy and muon content of \acp{EAS} using the fluorescence detectors and improved surface detectors of the Pierre Auger Observatory (now including an additional layer of scintillator allowing better electromagnetic to muonic component separation and even direct muon detection in a sub part of the array) will enable studies of the observed discrepancies in a non-degenerated way. Since the muon number depends on both the energy and the mass of the cosmic ray, independent measurements of these two parameters are a key element to quantify precisely the muon deficit in simulations. For instance, the radio extension of the Pierre Auger Observatory~\cite{PierreAuger:2021ece} or the GRANDProto300~\cite{Decoene:2019sgx} experiment will add new measurements for both the mass and the energy, testing a new technology that could replace the fluorescence measurements which are limited by their duty cycle. These measurements suffer from less theoretical uncertainties and can reach an energy resolution of about $10\%$~\cite{Glaser:2017fgg}. Thereby, they will also have direct impact on the resolution of muon measurements, providing high-precision data to study multi-particle production in \acp{EAS}.

The ability to measure the number of muons at the ground, on an event-by-event basis, with high statistics will allow to study the distribution features in a more in-depth way. For instance, the fluctuations of this distribution have been shown to be mainly connected to the energy partition of the first interaction~\cite{Cazon:2018gww}. Moreover, the muon number distribution for proton-induced showers exhibits a quasi-exponential tail for showers with low muon content. In Ref.~\cite{Cazon:2020jla}, it has been shown that the slope of this tail has a direct link with the high energy tail of the $\pi^{0}$ production cross section of the first ultra-high-energy interaction.

New techniques based on neural networks also provide new insight on the data which allow to extract direct information on hadronic interactions using correlations between different observable (e.g., the multiplicity and neutral pion fraction distributions extracted from the \xmax-$N_\mu$ correlation in proton induced showers~\cite{Goosthesis}). The hybrid approach will also allow simultaneous measurements of the longitudinal profile of the electromagnetic and muonic shower components, i.e., the shower maxima \xmax and \xmumax, which enables further insights in the inner degrees of freedom of \acp{EAS}, the former being mostly linked to the first interactions while the latter is driven by the full hadronic shower evolution~\cite{Ostapchenko:2016bir} and thus by pion-air interactions which are hardly accessible in laboratory experiments. For instance, \xmumax is very sensitive to the diffractive mass distribution which has been set in the models to the same value than in the case of proton interactions because of the lack of experimental data~\cite{Pierog:2015ifw}. But with a measurement of muons from air showers with a good timing resolutions, a good muon production depth measurement could be achieve to constrain this fundamental parameter. A better understanding of \xmumax will further reduce the uncertainties on the theoretical prediction of \xmax and thus on the mass composition of \acp{UHECR}~\cite{Ostapchenko:2019ubd}.

Moreover, the IceCube Neutrino Observatory is able to measure the muon content in \acp{EAS} at two vastly different energies using its surface and deep ice detectors. The simultaneous in-ice high-energy ($>$ few $100\,\mathrm{GeV}$) muon measurements and the estimation of the GeV muon content at the surface provide unique tests of hadronic interactions in the forward region and can constrain hadronic interaction models based on their predicted muon energy spectra. New surface detector extensions of IceCube will become operational within this decade \cite{Haungs:2019ylq}. These include new scintillator and radio antenna arrays, which will help to separate the GeV muon content in air showers, reduce systematic uncertainties of the current muon measurements, and extend the measurements towards higher cosmic ray energies. The radio array will allow an independent measurement of the radio emission of the \ac{EAS}, providing calorimetric measurements of the shower energy and measurements of \xmax, enabling multi-hybrid event detection in a unique phase space.

Indirect measurements of the muon spectrum in \acp{EAS} can also be performed by many experiments, using the zenith angle evolution of various experimental observables. As shown in Ref.~\cite{Espadanal:2016jse}, not only the muon number at the ground will have a strong evolution with the shower inclination, due to its attenuation through the atmosphere, but also the maximum of the muon production depth, the \xmumax evolution with zenith angle, will be differently affected depending on the muon energy spectrum. These measurements will thereby yield complementary data that is sensitive to the muon spectrum, providing additional tests of hadronic interaction models.

Furthermore, improved analysis techniques based on machine learning approaches~\cite{PierreAuger:2021nsq} are expected to exclude, or strongly constrain models of muon production in hadronic interactions throughout the next decade. This combination of new but well understood experimental methods and new analysis techniques, will lead to very precise measurements of the muon component in air showers and subsequently to an understanding of the origin of the Muon Puzzle. 

\subsubsection[Upcoming collider measurements]{Upcoming collider measurements}

The \ac{LHC} will take data with high-luminosity beams in the coming decade at 14\,TeV in \pp collisions, while measurements at even higher energies of 28\,TeV are only expected in the 2040s~\cite{FCC:2018bvk}. These runs will primarily improve the accuracy of charm and bottom production cross sections, which play an important role as a background for astrophysical neutrino searches. Future studies of unflavoured hadrons will benefit indirectly from the precise calibrations of the experiment that will become possible. In addition, LHCf plans to study strangeness production at zero-degree angles based on the decay $K^0_S \to 2\pi^0 \to 4\gamma$ with upgraded detectors~\cite{LHCf:2021qrz}. 

Of particular importance for air shower physics, and complementary to \ac{EAS} measurements, are also the approved plans to accelerate oxygen beams in order to measure \pO and \OO collisions in 2023 \cite{Citron:2018lsq,Bruce:2021hii}. The most common interaction in an air shower is $\pi$-N for which \pO collisions are an excellent reference. Current state-of-the-art models show considerable variance in their predictions of hadron production in \pO collisions, despite being tuned to \pp data, which reflects the theoretical uncertainties in extrapolating hadron production from a \pp reference system to a proton-ion collision. Together with the essential direct measurements in \pO, the study of both \pp and \pPb data is important to potentially detect simple scaling laws for production cross sections and the hadron composition in hadron-ion collisions. A model variance of $20\%$ is currently found in \pO collisions, which is expected to be strongly reduced with the upcoming \pO data. The shape of the hadron rapidity spectrum depends on the pomeron approach that is used in the hadronic models, and measurements over a wide pseudorapidity range are able to discriminate between the two main approaches in use (see Ref.~\cite{Albrecht:2021cxw} for details). In the forward region, yields of identified hadrons, pions, kaons, and protons, will be studied by LHCb in \pp collisions at $13\,\mathrm{TeV}$ and \pPb collisions at 8.16\,TeV, where other experiments lack particle identification capabilities. 

The \ac{LHC} experiments, in particular LHCb and LHCf, will determine in the coming decade whether the universal strangeness enhancement seen by ALICE \cite{ALICE:2016fzo} at mid-rapidity is also present in the forward region, by studying the hadron composition as a function of the track density in the event. A previous study by CMS \cite{CMS:2019kap} was not conclusive on this point, since the experimental uncertainties at the level of $20\%$ did not allow to detect the small effect from strangeness enhancement. LHCb will study beam-gas interactions with its upgraded SMOG2 device that allows for higher gas densities and more target gasses, including nitrogen and oxygen \cite{Barschel:2020drr}. With LHCb in fixed-target mode, it will be possible to study hadron production at $\sqrtsnn = 115\,\mathrm{GeV}$ at mid-rapidity $-2.5 < \eta_\mathrm{cms} < 0.5$ in the nucleon-nucleon center-of-mass frame.

In addition, the FASER~\cite{Feng:2017uoz,FASER:2018ceo, FASER:2018bac, FASER:2018eoc} and FASER$\nu$~\cite{FASER:2020gpr, Abreu:2019yak, FASER:2021mtu} experiments will perform measurements of particle production in the far-forward region at the \ac{LHC}, at pseudorapidities of $\eta_\mathrm{cms}> 7$. As shown in \cref{fig:eta-spectra}, this is the main rapidity range relevant for particle production in \acp{EAS}. Shielded by $100\,\mathrm{m}$ of rock and concrete from the ATLAS interaction point, FASER and FASER$\nu$ will be able to measure lepton fluxes up to TeV energies and higher. Thereby, these experiments will provide estimates of the $K/\pi$ ratio as a probe of forward strangeness production in hadronic interactions~\cite{Anchordoqui:2022fpn}, for example. In addition to the current efforts to measure particle production in the forward region at the \ac{LHC}, the proposed \acf{FPF} could provide further important data to test hadronic interactions in the forward region, as discussed in \cref{sec:FuturePartPhysUHECR}.

The complementary measurements of multi-particle production in \acp{EAS} and at the \ac{LHC} will strongly constrain hadronic interaction models in a large phase space. This process is already underway and in 10 years a large variety of high-statistic data will be available, yielding stringent constraints, leaving very little room for the interpretation of the data. Thus, these interdisciplinary efforts can be expected to reveal the origin of the Muon Puzzle within the next decade, opening a new era for particle physics measurements with \ac{EAS} observatories, as discussed in \cref{sec:FutureHadronicInteractions}.

\subsubsection{Searches for macroscopic dark matter and nuclearites}
\label{sec:macro}

In addition to the \ac{SHDM} the scenarios discussed in \cref{sec:PartSHDM}, macroscopic dark matter particles (macros) represent a broad class of candidates that provide an alternative to conventional particle dark matter. There is considerable evidence for dark matter \cite{Zyla:2020zbs}, and a wide range of macro masses, M, and cross sections, $\sigma_m$, that is not excluded yet could potentially still provide the entire observed dark matter in the Universe (see e.g., Ref.~\cite{Starkman:2020sbz} for a comprehensive review).

\begin{figure}[b]
    \centering
    \includegraphics[width=0.73\textwidth]{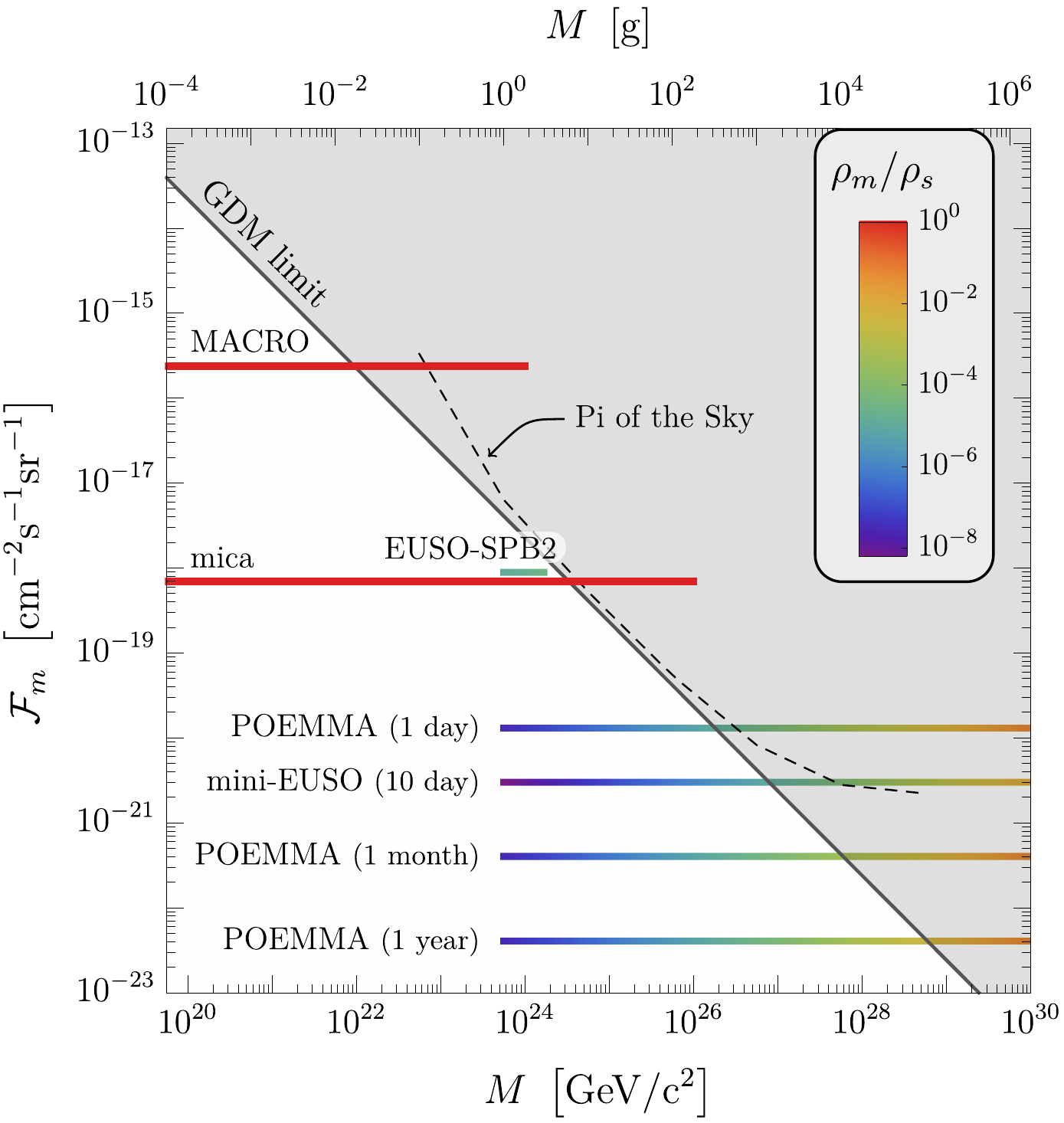}
    \caption{The projected $90\%$ confidence level upper limit on the macro flux, $\mathcal{F}_m$, as a function of the macro mass, $M$, and the macro density, $\rho _m$, resulting from null detection over different time spans of acquired data for \acs{POEMMA}~\cite{POEMMA:2020ykm}, \acs{EUSO}-SPB2~\cite{Paul:2021bhh}, Mini-\acs{EUSO}~\cite{Mini-EUSO:nuclearites}, and other experiments~\cite{Price:1988ge, MACRO:1999ijn, Piotrowski:2019}. The \acf{GDM} limit is indicated for comparison.}
    \label{fig:macro_parameter_space}
\end{figure}

Over the next decade, several experiments will have the ability to probe more regions of macro parameter space. In particular, bolide observation experiments are poised to examine a large chunk of this parameter space~\cite{Sidhu:2019fgg} because macros with sufficiently large $\sigma_m$ will produce a distinct luminous trail across the sky. Camera networks specifically designed for observing macros and interstellar meteoroids~\cite{Abe:2021ugf, DIMS:2021ipc} will likely probe a similar region of parameter space. Optical observations have been already used on setting limits on the flux of macros with strange-quark matter density, exemplified by nuclearites~\cite{Piotrowski:2019}. Under the right circumstances, macros could even initiate unique lightning strikes~\cite{Starkman:2020sbz,Cooray:2021dvp}.

\ac{UHECR} experiments have the potential to probe a unique part of the parameter space~\cite{POEMMA:2020ykm,SinghSidhu:2018oqs}. In contrast to relativistic cosmic rays, a macro would move much slower and will not generate an air shower. One caveat of current \ac{UHECR} experiments is that existing (and possibly future) cosmic ray detectors would require software or hardware accommodations to detect the more slowly moving macros. Such events would not currently be flagged by most of the existing \ac{UHECR} experiments because macros move much more slowly than relativistic cosmic rays, with the exception of the (Mini) \ac{EUSO}~\cite{Mini-EUSO:nuclearites}.

The detection of a luminous trail from macros would also shed some light on the light emission mechanism involved. Recent theoretical works \cite{SinghSidhu:2018oqs,Anchordoqui:2021xhu} suggest, that the intensity of a trail for nuclearites may be much lower than described in Ref.~\cite{DeRujula:1984axn}. In such a case, a different macro candidate, allowing for a larger cross sections is needed, for example so-called dark-quark-nuggets~\cite{Bai:2018dxf}. Here, the larger cross sections are obtained by allowing densities much smaller than the nuclear density. \cref{fig:macro_parameter_space} shows the expected sensitivities as a function of the macro mass, $M$, for Mini-\acs{EUSO}~\cite{Mini-EUSO:nuclearites}, planned orbital \acs{POEMMA}~\cite{POEMMA:2020ykm} (see also \cref{sec:POEMMA}), and constructed air-borne \acs{EUSO}-SPB2~\cite{Paul:2021bhh} \acs{UHECR} experiments, estimated using the procedure outlined in Ref.~\cite{Paul:2021bhh}, with the macro densities in the range of $10^6 < \rho _m   / \mathrm{(g/cm^3)} < 10^{15}$ and the nuclear density $\rho _s \sim 3.6 \times 10^{14}\  \mathrm{g/cm^3}$.
\fakesection{Astrophysics at the Energy Frontier}
\vspace{3cm}
{\noindent \LARGE \textbf{Chapter 4}}\\[.8cm]
\textbf{\noindent \huge Astrophysics at the Energy Frontier:}\\[3mm]
\textbf{\LARGE Pinpointing the most extreme processes in the Universe}
\label{sec:Astrophysics}
\vspace{1cm}


The evolving observational picture motivates new theoretical frameworks for understanding the origins of \acp{UHECR} and their journey through the cosmos. Answering the outstanding questions of the \ac{UHECR} picture will require the enhanced capabilities of a new generation of \ac{UHECR} experiments, as well as leveraging insights brought about by continued progress in supporting areas of astrophysics and the emerging multi-messenger landscape. The high-energy astrophysics community remains abreast with the evolving observational picture and has developed a wide variety of new exciting models that will be further tested by the data collected over the next decade.

\subsection[Open questions in UHECR astrophysics]{Open questions in UHECR astrophysics: The quest for a comprehensive interpretive framework}

The observations detailed in the previous chapters (\ac{UHECR} spectrum, mass composition and arrival directions) are central to identifying the cosmic-ray sources, and to understanding the physical processes particles undergo. Multi-wavelength and multi-messenger observations of secondary gamma-rays and neutrinos as well as associated gravitational waves also nourish the interpretations and are central to validate them. This section will summarize the tentative comprehensive picture that is emerging today and the many open questions that remain~\cite{AlvesBatista:2019tlv}.


\subsubsection{Galactic to extragalactic transition}

Cosmic rays below $10^{16.5}\,$eV are likely created and contained in the Galaxy \cite{Strong:2007nh,Fermi-LAT:2013iui,Amato:2021yyv}, and the large-scale anisotropy measurements from Auger imply that cosmic rays above $8\times 10^{18}$\,eV originate in extragalactic sources \cite{PierreAuger:2020fbi}. Therefore, a transition between Galactic to extragalactic components should happen somewhere within these two decades in energy. This transition region is a well of information, holding the key to identifying the {\color{red} highest energy} Galactic cosmic-ray sources, and to understanding the operating acceleration mechanisms. The energy at which the extragalactic component(s) emerges, and the exact spectral shape and mass composition around EeV energies are essential information to understand the injection, acceleration, interactions and magnetic deflections experienced by \acp{UHECR}. 

A possible picture that has acquired coherence over the past years is that the transition happens at the second knee, around $3$--$4$ $\times 10^{17}$\,eV (see e.g., Refs.~\cite{Haungs:2015ema,Globus:2015xga, 2019EPJWC.21004003K,Kaapa:2021Fk,Mollerach:2018lkt}). This is supported by dipole anisotropy data and the spectra of different mass groups (see \cref{sec:EnergySpectrum}). In particular, the iron spectrum cuts off in the range of $2$--$6$ $\times 10^{17}\,$eV, which can be interpreted as the signature of the end of the Galactic contribution \cite{Candia:2003dk}. On the other hand, the emergence at $\sim$\,$6\times 10^{17}\,$eV of a lighter component with a low level of anisotropy could signify the emergence of an extragalactic component since at these energies, lighter nuclei originating from the Galaxy should exhibit some level of anisotropy~\cite{Giacinti:2011ww}. For a lighter extragalactic component, anisotropy may only emerge at higher energies due to the distribution of \ac{UHECR} sources and magnetic deflection. The emergence of the dipole feature at $E \geqslant 8$\,EeV appears consistent with this picture. 

In the above framework, the nature of the ankle feature at $5\times 10^{18}\,$eV is still to be understood. It could be the signature of propagation effects in a single (intermediate nuclei) extragalactic component, or a cross-over region between two extragalactic source populations. Notably, a combined fit of Auger measurements of the \ac{UHECR} energy spectrum and composition across the ankle seems to suggest the presence of two extragalactic components, though an intermediate-mass galactic component might also be present below the ankle~\cite{PierreAuger:2021mmt}. 
\textcolor{red}{Assuming that the highest energy Galactic cosmic rays are of intermediate or heavy mass, more accurate measurements of the energy spectra and anisotropies of the proton and Helium fluxes at lower energies down to the knee region will complete the understanding of the Galactic to extragalactic transition.}
Also, the secondary neutrino and gamma-ray fluxes expected in these scenarios, for each source population model, will provide concrete constraints. 

\subsubsection{Clues from the energy spectrum}

Above the ankle region, the measurement of the flux first provides the energy budget that the population of the highest energy cosmic-ray sources have to supply: $\dot{{\cal E}}_{\rm UHECR}\sim 0.5\times 10^{44}\,$erg\,Mpc$^{-3}$\,yr$^{-1}$ at $E=10^{19}\,$eV \cite{Katz:2008xx}. 
The steep decline in flux above about 30\,EeV is reminiscent of \ac{GZK} cutoff \citep{Greisen:1966jv,Zatsepin:1966jv}. A similar cutoff could however be produced by a maximum acceleration energy $E_{\rm max}$ at the source, and the interpretation is still being debated. The detection of particles at energies above $10^{20}\,$eV implies 1) that sources have to be able to accelerate particles up to these energies, and 2) that the sources of these particles lie within a few hundreds of megaparsecs, as they would have experienced severe energy losses if they had travelled from further away. Criterion 1) can be further translated into a necessary condition on the source parameters, using upgraded Hillas criteria.

\subsubsection{Clues from the mass composition}

The latest composition measurements at the highest energies reported by the Auger Observatory and Telescope Array (see \cref{sec:Auger_DesignAndTimeline} and \cref{sec:TA_DesignAndTimeline}) point towards a mass composition of \acp{UHECR} that evolves from a proton dominated composition at a few EeV toward an intermediate nuclei dominated composition at around 50\,EeV. 

\ac{UHECR} source models, in which a heavy composition arises at the highest energies due to a combination of a low proton maximum acceleration energy (around 10~EeV) and $Z$ times higher maximum energies for heavier elements (present in a slightly higher abundance than Galactic), have been shown to reproduce the composition trends observed by Auger~\citep{Allard:2008gj,Aloisio:2009sj}, once intergalactic propagation is accounted for.
The problem is then shifted to finding powerful sources that inject mainly these low abundance elements and let them escape from the acceleration site.

Heavy or intermediate nuclei dominated injection models at the source require either an initial metal-rich region, or an efficient nucleosynthesis in the accelerating outflow. Moreover, because the acceleration sites are usually rich in baryons and intense in radiation, the escape and survival of nuclei from these regions is not obvious. Many works have shown the difficulty to overcome these problems in \ac{AGN}, clusters, and \acp{GRB} \cite{Pruet:2002hi,Lemoine:2002vg,Wang:2007xj,Murase:2008mr,Murase:2011cy,Horiuchi:2012by}. On the other hand, it has been recently shown that many novel transient source models, several involving stellar cores (see e.g., Refs.~\cite{Fang:2012rx, Kotera:2015pya,Globus:2015bko,Guepin:2017abw, Biehl:2017hnb,Bhattacharya:2021cjc,Ekanger:2022tia}), could be natural candidate sites for such injection, and that accelerated nuclei could successfully escape their source environment.

\subsubsection{Clues from arrival directions}

The interpretation of arrival directions of \acp{UHECR} in the sky is intricate, and intimately linked to poorly understood magnetic fields in the Universe. Intervening magnetic fields deflect charged \ac{UHECR} trajectories, causing spatial and temporal (for transient sources) decorrelations. The impact of galactic and extragalactic magnetic fields and the related challenges are discussed in \cref{subsec:magneticfields}. 

The observed hints of small-scale anisotropy at energies beyond the \ac{GZK} cutoff, remain insufficient to draw conclusions as to the sources of \acp{UHECR} with available data (see \cref{sec:Anisotropy}). 
In the future, studies \cite{Blaksley:2013eho,dOrfeuil:2014qgw} show that even for the most unfavourable composition scenarios (with e.g., no protons accelerated to the highest energies), an increase in statistics should allow for the measurement of a significant anisotropy signal, assuming the sources to follow the spatial and luminosity distribution of the large scale structures. In the ankle region ($E\gtrsim 5\,$EeV), where the sources are numerous enough to imprint a clustering pattern in the sky, and hence where the anisotropy signal should not be dominated by the clustering of events around individual sources, increased statistics can also allow for efficient source population discrimination~\cite{Oikonomou:2014zva}. 

Another information given by the distribution of the arrival directions is the absence of multiplets, namely cosmic ray events arriving with little angular separation in the sky. This lack can be used to constrain the apparent number density of sources to $n_0>10^{-5}\,$Mpc$^{-3}$, if cosmic rays are protons \cite{Kashti:2008bw,Takami:2011nn}, a simple evaluation leading to $n_0\sim 10^{-4}\,$Mpc$^{-3}$ \cite{Takami:2011nn}, and models with $\bar{n} < 10^{-5}~{\rm Mpc}^{-3}$ are strongly disfavoured for any chemical composition \cite{PierreAuger:2013waq} as long as average deflections above $70$~EeV do not exceed $30^{\circ}$.

\subsubsection{Transient vs. steady sources}

The possible candidate sources can be split into two categories: steady and transient sources, which lead to different observable signatures. 
A source can be categorized as steady if its emission timescale is longer than the spread in the arrival time of their \acp{UHECR} \citep{Takami:2011nn,Murase:2008sa,Waxman:2008bj}. In this case, the arrival directions of \acp{UHECR} can directly trace and  constrain the sky distribution of their sources, in conjunction with other neutral messengers like photons, neutrinos and gravitational waves. 

The spread in arrival time is caused by magnetic deflections of charged cosmic rays in Galactic and intergalactic media, which can be quantified as $\delta t\approx 10^5\,(l/100 \,{\rm Mpc})\left( \alpha/2^\circ\right)^2\,{\rm yr}$ \cite{Kotera:2008ae},  for a propagation distance $l$ and a total deflection angle $\alpha$. The time delay is noticeable if the source is of transient type: for these sources, one does not expect to observe counterparts to \acp{UHECR}. The distribution of events in the sky should however follow closely the large scale structure with a bias which could help discriminate the source populations~\cite{2011A&A...528A.109K}. 

In terms of sheer energy budget arguments, powerful transients are highly promising sources, as they can inject their huge amount of energy over short timescales. The increase in luminosity can lead to enhanced cosmic-ray acceleration, with subsequent particle interactions and production of secondary multi-messenger emissions. Moreover, anisotropy, source-density, energetic and magnetic-structure arguments strongly challenge steady-source scenarios for \ac{UHE} cosmic rays with light composition~\cite{PierreAuger:2012gro, Fang:2016ewe,Palladino:2019hsk}. A bright transient source observed in \ac{UHE} neutrinos with solid electromagnetic counterparts would enable an immediate identification of the source with clear evidence of \ac{UHECR} acceleration~\cite{GKO22}. Ultra-high-energy gamma-ray transients would also be expected~\cite{Murase:2009ah}.


\subsection{Challenges in identifying the sources of UHECRs}

\subsubsection[General considerations for UHECR acceleration]{General considerations for UHECR acceleration}


\begin{figure}[!ht]
    \centering
    \includegraphics[width=0.7\textwidth]{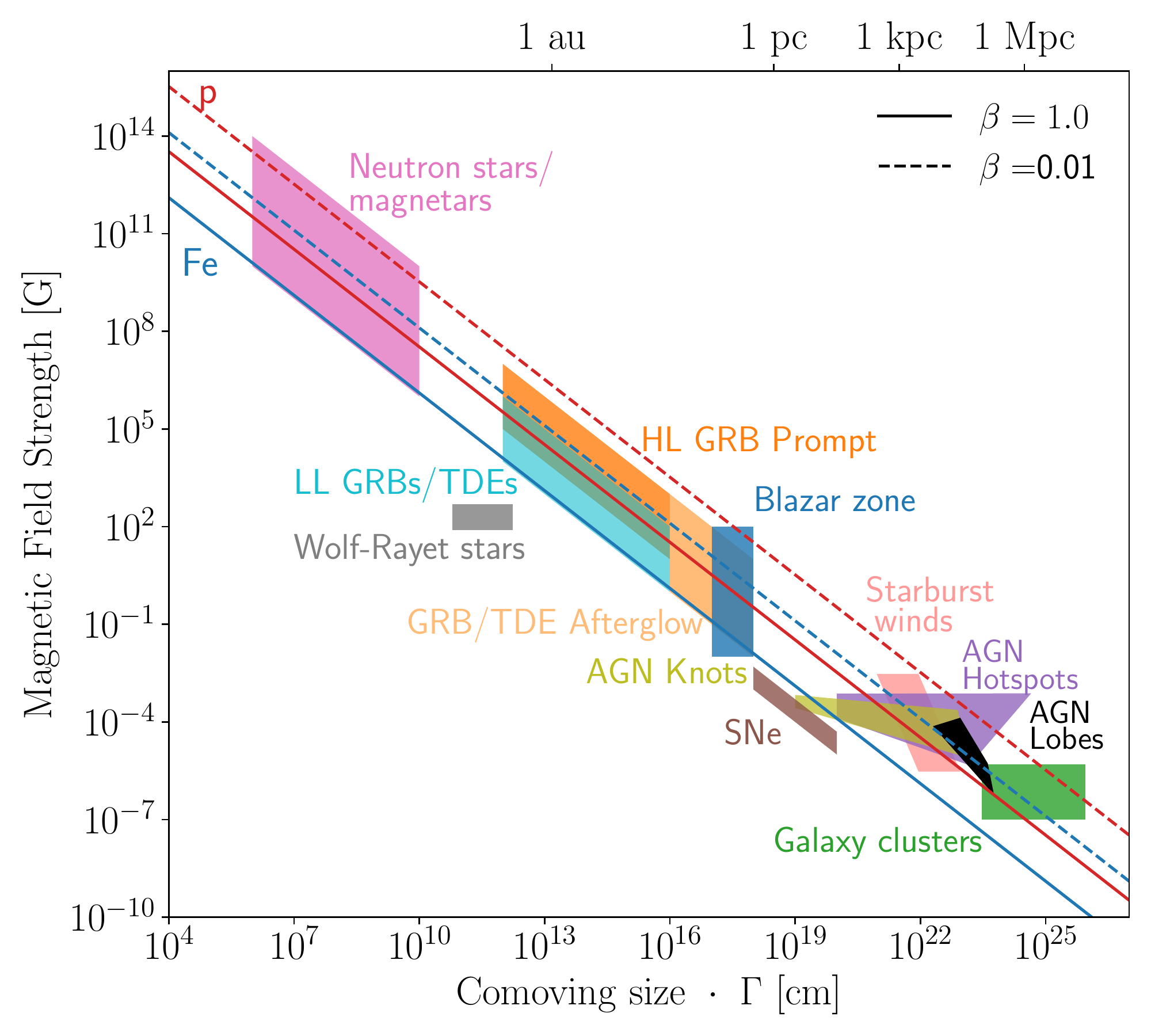}
    \caption{Hillas Plot, adapted from~\cite{AlvesBatista:2019tlv}. Source classes are shown as a function of their characteristic size and magnetic field strength. Source classes that lie to the right of the solid diagonal lines can confine $10^{20}$~eV iron (blue) and proton (red) nuclei respectively for sources with bulk outflows with velocity equal to the speed of light ($\beta = c$). Dashed lines illustrate the condition for sources with lower expansion velocity, namely $\beta = 0.01c$.}
    \label{fig:hillas}
\end{figure}


One of the most fundamental questions surrounding the origin(s) of \acp{UHECR} concerns how they attain their energies. Astrophysical models typically invoke some form of particle acceleration, a phenomenon that is as mystifying as the \acp{UHECR} themselves {\color{red} (detailed discussions of select acceleration mechanisms are provided in Sec. \cref{subsec:acctheory})}. The highly conductive environments of astrophysical plasmas make it difficult to maintain large electostatic fields in most cases. Instead, the necessary electric fields are generated through the bulk motions of magnetised plasmas ($-\vec{\beta} \times \vec{B}$, where $\vec{\beta}$ is the bulk velocity (in units of $c$) of the flow and $\vec{B}$ is the magnetic field). Under these circumstances, the maximum energy attainable by a particle with charge $Z e$ moving through an acceleration zone of size $R$ (
comoving size) is $E_{\rm max} = Z e \beta BR$~\cite{Matthews:2020lig}. Allowing for relativistic flows and inefficiencies in the acceleration process introduces a factor of the bulk Lorentz factor, $\Gamma$, and an efficiency factor $\eta$ such that the maximum energy is expressed as $E_{\rm max} = \eta^{-1} Z e \beta B R \Gamma$~\cite{AlvesBatista:2019tlv,Achterberg:2001rx,Bell:2019nnf}. This expression, commonly referred to as the Hillas criterion~\cite{Hillas:1984ijl}, imposes certain requirements on the characteristics of cosmic accelerators in order to achieve ultra-high energies. \cref{fig:hillas} provides an example of a Hillas diagram that plots the characteristic sizes of various candidate accelerators versus their magnetic field strengths in comparison with values of $B R$ required in order to accelerate protons and iron to $10^{20}$\,eV. As shown in the plot, a number of proposed source classes may possess the characteristics necessary to accelerate cosmic rays to the highest energies. However, meeting the Hillas criterion does not guarantee that a cosmic accelerator will be capable of accelerating cosmic rays to ultra-high energies. Ultimately, whether a given source is capable of producing \acp{UHECR} depends on energy losses within its environment and the details of the acceleration mechanism(s) at work. 

    
A necessary condition to accelerate \acp{CR} in a particular source environment is sufficiently large size and magnetic field strength as to confine the \acp{CR}~\cite{Hillas:1984ijl}. The maximum energy that can be achieved in a source of radius $R$ and magnetic field strength $B$ is $E_{\rm max} = \beta e B R \Gamma$, where $\beta$ is the velocity of the shock in units of the speed of light, $c$, and $\Gamma$ is the Lorentz factor of the motion of the emission region. Source classes that have sufficiently high values of $\beta R B \Gamma$ as to accelerate \acp{CR} to very high energies are shown in \cref{fig:hillas}. Those source classes that reside above the diagonal lines can plausibly accelerate \acp{CR} to ultra-high energies. 
    
\begin{figure}[!ht]
    \centering
    \includegraphics[width=0.7\textwidth]{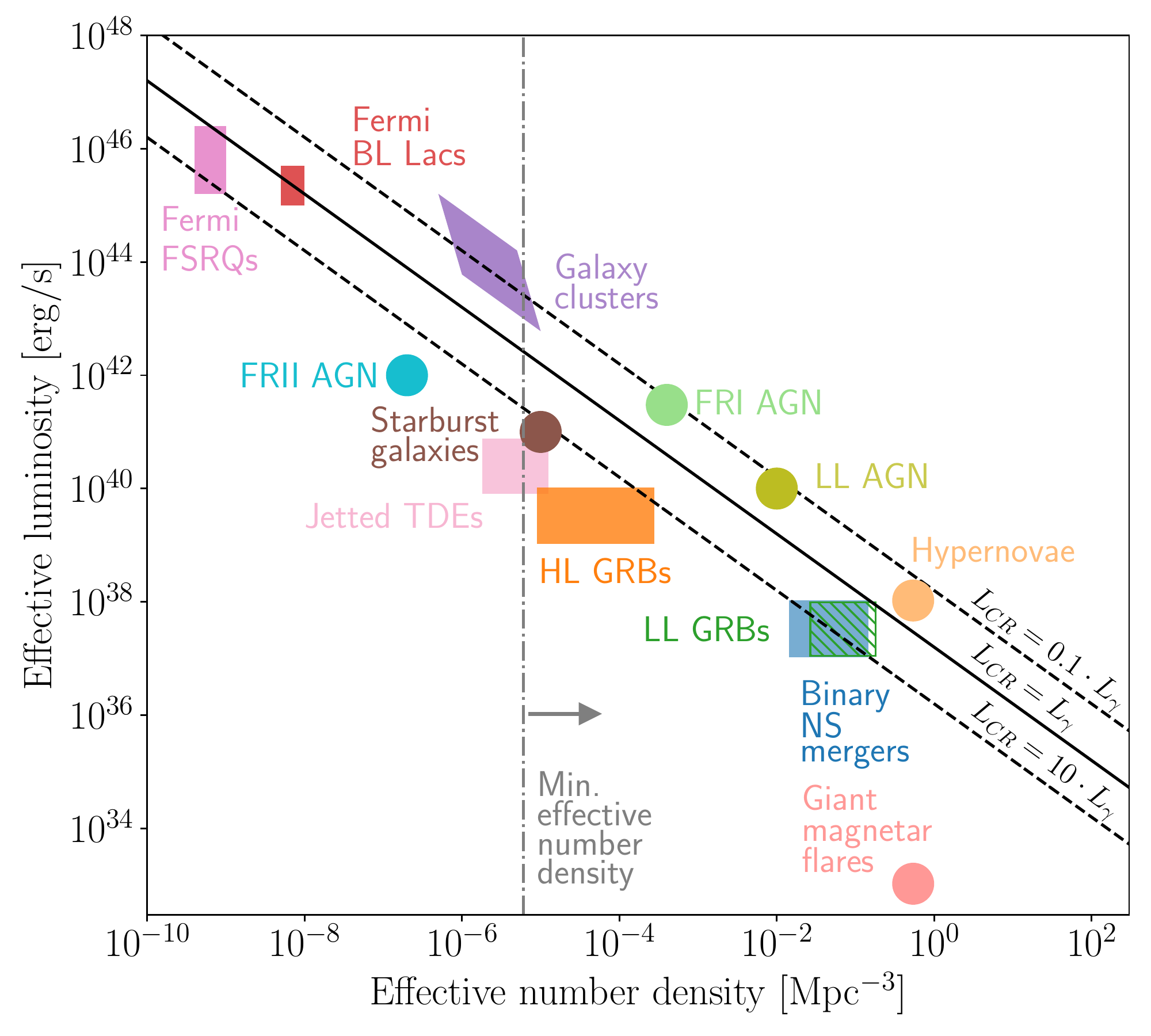}
    \caption{Characteristic source luminosity versus source number density for steady sources, and effective luminosity versus effective number density for transient sources assuming a characteristic time spread $t= 3 \times  10^5$\,yr. The effective number density for bursting sources is only valid for the assumed value of $t$, which corresponds to mean extragalactic magnetic-field strength of 1\,nG. Stronger magnetic fields would imply larger $t$ and hence, larger effective number density. The black solid line gives the best-fit luminosity density of \ac{UHECR} sources estimated in Ref.~\cite{PierreAuger:2016use}. Dashed lines bound the parameter space in which sources have luminosity density in the range 0.1 - 10 times the nominal \ac{UHECR} source luminosity density. The grey dashed line indicates the minimum \ac{UHECR} source number density estimated in Ref.~\cite{PierreAuger:2013waq}.}
    \label{fig:energy_budget}
\end{figure}

A clue to the origin of \acp{UHECR} comes from the measured diffuse intensity which can be converted to the UHECR energy production rate~\cite{Murase:2018utn,Jiang:2020arb} and compared to the emissivity of different source populations at various wavelengths. This allows to estimate whether a particular source population has sufficient power density as to produce the observed \acp{UHECR}. \cref{fig:energy_budget} shows different source classes in terms of their measured number density and characteristic luminosity. Source classes to the right and above the diagonal lines have sufficient emissivity as to power the observed \acp{UHECR}. An additional clue to the origin of \acp{UHECR} comes from the observed clustering of the arrival directions. The fact that there is no significant small scale clustering of the arrival directions above 70\,EeV disfavours rare source classes such as flat spectrum radio quasars as the sole sources of \acp{UHECR}~\cite{PierreAuger:2013waq}. This, lower bound on the source number density is, however, sensitive to the deflections suffered by the \acp{UHECR}. 

\subsubsection{Potential astrophysical source classes}\label{subsec:SourceCandidates} 

\paragraph{Gamma-ray bursts}

\acp{GRB}, bursts of MeV photons lasting a few seconds, are the most powerful transient sources in the Universe. \acp{GRB} dissipate kinetic energy in the form of relativistically expanding wind, a \emph{fireball}, whose inferred characteristics are believed to fulfill the requirements for acceleration of particles to $\sim$\,$ 10^{20}$~eV~\cite{Waxman:1995vg,Vietri:1995hs}. \acp{LLGRB} are less relativistic but more numerous than classical \acp{HLGRB}. They, as well as transrelativistic supernovae, are also thought to fullfill the requirements of acceleration of \acp{CR} to ultra-high energies~\cite{Murase:2008mr,Murase:2006mm,Zhang:2017moz,Zhang:2018agl}. The relatively heavy inferred elemental composition of UHECRs~\cite{Bellido:2017cgf,PierreAuger:2016qzj, PierreAuger:2014gko}, is generally easier to reconcile with  \acp{LLGRB} and transrelativistic SNe than with classical \acp{HLGRB} due to the stronger radiation fields in the latter environments, in which heavier nuclei are more likely to experience photodisintegration (see e.g., Refs.~\cite{Globus:2015bko,Zhang:2017moz,Globus:2014fka,Baerwald:2014zga,Boncioli:2018lrv,Bustamante:2014oka}).


\paragraph{Jetted active galactic nuclei}

\ac{AGN} with relativistic jets are one of the most popular candidate source classes of UHECRs. The relativistic jets, which can extend to $\sim$\,Mpc scales, host several sites of shocks where the product of magnetic field and size of the region may be sufficiently large as to allow UHECR acceleration. Proposed sites of UHECR acceleration include the termination shocks which are responsible for bright hotspots observed in radio galaxy jets, the giant radio lobes and other more compact but more highly magnetised regions closer to the base of the jet~\cite{Murase:2011cy,Matthews:2020lig}. Recent phenomenological studies have shown that UHECR observations are consistent with a jetted-\ac{AGN} origin of the bulk of UHECRs under different scenarios including shear acceleration, generally based on the idea of re-acceleration of Galactic \acp{CR}~\cite{Rodrigues:2020pli, Caprioli:2015zka,Kimura:2017ubz}.

\paragraph{Tidal disruption events}

It has been argued that sources that satisfy the minimum luminosity requirement from the leading candidate classes, namely \acp{GRB} and jetted \ac{AGN}, are not sufficiently prevalent inside the \ac{GZK} horizon as to supply the observed UHECR flux, leading to the need to consider other types of transient events~\cite{Farrar:2008ex}. Though this argument depends on the elemental composition of the UHECRs, it is in general true that the power requirement is a hurdle for most theoretical models. An alternate source population that has been suggested to be able to overcome these constraints are tidal disruption events that lead to the formation of an accretion disk and jet around a supermassive black hole~\cite{Farrar:2014yla}. Only a handful of jetted \acp{TDE} are known to date. Given the relatively low inferred rate of jetted \acp{TDE}, most studies conclude that whether they can satisfy the energy-budget constraint depends intricately on the relation between the \ac{TDE} radiative luminosity and UHECR luminosity~\cite{Guepin:2017abw,Zhang:2017hom}. Interestingly, \acp{TDE} have recently been associated with high-energy neutrinos~\cite{Stein:2020xhk} sparking renewed interest in understanding the multi-messenger role of these extreme transient phenomena. 

\paragraph{Starburst galaxies}

Starburst activity is an episodic phenomenon of extraordinarily high star-formation activity in a fraction of galaxies, which can be inferred from their having infrared luminosities that are much higher than those typically observed in normal galaxies. Starburst galaxies are observed to drive large-scale magnetised outflows which have been proposed as possible sites of \ac{UHECR} acceleration~\cite{Anchordoqui:1999cu}. No consensus has been reached on this scenario, with some authors concluding that the properties of the wind are not sufficient to accelerate particles to $10^{20}$~eV~\cite{Romero:2018mnb, Anchordoqui:2018vji, Anchordoqui:2020otc}. The Auger collaboration has reported the observation of an excess of \acp{UHECR} with respect to background expectations from nearby starburst galaxies~\cite{PierreAuger:2018qvk}. Such anisotropy, if established, does not necessarily indicate \ac{UHECR} acceleration in starburst winds, but may indicate \ac{UHECR} acceleration in stellar explosions which occur at higher rates in starburst galaxies than in normal ones. However, as discussed in Ref.~\cite{Anchordoqui:2019ncn}, it is probable that the higher rate of stellar explosions in nearby starbursts cannot fully compensate for the difference in number density between starburst and normal galaxies. This means that if stellar explosions were the primary mechanism driving \ac{UHECR} acceleration, the hints of anisotropy observed by Auger and \ac{TA} should correlate with the local matter distribution rather than with nearby starbursts.

\paragraph{Galaxy clusters}

Galaxy clusters are the largest gravitationally bound objects in the Universe. Despite relatively moderate inferred magnetic field strengths, they may be able to confine or accelerate particles to $10^{20}$~eV due to the extremely large size. They are generally thought to be calorimetric environments for high energy \acp{CR}, and host many of the other candidate source classes including jetted \ac{AGN}. They are therefore plausibly the sources of UHECR and high-energy neutrinos simultaneously~\cite{Kang:1996rp,Murase:2008yt,Kotera:2009ms,Fang:2017zjf}. 

\paragraph{Pulsars and Magnetars}

Despite being very compact, the extremely large magnetic fields that are inferred for pulsars and magnetars mean that they may be able to accelerate particles and nuclei to $10^{20}$~eV. As the collapsed cores of massive stars, pulsars and magnetars have the appealing feature of being embedded in environments that are highly enriched in heavy elements and are thus thought to be able to supply \acp{UHECR} consistent with the elemental abundances inferred from the most recent observations~\cite{Fang:2012rx}.

\subsection{UHECR propagation through the Universe}
\label{sec:UHECR_transport}

Once \acp{UHECR} are generated, they must navigate a universe replete with radiation and magnetic fields. Encounters with these phenomena significantly influence observable properties of \acp{UHECR} (see \cref{sec:CurrentStatus}) and thus, substantially impact the interpretation of their origin(s).


\subsubsection{Interactions with the extragalactic background light}

The Universe is awash in radiation from all light-emitting processes that have occurred throughout its history (collectively referred to as the \ac{EBL}; for a recent review, see e.g., Ref.~\cite{Cooray:2016jrk}). \acp{UHECR} primarily interact with the \ac{CMB} component of the \ac{EBL}, with the \ac{IR} -- \ac{UV} components making modest contributions. These interactions constitute the dominant source of energy loss for \acp{UHECR} after they leave their sources and propagate to Earth. 

%
For \ac{UHECR} protons, the most relevant interactions are Bethe-Heitler pair production at lower energies ($E \gtrsim 10^{18}$\,eV) and photo-pion production at higher energies ($E \gtrsim 10^{19}$\,eV). While the threshold energy for photo-pion production is $\epsilon \sim 145$\,MeV (where $\epsilon$ is the energy of the photon in the proton rest frame), the cross section for the process is dominated by the $\Delta\left(1232\right)$ resonance. Heavier baryon resonances appear at higher energies, as well as multi-pion production.

For heavier nuclei ($A > 1$), the interactions with \ac{EBL} photons are somewhat more complex due to the presence of multiple nucleons. As with protons, \ac{UHECR} nuclei engage in Bethe-Heitler pair production and photo-pion production interactions near their respective energy thresholds ($\epsilon \sim 1$\,MeV and $\epsilon \sim 145$\,MeV, respectively). At energies between these thresholds, \ac{UHECR} nuclei undergo photodisintegration, a process in which a nucleus absorbs an impinging photon and subsequently fragments into an excited daughter nucleus and one or more nucleons. The dominant process for photodisintegration is the giant dipole resonance at photon energies of $\sim 10$--$30$\,MeV, which mainly triggers single-nucleon emission. At higher photon energies, multi-nucleon channels can also be triggered, as well as the quasi-deuteron process. Ultimately, the energy losses for \ac{UHECR} nuclei are dominated by photodisintegration~\cite{Allard:2011aa}. 

The energy losses resulting from the above interactions impact \ac{UHECR} observations in a number of ways. For instance, the \ac{UHECR} energy spectrum will exhibit features at energies relevant for the various interaction processes. The most famous of these features is a cutoff at the highest energies due to attenuation of the \ac{UHECR} flux (see e.g., Refs.~\cite{Greisen:1966jv, Zatsepin:1966jv, PhysRev.180.1264, Puget:1976nz, Epele:1998ia, Stecker:1998ib}). Models that additionally consider \ac{UHECR} interactions in the regions surrounding their sources have been shown to reproduce the ankle feature of the \ac{UHECR} energy spectrum, as well as the evolution in \ac{UHECR} composition with energy~\cite{Unger:2015laa}. Finally, the energy losses limit the distances over which \acp{UHECR} can travel from their sources without suffering significant attenuation, the so-called horizon distance. The horizon distances range from $\sim$\,few to tens of Mpc for intermediate-mass nuclei (e.g., He, C\,N\,O, Si) up to $\sim 250$\,Mpc for protons and iron nuclei. Within these distances, the distribution of matter in the Universe is anisotropic~\cite{Huchra:2011ii}, which should be reflected in the sky distribution of \acp{UHECR} if they do originate from astrophysical sources. 

Due to their significant impacts on \ac{UHECR} observations and implications for their interpretation, efforts to model \ac{UHECR} interactions continue to this day. Aside from being a source of energy loss for \acp{UHECR}, the above processes will also give rise to secondary particles, such as photons and neutrinos, providing a means for studying \acp{UHECR} and their sources through multi-messenger observations. Efforts to precisely model \acp{UHECR} interactions within sources and through propagation have resulted in the release of several publicly-available numerical codes. The SOPHIA Monte Carlo event generator is designed for modeling photo-hadronic interactions in a variety of astrophysical settings, making use of the full photo-pion production cross section and treating resonance excitation and decay, direct single-pion production, and diffractive and non-diffractive multi-particle production~\cite{Mucke:1999yb}. For \ac{UHECR} propagation, CRPropa~\cite{AlvesBatista:2016vpy} is designed for efficient calculations of the energy losses due to interactions with the \ac{EBL} and the associated secondary photon and neutrino production. The latest version (CRPropa 3) also provides functionality for 3D and 4D (including energy losses arising from cosmological redshift) propagation simulations through magnetic fields.

\subsubsection{Charged-particle propagation through magnetic fields}\label{sec:magnetic_fields}



\begin{SCfigure}
    \begin{minipage}[!htb]{0.55\columnwidth}
    \centering
    \includegraphics[width=\columnwidth]{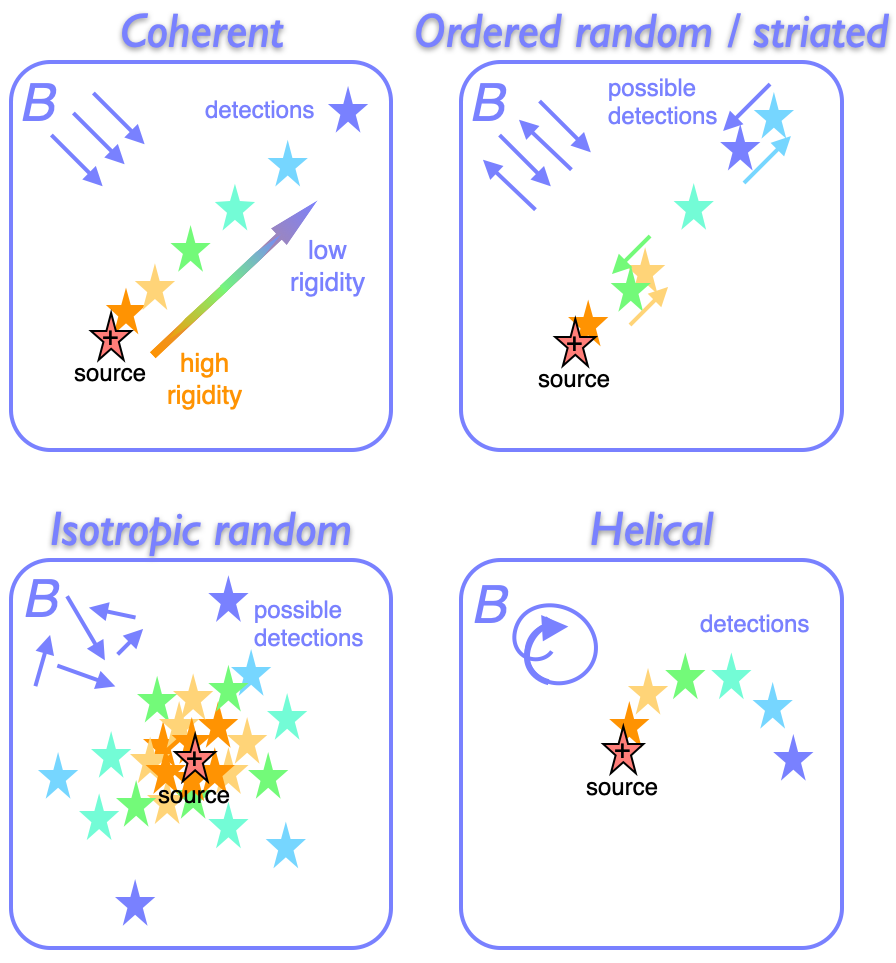}
    \end{minipage}%
    \caption{Illustration of the effects of different magnetic field components on the observed arrival directions of \acp{UHECR}. As shown in Ref.~\cite{Jaffe:2019iuk}, the traditional radio tracers of \acp{GMF} naturally divide the fields into regular and random components that can further split into coherent, isotropic random, anisotropic random (striated), and helical. For \acp{UHECR}, each component leads to a different deflection on the sky. For a source position (red star) emitting positively charged \acp{UHECR}, the coherent component (upper left) causes a systematic shift in the arrival direction as a decreasing function of rigidity (orange to blue).  The striated component (upper right) mixes these deflections along the same line on the sky.  The isotropic random component (lower left) causes a scatter in all directions, and the helicity (lower right) produces a curved set of deflections.}
	\label{fig:gmf_cartoon}
\end{SCfigure}

This section discusses how astrophysical magnetic fields influence \ac{UHECR} trajectories. Later sections will discuss how future \ac{UHECR} observations will be used to study cosmic magnetism.
For these purposes, it is instructive to introduce the conceptual components of magnetic fields that can be separated by different astronomical observations:
a coherent component pointing in a single direction through a large volume (also known as the mean field); an isotropic random component pointing stochastically in all directions equally; an anisotropic random component (sometimes referred to as striated fields) that has a constant orientation but changes direction stochastically. The helicity of the field (a topological property of it rather than a component) can also be probed by different combinations of certain observables.
Each magnetic field component plays a unique role in determining the observed arrival directions of \acp{UHECR} with respect to their sources (see~\cref{fig:gmf_cartoon}).
The coherent component of the field causes a deflection of the particle path that increases with decreasing rigidity\footnote{The rigidity of a particle with charge $Z\,e$ and momentum $p$ (energy $E$) is ${\mathcal R}=p\,c/(Z\,e) \simeq E/(Z\,e)$.
}, so that the arrival direction of the UHECR does not point back to where the source is located, but to a systematically shifted direction.  The deflections due to the stochastic components of the magnetic field cause a scatter in the arrival directions of \acp{UHECR} from a given source.
All field components are therefore important to include quantitatively to interpret UHECR hot spots, anisotropies, and correlations with tracers of large-scale structure.

As detailed in \cref{sec:Anisotropy} of this report, new
air-shower data collected in the last decade radically improved our
understanding of ultrahigh-energy cosmic rays.  A dipolar anisotropy
in the arrival directions of cosmic rays above 8~EeV was discovered
with high significance establishing the extragalactic origin of these
particles~\cite{PierreAuger:2017pzq,PierreAuger:2018zqu}. Moreover,
several tantalizing indications for small- and intermediate-scale
anisotropies are currently under scrutiny~\cite{TelescopeArray:2014tsd,PierreAuger:2018qvk,TelescopeArray:2021gxg,TelescopeArray:2021dfb}.


However, the astrophysical interpretation of these observations
depends on assumptions about the deflections of ultrahigh-energy cosmic
rays in the \ac{GMF} and \ac{IGMF}.  For instance, the strength and direction of
the detected dipolar anisotropy of cosmic-ray arrival directions is
expected to reflect the large-scale anisotropy of nearby extragalactic
cosmic-ray sources. But, due to the coherent deflection and partial
randomization of the arrival direction in the intervening magnetic
fields between the sources and Earth, a definite attribution of the
origin of the dipole requires a knowledge of the structure of the
coherent \ac{GMF} as well as the strength of the random component of the
\ac{GMF} and \ac{IGMF}~\cite{Ding:2021emg,Allard:2021ioh,Harari:2010wq,Globus:2017fym,diMatteo:2017dtg,Wittkowski:2017nfd,Globus:2018svy,Mollerach:2019wne,Eichmann:2020adn,Mollerach:2021ifa}.

\begin{figure}[!htb]
  \centering
  \includegraphics[height=5cm]{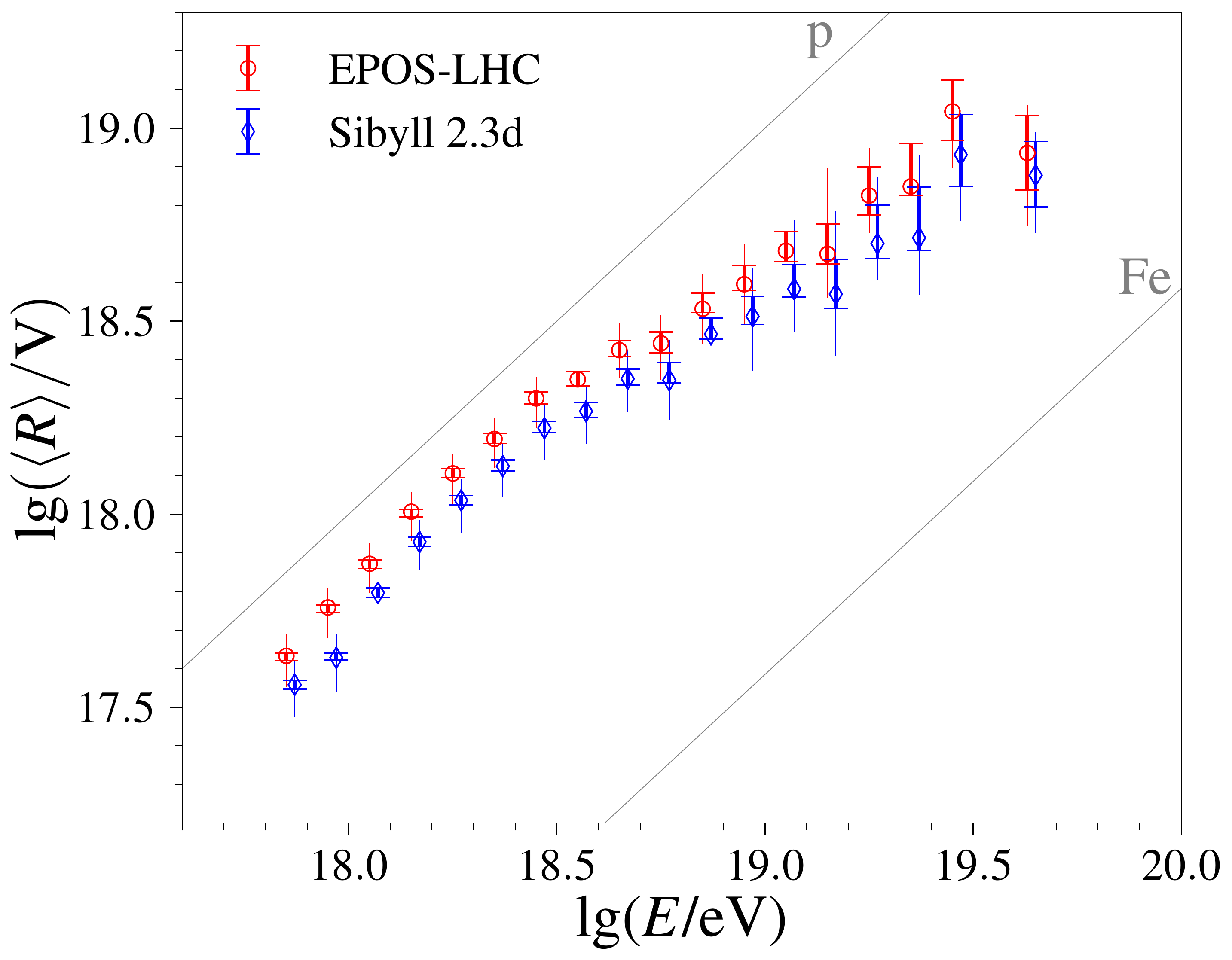}
  \includegraphics[height=5cm]{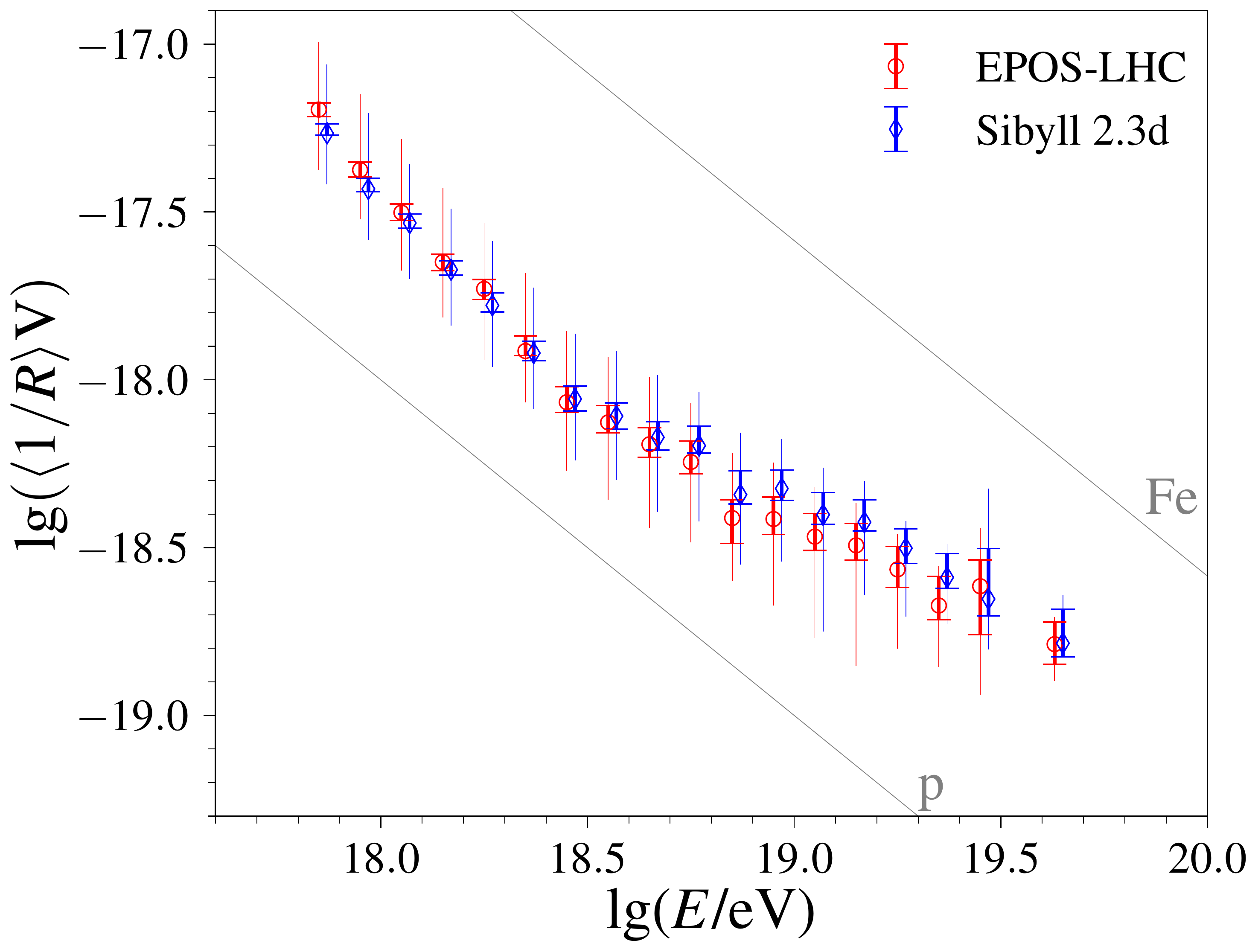}

  \caption{The average rigidity $\langle R \rangle$ (left) and $\langle 1/R \rangle$ (right) estimated from the \xmax distributions measured by the Pierre Auger Observatory~\cite{Bellido:2017cgf} using \eposlhc and \sibyll{2.3d} hadronic interaction models. Shown error bars denote statistical and total uncertainties. The inverse of the rigidity $\langle 1/R \rangle$ is proportional to the magnetic deflection angle.}
  \label{fig:rigidity}
\end{figure}

\begin{figure}[!htb]
  \centering
  \includegraphics[clip,width=0.28\linewidth]{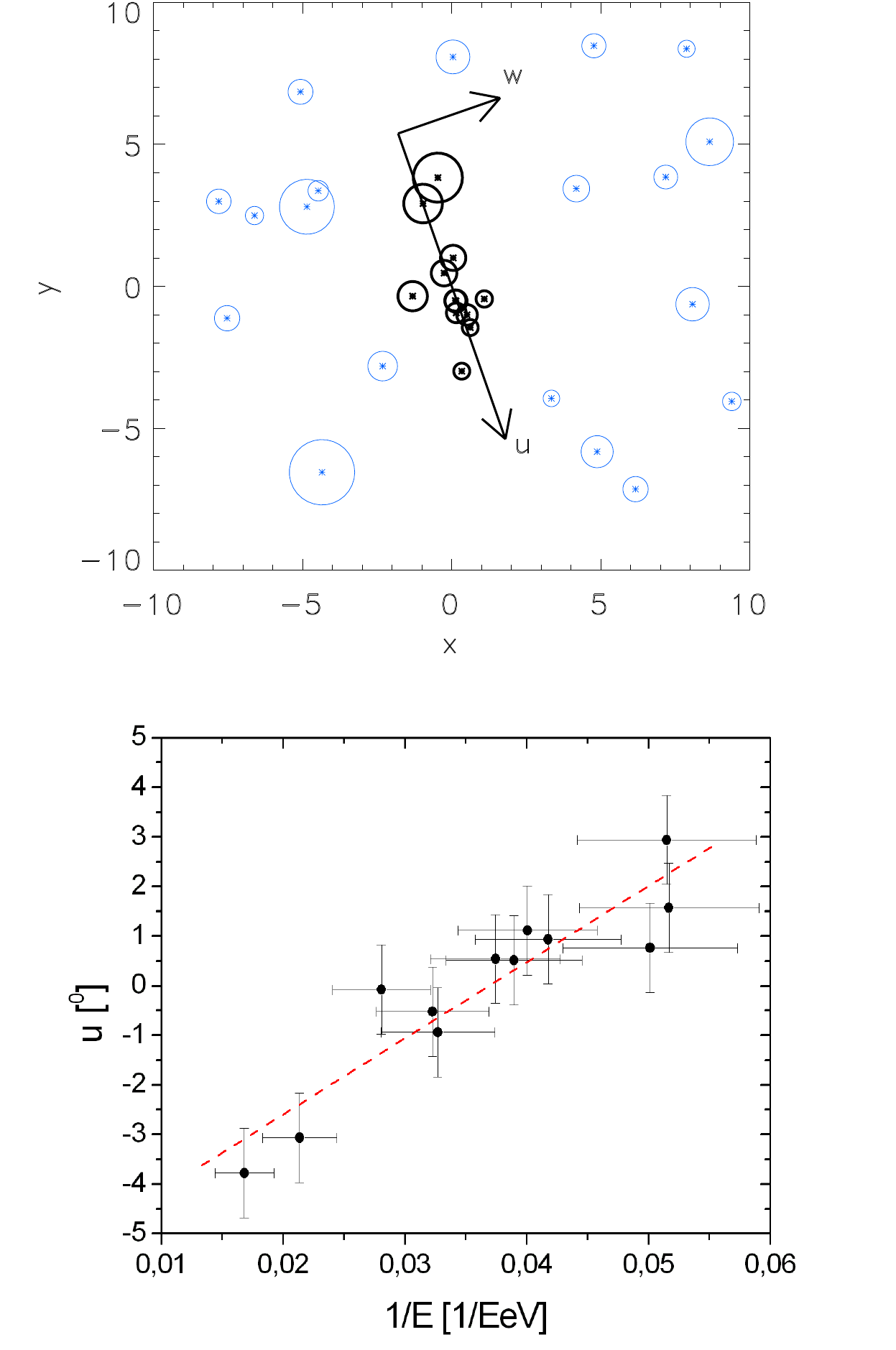}\includegraphics[clip,rviewport=0 -0.2 1 1,width=0.7\linewidth]{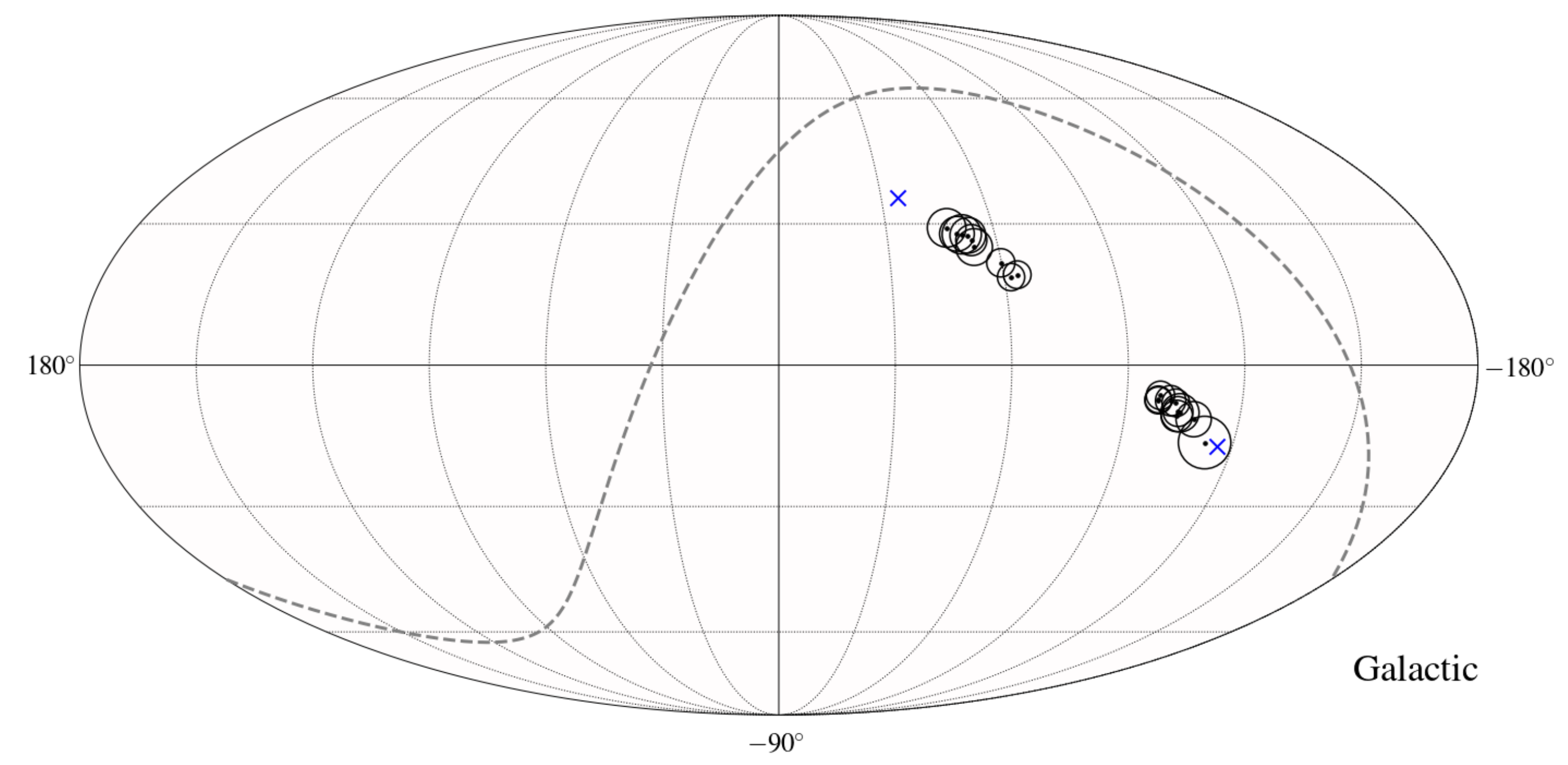}
  \caption{{\itshape Left}: Simulated magnetically-induced aligned of
    cosmic rays. The top panel shows the sky view with background
    events in light blue and source events in black. The size of the
    circles proportional to the energy of the cosmic ray. The lower
    panel illustrates the energy-angle correlation of cosmic rays
    along the $u$-axis shown in the upper
    panel~\cite{PierreAuger:2011mup}. {\itshape Right}: Two candidate
    multiplets reported by the Pierre Auger Collaboration
    \cite{PierreAuger:2020zii} above an energy threshold of
    $E=40$~EeV. The cross denotes the inferred infinite-rigidity
    source position and the size of the circles encode again the
    energy of individual events.}
  \label{fig:magpatterns}
\end{figure}

Even at ultra-high energies, deflections are expected to be
non-negligible, since the observed evolution of the average mass (and
thus charge) of cosmic rays at Earth leads to an energy-dependent
average rigidity that increases only slowly with energy (see \cref{fig:rigidity}). In the
quasi-ballistic regime, angular deflections in the \ac{GMF} are about
$
\delta \approx (1-5)^\circ / \left({\mathcal R}/(10^{20} \, \text{V})\right)$ depending
on the position of the sky (see e.g.,~Fig.\,3 in Ref.~\cite{IceCube:2015afa}).
Therefore, the correlation of the observed small- and
intermediate-scale anisotropies with astrophysical sources is not
straight-forward without taking into account these
deflections~\cite{Harari:2000az,Harari:2002dy,Dolag:2008py,Giacinti:2009fy,Takami:2009qz,Keivani:2014kua,Smida:2015kga,AlvesBatista:2017vob,Farrar:2017lhm,TelescopeArray:2021cvw,deOliveira:2021ckh}. And
similarly, multi-messenger studies of the cross-correlation of the
arrival directions of neutral particles, in particular high-energy
neutrinos, and \acp{UHECR} are challenged by the possibility of large
angular deflections of cosmic-ray
nuclei~\cite{IceCube:2015afa,ANTARES:2019ufk,Palladino:2019hsk,Carpio:2015ewa,Resconi:2016ggj,Carpio:2016wec}.

The most direct way to connect the sources of \acp{UHECR} with deflections
in the GMF would be magnetically-induced patterns in the arrival
directions of cosmic rays~\cite{Golup:2009cv}. As can be seen in the
left panel of \cref{fig:magpatterns}, such patterns arise if a
spectrum of energies is emitted from a source of identical
particles. More complicated signatures are expected for sources
emitting a mixed composition. So far, the search for
magnetically-induced patterns~\cite{PierreAuger:2011mup,PierreAuger:2020zii,PierreAuger:2014tos, TelescopeArray:2020acv}
has not yet resulted in a significant detection. Two candidate
cosmic-ray multiplets from the Pierre Auger Observatory are shown on
the right panel of \cref{fig:magpatterns}.

\begin{figure}[!t]
    \centering
  \includegraphics[width=0.72\linewidth]{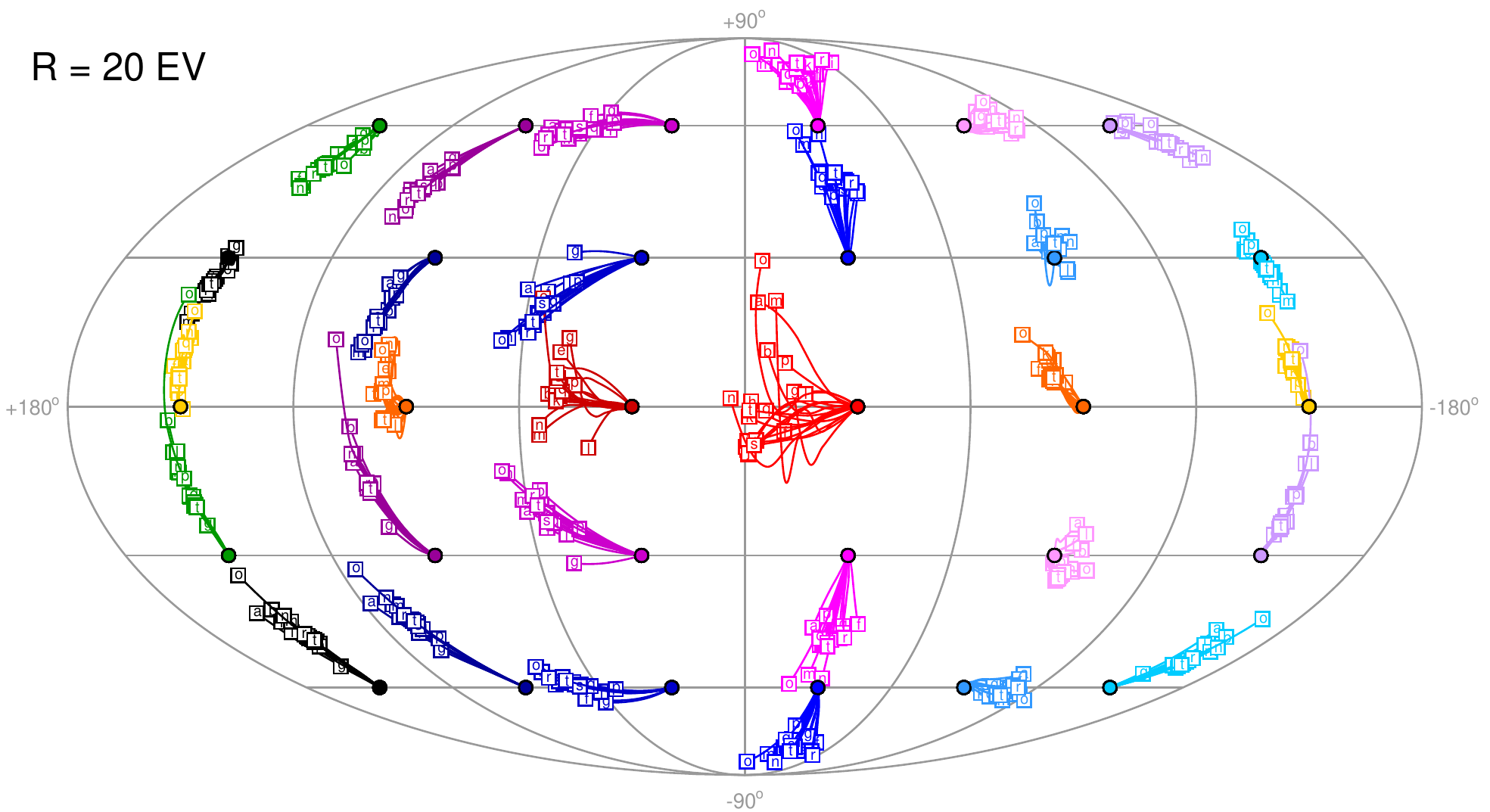}
  \caption{Backtracking of charged particles through the Galaxy
    starting from a regular grid of initial directions (dots).  The
    resulting directions outside of the Galaxy for particles with a
    rigidity of 20~EV are denoted by squares and the lines connecting
    the initial and final positions were constructed by performing
    backtracking at higher rigidities. Each of the letters (a)-(t)
    denotes a different GMF model that describes the sky maps of
    Galactic synchrotron emission and the rotation measures of
    extragalactic radio sources~\cite{Unger:2019xct}.}
\label{fig:defl}
\end{figure}

\subsubsection{Effects of Lorentz invariance violation}

\subparagraph{GZK limit} Assuming Lorentz invariance, ultra-relativistic nuclei can undergo photonuclear interactions with \ac{CMB} and \ac{EBL} photons such as pair production~${^AZ} + \gamma \to {^AZ} + e^+ + e^-$, photodisintegration e.g.,~${^AZ} + \gamma \to {^{A-1}Z} + n$, or pion production e.g.,~$p + \gamma \to p + \pi^0$ or~$p + \gamma \to n + \pi^+$.  These set a limit on the energy with which nuclei from cosmologically large distance can reach us, known as \ac{GZK} limit.  A positive~$\delta_\text{hadrons}^{(n)}$ (see \cref{eq:liv_delta}) can make such interactions kinematically forbidden, altering the resulting UHECR energy spectrum and mass composition at Earth.  Comparisons between propagation simulations and Pierre Auger Observatory data indicate that~$\delta_\text{hadrons}^{(0)} < 10^{-19}$, \ $\delta_\text{hadrons}^{(1)} < 10^{-38}\,\mathrm{eV}^{-1}$, and~$\delta_\text{hadrons}^{(2)} < 10^{-57}\,\mathrm{eV}^{-2}$ at the $5\sigma$~C.L.\ \cite{PierreAuger:2021tog}.

\subparagraph{Photon absorption}  Assuming Lorentz invariance, the secondary gamma rays produced during \ac{UHECR} interactions would be quickly absorbed by electron--positron pair production with \ac{CMB} and radio background photons, $\gamma + \gamma \to e^- + e^+$.  A negative~$\delta_\gamma^{(n)}$ could prevent this process, allowing such gamma rays to arrive intact at Earth.  In the assumption that there is a fraction of protons among the highest-energy cosmic rays, the non-detection of such gamma rays would imply that~$-\delta_\gamma^{(0)} < 10^{-21}$, \ $-\delta_\gamma^{(1)} < 10^{-40}\,\mathrm{eV}^{-1}$, and~$-\delta_\gamma^{(2)} < 10^{-58}\,\mathrm{eV}^{-2}$ \cite{PierreAuger:2021tog}, but it is still not known whether any such protons are present.



\subsection[The next decade and beyond]{The next decade and beyond: Charged particle astronomy and future \ac{SHDM} searches}

\subsubsection{Nuclear composition}

A major requirement for next-generation UHECR detectors is a precise measurement of the elemental composition of \acp{UHECR} up to $10^{20}$~eV. A key observation will be a measurement of the iron fraction up to $10^{20}$ eV. Absence of Iron up to $10^{20}$~eV would rule out Galactic reacceleration scenarios and thus \ac{AGN} as sources of the observed UHECRs and favor stellar transients where typically the iron core collapses into a black hole, see e.g., Refs.~\cite{Rodrigues:2020pli,Zhang:2017moz,Kimura:2017ubz,Mbarek:2021bay}. In parallel, next-generation UHECR detectors with excellent composition sensitivity will be able to determine whether the observed UHECR composition is consistent with originating in an environment with elemental composition similar to that of Galactic CRs thus strongly limiting plausible UHECR acceleration sites. 

\subsubsection{Charged-particle astronomy}

The anisotropy in the arrival directions of \acp{UHECR} caused by the anisotropy of the source distribution is expected to be strongest at the highest rigidities due to the reduced propagation horizon and the reduced magnetic deflections. An \ac{UHECR} detector with event-by-event composition identification,
even at moderate \xmax resolution,
is extremely well suited for anisotropy studies where identifying a light component is sufficient and large exposure is essential. A detector with exposure $\sim$\,$10^5~{\rm km}^2~{\rm yr}~{\rm sr}$ above 40\,EeV will allow for $5\sigma$ independent observation of all currently reported $3\mbox{--}4\sigma$ anisotropy hints including the \emph{TA Hotspot} and the Auger UHECR-starburst correlation~\cite{Sarazin:2019fjz}. Alternatively, a next-generation \ac{UHECR} detector that will determine mass composition on an event-by-event basis will measure the energy spectra of individual species and perform anisotropy searches above a fixed rigidity. Distinguishing individual elements or mass groups would enable tomographic mapping of \ac{UHECR} source populations, which would leverage the different propagation lengths and amounts of deflection for nuclei of various species~\cite{AlvesBatista:2019tlv}.
For example, if CNO or silicon are identified, it would be possible able to scrutinize the closest extragalactic \ac{UHECR} sources since these elements must originate from sources $\lesssim$\,tens of Mpc away. 


\subsubsection{The cosmic-ray energy spectrum}


A key observable that can unveil the accelerators of UHECRs is the cosmic-ray energy spectrum of individual nuclear species or elemental groups (light, CNO-like, Si/Fe like). With such observables it will be possible to strongly limit the plausible scenarios for the origin of UHECRs to those that can reproduce the observed scaling of features of the spectrum across different species. For example a spectrum that escapes the UHECR sources following a simple Peters cycle results in very different observations than models with in-source photodisintegration in this respect. 

At present, it is difficult to determine the extent to which the differences in the \ac{UHECR} energy spectra measured by Auger and \ac{TA} result from astrophysical effects, such as different source populations in the different parts of the sky. A full-sky \ac{UHECR} observatory with exposure $10^5~{\rm km}^2{\rm yr}~{\rm sr}$ at 100\,EeV will provide a final verdict on whether the UHECR spectrum is different in the two hemispheres. A very precise measurement of the diffuse spectrum will further allow to identify the features expected at the highest energies from transient sources which necessarily contribute to a narrow range in energy for individual chemical species if UHECRs originate in rare transient sources such as \acp{GRB}~\cite{Miralda-Escude:1996twc}. 

The suppression at the end of the cosmic ray spectrum due to photopion interactions of protons and/or photodisintegration of nuclei interacting with the \ac{CMB} is established with significance $>20\sigma$ compared to a continuous power-law extrapolation~\cite{HiRes:2007lra,TelescopeArray:2012qqu,PierreAuger:2008rol}. However, alternative interpretations of the suppression feature are viable, for example the Auger \ac{SD} and \ac{FD} data are compatible with scenarios in which the flux suppression at the highest energies is due to accelerators running out of steam~\cite{AlvesBatista:2019tlv}. If the suppression in the \ac{UHECR} spectrum is due to the \ac{GZK} process, a slight upturn (recovery) is expected if the source spectra continue up to energies beyond 100\,EeV and there are \ac{UHECR} sources within a few tens of Mpc of the Galaxy. As such, a recovery in the \ac{UHECR} spectrum beyond $100$\,EeV would have implications for the maximum energies achievable by \ac{UHECR} accelerators, as well as the distribution of \ac{UHECR} sources in the Universe. Such a recovery would be detectable by a next-generation \ac{UHECR} detector with an exposure $10^5~{\rm km}^2{\rm yr}~{\rm sr}$ at $100$\,EeV~\cite{Anchordoqui:2019omw}.  

\subsubsection{Insights into magnetic fields from future \acs{UHECR} observations}

The next decade of observations at ultrahigh energies will benefit
from the increased detection area of the Telescope Array in the Northern
hemisphere.  After the completion of the \TAxFour
upgrade (see \cref{sec:TAx4}), the
array will match the acceptance of the Pierre Auger Observatory in the
South and equal-exposure full-sky studies of the large-scale
anisotropies will allow answering the question ``How isotropic can the
UHECR flux be?''~\cite{diMatteo:2017dtg}, and it will be possible to
learn about the role of magnetic fields in deflecting and smoothing
large-scale patterns in the arrival directions of cosmic rays.

The upgrade of the Pierre Auger Observatory,
AugerPrime (see \cref{sec:AugerPrime})~\cite{PierreAuger:2016qzd,PierreAuger:2021ece}, will enable
an event-by-event mass-estimate for every air shower detected. This
will provide a large data set in which it is possible to enhance
low-charge primaries and to study the aforementioned anisotropies as a
function of rigidity.

These upgrades have the potential to pave the way towards
charged-particle astronomy in the semi-ballistic regime, i.e., at
rigidities where the trajectories are significantly deflected by the
coherent \ac{GMF}, but not fully isotropized. The ``nuclear window to the
extragalactic universe''~\cite{Erdmann:2016vle} is expected to open
at around 20~EV. As illustrated in \cref{fig:defl}, at around
this rigidity the differences between the deflections predicted
by different models of the \ac{GMF} are small enough such that it is conceivable
to use even limited knowledge of the \ac{GMF} to aid in \ac{UHECR} source searches. And even in the worst-case scenario for the \ac{IGMF}, in which voids have $\sim$~nG  fields, deflections at around these rigidities would still be less than 15$^\circ$~\cite{AlvesBatista:2017vob}.

The new experimental developments of the next decade will be supported by advancements in
the algorithms to determine the cosmic-ray charge from air shower data
(see e.g., Refs.~\cite{PierreAuger:2021xnt,Stadelmaier:2021ajo}) and new analysis techniques for the simultaneous fits of magnetic fields and
UHECR sources
(e.g., Refs.~\cite{2018arXiv180104341S,Erdmann:2018cvz,Bister:2020rfv,Wirtz:2021ifo}).

If the data collected in the next decade corroborate the existence of \emph{hot spots} in the \ac{UHECR} sky, then their location and angular extend will provide important insights on the \ac{GMF} and \ac{IGMF}, as demonstrated in studies using the current indications for these intermediate-scale anisotropies, see e.g., Refs.~\cite{Bray:2018ipq, VanVliet:2021sbc,Neronov:2021xua}.

The next decade will hopefully see a new generation of large-aperture
observatories, at least one with event-by-event rigidity capabilities like the \acf{GCOS} (see \cref{sec:GCOS}), possibly making use of the next generation of fluorescence telescopes developed with \acf{CRAFFT}~\cite{Tameda:2019wmj} or the \acf{FAST} (see \cref{sec:FD_tech_development}) \cite{Fujii:2015dra}, and complemented by large full-sky aperture from space provided by \acf{POEMMA} (see \cref{sec:POEMMA}).
In addition, air-shower neutrino observatories like the proposed Giant Radio Array for Neutrino Detection (\acs{GRAND}) (see \cref{sec:GRAND}) could also provide large-aperture observations of cosmic rays. A large
aperture will be the key for an unequivocal discovery of anisotropies
and sources at the highest
energies~\cite{Blaksley:2013eho,dOrfeuil:2014qgw,Oikonomou:2014zva}, and an
event-by-event sensitivity to the cosmic-ray charge opens up the
possibility to use cosmic rays as a novel probe to study Galactic and
extragalactic magnetic fields.

\subsubsection{Super-heavy dark matter searches}
\label{sec:FuturePartSHDMFuture}


\cref{sec:PartSHDM} presented the current status of \ac{SHDM} searches with existing \ac{UHECR} experiments and the resulting constraints on the mass and lifetime of \ac{SHDM} particles and on the effective coupling constant of hidden gauge interactions. Searches for \acf{SHDM} will continue through the next decade and beyond with the upgraded and next-generation \ac{UHECR} experiments. Increased exposure and upgraded instrumentation will lead to either a serendipitous discovery of \ac{SHDM} or further constraints on \ac{SHDM} scenarios. 
%
%

Aside from the generic \ac{SHDM} constraints discussed in \cref{sec:PartSHDM}, considering various \ac{SHDM} production scenarios provides an avenue for exploring a broader parameter space. This section illustrates constraints that will be achievable with the upgraded and next-generation \ac{UHECR} experiments on a specific category of \ac{SHDM} production models, namely \emph{freeze-in} scenarios (see e.g., Refs.~\cite{McDonald:2001vt,Hall:2009bx,Bernal:2017kxu}). \cref{subsubsec:SHDMwCMB} will discuss the framework of \ac{SHDM} production by time-varying gravitational fields at the end of inflation and complementary constraints that will be achievable with future \ac{UHECR} and \ac{CMB} experiments.

%

Typical \ac{WIMP} scenarios assume that \ac{DM} is a thermal relic with a current abundance determined by the ``freeze-out'' condition balancing \ac{DM} annihilation with the expansion rate of the Universe. However, in order for freeze out to occur, \ac{DM} would had to have been in thermal equilibrium with the rest of the Universe, requiring the coupling with the visible sector to be $\gtrsim \mathcal{O}\left(10^{-7}\right)$ (see e.g., Ref.~\cite{Enqvist:2014zqa}). On the other hand, if the coupling with the visible sector is weaker than this level, \ac{DM} can be produced through the freeze-in mechanism ~\cite{McDonald:2001vt,Hall:2009bx,Bernal:2017kxu}. In freeze-in scenarios, \ac{DM} particles are produced by the decay or annihilation of visible-sector particles until the temperature of the thermal bath cools below the energy scale of the interaction between \ac{DM} and the visible sector~\cite{Bernal:2017kxu}. In this manner, \ac{SHDM} can be produced during the reheating period following inflation. During this period, the inflaton field decays, producing Standard Model particles that can annihilate via graviton exchange and produce super-heavy particles
~\cite{Garny:2015sjg}. The freeze-in abundance of super-heavy particles can reproduce the \ac{DM} abundance observed today provided that the reheating period is fast enough and that the energy scale of the inflaton is high enough.

\begin{figure}[tb]
\centering
\mbox{\includegraphics[width=0.38\textwidth]{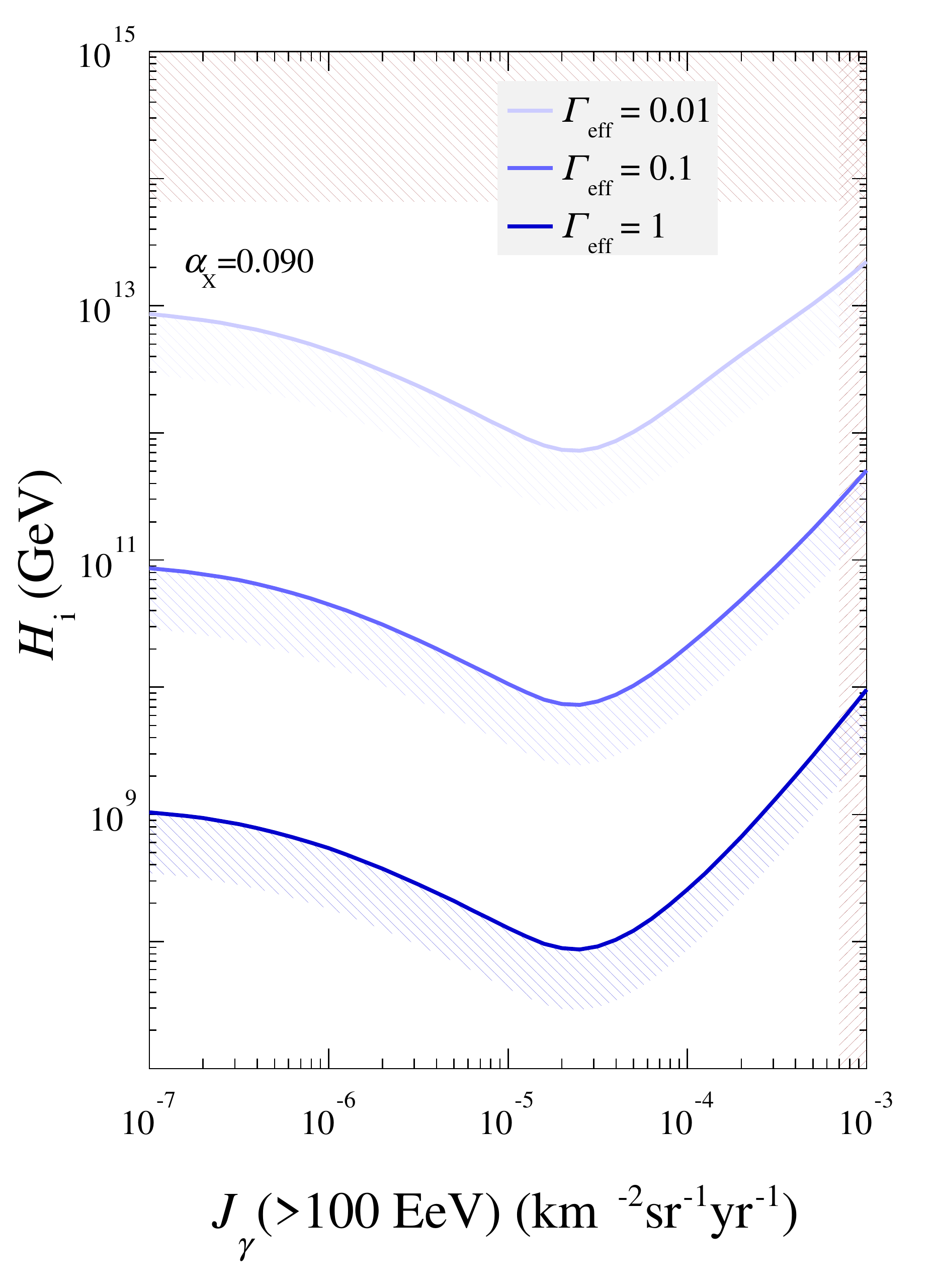}\hspace{-1em}
\includegraphics[width=0.64\textwidth]{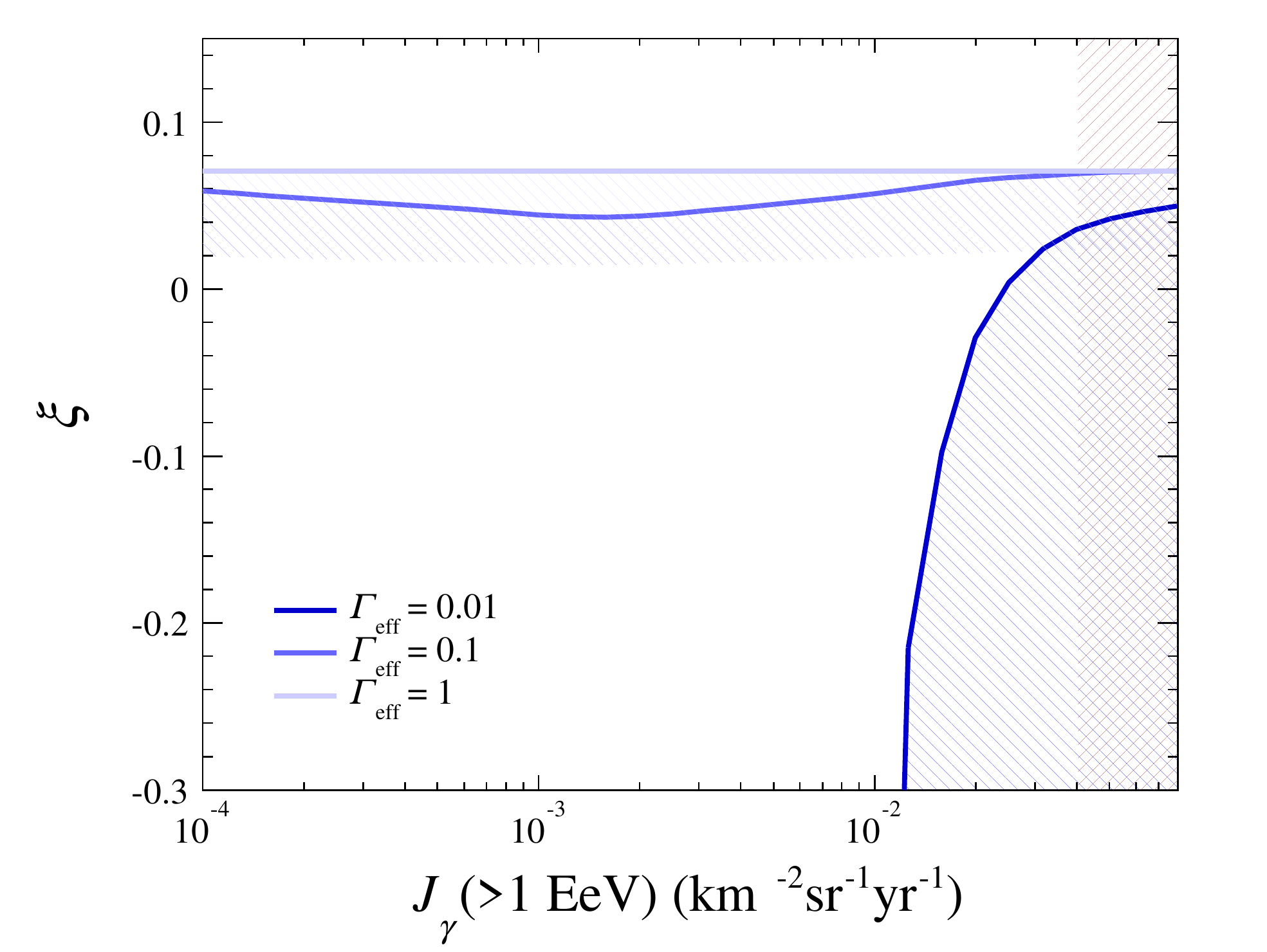}}
\caption{Left: Constraints on the Hubble rate, $H_{\mathrm{i}}$, as a function of the photon flux, $J_\gamma(E)$, for $E=100~$EeV. Right: Constrains on the non-minimal coupling, $\xi$ as  a function of $J_\gamma(E))$ for $E=1~$EeV. Figure adapted from Ref.~\cite{SHDM}.}
\label{fig:xiJ}
\end{figure}

Constraints on the flux of \ac{UHE} photons from \ac{UHECR} experiments (see \cref{sec:PartSHDM}) translate into constraints on the Hubble rate during the reheating period ($H_{\rm inf}$ and the duration of the reheating period (through the reheating efficiency parameter, $\Gamma_{\rm eff}$~\cite{ThePierreAuger:2022}.
The most recent constraints are shown in \cref{fig:xiJ} (left) for an energy threshold of $10^{20}~$eV. The viable regions are delineated for three different values of the reheating efficiency. The vertical dashed regions are excluded from the limits on $J_\gamma(>E)$, while the horizontal regions are excluded from the non-observation of tensor modes in the \ac{CMB}~\cite{Garny:2015sjg}. This demonstrates the complementarity between constraints provided by \ac{UHECR} experiments and those provided by \ac{CMB} experiments (see also \cref{subsubsec:SHDMwCMB}).



Next-generation \ac{UHECR} experiments with large exposures will be able to explore \ac{SHDM} freeze-in scenarios with sensitivites down to $J_\gamma(>E) \sim 10^{-4}~$km$^{-2}$~sr$^{-1}$~yr$^{-1}$ (e.g.,~\cref{fig:xiJ}). 
%
Such a sensitivity would allow for probes of the $(\Geff,\alpha_X)$ parameter space. Currently, regions of the $(H_{\mathrm{inf}},M_X)$ parameter space that reproduce the present-day relic abundance are excluded for $(\Geff\geq 0.01,\alpha_X\geq 0.10)$. 

Finally, it is important to assess the possible impacts of the Big Bang cosmology on other aspects of \ac{SHDM} models aside from particle production. In particular, the astronomical lifetime of the metastable vacuum of the Standard Model might be challenged in the cosmological context due to thermal fluctuations allowing the decay when the temperature was high enough, or due to large fluctuations of free fields generated by the dynamics on a curved background because of the presence of a non-minimal coupling $\xi$ between the field and the curvature of space-time. Requiring the electroweak vacuum not to decay yields constraints between the non-minimal coupling $\xi$ and the Hubble rate $\Hinf$~\cite{Markkanen:2018bfx}. The relationship can be established by formulating the Standard Model on a curved background. Propagating the stability bounds derived in the $(\xi,\Hinf)$ plane (for $\alpha_X=0.090$) into the $(\xi,J_\gamma(>E))$ parameter space yields constraints on the non-minimal coupling $\xi$~\cite{ThePierreAuger:2022}. Recent results are shown in \cref{fig:xiJ} (right) for $E=1$\,EeV. Values outside of the allowed range necessarily imply new physics different from that producing \ac{SHDM} particles in order to stabilize the Standard Model vacuum.

\subsection{Connections with other areas of physics and astrophysics}\label{sec:AstroOtherAreas}




\subsubsection{Synergies between future \acs{UHECR} searches for \acs{SHDM} and \acs{CMB} observations}\label{subsubsec:SHDMwCMB}

Searches for \acf{SHDM} will continue through the next decade and beyond with the upgraded and next-generation \ac{UHECR} experiments. Increased exposure and upgraded instrumentation will lead to either a serendipitous discovery of \ac{SHDM} or further constraints on \ac{SHDM} scenarios. At the same time, observations by other experiments will lead to complementary constraints on some \ac{SHDM} scenarios.
This section discusses the prospects of using \ac{CMB} observations to probe specific scenarios of \ac{SHDM} production by time-varying gravitational fields during the period following right after the end of the inflationary epoch.


In the standard paradigm of inflationary cosmology, the Universe undergoes a period of exponential expansion (inflation), which smooths out initial variations in density or temperature and reduces the curvature of space~\cite{Guth:1980zm,Linde:1981mu}. During this period, the Universe is completely dominated by the inflaton field, and the only density perturbations that exist are those that are generated due to fluctuations in the inflaton field. The rapid expansion of the background spacetime stretches these fluctuations to cosmological scales, laying the groundwork for them to become seeds of large-scale structure in the Universe~\cite{Guth:2013epa}. Other scalar fields present at the time of inflation will similarly obtain large values, $M_X \sim m_{\phi}$ (where $m_{\phi} \sim 10^{13}$~GeV is the mass of the inflaton), even if they couple only very weakly (or not at all) with other fields and do not couple to the inflaton~\cite{Chung:1998zb,Drees:2017iod}. Ref.~\cite{Chung:1998zb} proposed this scenario as a mechanism for generating \ac{SHDM}. In this mechanism, the very weak (or nonexistent) couplings of the \ac{SHDM} imply that it should be long lived, and its very large mass will prevent it from thermalizing, resulting in an abundance that depends only on the mass of the \ac{SHDM} and the behavior of the background spacetime. Ref.~\cite{Chung:1998zb} finds that in the range $0.04\leq M_X/\Hinf \leq 2$, where $\Hinf$ is of the order of the $m_\phi$, the \ac{SHDM} abundance is of the order of critical density, implying that the correct dark matter abundance can be achieved for particular values of $M_X$.

As noted earlier, the \ac{SHDM} gravitational production scenario is similar to the inflationary generation of gravitational perturbations that seed large-scale structure formation. Both processes will generation gravitational waves and contribute to the primordial gravitational wave background that, in turn, will induce a $B$-mode polarization pattern in the \ac{CMB}~\cite{Kamionkowski:1996zd,Zaldarriaga:1996xe,Kamionkowski:2015yta}. Thus, $B$-mode measurements by future \ac{CMB} experiments, such as CMB-S4~\cite{CMB-S4:2016ple}, will provide a search for \ac{SHDM} to complement the ongoing searches for \ac{SHDM} by current and future \ac{UHECR} experiments.

\begin{figure}[!ht]
\centering\includegraphics[width=0.80\textwidth]{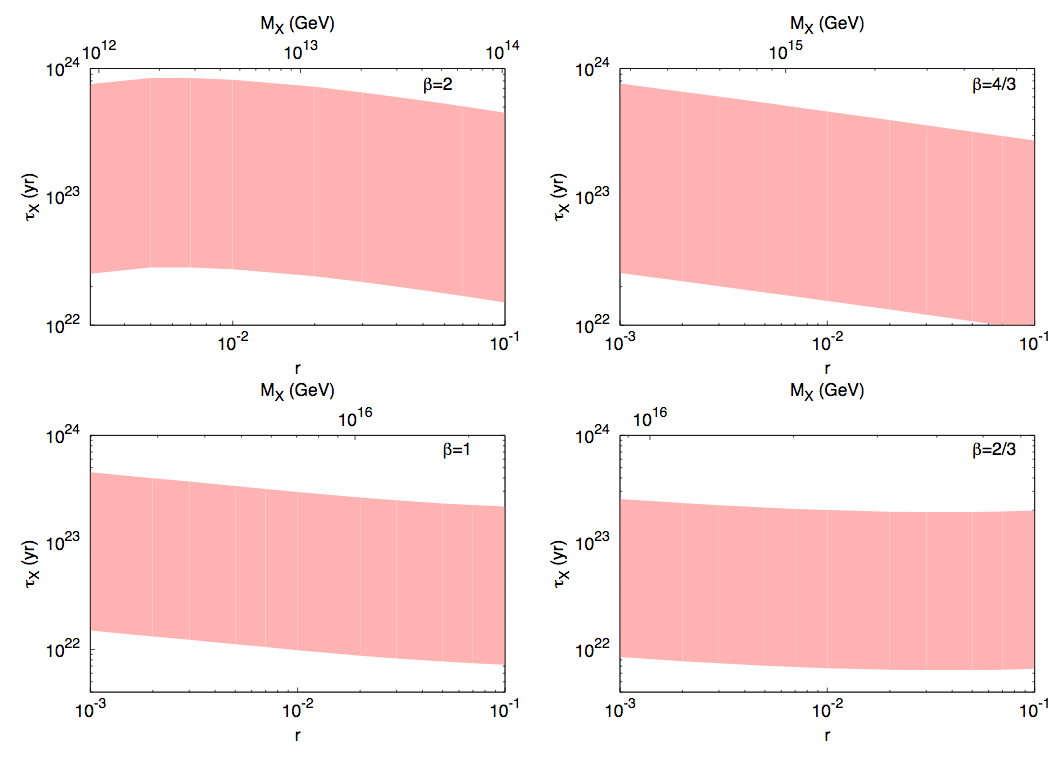}
\caption{\small{Regions of $(r,\tau_X)$ that will be accessible in the next decade within the framework of SHDM production by time-varying gravitational fields at the end of inflation (see text). From Ref.~\cite{Aloisio:2015lva}.}}
\label{fig:rtau}
\end{figure}

\subsubsection{Understanding particle acceleration in astrophysical sources}\label{subsec:acctheory}




Though particle acceleration is encountered in a myriad of astrophysical settings, acceleration to ultra-high energies is particularly illustrative, signifying the extremes of the phenomenon. While only the most powerful cosmic accelerators are capable of producing \acp{UHECR}, the questions of whether and how they do are deeply rooted in the processes by which they accelerate particles and the conditions present that may enable or prohibit acceleration to the highest energies. A variety of acceleration mechanisms have been proposed (for recent reviews, see e.g., Refs.~\cite{Matthews:2020lig,Marcowith:2020vho} and references therein), but crucial elements of the phenomenon remain unclear, challenging the development of a complete description. This section briefly summarizes the most commonly discussed acceleration mechanisms.


\paragraph{Fermi acceleration}\label{subsubsec:FermiAcc}

In Fermi acceleration, particles are accelerated through collisions with magnetic perturbations, or scattering centers, within a plasma. In the original description of this process that was proposed by Fermi, particles gain energy through collisions with scattering centers moving randomly at some speed $u_{\rm c}$ \cite{Fermi:1949ee}. With each encounter, particles gain or lose a fraction of their energy, depending on the orientation of the particle's velocity with respect to that of the scattering center. Head-on collisions in which particles gain energy are more likely to occur than rear-end collisions in which particles lose energy, resulting in a net gain in energy. On average, the energy gain per collision is $\propto \left(u_{\rm c}/c\right)^2$, and for this reason, this process is commonly referred to as stochastic acceleration or 2$^{\rm nd}$-order Fermi acceleration. 

In typical astrophysical scenarios, $u_{\rm c} \ll c$, and the particle must remain in the acceleration region for a long time in order to gain a substantial amount of energy through 2$^{\rm nd}$-order Fermi acceleration. As such, 2$^{\rm nd}$-order Fermi acceleration is relatively inefficient and unlikely to accelerate particles to ultra-high energies, particularly when accounting for energy losses in the source environment. One way for the Fermi acceleration process to be more efficient is for the scattering centers to move in the same general direction so that all of the collisions are nearly head on. In this case, the average energy gain scales as $\sim $\,$\left(u_{\rm c}/c\right)$, and the acceleration is a 1$^{\rm st}$-order process (or 1$^{\rm st}$-order Fermi acceleration) \cite{Parker:1958zza,1963ApJ...137..135W}. Such a scenario is naturally realized through collisionless shocks found in a variety of astrophysical systems, including candidate UHECR sources such as \acp{GRB}, \acp{AGN}, and galaxy clusters. 

Shock acceleration includes three basic processes (see e.g., Ref.~\cite{Marcowith:2016vzl}) of which the most commonly invoked to explain UHECRs is \ac{DSA}~\cite{1977ICRC...11..132A,1977SPhD...22..327K,Bell:1978zc,Blandford:1978ky}. In this process, turbulent magnetic fields on either side of the shock scatter particles scatter back and forth across front. With each shock crossing, particles gain a constant fraction of energy; hence, particles may reach very high energies as long as they remain in the acceleration region on long enough timescales to experience multiple shock crossings. The rate at which particles escape the acceleration region is also constant, which suggests that high-energy particles are as likely to remain in the acceleration region and reach even higher energies as lower-energy particles. As such, \ac{DSA} has the benefits of being relatively efficient and of being able to naturally produce power-law distributions that are similar to the measured CR spectrum.

While the above simplified picture of \ac{DSA} demonstrates its appeal as a model of particle acceleration, detailed studies of the process have highlighted several key elements that may ultimately determine whether it can accelerate UHECRs and in which source classes. In \ac{DSA}, the particles must already have superthermal energies in order to jump over the shock front and enter the acceleration process; however, the mechanism responsible for energizing (or injecting) these particles is as yet unclear (this is the so-called ``injection problem''). Other key elements of \ac{DSA} are related to the impacts of the accelerated particles themselves. If the acceleration process is efficient, it will produce a substantial population of accelerated particles that will exert a pressure and modify the shock structure~\cite{1977ICRC...11..132A,1981ApJ...248..344D,Drury:1983zz,1986MNRAS.223..353D}. Moreover, CRs can trigger various streaming instabilities that will generate the magnetic turbulence necessary to confine them to the acceleration region, transport them back-and-forth across the shock, amplify the turbulent magnetic field, and thereby determine the maximum attainable energy and the spectrum of accelerated particles~\cite{Bell:1978zc,Blandford:1978ky,2000MNRAS.314...65L,2004MNRAS.353..550B,Zirakashvili:2008bg,Reville:2008bp,Caprioli:2008st,Matthews:2017apu,Bell:2019uia,Malkov:2019pgq}. Numerical simulations have demonstrated that the conditions necessary for efficient \ac{DSA} can be realized in supernova remnants, allowing them to reach maximum energies of up to $\sim$\,$5 \times 10^{18}$~eV for Fe nuclei~\cite{Ptuskin:2010zn}. On the other hand, in ultrarelativistic shocks such as those expected in \ac{AGN} and \ac{GRB} jets, the time available for the CRs to generate the necessary magnetic turbulence is substantially limited due to the tendency of CRs to be overtaken by the shock in the upstream region and to be advected away from the shock in the downstream region. As such, DSA in ultrarelativistic shocks is expected to be inefficient, and the maximum energy predicted by the Hillas condition would be unattainable~\cite{Lemoine:2009vr,Reville:2014mta,Bell:2017zzx}. Thus, if \ac{AGN} or \ac{GRB} jets are the sources of UHECRs, then they either (1) accelerate particles in mildly relativistic shocks (e.g., \ac{GRB} internal shocks) or similar flow discontinuities (e.g., the boundary of the sheath of structured \ac{AGN} jet~\cite{Mertens:2016rhi}) or (2) accelerate particle via some alternative mechanism.

\paragraph{Unipolar induction}

The most straightforward and efficient mechanism for accelerating particles is through direct acceleration by persistent electric fields. Due to the high conductivity of astrophysical plasmas, such electric fields can only be realized in particular circumstances. One such instance is that of unipolar induction~\cite{1832RSPT..122..125F} by a rapidly rotating magnetized object, such as a neutron star~\cite{Fang:2012rx,Gunn:1969ej,Blasi:2000xm,Arons:2002yj,Fang:2013cba} or a black hole magnetosphere~\cite{Boldt:1999ge,Boldt:2000dx,Neronov:2009zz}. 

As with most astrophysical plasmas, neutron stars are excellent conductors and electrons and ions within the star redistribute themselves so that the internal electric field vanishes in the corotating frame, with electrons collecting at the poles and ions at the equator~\cite{Cerutti:2016ttn}. In the fixed lab frame, the charges create an electric field that balances the Lorentz force of the magnetic field and leads to an electrostatic potential that extends beyond the surface of the neutron star. Beyond the light cylinder radius, plasma can no longer corotate with the neutron star as this would require velocities greater than the speed of light. As a result, magnetic field lines that would extend beyond the light cylinder radius become open field lines that generate a relativistic wind. Voltage drops in the wind region can accelerate particles to high energies while avoiding catastrophic losses that would occur within the pulsar magnetosphere due to curvature radiation~\cite{Arons:2002yj}. These voltage drops are of order $\Phi = \Omega^2\mu/c^2$ ($\Omega$ is the angular velocity of the pulsar, $\mu = BR^3_{\ast}/2$ is the magnetic dipole moment, and $R_{\ast}$ is the radius of the pulsar), leading to energies for particles with charge $Z$ of $E\left(\Omega\right) = Ze\Phi\eta \sim 3 \times 10^{21} Z \left(\eta/0.1\right) \left(B/2 \times 10^{15}\,{\rm G}\right) \left(R_{\ast}/10\,{\rm km}\right)^3 \left(\Omega/10^4\,{\rm s}^{-1}\right)^2$\,eV, where $\eta$ is the fraction of the voltage experience by the particles as the travel through the wind region~\cite{Kotera:2011cp}. Thus, achieving \acp{UHE} is possible, but would require a very rapidly spinning magnetar (pulsar with magnetic field strengths on the order of $10^{15}$\,G). Typical magnetars spin much more slowly (spin periods of on the order of $1$--$10$\,s). Newborn magnetars do spin at much faster rates ($\sim$\,$ 100$--$300$\,s$^{-1}$)~\cite{Lander:2019guk}, though it remains a question as to whether they can reach high enough spin rates to produce the highest-energy cosmic rays\footnote{However, newborn pulsars may contribute to the population of galactic cosmic rays~\cite{Fang:2013cba}.}. The degree to which accelerated particles experience energy losses as they escape the pulsar wind is also a question that must be addressed by this scenario.

\paragraph{Magnetic reconnection}

Most, if not all, of the potential astrophysical sources presented in \cref{subsec:SourceCandidates} contain regions in which the energy contained in magnetic fields greatly exceeds that of the plasma~\cite{Blandford:2017chu}. Magnetic reconnection has garnered much interest because it provides a natural mechanism for transferring magnetic energy to the plasma, a necessary condition in order to power emission in these sources. 

Magnetic reconnection occurs in compact regions of converging flows in which the magnetic field topology abruptly changes~\cite[for detailed reviews, see e.g.,][]{Marcowith:2020vho,Kagan:2014hea}. In the original theoretical description proposed by Peter Sweet and Eugene Parker~\cite{1958IAUS....6..123S,Parker:1957abc}, a current sheet develops within region, for which the density becomes very large due to the compactness of the region. In such a situation, the electrical resistivity builds up to the point where the magnetic field decouples from the plasma, allowing field lines to diffuse and reconfigure so that they form a new topology. Magnetic tension acting on the reconfigured field lines forces the plasma out of the region in the form of exhausts; thus, the magnetic energy of the inflowing plasma is converted to kinetic energy of outflowing particles. While this picture assumes a collisional plasma, reconnection can also occur in collisionless plasmas, though factors other than the resistivity will drive the reconnection process (such as, electron inertia in a two-fluid model~\cite[see e.g.,][]{Melzani:2014jsa}).

The Sweet-Parker description of magnetic reconnection is quite effective in illustrating the phenomenology of the process; however, the reconnection rates it predicts are too low to explain observed phenomena in which reconnection is expected to play a role (i.e., solar flares~\cite[see e.g., ][]{Kagan:2014hea}). As such, a central focus of theoretical studies of magnetic reconnection is to determine how fast reconnection can occur~\cite{Matthews:2020lig}. Turbulent fluctuations can lead to the formation of many smaller reconnection sites along the current sheet~\cite[e.g.,][]{Lazarian:1998wd}. Tearing or plasmoid instabilities can fragment the current sheet into several magnetic islands~\cite[e.g.,][]{Loureiro:2007gv,2010PhRvL.105w5002U,DelZanna:2016uiq}. Both scenarios effectively decrease the transverse length scales over which reconnection takes place, increasing the reconnection rate.

While the descriptions of magnetic reconnection provided above focus on non-relativistic models, such models can be generalized to the relativistic regime~\cite{Lyubarsky:2005zt}, which is more favorable to efficient particle acceleration~\cite[see e.g.,][]{Sironi:2014jfa,Werner:2014spa,Guo:2014via}. Particle acceleration in reconnection scenarios can occur via several mechanisms~\cite[for review, see e.g.,][]{Blandford:2017chu}. The current sheets that develop during magnetic reconnection events provide electric fields that directly accelerate particles~\cite[e.g.,][]{1996ApJ...462..997L,Kirk:2004ap}. The converging flows inherent in magnetic reconnection events present a situation that is analogous to shocks or colliding scattering centers; hence, Fermi-like acceleration may occur~\cite[e.g.,][]{Sironi:2014jfa,deGouveiaDalPino:2003mu,2012MNRAS.422.2474D,deGouveiaDalPino:2013qng,Guo:2015ydj} (for discussion of Fermi acceleration, see \cref{subsubsec:FermiAcc}).

\paragraph{Future Progress in Acceleration Theory}

{\color{red} Detailed studies of plasma phenomena are key to revealing the physical processes connected with acceleration mechanisms.} 
%
Recent progress in kinetic and \ac{MHD} simulations have significantly advanced models of the aforementioned acceleration mechanisms, as well as enabled investigations into other mechanisms, including one-shot/shear acceleration, stochastic acceleration via turbulence, and wakefield acceleration. Future theoretical studies via dedicated numerical simulations together with detailed multi-messenger modeling and observations of candidate UHECR sources will provide crucial information that will reveal the origin(s) of \acp{UHECR} and the extremes of cosmic particle acceleration.

\subsubsection{Magnetic fields}\label{subsec:magneticfields}

Magnetic fields are ubiquitous in the Universe and exist on scales ranging from planets and stars up to galaxy clusters.
Despite the clear indications for the existence of magnetic fields in the large-scale structure of the Universe, it is not clear how they originated. Some local astrophysical process could have given rise to them, or they could have had a cosmological origin, through a global process such as Inflation or phase transitions (e.g., QED or QCD) in the early universe.
An evidence in favor of the cosmological scenarios would be the observation of  magnetic fields in cosmic voids.
Given the central role played by magnetic fields in the evolution of galaxies, it is important to understand how, where, and when the first magnetic fields were created.
The understanding of how, where, and when the first magnetic fields were created is of fundamental importance to many aspects of modern-day astrophysics. They play a major role in the evolution of galaxies, they could affect the synthesis of elements after the Big Bang, they could leave imprints on the cosmic microwave background distribution, and they are essential to describe the motion of charged cosmic messengers.

\paragraph{The Galactic magnetic field}
\label{sec:galactic_mag_fields}

The study of the Galactic magnetic field is a notoriously challenging
task, as described in Ref.~\cite{Jaffe:2019iuk}.  The observable signatures are degenerate with other quantities such as different particle distributions.  Three traditional observables remain among the best probes available for large-scale \ac{GMF}:  starlight polarization, Faraday \ac{RM}, and synchrotron emission. These observables can be simulated by numerical observations of model galaxies, as has been done in Refs.~\cite{broadbent,Jansson:2012pc}.  A number of models have been fit to some of the data, but the status of such studies today remains uncertain due to degeneracies in the parameter space of the components of the \ac{ISM}.  See various reviews for a summary of the current status of Galactic and extragalactic magnetic field studies \cite{Jaffe:2019iuk,Beck:2013bxa,Farrar:2014hma,2015A&ARv..24....4B,haverkorn:2014}. 

Based on observations that are available today, there are several global Galactic magnetic field models that can all fit some of the data, and there remain degeneracies among them \cite{Jaffe:2019iuk,Unger:2017kfh}.  
However, there is broad agreement on several features of the \ac{GMF} (for a detailed review, see e.g.,~\cite{haverkorn:2014} and references therein):
\begin{itemize}
\item in the disk of the Galaxy, the field follows an axisymmetric spiral (but the pitch angle is uncertain \cite{Unger:2019xct,2018arXiv180104341S}) with a total strength of about 6\,$\mu$G;
\item the total field strength is dominated by the turbulent component with a highly variable coherence length from parsecs to $\approx$\,100\,pc scales, see e.g., Ref.~\cite{Jaffe:2013yi};
\item the field likely extends to at least a few kpc above the Galactic disk (see e.g., Ref.~\cite{Orlando:2013ysa});
\item the coherent component reverses several times at large scales ($\gtrapprox$\,1\,kpc) in the Galactic midplane (see e.g., Ref.~\cite{Brown:2007qv});
\item an x-shaped vertical component is seen in almost all external galaxies observed with sufficient sensitivity \cite{Beck:2013bxa}, supporting hints of such a feature in the Milky Way (see e.g., Ref.~\cite{Jansson:2012pc}).
\end{itemize}

\ac{SKA}~\cite{Braun:2015zta,SKAMagnetismScienceWorkingGroup:2020xim} will provide an order of magnitude more pulsars in the Galaxy than currently observable and precise parallax distance measurements out to tens of kpc \cite{Smits:2011zh}, i.e., reaching even to the opposite side of the Galaxy. These measurements will provide probes of the 3D magnetic field at kpc scales across a large fraction of the Galactic disk. 
The mean \ac{RM} in a particular region probes the coherent field component, and the variance among the \acp{RM} in the region probes the stochastic components.  The low frequency Phase I will be coming online in 2023, though the full survey of Galactic pulsars will not be available until the end of the  decade at least.  But in the meantime, projects that are pathfinders for the \ac{SKA} are already taking data \cite{ASKAP:2007rlq,2010AAS...21547013G,vanHaarlem:2013dsa,Jarvis:2017aml}.
\ac{SKA} and its pathfinders will also map external galaxies at high resolution and sensitivity for both diffuse synchrotron emission and background \acp{RM}. Such studies provides insight into cosmic rays and magnetic fields in the disks and halos of galaxies similar to the Milky Way \cite{2015A&ARv..24....4B,2020A&A...639A.111S,2020A&A...639A.112K,Heald:2021wnt}. In turn, learning about these processes in other galaxies informs studies of the Milky Way, particularly through enabling probes of regions that are not visible from the inside. 
For instance, cosmic ray diffusion and streaming depend on the local magnetic field structure, which can be modeled on large scales ($\sim$\,1\, kpc) using the CHANG-ES polarization data.  On smaller scales, constraining the anisotropy in the turbulent component of the magnetic field can be done by measuring the correlation lengths of high angular resolution observations such as with the \ac{SKA} \cite{2015A&ARv..24....4B}. 

Mapping the local magnetic field in 3D will take another large step forward with the upcoming PASIPHAE survey \cite{2018arXiv181005652T}. This survey will cover 50\% of the sky beyond 30$^\circ$ from the celestial equator in both hemispheres, and measure starlight polarization out to 1--2\,kpc.  It will measure the orientation of the field toward 4 million stars observed in polarization. These measurements combined with Gaia distance and extinction information will provide a precise 3D map of magnetic fields in the nearest $\approx$\,2\,kpc.


By the end of the coming decade, the measurements described above will determine:
\begin{itemize}
\item whether the observed field reversals in the disk of the Galaxy are relatively local or whether they relate to the large-scale (more than a few kpc) structure of the Galaxy (pulsars;  \ac{SKA} pathfinders);
\item the strength of the coherent field component in the disk as a function of Galacto-centric radius and possibly spiral arm position (pulsars;  \ac{SKA} pathfinders);
\item the strength of the stochastic field components in the disk as a function of Galacto-centric radius and possibly spiral arm position (pulsars;  \ac{SKA} pathfinders);
\item the orientation of the magnetic field within 1--2\,kpc, accurate to within $\approx$\,10\,pc (dust and stars, Gaia and PASIPHAE);
\item the strength of all three field components as a function of Galacto-centric radius and height above the disk in the halos of external galaxies similar to ours viewed edge-on (external galaxies; \ac{SKA} pathfinders, \ac{SKA});
\item the locations of field reversals, if they exist, in the disks of external galaxies similar to ours (external galaxies;  \ac{SKA} pathfinders, \ac{SKA}).
\end{itemize}

Note, however, that the list above does not account for studies of \ac{UHECR} deflections that will be achievable with future measurements. For instance, the observation of multiplets confidently associated with specific sources will provide a crucial probe of the field strength in the halo of the Galaxy independent of other components of the \ac{ISM}, see \cref{sec:galactic_mag_fields}. Nonetheless, all of these data sets will need to be modeled simultaneously~\cite{Boulanger:2018zrk}.





While the first phase of \ac{SKA} results are expected toward the end of the next decade, some results will not arrive in full until beyond 2030. With the second phase, \ac{SKA} will ``potentially find all of the
Galactic radio-emitting pulsars in the \ac{SKA} sky which are beamed in our direction'' \cite{Keane:2014vja}. But some important features of the \ac{GMF} will remain to be determined, even with the phase two SKA. Without a 3D probe of the \ac{GMF} in the halo of the Galaxy where there are no pulsars or stars, the only measurements that will be possible are the average field components along the line of sight. The would be no way, for instance, to determine how high above the Galactic Plane the field extends, necessitating continued reliance models based on external galaxies seen edge-on. 


Even with \ac{SKA} measurements of Faraday \acp{RM} and distances to almost every pulsar in the Galaxy, the accuracy of the determination of the variation in the \ac{GMF} across spiral arm density waves in different components of the \ac{ISM} will be no better than of order 1\,kpc (based on current estimates for the density of pulsars remaining to be discovered). Again, models for the Milky Way will have to rely on high-resolution ($\approx$\,1\,pc) \ac{SKA} observations of nearby galaxies and high density background \ac{RM} sources through them.




In all cases, advances in theoretical work and numerical simulations will be crucial in using observations of other galaxies to  model the aspects of the \ac{GMF} that cannot be directly measured.

\paragraph{Intergalactic magnetic fields}

Knowledge of \acfp{IGMF} is presently limited. This is, in part, due to the lack of knowledge on how magnetic fields originated and how they evolved (see e.g., Refs.~\cite{Kulsrud:2007an,Vachaspati:2020blt} for reviews). \acp{IGMF} can be probed using a variety of techniques. Upper limits on primordial \acp{IGMF} can be obtained from \ac{CMB} measurements~\cite{Barrow:1997mj, Jedamzik:1999bm,Planck:2013pxb,Jedamzik:2018itu}. The magnetic field integrated along the line of sight can be obtained from Faraday tomography~\cite{Akahori:2014qja, Ravi:2016kfj,OSullivan:2018shr,OSullivan:2018adp, Vernstrom:2019gjr,Amaral:2021mly}, using polarized radiation from extragalactic sources with measured distances. Observations of synchrotron emission by a (known) distribution of relativistic electrons provide the field perpendicular to the line of sight~\cite{Vernstrom:2017jvh,2019Sci...364..981G, Locatelli:2021byc,Vernstrom:2021hru,Hodgson:2021rzd}. Lower bounds on \acp{IGMF} can be obtained using gamma-ray observations~\cite{Aharonian:1993vz,Plaga:1995ins}.

\begin{figure}[!ht]
	\centering
	\includegraphics[rviewport=0.05 0 1 1, width=0.465\textwidth]{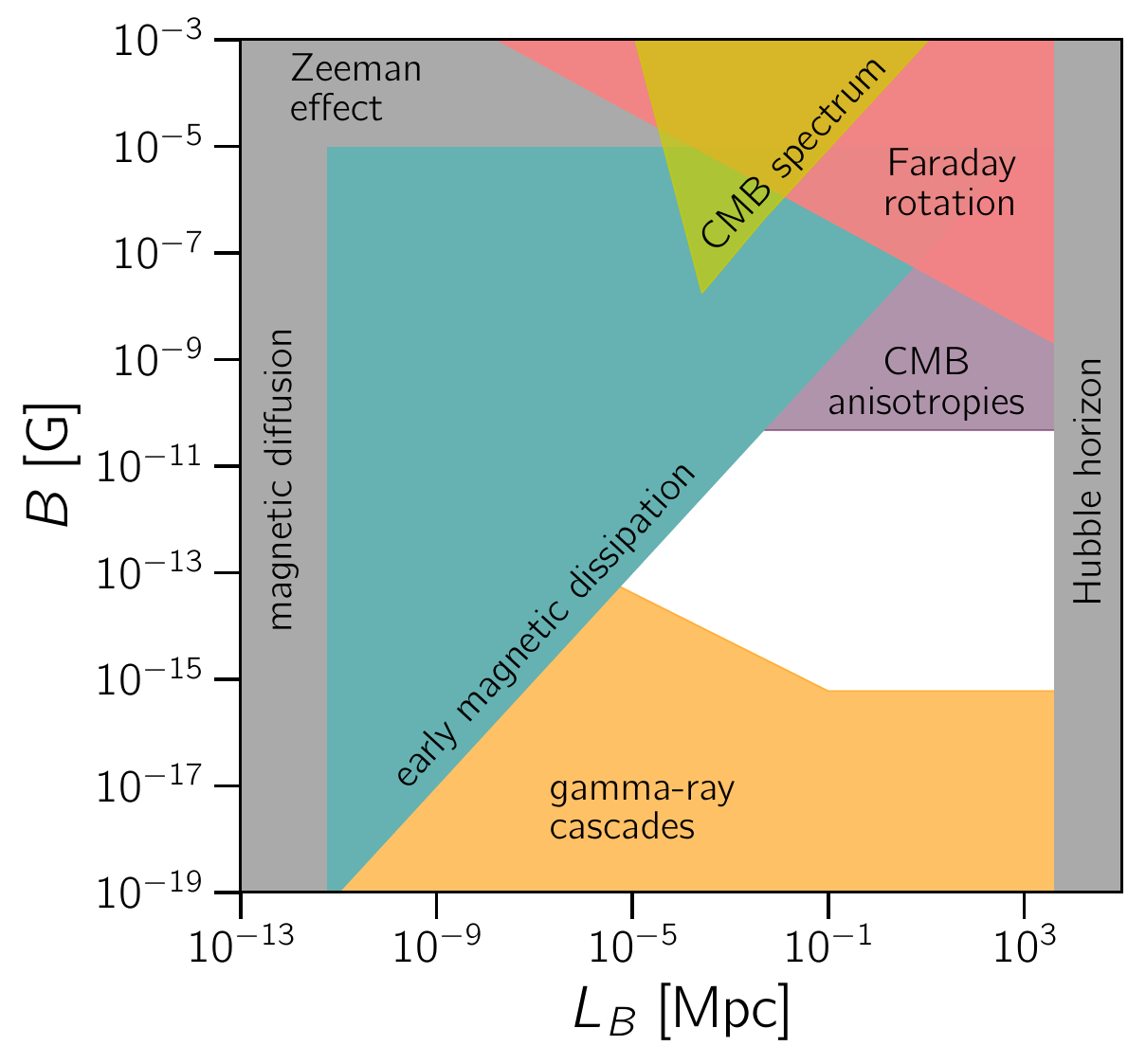}
	\includegraphics[rviewport=0 -0.1 0.95 1, width=0.525\textwidth]{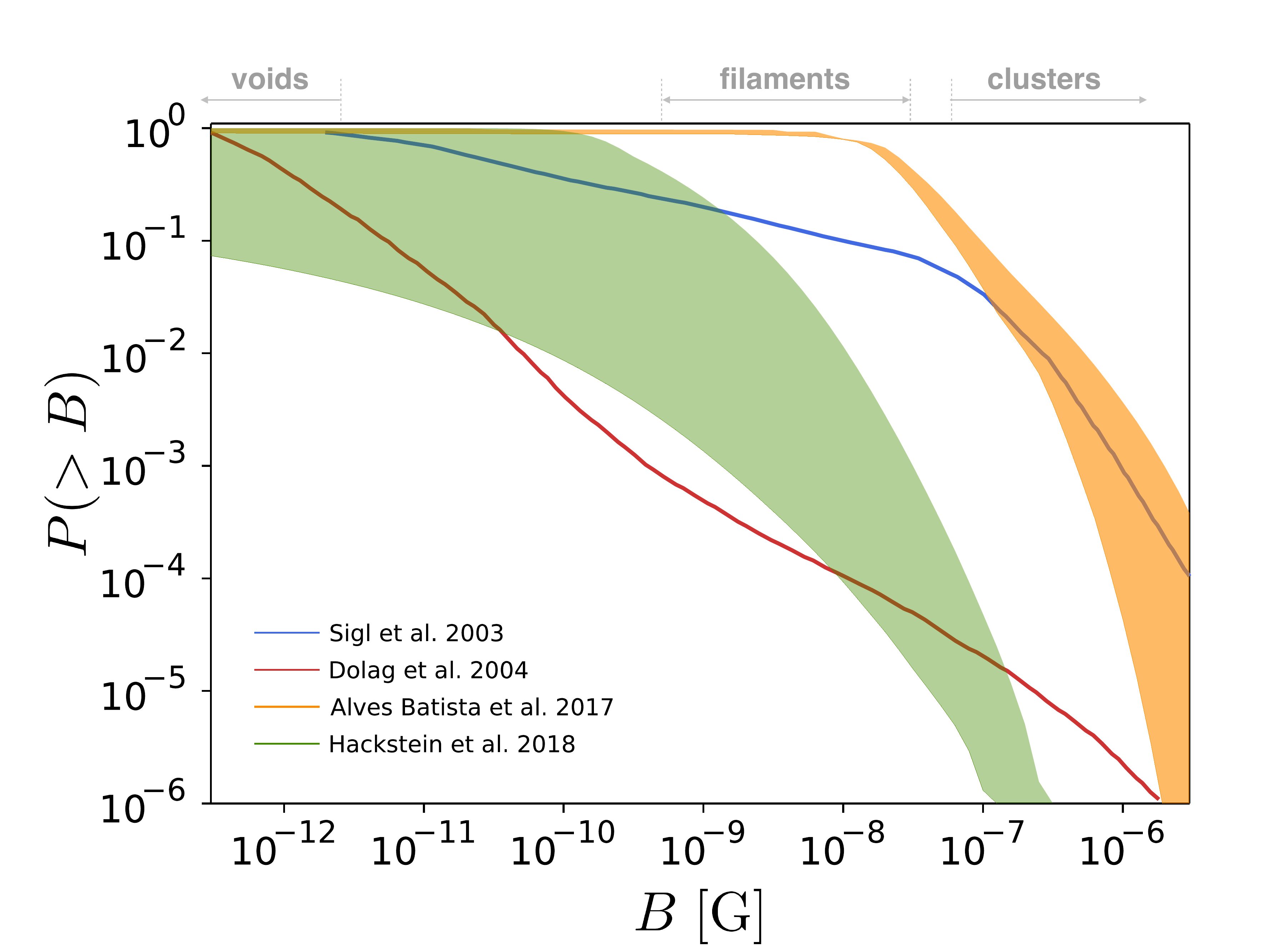}
	\caption{\textit{Left:} Compilation of the available IGMF constraints. The ``gamma-ray cascades'' bound is optimistic and loosely based on Ref.~\cite{Fermi-LAT:2018jdy}. Figure adapted from Ref.~\cite{Batista:2021rgm}. \textit{Right:} Cumulative volume filling factors (i.e., the inverse cumulative distribution function) for various models:  Sigl et al.~\cite{Sigl:2003ay}, Dolag et al.~\cite{Dolag:2004kp}, the upper-limit models by Alves Batista et al.~\cite{AlvesBatista:2017vob}, and the models by Hackstein et al.~\cite{Hackstein:2017pex}. The shaded bands encompass a whole family of models with different topological and spectral properties originated through various processes.}
	\label{fig:IGMFconstraints}
\end{figure}

\acp{IGMF} are present in various astrophysical sites. In galaxy clusters they can reach strengths of up to $\sim$\,$1 \; \mu\text{G}$ in the central regions~\cite{Ryu:2011hu, OSullivan:2020pll}. In filaments they are uncertain but believed to be weaker~\cite{Gheller:2019wlf, Vazza:2021vwy}, below $\sim$\,$10$ and~$100$\,nG~\cite{Locatelli:2021byc, Vernstrom:2021hru,Vazza:2017qge}. The picture is far from clear in cosmic voids. In the inner parts of these regions \acp{IGMF} could, in principle, not even exist if cosmic magnetic fields originated through some local astrophysical process. However, gamma-ray observations provide lower limits on the integrated \acp{IGMF} along the line of sight -- which is dominated by the contribution of the voids -- of $B \gtrsim 10^{-17}$--$10^{-15}$\,G~\cite{Neronov:2010gir, Tavecchio:2010ja,Dermer:2010mm,Finke:2015ona,Veres:2017aou,Fermi-LAT:2018jdy,AlvesBatista:2020oio}. This is generally supported by simulations studies~\cite{Vazza:2021vwy,Vazza:2017qge,Garaldi:2020xos, Martin-Alvarez:2020enk,Katz:2021iou,Mtchedlidze:2021bfy}. The constrained parameter space is summarized in \cref{fig:IGMFconstraints} (left).

The \ac{IGMF} uncertainties in cosmic voids are even more problematic considering that they fill about between 20\% and 80\% of the volume of the Universe, whereas galaxy clusters and filaments, together, fill the remainder of the volume, with clusters filling $\lesssim 10^{-3}$~\cite{Vazza:2017qge,Forero-Romero:2008svv}. Therefore, cosmic rays are more likely to be susceptible to the fields in voids. As a consequence, if they are highly magnetized and UHECR sources are not all local, understanding \acp{IGMF} is of utmost importance.

The coherence length of \acp{IGMF} is also poorly known and essential for understanding UHECR propagation, especially in the diffusive regime. In filaments and galaxy clusters they are generally bound by the size of these structures, but in voids they can assume any value from a fraction of a parsec up to the size of the observable universe~\cite{Vachaspati:2020blt,Batista:2021rgm}. The only existing limits are rather weak, in the range between 10~kpc and 100~Mpc~\cite{AlvesBatista:2020oio}.

The helicity of \acp{IGMF}, too, can significantly affect the propagation of UHECRs and their anisotropy~\cite{Kahniashvili:2005yp, AlvesBatista:2018owq}. This could have interesting implications for understanding the early Universe, since processes such as baryogenesis and leptogenesis can leave specific imprints in the helicity of \acp{IGMF} (see e.g., Ref.~\cite{Vachaspati:2020blt} for details on these connections).

Studies of UHECR propagation in the magnetized cosmic web generally rely on cosmological N-body simulations, in which a given volume is evolved from early times to the present according to magnetohydrodynamical prescriptions. Early works~\cite{Sigl:2003ay,Dolag:2004kp,Sigl:2004yk, Armengaud:2004yt,Das:2008vb} that studied the propagation of UHECRs in these cosmological volumes obtained seemingly contradictory conclusions regarding the prospects for identifying the sources of UHECRs. The situation did not improve with subsequent works, which showed that even the power spectrum of the seed magnetic field can have an impact on the deflections of UHECRs~\cite{AlvesBatista:2017vob,Hackstein:2017pex,Garcia:2021cgu}. The main source of these discrepancies is the different filling factors for each cosmological simulation, as shown in \cref{fig:IGMFconstraints} (right).


But the situation is not as dire as it may seem: even in the worst-case scenario wherein voids have $\sim$\,nG  fields, deflections of 50\,EeV protons from the majority of sources closer than 50\,Mpc would still be less than 15$^\circ$~\cite{AlvesBatista:2017vob}. Naturally, this also depends on the \ac{GMF} (see \cref{sec:galactic_mag_fields}).

\begin{figure}[t]
	\centering
	\includegraphics[width=0.49\columnwidth]{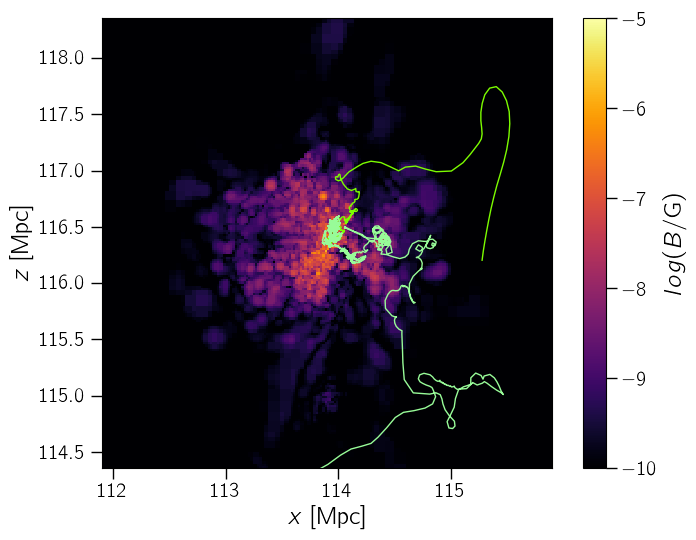}
	\includegraphics[width=0.49\columnwidth]{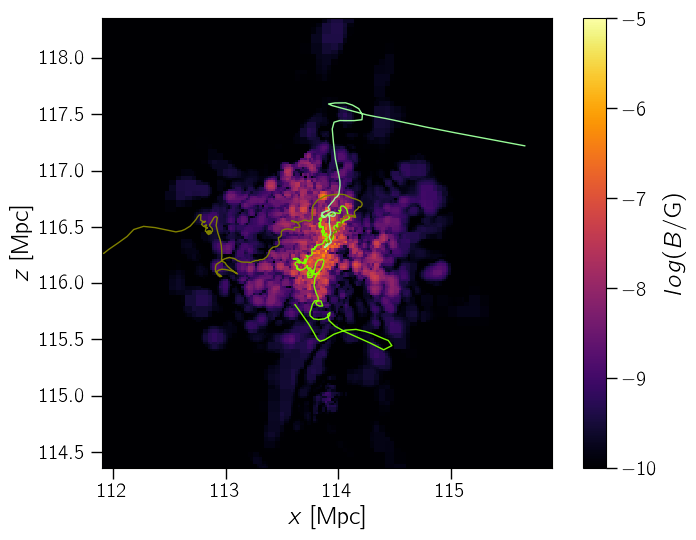}
	\caption{CRs with energies 0.1~EeV (left) and 1~EeV (right) escaping from the center of a galaxy cluster. The color scale indicates the magnetic-field strength. Figure taken from Ref.~\cite{AlvesBatista:2018kup}.}
	\label{fig:Bfields_trajectories}
\end{figure}

\acp{IGMF} also play an important role in determining the counterpart of \acp{CR} in other messengers by increasing the rate at which they can interact with the gas and radiation fields in a given environment, as shown in \cref{fig:Bfields_trajectories}, which may impact the energy spectrum and mass-composition of the observed \acp{CR} (e.g., Refs.~\cite{Unger:2015laa,Fang:2017zjf}). This results in the production of secondary particles such as neutrinos and gamma rays~\cite{Fang:2017zjf,Fang:2016amf, Hussain:2021kye}.
Furthermore, by confining \acp{CR} for a time comparable to the age of the universe, \acp{IGMF} also lead to magnetic horizon effects. Over larger scales, this depends on the distribution of \ac{CR} sources and the properties of the fields, such that it is unclear whether this effect could  be noticeable at energies above $\sim$\,$1$\,EeV~\cite{Globus:2007bi,Kotera:2007ca, Mollerach:2013dza,AlvesBatista:2014aky}.

In the next decade, upgrades of existing radio telescopes will deliver polarization surveys from which detailed rotation measure grids will be built~\cite{2010AAS...21547013G,Jarvis:2017aml, Lacy:2019rfe}. Observations of \acp{FRB} will likely play an important role in this scenario, potentially contributing to breaking the degeneracy between electron density and magnetic field, leading to better measurements of \acp{IGMF}~\cite{Akahori:2016ami,Hackstein:2019abb}. Nevertheless, the measurement of \acp{IGMF} in cosmic voids will remain a challenge.

New gamma-ray facilities that will start operating in the next decade such as \ac{CTA}~\cite{CTAConsortium:2017dvg,CTA:2020hii} might be able to improve the lower bounds on \acp{IGMF}, significantly reducing the parameter space shown in \cref{fig:IGMFconstraints} (left). There are also theoretical challenges that need to be overcome, some related to the difficulty of scanning the full parameter space of all relevant \ac{IGMF} properties~\cite{AlvesBatista:2021gzc}, others to the ongoing debate concerning the role of plasma instabilities on quenching gamma-ray cascades~\cite{Broderick:2011av,2012ApJ...758..102S, Broderick:2018nqf,Yan:2018pca,AlvesBatista:2019ipr}.

A particularly useful avenue to be explored is the potential of novel methods using, for example, multiple messengers~\cite{AlvesBatista:2020oio}, to mitigate \ac{IGMF} uncertainties and to \emph{measure} \acp{IGMF} (as opposed to only \emph{constraining} them). In this case, increasingly detailed cosmological simulations can be used as benchmarks to provide additional insights into the nature of \acp{IGMF}~\cite{Vazza:2017qge}.

On a longer timescale, facilities such as the \ac{SKA}~\cite{Braun:2015zta,SKAMagnetismScienceWorkingGroup:2020xim} and the next-generation Very Large Array (ngVLA)~\cite{Lacy:2019rfe} will reach unprecedented sensitivities and contribute to understanding \acp{IGMF}, delivering rotation measures that will compose tomographic maps of extragalactic magnetic fields. More constraints will come from gamma-ray observatories, combining data from ground-based facilities with observations by space-borne detectors such as the AMEGO~\cite{AMEGO:2019gny}, AMS-100~\cite{Schael:2019lvx}, GAMMA-400~\cite{Topchiev:2021uer}.
\fakesection{The evolving science case}
\vspace{3cm}
{\noindent \LARGE \textbf{Chapter 5}}\\[.8cm]
\textbf{\noindent \huge The evolving science case:}\\[3mm]
\textbf{\LARGE  Defining the new goals for the next decade}
\label{sec:EvolvingScienceCase}
\vspace{1cm}

From \cref{sec:CurrentStatus,sec:AccelSyn,sec:Astrophysics} it is clear that many of the original science goals for the current generation of \ac{UHECR} experiments have been met. However, it has also become clear that, in order to continue to progress toward answering the core questions of \ac{UHECR} physics, further upgrades to our instrumentation and analysis methods are required. This chapter outlines current and near future plans of the \ac{UHECR} community, and highlights new powerful analysis techniques which promise to illuminate the \ac{UHECR} flux in ways that were impossible until now.

\subsection{The upgraded detectors}
\label{sec:FutureDetectors}

Due to the outstanding progress that has made through the Pierre Auger Observatory, Telescope Array Project and the IceCube Neutrino Observatory, outlined in \cref{sec:GoBigOrHome}, it was clear that these established experiments should be further leveraged through detector upgrades and expansions. These upgrades are already well into (or finished with) the development and planning stage with both AugerPrime and \TAxFour in active deployment. Once completed, each of the following experiments will drive scientific discovery for the next 10-years and beyond.

\subsubsection[The AugerPrime upgrade of the Pierre Auger Observatory]{The AugerPrime upgrade of the Pierre Auger Observatory: 24/7 event-by-event mass sensitivity}
\label{sec:AugerPrime}
\begin{figure}[!htb]
    \centering
    \hspace{8mm}\includegraphics[height=3.05in]{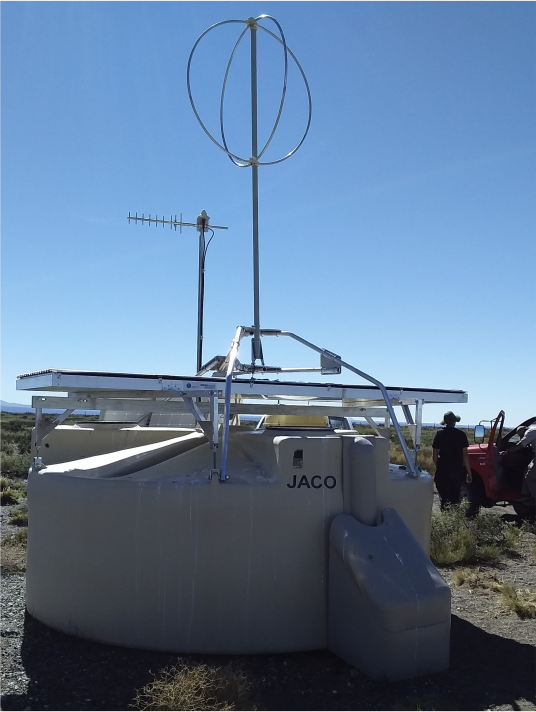}\hfill
    \includegraphics[height=3.05in]{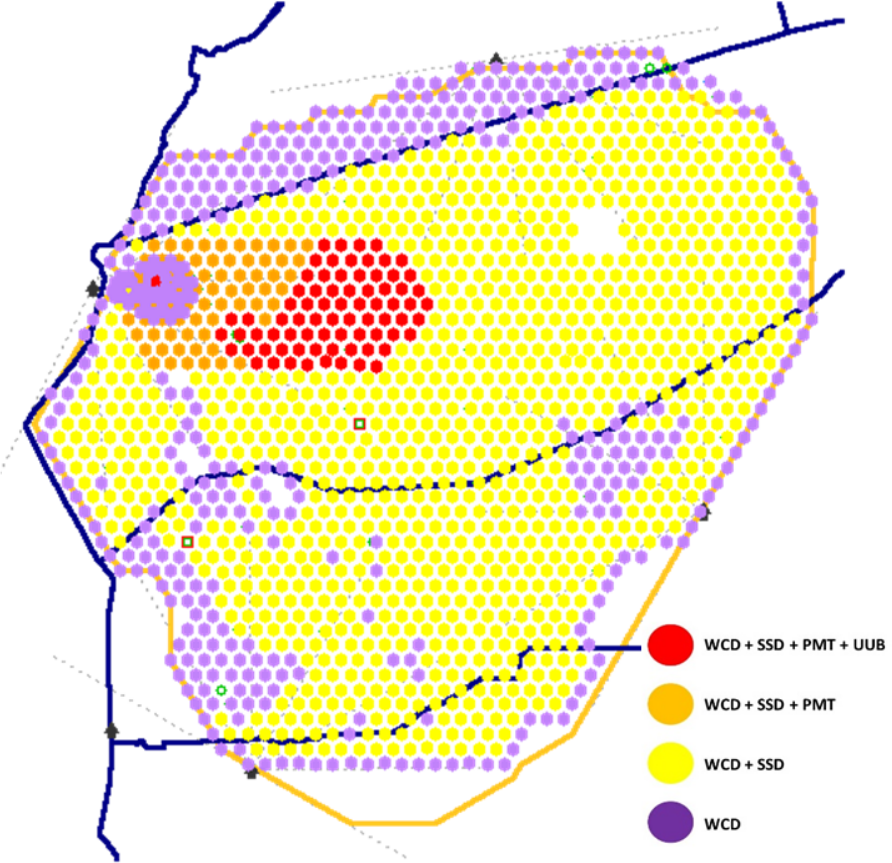}\hspace{8mm}
    \caption{Left: one of the AugerPrime SD stations. From top to bottom, the RD antenna, communication antenna, scintillation detector, and water-Cherenkov detector can be seen.  Right: deployment status of the AugerPrime SD array as of June~10, 2021 (see the text for details).}
    \label{fig:AugerPrime}
\end{figure}

The Pierre Auger Observatory is undergoing an upgrade process know as AugerPrime~\cite{PierreAuger:2016qzd, Castellina:2019irv}. 
In this upgrade, each \ac{SD} station (pictured in \cref{fig:AugerPrime}~left) is being enhanced with:
\begin{itemize}
    \item a $3.8\,\mathrm{m} \times 1.3\,\mathrm{m} \times 1\,\mathrm{cm}$~scintillation detector (\ac{SSD}) placed above the \ac{WCD} tank (excepting stations at the border of the array) to enhance the separation of the muonic and electromagnetic components in measured signals for vertical showers~\cite{PierreAuger:2021ccl};
    \item new, faster electronics and an additional \ac{PMT} with $1''$~diameter in the water-Cherenkov tank to extend its dynamic range~\cite{PierreAuger:2021nnx}; 
    \item an \ac{RD} antenna to detect the emission of inclined showers in the 30--80\,MHz frequency band, enabling a reconstruction precision comparable to that of \acp{FD} with a duty cycle comparable to that of \acp{SD} for inclined showers~\cite{PierreAuger:2021bwp,PierreAuger:2021ece};
    \item and a \ac{UMD} next to each SD-750 and SD-433 station consisting of three $\mathrm{10\,m^2}$~scintillation detectors buried at a depth of 2.3\,m~\cite{PierreAuger:2021nob}.
\end{itemize}
At the time of writing, \ac{SSD}s have already been deployed on 94\%~of the \ac{SD} stations (shown in yellow in \cref{fig:AugerPrime}~right), 150~of which have been equipped with \acp{PMT} (orange), with $\sim$\,$130$~already paired with an upgraded electronics boards (red). Additionally, 10~radio antennas, 7~of which have been so far equipped with read-out electronics, have been deployed along with 25~underground muon detectors.  
In spite of the delays due to the COVID-19 pandemic, the upgraded observatory will begin taking data in 2022 with the upgrade expected to be complete in 2023.
Once complete, the fully upgraded observatory is planned to operate until at least 2032. This upgrade assures that, for the time being, Auger will remain the leading observatory in operation. 
It also will provide an excellent site to cross-calibrate detector developments for the next generation of ground arrays.

\paragraph{Scientific capabilities}

Currently the energy scale of all air-shower measurements at the Observatory is pegged to \ac{FD} calorimetric energy measurements. These are affected by a $\pm14\%$ systematic uncertainty, which is mainly due to uncertainties in the absolute calibration of \ac{FD} telescopes, as well as uncertainties in the shower profile reconstruction, the fluorescence yield and in the aerosol content of the atmosphere \cite{Dawson:2020bkp}.
Once finished, the radio detector array will provide an absolute calibration of the energy scale from first principles independently of the FD measurements, with $\sim$\,$10\%$ systematic uncertainty \cite[][and refs.\ therein]{PierreAuger:2021bwp, PierreAuger:2021ece}.

For composition, by combining data from the \ac{WCD}s and \ac{SSD}s, which have different relative sensitivities to electrons/photons vs.~muons, the muonic content of air showers can be reconstructed. This is important as it represents a key observable for estimating primary masses on an event-by-event basis and for the pursuit of particle physics analyses.
Using a Fisher discriminant developed from all available data, AugerPrime will be able to distinguish between protons and iron showers with merit factors\footnote{The merit factor of an observable~$S$ between two elements~$i,j$ is defined as \(
    f = \frac{\left|\left<S_j\right> - \left<S_i\right>\right|}{\sqrt{\sigma^2(S_i) + \sigma^2(S_j)}}
\).} ranging from around~1.2 to~2.1 depending on the energy and zenith angle, after accounting for the resolution of the reconstruction~\citep[][Tab.~3.3]{PierreAuger:2016qzd}.

This increased mass sensitivity, particularly in the full duty cycle \ac{SD} is important as current Auger \ac{FD} data~\cite{Yushkov:2020nhr} show that at $E \gtrsim 2$\,EeV the composition becomes progressively heavier and less mixed as energy increases. 
However, available statistics quickly run out above the flux suppression (with only~35, 5, and~2 events above~$10^{19.6}$, $10^{19.8}$, and~$10^{20.0}$~eV respectively) so no statement can confidently be made at this stage about whether the trend to heavier compositions continues indefinitely. 
Indeed, in preliminary SD-based estimates \cite{ToderoPeixoto:2020rta} there are indications that the trend may be slowing down after a few tens of EeV\@.  
In particular, a non-negligible fraction of protons in the cutoff region cannot be excluded, which would have wide-ranging implications for the production of secondary neutrinos and gamma rays, the ability to locate \ac{UHECR} sources (as detailed below), and searches for new physics. 
Preliminary studies of techniques to extract composition information from \ac{SD} data using machine learning techniques are being performed \citep[][and refs.\ therein]{PierreAuger:2021nsq,PierreAuger:2021fkf}, but they are affected by large systematic uncertainties (for more see \cref{sec:comp10yr-MLM} and \cref{sec:10yearMass}). 
Thanks to the new detectors of AugerPrime, within five years of operation a proton fraction as low as~10\% will be detectable with $5\sigma$~statistical significance if such a component exists ~\citep[Fig.~8]{Castellina:2019irv}.

\begin{figure}
    \centering
    \includegraphics[width=0.32\textwidth]{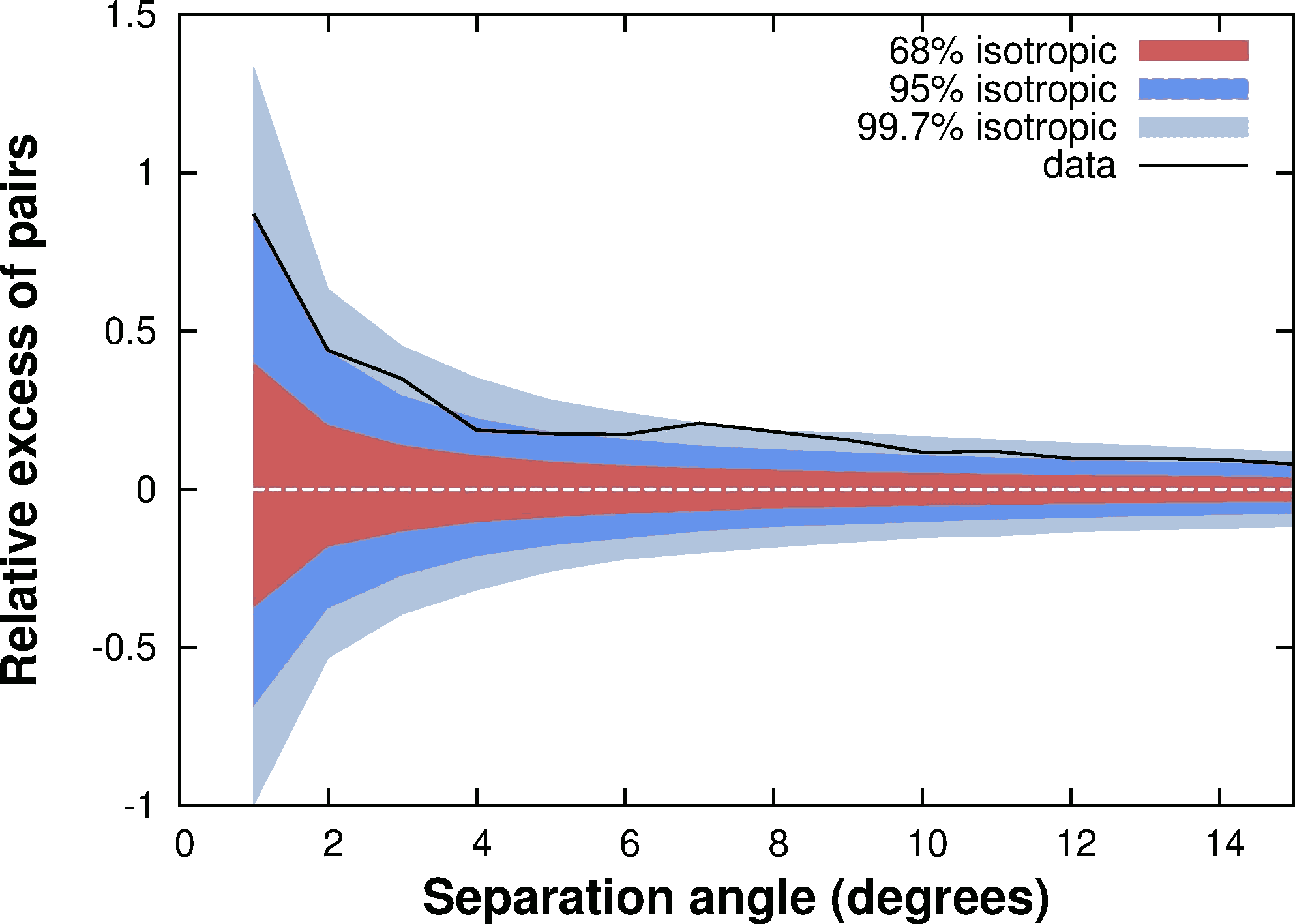}\hfill
    \includegraphics[width=0.32\textwidth]{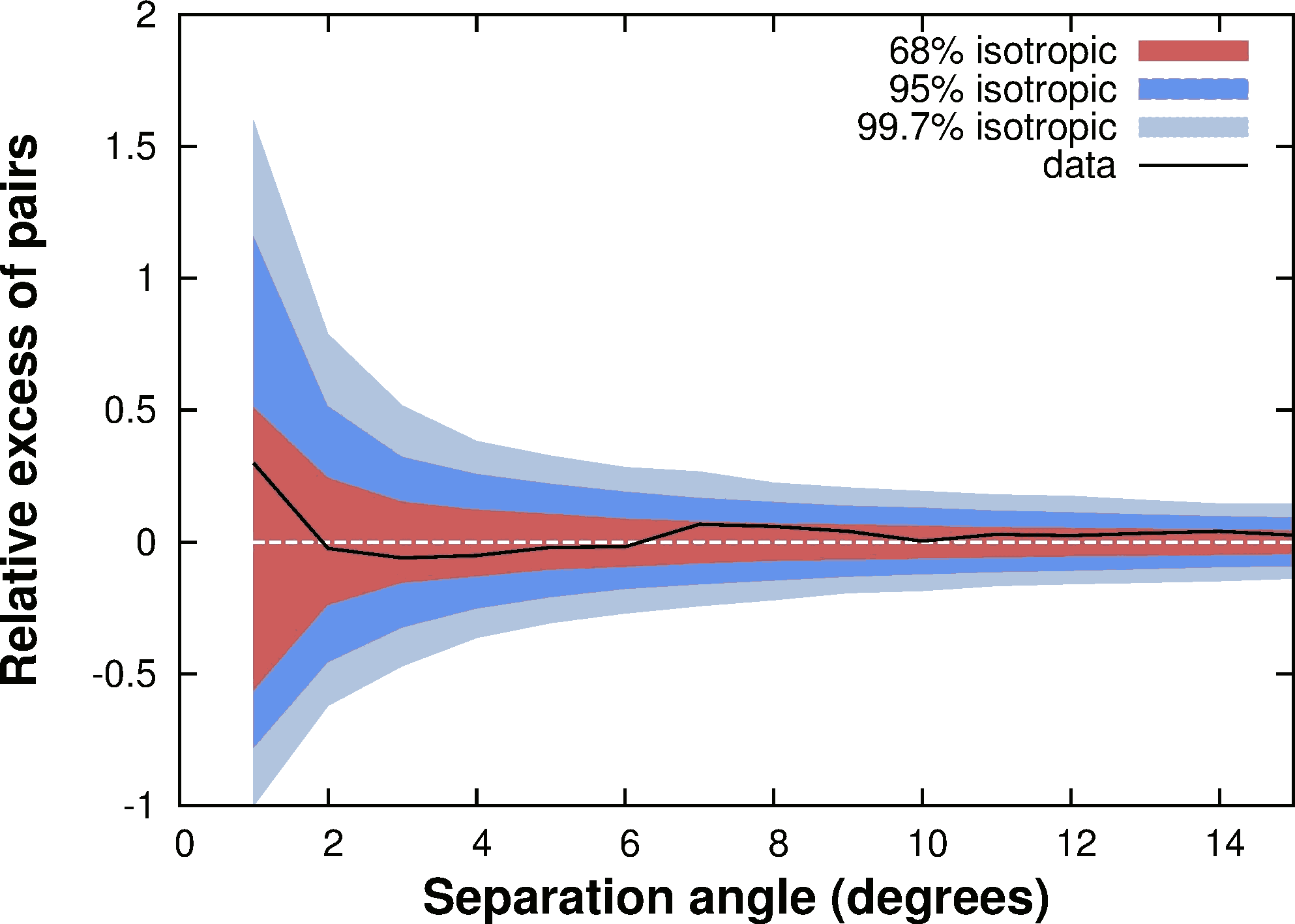}\hfill
    \includegraphics[width=0.32\textwidth]{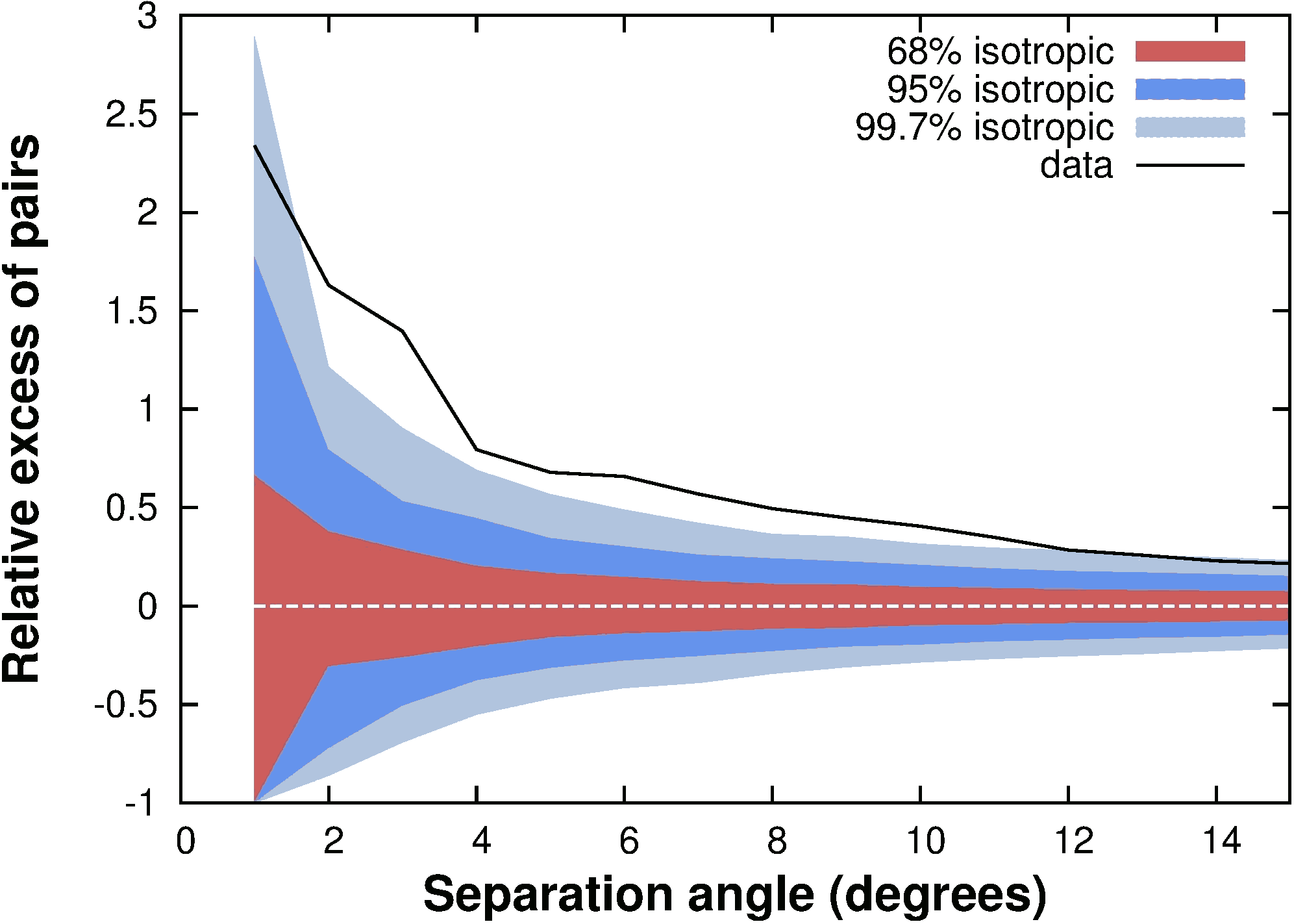}
    \caption{Relative excess of events within a radius~$r$ of \textit{Swift}-BAT \acp{AGN} in the simulated scenario of refs.~\cite{PierreAuger:2016qzd,Castellina:2019irv}, using (left)~all 454 events, (middle)~the 326 least proton-like events, and (right)~the 128 most proton-like events. }
    \label{fig:AGN}
    \vspace{-2mm}
\end{figure}

Beyond simply proton isolation, with estimates of the mass of primary cosmic rays on an event-by-event basis, AugerPrime will be able to study mass-dependent features in the distribution of UHECR arrival directions (see \cref{sec:10yearMass} and \cref{sec:Anisotropy_NextTenYears}). 
For a given energy, light nuclei are expected to undergo smaller deflections in intergalactic and Galactic magnetic fields than heavier ones, and hence to more closely track the distribution of sources. 
If a non-trivial fraction of cosmic rays at post-suppression energies are protons, AugerPrime will enable the construction of proton-enriched samples enhancing our sensitivity to possible anisotropies. 
For example, in \cref{fig:AGN} shows the results of a search for correlation with \textit{Swift}-BAT \acp{AGN} in a simulated scenario \cite{PierreAuger:2016qzd, Castellina:2019irv} where 10\% of cosmic rays with~$E \ge 40$\,EeV are protons, half of which originating from such AGNs. The improvement in sensitivity from being able to select the most proton-like events (right) with respect to using the whole data set without composition information (left) is striking.
Furthermore, the event-by-event composition information of AugerPrime will allow the statistical might of the full duty cycle of both the \ac{SD} and \ac{RD} to be used to confirm or refute the recently detected indication (from \ac{FD} data) that UHECRs are heavier on average at low Galactic latitudes as compared to higher Galactic latitudes \cite{PierreAuger:2021jlg}.

Even when not using the composition information from the new detectors, the continued operation of the Auger \ac{SD} array will further increase the available statistics sufficiently to confirm or refute the current indications of anisotropies.  For instance, as mentioned earlier in \cref{sec:Auger_DesignAndTimeline}, using a linear extrapolation, the indication of a correlation between UHECR arrival directions and the position of nearby starburst galaxies \cite{PierreAuger:2018qvk,PierreAuger:2021rfz} can be expected to reach $5\sigma$~statistical significance by the end of~2026 $\pm$ 2~years.

Also, as further outlined in \cref{MM_outlook10yr}, through the considerable increases to exposure, and likely increase in detection efficiency, AugerPrime will also allow for enhanced searches for UHE neutrino and gammma-ray fluxes. This will allow for current upper limits, which already are the most stringent available~\cite[][Fig.~10]{Castellina:2019irv}, to be lowered further~--- or perhaps to finally detect these phenomena. Either way, this will allow for further improvements to the constraints on models of UHECR sources \cite[][and refs.\ therein]{PierreAuger:2019ens} and on certain exotic scenarios (see e.g.,\cite{PierreAuger:2021tog, Anchordoqui:2021crl, SHDM}).

Last, but far from least, as outlined in \cref{sec:PartUHECROutlook}, the combination of information from different types of detectors and the resulting separation between the electromagnetic and muonic shower components is going to be of vital importance for probing hadronic interaction models in kinematic regimes not accessible to collider experiments \cite{Allen:2013hfa, Anchordoqui:2022fpn, FengSnowmass}.


\subsubsection[The TAx4 upgrade of the Telescope Array Project]{The \TAxFour upgrade of the Telescope Array Project: Massive exposure in the northern hemisphere}
\label{sec:TAx4}
\begin{wrapfigure}{r}{0.5\columnwidth}
    \centering
    \vspace{-11mm}
    \includegraphics[width=0.4\columnwidth]{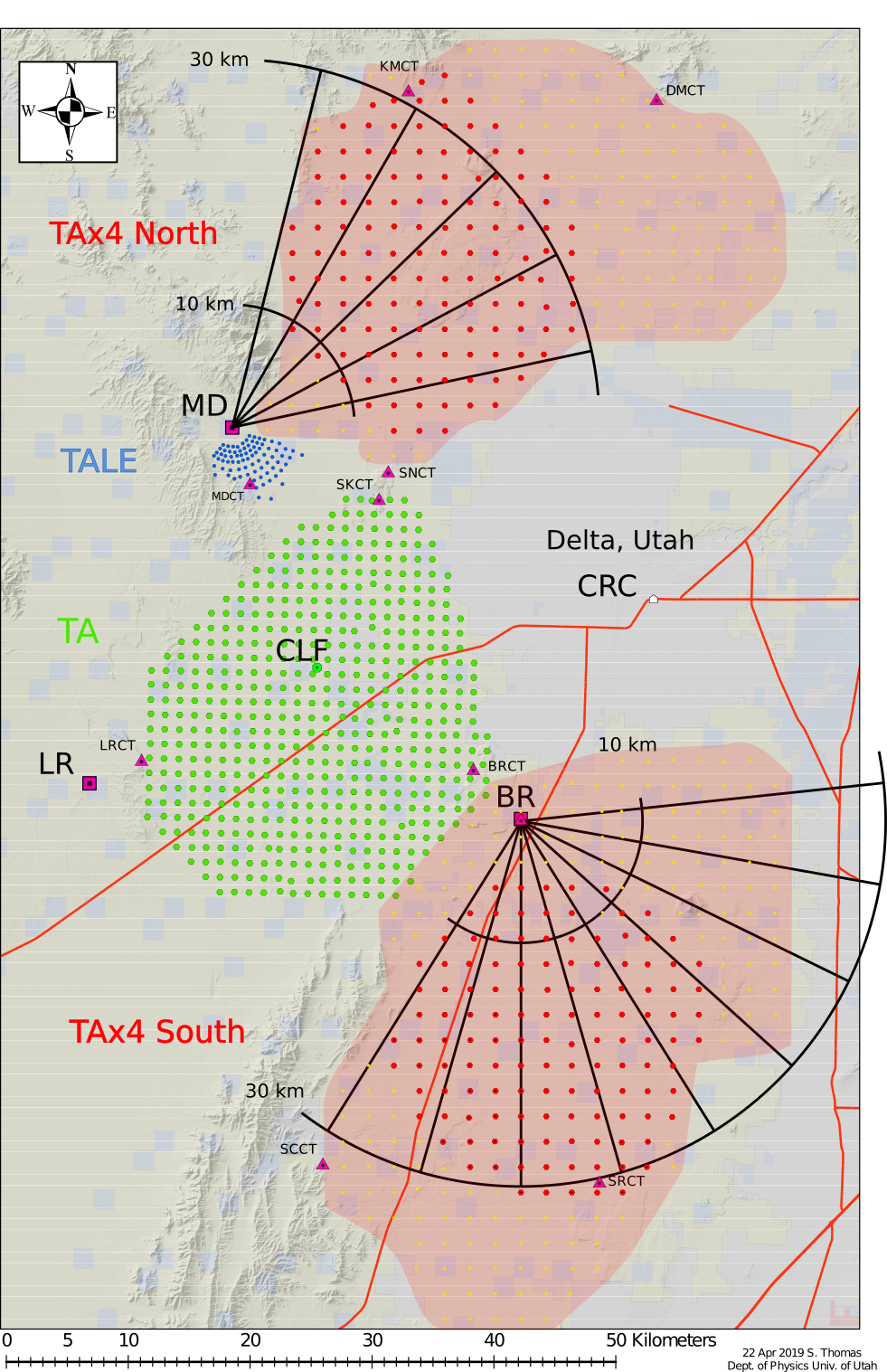}
    \caption{Map of \TAxFour. The new scintillation counters of \TAxFour\ are placed at 2.08 km spacing in two lobes to the northeast and southeast (red). The currently deployed \TAxFour\ \ac{SD} counters are shown with larger (red) dots. 12 new FD telescopes have been added to the MD and BR FD stations overlooking the new \ac{SD} lobes. The arcs mark the approximate extent of the coverage of the new telescopes up to $10^{18}$ and $10^{20}$\,\eV{}.}
    \label{fig:tax4-layout}
    \vspace{-5mm}
\end{wrapfigure}%

In 2014, the Telescope Array Collaboration reported an indication of an excess in the arrival directions of \ac{UHE} cosmic rays ($E>5.7\times 10^{19}$\,eV) just off the \ac{SGP} in the vicinity of Ursa Major \cite{TelescopeArray:2014tsd}.
To better understand this, the collaboration set about to expand the area of the \ac{SD} by a factor of four to $\sim$\,3000\,km$^{\rm 2}$ with the addition of 500 new scintillator detectors at a spacing of 2.08\,km. 
This upgrade, shown in \cref{fig:tax4-layout} has therefore been named \TAxFour{}.
The spacing was optimized to maximize aperture for detecting showers with $E>10^{19.3}$\,eV with full efficiency, while reducing the overall cost of the project. 
The first 257 of the new \TAxFour{} \acp{SD} were deployed in 2019 to maximize the aperture for hybrid events. 
To cover this new area, twelve new telescopes have already been added viewing 3-17$\,^{\circ}$ above the \TAxFour expansion detectors both to calibrate the scintillator array, with its new spacing, as well as to measure composition via hybrid measurement of events at the highest energies.
The deployment of the remaining \ac{SD} stations has been delayed due to COVID-19, however, plans are presently being explored on how to quickly complete the array, with the aim to complete the array in 2023.  
\paragraph{Scientific Capabilities}%
\TAxFour~\cite{Kido:2019enj} will increase the area of the surface of \ac{TA} from 700~km${}^2$ to $\sim$\,3000~km${}^2$, significantly accelerating the rate of data collection, especially at the highest energies. With this data it will be possible to more precisely observe anisotropy features, the energy spectra, and mass composition in the northern hemisphere at energies above $10^{19}$\,eV. The expansion of \ac{TA} composition data will come both from a further refinement of mass sensitive SD analyses applied to the new 3000~km${}^2$ surface array, and an increased hybrid aperture due to the addition of \ac{FD} sites observing the atmosphere over the newly instrumented northern and southern lobes of the \ac{SD}.

\begin{figure}
    \begin{minipage}[t]{0.48\columnwidth}
        \centering
        \includegraphics[width=\columnwidth]{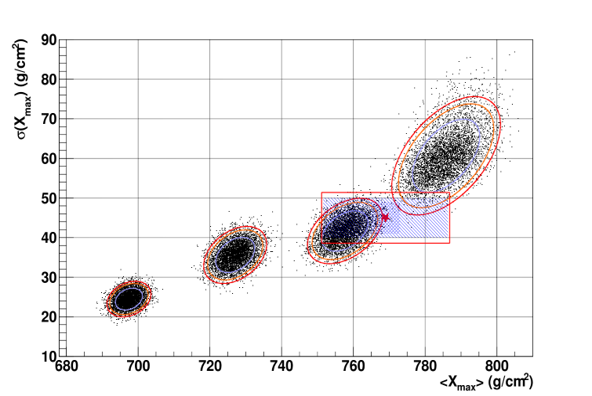}
        \caption{\small {Mean vs $\sigma$ of the \xmax{} distribution for $10^{19}<E<10^{19.2}$\,eV.
        The data, shown in the box, is for 9.5 years of \ac{TA} data.
        The ovals from from top right to bottom left show equivalent statistics for Monte Carlos simulations for p, He, N, and Fe.
        }}
        \label{fig:Comp19.1-9.5y}
    \end{minipage}
    \hspace{1mm}
    \begin{minipage}[t]{0.48\columnwidth}
        \centering
        \includegraphics[width=\columnwidth]{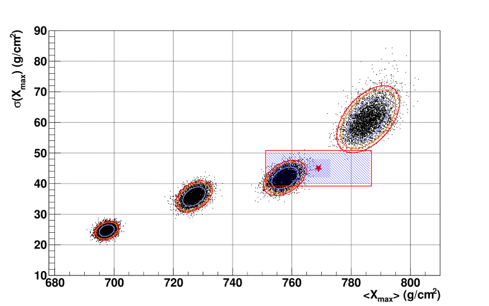}
        \caption{\small {Mean vs $\sigma$ of the \xmax{} distribution for $10^{19}<E<10^{19.2}$\,eV.
        The data, shown in the box, is for 9.5 years of TA data.
        The ovals from from top right to bottom left show Monte Carlos simulations for p, He, N, and Fe, now with statistics for 5 additional years of \TAxFour data.}}
        \label{fig:Comp19.1-9.5+5y}
    \end{minipage}\\[-1mm]
    \begin{minipage}[t]{0.48\columnwidth}
        \centering
        \includegraphics[width=\columnwidth]{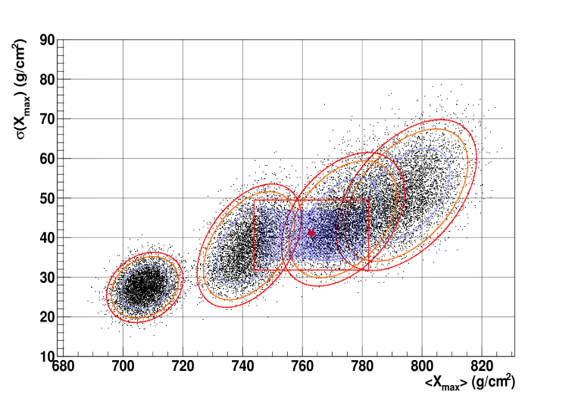}
        \caption{\small {Mean vs $\sigma$ of the \xmax{} distribution for $10^{19.2}<E<10^{ 19.4}$\,\eV{}.
        The data, shown in the box, is for 9.5 years of \ac{TA} data.
        The ovals from from top right to bottom left show equivalent statistics for Monte Carlos simulations for p, He, N, and Fe.}}
        \label{fig:Comp19.3-9.5y}
    \end{minipage}
    \hspace{1mm}
    \begin{minipage}[t]{0.48\columnwidth}
        \centering
        \includegraphics[width=\columnwidth]{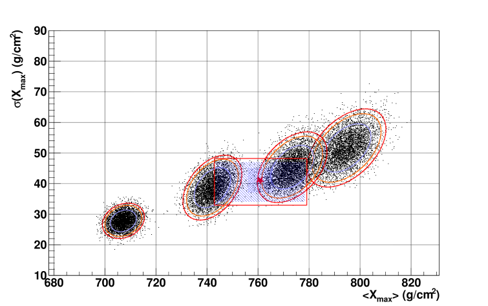}
        \caption{\small {Mean vs $\sigma$ of the \xmax{} distribution for $10^{\rm 19.2}<E<10^{\rm 19.4}$ eV.
        The data, shown in the box, is for 9.5 years of \ac{TA} data.
        The ovals from from top right to bottom left show Monte Carlos simulations for p, He, N, and Fe, now with statistics for 5 additional years of \TAxFour data.}}
        \label{fig:Comp19.3-9.5+5y}
    \end{minipage}
    \vspace{-2mm}
\end{figure}

The significance of the \emph{hotspot} after including the data collected through 2020 is about 5$\sigma$ pre-trial and 3.5$\sigma$ post-trial. While the original brightness seems to not be sustained, the growth of the significance is consistent with a linear trend. If the source is a single source and the significance continues to grow at the present rate, the experiment should have enough data by $\sim$\,2024 for a 5$\sigma$ post-trial observation.

Meanwhile, in the process of studying the energy difference in the high energy spectrum suppression observed by the Telescope Array versus that observed in the southern hemisphere by the Pierre Auger Observatory, the Telescope Array group found an additional bright spot with slightly lower energy ($E>4\times 10^{ 19}$\,eV) in the direction of the \ac{PPSC}. Like the \ac{PPSC} itself, the bright spot is somewhat spread out. The pre-trial significance of this is about 4.5$\sigma$. The penalty factor for this more diffuse spot is still being calculated, but additional high energy data will also be required to verify this as a source. If the rate of signal growth continues as anticipated from present data, this should be confirmed in the next few years.  

The Telescope Array hybrid measurement of cosmic ray composition examines the mean and width of the \xmax distribution. Analysis of the moments of these distributions are consistent with a light and mostly constant composition (protons and/or helium) for cosmic rays with energies greater than $\sim$\,10$^{\rm 18.2}$\,eV. However, for energies greater than 10$^{\rm 19.1}$\,eV the data set has limited statistics and the picture starts to get murky.  
\cref{fig:Comp19.1-9.5y} shows the distribution of the mean vs $\sigma$ of the \xmax distribution for 9.5 years of Telescope Array data in the energy range {10$^{\rm 19}$ $<$E$<$10$^{\rm 19.2}$\,eV} as compared to p, He, N, and Fe Monte Carlo simulations.  
In \cref{fig:Comp19.1-9.5+5y} the Monte Carlo has been updated to show the effect of adding five years of \TAxFour data.  
\cref{fig:Comp19.3-9.5y} and \cref{fig:Comp19.3-9.5+5y} show the same distributions for 
$10^{19.2}<E<10^{19.4}$\,eV.
The addition of five years of \TAxFour data should allow the hybrid composition measurement to extend to up $\sim$\,10$^{19.6}$\,eV.   

At the same time, The Telescope Array collaboration has been improving its machine learning programs to better determine the composition using only the SD data. This is especially important since the SD takes data with a nearly 100\% duty cycle. A \ac{BDT} analysis of 12 years of Telescope Array data also indicates a light unchanging composition (between p and He) for $10^{18}<E<10^{19.7}$\,eV.  
Meanwhile, Auger data from the southern hemisphere shows a composition which gradually becomes lighter from $10^{18}<E<10^{18.4}$\,eV and then proceeds to become heavier and moving towards nitrogen and larger nuclei at the highest energies.
The addition of \TAxFour{} data and continuous improvements to techniques will provide the statistical power needed to explore this potential difference.

There are a number of improvements in the spectrum measurement that will provide additional useful information about the sources and propagation of \ac{UHECR}s.  These include further spectral study of the \emph{hotspot} vs the rest of the sky, improved measurement of the \emph{instep} feature, and more detailed measurement of the declination dependence of the suppression in addition to more refined knowledge of the shape of the suppression itself. All of these require additional data to clarify the situation. For example, the spectral anisotropy in the \emph{hotspot} has a post-trial significance of $\sim$\,3.7\,$\sigma$.  
Additional data can make a large difference in understanding this potential source.  

\begin{SCfigure}
    \begin{minipage}[t]{0.32\columnwidth}
    \centering
    \includegraphics[width=\columnwidth]{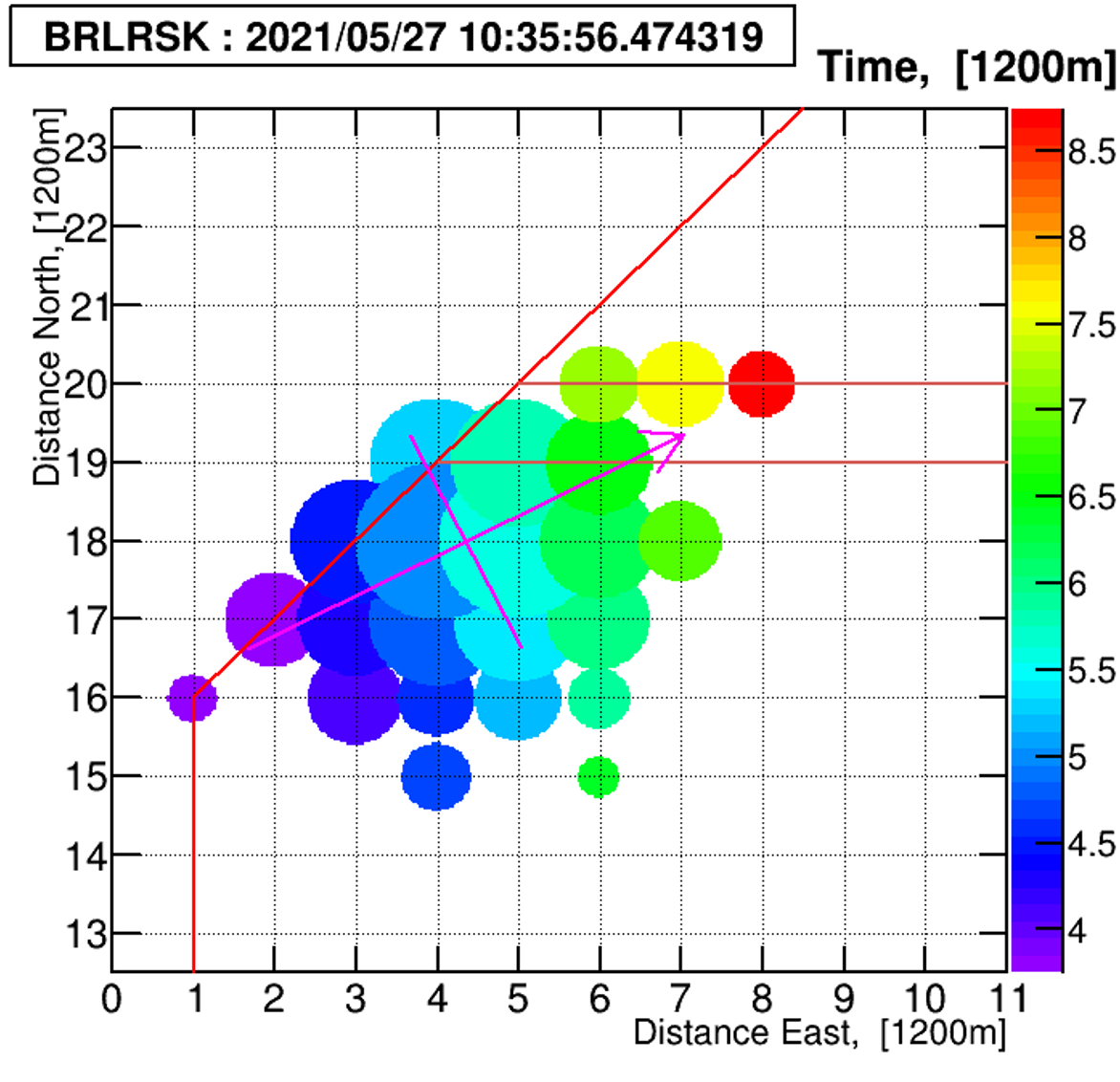}
    \end{minipage}%
    \hspace{1mm}
    \begin{minipage}[t]{0.32\columnwidth}
    \includegraphics[width=\columnwidth]{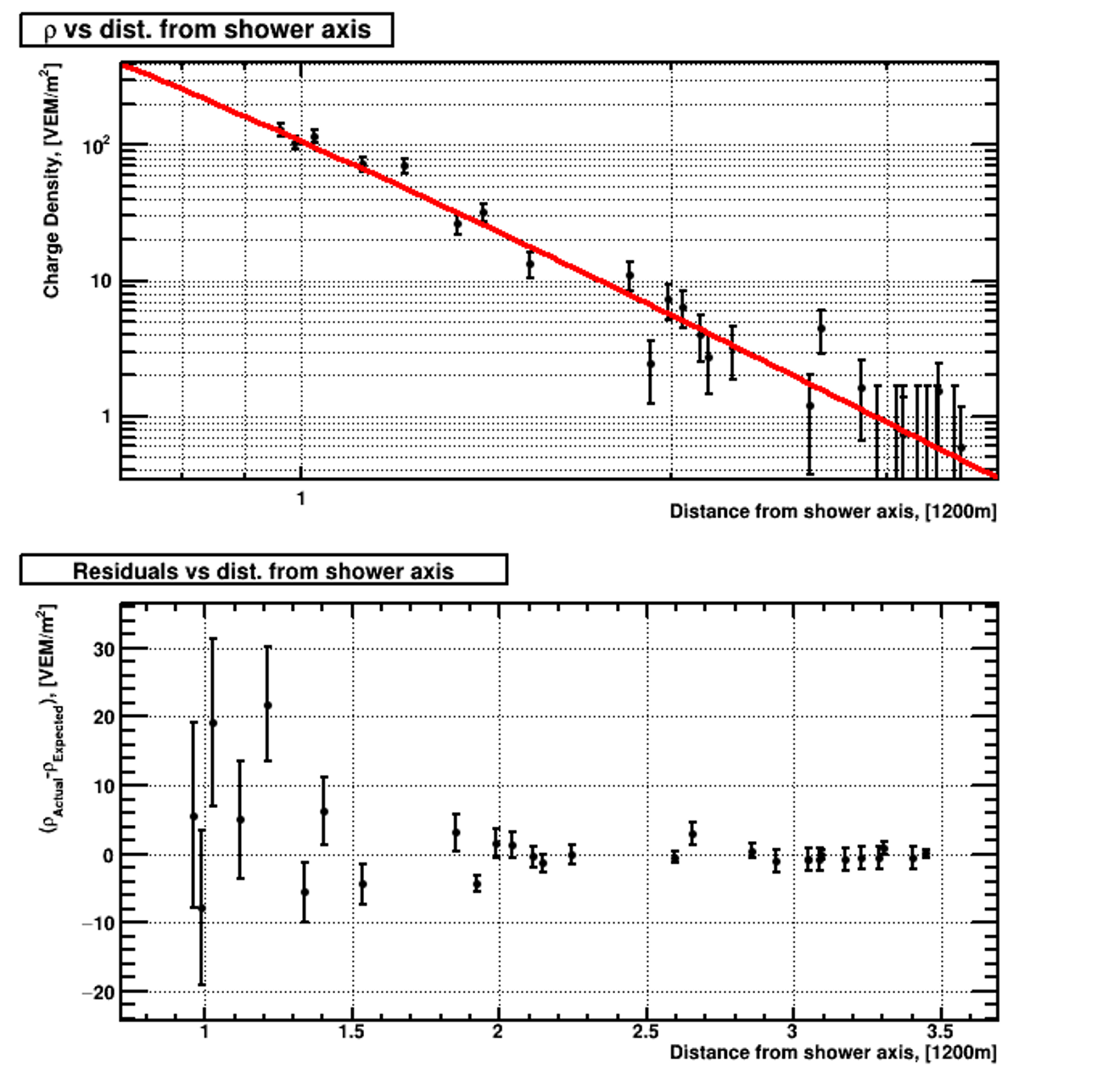}
    \end{minipage}%
    \caption{\small {\bf Left:} {\ac{SD} display of the highest energy event seen by \ac{TA}, at $10^{20.4}$\,eV. The circle size represents the \ac{SD} integrated signal, while the color represents the relative time. The shower core and direction are shown by the cross.} {\bf Right:} {The longitudinal profile of the event. The two counters closest to the core of the shower were saturated and are not included. The value of $S(800)$ is 530\,VEM/m$^2$.}}
    \label{fig:ta-highest}
    \vspace{-3mm}
\end{SCfigure}

Finally, In May 2021, the \ac{TA} \ac{SD} recorded the second most energetic cosmic ray event ever seen, making this event the most energetic seen in an \ac{SD}. This event, with an estimated energy of $10^{20.4}$\,eV, is a third again higher in energy than the next highest energy event observed by \ac{TA}, and gives reassurance that the highest energy event observed by the Fly's Eye Experiment at $10^{20.5}$\,eV was not an analysis artifact. An event display for this event is shown in \cref{fig:ta-highest}.

\subsubsection[The IceCube-Gen2 expansion of the IceCube Neutrino Observatory]{The IceCube-Gen2 expansion of the IceCube Neutrino Observatory: A unique lab for air showers}
\label{sec:IceCubeGen2}
\begin{figure}[!htb]
\centering
\includegraphics[width=0.98\textwidth]{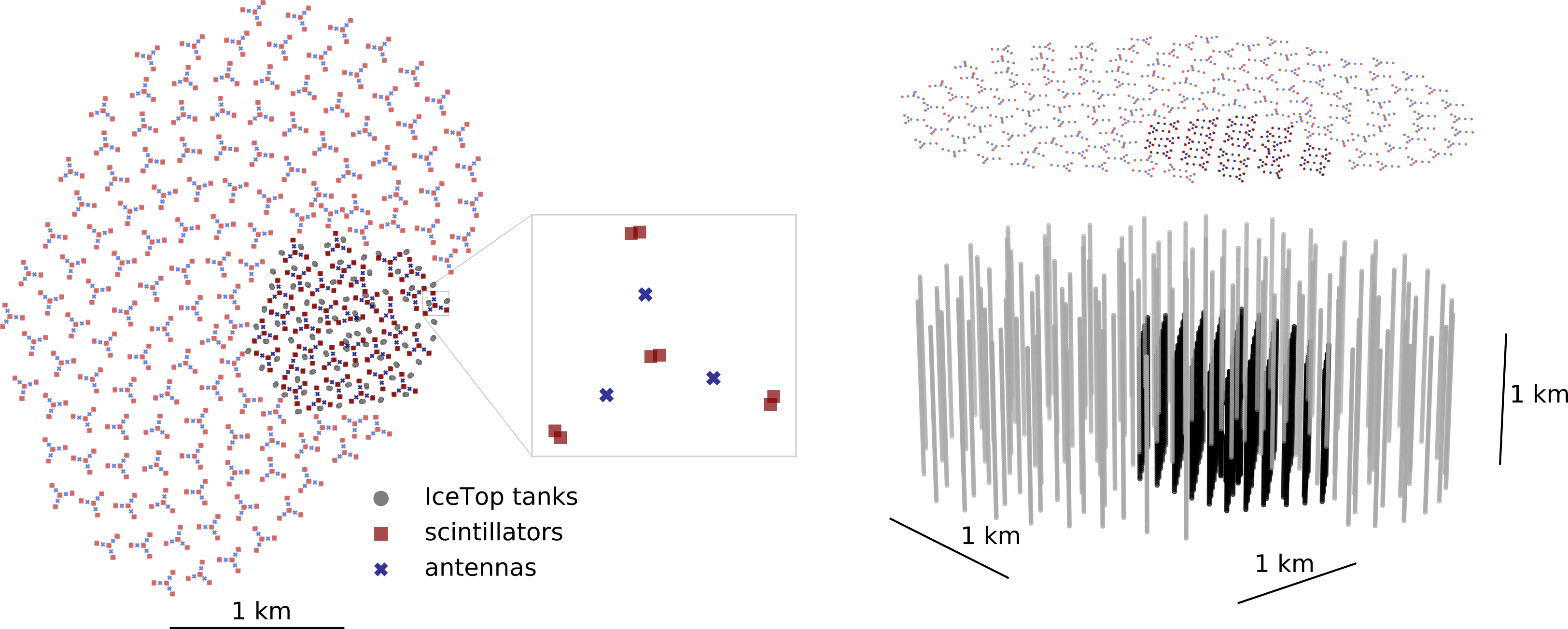}
\caption{\label{fig:icgen2_array} Layout of the IceCube-Gen2 surface array (left) and the in-ice deep optical array (right). The detectors of the IceTop enhancement and IceCube are shown in darker colors. }
\end{figure}

IceCube-Gen2 \cite{IceCube:2014gqr} is an envisioned next-generation extension of IceCube consisting of three sub-components: an $8\,\mathrm{km}^3$ in-ice array of \acp{DOM} optimized for high-energy neutrino astronomy;  a $\sim$\,$500\,\mathrm{km}^2$ radio array for EeV neutrino detection; and a $\sim$\,$6\,\mathrm{km}^2$ surface array instrumenting the snow surface above the in-ice array (see \cref{fig:icgen2_array}). The surface array consists of hybrid scintillation and radio antenna detectors and follows the station design of the IceTop surface enhancement~\cite{Haungs:2019ylq}, a prototype station of which is currently operating at the South Pole. The scintillator panels provide a low energy threshold and trigger for the radio antennas, which in turn allow higher-precision determination of the shower energy and \xmax. Because of the increased zenith-angle range acceptance, the geometric aperture for coincident surface and in-ice events will increase by a factor of $\simeq$\,30 over IceCube. The addition of IceAct air-Cherenkov telescopes can provide additional complementary measurements~\cite{IceCube:2021htd}. 

IceCube-Gen2 construction is planned over a period of 10 years, following the completion of the IceCube upgrade \cite{Ishihara:2019aao}. As with IceCube, data-taking can begin during the construction period, with the first surface array stations planned for installation in Project Year 4. Assuming a nominal construction project start date of 2025, IceCube-Gen2 will commence full operations in 2035.

\paragraph{Scientific Capabilities}
\label{sec:icgen2_science}

The surface component of IceCube-Gen2 is foreseen as a hybrid detector array capable of detecting air showers initiated by \acp{CR} of sub-PeV to a few EeV energies. Each surface station will consist of 8 scintillation detectors and 3 radio antennas placed above each in-ice detector string. Several additional surface stations will be placed between the IceCube and IceCube-Gen2 footprint to provide a uniform coverage between the future surface arrays (see \cref{fig:icgen2_array}). The large number of scintillation modules enables good sampling of the air shower footprint, a low detection threshold, and good reconstruction resolution. 

\begin{figure}
\centering
\includegraphics[width=.48\linewidth]{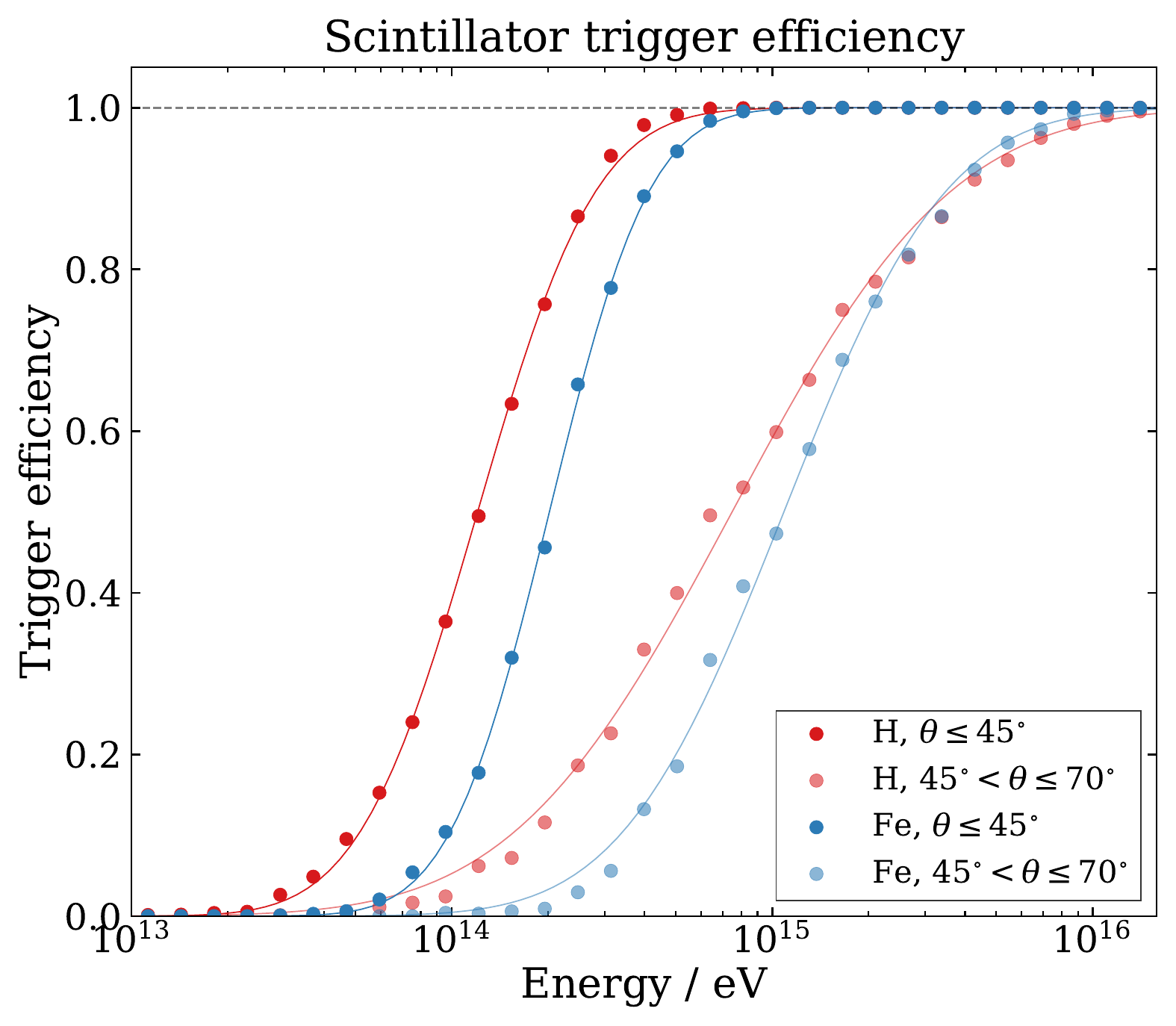}
\hfill
\includegraphics[width=.50\linewidth]{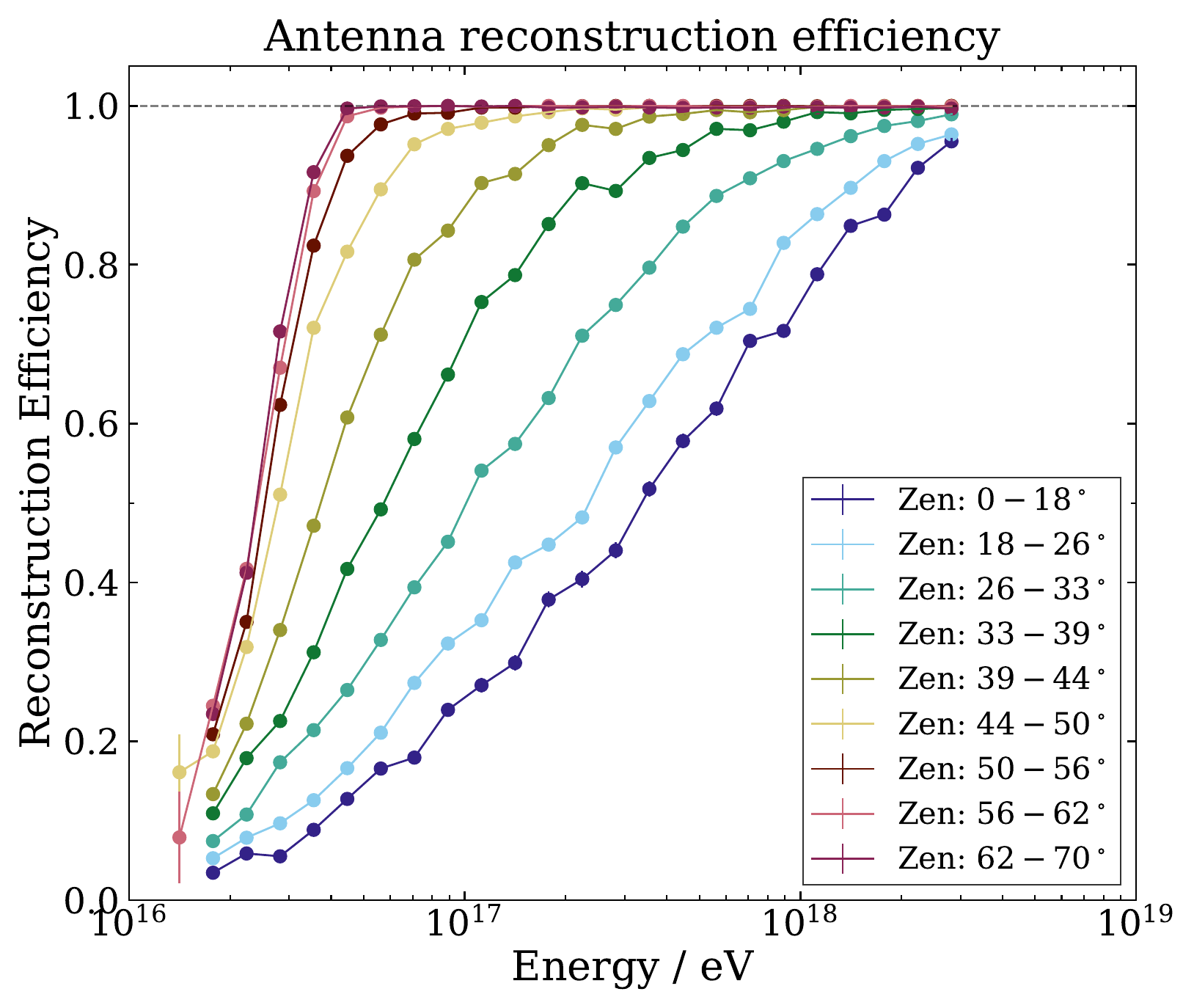}\vspace{-1mm}
\caption{Left: Scintillator array trigger efficiency vs.~primary energy for proton- and iron-induced showers. Right: Directional reconstruction efficiency for the radio antennas vs.~primary energy for different zenith angle ranges for 50/50 mix of p/Fe.}
\label{fig:icgen2_efficiency}\vspace{-4mm}
\end{figure}

\begin{wrapfigure}{r}{0.5\columnwidth}
\vspace{-3mm}
\centering
\includegraphics[width=0.5\textwidth]{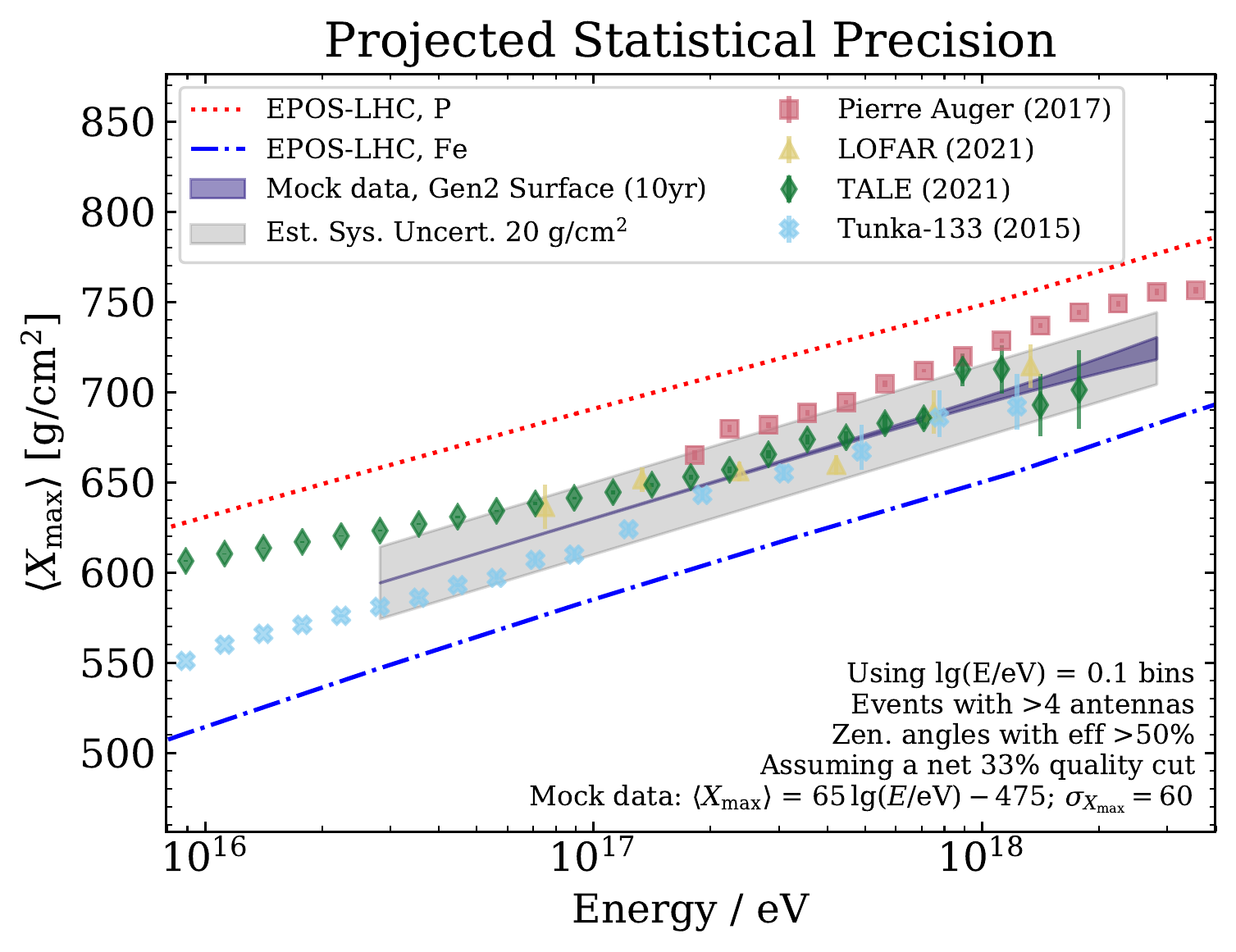}\vspace{-1mm}
\caption{\label{fig:icgen2_xmax} Projected precision for a measurement of \meanXmax for 10 years of mock data.}\vspace{-2mm}
\end{wrapfigure}

The trigger efficiency for proton- and iron-induced air showers (see left panel of \cref{fig:icgen2_efficiency}) indicates that the scintillator array alone will efficiently detect quasi-vertical air showers below PeV energies. This threshold will be also relevant for vetoing the atmospheric muons that constitute the main background for astrophysical neutrino searches. At a few tens of PeV energy and more inclined zenith angles, the radio array starts to be efficient as shown in the right panel of \cref{fig:icgen2_efficiency}. Measurement of the radio emission allows for a more precise reconstruction the energy of the \ac{CR} primary as well as the air-shower \xmax which correlated with primary mass. Hybrid measurements at \ac{CR} energies $\gtrsim$\,100\,PeV will allow for in-depth investigations of the transition region where extra-galactic sources are expected to begin to dominate the CR sky. After 10 years, IceCube-Gen2 will achieve a statistical precision in the \xmax radio measurements comparable to other experiments in this range, enhancing \ac{CR} primary mass determination (see \cref{fig:icgen2_xmax}). Improved measurements of the composition-dependent spectrum can improve the differentiation between different scenarios of the extra-galactic transition~\cite{Kampert:2012mx, Albrecht:2021cxw}. The increased coincident aperture will also allow more sensitive searches for PeV photons \cite{IceCube:2019scr} and improved methods for gamma-hadron separation.



The unique capabilities of this instrumentation are in the combined detection of the mainly electromagnetic component of an air shower at the surface and the high-energy muonic component in the ice. As discussed in \cref{sec:UHECRPartSyn}, atmospheric muons originate from hadronic cascades which above $\approx$\,10\,PeV energies cannot be reliably described by current interaction models; current model predictions underestimate the number of muons arriving at the surface (the Muon Puzzle, see also Refs.~\cite{EAS-MSU:2019kmv,Soldin:2021wyv,Cazon:2020zhx} and \cref{sec:UHECRPartSyn}). This effectively introduces an uncertainty in the interpretation of CR measurements that typically rely on air-shower simulations. 

\begin{SCfigure}
    \begin{minipage}[t]{0.5\columnwidth}
    \centering
    \includegraphics[width=\textwidth]{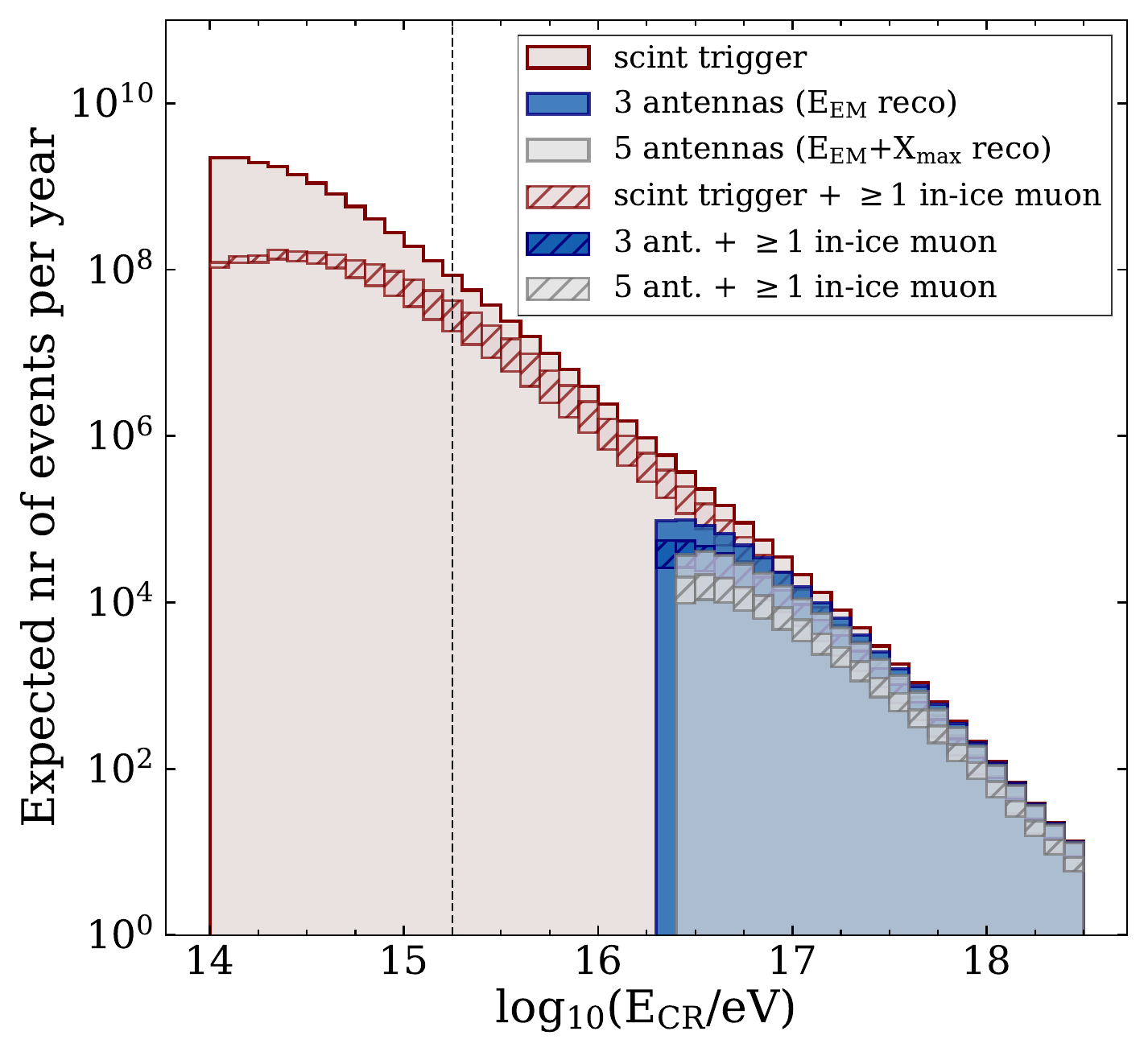}
    \end{minipage}%
    \caption{\label{fig:icgen2_event_cnt} Event rates for the IceCube-Gen2 surface array as a function of primary energy. Rates are shown at trigger level and for muon coincidences with the in-ice optical array, as well as for events at higher energies with radio signals suitable for reconstruction of shower energy and \xmax. The energy threshold for $>$\,99\% detection efficiency is indicated by the dashed line.}\vspace{-7mm}
\end{SCfigure}

The in-ice high-energy ($\gtrsim$ few 100\,GeV) muon measurements and the estimation of $\sim$~GeV muon content at the surface provide unique tests of hadronic interactions in the forward region and can constrain simulation models based on their predicted energy spectra. Preliminary studies combining IceTop and IceCube have recently shown internal inconsistencies in the description of GeV and TeV muons in state-of-the-art hadronic interaction models~\cite{IceCube:2021ixw}; improved analysis techniques are expected to strongly constrain models of muon production in hadronic interactions throughout the next decade. The extension to higher CR energies by IceCube-Gen2 will provide coverage of the region where the Muon Puzzle appears and enable overlap with data from the \ac{UMD} at the Pierre Auger Observatory~\cite{PierreAuger:2020gxz}. The increased aperture for coincident events also opens the possibility to study the angular dependence of the muon content. The estimated statistics for coincident measurements are shown in \cref{fig:icgen2_event_cnt} (for more details on the calculations, see~\cite{IceCube-Gen2:2021jce}). Together with the increased precision resulting from the enhanced air-shower reconstruction provided by the surface array and the improved in-ice calibration, this will contribute to the improvement of current hadronic interaction models at the intersection between cosmic-ray and particle physics~\cite{Albrecht:2021cxw, Dembinski:2021szp, Anchordoqui:2021ghd}. With this, air-shower data from many experiments can in turn be re-analyzed in the context of CR mass composition measurements.

The combined detection of atmospheric muons in relation to parent CR is also relevant at $\gtrsim$~PeV muon energies. These muons dominantly come from the decay of charmed and unflavoured mesons and are produced mainly by CRs of PeV to EeV energy~\cite{Fedynitch:2018cbl}, exactly in the range covered by the surface array of IceCube-Gen2. Due to the large aperture and given enough exposure, IceCube-Gen2 could make the first measurement of the prompt component of the muon spectrum. This will also constrain prompt neutrino production at the highest energies and contribute to a better understanding of the background estimates for astrophysical neutrino searches~\cite{IceCube:2015wro}.

The increased acceptance and statistics of the IceCube-Gen2 surface and in-ice arrays will also allow improved measurements of \ac{CR} anisotropy. The amplitude and phase of the \ac{CR} dipole feature can change for different mass groups of \acp{CR}, in particular, in the transition region of Galactic to extragalactic origin of \acp{CR}.  IceCube-Gen2 will perform precise measurements of these composition-dependent anisotropies in an extended energy range up to a few EeV. 

The unique measurements that can be provided by the surface and in-ice arrays of IceCube-Gen2 will improve the understanding of particle interactions in the air showers and boost current results in cosmic-ray physics. This in turn will provide essential information for the future analysis of multi-messenger data in conjunction with gamma-ray, neutrino, and gravitational wave observations~\cite{Schroder:2019rxp}.


\subsection[Computational advances]{Computational advances: Educated algorithms}
\label{sec:10yrComputation}
As outlined above in \cref{sec:FutureDetectors}, all leading experiments in the field are undergoing major upgrades, aiming to supplement their statistics, particularly at the highest energies, and enhance the quality of their data. These efforts will considerably increase the volume and complexity of the data. Because of this, their reconstruction and Monte Carlo codes are being updated and simultaneously adapted to run in multi-core architectures and heterogeneous clusters combining \acp{CPU} with accelerators (\acp{GPU}, \acp{FPGA}, \acp{DPU}, etc.). 

Most data processing and simulations are produced in \ac{HPC} centers from specific local groups using, predominantly, single-core architectures. However, processing the newly acquired data and simulations requires unprecedented \ac{CPU} time, which can only be overcome by parallel computation using multi-core architectures. Additionally, the growing number of machine learning algorithms employed in recent analyses and the simulation of the radio and Cherenkov emission of extensive air showers all call for accelerators (currently, mostly \acp{GPU})~\cite{Baack:2020jwb, c8}. Given the escalating demand for computing power, data production is progressively being transferred to distributed resources such as grid computing. Furthermore, the responsibility of producing extensive simulation libraries and data processing is being deployed to specific task groups that ensure the centralization and quality of the reconstructed data and Monte Carlo simulations. In parallel, work is being carried out to better coordinate cluster and grid production to foster more uniform data processing and management. 

The typical data output of most experiments is relatively modest compared to the \ac{LHC} experiments. However, the need for larger storage systems will increase in the coming years to accommodate future experiments. While the growth in the data output of the upgraded Pierre Auger Observatory and \ac{TA} experiments is expected to remain relatively modest, IceCube/IceCube-Gen2 will require increased resources. As another example, substantial computing resources will be required for the study of extensive air showers with the \ac{SKA} given the high density of radio antennas per event~\cite{Huege:2016jvc} (see \cref{sec:SKA} for more).

\subsubsection{The advent of machine learning methods}
To further drive the need for accelerators, such as \acp{GPU} machine learning is expected to be more and more of a critical component to UHECR analysis in the future. In particular, driven by recent developments in parallel computing, the large quantity of available training data, and the progress in the design and training of neural networks, \emph{deep learning} with \acp{DNN}, will predominantly shape the world of machine learning today and in the future~\cite{dlbook}. The success of deep learning based algorithms in computer vision and speech recognition~\cite{lecun_deep_2015} has led to first applications in many other fundamental sciences, including physics~\cite{ml_physics_review, dlfpr}.

In the era of \ac{MMA}, these technologies in particular provide promising tools to meet the upcoming challenges of analyzing ever-increasing amounts of data from large-scale astroparticle-physics experiments quickly and accurately.
Machine learning methods accelerate data processing and enable the design of analysis pipelines with very rapid response times, which is essential for \ac{MMA}. 
Speed is not the only advantage that machine learning brings to the table. The new technologies offer the opportunity to significantly improve present reconstruction methods and analysis techniques by identifying subtle patterns in the data that were previously inaccessible. That enables us to devise new strategies to analyze future data and re-analyze existing data, unlocking new opportunities in the field of data-driven knowledge discovery.

In recent years, first applications were developed to adapt machine learning-based analysis techniques in \ac{MMA}, including gamma-ray astronomy~\cite{iact_2019}, neutrino astrophysics~\cite{icecube_cnn}, gravitational-wave detection~\cite{ligo_2018}, and, as seen in \cref{sec:10yearSpectum} through \cref{MM_outlook10yr}, cosmic-ray observations.
So far, most progress has been made in the area of supervised learning and object reconstruction using \acp{CNN}, \acp{RNN}, and more traditional approaches like decision tree learning. Other developments using graph neural networks and generative models are about to unfold.

\subparagraph{Machine learning based event reconstructions}
In object reconstruction, algorithms are developed using an end-to-end approach which involves training the machine learning algorithms on large simulated data sets to infer physically significant quantities; one example is extraction properties of the primary particles given a particle footprint measured by surface-detector arrays. Specific examples of applications include event classification such as discriminating photons from hadrons~\cite{Carrillo-Perez2021}, or distinguishing signal from backgrounds~\cite{Erdmann:2019nie}, as well as reconstructing physics observables like the primary energy, arrival direction, and mass composition~\cite{Erdmann:2017str, IceCube:2019hmk, IceCube:2021ldz}. By applying the algorithms directly to the data, it has been demonstrated that \acp{DNN} are capable of reaching the performance of and even outperforming state-of-the-art results when compared to classical methods. There are examples of such applications from the Pierre Auger Observatory~\cite{Glombitza:2020yhw, carceller}, the Telescope Array~\cite{Kalashev:2020fzg, ta_ml}, and IceCube~\cite{icecube_cnn}, as outlined in \cref{sec:10yearMass} below. The increase in performance is looks to be particularly significant in the reconstruction of mass-sensitive observables and separating out the muonic component of showers, as these are exceptionally complex to extract from detector data. 

 

\subparagraph{Data-driven analysis strategies}
Beyond the reconstruction of physics observables, there have been initial steps towards sensor-close applications, like the denoising~\cite{Erdmann:2019nie, Rehman:2021nw} and unfolding~\cite{unfolding} of measured radio signals as well as the development of a real-time trigger stream~\cite{Zehrer_2021} for \ac{AMON}.
On top of the application to fundamental event reconstruction, approaches for high-level analyses have been developed, for example, by studying cosmic-ray propagation and source properties~\cite{Bister:2021arb}. 

Other approaches exploit the arrival directions of cosmic rays to obtain insights into their origin~\cite{Bister:2020rfv, Kalashev:2019skq} and explore algorithms on non-Euclidean surfaces. The results from simulations are encouraging. However, due to the large uncertainties in the simulated training data, arising, for example, from the modeling of the Galactic magnetic field, significant systematic biases propagate in the analyses. These are challenging to estimate and are so far not well controlled.

\subparagraph{Domain dependency and systematic biases}
Inadequate modeling in simulations can lead to systematic biases when applying models trained on simulations to measured data. This raises particular challenges for the application of machine learning in contexts where the existence of differences between simulations and data are well known and calibration using reference measurements is not possible. Aside from the challenges of modeling of the \ac{GMF}, the precise simulation of hadronic interactions in air shower physics is a major challenge in \ac{UHECR} research (see \cref{sec:AccelSyn}), which rely on simulation-trained algorithms.
In this context of so-called domain adaption, the first basic machine-learning techniques were developed for particle physics~\cite{dlfpr} and \ac{UHECR} observatories~\cite{Erdmann2018}. The results are promising, but more research is needed to better understand and exploit the potential of these techniques.

\subsection[Energy spectrum]{Energy spectrum: A fixed energy scale at higher resolution}
\label{sec:10yearSpectum}
As described in \cref{sec:EnergySpectrum}, it is clear that the overall picture has considerably improved in the last two decades.
The $\sim$\,$10^{5}$\,km$^2$\,sr\,yr of accumulated exposure has allowed for a precise measurement of the spectrum shape, to find the new instep feature (see \cref{fig:newFeature}), and to confirm beyond any doubt the suppression at the highest energies. 
The spectrum has been measured in different declination bands and the differences between the measurements performed in the Southern and Northern hemisphere by Auger and \ac{TA}, respectively, have been scrutinized.
The joint work has revealed an overall good agreement up to $10^{19}$\,eV and some evidence of potential differences in the two hemispheres at larger energies, which need further study. 
The TA-Auger working groups for the spectrum and mass composition have been proven to be very effective in constraining astrophysical models~\cite{PierreAuger:2020kuy}. 
However the interpretation of the suppression of the spectrum in terms of \ac{GZK} effect~\cite{Greisen:1966jv,Zatsepin:1966jv} and maximum acceleration at the sources is still uncertain, being limited by the lack of \ac{FD} data to address primary mass composition at highest energies (see \cref{sec:Astrophysics}).

\subsubsection{Improved exposure and resolution, improved astrophysical insights}
The Auger and TA collaborations are currently implementing an extension of the detection capabilities of the two observatories, aiming to increase the statistics and the sensitivity to primary mass composition at the highest energies. 
The \TAxFour project (see \cref{sec:TAx4}) is in the construction phase and is planned to increase the size of the observatory from 700\,km$^2$ to 2,800\,km$^2$ ($\sim$1,700\,km$^2$ in 2021). 
By 2030, the experiment will have accumulated an exposure $\sim$\,4 times larger than what \ac{TA} has collected so far. 
The Pierre Auger Observatory is also completing an upgrade, called AugerPrime (see \cref{sec:AugerPrime}).
This upgrade does not include an increase in aperture and thus the continued data taking will amount to a $\sim$\,$\sqrt{1.5}$ improvement of the statistical resolution of the spectrum by the end of the decade.
The Auger upgrade will instead bring a better understanding of the mass composition up to the most extreme energies which is crucial to understanding both particle- and astro-physics at the highest energies. 

The extension of the \ac{TA} array is extremely important to confirm, with high statistical significance, the declination dependence of the position of the spectrum steepening at the highest energies as shown in \cref{fig:decbands}.
Moreover, the increase in exposure will allow to significantly reduce the statistical fluctuations that could affect the comparison of the Auger and \ac{TA} spectra in the common declination band (see~\cref{subsec:AugerTAWG}). 
The higher statistics in \ac{TA} and the combination of \ac{WCD} and scintillators in AugerPrime will also allow to understand the systematics between the two experiments and to put a final word on the discrepancy between the spectra at the highest energies. 

\subsubsection{Understanding of the galactic/extragalactic transition}
As shown previously in~\cref{fig:all_spectra}, the spectrum measurements performed at the Auger and \ac{TA} observatories extend to lower energies, allowing for the coverage of almost the entire energy range in which \acp{CR} are studied through the detection of extensive air showers. The lowest energies are attained by analyzing the events dominated by Cherenkov light detected with special fluorescence telescopes that point at high elevation angles ~\cite{TelescopeArray:2018bya,PierreAuger:2021ibw}. For the \ac{SD}-based measurements, the energy threshold is lowered using denser arrays of \ac{SD} stations nested in the main array. Recently, Auger published the spectrum down to $10^{17}$~eV using an array with 750\,m spacing~\cite{PierreAuger:2021hun}. In the near future the region around the \emph{second knee} will be completely covered by both Auger, using an array with 433\,m spacing~\cite{PierreAuger:2021tmd}, and by TA, using \ac{TALE}-\ac{SD}~\cite{TelescopeArray:2021auw}. 
These SD-based measurements are important since they benefit from larger statistics and a more model independent reconstruction, unlike the \ac{FD} ones that must rely on simulations for the exposure calculation. However, the second knee will also be covered by the IceCube-Gen2 (see Section \ref{sec:IceCubeGen2})~\cite{IceCube:2014gqr} experiment at the South Pole, with complimentary methods which should further reduce global systematics and increase statistics.
This is important as a precise characterization of the spectrum at \ac{UHE} is crucial to the study of the transition from galactic to extra-galactic \acp{CR}.

\subsubsection{Better understanding of energy scales}
A further improvement in the understanding of the systematic uncertainties in the measurements performed by \ac{TA} and the Pierre Auger Observatory, in particular the ones affecting the energy scales, will be attained via several activities of cross-calibration. 
One method includes deploying Auger \ac{SD} stations at the \ac{TA} site~\cite{Auger:2017nlp,PierreAuger:2019mun,PierreAuger:2019vqa}. 
By operating an independent Auger hexagonal elementary cell within TA, the parameters extracted from \ac{TA} and Auger \ac{SD} reconstruction algorithms can be compared for the exact same showers. 
This may reveal some discrepancies in the energy determination of showers observed by the \ac{SD} of each experiment. 
The \ac{FAST}~\cite{Fujii:2015dra,Malacari:2019uqw} concept includes deploying an array of low-cost fluorescence detectors at both the TA and Auger sites.
Prototypes of the telescopes have already been deployed and have demonstrated the ability to reconstruct air showers based on the economical design. 
Another proposal includes a portable array of antennas that can be deployed at different ``host" experiments~\cite{Mulrey:2020oqe,Mulrey:2021fqw}. 
Cosmic rays can be measured with the radio array at each site, contemporaneously with the traditional cosmic-ray measurements of the host experiment, and the radiation energy for each event will be reconstructed. The radiation energy at each site can be directly compared, which in turn allows the hosts’ reconstructed cosmic-ray energy to be directly compared.  

At ultra-high energies, the calorimetric measurements of the electromagnetic content of air showers have historically been performed using fluorescence techniques~\cite{Baltrusaitis:1985mx,Abu-Zayyad:2000vin,PierreAuger:2015eyc,TelescopeArray:2012uws}.
However in recent years, the development of the radio technique has proven to be a viable method to directly access the calorimetric energy in the electromagnetic cascade as well~\cite{PierreAuger:2016vya}.
This method, discussed more completely in \cref{sec:RD_tech_development}, will allow for a second method to validate the energy scale of future experiments, therefore providing further information on the largest contribution to the systematic uncertainty affecting the measurement of the energy spectrum.
The measurements performed with \ac{AERA}~\cite{PierreAuger:2015hbf}, a set of radio detectors installed in the denser array of the SD at the Auger site, together with the measurements that will be performed in very inclined showers with the AugerPrime radio antennas, will be important  
for improving the understanding of a major systematic uncertainty.


One of the largest contribution to the uncertainty in the energy scale of the \ac{UHECR} observatories is related to the absolute calibration of the detectors, both for the fluorescence and radio detection techniques.
For both \ac{TA} and Auger, the uncertainty in the absolute calibration of the fluorescence telescopes is 10\% against the total uncertainty of 21\%~\cite{Abu-Zayyad:2011ugz} and 14\%~\cite{Verzi:2013ajy,Dawson:2020bkp}, respectively.
A new calibration system is being developed in Auger that consists of using a portable, calibrated light source mounted on a rail system is moved across the aperture of each telescope~\cite{PierreAuger:2021cwj}. The light source is an integrating sphere that is calibrated in a dedicated setup operated in the laboratory and its intensity is measured with a 3.5\% precision.  For the radio detection technique the typical uncertainty in the calibration of the overall gain (antenna and electronics) is about 9\%. The calibration in situ is performed using external radio sources, e.g., carried out by an octocopter as in the case of \ac{AERA}~\cite{PierreAuger:2017xgp}. An independent method using the background Galactic emission is being developed~\cite{Mulrey:2019vtz,PierreAuger:2012ker} which will allow to make cross-checks and has the advantage to provide a calibration stable over time. 

\subsection[Primary mass composition]{Primary mass composition: Toward event-by-event separation and the post-suppression picture}
\label{sec:10yearMass}
There are a few main goals of near-term and future projects with respect to primary composition. 
The first is to remove the ambiguity between mass composition and hadronic interactions through the collection of high-statistics air-shower observations with multiple observables with energies ranging from 100\,PeV (i.e., close to the center-of-mass energy of the \ac{LHC}) up to ultra-high energies \cite{Klages:2007zza,Thomson:2011gke, PierreAuger:2016crs, Ogio:2020fyf, Bergman:2019pnx,Holt:2019tja}.
The second is to gather enough composition data at the highest energies to constrain the mass picture above the suppression. 
A third would be to explore how mass information can be combined with arrival directions to probe the \ac{UHECR} sky at all energies with increased power.
These and other goals will be accomplished through a combination of upgraded detectors and new analysis techniques.


\subsubsection{Machine learning methods and mass composition}\label{sec:comp10yr-MLM}

As outlined in \cref{sec:10yrComputation}, machine-learning methods, and particularly DNNs, are beginning to be leveraged to reconstruct primary cosmic-ray mass to great effect. The current mass composition related machine-learning methods efforts of each of the major current observatories are described below, along with outlooks on how these methods should progress in the next 10-years.

\subparagraph{IceCube and IceTop}
The analysis of the combined data from surface IceTop and deep in-ice IceCube are well suited for the application of various machine learning methods. In recent years, several neural network and random forest methods were successfully applied to analyze the cosmic-ray data from both detector components. Recent technical developments show promising results for the future IceCube-Gen2 observatory to increase the usage of machine learning methods even further. Those methods include for example reconstructions using deep \acp{CNN} as well as \acp{GNN} and recurrent neural networks for filtering.\medskip

\subparagraph{Pierre Auger Observatory}
Two machine learning based algorithms have been developed with the goal of extracting mass composition information from the \acp{WCD} of the surface detector array. The first technique~\cite{PierreAuger:2021fkf} provides a direct reconstruction of \xmax{} with the SD using recurrent and convolutional neural networks which analyze the time-dependent signals detected by the \acp{WCD}. Though the network was trained using extensive simulation libraries, dependencies on the hadronic model were removed using hybrid events to validate the reconstruction and cross-calibrate it to the \xmax{} scale of the fluorescence measurements. When applied to data, the post calibration event-by-event \xmax{} resolution amounts to roughly $25$\,\gcm{} (see \cref{fig:ResolutionVsScience}) above a few EeV~\cite{PierreAuger:2021xnt}. This enables improved composition studies at the highest energies compared to those possible with classic SD analyses, for example the interpretation of the signal rise time~\cite{PierreAuger:2017tlx}. 
The second method~\cite{PierreAuger:2021nsq} aims to directly extract the muon signals recorded by each \ac{WCD} using recurrent neural networks as the total number of muons produced in a shower \nmu is strongly correlated with primary mass and is subject to lower shower-to-shower fluctuations than \xmax{}. In simulations it was found that the network was able to estimate the fraction of the total \ac{WCD} signal contributed by muons with a bias of less than 2\,\% and a resolution better than 11\,\%. 

Hadronic interaction model uncertainties in the muon production currently serve to limit the precision of SD-based composition studies as measuring the muon content of the shower would be the natural approach for ground-based detector arrays. In the case of \nmu, interpretation is particularly impaired at the highest energies where the statistical power of the SD is badly needed. SD-based reconstructions of \xmax{} suffer less from model uncertainties, their resolutions are limited by the need to cross-calibrate their reconstructions with the FD and the inherently lower sensitivity of \xmax{} itself. Through the addition of the \ac{SSD}, the AugerPrime upgrade currently underway offers an opportunity to improve the resolutions obtainable by both methods. This in turn will provide much-needed data to aid in improving hadronic interaction models and would provide the statistical power needed to constrain primary composition at energies higher than those reachable through the FD.

\subparagraph{Telescope Array} An analysis of \ac{TA} \ac{SD} data using a \ac{BDT} has been developed to measure CR composition~\cite{TelescopeArray:2018bep, Zhezher:2021qke}. The variables considered in the \ac{BDT} include \ac{SD} observables related to the shower LDF, the shower front thickness and curvature, and the shower muon content as observed by a combination of the number of peaks in \ac{SD} traces and upper/lower layer differences. The \ac{BDT} analysis results in a classifier variable that is calibrated using CORSIKA simulations with different HE interaction models. Single-species MC sets are reconstructed to give the average classifier value. Then the classifier value for the data is compared, after a bias correction, and a \meanlnA\ value is determined. The results have shown constant composition as a function of energy at about the helium level~\cite{TelescopeArray:2021gov}. This method, combined with the four-fold increase in SD statistics and the expanded hybrid aperture of \TAxFour will drastically increase mass composition statistics at \ac{TA}.

\subparagraph{Longitudinal profiles and machine learning methods} In addition to their application to \ac{SD} data, there may also be significant advantages in using these same machine learning methods to extract additional mass information from the profiles of showers. Though so far mostly untried, it has already been shown that there is significantly more primary mass information in the longitudinal profiles then that which can be provided by \xmax{} alone \cite{Andringa:2011zz}. From this it is clear that there is a good opportunity to apply similar machine learning methods as described above to increase the mass resolution of \ac{FD} only measurements. It can be expected that these methods will be experimented with in the next 10-years and may feature alongside the already proven \ac{SD} methods in the mass composition analyses of the next generation of detectors.\medskip

These developments, together with a more complete understanding of hadronic interactions at high energies, have the potential to determine the mass composition at the highest energies with unprecedented statistics and fidelity in the next 10 years.

\subsubsection{Mass composition and arrival directions}\label{sec:MEAD}

By combining the primary mass with the arrival direction and energy of each cosmic ray, charged-particle astronomy gains sensitivity in a way comparable to adding multiple wavelengths to optical astronomy. Additionally, with mass composition, the charge of primaries is also known, which when combined with a high-resolution energy reconstruction results in the availability of primary rigidity for analysis. When this is combined with modern magnetic field models, the possibility to perform charged-particle astronomy, even at energies below the flux suppression, is recovered as long as the rigidity of the evaluated component is above $\sim10$\,EV (see \cref{sec:magnetic_fields} for more). 

Currently, because the collaborations are on the cusp of meeting either the required statistical power with \ac{FD} methods, and/or the required mass resolution with \ac{SD} methods, there are many techniques currently under development which will come into their own in the next 10-years. It can therefore be expected that these types of studies will be central to \ac{UHECR} science in the next generation of experiments. As examples, in rough order of increasing complexity:
\begin{enumerate}[i),topsep=2pt]\setlength\itemsep{-0.2em}
  \item light-only anisotropy studies;\label{MassSplit}
  \item split sky mass studies;\label{SplitSky}
  \item mass composition sky-mapping;\label{CompMap}
  \item mass $+$ arrival direction $+$ energy spectrum combined fits;\label{GrandCombined}
  \item event-by-event magnetic field inversion.\label{GMFBackTrack}
\end{enumerate}\vspace{2mm}

Each of the above studies and methods will benefit greatly from the increased mass resolution and aperture afforded by the upgrades of both \ac{TA} and Auger. However, these types of analyses are already being carried out and are producing interesting results. An analysis in the vein of \ref{MassSplit} was recently carried out on \ac{SD} data in \cite{Zhezher:2021qug} which hinted at an excess of light events clustering near established hot spots. In \cite{PierreAuger:2021jlg} analyses of the types \ref{SplitSky} and \ref{CompMap} have already been performed on data as well. The result, illustrated in \cref{fig:aniGP}, hints that at energies above the ankle the mean mass of UHECR arriving from middle galactic latitudes is higher than that of UHECR arriving from other parts of the sky. The mean mass difference found is much larger than would be expected from current source and propagation models, leading to significant tension \cite{Allard:2021ioh}. 
An analysis in the vein of \ref{GrandCombined} has also been explored using simulations \cite{PierreAuger:2021oxo}, which shows promise in differentiating between source scenarios when the method is applied to data. An analysis of type \ref{GMFBackTrack} has not yet been performed and cannot realistically be carried until improved \ac{GMF} models, reconstruction methods and/or upgraded instrumentation are available. This is an area of intense study which and is outlined in \cref{subsec:magneticfields}.

In all of these studies, statistics have proven to be a limiting factor, yet all are also showing hints of results which if confirmed would have major impacts on the understanding of UHECR sources and propagation. In the next 10-years, \acp{SD} at the upgraded observatories will be able to add their considerable statistical might to these efforts and primary composition anisotropy studies will become more frequently leveraged to study the cosmic ray sky. This is already beginning as SD data is being reanalyzed using machine learning techniques expanding the composition sensitive aperture at the highest energies. This can only continue as the upgrades will increase the resolution of \ac{SD} methods to eventually enable the event-by-event study of mass composition as a function of arrival direction. With these advances the study of the UHECR sky as a function of rigidity will be a key component of results from the upgraded observatories and the next generation of detectors.

\subsubsection{Towards a model-independent measurement of composition}\label{sec:MassModelIndependent}

There are several possibilities to decrease the theoretical uncertainties on primary composition due to our limited understanding of hadronic interactions by using the data from air shower experiments.
Most importantly, the correct mass scale needs to be established for at least one mass-sensitive air shower observable (shower maximum, number of muons, muon production depth, etc.) and then transferred to all other observables via

\subparagraph{Analyses with low sensitivity to uncertainties in hadronic models} 
The best-known example of such analyses are nearly model-independent inferences on the evolution of \meanlnA with energy from the elongation rates of different shower observables~(see Figures~\ref{fig:LnA} and \ref{fig:lnA_SD_FD}).  However, recently a method based on the correlation between \xmax{} and particle density at ground~\cite{Younk:2012mp} was applied by Auger for constraining the spread of the masses in the primary beam~\cite{PierreAuger:2016qzj, Yushkov:2020nhr}. This study proved that near the ankle the composition is mixed and includes nuclei heavier than helium. As yet another example, a method to extract the proton-to-helium ratio~\cite{Yushkov:2016xiz} was applied in \ac{TA} to set lower p/He limits~\cite{Karpikov:2018itx}. With the higher statistics of the Auger data even stronger p/He limits should be possible. Input from these kind of analyses will help to better restrict hadronic interactions which in turn will lead to even smaller uncertainties in the determination of the mass composition. This will allow one to perform stricter tests of self-consistency of hadronic models.

\subparagraph{Self-Consistency} 
An example of the power of air-shower data to perform data-driven tests of the consistency of hadronic interactions and the inferred cosmic-ray composition is the analysis of the first two moments of the distribution of shower maximum~\cite{PierreAuger:2013xim, PierreAuger:2014sui} with which it could be shown that the \xmax{} values predicted by air-shower simulations with the hadronic interaction model \qgsii~\cite{Ostapchenko:2010vb} are incompatible with the data. Further examples which exploit the FD data make use of the fractional composition fits of \xmax{} distributions with simultaneous adjustments of the \meanXmax and \sigmaXmax scales~\cite{Blaess:2018msv} or proton-proton interaction cross-section~\cite{Tkachenko:2021bja}.
Even more powerful consistency checks are possible with the inclusion of ground-level particle densities, see e.g., Refs.~\cite{PierreAuger:2016qzj, Vicha:2020sfv, PierreAuger:2021xah}. 
Many of these self-consistency checks have been performed at low energies, where the current experiments collected a lot of events. 
Similar studies at \ac{UHE} will need much larger exposures for high-quality, event-by-event measurements of multiple mass-sensitive air showers observables, which will be provided through the upgrades.

\subparagraph{Cosmic Spectrometer} 
Another possibility for the study of composition at \ac{UHE} relies on the detection of point sources in the arrival directions of cosmic rays.
Recently there have been tantalizing hints with significances of up to 4.5~$\sigma$ for a clustering of cosmic rays at intermediate angular scales~\cite{PierreAuger:2018qvk, TelescopeArray:2014tsd}. 
If these \emph{hot spots} in the cosmic-ray sky are corroborated by future data, then the study of the arrival directions can open a window of opportunity to determine the cosmic ray composition without the use of hadronic interaction models. 
The location of the apparent image of the sources will be distorted by the \ac{GMF}~\cite{Farrar:2017lhm}, which acts as a particle spectrometer on the charged cosmic rays~\cite{Kachelriess:2005qm, Anchordoqui:2017abg, Erdmann:2018cvz}.
An even more direct handle on the cosmic-ray composition could be provided by the discovery of multiplets of magnetically-aligned arrival direction of cosmic rays~\cite{PierreAuger:2011mup, PierreAuger:2014tos, TelescopeArray:2020acv, PierreAuger:2020zii}.
Both of these potential studies call for a large-exposure detection of cosmic rays with event-by-event mass sensitivity.

\subparagraph{Cosmic Mass Degrader} 
Another advantage of a high-statistics measurement of cosmic rays at ultra-high energies is that extragalactic photon fields limit the propagation distance of cosmic-ray nuclei.
Between 100\,EeV and 300\,EeV, the interaction length is largest for proton and iron particles. It is therefore possible (if the extra galactic cosmic-ray flux is dominated by ($>10$~Mpc) sources) that at these energies the particle beam arriving at our Galaxy consist of only iron and some protons, as intermediate mass primaries are efficiently photo-disintegrated~\cite{Allard:2011aa}. 
The observation of a bi-modal distribution of air shower observables, e.g., in the muon-number/shower-maximum plane, could set the mass scale for these two variables with high precision and without the need to resort to air shower simulations.

\subsection[Shower physics and hadronic interactions]{Shower physics and hadronic interactions: Beyond the Muon Puzzle}
\label{sec:FutureHadronicInteractions}
As described in \cref{sec:AccelSyn}, accurate measurements of extensive air showers in the atmosphere provide broad opportunities for interdisciplinary studies between modern astroparticle and high-energy particle physics. In this section, these synergies will be further explored in the context of upcoming and proposed air shower and collider experiments. In \cref{sec:FutureUHECRPartPhys} how future \ac{UHECR} observatories can inform particle physics will be discussed, while the impact of upcoming collider experiments on air shower physics will be described in \cref{sec:FuturePartPhysUHECR}

\subsubsection{Particle physics with UHECR observatories}
\label{sec:FutureUHECRPartPhys}

The main goal of future large-scale \ac{UHECR} experiments, either on the ground or in space, will be to increase the aperture to reach a sufficiently large number of events at the end of the energy spectrum to study the sources of cosmic rays. As a result, if even a small fraction of protons at the highest energies exists, event statistics will be sufficient to directly measure the proton-air cross section at these energies as has already been done at lower energy by the Pierre Auger Observatory~\cite{PierreAuger:2012egl} or Telescope Array~\cite{Abbasi:2015fdr,Abbasi:2020chd} (see \cref{sec:UHECRPartSyn}).

\begin{figure}[tb]
    \centering
    \includegraphics[width=0.8\textwidth]{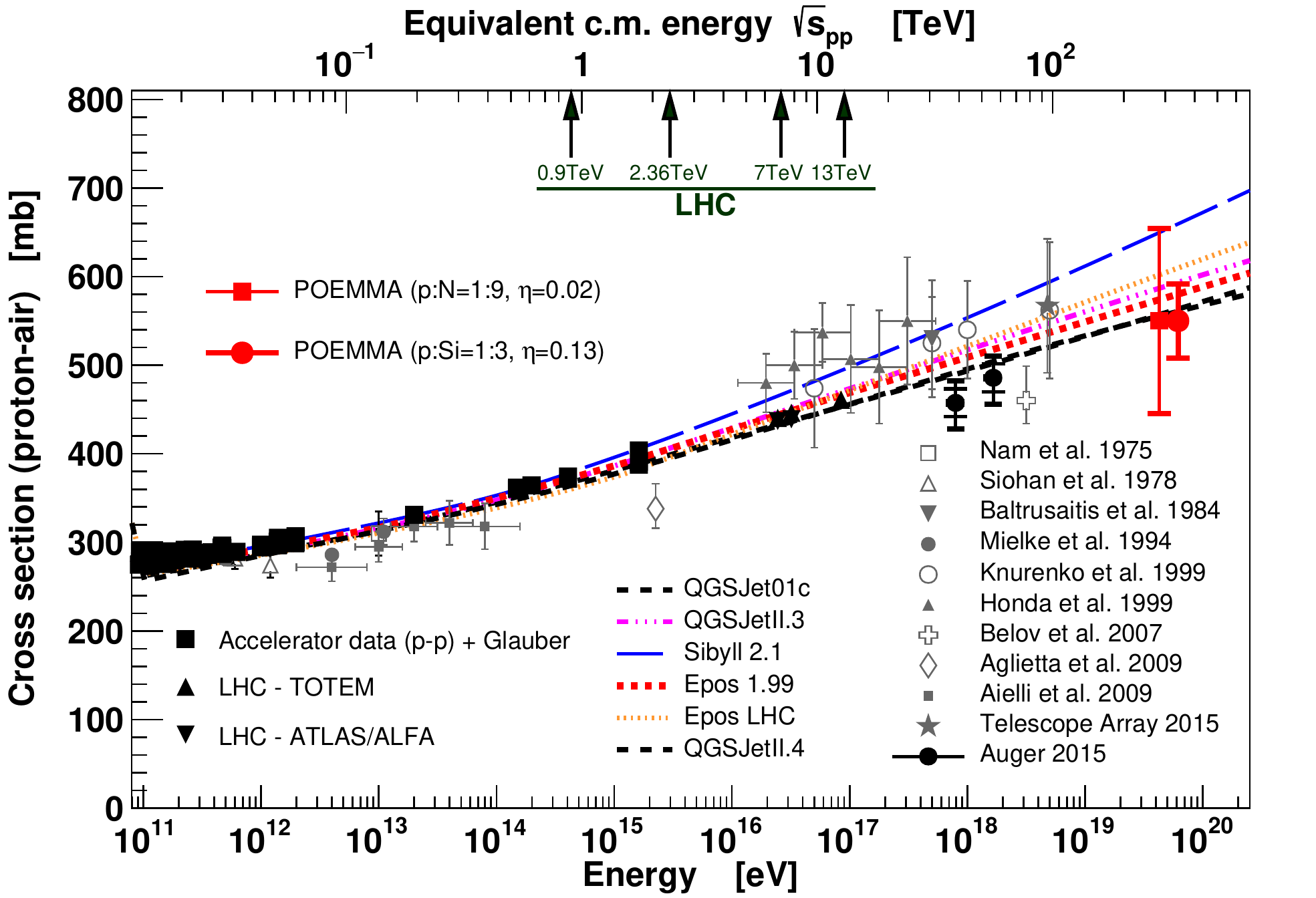}
    \caption{Potential of a measurement of the proton-air cross section with \ac{POEMMA}~\cite{Anchordoqui:2019omw}. Shown are also current model predictions and a complete compilation of accelerator data, converted to a proton-air cross section using the Glauber formalism. The expected uncertainties for two composition scenarios (left p:N=1:9, right p:Si=1:3) are shown as red markers with error bars. The two points are slightly displaced in energy for better visibility.}
    \label{fig:POEMMA}%
\end{figure}

The p-Air inelastic cross section is extracted from the tail of the $X_\mathrm{max}$ distribution using a fraction $\eta$ of the events with the largest $X_\mathrm{max}$. Depending the fraction of protons compared to nitrogen or silicon, different fraction $\eta$ can be used, leading to results with different precision. An example is given in \cref{fig:POEMMA} which shows a feasibility study for \ac{POEMMA}~\cite{Anchordoqui:2019omw} (see also \cref{sec:POEMMA}). With a ratio p:N of 1:9, only $2\%$ of the events can be used leading to much larger error bars, compared to a p:Si ratio of 1:3 which would allow the use of $13\%$ of the measured events at $E=10^{19.6}\,\mathrm{eV}$, equivalent to a center-of-mass energy of $283\,\mathrm{TeV}$. This measurement would further extend previous cross section measurements by \ac{UHECR} experiments into a phase space far beyond current or future colliders, at least for the next $\sim$\,$50$ years.

As previously discussed in \cref{sec:PartUHECROutlook}, it can be expected that the muon production in air showers will be precisely quantified within the upcoming decade. At the same time, precise measurements of multi-particle distributions in the forward region at the \ac{LHC} will become available and put strong constraints on the hadronic interaction models. With this wealth of data, it is expected that the currently missing ingredient(s) in recent hadronic interaction models will be found and that future models will become reliable tools to fully exploit air shower data even at higher energies (validated by self-consistency checks in hybrid \ac{EAS} measurements). In turn, this will provide essential information for the future analysis of multi-messenger data in conjunction with gamma-ray, neutrino, and gravitational wave observations~\cite{Schroder:2019rxp}. If, however, \ac{LHC} data can be fully reproduced but the Muon Puzzle remains unsolved, the quality of the \ac{EAS} data will allow for tests of \ac{BSM} physics scenarios, using either the bulk properties of the data or using tails of certain distributions, like the muon number and the \xmax distributions. In both cases, a new era of high-precision particle physics studies with accurate air shower data will be opened.

In order to fully realize the physics potential of muon measurements with large-scale \ac{UHECR} observatories, future experiments should be equipped with radio antennas to measure the shower energy very precisely, and buried or shielded muon detectors with high spatial and time resolution and large collection area. Shielding is needed to have a clean muon signal without contamination from photons or electrons. This, and high resolution in time is needed to make full use of the information in the muon production depth. These ideas already influenced the design of \ac{GCOS}~\cite{Horandel:2021prj} or the \ac{GRAND}~\cite{GRAND:2018iaj} sub-array with particle detectors, full described later in \cref{sec:GCOS} and \cref{sec:GRAND}. With a very high event statistic and accurate models, including rare high-energy physics phenomena like particle physics event generators, such as \textsc{Pythia} \cite{Sjostrand:2006za, Sjostrand:2007gs, Sjostrand:2014zea, Sjostrand:2019zhc, Sjostrand:2021dal} (see also the contribution to Snowmass 2021 on high-energy \ac{MC} event generators~\cite{GeneratorsWhitepaper}), standard model predictions could be tested at energies much higher than any at current or future accelerator. In particular, the production of heavy hadron flavors can be tested which will carry an increasingly significant part of the energy.

The extension of the IceCube Neutrino Observatory to higher cosmic ray energies by IceCube-Gen2 \cite{IceCube:2014gqr} (see also \cref{sec:IceCubeGen2}) will provide coverage of the region which overlaps with data from the Pierre Auger Observatory, enabling combined studies of the atmospheric muon fluxes. The increased aperture of IceCube-Gen2 for coincident events also opens the possibility to study the angular dependence of the muon content in \acp{EAS}. Together with an increased precision resulting from the enhanced air shower reconstruction provided by the surface array and the improved in-ice calibration, this will contribute to further tests of the improved hadronic interaction models. With this, air shower data from many experiments can in turn be re-analyzed in the context of cosmic ray mass composition measurements more reliably, or, if some discrepancy remains, it could potentially lead to the discovery of more exotic particle phenomena.

The combined detection of atmospheric muons in relation to initial cosmic ray is also relevant at $\sim$\,$\mathrm{PeV}$ muon energies. These muons dominantly originate from decay of charmed and unflavoured mesons and are produced mainly in air showers at PeV to EeV energies~\cite{Fedynitch:2018cbl}, exactly in the range covered by the surface array of IceCube-Gen2. Due to the large aperture and given enough exposure, IceCube-Gen2 could make the first measurement of the prompt component of the muon spectrum. This will also constrain prompt neutrino production at the highest energies and contribute to a better understanding of the background estimates for astrophysical neutrino searches~\cite{IceCube:2015wro, Soldin:2018vak}. An in-depth discussion of astrophysical neutrino searches in a multi-messenger context can be found in complementary contributions to Snowmass 2021 on high-energy and ultra-high-energy neutrinos~\cite{UHEnuWhitepaper}, and multi-messenger astronomy and astrophysics~\cite{MMWhitepaper}.

\subsubsection{Measurements at the high-luminosity LHC and beyond}
\label{sec:FuturePartPhysUHECR}

Measurements at collider experiments provide important complementary information which is crucial for the understanding of particle interactions in air showers, as discussed in \cref{sec:AccelSyn}. Existing measurements from the \ac{LHC}, as well as data from the upcoming \ac{HL-LHC} run, will play a crucial role in understanding the origin of the Muon Puzzle, for example.

In the future, the synergies between astroparticle and high-energy physics could be further exploited with the proposed \acf{FPF} at the \ac{HL-LHC}~\cite{Anchordoqui:2021ghd}. The \ac{FPF} is proposed to be located several hundred meters from the ATLAS interaction point, shielded by concrete and rock, and it will host a variety of experiments to uniquely probe physics in the far-forward region. As discussed in-depth in a dedicated contribution to Snowmass 2021 \cite{FengSnowmass}, measurements of leptons with the proposed experiments at the \ac{FPF} can provide important information about multi-particle production in hadronic interactions in the far-forward region. This will further improve the modeling of high-energy hadronic interactions in the atmosphere. The construction is proposed to take place from 2026 to 2028, in order to install support services and the proposed experiments starting in 2029, and to take data not long after the beginning of Run 4 at the \ac{HL-LHC}. 

Another interesting proposed option to measure particle production in the forward region of an \ac{HL-LHC} interaction point is the construction of a dedicated \ac{VFHS} \cite{Albrow:2018kxz}. Such an experiment would enable measurements of the charged hadron production in hadron-hadron collisions with longitudinal momentum fraction, i.e., Feynman-x, between $0.1$ and $0.9$. Hence, the \ac{VFHS} could potentially also yield important information about forward multi-particle production in hadron interactions on order to further improve hadronic interaction models. 
 
Once the hadronic interaction models can successfully describe all details (i.e., various observables and their correlations) of the air shower development at ultra-high energy ($\tilde 100\,\mathrm{TeV}$ center-of-mass energy), they will become reliable tools for the development of the proposed \ac{FCC} and associated experiments. In order to study both the background of secondary particle production associated with the production of rare but relevant high-energy physics phenomena (e.g., Higgs or Top production, \ac{BSM} physics, \emph{etc.}) and the detector response, models are required that are able to generate hadronic interactions under conditions that can not be tested in man-made experiments but which occur in extensive air showers (e.g., high energy, meson projectiles, forward particle production). The best models for the \ac{FCC} development should be tested against air shower data of high precision to be validated at the energy of the \ac{FCC}. The models used for \ac{EAS} simulations are already used in tools like \textsc{Geant4} \cite{GEANT4:2002zbu, Allison:2006ve} for other direct cosmic ray experiments like DAMPE~\cite{Jiang:2021cit}, for example, where the cascade energy generated in the calorimeter goes beyond the energy range of traditional hadronic models used in \textsc{Geant4} for the \ac{LHC}. These developments will further be extended into the \ac{FCC} era.

\subsection[Anisotropy]{Anisotropy: {\color{red} Bringing sources into focus}}
\label{sec:Anisotropy_NextTenYears}
Taken globally, the existing \ac{UHECR} data indicate that cosmic ray deflections in the intervening magnetic fields are typically large -- too large to allow for a direct identification of sources via small-scale clustering, with the currently available statistics, but apparently not large enough to completely isotropize the \ac{UHECR} flux -- as indicated by the observed large-scale dipole anisotropy and the interesting hints of anisotropies at intermediate scales.  To extract more information about \ac{UHECR} sources from \ac{UHECR} anisotropies, two advances are underway:  further increasing statistics, especially in the northern hemisphere with the \TAxFour upgrade, and adding event-by-event information on the charge, $Z$, of each \ac{UHECR} with the AugerPrime upgrade.   

\subsubsection{Improving statistics}
Regarding statistics, at the time of writing, the Telescope Array detector is currently undergoing the major upgrade to \TAxFour which will increase its effective area by a factor of $\sim$\,$4$ \cite{TelescopeArray:2021dri}, with about half of the planned detectors having already been deployed and taking data. 
The important goal of this extension is to discover intermediate-scale anisotropies of the \ac{UHECR} flux at the highest energies by significantly increasing the number of detected events. This will also boost the accuracy of the combined full-sky \ac{TA} and Auger Observatory analyses as the relatively small statistics of events in the northern hemisphere is the main limitation at present.   Continuing operation of Auger should yield a significance level of $5\sigma$ for the Centaurus region excess by the end of 2025 ($\pm2$ calendar years), possibly preceded by a similar significance milestone in the correlation with the starburst catalog, if those excesses 
continue to grow.   And with the merged data sets of \ac{TA} and Auger, measuring the energy dependence of the dipole anisotropy, identifying or placing limits on a quadrupole or higher component, and separating the Galactic and extragalactic dipoles, should all become feasible.



\subsubsection{Composition-enhanced anisotropy searches}
On the Auger side, much more impactful than merely the growth of statistics will be the full deployment of the upgraded capabilities of the \ac{SD} array, i.e., AugerPrime \cite{PierreAuger:2016qzd}. With the upgrade, AugerPrime will be able to disentangle the electromagnetic and muonic components of the air showers registered by the surface detector on an event-by-event basis, allowing to have mass-sensitive parameters for each SD event.  Additionally, the radio detector array will provide composition constraints for large-zenith angle events.  Taking data steadily from 2023, AugerPrime should collect enough events by the end of the decade with individual events' rigidities determined (with some uncertainty), to map the composition anisotropy and possibly reveal a component of low-$Z$ \acp{UHECR} which should be particularly useful for source identification. 

The data from the Phase 1 of the Auger Observatory indicate that the composition becomes heavier with increasing energy \cite{PierreAuger:2014sui, PierreAuger:2014gko, PierreAuger:2017tlx}. However, these results do not rule out a fraction of light nuclei at the highest energies, which can be expected assuming there is a diversity of source types.  Indeed, some analyses already suggest the presence of a light or proton-like component, see e.g., Ref.~\cite{Muzio:2019leu}.  With AugerPrime, it will be possible to identify the subset of events which are candidates to be protons or light nuclei and thus the easiest events to use for anisotropy studies, given that (for a given energy) those are the ones least deflected by the Galactic and extragalactic magnetic fields.   This important new capability of AugerPrime will enable the entire accumulated Auger Phase 1 data set to be retroactively tagged by mass-composition estimators on an event-by-event basis, using machine-learning techniques \cite{PierreAuger:2021fkf} and an approach based on the concept of air-shower universality \cite{Bridgeman:2017rcv}, calibrated with events detected with AugerPrime.  
\begin{figure} [b]
\centering
\includegraphics[width=0.95\textwidth]{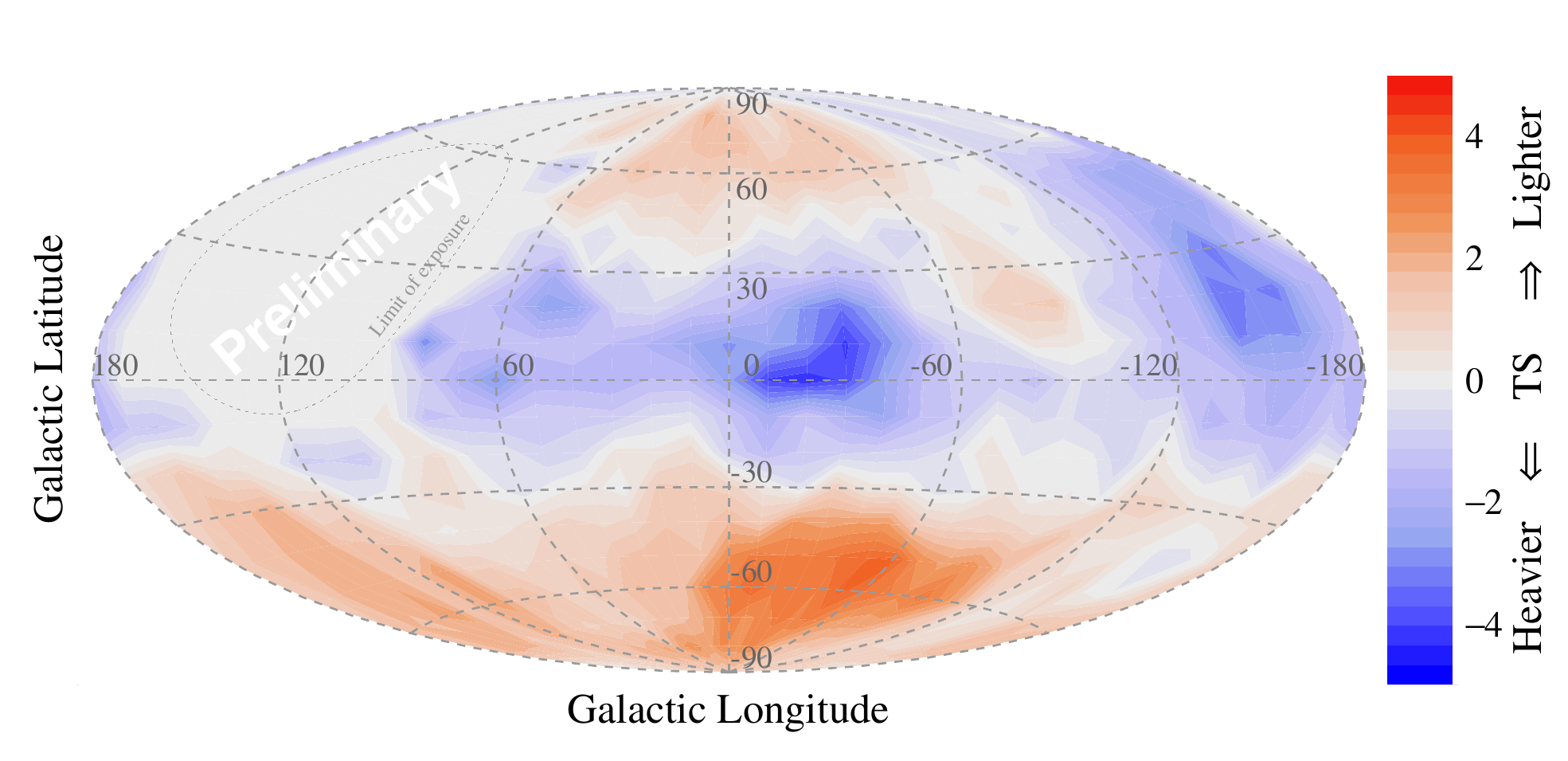}
\caption {Map showing the relative cosmic-ray composition detected by the Pierre Auger Observatory above $10^{18.7}\,$eV with the \ac{FD}, in Galactic coordinates. From Ref.~\cite{PierreAuger:2021jlg}.}
\label{fig:aniGP}
\end{figure}

The discovery of differences in arrival directions for particles of different species is a tantalizing goal. All the anisotropy searches will benefit from having mass-composition proxy on an event-by-event basis, by re-performing the same analyses with events likeliest to have high rigidity. 
For example, if 
the Centaurus region excess is real and all the excess events have high rigidity,  standing over a low-rigidity background, 
being able to reject 22\% of the heaviest events in the sample would already yield a $5\sigma$ significance with the current statistics.
Moreover, composition information will allow the Auger Collaboration to see if there is evidence of a Peters' cycle structure (maximum energy achievable at the accelerator depending on the rigidity) in the energy evolution of the excesses. Furthermore, Auger will perform combined analyses such as in Ref.~\cite{PierreAuger:2021oxo}, simultaneously fitting the energy spectrum, the arrival directions of the events and the mass-composition estimators; this combined analysis has proven to have a much better sensitivity to distinguish between different catalogs of source candidates, even with the much lower statistics available for composition information from the fluorescence detector.  With the event-by-event mass-composition estimator, Auger will also update the search for \emph{multiplets}, i.e., sets of events that show a correlation between their arrival direction and the inverse of their rigidity \cite{PierreAuger:2020zii} as expected if they come from a common source. Discovering such multiplets will give extremely valuable information on the \ac{GMF}, since the deflection as a function of rigidity will be fully determined with no further assumptions as needed for most probes of the \ac{GMF}. {\color{red} In parallel, progress in our understanding and modeling of the \ac{GMF} will help bring \ac{UHECR} sources into focus.}

Finally, by having a mass-composition estimator with the statistics of the surface detector, Auger will be able to test independently the $3.3\sigma$ anisotropy laying along the galactic plane which depends on the mass of primary cosmic-rays, using the events registered by the fluorescence detector (a dataset which is an order of magnitude smaller) \cite{PierreAuger:2021jlg}. This hint of anisotropy, which was detected with events with an energy above $10^{18.7}\,$eV and a galactic latitude splitting at $|b| = 30^\circ$, seems to indicate that the events detected in the on-plane region are heavier than the ones in the off-plane one (see \cref{fig:aniGP}). This effect could be caused by the \ac{GMF}, if sources are extragalactic, non-homogeneously distributed and the \ac{UHECR} composition is mixed \cite{PierreAuger:2021jlg}.

\subsection[Neutral particles]{Neutral particles: Improved sensitivity and game-changing detection}
\label{MM_outlook10yr}
As a result of the developed strategies to detect neutrinos, photons, and neutrons with the Pierre Auger Observatory, as well as of the increased statistics, significant improvements can be expected in the next decade to the upper limits that are to be deduced in case that no candidate events are found. This applies to diffuse fluxes, to specific source directions and candidates, as well as to a variety of transient events. 

Many more opportunities to find EeV neutral particles  will come with the vastly increasing number and better localization of detected sources of gravitational waves with the network of LIGO-Virgo-KAGRA interferometers; the increased number of detected sources of TeV-PeV photons with gamma-ray telescopes such as \ac{CTA}, and most likely of 100\,TeV - PeV neutrinos with IceCube and possibly other neutrino telescopes in correlation with these sources. 

\subsubsection{Cosmogenic and astrophysical photons and neutrinos}

\begin{figure}[t!]\centering
\includegraphics[width=0.9\linewidth]{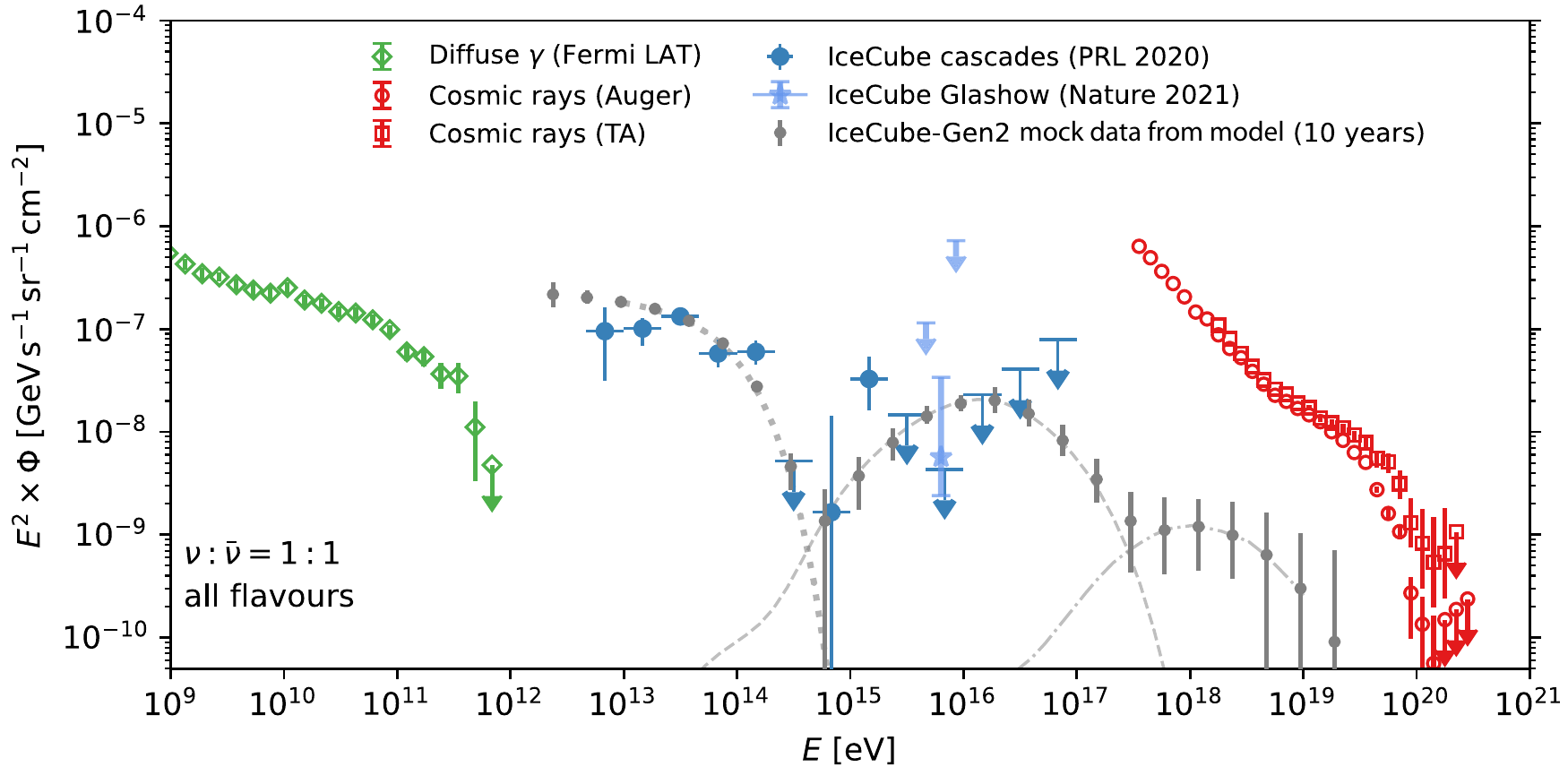}
\caption{A fiducial model for the flux of neutrinos which illustrates the qualitative range of reasonable possibilities. 
This model consists of three components: 
1) a \ac{UHECR}-produced peak at $10^{16}$\,eV giving the best-fit to the high-energy astrophysical neutrino flux consistent with \ac{UHECR} data from Auger and IceCube, taken from \cite{Muzio:2021zud}; 
2) a peak at $10^{18}$\,eV due to \ac{GZK}-produced neutrinos assuming a 10\% proton fraction above 30\,EeV, taken from \cite{vanVliet:2019nse};
and 3) a low-energy component of neutrinos produced by some non-\ac{UHECR} sources, tuned to give the best-fit to the low-energy astrophysical neutrino data. 
The shown points for IceCube-Gen2 are mock data for this model for 10 years of combined optical and radio measurements.
A number of other plausible models for the astrophysical neutrino flux based on specific astrophysical source types are explored e.g., in Refs.~\cite{Biehl:2017hnb,Fang:2017zjf,Zhang:2018agl,Rodrigues:2020pli}.}
\label{fig:UHECR-nu}
\end{figure}

A possible scenario of neutrino production is shown in Fig.~\ref{fig:UHECR-nu}. It assumes the generation of astrophysical neutrinos directly at the sources, and of cosmogenic neutrinos from \ac{UHE} proton interactions with the \ac{CMB}. The strong dependence of the cosmogenic photon and neutrino fluxes on the \ac{UHECR} composition at the highest energies, will allow for an estimation of the primary composition in case cosmogenic fluxes are observed. In fact, present data of the Auger Observatory allow the possibility of a subdominant proton component in the \ac{UHECR} flux sticking out to the highest energies. Conservative extrapolations of the sensitivity of the Pierre Auger Observatory to EeV cosmogenic neutrinos, lead to the conclusion that a fraction of protons at a level of $10\,\%$ at the highest energies will allow detection of cosmogenic neutrinos, unless the cosmological source evolution is softer than what is expected from star formation \cite{PierreAuger:2019ens}. In combination with direct composition measurements at the highest energies, searching for \ac{UHE} cosmogenic neutrinos provides the opportunity to also constrain the \ac{UHECR} source evolution more sharply than is possible at present (see e.g.,~\cite{Heinze:2015hhp}).neutrino

The same arguments do also apply to the production of cosmogenic photons.
Just as with cosmogenic neutrinos, a substantial proton flux will lead to higher fluxes of photons \cite{Muzio:2019leu}, and consequently, a non-observation of photons strongly constrains the proton fraction~\cite{Berezinsky:2016jys, Supanitsky:2016gke}. In this case, because of photon-photon interactions, the intensity of the flux is strongly influenced by the {\em local} source distribution, decreasing as the local source density decreases. Similarly as to neutrinos, presently existing upper bounds to cosmogenic photons start to enter into the parameter space of \ac{GZK}-expectations \cite{PierreAuger:2021mjh,TelescopeArray:2020hey}, provided the proton fraction is sufficiently high at the highest energies. Due to their limited horizon, cosmogenic photon fluxes are rather insensitive to the cosmological source evolution but, other than neutrinos, probe the local Universe, often expressed in terms of negative evolution parameters. This example demonstrates the complementarity of the two messengers. 

The bounds on neutrino and photon fluxes will become stronger in the next decade, because of more statistics becoming available and of improved analysis techniques being developed. Extrapolation of the limits obtained so far by the Pierre Auger Observatory lead to improvements by a factor of $\sim$\,2 for neutrinos and 3 for photons with respect to those shown in ~\cref{fig:nu_panorama}. 
These are very conservative estimates because they ignore all the upgrades that are being deployed, which will help to improve the selection capabilities of the Auger Observatory to detect these particles. 

\subsubsection{Neutrons} 
Similarly as with \ac{UHE} photons and neutrinos, the search for \ac{UHE} neutron point sources will benefit from increasing statistics and improved techniques becoming available in the next decade. This will enable more sensitive searches for transient galactic sources\cite{PierreAuger:2014tey}, and push down the bounds on the neutron energy flux to a factor of about 100 below those expected from a $1/E^2$ extrapolation of TeV $\gamma$-spectra from galactic sources \cite{PierreAuger:2019fdm}. 

\subsubsection{Follow-up observations \& transient events} 
The next decade of multi-messenger observations will strongly benefit from the progress in gravitational wave detection. The enhanced sensitivity that is being reached as existing gravitational wave detectors are optimized and as new ones come into play, should increase the rate of events to follow by very large factors. The example of the neutron star merger GW170817 has impressively demonstrated the science potential of follow-up observations of \ac{UHE} neutrinos (and photons), with their upper bounds being close to expectations from models of jet formation \cite{ANTARES:2017bia}. In the future, the rate of \ac{GW} event observations will vastly increase, which promises a rich science harvest. 
Different from imaging $\gamma$-ray telescopes pointed to specific regions of the sky, and complementary to $\gamma$-ray arrays observing the sky at lower energies, observatories such as IceCube, Auger and the Telescope Array, provide continuous coverage of a large part of the sky and thus initiate automated neutrino (and photon) searches upon \ac{GCN} alerts, and also contribute by sending alerts. Besides analysing individual events, stacking analyses, such as those started by the Auger Collaboration \cite{PierreAuger:2021asv, PierreAuger:2021oks}, will allow to push down the neutrino bounds in direct proportion to the number of detected \ac{GW} events.

\subsubsection{Indirect information on neutral particles from UHECR measurements}

One of the most important developments that can be expected to take place in the following decade, specifically in \ac{UHECR} measurements, is a more precise determination of composition of \ac{UHECR} on an event-by-event basis, with the goal to enable composition enhanced anisotropy studies, particularly at the highest energies. The Auger upgrade AugerPrime \cite{PierreAuger:2016qzd}, is mostly designed with this as a main objective. 
Through it, an increase in statistics by at least an order of magnitude will be achieved, which will also allow a better establishment of the average composition and, in particular, that of the highest-energy particles. A more accurate determination of \ac{UHECR} primary mass will open new possibilities to select samples of particles with enriched rigidity from a large fraction of the sky, for which the anisotropy signals are likely to be enhanced and easier to be detected. The study of composition-driven anisotropies will be crucial in further constraining the sources of cosmic rays, and the secondary fluxes of neutrinos and photons that could arise from their interactions with matter and/or radiation.

While mass measurements are already giving an increasingly clearer picture of the composition becoming heavier as the energy rises in the 3 to 50\,EeV range, there are no measurements at the highest energies, yet. Composition inference has been achieved with combined fits of the spectrum and measurements of the average $X_{\rm max}$ and its fluctuations under the hypothesis of a rigidity limited acceleration at sources (Peters' cycle) which predicts heavier components at the highest energies~\cite{PierreAuger:2016use}. However, in case that the acceleration mechanism is more complex than Peters' cycle hypothesis, and/or if the sources of \ac{UHECR} are not of a unique type, a very different composition beyond 80\,EeV could be expected and few constrains on composition could be obtained from the scarce data that is available today. As an example, if a component of protons exists at the highest energies, even if it has a small fraction of order $10\%$, the possibilities of doing \ac{UHECR} astronomy will be notably enhanced and possible sources may be imaged with the cosmic rays, besides obtaining invaluable information about the intervening magnetic fields. This in turn will allow \ac{UHECR} to be finally added to the list of `messengers' available for multi-messenger studies of astrophysical sources and processes.

With the upgraded \ac{UHECR} detectors and the increase in statistics, all searches for anisotropies with \ac{UHECR} can be expected to improve to the level of providing further and more precise tests of indications of correlations with potential candidate sources or localized excesses which have not yet reached a high enough statistical significance. 
Moreover, the wealth of observational data probing the Galactic magnetic field is expected to be increased by more than an order of magnitude over the next decade by upcoming instruments, in particular the \ac{SKA} and its pathfinders and surveys. These observations will significantly reduce the uncertainty on the 3D magnetic field, both locally and throughout the Galactic disk, providing information about the magnitude of the coherent and stochastic field components, as well as their overall orientation. This will significantly reduce uncertainties in modeling the \ac{GMF}, and enable much more robust correlations between \ac{UHECR} events and neutral messengers. For a more detailed discussion of the current and future status of the Galactic magnetic field see \cref{sec:galactic_mag_fields}. 

All these observations, in combination with other multi-messenger observations, are expected to further constrain \ac{UHECR} acceleration and the astrophysical sources where it takes place, giving also a clearer picture of their spatial distribution about the Earth. This will have a direct impact in constraining the fluxes of \ac{UHE} photons and neutrinos that could be expected at the Earth.
\fakesection{Instrumentation road-map}
\vspace{3cm}
{\noindent \LARGE \textbf{Chapter 6}}\\[.8cm]
\textbf{\noindent \huge Instrumentation roadmap:}\\[3mm]
\textbf{\LARGE  A strategy for the next generation of UHECR experiments}
\label{sec:TNG}
\vspace{1cm}

While the upgrades of the current generation of cosmic-ray air shower arrays are essential for progress in the next decade, these experiments are too limited in their exposure to solve some of the key science questions of \ac{UHECR}.
Therefore, a new generation of experiments is required featuring an order of magnitude higher aperture to identify the sources of \ac{UHECR}, study the particle physics of air showers at the highest energies, search for ZeV particles and \ac{BSM} physics.
Building on recent and ongoing technology and computational developments, three future \ac{UHECR} experiments expected to be operational in the next decade will complement each other in achieving the various \ac{UHECR} science goals.

\begin{figure}[t]
  \centering
      \includegraphics[width=\textwidth]{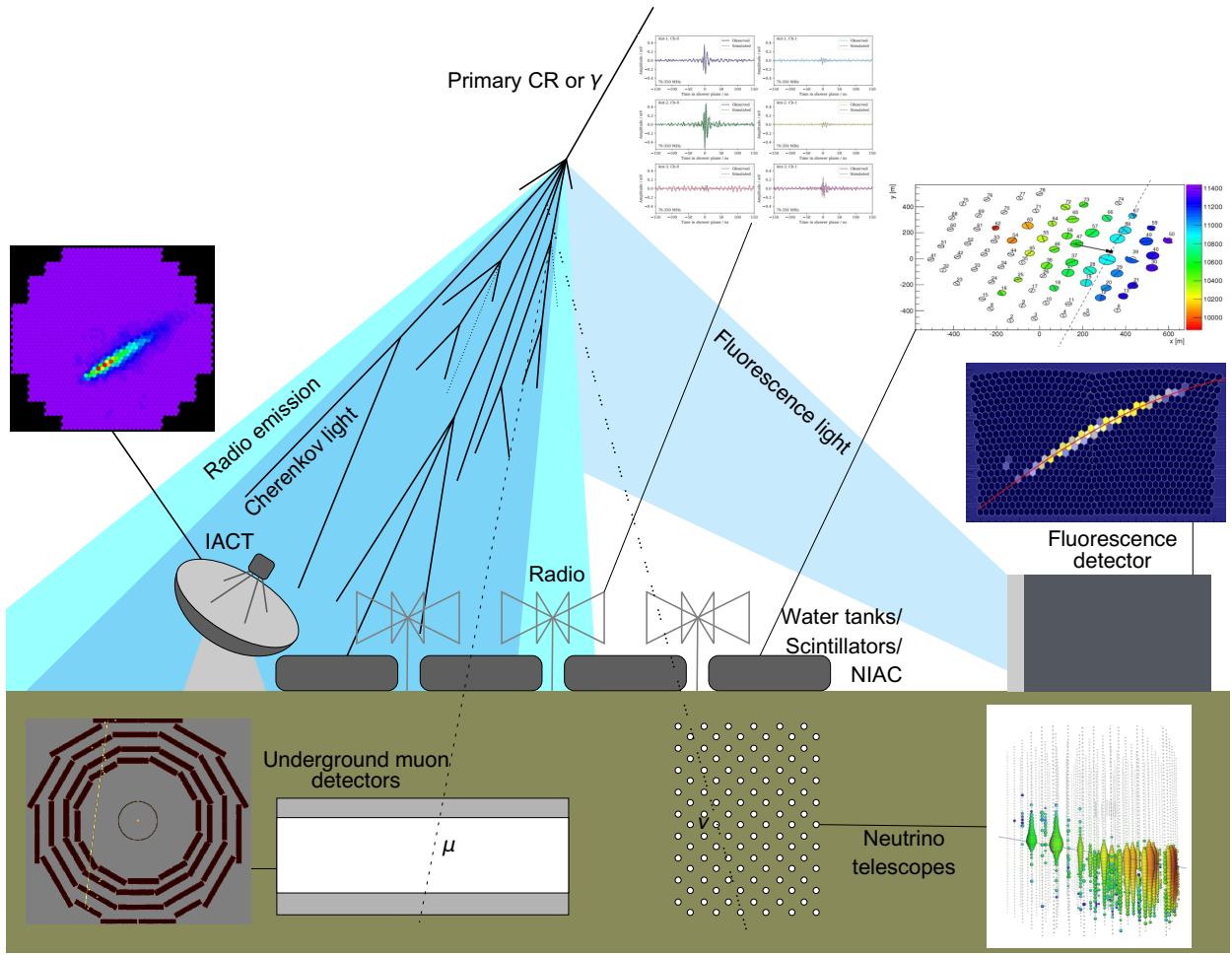}
  \caption{Schematic of indirect CR detection methods for \ac{EAS}. Surface and underground particle detectors measure electromagnetic particles and muons. Imaging (IACT) and non-imaging (NIAC) air-Cherenkov detectors as well as radio antennas provide a measurement of the electromagnetic shower component when located in the footprint of the shower, while fluorescence light detectors can observe the shower development from the side (pictures from Refs.~\cite{IceCube:2021epf,Hussain:2013ui,Radel:2017ule,showerSchematics_photo1,showerSchematics_photo2,showerSchematics_photo3}}
  \label{eas_detection_technique}
\end{figure}

\subsection{Technological development for the future}
\label{sec:tech_development}

Various techniques are used for the detection of cosmic-ray \acp{EAS}, which measure different observables of the air showers. They each have their advantages depending on needs.
The following paragraphs summarize recent technology progress and ongoing developments.
These build the foundation for the next generation of \ac{UHECR} experiments.
Developments are ongoing regarding all detection techniques (\cref{eas_detection_technique}), making use of \acp{SiPM} as well as recent electronics advances. 
Substantial progress has been achieved in the last decades especially regarding the digital radio technique for air showers, which has matured to a level that it will play a major role in the next generation of arrays. 
Moreover, the established technique of fluorescence detection has been made ready for space.

Some of the science goals of the next generation require huge exposure, but have less strict requirements regarding the accuracy of the energy and mass of the primary particles. 
These science goals will benefit from technology development making techniques such as fluorescence or radio detection cost-effective for huge ground arrays or ready for space.
Other science goals require higher accuracy for the rigidity of the primary particle than achievable by any single technique standalone. 
These science goals will benefit mostly from the improvements in particle detectors for surface arrays that allow for measurements of the electromagnetic and muon particles and can be combined with a calorimetric measurement technique such as the simultaneous air-fluorescence, air-Cherenkov, or radio measurement of the same air showers.

\subsubsection[Surface detectors]{Surface detectors: more mass sensitivity}
\label{sec:SD_tech_development}

\paragraph{Current Picture and Status}\label{sec:ParticleDetectors_CurrentStatus}

Indirect measurements of cosmic rays are usually performed using particle detectors deployed on the ground. 
These detectors are covering large surfaces depending on the energy range on interest and reach areas of up to 3000\,km$^2$ as in the case of Auger (for more details see~\cref{sec:FutureDetectors}). 
The size of the \ac{EAS} footprint on the ground depends on the energy of the primary cosmic ray and on the amount of matter traversed by the air shower in the atmosphere. With an energy of 10\,EeV, a vertical cascade would produce a footprint with a diameter of about 10\,km while at around 1\,EeV the footprint could extend to more than 3\,km. 
To be able to properly sample the particles on the ground the arrays need to be dense enough (distance between detectors smaller than 2\,km to have at least 4~detectors triggered at 10\,EeV) and in the same time to cover a sufficiently large area to compensate for the strong decrease of the flux.
With foreseen ground arrays extending over hundreds of kilometers in diameter (\cref{sec:GCOS}) or being placed at hardly accessible locations (\cref{sec:IceCubeGen2}), the particle detectors need to be very robust, and with a very low need for maintenance.

The main components of the particles that are reaching the ground are the electrons, positrons, muons, anti-muons, and photons. To obtain a good resolution on the mass composition of the primary particles, a sufficient separation between the electromagnetic and muonic components needs to be achieved by the surface detectors.
Moreover, a very good dynamic range is required to cover the signal produced by more than 1000~particles/m$^2$ close to the core of air showers as well as smaller signal produced by just one muon far from the shower axis.

The effective area of the individual particle detectors needs to be large enough to be able to measure the signals at certain distances with statistical fluctuations of less than 10 to 15\,\%. 
The number of particles decreases with increasing distance to the shower axis. For air showers at 10\,EeV, there are around 2 particles/m$^2$ at 1000\,m. 
This number decreases with decreasing energy and increasing zenith angle.

The main observatories currently operating, Auger~\cite{PierreAuger:2015eyc}, \ac{TA}, and IceCube~\cite{IceCube:2016zyt}, are employing very simple and robust detectors to measure the particles at ground: containers filled with water/ice to measure the Cherenkov light and plastic scintillation detectors. 
Each of these detectors has been constructed to be independent, equipped with their own electronics processing local triggers, solar panels and batteries, GPS receivers for timing and radio antennas for data transmission and communication~\cite{PierreAuger:2004naf}. 
The Cherenkov light produced in the water/ice and reflected on the sides of the detectors is observed by \acp{PMT} optically coupled to the water/ice, while the scintillation light is usually collected and guided via wavelength shifting optical fibers and then read out either by solid-state photosensors (multi-pixel photon counters, MPPCs) or by \acp{PMT}.

\paragraph{The near future - 10 year outlook}\label{sec:ParticleDetectors_10Year}

Part of the limitations of the simple detectors that are containing just one optical volume are related to the difficult task of separating the muonic from the electromagnetic components of the air showers. 
While modern techniques based on deep neural networks are improving upon the separation, it is still not clear that they will provide the needed resolution for the determination of the maximum of the shower development or the number of muons in air showers (the main variables sensitive to the composition of \acp{UHECR}).

The Pierre Auger collaboration is currently deploying the AugerPrime upgrade~\cite{PierreAuger:2016qzd} aimed at better understanding the physics of air showers and separating the \ac{EAS} components. 
Scintillators are placed on top of the water Cherenkov detectors delivering an alternative measurement of the particles arriving on the ground: in the water, all particles are measured with photons dominating the signals close to the \ac{EAS} axis and muons dominating at larger distances and inclined events; in scintillators, the signal is mainly produced by the charged particles. 
This double measurement at the same location will make it possible to differentiate the \ac{EAS} components and enhance the capability of the surface detector to provide the sensitivity to measure the composition of \acp{UHECR}. 
Another important upgrade of the Auger surface detector is the deployment of buried scintillators on a smaller area to directly measure the high energy muons. 
The particle detectors of AugerPrime, the main infrastructure in the array, will be operated for at least the next 10 years and will provide a deep insight into the \ac{EAS} physics and about \acp{UHECR}.

A similar upgrade is planned to be deployed at the South Pole for the surface detector of IceCube, IceTop~\cite{IceCube:2012nn}. 
On top of the IceTop array comprised of ice-Cherenkov detectors, an array of scintillators similar to the AugerPrime ones will be placed (complemented by surface radio antennas)~\cite{Haungs:2019ylq,Schroder:2019suq}. 
While this upgrade of IceTop is aimed at reducing the systematic uncertainties produced by the snow accumulation, it will also be used to enhance the sensitivity to mass composition. 
In the \ac{EAS} measurements with IceTop, a crucial role is played also by the in-ice detectors which, similarly to the underground muon detector in Auger, measure the high-energy muons as well as muon bundles, relevant for understanding the high-energy hadronic interactions. 
An extension of the surface area of the IceTop array with scintillators, in the framework of the IceCube-Gen2 extension, is foreseen to increase the effective area and reach higher cosmic-ray energies.

Particle detectors deployed on a large surface is the only way to have the largest possible statistics at the highest energies with ground detectors (due to the 100\% duty factor and the large spread of particles on the ground covering several kilometers ). 
To increase the exposure, \ac{TA} is increasing  the area covered by the scintillator array by a factor of four in the following years.

\paragraph{The next generation - recommendations for 10-20 years}\label{sec:ParticleDetectors_20Year}
\begin{figure}[t]
\subfigure[ ]{\includegraphics[width=0.699\textwidth]{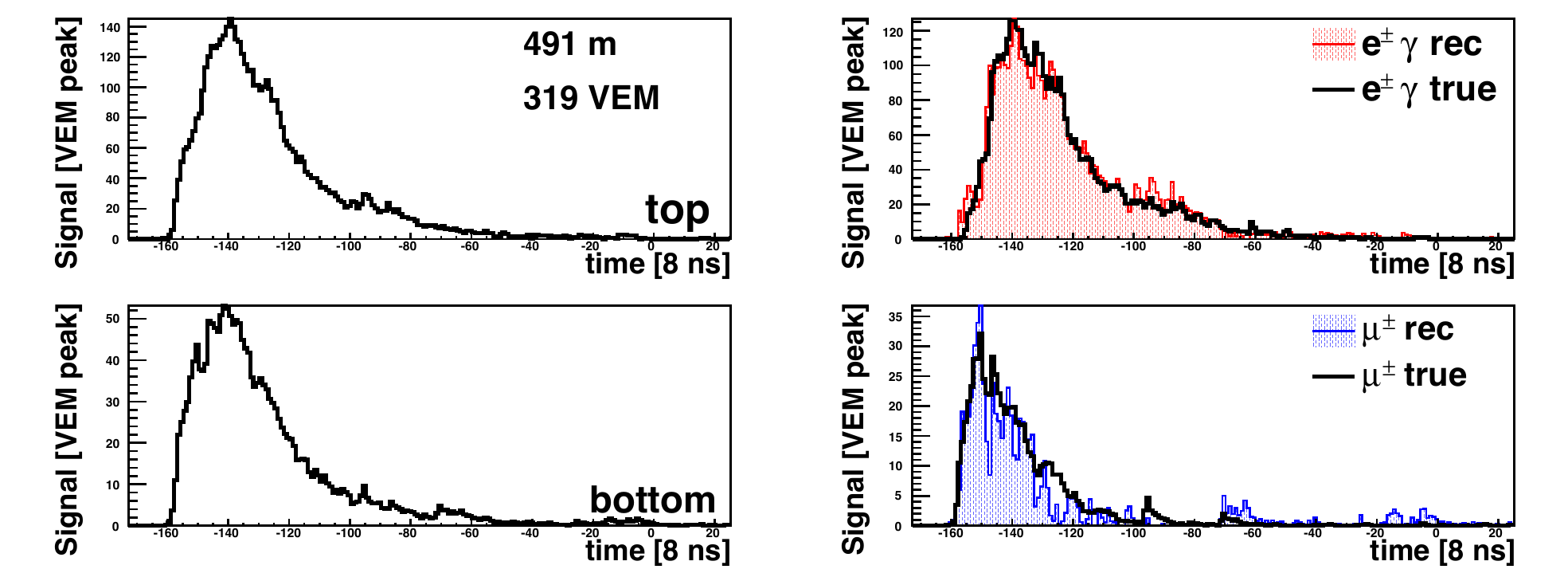}}
\subfigure[]{\includegraphics[width=0.28\textwidth]{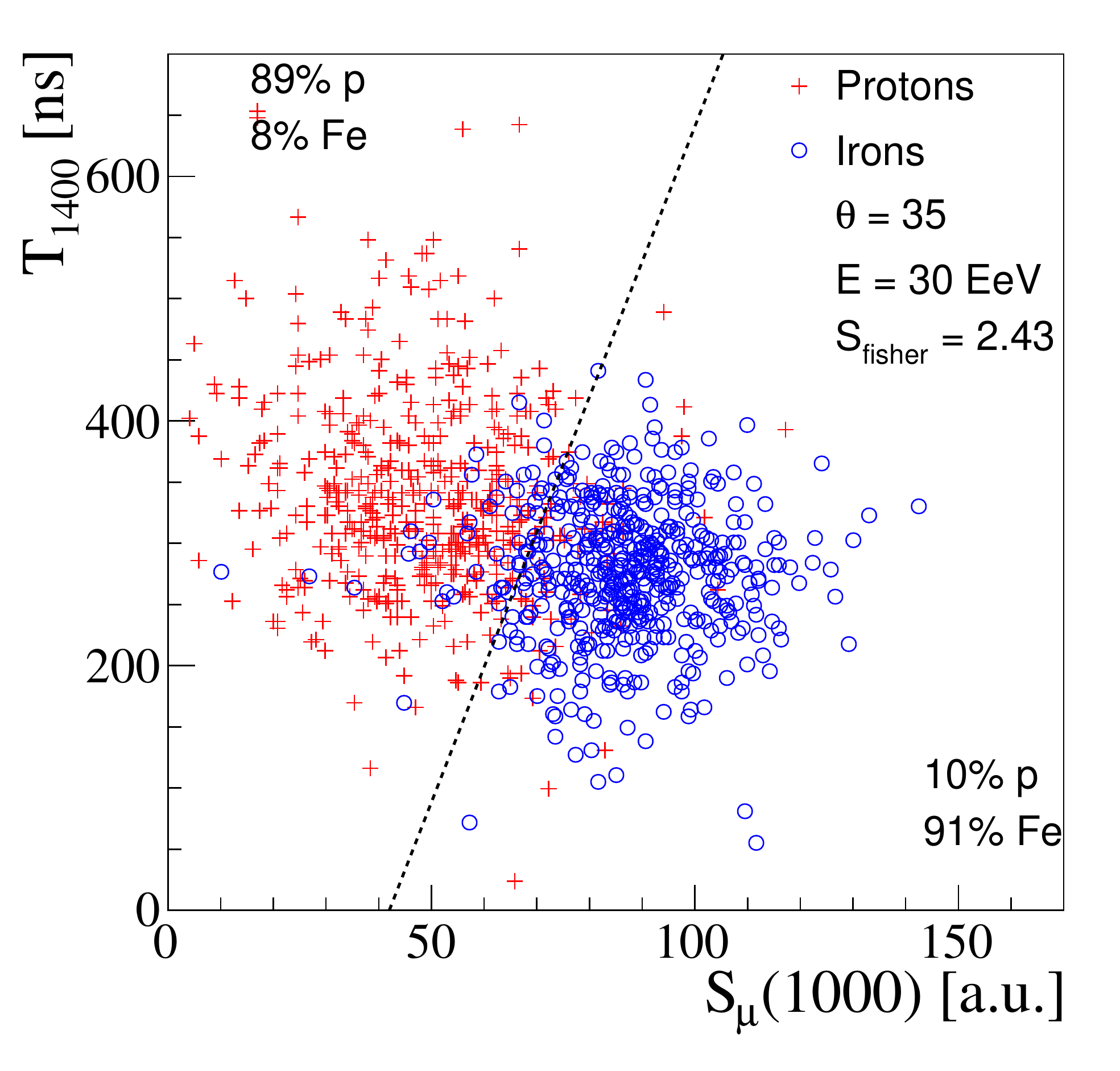}}
\caption{Example of the performance of a layered surface detector from Ref.~\cite{Letessier-Selvon:2014sga} simulated with the size of the Auger tank and two optical superimposed layers. (a) Traces in the upper and lower optical layers (left panels). Reconstructed time distributions of muons and electromagnetic components compared to the true distributions (right panels). (b) Example of a p-Fe separation based on two event variables sensitive to the \xmax ($T_{1400}$) and muonic signal $S_{\mu}(1000)$. The Fisher significance is extremely good, 2.43., allowing more than 80\,\% proton separation with less than 10\,\% Fe contamination.}
\label{fig:lsd}
\end{figure}
In the next 10 years, with the enhancement of particle detectors and modern analysis techniques, the current observatories will probably reach a resolution on the muon numbers at a station level of around 20 to 25\% by combining different type of detectors, which will translate to about 10 to 15\,\% resolution at event level (depending on the number of stations participating in the events).  
They are also expected to achieve a resolution on \xmax similar to fluorescence-detector measurements (better than 30\,g/cm$^2$). 
The next generation of ground detectors will have to improve upon this to provide a better resolution and cover huge areas for the measurement of the low flux of cosmic rays at the highest energies.

Given the steep lateral distribution function of the particles on ground (Moli\`ere radius of about 100\,m), the particle detectors will have to be large enough to provide enough statistics, i.e. of the order of tens of square meters and ideally not flat as simple scintillators (note that their effective area is halved at zenith angle of $60^\circ$ with respect to vertical). 
Therefore a water-Cherenkov detector, which is a 3D type of detector, is the natural solution. 
One of the limitations of these detectors is their time response, for example the decay time of the light in the Auger tank is about 60\,ns and is caused by the refection losses of photons in the tank. 
To shorten the decay for a better determination of the single-particle peaks a tank with black inner walls could be a more suitable choice, for which part of the interior of the detector is absorbent.
By this choice a decay time as low as 30\,ns without a substantial loss in the detection efficiency can be achieved.

Improving the decay constant might help in the determination of the muon number using deep neural networks, however, it is clear that the separation of the electromagnetic and muonic components is also required to achieve the best resolutions. 
One of the proposed solution is a layered~\cite{Letessier-Selvon:2014sga} or nested surface detector designed based on the energy deposit of particles in water. 
The majority of the electrons and positrons that reach the ground have an energy of around 10\,MeV and thus will be absorbed within a few cm, the photons will deposit their energy within more or less one radiation length ($<40\,$cm), while muons will traverse the entire water volume and produce Cherenkov photons all along their path. 
By separating the optical volumes in two pieces to enhance the difference between the signals from the different components, a layered or nested surface detector can provide very good resolution on the separation of the \ac{EAS} components at a individual station level (see~\cref{fig:gcos-wcd}) and can be a very good solution for next generation surface arrays.

\subsubsection[Fluorescence and Cherenkov detectors]{Fluorescence and Cherenkov detectors: more coverage for less price}
\label{sec:FD_tech_development}
Air showers are visible in clear nights by their \ac{UV} emission due to atmospheric fluorescence and Cherenkov light. 
The corresponding detection techniques are the backbone for any cosmic-ray physics that requires a high accuracy for the shower energy and for \xmax.
Although the technique is mature and high-quality, recent progress was achieved in making the technique more cost-effective, exploiting progress in fast timing and the development of \acp{SiPM}.
Exemplary projects of such technology development, that each will likely come of use in at least one of the future \ac{UHECR} detectors, are presented in this section. 
The selection of specific projects, such as \ac{FAST} for the fluorescence technique, is done for the purpose of readability, and does not indicate a preference over complementary R\&D projects such as \ac{CRAFFT}~\cite{Tameda:2019wmj}. 

\paragraph{Fluorescence detector Array of Single-pixel Telescopes (FAST)}
The \acf{FAST}\footnote{https://www.fast-project.org} features compact \ac{FD} telescopes with a smaller light-collecting area and far fewer pixels than current-generation \ac{FD} designs, leading to a significant reduction in cost~\cite{FAST:2021bkj,Malacari:2019uqw,Fujii:2015dra}.
Although \ac{FAST} features only four pixels, it is possible to extract timing information from each individual \ac{PMT} off the traces 
to reconstruct energy and \xmax values, resulting in comparable resolutions to conventional \acp{FD}.
\ac{FAST} is capable of providing a cost-effective method to achieve a calorimetric energy determination and a mass composition sensitivity for future ground array.

\begin{figure}[bt]
\centering
\includegraphics[width=1.0\linewidth]{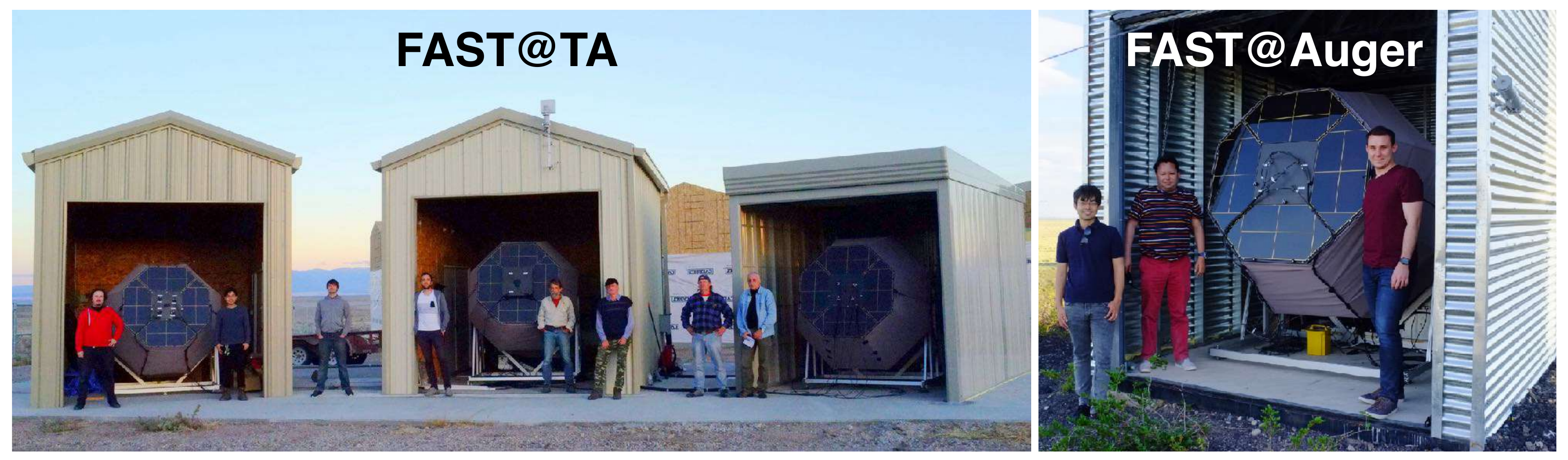}
\caption{The prototypes of the \acf{FAST} installed at the Telescope Array Experiment and Pierre Auger Observatory, dubbed FAST@TA and FAST@Auger, respectively.}
\label{fig:fast}
\end{figure}

In the \ac{FAST} design, a 30$^{\circ}$ $\times$ 30$^{\circ}$ \ac{FoV} is covered by four 20\,cm \acp{PMT} at the focal plane of a compact segmented mirror of 1.6\,m diameter~\cite{FAST:2017vfb} (see also \cref{fig:gcos-fast}). 
Its smaller light-collecting optics, smaller telescope housing, and fewer number of \acp{PMT} significantly reduces its cost. 
As shown in \cref{fig:fast}, three full-scale \ac{FAST} prototypes dubbed FAST@TA were installed at the \ac{TA} site for a concept validation, and an identical \ac{FAST} prototype dubbed FAST@Auger was also installed at the Auger site for a cross-calibration of energy and \xmax scales.
An automated all-sky monitoring camera is used to record cloud coverage and atmospheric transparency to reduce these uncertainties~\cite{FAST:2020blj}.

\begin{figure}[bt]
  \centering
  \subfigure[Reconstructed \xmax distributions]{\includegraphics[width=.49\linewidth]{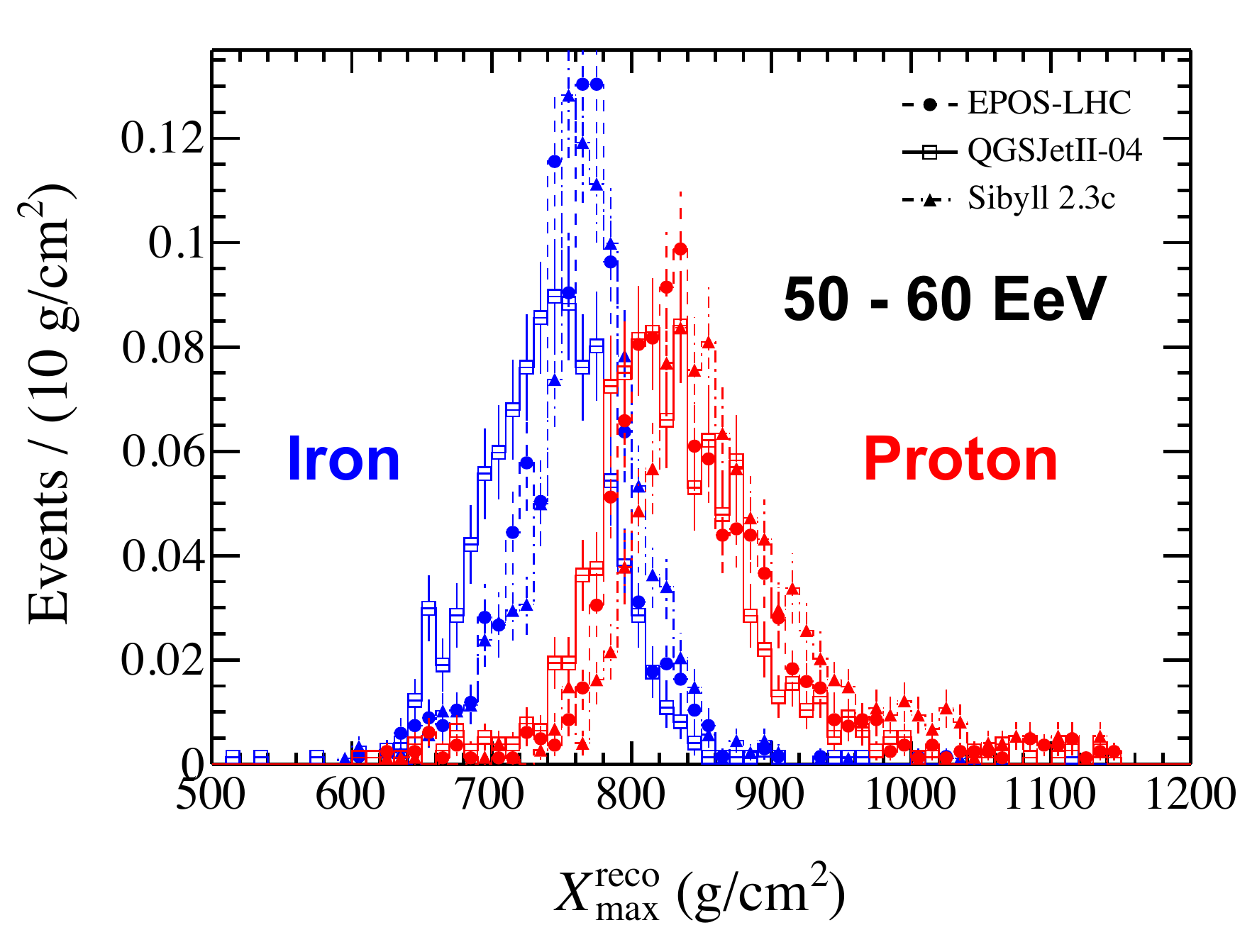}}
  \subfigure[Preliminary result with FAST@TA]{\includegraphics[width=0.43\linewidth]{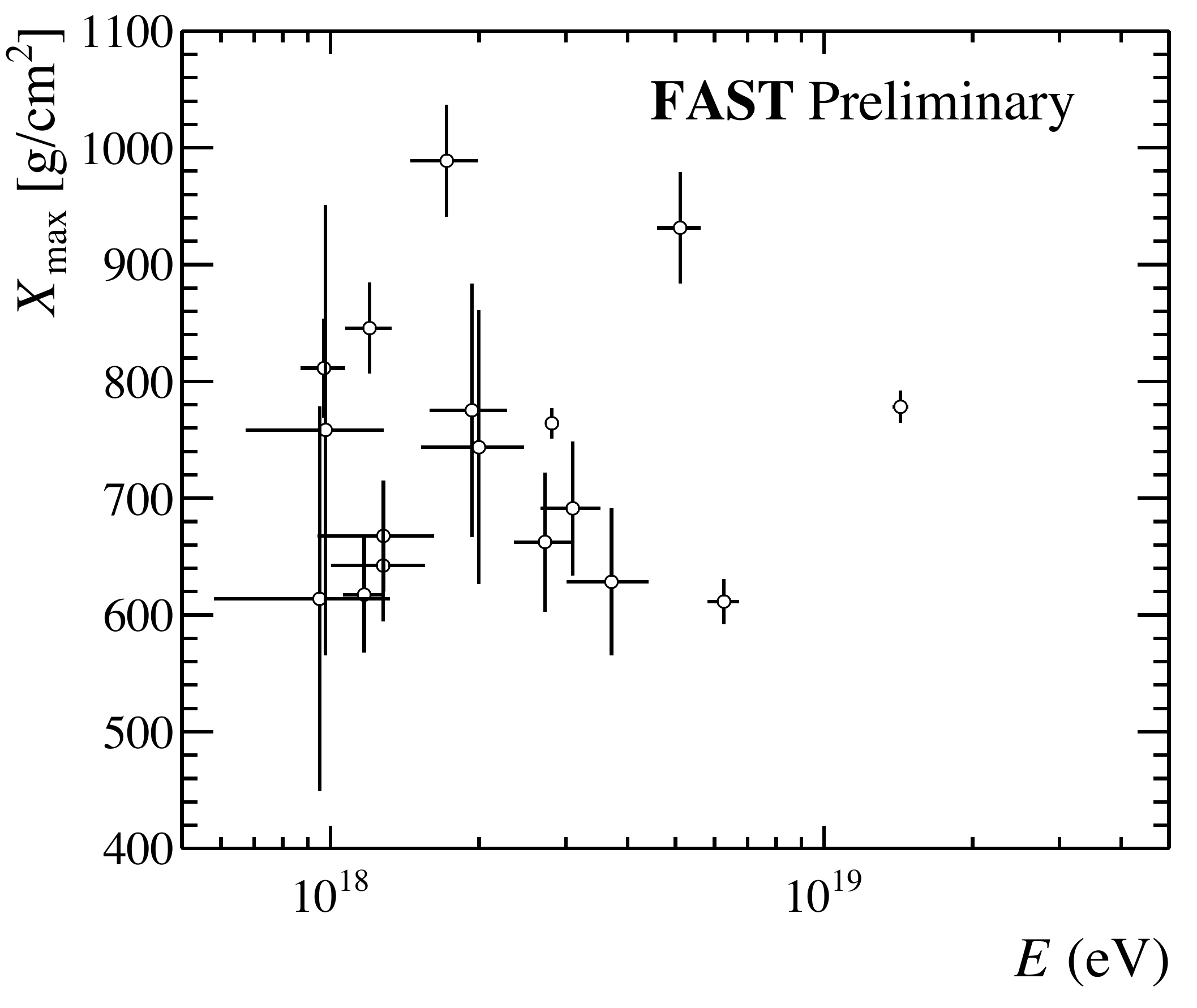}}
  \caption{(a) Reconstructed \xmax distributions estimated by the \ac{MC} simulation with an energy range from 50\,EeV to 60\,EeV, (b) Preliminary result of Energy and \xmax reconstructions by the FAST@TA prototypes~\cite{FAST:2021bkj}.}
  \label{fig:xmax}
\end{figure}

\cref{fig:xmax}(a) shows the expected \xmax distribution with an energy range from 50\,EeV to 60\,EeV by \ac{FAST}  evaluated by a detailed detector simulation~\cite{Malacari:2019uqw} and a neural network reconstruction~\cite{Justin:2021phd} using proton and iron primaries
with three hadronic interaction models (\eposlhc, \qgsii and \sibyll{2.3c})~\cite{Heck:1998vt}. The expected resolutions of \ac{FAST} are 8\,\% in energy and 30\,\gcm on \xmax around 50\,EeV. Analyzing 224 hours of data measured by FAST@TA, significant signals of 17 showers were found in time coincidence with the \ac{TA} fluorescence detectors. \cref{fig:xmax}(b) shows preliminary energy and \xmax values reconstructed by the FAST@TA prototypes. This result demonstrates the calorimetric energy determination and the mass composition sensitivity by the \ac{FAST} prototypes from field measurements.

\subsubsection{Air Cherenkov technique}
\label{sec:air-cherenkov}
Incoming \acp{CR} create \acp{EAS}, that also produce Cherenkov light in the atmosphere, as shown in \cref{eas_detection_technique}.  Detection of this Cherenkov light is a powerful tool in the study of both gamma-rays (which will not be discussed in this paper) and charged \acp{CR}.  Detectors designed for air-Cherenkov detection can be used independently or in conjunction with other \ac{EAS} detection techniques to study both the energy and mass composition of primary \acp{CR} incident on the atmosphere.  Air-Cherenkov detection of \ac{EAS} can be divided into imaging and non-imaging techniques.

\paragraph{Non-imaging air Cherenkov detection of cosmic rays}
\ac{NIAC} detectors are arranged into ground-based arrays which sample the lateral distribution of \ac{EAS}-produced Cherenkov light at the ground.  These detectors often consist of large Winston cones facing up toward the night sky, which collect the light and concentrate it into a \ac{PMT} for measurement.  The primary particle information is then reconstructed using techniques similar to those of the ground-based charged particle detectors: this involves fitting the data to an expected Cherenkov-light lateral distribution function (originally worked out in Refs.~\cite{Hillas:1982wz, Patterson:1983qj}), which then allows for the extraction of the calorimetric energy and depth of shower maximum of the air showers.  

The NIAC technique has been successfully performed several times, including using the AIROBICC detectors at HEGRA \cite{Karle:1995dk} and CASA-BLANCA in Utah \cite{Fowler:2000si}.  More recently the Yakutsk array \cite{Ivanov:2014tra}, Tunka-133/Taiga array \cite{Prosin:2021nad} and the \ac{NICHE}~\cite{Omura:2021nkh} are utilizing hybrid detection of \acp{CR} using different detection techniques to reach ultra high energies. 

\paragraph{Imaging air Cherenkov detection of cosmic rays}
\acp{IACT} collect the air-Cherenkov light produced by \acp{EAS}, historically using very large mirrors and/or lenses.  The light is then measured using a multi-pixel camera consisting of high-speed photon detectors.  The images produced using this technique are related to the shape of the \ac{EAS} in the atmosphere, which is dependent on both the energy and composition of the incident primary particle (in addition to atmospheric properties).  To analyze the images produced in the cameras, \acp{IACT} use either several parameters devised by Hillas \cite{1985ICRC....3..445H} or state-of-the-art machine learning algorithms.  Either of these methods provides information related to the \ac{EAS} geometry including the composition-sensitive depth of the shower maximum, in addition to the primary energy.  

To increase the accuracy of these measurements, multi-telescope observations are used to record the Cherenkov light from the \acp{EAS} from multiple perspectives.  For example, the \ac{HESS} \cite{Forster:2014rra} includes 5 telescopes, and the \ac{CTA} \cite{Gueta:2021vrf} plans to include more than 100 telescopes divided between two arrays: one in the northern hemisphere, the other in the southern hemisphere. Although these observatories in particular were designed to measure the gamma-ray flux from stellar objects, they are also able to measure the diffuse \ac{CR} flux and mass composition, as discussed in Refs.~\cite{Jankowsky:2020baz,VERITAS:2018gjd}.

\begin{figure}[!htb]
  \centering
    \includegraphics[height=7.cm]{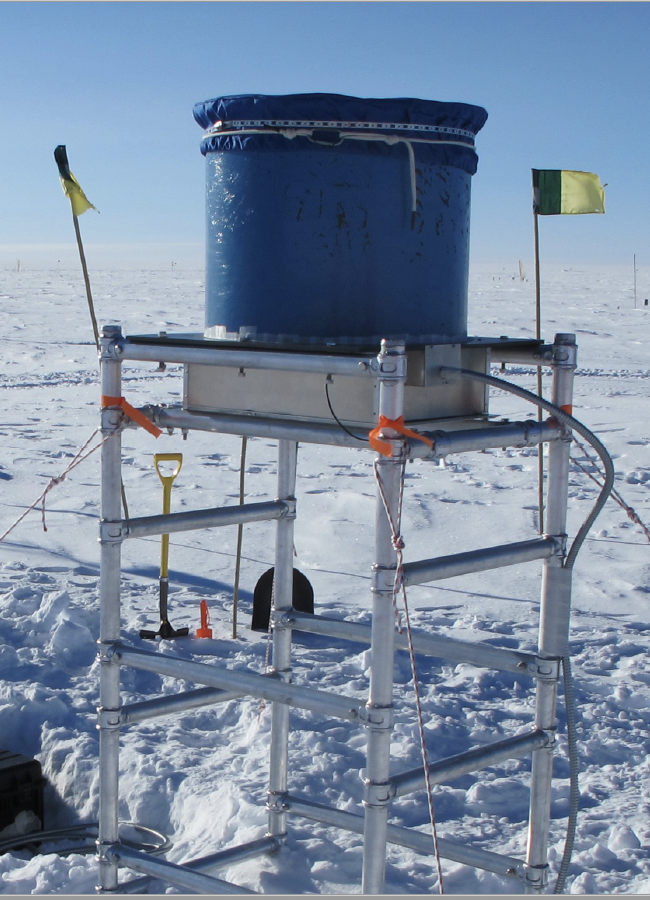}
    \includegraphics[height=7.cm]{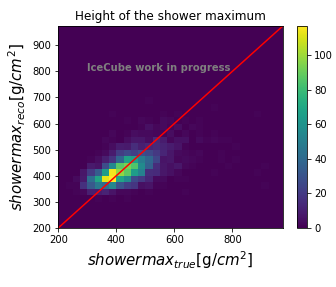}
  \caption{(left) IceAct prototype detector at the South Pole. (right) Preliminary air shower maximum reconstruction using machine learning~\cite{larissa_paul_2022_6354743} with IceAct.}
  \label{IceAct_fig}
\end{figure}

Furthermore, several existing observatories are presently planning upgrades to include compact \acp{IACT} utilizing cost-efficient \ac{SiPM} camera designs. For example, two prototype \emph{IceAct} \acp{IACT}~\cite{IceCube:2021htd} are providing a low-energy \ac{CR} enhancement for the IceCube Neutrino Observatory at the South Pole. These telescopes have a fixed pointing and a wide \ac{FoV} and have been operational in a stable configuration since 2019.  IceAct measures the air Cherenkov portion of the \ac{EAS} in stereo configuration and in hybrid mode together with IceTop/IceCube.  These two IceAct prototypes, as shown in \cref{IceAct_fig}, will be able to extend the most recent composition and energy spectrum measurements from IceTop and IceCube \cite{IceCube:2019hmk} from a few PeV down to $\sim$\,50~TeV \cite{IceCube:2021htd} to cross the transition region from galactic to extra-galactic cosmic rays.  An array of 4 stations with 7 IceAct telescopes each is planned for IceCube-Gen2, which will increase the sky coverage and number of events at higher energies, providing a new handle on the \ac{UHECR} composition by directly measuring the air shower maximum, which is shown in right side of \cref{IceAct_fig}, and the energy spectrum for IceCube.

Imaging and non-imaging techniques can also be combined.  For example, \ac{NICHE} and \ac{TALE}~\cite{TelescopeArray:2018bya} can work together to study \acp{CR} at energies above 1\,PeV~\cite{Bergman:2020pjh}.  In this case, the detection threshold of \ac{TALE}, which was designed as an air-florescence telescope, is extended by utilizing Cherenkov-dominated events, which essentially turns \ac{TALE} into an \ac{IACT}.  Similarly, Cherenkov dominated events are used in the fluorescence detector extension \ac{HEAT}~\cite{Novotny:2021lfu} of the Pierre Auger Observatory to provide the \ac{CR} energy spectrum and the shower maximum above $10^{16.5}$\,eV.

\subsubsection[Radio detectors]{Radio detectors: the multi-hybrid perspectives of a new orthogonal measurement technique}
\label{sec:RD_tech_development}
Radio detection of \acp{EAS} has proven its competitiveness with other detection techniques over the last decade~\cite{Huege:2016veh,Schroder:2016hrv}. 
Digital antenna arrays have demonstrated that they can deliver an accurate measurement of the arrival direction, electromagnetic shower energy, and depth of the shower maximum, \xmax. 
In combination with muon detectors, radio antennas can provide a path to around-the-clock measurements of the rigidity of the primary particle.

The threshold of the radio technique is around $10^{16}$ to $10^{18}$\,eV depending on the frequency band and density of the antenna array, and also depending on the detector elevation and the strength and orientation of the magnetic field relative to the \ac{CR} arrival directions. 
While full-sky coverage requires an antenna spacing of the order of $100$\,m, sparse arrays with spacing of a kilometer or more still enable full efficiency for very inclined showers~\cite{PierreAuger:2021ece}. 
Therefore, the radio technique is suited for a large variety of different use cases and will play a role in many of the future ultra-high-energy astroparticle observatories.
This section provides an overview over the state-of-the-art and future developments regarding various aspects.

\paragraph{Theory and simulations of radio signals from particle showers}

One of the main reasons for the success of the radio detection technique is the detailed understanding of the radio-emission physics achieved in recent years~\cite{Huege:2016veh}. 
Due to the interplay of the emission mechanisms relevant to \acp{EAS}, the dominant geomagnetic emission and the subdominant charge-excess or Askaryan emission, and Cherenkov time compression, the radio signal on ground has a more complicated structure than the particle footprint.
Nonetheless, this feature-rich radio signal has been mastered in recent radio projects because of substantial progress in the theoretical understanding and the availability of state-of-the art simulation tools.

The ``work horse'' for the calculation of radio emission from particle showers is the calculation of the emission from every single electron and positron in a particle shower using classical electrodynamics in a ``microscopic'' Monte Carlo simulation approach. 
These calculations make no assumption on the underlying \emph{emission mechanisms} by the air-shower particles; they thus directly and unambiguously predict the absolute signal strength. 
The CoREAS~\cite{Huege:2013vt} and ZHAireS~\cite{Alvarez-Muniz:2010wjm} simulation codes, two independent programs implementing numerically different approaches, have in particular been successful in simulating the radio emission from air showers in a vast variety of applications and for frequencies from $30\,$MHz to $4.2\,$GHz~\cite{Huege:2016veh}. 
Comparisons between the two codes~\cite{Gottowik:2019yih} and with lab-experiments~\cite{T-510:2015pyu} have illustrated the ability to predict the absolute strength of the emission correctly, including details such as the (small) degree of circular polarization in the mostly linearly polarized radio signal~\cite{Scholten:2016gmj}.

While tremendously successful, these microscopic simulations suffer from the problem that they are very computing-intensive. 
Several strategies are being followed to keep computing feasible in light of increasing need for simulation accuracy and level of detail for next-generation experiments:

\textbf{Thinning:} At energies of $10^{17}$~eV and higher, particle thinning algorithms are applied which, however, lead to an overestimation of coherent radio emission at high frequencies. Especially at energies well beyond $10^{18}$~eV, this \emph{thinning noise} starts to dominate over Galactic noise even at frequencies of 30--80~MHz and thus introduces problems in simulation-based analyses. 
Strategies will need to be worked out to minimize or compensate for the impact of thinning artifacts in simulations at the highest energies.

\textbf{Parallelization:} 
Parallelization of the simulations using MPI is already possible with CoREAS \cite{Huege:2013vt} and effectively solves the problem of long computation times for UHE showers (but of course not total computing requirements). Another area with potential is the parallelization on \acp{GPU} in the context of the CORSIKA~8 project~\cite{Karastathis:2021akf}. 
\ac{GPU} parallelization will also improve the energy efficiency of the simulations, thereby reducing their ecological impact.

\textbf{Accurate approximations:} For \emph{top-down} analysis approaches such as the ones described in \cref{sec:CurrentStatus:RadioXmax}, and in particular for future dense radio arrays such as \ac{SKA}~\cite{Huege:2015jga}, computing requirements for simulations constitute a limiting factor.
Building on the experience of -- yet less accurate -- ``macroscopic'' calculation approaches~\cite{Scholten:2007ky,Scholten:2017tcr}, efforts have been made to investigate approaches to exploit \emph{universality} in the radio emission from particle showers to calculate the radio signals from a desired air shower using a reference or template shower~\cite{Butler:2019kde,Chiche:2021iin}. 

Further work is also envisioned for the application of the radio technique to very inclined air showers as well as cross-media showers. 
The former have been measured by \ac{AERA}~\cite{PierreAuger:2018pmw} and are the focus of the Radio Detector component of the ongoing AugerPrime upgrade~\cite{PierreAuger:2021bwp,PierreAuger:2021ece}, the potential radio component of \ac{GCOS}, and \ac{GRAND}~\cite{Alvarez-Muniz:2018bhp}. 
Simulations for these very inclined geometries will need to be validated in depth, in particular because refractive effects in the atmosphere and potentially also ground reflections start to play a role~\cite{Schluter:2020tdz}. 
The existing codes cannot be easily adapted to simulate these and other complex scenarios, such as cross-media showers important for in-ice radio detection~\cite{DeKockere:2021qka}, but CORSIKA~8 will allow the flexibility to perform such simulations.

\paragraph{Radio Energy} \label{sec:CurrentStatus:RadioEnergy}

Radio measurements are well suited for doing electromagnetic energy reconstruction.
Radio emission is produced primarily by the \ac{EAS} electromagnetic component, and as discussed above, can be calculated from first principles. Furthermore, the measured radio signal is integrated over the entire air shower, so measurements can be used to perform calorimetric energy reconstructions~\cite{Huege:2016veh}. 

In the last decade, a number of approaches have been used to reconstruct cosmic-ray energy using radio measurements. The \ac{LOPES} and Tunka-Rex used the method of determining the signal strength relative to a characteristic distance from the shower axis in the shower plane at which shower-to-shower fluctuations are minimized.  This method achieved a resolution of better than 15\% for Tunka-Rex~\cite{Kostunin:2015taa}, and better than 20\% for \ac{LOPES}~\cite{LOPES:2014bps}.  The \ac{LOFAR} and later Tunka-Rex have used an approach which directly compares the measured signal strength in each antenna to the strength predicted by CoREAS simulations, achieving a resolution of 15\% or better~\cite{Buitink:2014eqa,Mulrey:2020oqe,Bezyazeekov:2018yjw}.

\begin{figure}%
    \centering
    \includegraphics[width=8.0cm]{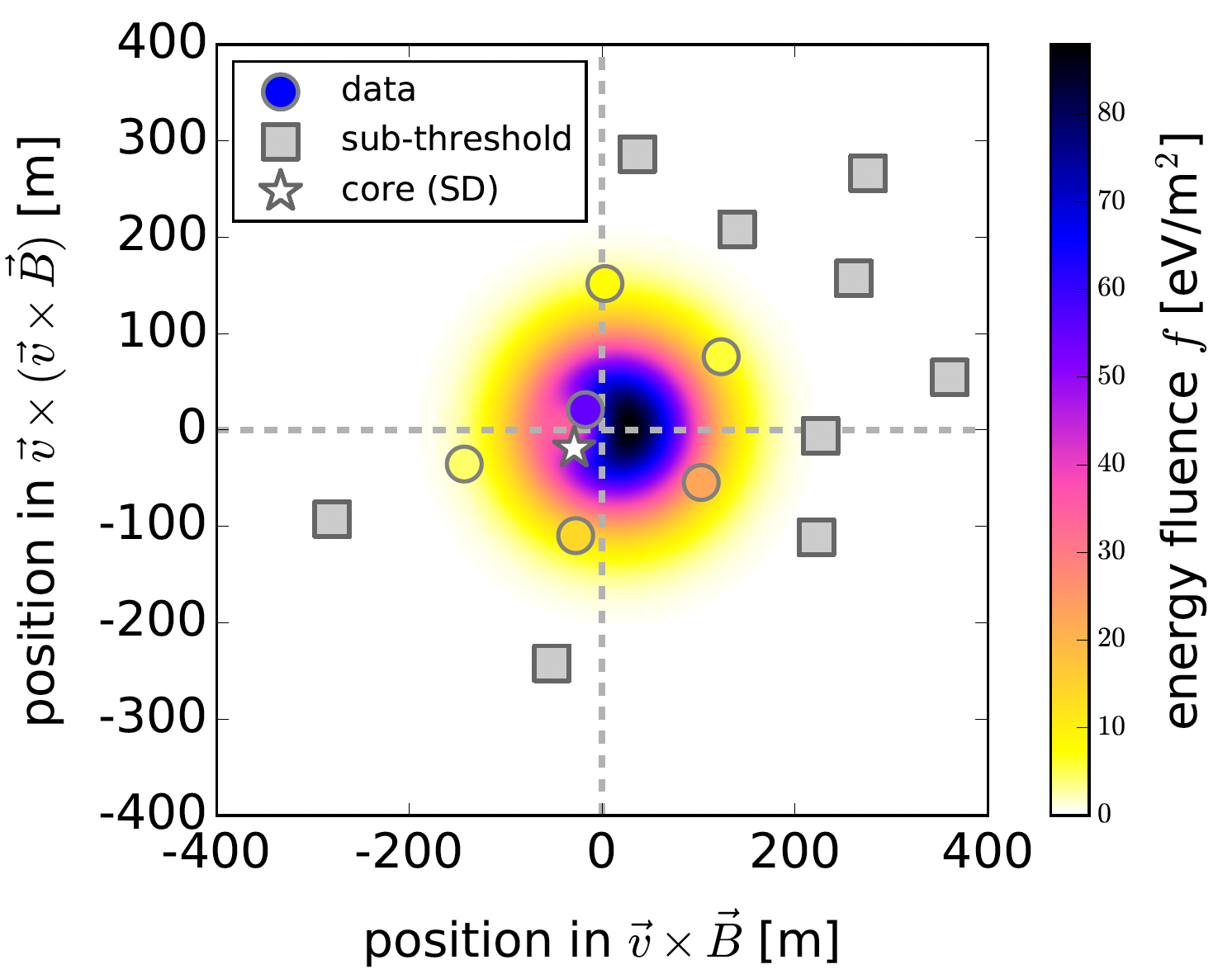} 
    \hfill
    \includegraphics[width=6.5cm]{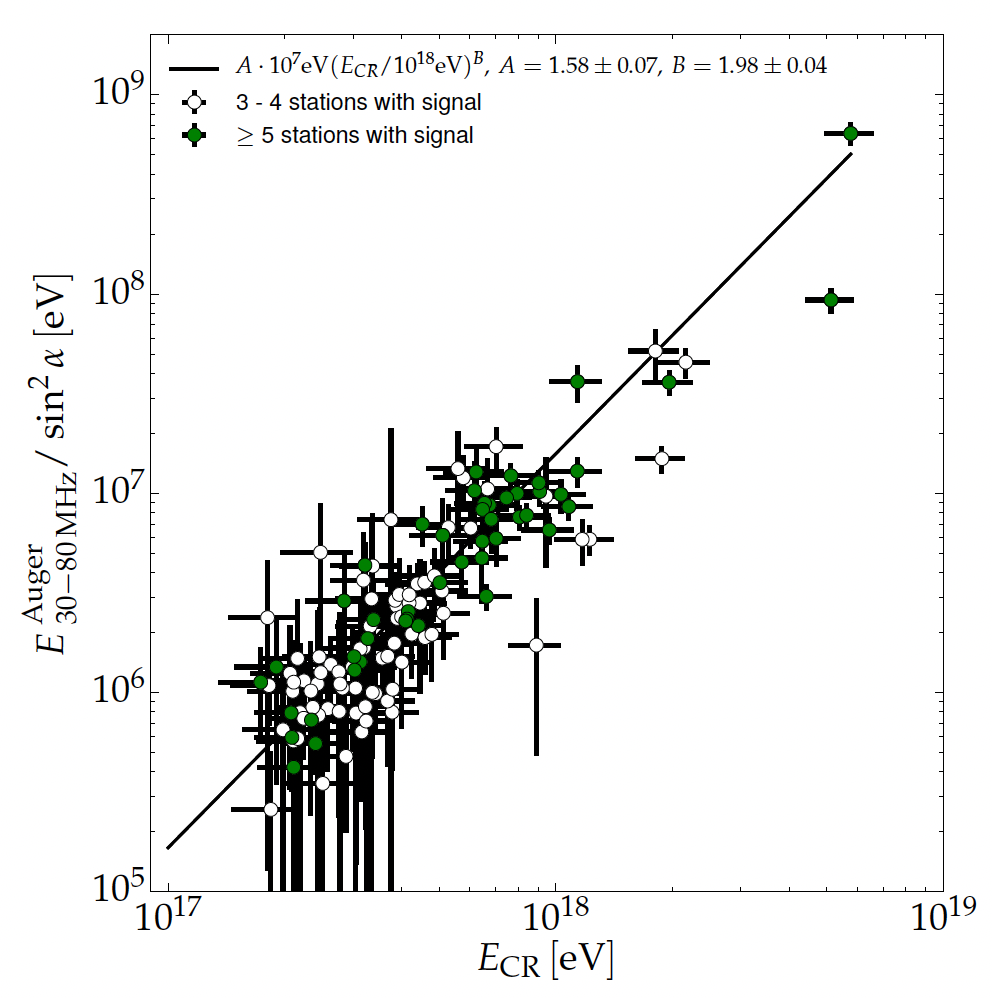}
    \caption{Left: Energy fluence footprint for an extensive air shower with an energy of $4.4\times 10^{17}$~eV detected by \ac{AERA}, and positions of the \ac{AERA} antennas. Right: Correlation between the radiation energy measured with \ac{AERA} and the total cosmic-ray energy as determined by the Auger surface detector~\cite{PierreAuger:2016vya}}%
    \label{fig:energy}%
\end{figure}

Another technique focuses on determining the total energy radiated by the air shower in the form of radio emission, or the radiation energy, which scales quadratically with the electromagnetic energy of the air shower~\cite{Glaser:2016tng}.  The geomagnetic emission strength scales with the absolute value of the local geomagnetic field and the sine of the angle between the shower axis and the geomagnetic field. There are also second-order effects from the influence of the atmospheric density on the shower development and the relative charge excess contribution. 
\ac{AERA} has fit the measured energy fluence at different antenna positions to a 2D \ac{LDF}~\cite{Nelles:2014xaa,PierreAuger:2016vya}. 
When integrated, this yields the radiation energy of the shower. 
An example of the energy fluence map for an \ac{AERA} event is shown in the right panel of \cref{fig:energy}. The left panel of \cref{fig:energy} shows the correlation between the radiation energy measured with \ac{AERA} and the total cosmic-ray energy as determined by the Auger \ac{SD}~\cite{PierreAuger:2016vya}.

A strong prospect for energy reconstruction in the future is the use of broadband radio signals, rather than the traditional $30-80$~MHz bandwidth currently used by most experiments.  The spectral shape of the signal can be used to determine the distance of an antenna to the shower core.  The amplitude of the signal can then be directly related to the radiation energy in the shower.  Tunka-Rex demonstrated this principle, using the core position as determined by the Tunka-133 air-Cherenkov array~\cite{Tunka-Rex:2016gcn}. The \ac{ANITA} and the \ac{ARIANNA} have shown that the radiation energy can be reconstructed with with a single antenna station even without external information on the shower geometry~\cite{Schoorlemmer:2015ujs,Welling:2019scz}.  The \ac{GRAND} experiment will use antennas in the $50-200$ MHz bandwidth, and the radio installation at IceTop~\cite{Schroder:2018dvb} and the \ac{SKA}~\cite{Buitink:2021pkz} will measure \acp{CR} between $50-350$~MHz.  This will also be highly relevant for \ac{GCOS}, where an energy resolution of about $10\,\%$ will be required to fully investigate features of the energy spectrum~\cite{Horandel:2021prj}.

The ability of any of these techniques to produce a valid energy scale relies on the absolute calibration of the antennas, which can be determined by using an external reference source. One effective reference source is the background Galactic emission.  When calibrating the antennas using this technique, the systematic uncertainty on the energy reconstruction has been shown to be $14\,\%$, with the dominating contribution being the uncertainty on the models used to predict the background Galactic emission~\cite{Mulrey:2019vtz,PierreAuger:2012ker}.  The remaining contributions to the absolute scale uncertainty can be reduced to less than about $8\,\%$, and the performance of the antennas promises to be stable over time. In summary, the radio detection technique produces energy reconstructions with absolute scale uncertainties competitive with other techniques already today. 
Efforts will be made to further reduce the uncertainty on the antenna calibration in the future which brings in reach a precision of individual events as well as an absolute accuracy for the energy of better than $10\,\%$. 

\paragraph{Measuring the depth of shower maximum with Radio} \label{sec:CurrentStatus:RadioXmax}

\begin{figure}[!ht]
  \centering
  \includegraphics[height=8.5cm]{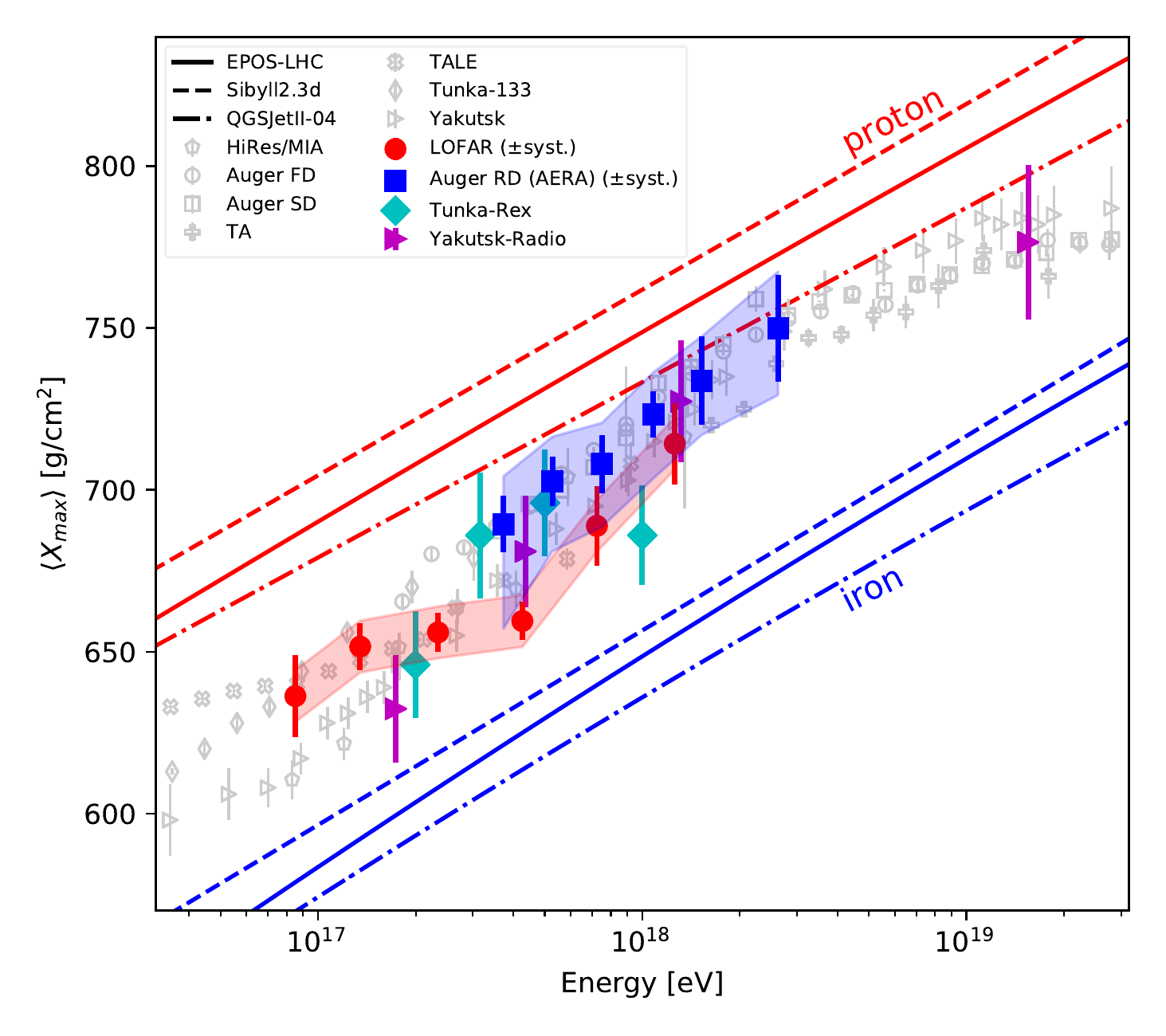}
  \caption{Measurements of the mean of the \xmax distribution by radio experiments (\ac{AERA}~\cite{PierreAuger:2021rio}, \ac{LOFAR}~\cite{Corstanje:2021kik}, Tunka-Rex~\cite{Bezyazeekov:2018yjw}, and Yakutsk-Radio~\cite{Petrov:2020edv}) and compared to world data (Auger \ac{FD}~\cite{Yushkov:2020nhr} and SD~\cite{ToderoPeixoto:2020rta}, HiRes/MIA~\cite{HiRes-MIA:2000ook}, \ac{TA}~\cite{TelescopeArray:2018xyi}, \ac{TALE}~\cite{TelescopeArray:2020bfv}, Tunka-133~\cite{TAIGA:2021fgx}, and Yakutsk~\cite{Knurenko:2019oil}). The statistical uncertainties are plotted as error bars and for radio the systematic uncertainties as bands if available. The results are compared to predictions from CORSIKA air shower simulation for multiple hadronic interaction models (lines) for proton (red) and iron (blue) mass compositions~\cite{Yushkov:2020nhr}.}
  \label{fig:Radio_Xmax_CombinedPlot}
\end{figure}

The radio signal as measured on the ground is sensitive to \xmax manifesting predominantly in a change of shape of the radio emission footprint. Early work on this was done by Allan in 1971~\cite{ref:radio_xmax_Allan1971,ref:radio_xmax_AllanICRC71} relating the footprint width to \xmax, but it wasn't until the arrival of fast digital data acquisition that \ac{CR} radio arrays became an effective way to study \xmax.

Measurements by \ac{LOPES}~\cite{LOPES:2012xou,Palmieri:2013kvf} of the slope of the \ac{LDF} demonstrated the feasibility of radio \xmax measurements, but did not yet reach a competitive resolution. A similar method was also used by the Yakutsk radio array~\cite{Petrov:2020edv}.
Understanding of the geomagnetic and charge excess emission mechanisms with 2-dimensional \ac{LDF} parametrization functions~\cite{Nelles:2014xaa,Glaser:2018byo,Tunka-Rex:2015zsa} improved on this. In addition to \ac{LDF} parametrizations also the slope of the frequency spectrum~\cite{Jansen:2016sjo,Canfora:2021xkh} and the shape of the shower wave front~\cite{Apel:2014usa} were investigated for \xmax reconstruction but were limited in practice by core position resolution and understanding of the antenna response. The highest resolution has been achieved only in the past few years by matching measured radio signals to signals from dedicated sets of CORSIKA/CoREAS full Monte-Carlo air shower simulations for each measured \ac{EAS}~\cite{Buitink:2014eqa}. Recent advancements such as from including time-varying atmospheric conditions~\cite{Mitra:2020mza} into the simulations have improved \xmax resolutions further, circumventing the uncertainties previously encountered in the averaged \ac{LDF} parametrization models for \xmax. Results by Tunka-Rex \cite{Bezyazeekov:2018yjw}, \ac{LOFAR}~\cite{Corstanje:2021kik}, and \ac{AERA}~\cite{PierreAuger:2021rio} have shown resolutions up to $15-25$~g/cm$^2$ can be achieved with similar implementations of this method. 

Recent efforts by \ac{LOFAR}~\cite{Corstanje:2021kik} and \ac{AERA}~\cite{PierreAuger:2021rio} have also performed detailed studies to quantify systematic uncertainties on radio \xmax measurements, including direct comparisons to fluorescence \xmax measurements at \ac{AERA} showing radio and fluorescence measurements to be fully compatible. An overview of radio \xmax measurements with statistical uncertainties (bars) and systematic uncertainties (bands) is shown in \cref{fig:Radio_Xmax_CombinedPlot} superimposed on \xmax data from optical Cherenkov and fluorescence light measurements. This highlights that the radio technique has already shown to be competitive in mass composition studies even with small sparse arrays.


\paragraph{Interferometric measurements of extensive air showers} \label{sec:CurrentStatus:RadioInterferometry}
Interferometric techniques for the detection of extensive air showers make use of not only the amplitude but also the phase information of the received radio emission. 
By combining the waveforms recorded by several receiving antennas into a single, directed beam, the signal-to-noise ratio can be increased by $\lesssim\sqrt{N_\mathrm{ant}}$ while anthropogenic radio-frequency-interference (RFI) which is typically emitted by sources at, or close to the horizon, is suppressed. 
Examples for the successful application of interferometric measurements with the aforementioned objectives are \ac{ANITA}~\cite{ANITA:2008mzi} or \ac{LOPES}~\cite{LOPES:2021ipp}, which both featured the required (sub-)nanosecond time synchronization~\cite{Schroder:2010sa}.

\begin{figure}[t]
  \centering
  \includegraphics[width=0.6\linewidth]{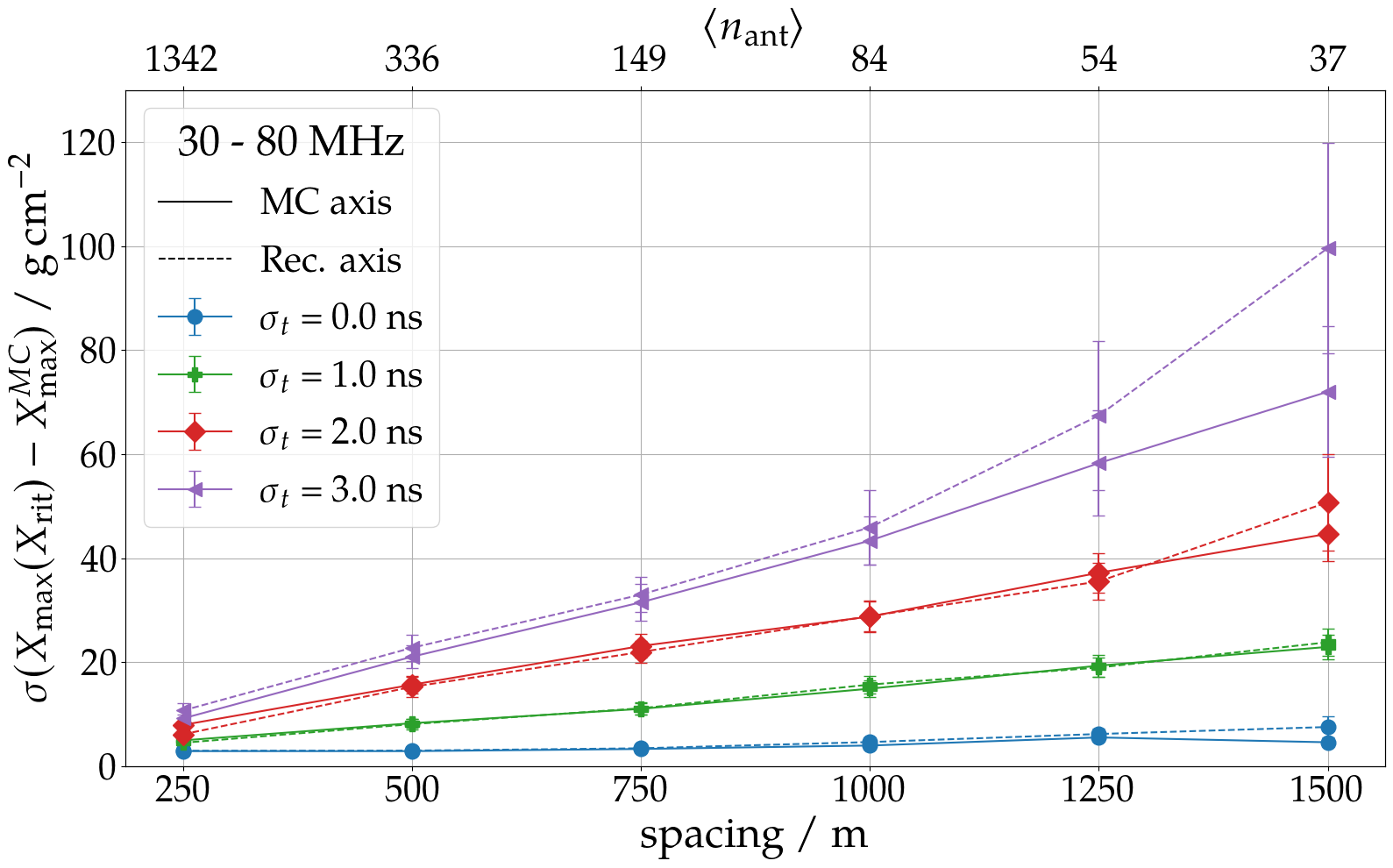}
  \caption{Achievable \xmax resolution for different time-synchronisation accuracy and antenna multiplicity (i.e., antenna spacing) scenarios. From Ref.~\cite{Schluter:2021egm}.}
  \label{fig:interferometry}
\end{figure}

A novel algorithm to reconstruct the depth of the shower maximum \xmax with beamforming has proven to achieve exquisite accuracy on ideal simulations~\cite{Schoorlemmer:2020low}. 
However, current air-shower antenna arrays do not fit the requirements in terms of time synchronization accuracy and antenna multiplicity~\cite{Schluter:2021egm}, see \cref{fig:interferometry}. 
In the near future, astronomical observatories such as \ac{SKA}~\cite{Huege:2015jga} or OVRO-LWA~\cite{Plant:2021iqx} promise great potential to employ interferometric measurements to lower their energy threshold and reconstruct \xmax. 
If proven applicable with measured data, this novel \xmax reconstruction would be extremely valuable to 
enable accurate \xmax reconstruction for very inclined air showers with sparse, large aperture antenna arrays.
The rapid development of wireless communication~\cite{9076025} might enable a sufficiently accurate time synchronisation for such large-scale antenna arrays of independent detectors.


Beamforming on the trigger level is currently tested by radio in-ice experiments for neutrino detection~\cite{Allison:2018ynt,RNO-G:2020rmc} which exploit their particular vertical detector geometry. 
Very fast online data processing or the focus to certain regions in the sky (e.g., positions of candidate sources of ultra-high energy gamma rays~\cite{Escudie:2019tlt} or a target mountain range in searches for tau neutrinos~\cite{Hughes:2020ghq}) might enable interferometric triggers also for air showers arrays.


In the next decade, if proven applicable to data of sparse radio air-shower array, interferometric methods exhibit great potential to empower the scientific capabilities of large-scale experiments such as \ac{GRAND}, \ac{GCOS}, or the surface array of IceCube-Gen2.

\paragraph{Autonomous radio-detection of extensive air showers} 
Autonomous radio detection of air showers can be defined as the set of hardware and software processes allowing for the detection and identification of air showers using solely information from radio antennas. 
For giant arrays such as the $200,000$\,km$^2$ of the planned \ac{GRAND} project~\cite{Alvarez-Muniz:2018bhp}, this is essential for obvious technical and financial reasons. 
Yet for hybrid setups combining the radio technique with an array of particle detectors, autonomous radio-detection also has advantages: the larger radio footprint of inclined showers allows for an improved efficiency for radio-detection if either a self-trigger or continuous buffering is in place (see e.g., Ref.~\cite{Huege:2016veh}), whereas the increased absorption in the atmosphere of the electromagnetic component for these showers~\cite{Zas:2005zz} affects the efficiency of ground arrays of particle detectors. 
This is even more important for muon-poor showers, such as the ones induced by $\gamma$-rays.

Yet the detection --- and even more the identification --- of air showers from faint radio signals of duration $\lesssim$100\,ns is challenging.
Dedicated efforts have been initiated over the last decade by various experiments, taking advantage of specific signatures of air-shower radio signals to distinguish them from thermal and  anthropogenic background (e.g., transient pulses from RFI sources):

\ac{ARIANNA} benefited from the very limited anthropogenic noise of the Ross Ice-Shelf to reach an event rate as low as $10^{-3}$\,Hz with a basic trigger condition (causal coincidence between 2 antennas with signal-over-threshold). An additional offline treatment, based on the adjustment of template signals (built from simulated air showers) to recorded pulses, allowed to identify 38 cosmic-ray candidates~\cite{Barwick:2016mxm}.

The \ac{ANITA} balloon probe used interferometry followed by dedicated   analysis tools to identify cosmic-ray events from the billions of radio signals recorded during its four fights above Antarctica~\cite{ANITA:2010ect,ANITA:2010hzc,ANITA:2018vwl,ANITA:2019wyx}. 
Eventually a few tens of cosmic-ray events could be selected in the whole \ac{ANITA} dataset through an additional selection on signal polarity (positive) and polarization (horizontal). The pioneering work of \ac{ANITA} will be followed by the \ac{PUEO}~\cite{PUEO:2020bnn}, a next generation balloon-borne radio detector, and could be adapted to in-ice experiments~\cite{IceCube-Gen2:2020qha, RNO-G:2020rmc}. 

Outside polar areas, anthropogenic noise is much higher: in \ac{AERA}, an average 15\,kHz trigger rate was measured on antennas running in self-trigger mode~\cite{Huege:2019ufo}. Advanced trigger methods were investigated within this prospective experiment, (see e.g., Ref.~\cite{Schmidt:2011vue}), and self-triggered radio events were identified as air showers using information from the Auger Surface Detector~\cite{PierreAuger:2012gwg}.
Efforts towards self-triggering were eventually halted given the adverse background conditions observed at the \ac{AERA} site and the easy availability of an external trigger provided by the other Auger detectors. 

The \ac{TREND} experiment was a fully autonomous array of 50 antennas deployed in a remote valley of the TianShan mountains in China.
Dedicated (offline) selection algorithms were developed, based on distinct characteristics of air showers (e.g., brief pulses, limited curvature of the wavefront) and background pulses (e.g., clustering in time or direction). 564 air shower candidates were selected out of the $7\cdot10^8$ events recorded in 314 live-days, with an estimated $\sim$\,80\% purity~\cite{Charrier:2018fle}. 
This positive result was mitigated by the low value ($3\,\%$ only) of \ac{TREND} air detection efficiency~\cite{Charrier:2018fle}, mostly due to detector instability. 
Nonetheless, in an earlier analysis, several \ac{TREND} radio events were found to be in coincidence with a 3-units particle detector~\cite{Ardouin:2010gz}. 

These various results show that a large set of analysis tools can be developed (online or offline) to allow for an efficient identification of air showers --- even though further developments are needed to optimize the efficiency when keeping the purity high.
The GRANDProto300 experiment~\cite{Decoene:2019sgx,Zhang:2021Tk}, presently being deployed in a radio-quiet site in the Gobi desert, could be the next step on this path (see \cref{sec:GRAND}).


\paragraph{Future developments}
While the radio technique is ready to play a significant role in the design of future experiments, it has not yet reached its feasible limits. 
With appropriate R\&D regarding the calibration and analysis techniques, the radio method may achieve a measurement precision and absolute accuracy for the energy and for \xmax even higher than that of the leading optical methods today. 

However, even with perfect \xmax resolution, the accuracy for the mass of an individual shower is statistically limited by shower-to-shower fluctuations. 
Overcoming that limit requires the addition of further mass-sensitive parameters. 
A straight forward approach is exploiting the high mass-separation power of the muon number by combining radio and muon measurements~\cite{Holt:2019fnj} in hybrid arrays such as Auger, IceCube-Gen2, or \ac{GCOS}.

Another approach to further increase the accuracy for the primary particle can be to utilize the yet unexploited richness of features in the radio signal which contain information on the on the shower development beyond the simple position of the shower maximum. 
Methods, such as near-field imaging or the reconstruction of the width parameter $L$ of the shower profile, can be explored at the ultra-dense \ac{SKA}-low array (see \cref{sec:SKA}). 
The lessons learned with \ac{SKA} can then be transferred to other radio arrays.
One promising way to exploit the additional information contained in the radio signals also with sparser arrays is machine learning.

Consequently, employing machine-learning techniques to digital radio arrays is another promising area of future R\&D. 
Neural networks have already been trained to recognize air-shower particles against background~\cite{Shipilov:2018wph,Erdmann:2019nie,Rehman:2021oby}, which can lower the detection threshold and increase the reconstruction accuracy.
Thanks to the accurate simulation tools available for training, it is probable that machine-learning techniques can also be applied to high-level event reconstructions such as the energy and \xmax, which will further increase the impact of the radio technique on the field of ultra-high-energy cosmic-ray physics.

In summary, the radio technique has reaches sufficient maturity to play a major role in the next generation of ground-based air-shower arrays.
With further R\&D applied, the accuracy of the radio technique for the energy and mass of the primary particles is likely to surpass today's accuracy of other state-of-the-art detection techniques.

\subsubsection[Space based detectors]{Space based detectors: the final frontier}
\label{sec:SBD_tech_development}
The detection of \ac{UHECR} from space poses several technical challenges, mostly related to the constraints of volume, mass and power typical of space-borne detectors, with the  compounded requirements of large optics and focal surface need to observe the UV emissions of atmospheric showers. The large optics is needed  to lower the threshold of \ac{UHECR} shower reception to $\simeq$\,$10^{19} eV $ and overlap the measured \ac{UHECR} spectrum from space with those taken with  ground based observatories. Furthermore, the low particle flux at these energies requires a high field of view (f.o.v.) and a large focal surface to gather enough events for a meaningful statistics. In addition, the signal/noise ratio in a given pixel decreases with the size of the area it is observing (since more background photons hit the same pixel), thus requiring  a highly-pixelated focal surface. The high readout speed ($\simeq$\,$\mu s$) associated to the shower development in the atmosphere completes the bill of requirements for a space-borne detector. Next generation detectors and associated electronics capable of faster readout speeds ($\simeq$\,$10$\,ns) for the observation of direct Cherenkov light are also being developed.

In the last decades, development of the technologies and production techniques has   
resulted in the convergence on specific systems capable of meeting these requirements, offering ample margin of improvement for future missions. Most of these systems have flown on space-born detectors (Tatiana \cite{Sadovnichy2011}, TUS \cite{Klimov:2017lwx}, Mini-EUSO \cite{Bacholle:2020emk}), or on balloon-borne detectors (EUSO-Balloon \cite{Adams:2015pec,Abdellaoui:2019qmg,Adams:2022oko}, EUSO-SPB1 \cite{Bacholle:2017dye}, EUSO-SPB2 \cite{Kungel:2021anx}, launch foreseen in 2023) and are thus in various stages of Technical Readiness Level. 

\paragraph{Optics}

\begin{figure}[ht]
\centering
\includegraphics[width=0.5\textwidth]{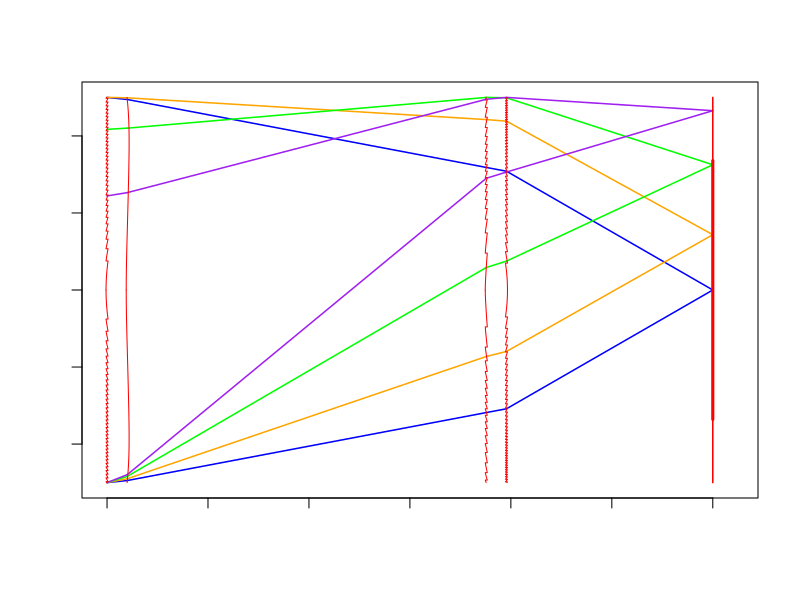}
\includegraphics[width=0.4\textwidth]{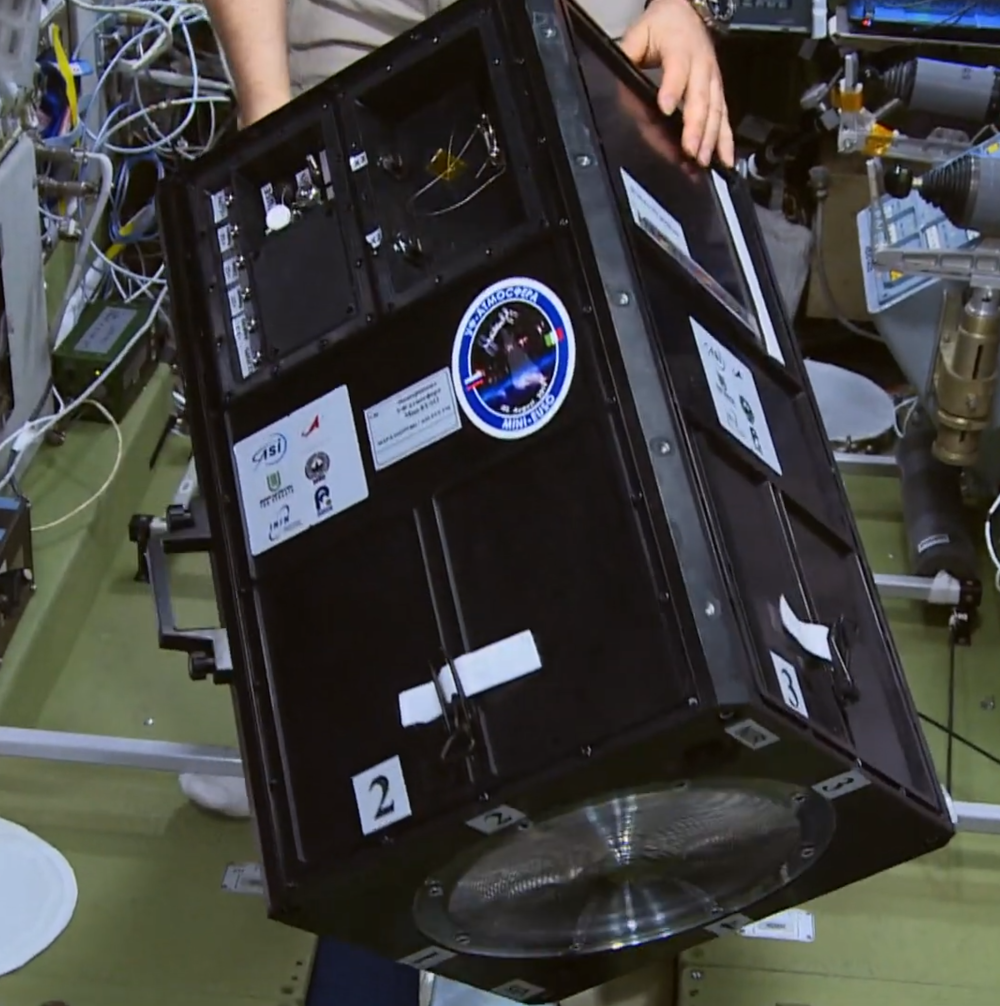}
\caption{Left: Mini-EUSO optical system design \cite{Bacholle:2020emk}. The light enters from left of the picture, crosses the two  lenses  and reaches the focal surface  to the right (focal length 30 cm). The  lines show the different paths followed by light impinging  on the detector with different angles of incidence and being focused on different points of the focal surface: $30^o$ (purple), $20^o$ (green), $10^o$ (yellow), $0^o$ (blue). Right: Picture of the Mini-EUSO telescope on board the \ac{ISS}, prior to installation on the UV transparent window of the Zvezda module. The front lens is in the bottom of the picture.}
\label{MiniEUSO_optics}       
\end{figure}

\begin{figure}[ht]
\centering
\includegraphics[width=0.9\textwidth]{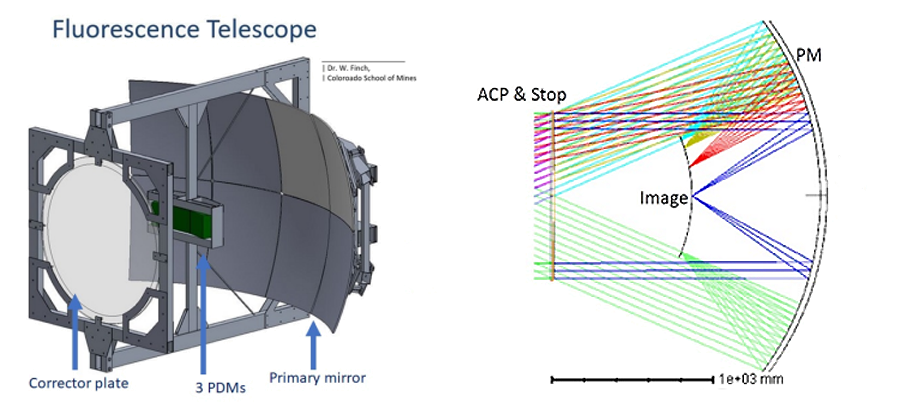}
\caption{Left: EUSO-SPB2 fluorescence telescope design (model by W. Finch). Right: optics  raytracing model (by P. J. Reardon).
The Primary Mirror   is on the right of the pictures, the Achromatic Corrector Plate  on the left, and the focal surface (3~\acp{PDM}) at the centre. Note the curvature of the Schmidt-design focal surface. Adapted from Ref.~\cite{Kungel:2021anx}.}
\label{SPB2_optics}       
\end{figure}

Optics design for \ac{UHECR} detection fall in two broad categories: lens (refractive) and mirror (reflective). 
Refractive optics  usually employ Fresnel lenses, which allow a lower mass and higher robustness to launch vibrations. The presence of the various transition surfaces (four in two-lenses systems) and of the grooves of the Fresnel structure result in a lower efficiency than reflective systems. Furthermore, the refraction of photons suffers from  wavelength dispersion, requiring a diffractive lens to compensate this effect (since the two phenomena have opposite frequency dependence). The refractive design  has the advantage of better protecting the focal surface from the harsh environment of space (atomic oxygen, low energy electrons...) and of usually being more easily deployable in space. 
\cref{MiniEUSO_optics} shows  an example of refractive optics used in the Mini-EUSO detector: it consists of two, 25~cm diameter, Fresnel lenses with a wide field of view (44$^{\circ}$ on the focal surface). Poly(methyl methacrylate) - PMMA - is used to manufacture the lenses with a diamond bit machine. In this way it is possible to have a  light (11~mm thickness, 0.87~kg/lens), robust and compact design well suited for space applications.  The effective focal length of the system is 300~mm, with a Point Spread Function  of 1.2~pixels, of the same dimension as the pixel size of the \acp{MAPMT}.

Reflective optics telescopes have the advantage of providing  a high efficiency and of being wavelength  independent. To increase the field of view, a Schmidt optics may be employed, in which case a refractive corrector plate is employed. Disadvantages of these systems are the higher positioning requirements, the occultation by the focal surface  and its higher exposure to space and the resulting day-night thermal  fluctuations.  In \cref{SPB2_optics} is shown the design of the fluorescence detector of the EUSO-SPB2.  A corrector lens is located in front of the detector and acts as an entrance pupil. Light is then reflected by the mirror, composed of six  segments with an overall  field of view of $36^{\circ} \times 12^{\circ}$ to the focal surface, composed of three \acp{PDM} (6912 pixels). 

\paragraph{Focal Surface detectors}

\begin{wrapfigure}{r}{0.48\columnwidth}
\vspace{-10mm}
\centering 
\includegraphics[width=0.48\textwidth]{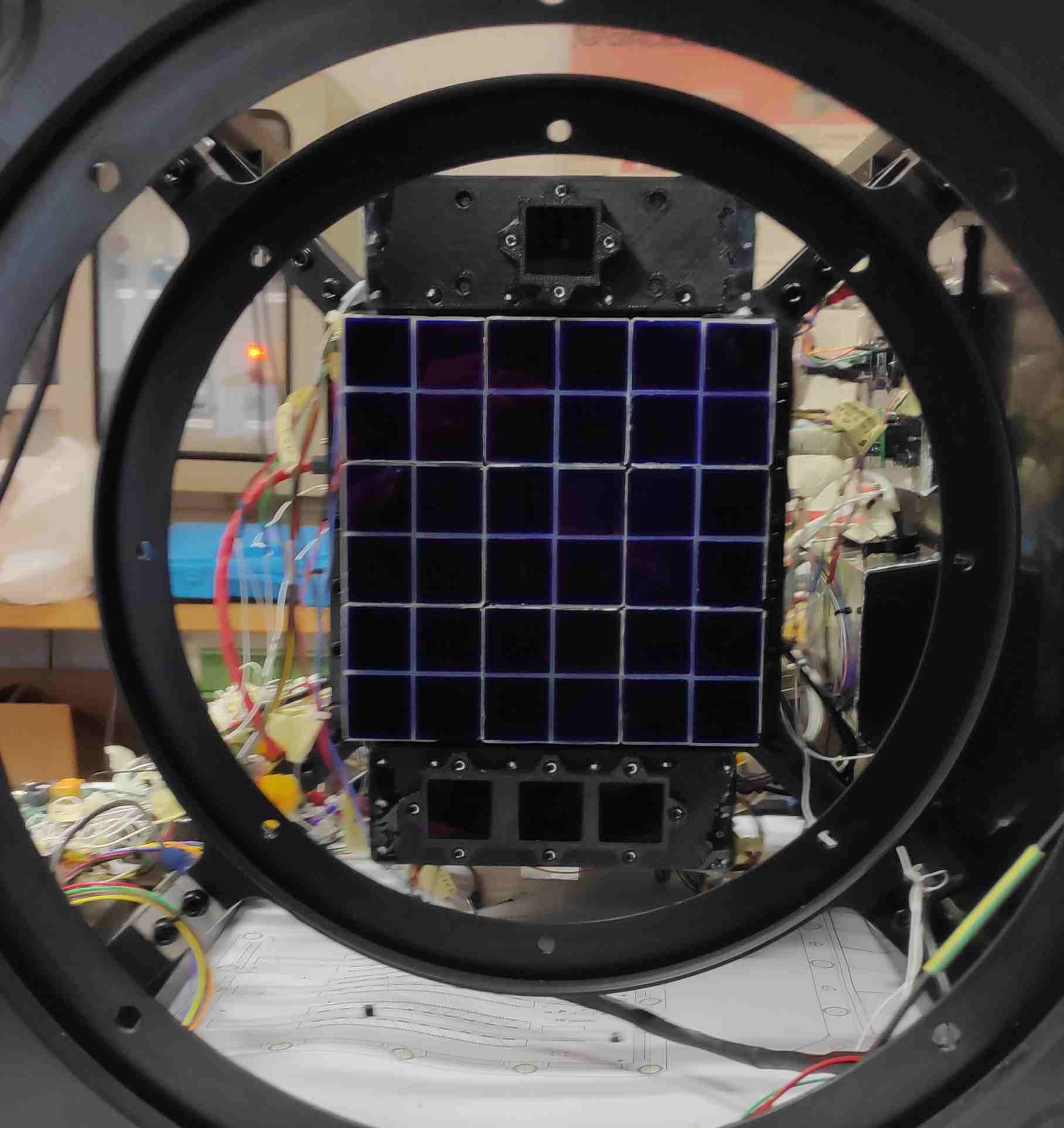}
\caption{Mini-EUSO focal surface. The Photo Detector Module (\ac{PDM}) is composed by 36 \acp{MAPMT}, each with 64 independent channels (2304 total pixels) and arranged in groups of four (an Elementary Cell, EC). On top of the \ac{PDM} is a 64 channel \ac{SiPM}, at the bottom of the \ac{PDM} are two light sensors and a single-pixel \ac{SiPM}.}
\label{fig_FS}     
\vspace{-5mm} 
\end{wrapfigure}
Multi-anode and silicon photomultipliers are currently the most promising candidates as Focal Surface detectors. 

\subparagraph{Multi-Anode Photomultiplier (MAPMT)} technology has been successfully tested both on balloons and in space (TUS, Mini-EUSO), making them a reliable and scalable detector. These detectors are robust, resistant to launch vibration, radiation and temperature changes. The main disadvantage is the high ($\simeq$\,$1000 ~V$) voltage needed and the higher mass of these detectors. 

\acp{MAPMT} focal surfaces (\acp{PDM}) have been extensively employed in several detectors: in the ground telescope of EUSO-TA \cite{Abdellaoui:2018rkw}, in the first two balloon flights, EUSO-Balloon \cite{Adams:2015pec,Abdellaoui:2019qmg,Adams:2022oko}  and  EUSO-SPB1 \cite{Bacholle:2017dye}, and in Mini-EUSO \cite{Bacholle:2020emk}.
A more complex setup, involving three \acp{PDM} side-by-side, will be used in the upcoming EUSO-SPB2 flight \cite{Adams:2017fjh}.  Each \ac{PDM} consists of a matrix of 36 \acp{MAPMT} (Hamamatsu Photonics  R11265-M64), arranged in an array of 6$\times$6 elements. Each \ac{MAPMT} consists of 8$\times$8 pixels, resulting in a total of 2304 channels (\cref{fig_FS}). They are powered by a low-power consumption Cockroft-Walton high voltage power supply (HVPS) and read-out by  Spaciroc-3 ASIC \cite{Blin:2018tjp}.  Each Spaciroc-3 handles in parallel 64 independent channels and thus preamplifies and digitizes the photoelectron signals coming  from a single \ac{MAPMT}. The \acp{MAPMT} are operated in photon counting mode  to minimize the contribution of the integrated noise, with readout occurring on the order of $\mu s $ time scale. 
\ac{MAPMT} technology  is thus mature enough to be scaled in number and  employed  in a  large focal surface for detection of atmospheric showers of UHECR such as that of a POEMMA-like mission.

\subparagraph{\Acfp{SiPM}} have been rapidly establishing as a reliable technology in a number of fields, ranging from high energy physics to medical applications. 

In space, they have been first flown on the inside of the \ac{ISS} in 2005 as part of the LAZIO-SIRAD mission \cite{Casolino:2021pss} as scintillator readout. A 8x8 multipixel \ac{SiPM} (Hamamatsu) is also being flown in the Mini-EUSO detector (it is visible in Figure \cref{fig_FS} on top of the \ac{MAPMT} Focal Surface). A wide-area \ac{SiPM} focal surface of 512 pixels (Hamamatsu S14521-6050AN-04) is employed as the Cherenkov detector camera of the SPB2 payload \cite{Otte:2019lbq}. In this case the  field of view of the Schmidt bifocal optics is $12.8^{\circ}$$\times$$6.4^{\circ}$\footnote{The bifocal design results in a double image, with spots offset by 12~mm to detect Cherenkov light but reject single hits coming from direct hits of cosmic rays.}.    

The limiting  factor in the adoption of \ac{SiPM} on a wide focal surface in space is mostly related to the high temperature dependence of the gain and their sensitivity to ionizing radiation. The former effect can be offset (up to a limit) by voltage-dependent temperature compensation and the latter by shielding the focal surface as much as possible (but this makes the design of reflective optics more challenging). Overall, the  long-term durability of \ac{SiPM} in open space needs still further studies. Various efforts to raise the TRL of these systems are currently taking place all over the world. This development is temporally consistent with the employment as Cherenkov detectors in a \ac{POEMMA}-like mission. 

Overall, the recent development of detectors and their adoption in various balloon- and space-borne experiments have advanced the electronics and optics technology  to the point that a large detector in space (either on a free-flyer or on a space station) is a concrete possibility.

\subsection[The computational frontier]{The computational frontier - Harder, Better, Faster, Smarter}

In forthcoming years, it is foreseen that the use of machine learning in a wide range of applications 
will be fully established and consolidated. However, work will be required to include these new techniques in standard codes and to achieve good performance at large scales and in \ac{HPC} centers with heterogeneous resources and architectures.
Currently, the \ac{CVMFS}~\cite{cvmfs} is being used for software distribution using container technology. 
Adapting machine learning techniques to this environment is a forthcoming challenge that may require new tools.
As deep learning relies on the availability of \acs{GPU} resources that are not widely available in most university computing clusters, the pay-per-use of commercial cloud computing centers are a possible solution for the near future. The utility of cloud resources in astroparticle physics has  already been explored by the IceCube collaboration for simulation to analyze the performance, usability, and running costs~\cite{cloud_1_2020,cloud_2_2020,cloud_2021}.

Over the next decade, data complexity will undoubtedly continue to increase, and several aspects need to be considered, namely:
\vspace{-3mm}
\subparagraph{$\bullet$ Portability and compatibility}
It is vital to ensure that key reconstruction software and Monte Carlo codes can fully retain functionality in the face of the swift evolution 
of the adopted operating systems, inherent software, and respective compilers.
Moreover, Monte Carlo simulations are typically done in a chain of several programs using different programming languages. 
In particular, the most relevant codes for the simulation of \acp{EAS}, which were developed and extensively
refined over several decades, were written in FORTRAN and have a very rigid structure that is becoming harder to adapt to the new needs and interface with new software. 
Furthermore,  there are fewer people with a deep knowledge of these codes, making the path forward uncertain.  
In this sense, the CORSIKA~8 project~\cite{c8} aims at providing a modern framework and more realistic simulations in  C++ and Python, replacing the previous FORTRAN code. 
Backward compatibility is another ongoing challenge. One example is the transition from Python 2 to Python 3, which may have led to significant changes in some programs and introduced a maintenance burden to prevent code obsolescence.
\vspace{-3mm}
\subparagraph{$\bullet$ Code modularity}
Best efforts have been devoted to the development of modular codes in which new features
can be easily added.
Nonetheless, given the extent and complexity of the current codes, new users are 
experiencing more difficulties getting acquainted with the whole data processing workflow. 
Effort is required to provide a transparent framework connecting all stages of data reconstruction 
and production of simulations. 
Pipeline frameworks should be envisaged for all standard types of data reconstruction.  
\vspace{-3mm}
\subparagraph{$\bullet$ Data management, distribution, and integrity}
Data management and distribution must be optimized for future experiments while securing data integrity.
Data volumes and complexity will continue to increase.
From the user's perspective, all available data and metadata should be organized into databases, allowing users to straightforwardly locate data of interest and automatically transfer it from its physical storage element. The massive data transfer
should be made more reliable for the administrators, ensuring data preservation, since currently there are high
chances of data loss or corruption. A desirable feature would be the direct communication between the
distribution systems \ac{NFS}, Lustre, dCache, or a system similar to the \ac{iRODS} or other data management software, from which users prefer to download their data.
On the latter, the issue of latency has to be addressed, particularly for accessing data recorded on tape.
\vspace{-3mm}
\subparagraph{$\bullet$ Bursts of heavy data processing}
The number of Multi-Messenger Alerts will increase with the first light of several new detectors.
Experiments must endow their computing resources for frequent bursts of heavy data processing.
\vspace{-3mm}
\subparagraph{$\bullet$ Quantum Computing}
Quantum computing is no longer a distant reality, and there are plans for both IceCube and IceCube-Gen2 to potentially use it to calculate neutrino transport, interactions, and event generation~\cite{Arguelles:2019phs, Bauer:2019qxa, Wei:2019rqy}.

\subsubsection{Machine learning in the future}
As discussed earlier in \cref{sec:10yrComputation}, recent pioneering work in applying machine learning methods to astroparticle physics challenges has been accomplished, revealing the enormous potential of the new technology.
These initial approaches, however, do not nearly represent the full spectrum of possible applications. In the next decade, machine learning will spread into many more areas of cosmic-ray research.

It further has to be emphasized that not one and the same machine learning algorithm can be applied to all tasks and challenges. Instead, the new technology offers various new tools and methods to be designed and adapted to the respective application.
To ensure applicability for different data types and symmetries, network architectures beyond \acp{CNN} and \acp{RNN} will become established~\cite{Benato:2021olt}. 
Possible candidates include \acp{GNN}~\cite{geometric_dl} and transformer networks~\cite{vaswani2017attention} as they are very flexible. In addition, architectures will be extended to predict reconstruction uncertainties at the event level.

\subparagraph{Improved reconstructions and sensor-close applications}
Given the success of the studies performed so far, it is clear that further progress will be made in the field of event reconstruction, signal de-noising, and unfolding. 
Incorporation of data from recent upgrades into machine learning models will further
improve results.
For example, the AugerPrime upgrade includes both new detector components and enhancements to the existing electronics, which improves the sampling rate by a factor of three.
Especially promising is the potential for precise reconstruction of the cosmic-ray mass at the event level using the radio and surface scintillator detectors.
In addition to leading to more precise composition measurements, event-by-event composition information would open up entirely new prospects in the field of anisotropy studies and source identification.

Another important step is the development of machine learning algorithms close to the sensors. Some very first steps in the context of air-shower signal detection and de-noising with radio antennas were already made. In the next ten years, more research can offer intelligent triggering solutions for future large-scale projects like \ac{GRAND}. It is also worth noting that the \acs{EUSO} Super Pressure Balloon 2 Experiment~\cite{wiencke2019extreme} will employ a convolutional neural network to prioritize triggered events for download via the bandwidth-limited telemetry~\cite{Filippatos:2021noz}. The network is trained prior to flight using a combination of simulations and data from similar experiments.

\subparagraph{Generative models and domain adaption}
Inspired by the recent progress in computer science~\cite{goodfellow2014generative} and its application in particle physics~\cite{Paganini:2017hrr, Erdmann2018}, it is likely that generative models will support and accelerate physics simulation in the future. These changes could well go hand in hand with differentiable programming for physics simulations. 
Since approaches to domain adaptation are intertwined with developments in unsupervised learning and generative models, further advancement in one field will likely also advance the other.
The expected progress in these areas would have the potential to facilitate domain-robust algorithms and techniques that help to locate and reduce discrepancies in simulations and measured data.

To meet the future challenges of large-scale experiments, it will be more and more critical to supplement physical analysis workflows with machine learning. Therefore, it is crucial to find synergies between machine learning and physics in the near future, to establish a solid foundation for data-driven and data-intensive applications in the long run.
Machine learning will be a key to finding structures that can not be resolved with conventional methods in the ever more comprehensive, detailed, and multidimensional event data. 
The adaption and development of new algorithms tailored to the particular application, the data structure, and the scientific requirements will be key to deepening our understanding of \acp{UHECR} and our cosmic environment.

\subsubsection{Computational infrastructure recommendations}
\label{sec:ComputationFutureInfrastructure}

\subparagraph{Accelerator clusters}
\ac{UHECR} physics is facing technical challenges in connection to Big Data. The upcoming experiment detector upgrades give a first impression of this. Similar developments are taking place in particle physics with the significant increase in the data rate with the High Luminosity \ac{LHC} upgrade.
There are strong synergies between \ac{HEP} and \ac{UHECR} communities.  
The computing infrastructure, for instance, is similar for the two communities.
Multi-processing and the implementation of accelerator clusters are critical for the future, especially with regard to machine learning.
While the evaluation of machine learning models can in principle be performed with \acp{CPU}, the training of algorithms requires accelerators (currently \acp{GPU}) and the evaluation is greatly aided by accelerators (both \acp{GPU} and \acp{FPGA}). 
Estimating the requirements for training neural networks, it is to be expected that for competitive research, each scientist must be granted access to a cluster that provides several \acp{GPU}. Nowadays, this is not the case since almost no computing center features (sufficient) \acp{GPU}. It should be noted that the development of data-driven algorithms requires multiple \acp{GPU} per scientist.
In addition, the need for a vast amount of Monte Carlo simulations to train machine learning models will require computational resources. To ensure optimum utilization of resources, fast storage solutions like SSDs or computational storage devices and intelligent caching will be required.

In addition to new computing infrastructures and the preparation and preservation of data in public data centers, analyses of the data with data-driven methods are essential. Since instruction is such methods is still not common in most physics curricula, investments in education and the organization of dedicated schools workshops and conferences are indispensable. Ideally, these programs should become available to students from undergraduate to doctoral levels.
Moreover, greater recognition of software contributions will be essential to ensure quality, sustainable software. 
It is notable that, relatively recently, dedicated projects have begun 
devoted to computation research, development, and education, such as the NSF-supported Institute for Research and Innovation in Software for High Energy Physics (IRIS-HEP)~\cite{iris-hep}.

\subparagraph{Quantum computing}
The possibility to use more and faster classical computing power has been the core resource in many advancements in experimental high-energy physics.
Quantum computing technologies promise to revolutionize these computational approaches.
In the near future, one of the main tasks will be to investigate if the most difficult and prohibitive computing problems in astroparticle physics can utilize quantum computers rather than traditional integrated circuits.
The first applications of these new resources are already used in high energy simulation~\cite{Bauer:2019qxa} and reconstruction~\cite{Wei:2019rqy}.
Possible opportunities in astroparticle physics include challenges where the represented phase-space can be larger than a classical system can handle in reasonable times, as in systems of a large number of coupled differential equations.
The applications in detector simulations are also promising. For example, photon propagation from the source through a medium to the detector can be modeled with a classical path integral~\cite{collin2018pathintegrals}. The challenge imposed by such large-phase spaces could be elegantly addressed using quantum computing techniques.
Quantum computing can also be applied during the reconstruction, classification, and event selection by using quantum machine learning techniques~\cite{quantum_ml_2014,quantum_ml_2017} that may yield great enhancements for future (complex multi-component) \ac{UHECR} observatories.


\subsection{UHECR science: The next generation}

For the next generation experiments, an essential requirement is a significant increase of the exposure because the current generation of experiments is limited by statistics at the highest energies.
Moreover, at least one of the next generation experiments needs to feature a high accuracy for the mass and, thus, rigidity of the primary particles because it is unknown whether the \ac{UHECR} mass composition is pure or mixed at the highest energies, e.g., by containing a small fraction of protons next to heavier nuclei. The proposed experiment in this category is \ac{GCOS}.
Such an experiment with high accuracy for the rigidity needs to be combined by experiments maximizing the total exposure for \ac{UHECR}. 
Exploiting the synergies given by the multi-messenger approach, some experiments planned primarily for neutrino detection at the same time can deliver a huge exposure for cosmic rays at the highest energies.
The proposed experiments in this category are \ac{POEMMA} in space and \ac{GRAND} on ground, which are the ideal complements to \ac{GCOS} to cover the \ac{UHECR} science case of the next decade.

\subsubsection{POEMMA -- highest exposure enabled from space}
\label{sec:POEMMA}
\ac{UHECR} measurements from space-based experiments vast atmospheric volumes that contain the \ac{UHECR} and \ac{UHE} neutrino \ac{EAS} development, which are viewed using wide \ac{FoV}, large optical systems to image the \ac{EAS} air fluorescence signal developments. This results in large effective geometry factors, even assuming a conservative 10\% duty cycle for the observations. For example, a telescope with a full \ac{FoV}$=45^\circ$ from a 525\,km orbit translates to viewing a large, nearly $10^6$ km$^2$ area of the ground. This leads to unique sensitive to Earth-emergent tau neutrinos by observing the optical Cherenkov signal from the upward-moving \ac{EAS}. The different nature of the signals and their development, $300~{\rm nm} \lsim \lambda \lsim 500~{\rm nm}$ and 10's of $\mu s$ timescales for air fluorescence (AF) and $300~{\rm nm} \lsim \lambda \lsim 1000~{\rm nm}$ and 10's of $ns$ timescales for optical Cherenkov (OC), drive the design of the photo-detection instrumentation used in such an experiment.

\paragraph{Design and Timeline}
\label{sec:POEMMA_DesignAndTimeline}

\begin{figure}[!htb]
\begin{center}
    \includegraphics[width=0.99\textwidth]{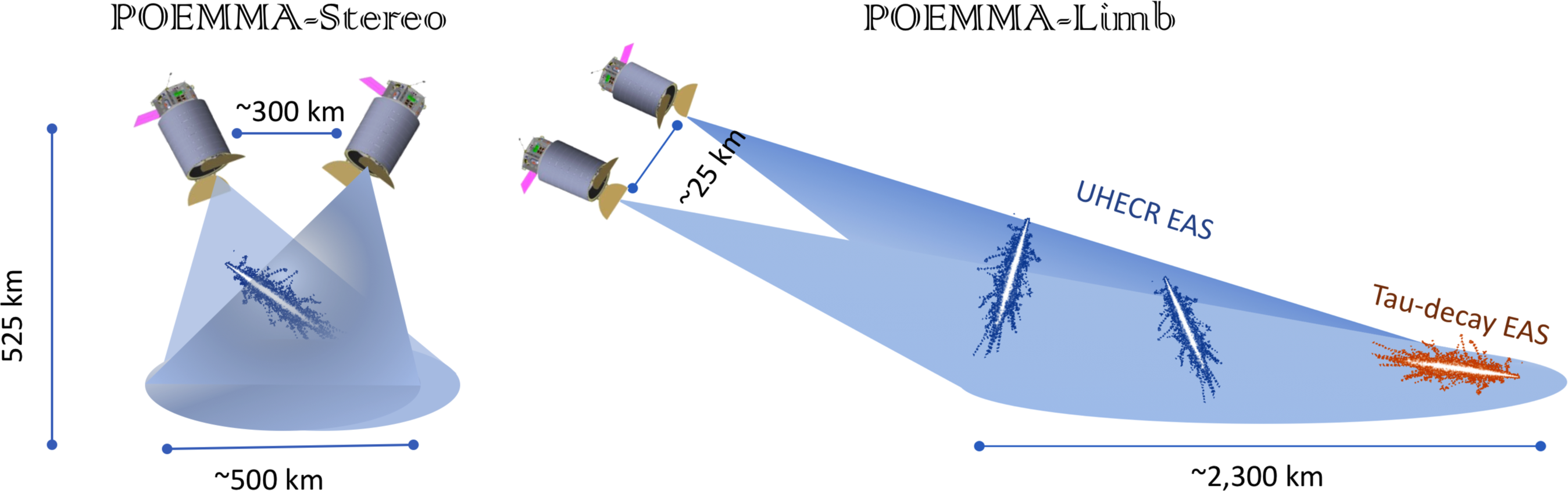}
 \end{center}
\vspace{-0.3cm}
  \caption{The \ac{POEMMA} science modes. Left: \ac{POEMMA}-Stereo where the spacecraft are separated and viewing a common atmospheric volume to measure the fluorescence emission from \ac{EAS}. Right: \ac{POEMMA}-Limb where the instruments are tilted to view near and below the limb of the Earth for optical Cherenkov light from upward-moving \ac{EAS} induced by tau neutrino events in the Earth. From Ref. \cite{POEMMA:2020ykm}
  }\label{POEMMAmodes}
\end{figure}

\begin{figure}[!htb]
\begin{center}
    \includegraphics[width=0.44\textwidth]{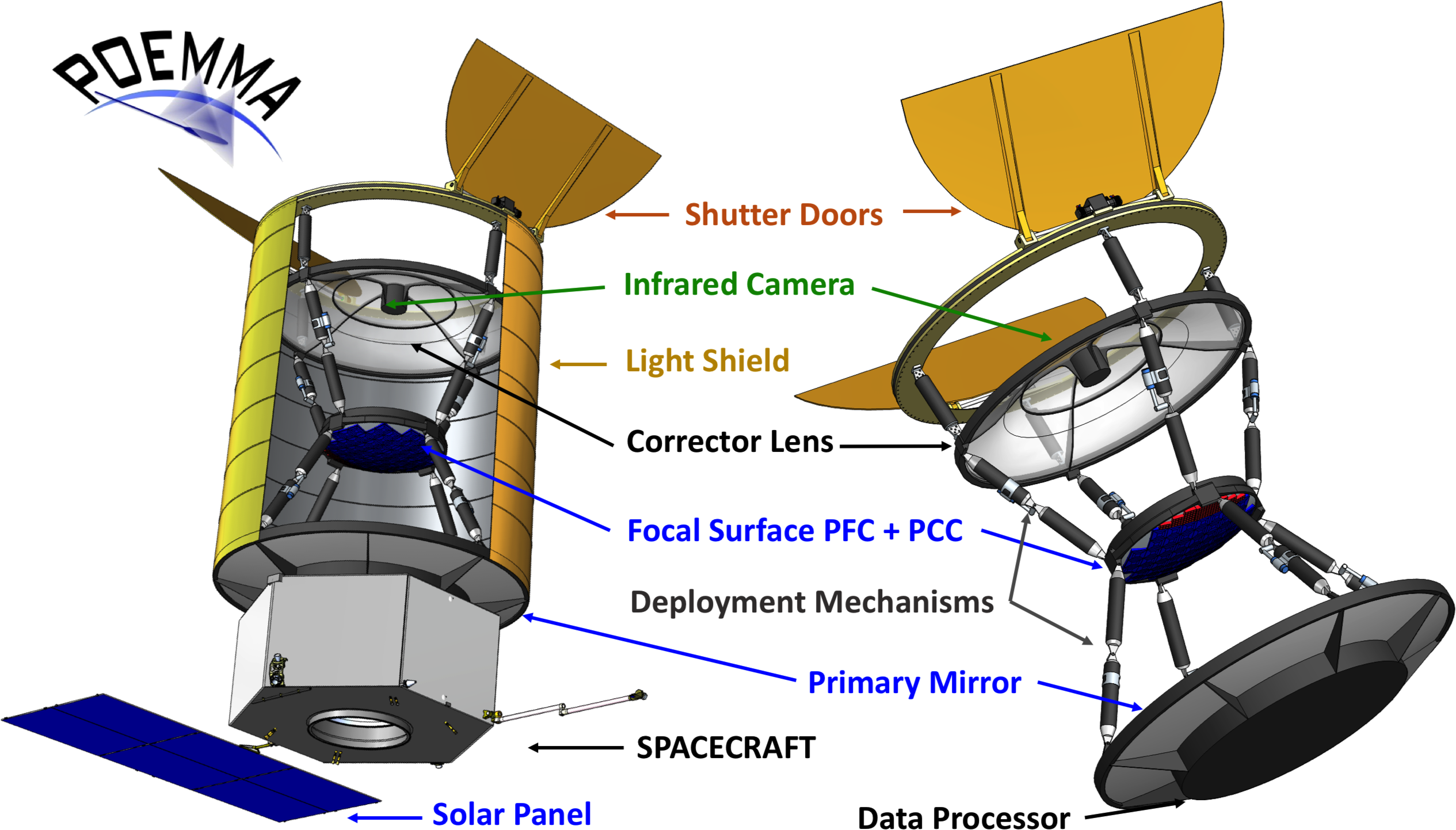}
    \includegraphics[width=0.54\textwidth]{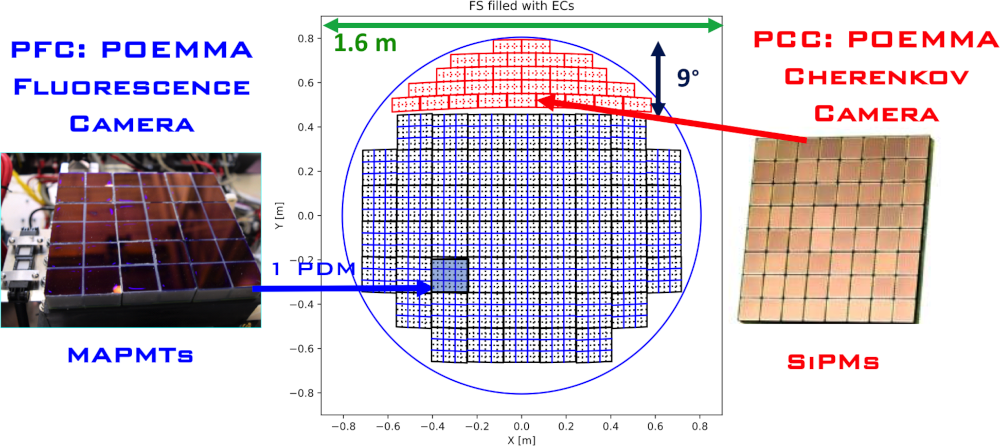}
 \end{center}
\vspace{-0.3cm}
  \caption{Left: Schematics of a \ac{POEMMA} satellite  and the Schmidt telescope consisting of a 4-m diameter primary mirror, 3.3-m diameter corrector plate, and 1.6-m diameter focal surface comprised of 126,720 pixels in the \ac{PFC} and 15,360 pixels in the \ac{PCC}  Several components are detailed in the schematic including infrared cameras which will measure cloud cover within the 45$^\circ$ full \ac{FoV} of each telescope during science observations.
  Right: The layout of the hybrid focal plane of a POEMMA Schmidt telescope. The majority of the area is comprised of \ac{PFC} MAPMT modules with a UV filter to record the 300$-$500 nm air fluorescence light in 1\,$\mu$s snapshots. The \ac{PCC} is comprised of \ac{SiPM} pixels whose 300$-$1000 nm wavelength response is well-matched to that from the \ac{EAS} optical Cherenkov signals and are recorded with 10\,ns cadence. From Ref.~\cite{POEMMA:2020ykm} }
\vspace{-3 mm}
  \label{POEMMAfp}
\end{figure}

Designed as a NASA Astrophysics Probe-class mission, the \acf{POEMMA} observatory is currently the most capable space-based experiment proposed to identify the sources of \acp{UHECR} and to observe cosmic neutrinos both with full-sky coverage. \ac{POEMMA} consists of two spacecraft that co-view EAS while flying in a loose formation, separated by 300\,km, at 525 km altitudes at 28.5$^\circ$ inclination. \cref{POEMMAmodes} illustrates the two science modes of POEMMA: POEMMA-Stereo optimized for UHECR AF stereo observations and POEMMA-Limb optimized for $\nu_\tau$-induced OC detection. Each spacecraft hosts a large Schmidt telescope with a \ac{FoV} of $45^\circ$ and with a novel focal plane optimized to observe both the isotropic near-UV fluorescence signal generated by \ac{EAS} from UHECRs and \ac{UHE} neutrinos and forward beamed, optical Cherenkov signals from \ac{EAS}. A \ac{POEMMA} focal plane is shown in Fig.~\ref{POEMMAfp}. The iFoV (or pixel \ac{FoV}) of $0.084^\circ$ yields high accuracy of the \ac{EAS} reconstruction from the stereo fluorescence technique and large field-of-view from LEO. For POEMMA, this leads to the ability to accurately reconstruct the development of the \ac{EAS} with $\lsim 20^\circ$ angular resolution, $\lsim 20\%$ energy resolution, $\lsim 30$\,\gcm \xmax resolution \cite{Anchordoqui:2019omw}. This performance yields excellent sensitivity for all neutrino flavors for \ac{UHE} \ac{EAS} that begin deeper in the atmosphere and well separated from the dominant UHECR flux. 
The high-statistics ($\ge 1,400$ UHECR events in a five-year mission) full-sky UHECR measurements above 20\,EeV using the stereo air fluorescence technique would provide a major advance in discovering the sources of UHECRs \cite{POEMMA:2020ykm,Anchordoqui:2019omw}. POEMMA also provides unique sensitivity to \ac{UHE} cosmic neutrino searches using stereo air fluorescence measurements, and an Earth limb-pointed mode to observe \ac{VHE} Earth-interacting cosmic tau neutrinos using the beamed optical Cherenkov light generation from \ac{EAS} for $E_\nu \gsim$ 20\,PeV \cite{Reno:2019jtr, Venters:2019xwi}. \cref{POEMMAmodes} illustrates the two science modes of POEMMA.

\paragraph{Scientific Capabilities}
\label{sec:POEMMA_ScientificCapabilities}

\begin{figure}[!htb]
\begin{center}
  \includegraphics[width=0.54\textwidth]{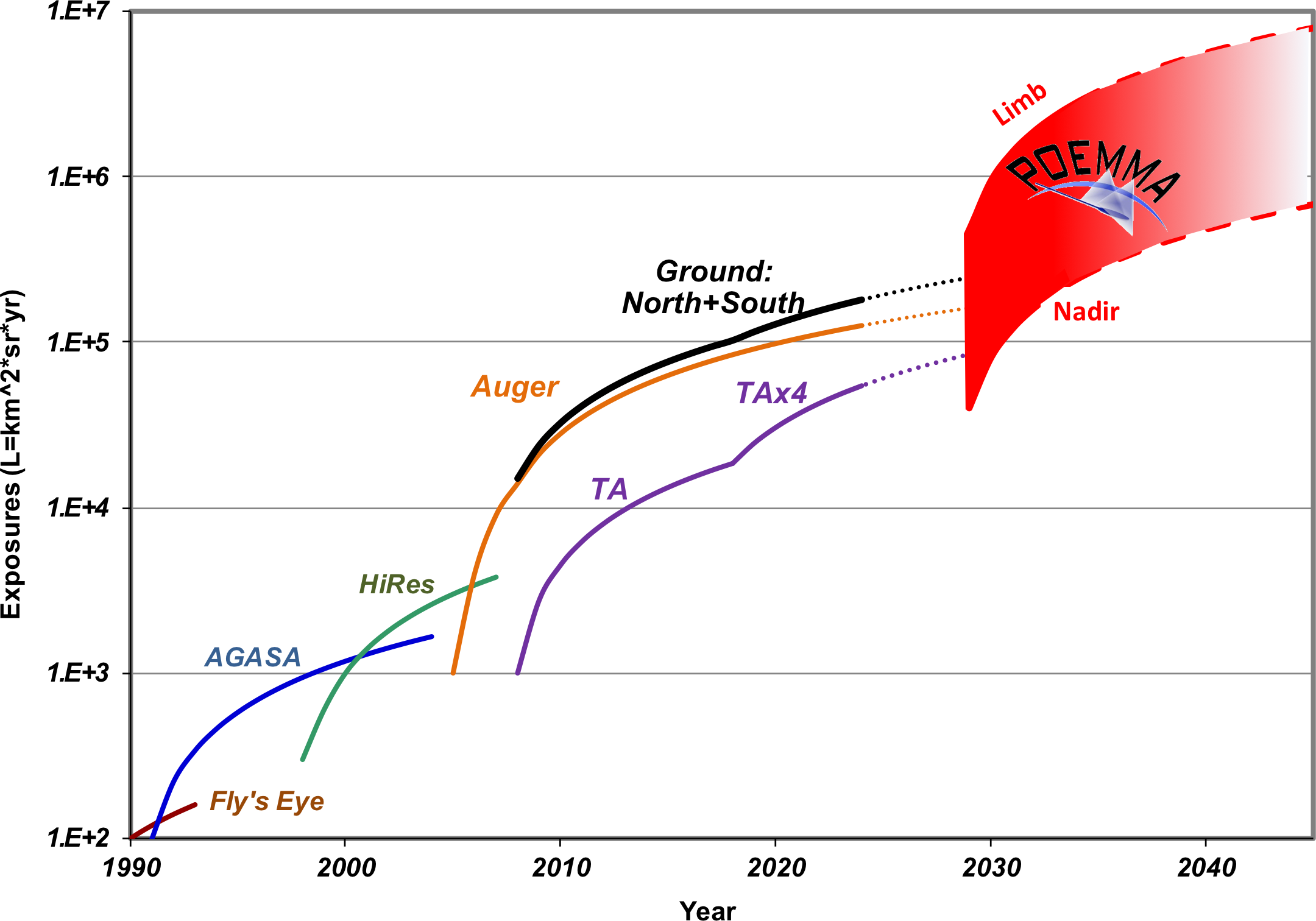}
  \hspace*{2 pt}
   \includegraphics[width=0.42\textwidth]{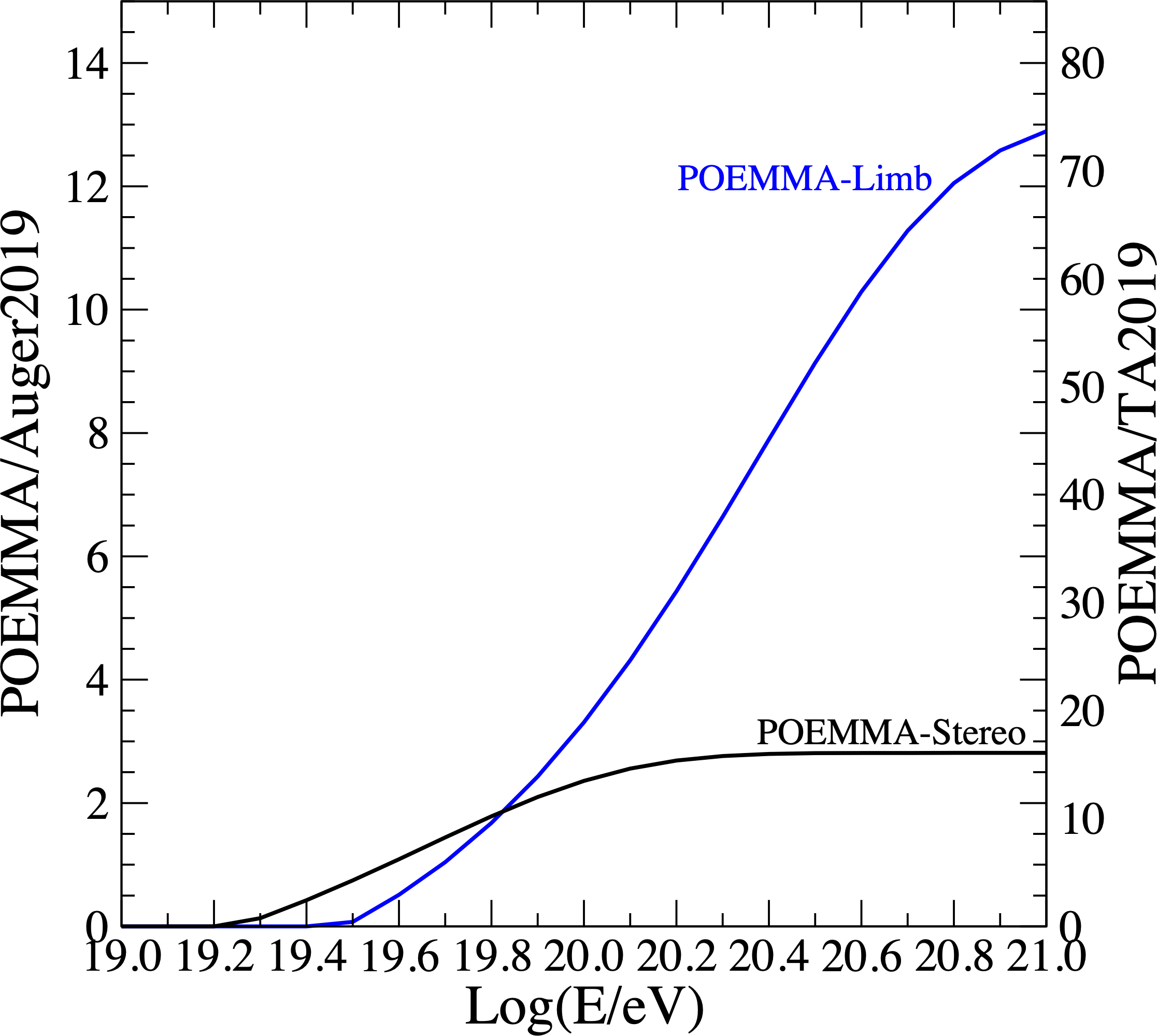}
\end{center}
    \caption{Left: The anticipated \ac{UHECR} exposure growth vs operation time for \ac{POEMMA}  compared to other \ac{UHECR} experiments. The \ac{POEMMA} band is defined by nadir-pointing stereo fluorescence measurements (lower) versus limb-pointing \ac{UHECR} measurements (upper). Right: The comparison of 5-year \ac{POEMMA} exposure versus \ac{UHECR} energy in terms of the Pierre Auger Observatory and Telescope Array exposures reported at the
    2019 ICRC. From Ref.~\cite{POEMMA:2020ykm}.}
\label{POEMMAexposure}

\begin{center}
    \includegraphics[width=0.3\textwidth]{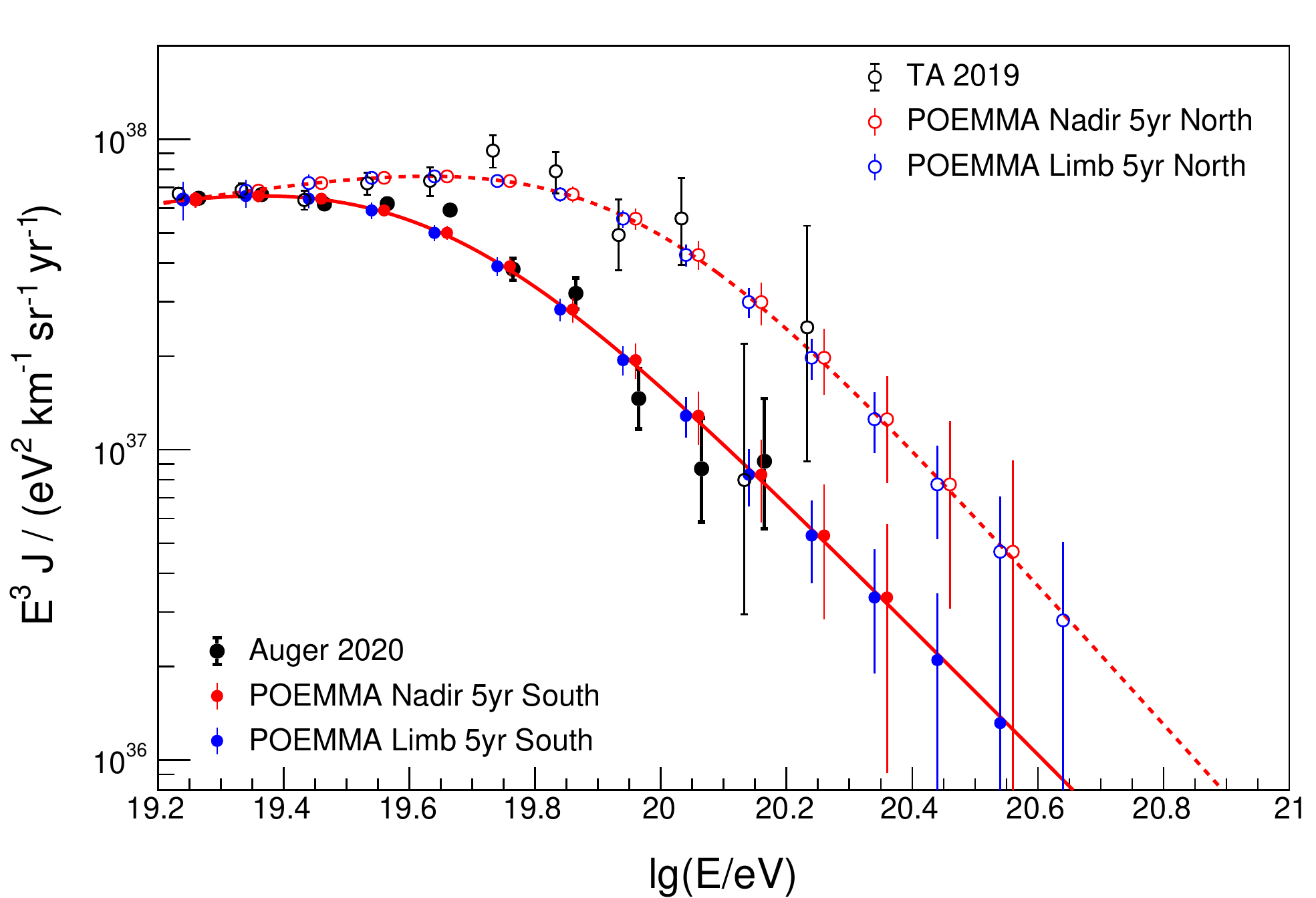}
      \hspace*{2 pt}
   \includegraphics[width=0.68\textwidth]{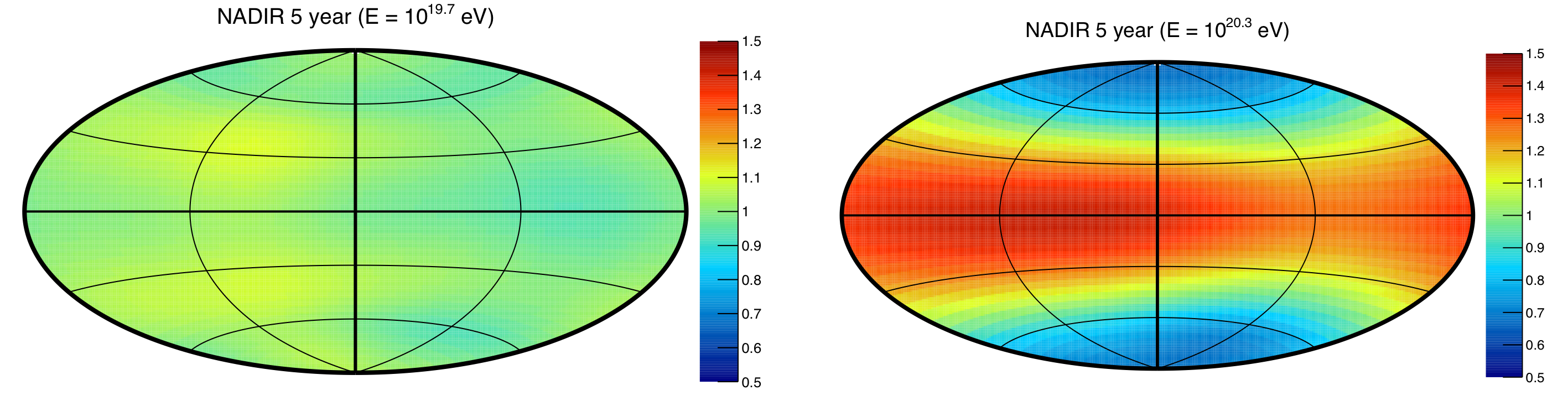}
 \end{center}
\vspace{-0.3cm}
  \caption{Left to right: The simulated \ac{POEMMA} spectra extrapolated from and compared to the Auger 2020 spectrum (black dots and
solid line) \cite{PierreAuger:2020kuy} and the extrapolation and comparison to the TA 2019 spectrum (black open circles and dotted line) from Ref.~\cite{Ivanov:2020rqn}
for the POEMMA-Stereo (red) and POEMMA-Limb (blue) observational modes, for UHECRs above 16\,EeV.  The sky exposure of POEMMA-Stereo UHECR observations in declination versus
right ascension at 50\,EeV and 200\,EeV, with the color scale denoting the exposure variations for a 5-year mission.
From Ref.~\cite{POEMMA:2020ykm}.}
\vspace{-3 mm}
  \label{POEMMAspectra}
\end{figure}

\begin{figure}[!htb]
\begin{minipage}[t]{0.54\textwidth}
\captionsetup{width=0.98\columnwidth}
\begin{center}
    \includegraphics[width=0.99\textwidth]{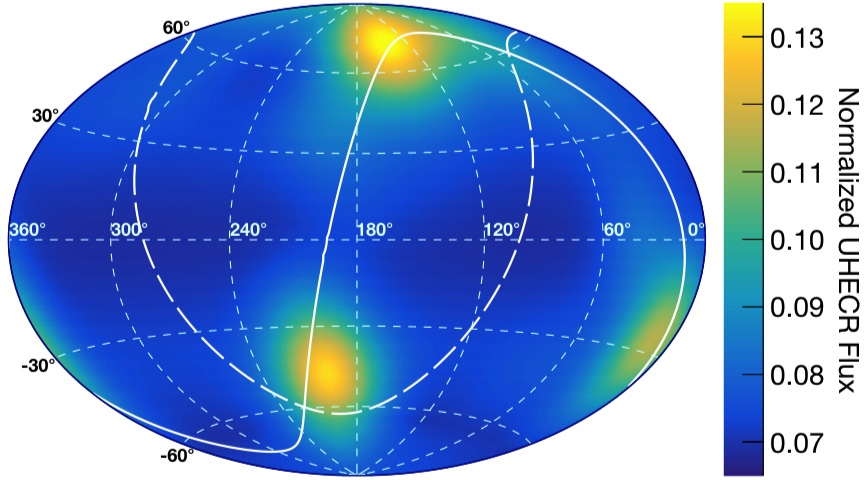}
  \caption{The equatorial coordinate sky maps of simulated POEMMA UHECR measurements for different astrophysical catalogs using the best fit parameters reported by the Auger collaboration \cite{PierreAuger:2019phh} using starburst galaxies with 11\% anisotropy fraction and 15\% angular spread of the arrival directions. From Ref.~\cite{POEMMA:2020ykm}.          \label{POMMAsbMAP}
}
\end{center}
\end{minipage}
\begin{minipage}[t]{0.44\textwidth}
\captionsetup{width=0.98\columnwidth}
\begin{center}
       \includegraphics[width=0.99\textwidth]{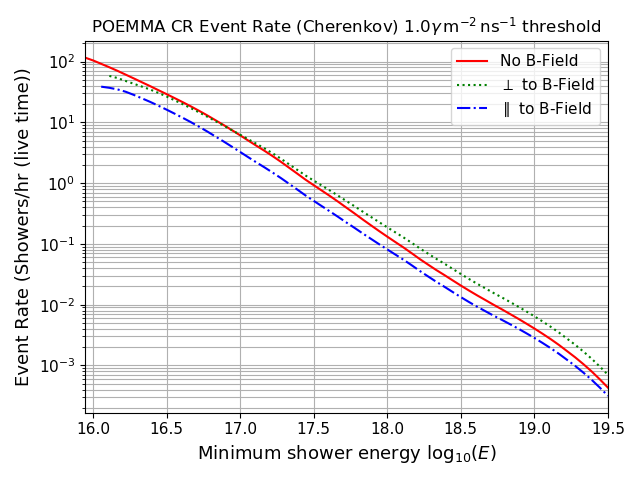}
  \caption{The simulated integral cosmic ray event rates (above a threshold $E$) for observing over-the-limb events via the optical Cherenkov signal for \ac{POEMMA}. 
  From Ref.~\cite{Cummings:2021bhg}.          \label{VHECRrate}
}
\end{center}
\end{minipage}

\end{figure}

\cref{POEMMAexposure} illustrates the gains in exposure using \ac{POEMMA} space-based UHECR measurements.  Assuming 5 years of POEMMA-Stereo operation, the total exposure is simulated to be $\sim$\,$8 \times 10^5$\,km$^2$\,sr\,yr with precision measurements of \acp{UHECR} above 40\,EeV: energy resolution of $< 20$\%, an angular
resolution of $\le 1.5^\circ$ above 40\,EeV; and a \xmax resolution of $\le 30$\,\gcm. This performance allows for the statistical identification of proton, helium, nitrogen, and iron mass groups in a mixed UHECR composition to $\lsim 20\%$, assuming minimum statistics of $\sim$\,$100$ events~\cite{Krizmanic:2013tea}. 
\cref{POMMAsbMAP} shows the source sky map obtained by the starburst galaxy hypothesis regarding the astrophysical distribution of \acp{UHECR} \cite{POEMMA:2020ykm} motivated by Auger results and correlation analysis with a similar catalog \cite{PierreAuger:2019phh}. The results demonstrate the need for full sky coverage with precision UHECR measurements for definitively identifying the astrophysical sources of \acp{UHECR}.

In POEMMA-Limb mode, \ac{POEMMA} will perform UHECR measurements with a significant gain in exposure at the highest energies, but at a cost of increased UHECR detection energy threshold and reduced precision on the \ac{EAS} measurements. Thus in the case of a recovery of the UHECR spectrum is observed above 100\,EeV, POEMMA-Limb mode yields increased exposure with good angular and energy resolution, with the capability to distinguish proton for iron primaries \cite{Anchordoqui:2019omw, POEMMA:2020ykm}, and while simultaneously searching for $\nu_\tau$-induced \ac{EAS} events using the optical Cherenkov channel \cite{Reno:2019jtr, Venters:2019xwi,POEMMA:2020ykm}. 
The POEMMA UHECR measurement performance demonstrated in in both the POEMMA-Stereo and POEMMA-Limb simulations also expands the sensitivity for \ac{UHE} neutrinos, \ac{UHE} photons, and measurement of the proton-air cross section at $\sqrt{s} = 450$\,TeV \cite{Anchordoqui:2019omw} and also provides exceptional sensitivity to the detection of \ac{SHDM} from decay or annihilation into \ac{UHE} neutrinos \cite{Guepin:2021ljb} or \ac{UHE} photons \cite{Anchordoqui:2019omw}.

For neutrino observations, the \ac{POEMMA} telescopes can easily slew in both azimuth and zenith ($90^{\circ}$ in $\sim$\,$8$ minutes), yielding unprecedented follow-up on transient astrophysical events by tracking sources as the move across the sky \cite{Venters:2019xwi,POEMMA:2020ykm}.  The separation of the \ac{POEMMA} spacecraft can also be decreased to $\sim$\,25\,km to put both telescopes in the upward-moving \ac{EAS} light pool for each event, thus reducing the neutrino detection energy threshold. The orbital period of the \ac{POEMMA} telescopes is 95 minutes, providing able to achieve full-sky coverage for both UHECR and \ac{EAS} neutrino sources.

The optical Cherenkov measurement ability of \ac{POEMMA} also extends to measuring above-the-limb \acp{VHECR} with energies for $E_{CR} \gsim 10$\,PeV \cite{Cummings:2021bhg}. \cref{VHECRrate}, presents a calculation of the \ac{VHECR} event rate versus energy threshold for POEMMA, showing rates between 30$-$100 per hour of livetime, dependent on the assumption of the orientation of the \ac{EAS} to the local geomagnetic field.  Thus, the \acp{VHECR} provide a source of optical Cherenkov \ac{EAS} signals to demonstrate the space-based measurement of these signals while searching for neutrino induced signals.  

The Astro2020 decadal review has recommended that  NASA Astrophysics Probe-class missions be implemented and how this will be done remains to be seen.  However, if POEMMA was one of the first Probe missions, the earliest POEMMA would launch would be in 2030.  In the meantime, the ESUO-SPB2 ultra-long duration balloon experiment is planned to fly in 2023 and will use similar instrumentation, including a dedicated Cherenkov telescope using \acp{SiPM} to search for the cosmic tau neutrino flux viewed below the Earth's limb while measuring the \acp{VHECR} flux using the EAS optical Cherenkov signals view from above the limb. This \acp{VHECR} measurement capability has motivated two potential future missions, the Terzina SmallSat mission that will use a Schmidt telescope with an effective area of 0.2\,m$^2$, and $8^\circ \times 2^\circ$ \ac{FoV} with an iFoV$ = 0.125^\circ$ in a sun-synchronous orbit.  Another is the Wide-Angle Telescope-Transformer (WATT) that is built on the success of the \ac{mini-EUSO} mission \cite{Bacholle:2020emk} and recent work based on K-EUSO \cite{Fenu:2021wub} using a larger \ac{mini-EUSO} type design with 40-cm diameter outer lens and a \ac{FoV} of $60^\circ$. These experiments, along with EUSO-SPB2, will also provide a wealth of data on the impact of \acp{VHECR} backgrounds, dark-sky background, and atmospheric refraction on  \cite{Venters:2019xwi} detection of $\nu_\tau$-induced \ac{EAS} events.

\subsubsection{GRAND -- highest exposure from ground by a huge distributed array}
\label{sec:GRAND}
\begin{figure}[!htb]
\centering
\includegraphics[width=4in]{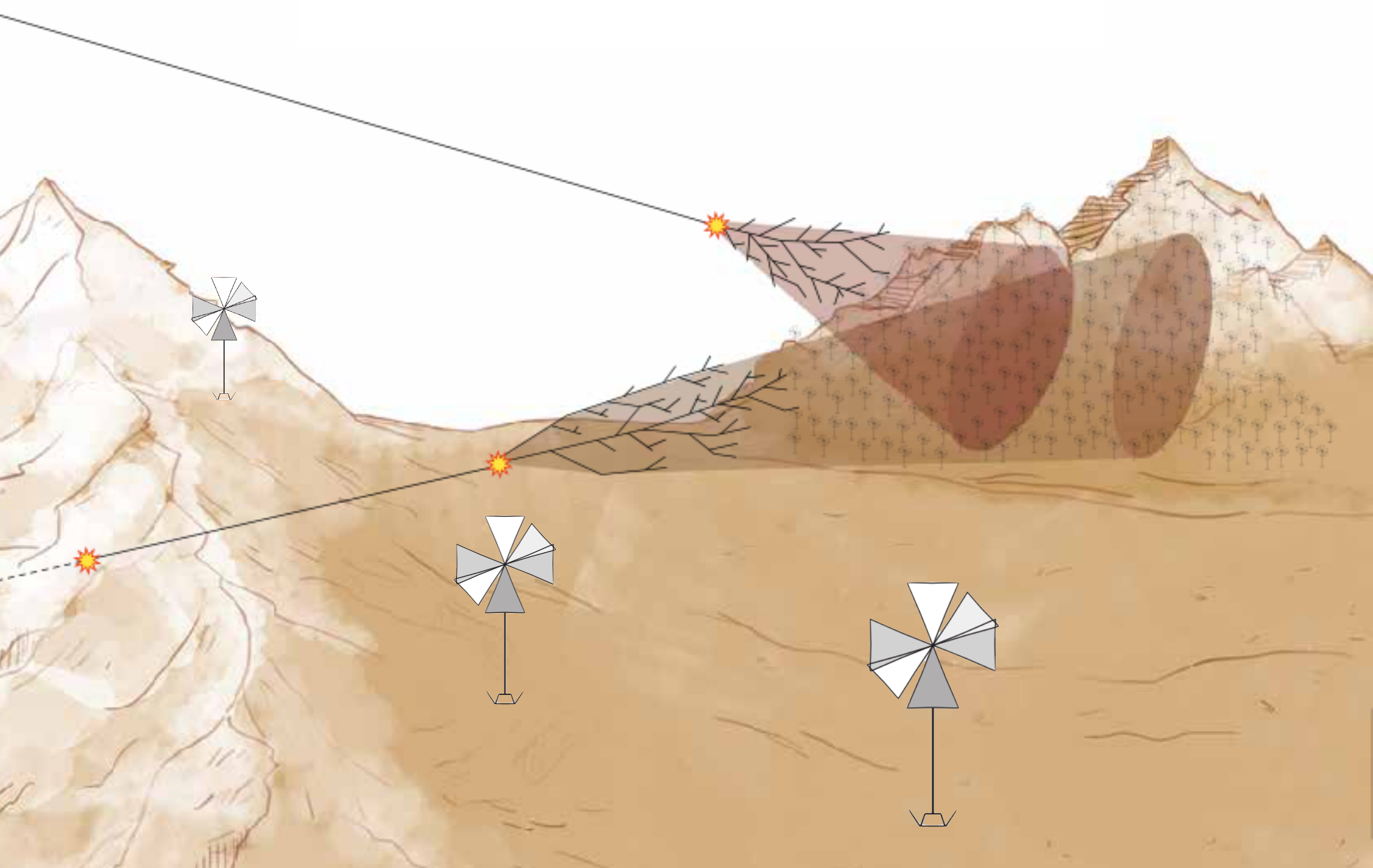}
\caption{A schematic of the \ac{GRAND} design \cite{GRAND:2018iaj}.
Arrays of radio antennae on the side of mountains can detect inclined extensive air showers from Earth-skimming and mountain passing tau neutrinos (lower trajectory) and inclined \ac{UHECR} events (higher trajectory).}
\label{fig:schematic}
\end{figure}

\paragraph{Design and Timeline}\label{sec:GRAND_DesignAndTimeline}
The \acf{GRAND} is a proposed experiment to detect the most energetic cosmic particles: neutrinos, cosmic rays, and gamma rays \cite{GRAND:2018iaj,Kotera:2021hbp}.
When high-energy particles interact in the atmosphere, an \ac{EAS} is produced and the Earth's magnetic field causes a separation of charge within the shower.
This charge separation leads to a coherent radio signature in the $\sim$\,10--100\,Hz range lasting $\sim$\,$100$\,ns.
The amplitude for the radio wave is large enough to be detected for air showers with $E\gtrsim10^{16.5}$\,eV \cite{Huege:2016veh,Schroder:2016hrv}.
GRAND will use a large number of very spaced-out radio antennas to detect these short showers, see \cref{fig:schematic} for a schematic of the process.
This technique builds on work done by previous radio arrays such as \ac{AERA}~\cite{PierreAuger:2012ker,PierreAuger:2018pmw}, CODALEMA~\cite{Ardouin:2009zp,Charrier:2012zz}, LOFAR~\cite{Nelles:2014dja,Corstanje:2014waa,Buitink:2016nkf}, TREND~\cite{Ardouin:2010gz,Charrier:2018fle}, and Tunka-Rex~\cite{Prosin:2016jev}.
As \ac{GRAND} focuses on inclined showers which are spatially much larger than vertical showers, the array of detectors can be more diffuse, allowing for larger arrays increasing the effective area.
The ultimate design for \ac{GRAND} is 10--20 locations around the globe containing 10,000-20,000 antennas over areas of 10,000-20,000\,km$^2$ each.
Due to the relatively straightforward scalability, numerous benchmark steps are in place as the design and construction progresses \cite{Martineau-Huynh:2021oqi}.

The precursor to \ac{GRAND} was the TREND array composed of 50 antennas over 1.5\,km$^2$ in the Tianshan mountains in China.
The goal of this array was to develop self-triggering capabilities of a large radio array.
After taking data for 314 days in 2011 and 2012, with only a relatively simple setup and cuts, TREND successfully self-triggered and identified 564 air shower candidates consistent with the expected flux up to $E=$ few$\times10^{18}$\,eV \cite{Charrier:2018fle}.
TREND's detection was with an efficiency of only 3\,\% due mostly to dead time of hardware which is expected to significantly improve in future iterations.
In addition, the detector design and lack of polarization information did not allow for an optimal analysis; this knowledge will be carried over into next-generation detectors.

TREND's success is to be followed up with GRANDProto300 (GP300) designed to reproduce TREND's self-triggering success, bring the efficiency close to $\sim$\,$100\,\%$, and build a larger array to provide measurements of cosmic rays up to $E=$ few$\times10^{18}$\,eV \cite{Decoene:2019sgx}.
Due to delays, site selection is still on-going, but the array is being deployed over 2021--2023.
After GP300, a large unit of 10,000 antennas will be deployed over an area of 10,000 km$^2$, likely in China, estimated to begin deployment in 2026.
This will allow for the final testing of the electronics and the detector design before deploying 10--20 more such arrays at other sites for the final distributed design with a total of 200,000 antennas over a total of 200,000 km$^2$ which should be rolled out in the early 2030s.

\paragraph{Scientific Capabilities}
\label{sec:GRAND_ScientificCapabilities}
\ac{GRAND} aims to be a state-of-the-art neutrino experiment measuring Earth-skimming topologies \cite{Feng:2001ue}, while also achieving leading measurements of \acp{UHECR} and high energy gamma rays.
The secondary physics case is also very rich including fast radio bursts, epoch of reionization, multi-messenger studies within a single experiment, among others. 
Below, the UHECR science case of \ac{GRAND} is given special focus.

As the various subarrays will be designed to target horizontal showers in order to maximize efficiency for detecting Earth-skimming tau neutrinos, \ac{GRAND} will be fully efficient for UHECR showers in the zenith angle range $[65^\circ,85^\circ]$ and for shower energies $\gtrsim$\,$10^{18}$\,eV.
This gives rise to a 100,000 km$^2$ sr effective area which is enough to match the current global accumulated exposure in $\sim$\,$1$\,year.
Moreover, simulations indicate that \ac{GRAND}'s effective area is actually $\sim$\,$5\times$ larger when considering events with shower cores just outside the instrumented area \cite{Kotera:2021hbp}.
The advance in statistics is \ac{GRAND}'s primary advantage to UHECR physics over existing measurements, but \ac{GRAND} will also be the only single experiment with full-sky coverage.
The impact of the advantages on the physics goals are discussed below.

The primary UHECR physics goals of \ac{GRAND} are to measure and characterize the spectrum, to identify the sources of \acp{UHECR} via anisotropy searches, and to determine the composition of \acp{UHECR} and its evolution as a function of energy.

First, while measurements of the spectra exist from Auger \cite{PierreAuger:2020qqz,PierreAuger:2020kuy} and TA \cite{TelescopeArray:2015dcv}, there is some discrepancy among them \cite{Abbasi:2018ygn}.
GRAND will address this discrepancy in two ways.
The increased statistics will make for a precise measurement both below the break at $10^{19.5}$\,eV as well as above it where the statistics of existing measurements start to fall off.
This will also allow for a more precise measurement of the break energy itself.
In addition, since \ac{GRAND} will have significant exposure in both the Northern hemisphere where TA is located and in the Southern hemisphere where Auger is located, it will allow for a measurement of the flux in both regions with the same systematics which will help to understand the difference between these measurements.

\begin{figure}[!htb]
\centering
\includegraphics[width=3.5in]{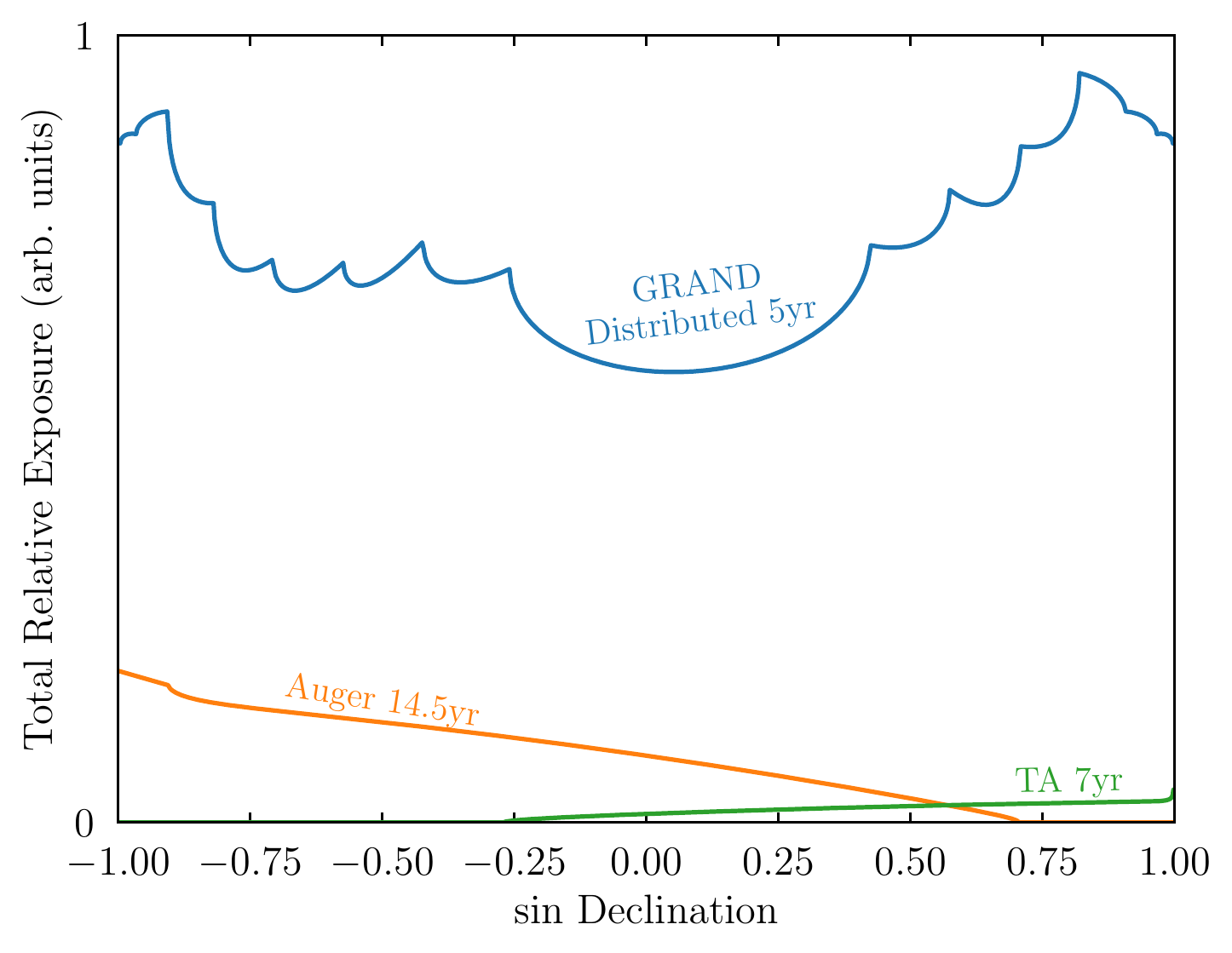}
\caption{The relative integrated exposure as a function of declination for Auger, TA, and a distributed \ac{GRAND} array \cite{Kotera:2021hbp}.}
\label{fig:GRAND exposure}
\end{figure}

Second, identifying the sources of \acp{UHECR} represents one of the chief ultimate goals of UHECR physics and \ac{GRAND}'s high-statistics measurements will provide an excellent opportunity to do that.
Due to the fact that \acp{UHECR} bend in intervening extragalactic and Galactic magnetic fields, learning about the source distribution is best done with a large field of view experiment.
As shown in \cref{fig:GRAND exposure}, \ac{GRAND} distributed across the planet will have roughly uniform exposure across the whole sky.
This will allow for the most efficient reconstruction of large-scale anisotropies \cite{Billoir:2007kb,Denton:2015bga,Deligny:2017wbx}.
Moreover, due to \ac{GRAND}'s large exposure, it will see more high-rigidity events with less bending by magnetic fields allowing for an increased power to identify correlations with catalogs of potential sources.
Finally, it is important to note that \ac{GRAND} will be well suited to test existing hints of large-scale anisotropy from TA and Auger due to significant exposure in each hemisphere \cite{TelescopeArray:2014tsd,PierreAuger:2016gkp,PierreAuger:2017pzq}.

Third, the composition of \acp{UHECR} seems to evolve from lighter elements at lower energies to heavier elements at higher energies \cite{PierreAuger:2014sui,Abbasi:2014sfa}, although the composition of cosmic rays at the highest energies is somewhat uncertain due to some slight tensions between the existing data sets.
Moreover, the conclusions depend on the details of the analyses as well as the hadronic shower model used, see e.g., Ref.~\cite{Arsene:2021inm}.
Thus additional higher statistics measurements are needed to resolve this picture.
The composition of \acp{UHECR} is extracted from \xmax which provides information on the depth of the shower; lighter elements propagate deeper into the atmosphere than heavier elements with the same energy.
In addition, fluctuations in the depth of the shower, \sigmaXmax provides information as well since lighter elements tend to have larger fluctuations.
Together these can provide an estimate of the composition of \acp{UHECR} as a function of energy.
A radio version of \xmax has been developed and GRAND can measure it statistically at $\sim$\,65\,\gcm for protons and $\sim$\,$25$\,\gcm for iron which is comparable to the shower-to-shower fluctuation size \cite{Decoene:2021ncf,Decoene:2021atj}.
GRAND's measurement of \xmax will provide key additional information to improve our understanding of the composition of \acp{UHECR}.
GRAND will have good enough \xmax resolution and will be able to apply it to higher energies than existing measurements due to the improved statistics allowing for a clearer determination of the evolution of the composition.
This information will then further propagate out into helping understand anisotropies and identifying sources.

Beyond making measurements of \acp{UHECR} directly, \ac{GRAND} will measure the \ac{UHE} neutrino flux.
This flux could contain contributions coming directly from the sources, or even be dominated by them (see, e.g., Ref.~\cite{Rodrigues:2020pli}). 
The sources of UHE neutrinos may be the same ones behind the TeV--PeV neutrinos seen by IceCube, or different ones.
There is one guaranteed contribution to \ac{UHE} neutrino flux, which is from \acp{UHECR} interacting with the cosmic microwave background \cite{Greisen:1966jv,Zatsepin:1966jv}.
Since \acp{UHECR} with $E\gtrsim10^{19.7}$\,eV only come from relatively nearby, \ac{GRAND}'s measurement of the neutrino flux which depends on the total UHECR flux provides key information about the redshift evolution of sources as well as the UHECR composition \cite{AlvesBatista:2018zui,Moller:2018isk}.

In summary, \ac{GRAND} will be a state-of-the-art large-scale \ac{UHE} astroparticle experiment.
It will benefit from the distributed nature of radio arrays which have already demonstrated the key benchmark of self-triggering.
The physics case within \acp{UHECR} alone is broad and will also provide leading measurements of neutrinos, gamma rays, and a number of other secondary physics cases.

\subsubsection{GCOS -- accuracy for ultra-high-energy cosmic rays}
\label{sec:GCOS}
\paragraph{Design and Timeline}
\label{sec:GCOS_DesignAndTimeline}
\begin{wrapfigure}{r}{0.7\textwidth} 
  \vspace{-5mm}
  \includegraphics[width=0.7\textwidth]{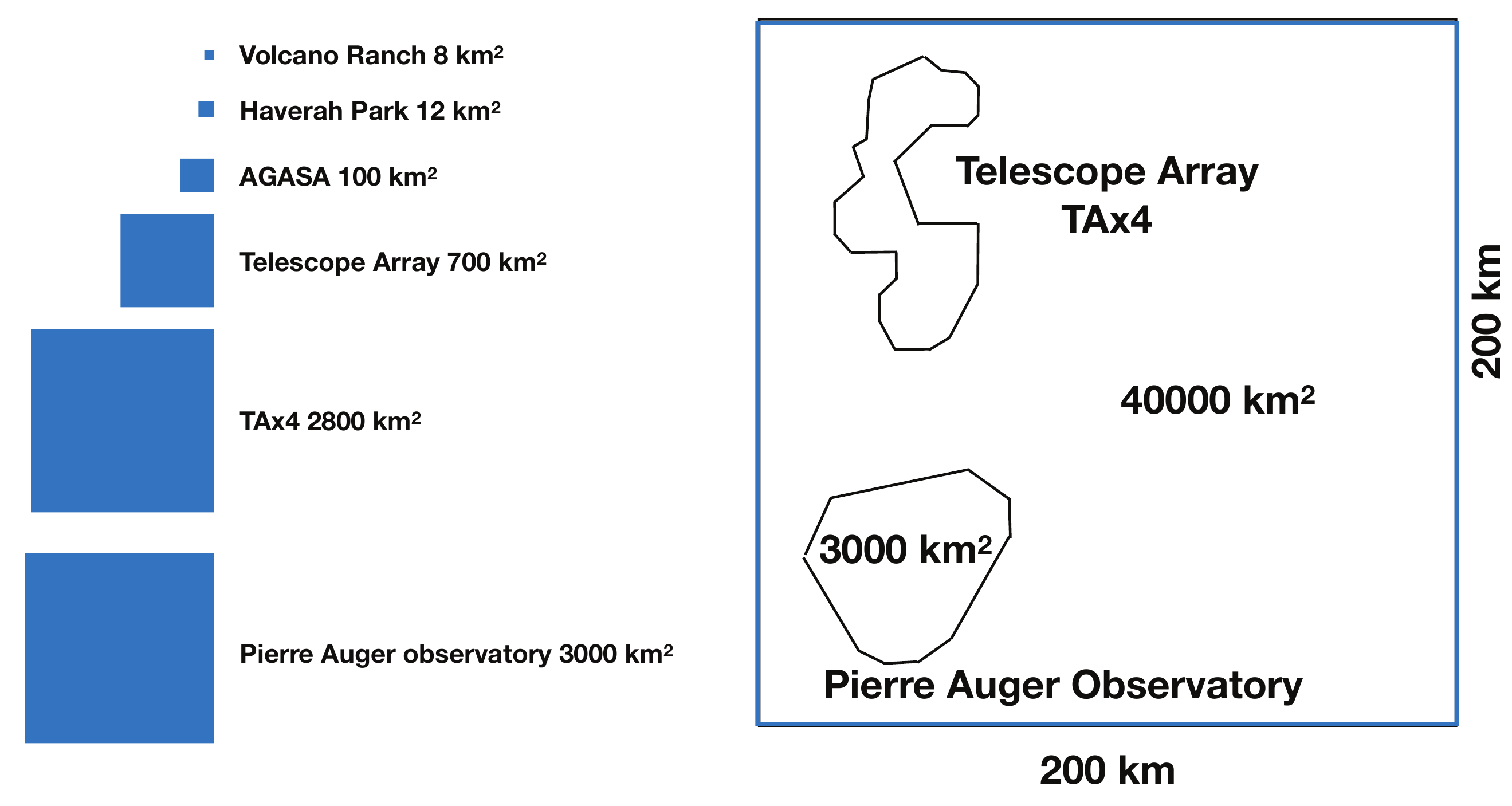}
  \caption{Illustration of total collection area for existing installations and GCOS.}
  \label{fig:gcos-area}
  \vspace{-2mm}
\end{wrapfigure}
The \acf{GCOS} is in the design phase with the final detection concept and setup being worked out.  More detailed considerations can be found in Ref.~\cite{Horandel:2021prj} and references therein.
With the goal of reaching  an exposure of  at least $2\times 10^5$\,km$^2$\,yr in a period of 10 years and full-sky coverage a set of surface arrays with a total area of about 40000\,km$^2$  is anticipated as shown in \cref{fig:gcos-area}. 
Identification of the ultra-high-energy particle sources will require a good angular resolution.
Assuming a detector spacing of the order of $1.6-2$\,km an angular resolution $<0.5^\circ$ is realistic.
For a determination of the fine structures in the energy spectrum, \ac{GCOS} is expected  to provide an energy resolution around $10-15\,\%$.  
In particular, in regions of a steeply falling energy spectrum, as e.g., at the highest energies a good energy resolution is important to restrict spill over of measured events to higher energies to an acceptable amount.
Good energy resolution is also important to identify and investigate transient sources.

\begin{figure}[!htb]
\centering
  \vspace{-5mm}
  \includegraphics[width=0.6\textwidth]{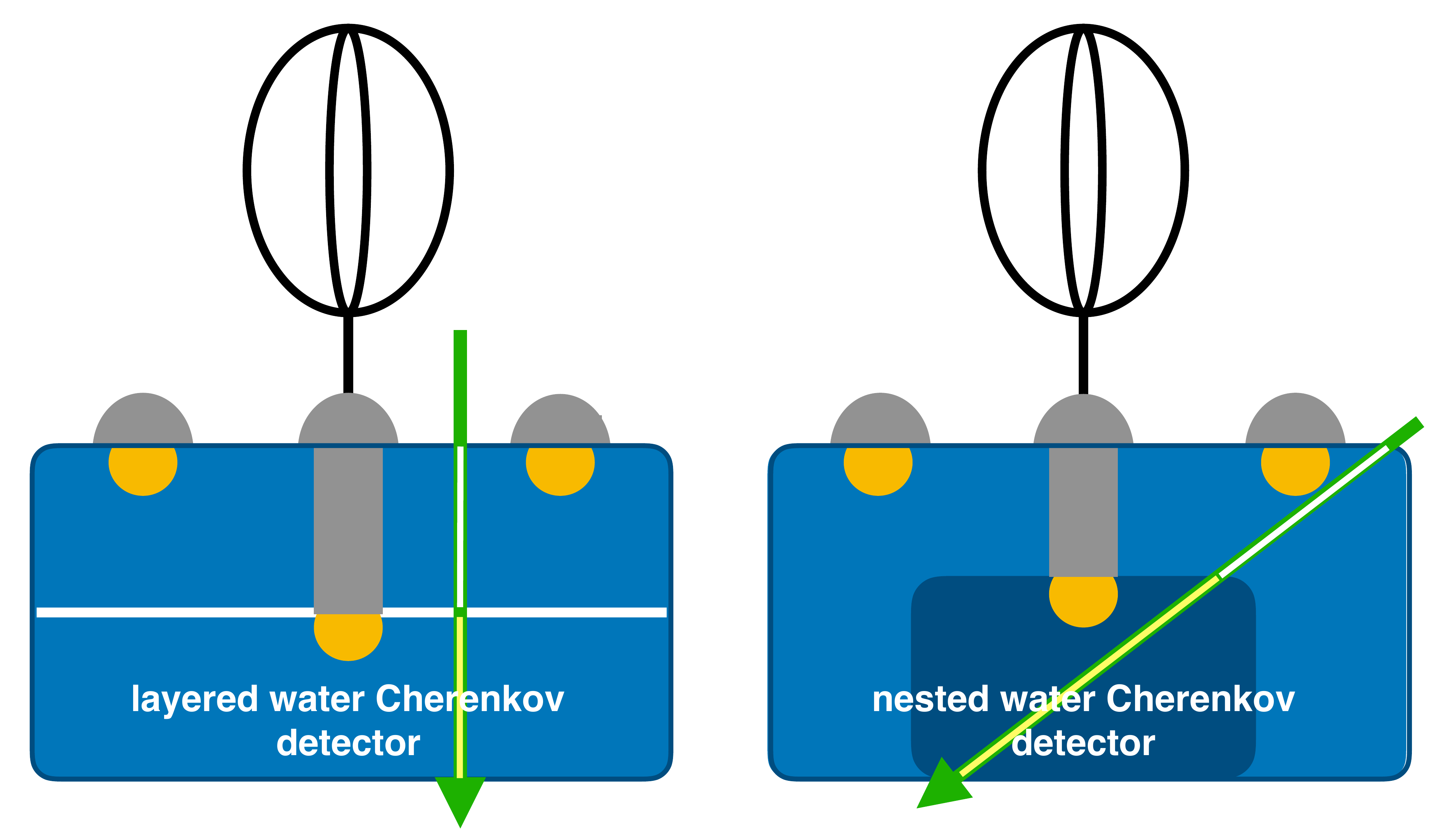}
  \vspace{-5mm}
  \caption{Detection concepts, using a layered (left) and a nested (right) water Cherenkov detector with a radio antenna on top.}
  \vspace{-2mm}
  \label{fig:gcos-wcd}
\end{figure}
Another important requirement for the \ac{GCOS} design is to have the capability to identify the mass, and ultimately the rigidity of each ultra-high energy particle measured.
This requires a good measurement of the atmospheric depth of the shower maximum \xmax{} and of the ratio of the electromagnetic to muonic particles in an air shower. Both quantities depend only logarithmically on the atomic mass of the primary particle. 
Ultimately, \ac{GCOS} will need to have excellent rigidity resolution. Since $R=E/Z$, this will require simultaneously good energy resolution of the order of $\sim$\,10\,\%--15\,\% and good mass resolution with $\Delta\ln A <0.8$ for individual showers. This will allow to distinguish at least five mass groups for the elemental composition, about equally spaced in $\ln A$ (p, He, CNO, Si, Fe). 
The charge $Z$ can only be derived indirectly, assuming $Z\approx A/2$, with the obvious exception for protons with $A/Z=1$.
This provides the foundation to find and study sources, but also to study particle physics and fundamental physics at extreme energies.
To achieve such a mass-resolution requires to measure the depth of the shower maximum with an accuracy better than 20~\gcm{} and a resolution for the measurement of the muonic shower content of the order of 10--15\,\%.

\begin{figure}[t]
\centering
  \includegraphics[width=0.4\textwidth]{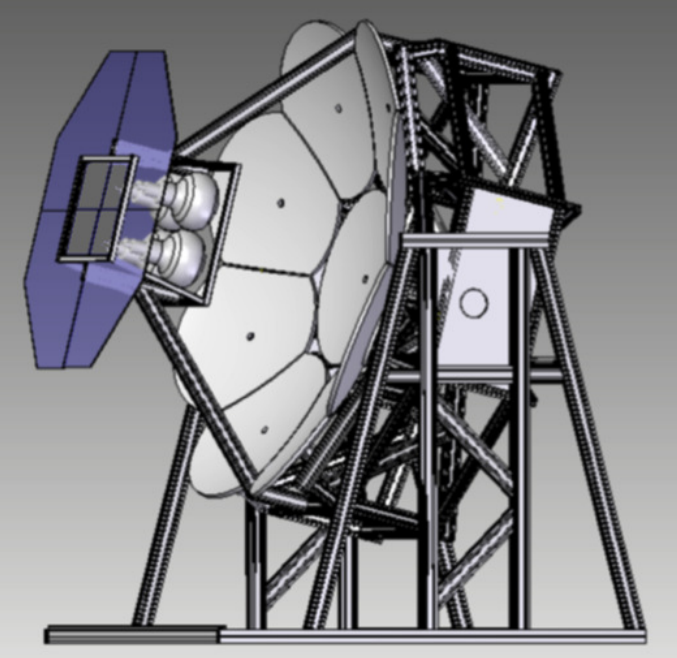}
  \vspace{-2mm}
  \caption{Concept of a fluorescence telescope frame, showing four \acp{PMT} at the focus of a 1.6~m diameter segmented mirror. The support structure is made from aluminium profiles. The UV band-pass filter can be seen attached to the periphery of the camera box.}
  \vspace{-3mm}
  \label{fig:gcos-fast}
\end{figure}
Different detection concepts are at hand. They need to be optimized to reach the targeted physics case.
Fluorescence detectors  provide a calorimetric measurement of the shower energy and a direct and almost model-independent measurement of \xmax{}. However, they have only a limited duty cycle ($\sim$\,$15\,\%$) due to constraints on atmospheric transparency and background light conditions.
An alternative with almost 100\,\% duty cycle is the use of radio antennas in a frequency range where the atmosphere is transparent to radio waves. Such detectors require radio-quiet regions.
The classical approach of a particle detector ground array has no restrictions with respect to radio interference or background light and the particle type is inferred from the ratio of secondary particles on the ground. 
The conversion from measured signal ratios (or to a lesser extend from \xmax{}) to the nature of the incoming particle is based on Monte Carlo simulations. The accuracy will depend on the progress in matching  the air-shower simulations with data. 

In order to determine the mass of each incoming cosmic ray  with a detector array one typically measures two shower components simultaneously, mostly the electromagnetic and muonic components are used.
A cost effective approach is the use of layered water Cherenkov detectors. A big water volume is read out through optically separated segments as illustrated in \cref{fig:gcos-wcd}. If enhanced electron-muon separation is also desired for horizontal air showers (e.g., for neutrino detection) a possible design could be a nested detector. 
Radio antennas on top of a particle detector are a very promising concept also for \ac{GCOS}.
They provide a calorimetric measurement of the electromagnetic shower component with high precision ($\sim$\,$10\,\%$). In particular, this will allow to measure the electron-to-muon ratio for horizontal air showers in combination with a water Cherenkov detector. In addition, the radio technique can be used to calibrate the absolute energy scale of a cosmic-ray detector.
Cost-effective fluorescence detectors, as shown in \cref{fig:gcos-fast}, could be included  to measure the calorimetric shower energy and the depth of the shower maximum or a large stand-alone array of fluorescence detectors could be an alternative option for \ac{GCOS}.

\ac{GCOS} is at present in the phase of concretizing the science case and studies are being conducted to optimize a suitable detector design. 
The performance of existing detector systems, in particular the upgrade of the Pierre Auger Observatory gives concrete and proven design examples to achieve the needed resolution for the depth of the shower maximum and the electron-to-muon ratio. The final design of \ac{GCOS} will  depend on the results which will be obtained with the Auger Observatory and the Telescope Array in the coming decade. New findings concerning the elemental composition and the arrival direction of cosmic rays at the highest energies will influence the ultimate science case, and, thus the design of the observatory. A promising approach towards a full-scale \ac{GCOS} observatory could be to gradually increase the aperture of the existing arrays. For example, findings with an aperture of a few times the aperture of the current Auger observatory will improve the understanding about the highest-energy cosmic rays, and, thus will clarify the design goals (with respect to mass resolution and direction resolution capabilities) for an even larger observatory.
Prototype detectors are expected to be build after 2025. The construction of \ac{GCOS} at multiple sites is expected to start after 2030 with an anticipated operation time of twenty years.

\paragraph{Scientific Capabilities}
\label{sec:GCOS_ScientificCapabilities}
We are living in a golden epoch in Astrophysics where we have witnessed the birth and the first steps in the development of multi-messenger astronomy. Our understanding of the high-energy Universe has significantly expanded and progressed thanks to observations obtained recently with different messengers in a broad range of energies. The objective of \ac{GCOS} is to conduct precision multi-messenger studies at the highest energies. \ac{GCOS} will be designed to have good sensitivity to measure charged cosmic rays, gamma rays, and neutrinos, thus, being able to address the following {\bf scientific questions}:

Nature is providing particles at enormous energies, exceeding $10^{20}$\,eV -- orders of magnitude beyond the capabilities of human-made facilities like the \ac{LHC}.
At the highest energies the precise particle types are not yet known, they might be ionized atomic nuclei or even neutrinos or photons. Even for heavy nuclei (like, e.g., iron nuclei) their Lorentz factors $\gamma=E_{\rm tot}/m\,c^2$ exceed values of $10^9$.

The existence of such particles imposes immediate questions, yet to be answered:
\begin{itemize}[topsep=2pt]\setlength\itemsep{-0.2em}
    \item  What are the physics processes involved to produce these particles?
    \item Are they decay or annihilation products of dark matter \cite{Alcantara:2019sco,Aloisio:2015lva}?
\end{itemize}

If they are accelerated in violent astrophysical environments:
    \begin{itemize}[topsep=2pt]\setlength\itemsep{-0.2em}
        \item How is Nature being able to accelerate particles to such energies?
        \item What are the sources of the particles?
        \item Do we understand the physics of the sources?
        \item Is the origin of these particles connected to the recently observed mergers of compact objects -- i.e., the gravitational wave sources~\cite{LIGOScientific:2017ync, Lipari:2017qpu, Branchesi:2016vef, Gergely:2007ny, Gergely:2008dw, Gergely:2010xr, Tapai:2013jza}?
    \end{itemize}
    
The highly relativistic particles also provide the unique possibility to study (particle) physics at its extremes:
\begin{itemize}[topsep=2pt]\setlength\itemsep{-0.2em}
    \item  Is Lorentz invariance (still) valid under such conditions
\cite{Klinkhamer:2008ss, Klinkhamer:2017puj, Aloisio:2000cm, Cowsik:2012qm, Martinez-Huerta:2017ulw, Trimarelli:2021Ul, PierreAuger:2021tog}?
    \item How do these particles interact?
    \item Are their interactions described by the Standard Model of particle physics?
\end{itemize}
When the energetic particles interact with the atmosphere of the Earth, hadronic interactions can be studied in the extreme kinematic forward region (with pseudorapidities $\eta>15$):
\begin{itemize}[topsep=2pt]\setlength\itemsep{-0.2em}
    \item What is the proton interaction cross section at such energies ($\sqrt{s}>10^5$\,GeV)?
\end{itemize}

A key objective of \ac{GCOS} will be to {\bf identify and study the sources of \ac{UHE} particles}.
Even in the most conservative case the high-energy sources are amazing objects that challenge our view and constitute unique laboratories to test the fundamental laws of physics.
This is already of great interest in the most conservative case, let alone the case of exotic phenomena of new physics, which obviously represents an exciting additional possibility.
A key component for a science case will be to be able to backtrack charged cosmic rays in the Galactic and extra-galactic magnetic fields. This requires detailed knowledge of the structure of cosmic magnetic fields. Corresponding models are being developed in parallel to the \ac{GCOS} hardware.
To conduct charged-particle astronomy it is also desired to have
a large exposure to reach high rigidity values and the ability to determine the charge for each measured cosmic ray.
If the knowledge about cosmic magnetic fields will allow correction for deflections on the $10^\circ - 20^\circ$ scale, this would dramatically improve the ability to backtrack the particles and conduct particle astronomy.
Model scenarios will need to consider in detail the physics of various sources, the acceleration mechanisms taking place, the physics which governs the escape of particles from the source region, the particle propagation through intergalactic and interstellar space until their interactions with the atmosphere of the Earth. Ideally, full end-to-end simulations will be prepared for different source classes, such as \acp{AGN}, gamma-ray bursts, and gravitational-wave sources. Such simulations will yield quantitative estimates for the measurable quantities.

\subparagraph{Multi-messenger connections:} \ac{GCOS} will be capable of detecting neutrinos and photons, greatly enriching its science case. In the multi-messenger era, it will be an important partner to search for neutral ultra-high-energy particles associated with transient events such as mergers of compact binaries, tidal disruption events, and gamma-ray bursts, among others, providing insights into the most energetic processes in Nature.  In addition, \ac{GCOS} will either measure or constrain the fluxes of cosmogenic neutrinos and photons, consequently improving our understanding about ultra-high-energy cosmic-ray sources.
Three messengers are ``inextricably'' tied together (cosmic rays, gamma rays, high-energy neutrinos) and provide complementary information about the same underlying physical phenomena.

\ac{GCOS} will also be able to address complementary science cases. They include in particular:
\subparagraph{Dark Matter searches:}
For many decades, the favored models characterized \ac{DM} as a relic density of \acp{WIMP}. Despite the fact that a complete exploration of the WIMP parameter space remains the highest priority of the \ac{DM} community, there is also a strong motivation to explore alternatives to the \ac{WIMP} paradigm. \ac{DM} could manifest itself by an excess of photons and neutrinos at high energies. Thus, it will be crucial that \ac{GCOS} will have photon and neutrino-detection capabilities to constrain, e.g., the flux of photons and neutrinos from certain regions, such as the Galactic center.

\subparagraph{Fundamental physics and quantum gravity:}
Ultra-high-energy particles can be used as probes of fundamental physics and quantum gravity. 
The data can be used to search for \acp{LIV} in the nucleon or photon sector.
Another important aspect are effects of \ac{LIV} on air showers. The main idea is that modified decay rates of neutral and/or charged pions and muons can change the shower characteristics, such as the muon content and \xmax{}.

\subparagraph{Particle physics:} 
One of the experimental challenges in determining the mass of cosmic rays from air shower measurements is the degeneracy between the mass of the incoming particles and hadronic interactions. Thus, hybrid measurements of air showers are mandatory for \ac{GCOS} to verify hadronic interaction models. Air shower data are also used to measure the cross sections for proton-air and proton-proton collisions at center-of-mass energies far above values reachable at accelerators. 

\subparagraph{Geophysics and atmospheric science:}
\ac{GCOS} will also be able to address scientific questions from the areas of geophysics and atmospheric science. 
An example is the study of ELVES which are a class of transient luminous events, with a radial extent typically greater than 250~km, that occur in the lower ionosphere above strong electrical storms.
Radio antennas allow detailed insights into the spatial and time structure of the development of lightning strokes in the Earth's atmosphere.

\subsubsection{Complementary experiments}
\label{sec:otherExperiments}
A number of experiments will complement the science of the three major projects mentioned above. 
Although not of the same large-scale scope regarding cosmic-ray physics, they will still make unique contributions to specific scientific questions of cosmic-ray particle and astrophysics.

At very high cosmic-ray energies reaching up to the presumed transition form Galactic to extragalactic cosmic rays, these are, in particular, to arrays dedicated primarily to the observation of TeV to PeV gamma rays: LHAASO~\cite{LHAASO:2019qwt}, which  recently started operation in the northern hemisphere in China, and SWGO \cite{Hinton:2021rvp}, which is planned in the southern hemisphere in South America. 
Further progress in this lower energy range will result from new analyses of data from multi-detector experiments such as TAIGA~\cite{TAIGA:2021fgx} or KASCADE-Grande~\cite{Apel:2010zz}, once better hadronic interaction models will be available.

The importance of building SWGO~\cite{SWGO:LoI191, SWGO:LoI226} must be fully recognised, and a detailed description of it is only omitted due to the scope of this whitepaper, specifically that it focuses on the highest cosmic-ray energies.
That said, below a few experimental activities that are relevant in the ultra-high-energy range in addition to the major large-scale projects are briefly described.

\paragraph{The Cosmic Ray Extremely Distributed Observatory (CREDO)}


\acf{CREDO} Collaboration (see Ref.~\cite{CREDO:2020pzy} and the topical references therein) asks under which circumstances and with which conditions \acp{CRE} can reach the Earth and be at least partly detected with the available or possible infrastructure. 

This approach is equivalent to looking for both small and large scale cosmic ray correlations in space and time domains, embracing the whole cosmic ray energy spectrum and all primary types. Within such a general approach the list of \ac{CRE} scenarios and interdisciplinary opportunities to be studied/pursued using \ac{CREDO} cannot be closed, so one rather considers \ac{CREDO} an infrastructure capable of hosting a wide scientific program, not just one individual project. An example of astrophysical scenario presently being studied within CREDO include the simulations and search for \textit{air shower walls}, a specific class of \acp{CRE}, composed of thousands or even millions of photons of energies from MeV to EeV, which are expected to be created due to the interactions of ultra-high energy photons with the magnetic field of the Sun \cite{CREDO:2018ghi}. 

Another promising science goal concerns the bursts of 0.1\,PeV air shower events recently recorded by one of the CREDO facilities~\cite{CREDO:2022qpa}. This case illustrates one of the observation strategies being implemented specifically with \ac{CREDO}: \textit{"the quest for the unexpected"} or (simply) \textit{"fishing"}. The fishing strategy helps not to miss the breakthrough observations despite the (yet) missing theoretical background: with no a priori assumed scenario (it is reasonable to assume that the UHECR community might not yet be aware of all the scenarios being realized in nature) one can still analyze the data to search for statistically significant signal excesses or anomalies which would provide a valuable input for theoretical considerations. 

The optimum \ac{CRE}-oriented experimental strategy should be based on forming an openly inter-operable alliance/network of observatories, experiments and individual detectors sensitive to cosmic signal (including also e.g., muons in underground or underwater facilities, radiation detected in CCD/CMOS pixel cameras used e.g., in classical astronomy, off-beam measurements in particle accelerators) that would enable both historical data analyses and CRE-candidate alerts, inevitably adopting front-end AI and big data technologies.

\paragraph{The Latin American Giant Observatory (LAGO)}
\acf{LAGO}, is a project conceived in 2006~\cite{Allard:2008zz} to detect the high energy component of \acp{GRB}, with typical energy of primaries $E_p \gtrsim 20$\,GeV, by installing $10$\,m$^2$--20\,m$^2$ \acp{WCD} at very high altitude sites across the Andean ranges. From this initial aim, \ac{LAGO} has evolved toward an extended astroparticle observatory at a regional scale, currently operating \acp{WCD} and other particle detectors in ten countries in LA, Argentina, Bolivia, Brazil, Chile, Colombia, Ecuador, Guatemala, Mexico and Peru, together with the recent incorporation of institutions from Spain. \ac{LAGO} is operated by the \ac{LAGO} Collaboration, a cooperative and non-centralized collaboration of 29 institutions.

The \ac{LAGO} detection network consists of single or small arrays of astroparticle detectors installed in different sites across the Andean region~\cite{Sidelnik:2015yky}. The detection network spans a region from the south of Mexico, with a small array installed at Sierra Negra ($4550$\,m a.s.l.), to the Antarctic Peninsula, with the recent installation of two \acp{WCD} at the Marambio Base (Arg., 200\,m a.s.l.), and at Macchu Picchu Base (Perú, 10\,m a.s.l.), mainly oriented for Space Weather studies and monitoring~\cite{Suarez-Duran:2015pas, Dasso:2015lfq, Dasso2019}.

The network is distributed over similar geographical longitudes but a wide range of geographical latitudes and altitudes. By combining simultaneous measurements at different rigidity cutoffs from regions with differing atmospheric absorption \ac{LAGO} is able to produce near-real-time information at different energy ranges of, for example, disturbances induced by interplanetary transients and long term space weather phenomena.

\ac{LAGO} has three main scientific objectives: to study high energy gamma events at high altitude sites, to understand space weather phenomena through monitoring the low energy cosmic rays flux at the continental scale, and to decipher the impact (direct and indirect) of the cosmic radiation on atmospheric phenomena. These objectives are complemented by two main academic goals: to train students in astroparticle and high energy physics techniques, and foremost, to support the development of astroparticle physics in Latin America~\cite{Asorey:2016pru}.

Scientific and academic objectives are organized in different programs and are carried out by the corresponding working groups. \ac{LAGO} programs cover several aspects of the project, from the installation, calibration and operation of the detectors to the search for pathways to transfer data from remote sites.

\paragraph{The Square Kilometer Array (SKA)}
\label{sec:SKA}

\acf{SKA-low}~\cite{Huege:2015jga}, currently under construction, will have a dense array of about $60{,}000$ radio antennas on an area of about $0.5$\,km$^2$ that will be able to measure air showers with energies between roughly $10^{16}$--$10^{18}$\,eV in a wide radio frequency bandwidth of 50--350\,MHz.

The \ac{SKA} will be able to study radio footprints in unprecedented detail with an expected \xmax resolution below $10$\,g/cm$^2$ \cite{Buitink:2021pkz}. 
Additionally, sensitivity to the width of the longitudinal shower profile ($L$)~\cite{Andringa:2011zz} with percent-level precision will enable the SKA to test hadronic interaction models regarding their predicted correlations of \xmax and $L$.
Once hadronic interaction models have been improved to accurately predict the $L$ parameter for different primary particles, this provides a complementary way to increase the mass sensitivity beyond the sensitivity achievable by a perfect \xmax resolution.

Therefore, the SKA may be able to differentiate between air-shower physics and astrophysical models, and in addition reduce systematic uncertainties on \xmax and mass composition measurements. 
The ability to perform interferometric measurements with SKA will have great potential to achieve even higher sensitivity to the mass composition \cite{Schoorlemmer:2020low,Schluter:2021egm} and might even be sensitive to the 3-dimensional profile of the shower emission regions, potentially enabling deeper studies of air shower physics. 
Finally, complementing \ac{SKA}'s measurements of electromagnetic shower content and \xmax{} with muon-sensitivity via a suitable particle detector array could unlock further potential in high-accuracy studies of mass composition and hadronic interaction physics in the energy range of the Galactic-to-extragalactic transition.

\subsection{The path to new discoveries}


\subsubsection[Energy spectrum]{Energy Spectrum: Characterizing the rarest particles}
\label{sec:nextgen_spectrum}

\begin{figure}[thb]
    \centering
    \includegraphics[width=0.99\textwidth]{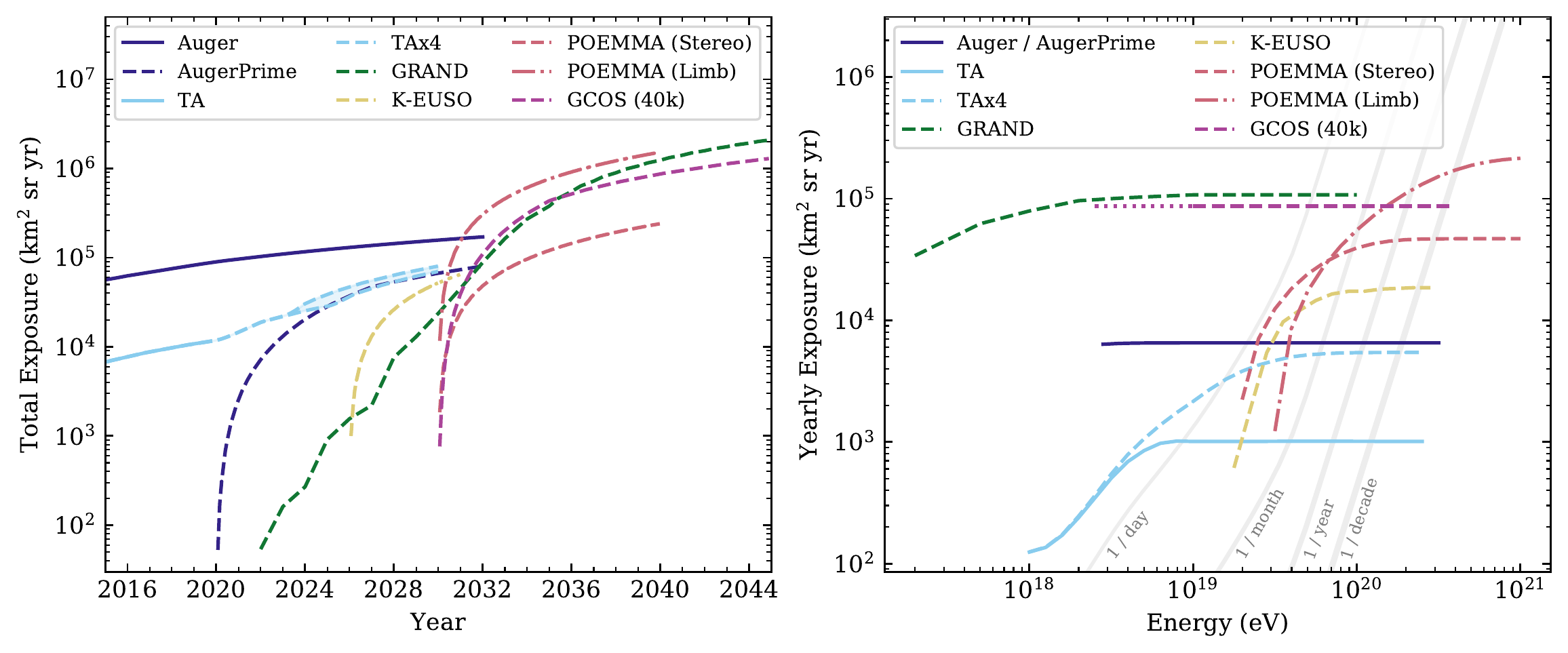}
    \caption{Left: The exposure to cosmic rays near the suppression region (50\,EeV) is shown as a function of time for Auger \& Auger-Prime (SD-1500), \ac{TA} \& \TAxFour, \ac{GRAND}, K-EUSO~\cite{Klimov:2022jzk}, \ac{POEMMA} in stereo-mode, and \ac{GCOS}. The exposure for \ac{POEMMA} in limb-mode is shown for 300\,EeV. A band is shown to indicate the exposure for various deployment schedules for \TAxFour. Right: The effective aperture of the experiments are shown as a function of energy. The gray lines indicate the yearly exposure that is required for an experiment to observe the indicated event rate, according to the flux model given in Ref.~\cite{PierreAuger:2021ibw}. In both panels, currently operating experiments are shown in solid lines and future experiments/upgrades are shown in dashed lines.}
    \label{fig:future_exposure}
\end{figure}

The contribution of Auger and \ac{TA} to the understanding of \acp{UHECR}' nature is certainly very remarkable and will be significantly improved in the next decade via the respective upgrades to the observatories. At ultra high energies, cosmic-rays are of extra-galactic origin~\cite{PierreAuger:2021dqp} and there are strong hints of correlation of the arrival directions at intermediate angular scales with some known catalogue of sources, see \cref{sec:Anisotropy}~\cite{PierreAuger:2018qvk} and of a clustering of events in the northern hemisphere~\cite{TelescopeArray:2014tsd}.  The improved sensitivity to the primary mass and the  increase in exposure that will be available with AugerPrime and \TAxFour will possibly lead to a $5\sigma$ discovery in the anisotropy searches at intermediate angular scales. 
Such result, probably achievable within the next decade, will confirm the feasibility of more detailed studies on the properties of single (or groups of) sources like the spectral shape.
A limit in such studies is the relatively small statistics achieved with both Auger and \ac{TA} at the highest energies.
The all particle spectrum must be subdivided in mass groups and in different sources, further reducing the statistics. This will require a large increase in exposure that will be only provided by the next generation experiments.

\subparagraph{Leveraging the spectral shape at the highest energies}
The limited statistics of Auger and \ac{TA} at the highest energies, 
hinders a proper characterization of the shape of the all-particle spectrum above the flux suppression and therefore to discover possible new spectral features in this range (like recently happened with the {\it instep} at $10^{19}$\,eV). 
A significant increase of the exposure is therefore needed.
Measuring precisely the spectrum at these extreme energies with high statistics is of fundamental importance to understand the maximum energy achievable by accelerators as the continuation of the very steep decay of the flux far above the suppression will confirm the end of the cosmic-ray spectrum. A large exposure would also allow to explore the spectrum above few times $10^{20}$~eV, where only upper limits to flux are currently available. A new hardening in the flux suppression of the energy spectrum could indicate the presence of a local source capable of accelerating particles at such high energies~\cite{Armengaud:2004yt,Taylor:2011ta,Gonzalez:2021ajv} and would provide new insights in the understanding of the mechanisms responsible for the acceleration of the highest-energy \acp{CR}~\cite{Hillas:1985is}.
A \emph{recovery} of the spectrum above $10^{20}$~eV has been moreover predicted~\cite{Scully:2008jp} in the context of \ac{LIV} allowing to test the frontier of particle acceleration in the Universe, and new physics as well~\cite{Colladay:1998fq, AmelinoCamelia:2000zs, Torri:2019gud}. 
In such kind of studies, a significant increase of the sensitivity is obtained by adding information on mass composition~\cite{Aloisio:2000cm}.
The combined fit of the spectral shape and of the composition has been used by the Auger collaboration to set stringent limits on the \ac{LIV} amplitude~\cite{PierreAuger:2021tog}. A significant increase in statistics together with an improvement in mass sensitivity for future observatories will be extremely beneficial to improve such limits.
Finally, the combination of high statistics, mass sensitivity and anisotropy will be of extreme value also to constrain production models in a similar way to what already done by Auger~\cite{PierreAuger:2016use, PierreAuger:2021oxo} and \ac{TA}~\cite{Bergman:2021djm}.

To significantly increase the statistics at the highest energies, a few possible detector concepts have been so far proposed: in particular \ac{GCOS} (see \cref{sec:GCOS_DesignAndTimeline}), \ac{GRAND} (see \cref{sec:GRAND_DesignAndTimeline}), K-EUSO~\cite{Klimov:2022jzk} and \ac{POEMMA} (see \cref{sec:POEMMA_DesignAndTimeline}). The first two are giant ground-based arrays while the last two are space-based 
telescope.  Although such projects are all in the development phase, they will bring a substantial increase in the exposure, as shown in \cref{fig:future_exposure} especially at the most extreme energies. 
The current baseline design for \ac{GCOS} calls for a ground-based observatory spanning $\sim$\,40,000\,$\mathrm{km}^2$
for which several detector designs are being studied, that will allow to obtain a yearly exposure of $\sim$\,100,000\,km$^2$\,sr\,yr. 
\ac{GRAND} will be a 200,000\,$\mathrm{km}^2$ radio observatory on ground with a $\sim$\,100,000\,$\mathrm{km}^2\ \mathrm{sr}\ \mathrm{yr}$ yearly exposure in the zenith angle range 65--80$^\circ$. Its prototype is currently being deployed in China.
K-EUSO is a mission aiming at the deployment of a single detector on the Russian section of the \ac{ISS}. It will be the first, from 2026, to measure cosmic ray from space through the fluorescence technique.
\ac{POEMMA} aims at the deployment of two fluorescence telescopes in space to operate either in stereo mode pointing straight down to the Earth, for the detection of cosmic rays, or tilted toward the horizon mainly for the detection of neutrino events. Depending on the tilting angle, \ac{POEMMA} can achieve at least $\sim$\,46,000\,$\mathrm{km}^2\,\mathrm{sr}\,\mathrm{yr}$ per year in nadir mode at $10^{20}$\,eV which can become over 200,000\,$\mathrm{km}^2\,\mathrm{sr}\,\mathrm{yr}$ each year around $10^{21}$\,eV when pointing toward the horizon. Space based configurations can moreover achieve a very uniform exposure on both hemispheres.
For comparison the yearly exposures of Auger and \TAxFour is between 5000 to 7000\,km$^2$\,sr\,yr.

Despite the strong specificity of the single concepts, the experimental techniques are inherited from the developments of currently operating detectors and the details will be defined in this decade also following the results from the operation of Auger and \ac{TA}.
The present generation of experiments, Auger and \ac{TA}, is in any case going to lead the field at least for this decade, until the future generation of experiments will take over in the first half of the 2030s.

\subsubsection[Mass composition]{Mass composition: The 20-year picture}
A significant increase of the exposure is required for collecting sufficient statistics at extreme energies with composition-sensitive detectors. 
This is critical as these are the energies where the rigidity of the primary particles might reach values $>$\,$20$~EV which would also for more straightforward identification of point sources~\cite{Erdmann:2016vle, Unger:2017kfh}.
New experiments building on novel detection technology, such as \ac{POEMMA} (see \cref{sec:POEMMA}), \ac{GRAND} (see \cref{sec:GRAND}), and \ac{GCOS} (see \cref{sec:GCOS}) with apertures of more than one order of magnitude larger than that of Auger, are currently in the design stage.
Also technology approaches such as \ac{CRAFFT}~\cite{Tameda:2019wmj} and \ac{FAST}~\cite{Fujii:2015dra, Malacari:2019uqw} are under evaluation, and may be integrated in \ac{GRAND} or \ac{GCOS} sites or added as additional sites to further increase the exposure.
The design of these observatories will benefit from the knowledge that will be gained in the next decade with the data of AugerPrime and \TAxFour.

\begin{figure}[!htb]
  \centering
  \includegraphics[height=6.5cm]{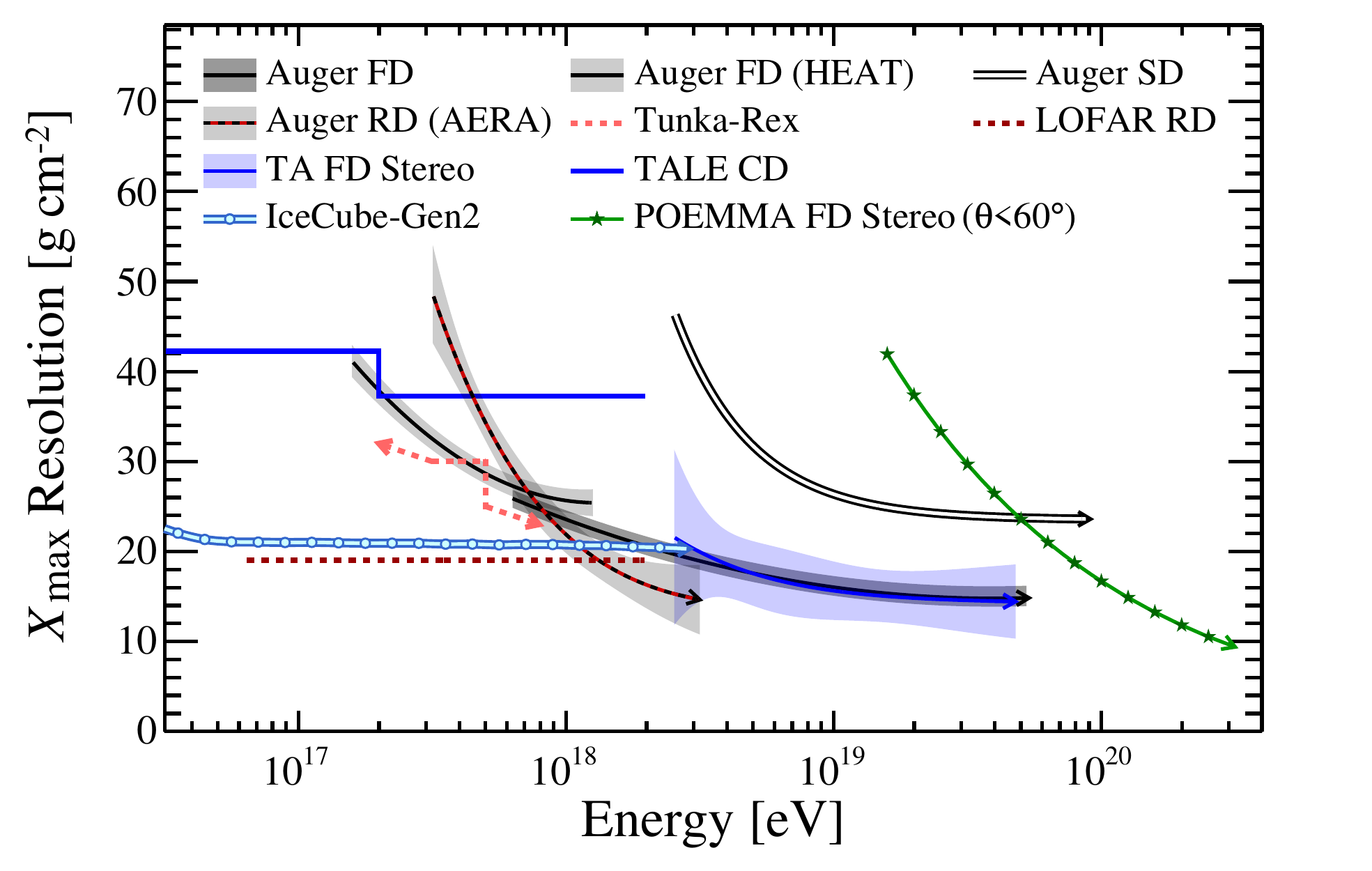}
  \caption{The \xmax{} resolutions reported by current experiments and estimated for some future observatories \cite{Bergman:2021djm, TelescopeArray:2020bfv, Bellido:2017cgf,  PierreAuger:2021rio, Corstanje:2021kik, Bezyazeekov:2018yjw, PierreAuger:2021xnt,  Anchordoqui:2019omw, private:AlanColeman}. Reported data points have been interpolated for Auger RD, \ac{TA} Stereo and \ac{POEMMA}. For \ac{LOFAR}, Tunka-Rex, and \ac{TALE}, only average resolutions for energy ranges were reported.
  }
  \label{fig:ResolutionVsScience}
\end{figure}

Depending on the methods and designs of the next generation of detectors, there are different types of composition related studies which can be pursued. Generally, these can be sorted into two groups. Those which, by their nature, require an event-by-event sensitivity to the mass group of the primary, and those which can be done through the analysis of observables with a moderate mass sensitivity. A non-exhaustive list of both types of studies follows:
\begin{enumerate}
    \item Moderate resolution mass composition analyses:
    \begin{itemize}[topsep=2pt]\setlength\itemsep{-0.2em}
        \item constraints on the \xmax{} and \nmu{} mass scales;
        \item \xmax{}/$\ln A$ moment and elongation rate studies;
        \item fitted mass group component fractions and energy spectra;
        \item mass composition anisotropy studies from \cref{sec:MEAD} of types \ref{MassSplit} to \ref{GrandCombined};
        \item \ac{GZK}/Photo-Disintegration/Peters-Cycle cutoff scenario differentiation;
        \item constraints on acceleration scenarios and composition at source;
        \item constraints on cosmogenic \ac{UHE} neutrino and photon fluxes;
        \item UHE neutrino and photon searches.
    \end{itemize}
    \item Event-by-event mass composition analyses:
    \begin{itemize}[topsep=2pt]\setlength\itemsep{-0.2em}
        \item generally higher fidelity versions of the studies above;
        \item mapping of individual mass groups;
        \item event-by-event \ac{GMF} inversion and source identification;
        \item better proton/air cross section measurement;
        \item determination of the \xmax{} and \nmu{} mass scales;
        \item expanded searches for new physics at ultra-high energies.
    \end{itemize}
\end{enumerate}

\begin{figure}[!htb]
  \centering
  \includegraphics[height=6.7cm]{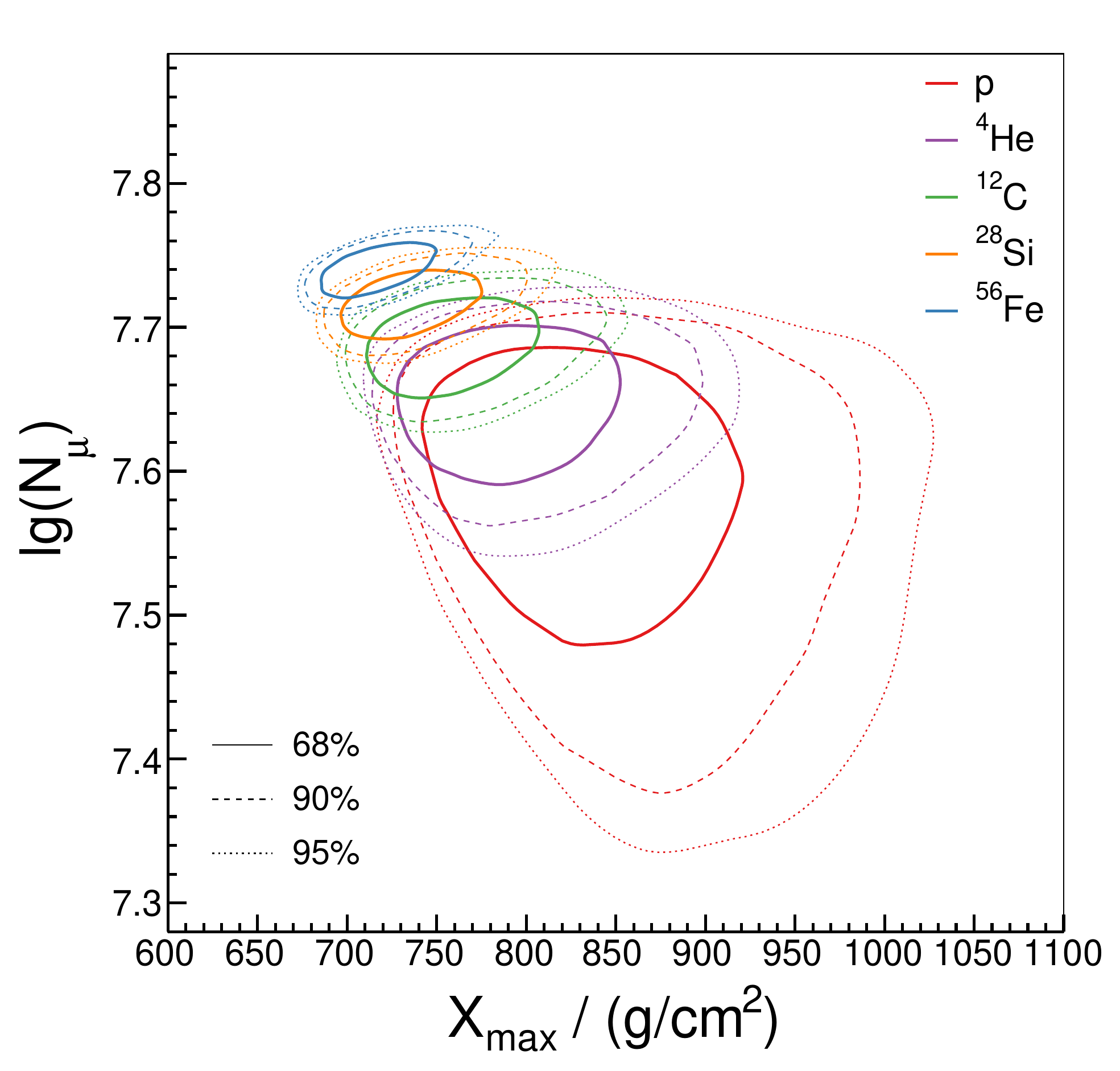}%
  \includegraphics[height=6.7cm]{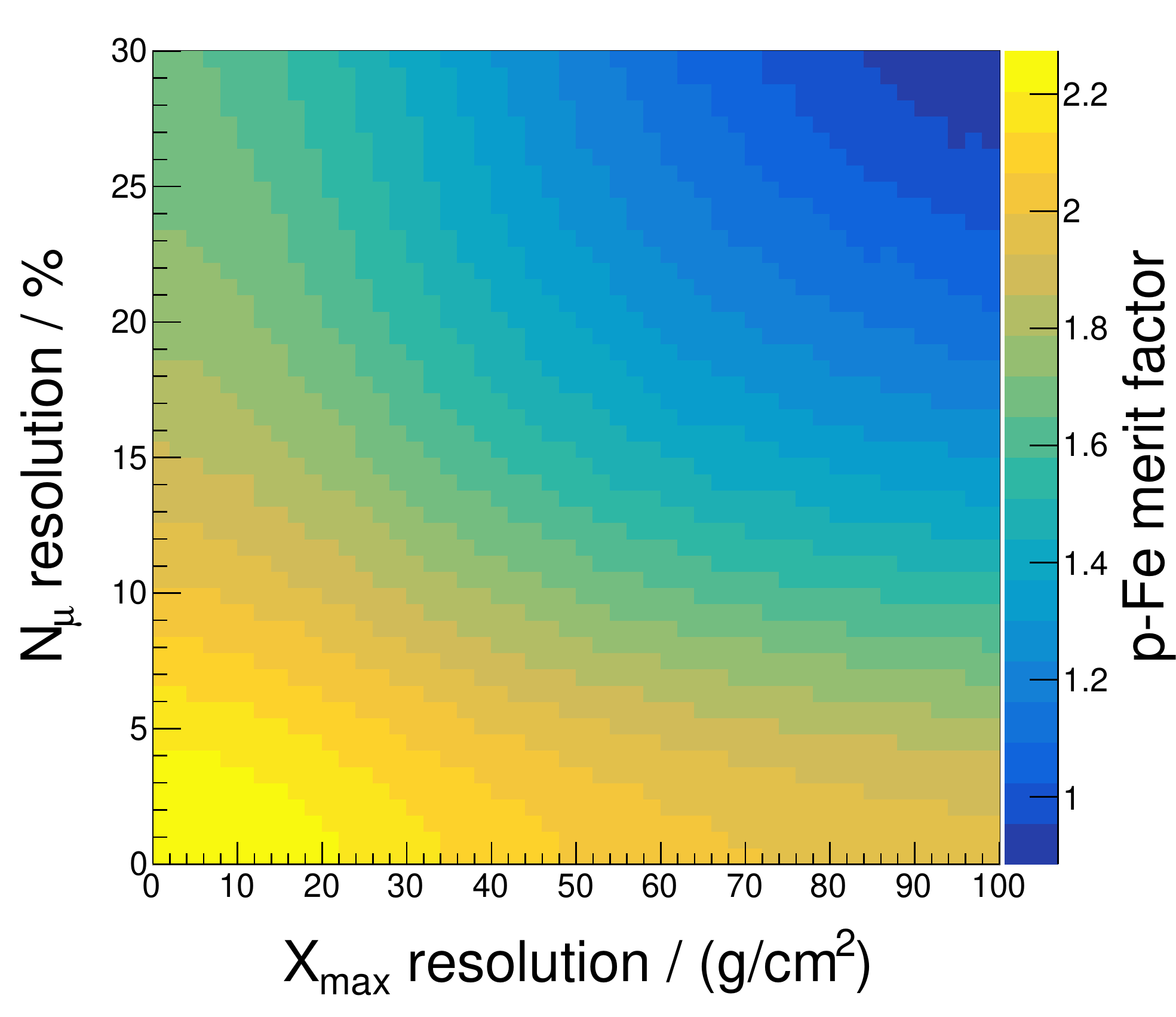}
  \caption{{\itshape Left}: shower maximum, \xmax, and logarithmic
    muon number, $\lg(\nmu{})$, for five cosmic-ray
    elements for \sibyll{2.3d} at 10\,EeV. The lines show the contours
    containing 68, 90 and 95\% of all showers. {\itshape Right}:
    ``merit factor'' which illustrates degree to which proton and iron showers can be distinguished,
    $\text{mf} = (\mu_\text{p} -
    \mu_\text{Fe})/\sqrt{\sigma_\text{p}^2 + \sigma_\text{Fe}^2}$
    for a given experimental resolution on \xmax and $\lg(\nmu{})$. $\mu$
    denotes the mean and $\sigma$ is the standard deviation of the
    mass observable, which in this case is the linear combination of
    \xmax and $\lg(\nmu{})$ maximizing $\text{mf}$. The merit factors
    for a single observable are close to the values shown for a large
    resolution of the other one, i.e., the top row for \xmax and the
    right column for $\lg(\nmu{})$.}
    \label{fig:MassResByObservable}
\end{figure}

The vast majority of composition studies which informed the review in \cref{sec:CurrentStatus} are of the first variety. This is because currently, \xmax{} is the most sensitive mass related parameter available for composition studies. As can be seen in \cref{fig:ResolutionVsScience}, the current and future resolutions on \xmax{} are already on the order of 20\,\gcm{} or better, which, as can be seen on the right of \cref{fig:MassResByObservable}, only marginally contributes to lowering the overall mass resolution. This is because the location of \xmax{} is subject to large shower-to-shower fluctuations for any given primary and energy. This is clearly illustrated on the left of \cref{fig:MassResByObservable} where a significant overlap in the \xmax{} distributions of adjacent mass groups is visible, with protons and iron overlapping at the 1.5\,$\sigma$ level. This means that an event-by-event discrimination between mass groups is challenging with \xmax{} alone.

The reconstruction of \nmu{} promises a much higher event-by-event mass resolution due to the much lower shower-to-shower fluctuations in the number of muons produced in a shower for any given primary and energy. This increase in sensitivity is clearly visible in the separation of elements along the ordinate of \cref{fig:MassResByObservable}~(left). 
However, there is still significant overlap in the distributions meaning that high certainty event-by-event mass reconstruction is still not obtainable with \nmu{} alone. As was discussed in \cref{sec:tensions}, the interpretation \nmu{} is complicated by high uncertainties in the muon scale due to problems with the current generation of hadronic interaction models, leading to unreliable results as compared to \xmax{} related studies.

The clearest path to event-by-event primary mass reconstruction lies in a high resolution independent reconstruction of both \xmax{} and \nmu{} coupled to a high resolution energy reconstruction. Right now, the uncertainties in hadronic interaction models serve as an effective barrier to decoupling the reconstructions and interpretations of \xmax{} and \nmu{}. Unfortunately, the current low event-by-event primary mass resolution of \ac{UHECR} events also serves to hinder progress on refining hadronic interaction models due to the large uncertainties it creates in the constraints \ac{UHECR} events can provide at the highest energies. This leads to a difficult to resolve mass-hadronic model interdependency, which means an iterative approach will be necessary. However, once the heaviest mass group can be identified and a high resolution \nmu{} measurement can be made, very strong constraints on muon production will be available which should significantly contribute to solving the Muon Puzzle.

As stated in the list from the beginning of this section, both studies with moderate mass sensitivity and event-by-event mass-resolution can allow for significant progress on the most important questions currently being posed in \ac{UHECR} and \ac{UHE} particle physics. Event-by-event detection will always provide a superior resolution and stronger constraints than statistical methods can. However, less sensitive methods can have large impacts at the highest energies if sufficient statistics and \xmax{} resolutions are achieved. This is particularly true if the trend of an apparent purification of primary beams with energy continues as energy increases \cite{Yushkov:2020nhr}, or alternatively if the composition approximately bifurcates into distinguishable very heavy and very light components due to propagation effects on distant sources, the so-called `cosmic mass degrader' scenario described in \cref{sec:MassModelIndependent}. If either of these cases occur, then beyond cut-off energies, most composition-dependent questions on source types and acceleration/propagation scenarios can be likely answered with profile measurements alone. This provides a good target for large aperture \ac{FD} and space-based detectors, without assuming resolution gains from advanced profile reconstruction techniques. However, if both of the above scenarios prove false, and the composition of the flux at the highest energies is both heavy and significantly mixed, very large detectors with event-by-event mass sensitivity will be required to fill out the \ac{UHECR} picture. Additionally, making any new progress below $\sim$\,$40$\,EeV, will also require event-by-event mass reconstruction with large exposure as the limits of what can be done using statistical methods is being reached by the current generation of cosmic ray observatories. 
Understanding of the degree of mixing at the highest energies and therefore knowledge of the degree of mass resolution needed to progress at energies above the flux suppression will need to wait a few years until AugerPrime and \TAxFour have collected enough data to constrain 
the mass composition at ultrahigh energies.

\subsubsection[Anisotropy]{Anisotropy: Towards the discovery of the sources}

In the next decade, the Pierre Auger Observatory will make sufficiently precise measurement of the composition so that it will be clear whether a significant fraction of protons or other low-$Z$ cosmic rays persists to the highest energies. The measurement of the spectrum of such a component will be also carried out, giving information on how extended in energy this fraction is. 
This spectral/composition information will be complemented by the major increase of statistics of events in the northern hemisphere from the data to be collected by \TAxFour (see \cref{sec:TAx4}). 
Depending on the outcome of the upgrades, the essential attributes of future detectors will differ. However, by evaluating the different possibilities, definite perspectives can be drawn.

\subparagraph{Large-scale: going to full-sky coverage}
Regarding large-scale anisotropies, full-sky coverage with detectors having as close to identical energy calibration in the northern and southern hemispheres as possible, to avoid spurious effects from an inconsistent energy threshold, is essential to reconstructing the spherical harmonics -- the most basic characterization of anisotropy, yet presently out of reach.  The data sets of Auger and \ac{TA} have been combined by cross-calibrating the energies in the overlap region but the statistics in this region are limited and an accurate calibration correction should ideally be done for each energy bin.  Therefore an observatory capable of full-sky surveys, such as \ac{POEMMA} (see \cref{sec:POEMMA})~\cite{POEMMA:2020ykm} or, if ground-based, using the same technology in both hemispheres such as \ac{GRAND} and \ac{GCOS} (see \cref{sec:GRAND} and \cref{sec:GCOS})
\cite{GRAND:2018iaj, Horandel:2021prj}, will thus greatly reduce systematics in measuring large-scale structures in the \ac{UHECR} sky.  In addition, much larger statistics are needed.  To achieve a $5\sigma$ significance level for the dipole anisotropy in the energy bins $16<E/{\rm EeV}<32$ and $E>32\,$EeV requires double and triple the current Auger exposure, respectively; this must be achievable with the next generation observatories on a fast time scale if possible.

\begin{figure} [ht]
\centering
\begin{tabular}{cc}
\includegraphics[width=0.45\textwidth]{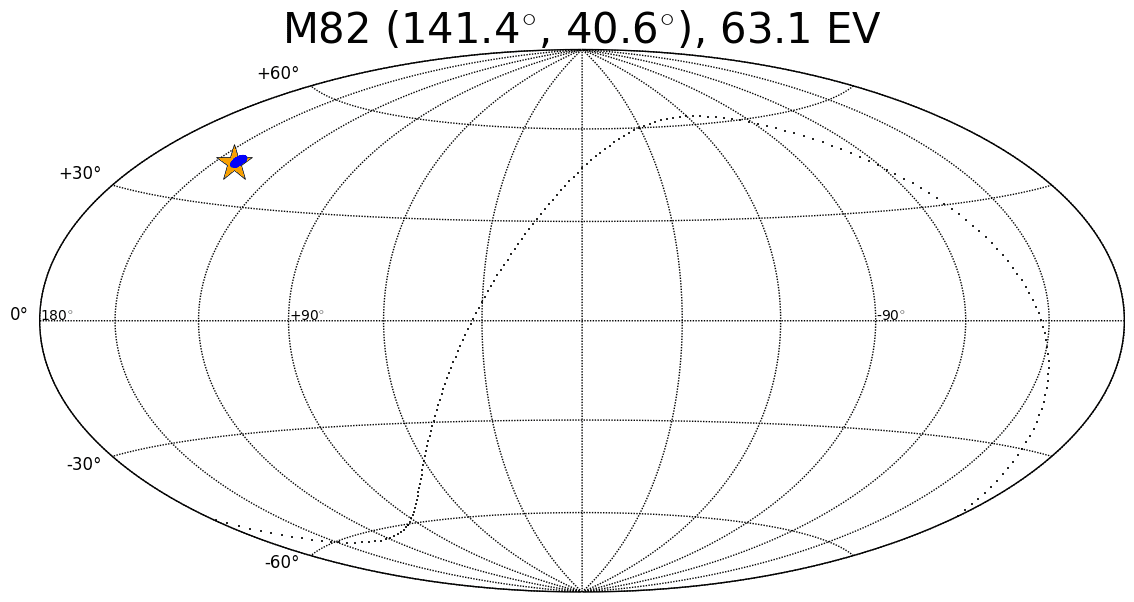} &
\includegraphics[width=0.45\textwidth]{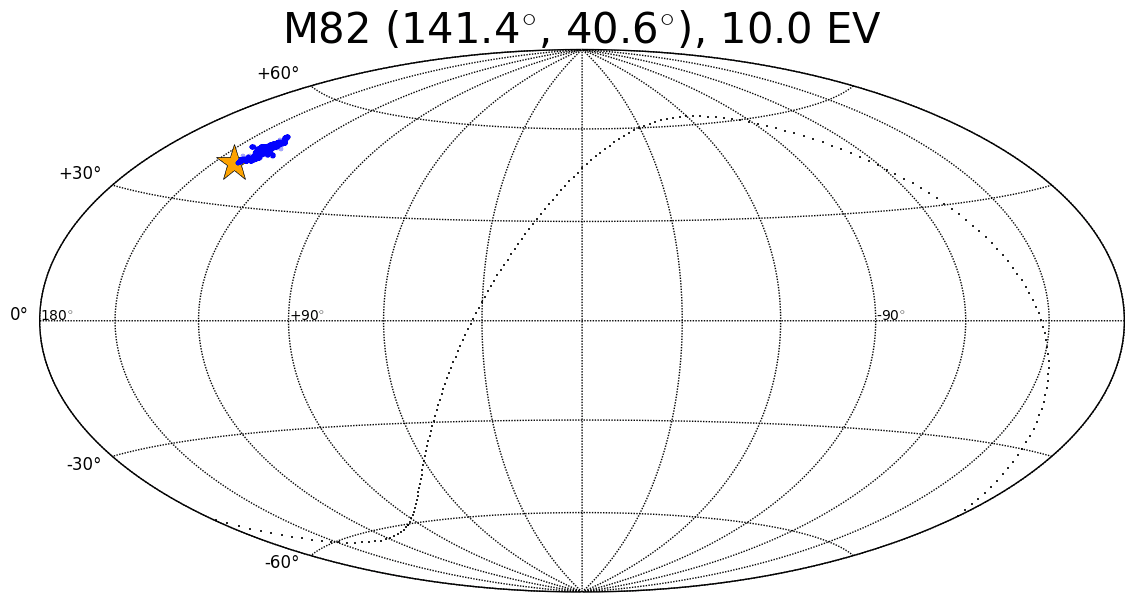}\\
\includegraphics[width=0.45\textwidth]{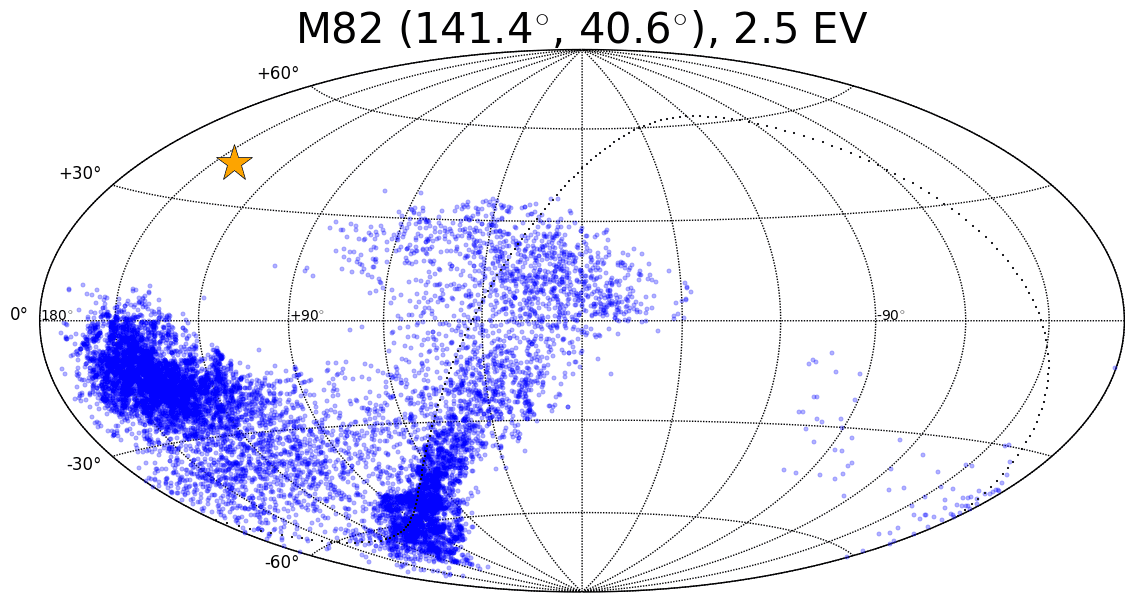} &
\includegraphics[width=0.45\textwidth]{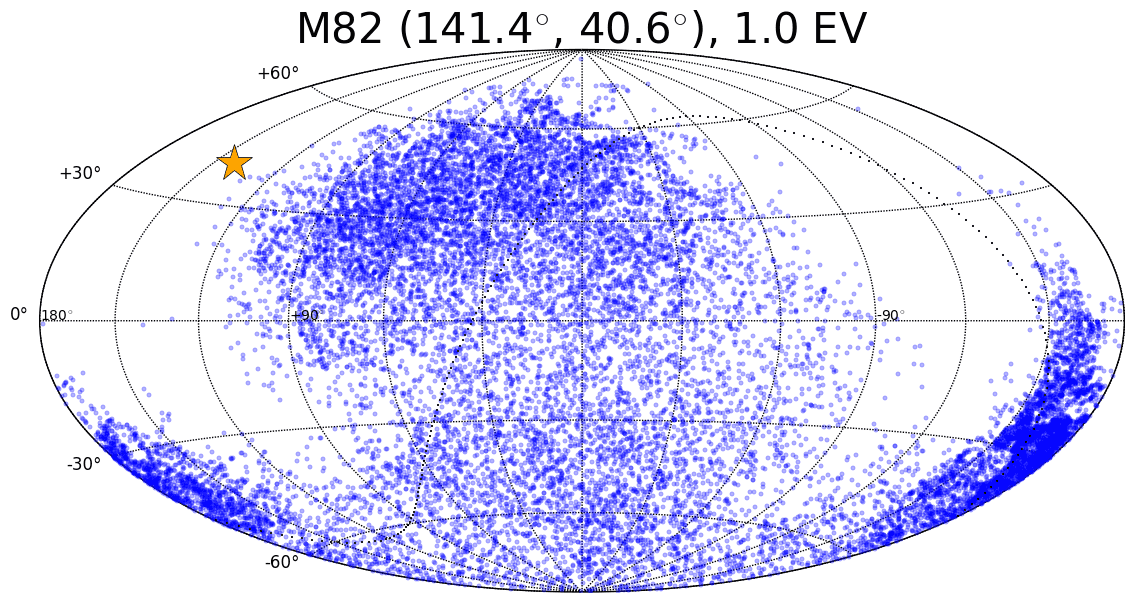}
 \end{tabular}
\caption {Skyplot showing the UHECR image of M82 for illustrative rigidities, after propagation through the JF12 magnetic field for a random component with a 30\,pc coherence length, from Ref.~\cite{Farrar:2017lhm}.}
\label{fig:M82}
\end{figure}

\subparagraph{Best-case scenario: A significant fraction of protons and Helium}
Regarding individual source discovery, current knowledge is consistent with two possibilities.  In the simpler scenario, a proton or light-nucleus component will be isolated at the highest energies and doing astronomy with charged particles, as long yearned for, could become reality.  In this case, next-generation observatories with larger apertures and a similar or better mass discrimination capability than achieved with AugerPrime, are needed to gather enough high-rigidity events to make a high-statistics skymap of their distribution.   Events arriving from the  hemisphere away from the Galactic Center (e.g., the \ac{TA} \emph{warm spot} candidates) surely experience smaller and less complicated deflections than \acp{UHECR} which have crossed the central region of the Galaxy, simply because the \ac{GMF} on average falls with distance from the Galactic center.   Thus restricting to high-rigidity events, individual sources should stand out over background, as seen in \cref{fig:M82} showing the image of M82, potentially the source of the \ac{TA} hotspot, for 4 different rigidities in the JF12 magnetic field model. 

Once sources are identified through their highest rigidity events, structure in lower rigidity events will constrain the \ac{GMF}. As the halo \ac{GMF} is better constrained in the anti-center direction, the great bulk of the halo \ac{GMF} becomes better determined as well, due to the approximate azimuthal symmetry of the Galactic halo.  This will calibrate, validate, or point to needed modifications of \ac{GMF} models.  Searches for individual sources in the hemisphere toward the Galactic Center (e.g., the Cen A region and other Auger over-densities) can then be interpreted with greater understanding and the anisotropy patterns toward the inner galaxy will further constrain the \ac{GMF} even in the central region of the Galaxy leading to a virtuous cycle of better-and-better ability to find sources.  

The shape of the rigidity spectrum from an individual source will indicate whether the \acp{UHECR} were produced by a transient or steady source, since the highest-rigidity events of a transient have already passed while the lowest-rigidity ones have not yet arrived, causing the spectrum to peak around some particular rigidity (which decreases, over thousand-year year timescales) rather than displaying the primary power-law behavior of the time-integrated spectrum~\cite{Waxman:1996zn}.

\subparagraph{The challenging scenario}
In the event that no clear proton-Helium component exists or it appears to be suppressed with growing energy, the path to the identification of \ac{UHECR} sources is similar but more demanding.  The dominant composition as determined by Auger has intermediate mass, say $Z\approx 6$, so the rigidity of a 60\,EeV particle is $\approx$\,10\,EV;  this is still high enough that deflections are small for sources away from the Galactic center, as illustrated for M82 in \cref{fig:M82}.
In this scenario, future observatories will ideally have still better mass discrimination and still higher statistics than needed for source discovery in the simpler case, because the lowest-$Z$, highest rigidity events remain the most powerful for finding sources, and mass-indicators tend to be sensitive to ln$A$. In this context, a much larger aperture than the existing detectors is even more valuable because the more complex the puzzle which must be unraveled, the more important it is to have large statistics, as will be the case with \ac{GRAND}, \ac{GCOS} and \ac{POEMMA} \cite{GRAND:2018iaj, Horandel:2021prj, POEMMA:2020ykm}. Moreover, \ac{GCOS} should be able to select events on rigidity and explore how rigidity-based selections can help to identify the sources.

In the end it is important to remind that the combination of neutrino, \ac{UHECR} and electromagnetic data --  multi-messenger astronomy using spectra and timing as well as arrival directions -- adds very powerful complementary information to the purely \ac{UHECR} anisotropy studies discussed here (see \cref{sec:NeutralParticles} and \cref{MM_outlook10yr} of this report). \textcolor{red}{\ac{UHECR} acceleration requires an unusual set of conditions that will necessarily impact gamma-ray and neutrino production. In conjunction with detailed studies of particle acceleration and transport (see discussion in \cref{subsec:acctheory}), observations in all three messengers will be crucial in determining whether candidate sources do produce \acp{UHECR} and in pinpointing the physical processes involved in their acceleration. Extending such studies to entire populations of candidate \ac{UHECR} sources will provide essential constraints on their contributions to the diffuse gamma-ray and astrophysical neutrino fluxes and the \ac{UHECR} spectrum.} Fully understanding \ac{UHECR} astrophysics is an ambitious goal, which we can hope to reach only by taking into account all the interconnections between different fields. Every step towards this, however, will be able to shed some new light on the physics of the most energetic particle accelerators in the universe. From this it is clear that in the next decades huge progress will be made, opening a new window for astrophysics.

\subsection[The big picture of the next generation]{The big picture of the next generation: Conclusions and recommendations}

Having the different scenarios in view, the ideal experiment of the next generation would combine huge exposure with the ability to measure the rigidity for every single cosmic-ray event. 
However, such an approach of one experiment for all scenarios and science goals will neither be economically attractive nor is it necessary. 
The community has proposed a small number of large-scale, but still feasible experiments which perfectly complement each other with their individual strengths and science goals.

In the ongoing decade, the Pierre Auger Observatory will remain the leading experiment with its AugerPrime upgrade in terms of exposure as well as accuracy for \ac{UHECR} primary mass. 
By its multi-hybrid design, it is also ideally suited to study the Muon Puzzle and, more generally, particle physics in \ac{UHE} air showers.
Auger in the southern hemisphere will be complemented by the \TAxFour upgrade of the \ac{TA} experiment. 
\TAxFour features a similar aperture as Auger, but in the northern hemisphere, though with a worse mass resolution than AugerPrime has. 
Nonetheless, it is essential, to have both experiments running in parallel for another decade, to harvest the science of the full sky coverage. 
In addition to their astrophysical goals, this high aperture will also serve the case of particle physics, e.g., for more precise measurements of the proton cross-section at the highest energies.

\begin{figure}[b]
    \centering
    \includegraphics[width = .99 \textwidth]{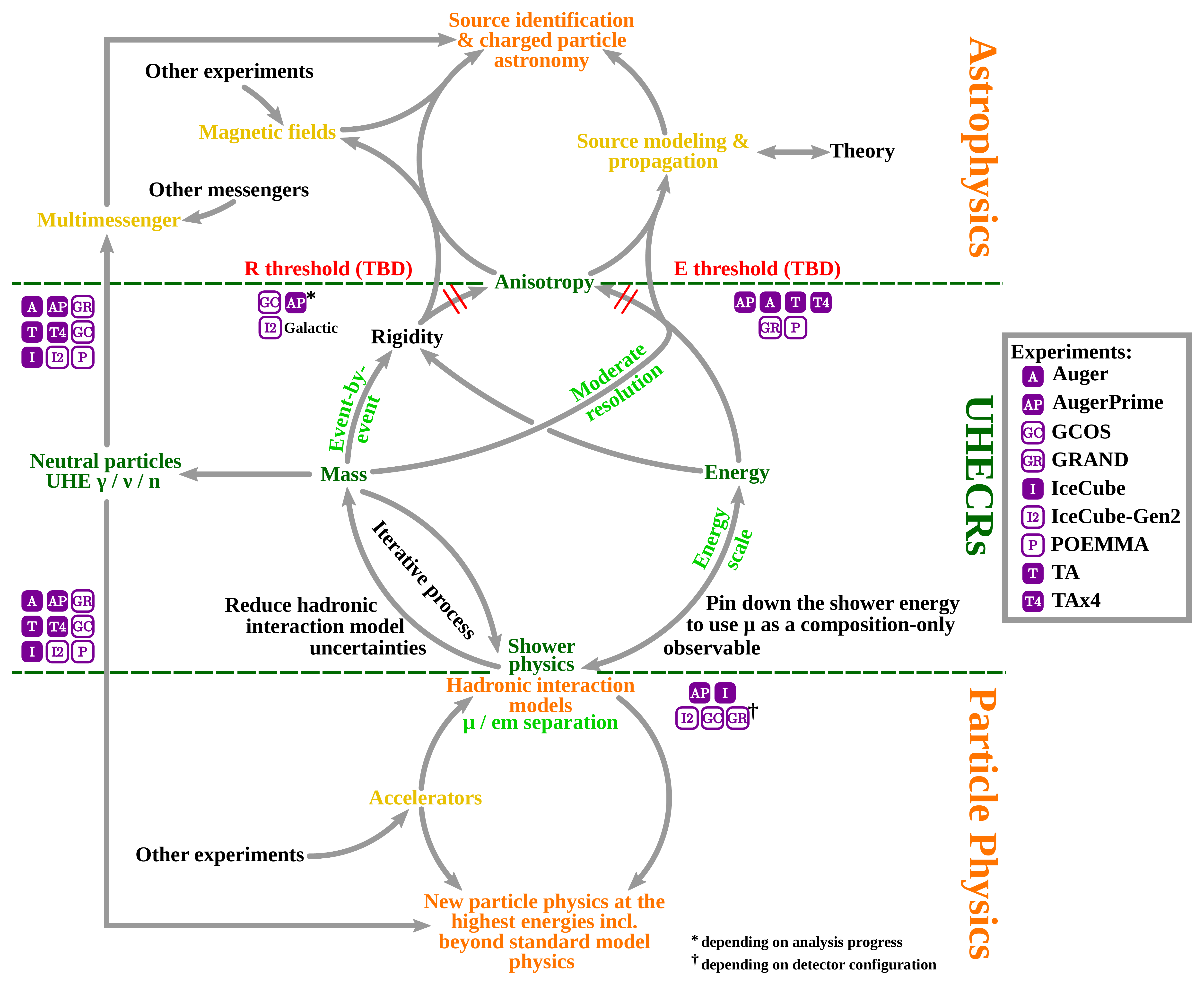}
    \caption{The main experiments of \ac{UHECR} physics in the coming two decades and how their particle and astrophysics goals interplay: Sufficient measurement resolution for the energy and mass of \ac{UHECR} is a prerequisite for any future cosmic-ray observatory. Currently, the mass resolution is hampered by an insufficient understanding of the hadronic interactions in air showers, but in addition to accelerator measurements, more accurate \ac{EAS} measurements themselves will provide the information needed to solve that puzzle. Once the hadronic interaction models are improved and compatible with data, they will provide the basis for the search of new particle physics on the one hand, and the ability to measure the rigidity of individual cosmic-ray particles with observatories featuring simultaneous muon and \xmax detection. This opens rigidity-enhanced anisotropy as additional way to search for the most energetic particle accelerators in the universe. This will complement the classical way of huge exposure observatories, which, by their unprecedented statistics, also will search for yet undiscovered ZeV particles and \ac{BSM} physics.}
    \label{fig:roadmap_with_experiments}
\end{figure}

Targeting somewhat lower energies up to EeV energies, IceCube and its extension IceCube-Gen2 play still a crucial role for the progress on \ac{UHECR}. By its combination of a surface array with a deep array measuring TeV and PeV muons, IceCube provides unique contributions to solve the Muon Puzzle and to study other unsolved problems in the physics of air showers, such as the production of prompt leptons. 
For this purpose, it is essential the surface array of IceTop is enhanced as planned by scintillators and radio antennas to deliver the maximum possible accuracy on the air showers producing the muons in the ice. 

Together with AugerPrime at higher energies, IceCube and IceCube-Gen2 thus provide the foundation to study and solve the puzzling discrepancies of state-of-the-art hadronic interaction models.
Improving these hadronic interaction models by utilizing muon and electromagnetic measurements of the same air showers at IceCube and Auger is a necessary foundation of both, deeper particle physics as well as astrophysics of \ac{UHECR} (\cref{fig:roadmap_with_experiments}).

Although Auger, and to some extent \ac{TA}, will keep leading the field of \ac{UHECR} physics in this decade, they fall short of statistics at the highest energies. 
Their exposure is insufficient to search for particles at ZeV energies or to resolve a larger set of individual cosmic-ray sources. 
These will be important goals of the next generation of experiments.

\begin{figure}[!htb]
    \centering
    \includegraphics[width = 0.5 \textwidth]{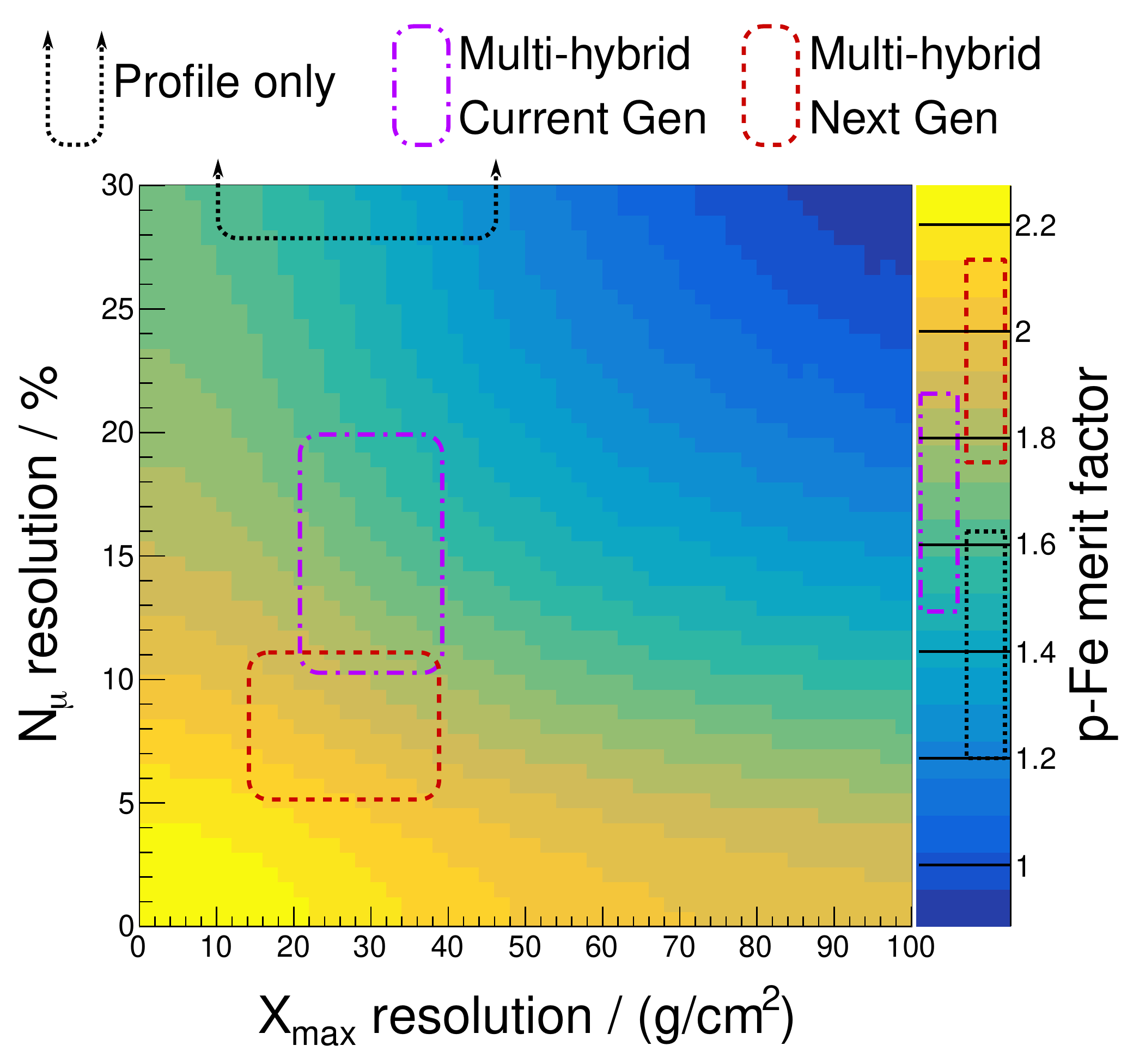}
    \caption{Estimated p-Fe merit factor ranges for various upgraded and next-gen type detector configurations based on reported designs and extrapolated from current methods. The profile-only measurements currently only leverage \xmax, however more mass information exists in the profiles than mass alone (ex.~$R$ and $L$ \cite{Andringa:2011zz}). The multi-hybrid values are derived from and refer to only \ac{SD} methods.}
    \label{fig:summary_FOM}
\end{figure}

With \ac{GRAND} on ground and \ac{POEMMA} in space, two experiments will aim at maximum exposure with a limited statistical mass resolution provided by \xmax (\cref{fig:summary_FOM}). 
They will provide the statistics needed to identify the accelerators of the most energetic particles in the universe and to search for particles at the ZeV scale and, thus, potentially new physics at the Energy Frontier.
Interestingly, these experiments have a very strong multi-messenger science case, as their primary science goal is to search for \ac{UHE} neutrinos using their huge exposure.
Therefore, their cosmic-ray science case can be realized for a relatively small additional effort, and yet provides guaranteed progress in \ac{UHECR} physics.

They will be complemented by \ac{GCOS}, which plans to provide per-event rigidity information by combining \xmax and the electron-muon ratio as mass sensitive parameters in a multi-hybrid surface array. 
While the hybrid-technology \ac{GCOS} approach is certainly more expensive than single-technology arrays, it seems to be the only way to achieve the necessary measurement accuracy required for all science goals that need a per-event mass separation (see previous section).
It is therefore essential that the huge exposure experiments, will be complemented by \ac{GCOS}, which will have an order of magnitude larger exposure than Auger and feature even better mass and energy resolutions than AugerPrime does today.

A particular feature of both \ac{GCOS} and \ac{GRAND} is their multi-site approach. 
It is therefore possible that these two experiments will not be completely distinct from each other, but share one or even a few common sites.
This will have various benefits, being it a reduction of cost by sharing infrastructure or the ability to cross-calibrate each others measurements.

\begin{figure}[!htb]
    \centering
    \includegraphics[width = 0.99 \textwidth]{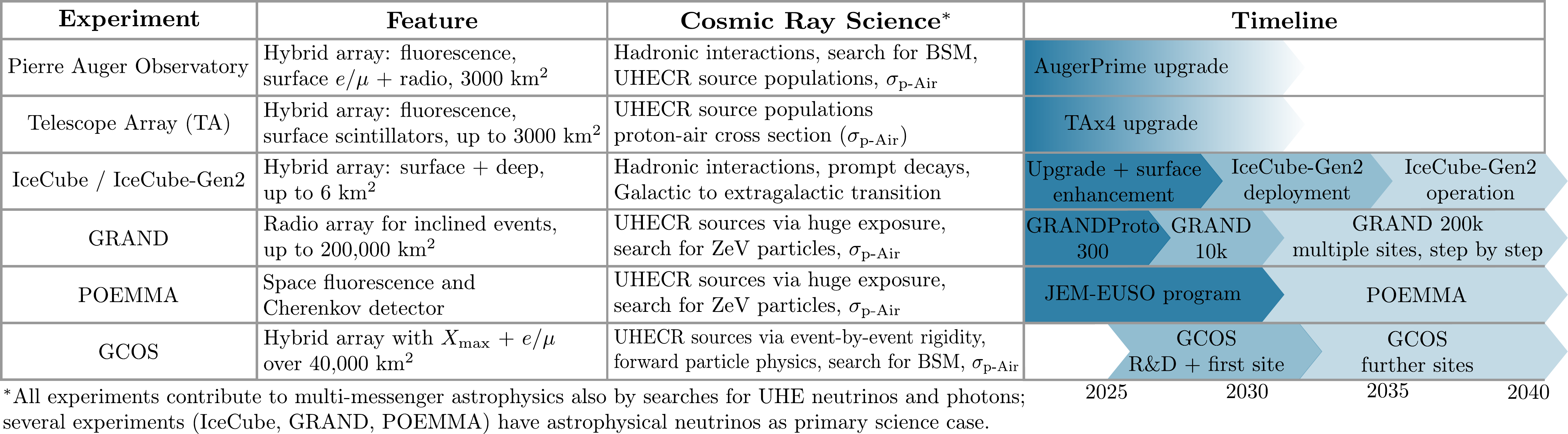}
    \caption{This table summarizes the major experiments that are expected to lead \ac{UHECR} physics in the next twenty years: Three current experiments with their ongoing and planned upgrades will be followed by three future experiments complementing each other with huge exposure and the ability to measure the rigidity of \acp{UHECR}.}
    \label{fig:summary_table}
\end{figure}

In summary, we reach the conclusions that are summarized in \cref{fig:summary_table} which lists the main and explicitly recommended experiments in \ac{UHECR} physics for the next twenty years.
There are further important experiments beyond those in the table, and some of them will have a leading or unique contribution to a specific science case (see \cref{sec:otherExperiments}). 
Our table should not be misunderstood as recommendation against such experiments, and some cosmic-ray experiments are simply not considered for the sole reason of focusing at lower energies than this white paper.
At the highest energies, the field of \ac{UHECR} will observe a transition to a new generation of large-scale experiments in the coming decade. 
Until that transition is made, it is essential to continue the currently upgraded observatories Auger and \ac{TA} into the next decade. 
Data taking at least until 2032 will be required to reach the full potential of the upgrades currently under construction. 

IceCube and IceCube-Gen2 have a special role because they target a lower energy range and are primarily neutrino detectors. 
Nonetheless, they provide unique cosmic-ray science regarding the Galactic-to-Extragalactic transition and the study of hadronic interactions, which is also important to interpret measurements at higher energies correctly.
It is thus highly important for the \ac{UHECR} community that IceCube and IceCube-Gen2 are equipped with a high-quality surface array that enables this unique cosmic-ray science.

The new generation of \ac{UHECR} experiments is expected to go online in the 2030's. 
This will be \ac{POEMMA} as first space experiment that will drive \ac{UHECR} science, and the \ac{GRAND} array that for little additional effort will also provide huge exposure for cosmic rays in addition to its neutrino science case.
These need to be complemented by \ac{GCOS}, the only next generation experiment featuring the accuracy for per-event rigidity. 
To maximize the outcome of \ac{GCOS}, it needs to be preceded by appropriate R\&D during this decade, e.g., by testing and calibrating \ac{GCOS} detectors at a sufficiently large scale at the Auger site for example. 

\textcolor{red}{On a final note, the processes of designing experiments and planning observations must expand to consider particularly timely concerns in order to fully realize the science return of these efforts. For instance, theory should be incorporated into these processes as it provides testable predictions that drive experimental design (e.g., \cref{fig:roadmap_with_experiments}). To that end, opportunities for closer collaborations between theorists and experimentalists should be explored. Furthermore, it is imperative that collaborations adopt inclusive, equitable, and accessible practices to empower all members of the scientific community to contribute at their fullest potential (see \cref{sec:Diversity} and \cref{sec:OpenScience}).} Lastly, as discussed in \cref{sec:CarbonFootprint}, it is critical that environmental concerns, and, in particular, the \COT{} footprint of their construction and operation be given weight while planning out the design and construction of these future detectors.

\fakesection{Broader scientific impacts}
\vspace{3cm}
{\noindent \LARGE \textbf{Chapter 7}}\\[.8cm]
\textbf{\noindent \huge Broader scientific impacts:}\\[3mm]
\textbf{\LARGE  Leveraging our infrastructure for other fields}
\label{sec:SuppSci}
\vspace{1cm}


Very large aperture fluorescence and Cerenkov telescopes with highly dynamic electronics allow for the detection of phenomena of an entirely different class and nature from \acp{UHECR}. The active fields of investigation which leverage the data of \ac{UHECR} observatories include astrobiology, lighting science, meteor investigation, dark matter, aurorae, and airglow observations among others. The reason for this is two-fold, first \ac{UHECR} experiments require extreme sensitivity in light intensity, with the possibility of detecting even single photons, and ultra-fast read-out electronics which can reach 100\,ns time resolutions. Second, 
the enormous extensions of the detector arrays, that reach thousands of square kilometers on ground, and potential footprints approaching millions of square kilometers for planned space experiments, allow for the direct monitoring of huge areas and atmospheric volumes. These factors together mean that \ac{UHECR} observatories often times meet or exceed the capabilities and sensitivities of experiments dedicated to the above fields of study for certain analyses. By acknowledging this reality, and keeping it in mind for the design of the next-generation detectors, the science reach of \ac{UHECR} experiments has been, and can further be, extended well beyond the realm of cosmic rays and related fundamental physics. In the following, a summary of some of the contributions that \ac{UHECR} experiments have provided in the past and might provide in the future is presented. 

\subsection{Astrobiology}
\label{sec:astrobiology}

The search for life beyond Earth requires an understanding of life itself as well as the nature of the environments that support it.
In particular, a key environmental factor to consider is the level of background ionizing radiation. As explained earlier, when cosmic rays interact with planetary atmospheres or the surfaces of small bodies such as moons, asteroids or comets, they initiate extensive showers of secondary particles. Through these showers, cosmic rays can lead to a host of interesting effects potentially relevant for habitability \cite{2011AsBio..11..551D, Griessmeier:2015zna, griessmeier2016}, such as:
\begin{itemize}[topsep=2pt]\setlength\itemsep{-0.2em}
\item the modification of the atmospheric chemistry,
\item an influence on atmospheric lightning,
\item the production of organic molecules within the atmosphere or at planetary surfaces, 
\item the destruction of stratospheric ozone, 
\item the sterilization of planetary surfaces and environments, 
\item the degradation of biosignatures.
\end{itemize}
However, beyond their impact on the habitability of different environments, cosmic rays also directly influence the path life takes once it appears and can even have a hand in its formation. The impact of \acp{UHECR} on clouds, their influence on lightning and their ability to obscure bioluminescence (a potential biosignature) are treated in separate sections. In this section, the direct influence of cosmic rays on living organisms, and their potential role in the emergence of life is outlined. In particular, the influence of cosmic rays on the growth and evolution of living organisms through their effects on mutation \cite{ferrari2009}, and how spin-polarized cosmic muons may induce enantioselective mutagenesis leading to the emergence of biological homochirality \cite{Globus:2020gud} are covered. 

\begin{figure}[!htb]
\centering
    \includegraphics[scale=0.27]{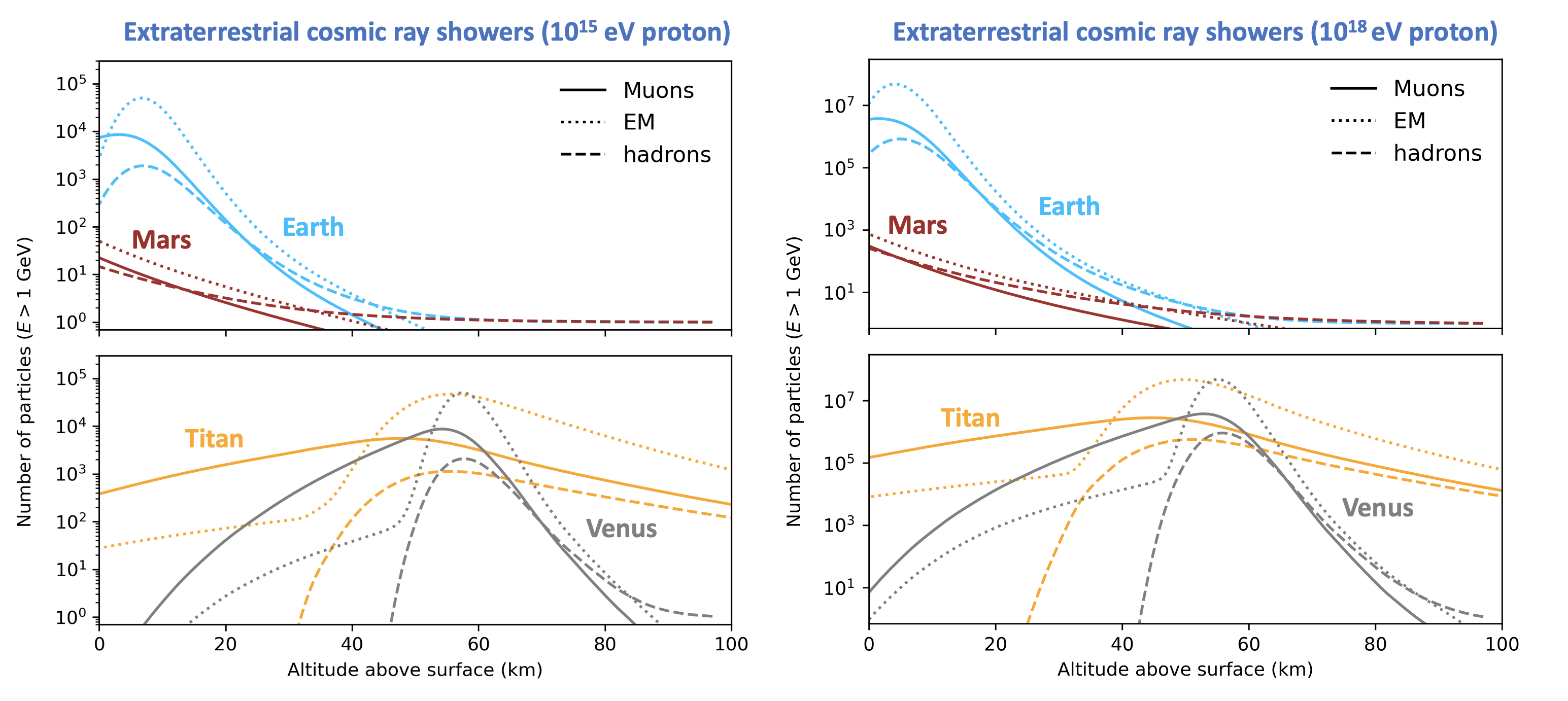}
\caption{Number of particles as a function of the altitude above the surface initiated by a proton at 1\,PeV (left) and 1\,EeV (right), on Earth (blue), Mars (red), Titan (yellow) and Venus (grey). On Venus, it is only for EeV proton that the secondary particles (muons) reach the surface. Therefore in certain environments only the showers induced by the highest energy particles might affect the evolution of life forms. Calculations from Ref.~\cite{globus2021}.
    }
    \label{CRfluxes}
\end{figure}

Cosmic rays have a direct influence on life because, even at modest intensity, these particles promote mutations and force the exploration of different evolutionary pathways which is necessary for adaptation of living organisms. Also, when the particle flux is high enough, it is destructive and can create sterile environments or apply a strong selective pressure.
There are two alternative modes of interaction of radiation with biopolymers: either directly via ionization, or indirectly via interactions with radicals produced by radiolysis of cellular water molecules \cite{kiefer1996}.
While rare and cataclysmic events such as supernovae, or (rarer) binary neutron star mergers, gamma-ray bursts, have been invoked as a limiting factor for life \cite{ellis1995,dar1998}, such events would not severely affect the majority of marine or underground life. Furthermore, these high energy events would have a dominant impact through their muons \cite{Thomas:2016fcp, melott2017} whose potential role in astrochemistry and astrobiology has been, so far, overlooked. As the number of secondary particles is proportional to the energy of the primary cosmic rays (see \cref{CRfluxes} for illustration), it is only underground and underwater that the effect of the \acp{UHECR} become relevant through their muon production. Even if the flux is smaller, if there was an elevated level of \acp{UHECR} for some period of time, it would have a greater impact, because densely ionizing radiation is more efficient in inducing damages in comparison with sparsely ionizing radiation. 

Muons are the only biologically significant cosmic radiation with energy sufficient to penetrate considerable depths, and they are, on average, spin-polarized \cite{Lipari:1993hd}. The mean energy of muons at the ground under contemporary conditions is $\sim$\,$4$\,GeV which is enough to penetrate a few meters of rock and several hundred meters of ice~\cite{globus2021}. In worlds with very dense atmospheres, such as  Titan and Venus, polarized muons dominate the radiation at altitudes around 50\,km (and interestingly this is the habitable layer in Venus clouds \cite{limaye2021venus}). The surface irradiation, comparable to that below 400\,m of rock, is negligible.

The fact that muons come from a decay involving the weak interaction have important consequences. Muons are, on average, spin-polarized as a consequence of parity violation. Interestingly, biological molecules also violate parity. Life has chosen one of two structurally chiral systems which are related by reflection in a mirror: DNA is made of sugar that all have the same chirality~\cite{takahashi2019}. The homochirality of organic molecules is a phenomenon only produced by life. Homochirality is thus a very unambiguous biosignature and its presence on an extraterrestrial body would be a powerful indicator of life~\cite{avnir2021}. It has been proposed that prebiotic chemistry produces both chiral versions of the molecular ingredients of life  and that at some stage in the early development of life, a small difference in the mutation rate has given a preference to one genetic polymer over its mirror-image~\cite{Globus:2020gud}.

The dynamics in mutation rate underlie evolution. If the organisms are subject to stress then the mutation rate become higher, so that the organisms are more likely to adapt. Cosmic radiation affects the mutation rate. Elevated level of \acp{UHECR} (from a local powerful accelerator such as a relativistic jets associated with core-collapse supernovae or a binary neutron star mergers) would increase the level of primary cosmic rays, and hence,  radiation doses due to secondary muons. Organisms living under rocks, under water and inside caves, which are well shielded from other forms of radiation such as ultraviolet light, are still subject to damage from muons~\cite{marinho2014}. Such shielded  environments are prime targets for the search of life in our solar system. For example, evidence has accumulated that subsurface liquid regions exist beneath the surface of Europa~\cite{zimmer2000Europa} and Enceladus~\cite{thomas2016enceladus}. Also recently, a 20-km-wide lake of liquid water has been detected in the martian undergound, at a depth of approximately 1.5\,km~\cite{orosei2018radar}. If microbial life started in similar hot springs~\cite{damer2020hot, teece2020biomolecules}, it is likely that after Mars’ geological death and the loss of its atmosphere, microbial life would have no longer be able to survive above ground, however it seems the Martian subsurface may have preserved the key ingredients to support life for hundreds of millions of years~\cite{checinska2019habitability}. Any punctuated elevated levels of muons may have an influence on chemistry and biology in these underground worlds. In the future, radiation damage experiments using laboratory techniques that mimic the interactions between cosmic muons and biomolecules would help to understand the role of secondary muons on the mutation rate and evolution of life. 

We hope this section demonstrated the importance of cosmic radiation in the origin and evolution of life. It is clear that through a better understanding the sources of \ac{UHECR} it will also become possible to have firmer grasp on historical exposure and fluctuation rate of our solar system to the \ac{UHECR} flux and thereby obtain a better understanding of the degree of influence \ac{UHECR} have had on the formation and evolution of life. To further explore these important ideas we strongly encourage future interdisciplinary workshops that would bring together biologists and cosmic ray physicists to discuss these important questions.

\subsection{Transient luminous events}
\label{sec:tle}

Atmospheric electricity drives a category of phenomena termed \acp{TLE}, which are ultimately associated with a parent thunderstorm, but can reach up to the lower edge of the ionosphere~\cite{Rodger1999,Pasko2011}. They include a variety of shapes and processes, ranging from upward leaders with streamer branches escaping from cloud tops (blue jets and gigantic jets), bunches of cold plasma streamers in the stratosphere and mesosphere (sprites), to large patches of diffuse emissions in the upper mesosphere (halos and ELVEs) above roughly 70\,km altitude where dielectric relaxation timescales suddenly drops. \ac{TLE} emissions include the near-IR to \ac{UV} range and radio signals, with typical light signal durations from a few hundreds of ms (gigantic jets), to tens of ms (blue jets and sprites), all the way down to 1\,ms (ELVEs)~\cite{Pasko2011}. Together with high energy emissions from lightning (or \acp{TGF}, see \cref{sec:tgf}), \acp{TLE} are a manifestation of the extraordinary impact of thunderstorms onto the Earth's atmosphere, and have as of yet unconstrained implications on atmospheric chemistry and the climate \cite{Vazquez2021}. In particular, low altitude \acp{TLE}, together with in-cloud streamer coronas and their associated UV and near-UV flashes, may exert a greater role in atmospheric chemistry than previously thought~\cite{soler2021}.

\ac{TLE} observations over the past decades have revealed the nearly global nature of these phenomena, which largely follow the distribution and seasonal modulation of lightning at a rate of a few \acp{TLE} every 1000 lightning flashes. According to the \ac{TLE}-dedicated \ac{ISUAL} global space mission~\cite{Chen2008}, ELVEs represent 80-90\% of all \acp{TLE}, occurring at a global rate of around once every 20 seconds. \emph{ELVEs} stands for Emission of Light and Very Low Frequency perturbation from \ac{EMP} Sources~\cite{GRL:GRL9448}), and are observed as rapidly-expanding (less than 1~ms) luminous circles of up to 300~km in diameter, and are caused by the interaction of an upward propagating \ac{EMP} with the lower edge of the ionosphere. ELVEs are associated with \acp{EMP} emission from powerful intra-cloud or cloud-to-ground lightning discharges~\cite{GRL:GRL53200}. Another phenomena, Sprites occur globally about once every two minutes. They are produced by the quasi-electrostatic field present in mesosphere, which has been enhanced by (mostly) positive parent cloud-to-ground lightning discharges with high charge moment change. Such a field causes the ignition of streamers at about 70\,km altitude, which then extend downward towards the cloud top over a few microseconds to a few tens of microseconds and then propagate upwards as diffuse emissions~\cite{pasko1998,cummer2006}, and are often accompanied by the diffuse ionized patch of a halo. Blue starters and blue jets ascend from cloud tops reaching 20--30 (starters) or 40--50 (jets) km altitude, emerging as a leader accompanied by bunches of streamers at its head~\cite{Wescott1998}. Similarly, the much rarer gigantic jets emerge above the thunderstorm top, but develop all the way to the ionosphere at about 90 km altitude \cite{Su2003}. This overall picture was supported by space missions, e.g., the current \ac{TLE} and \ac{TGF}-dedicated ASIM space experiment on-board \ac{ISS} \cite{Neubert2019}, as well as from ground networks of \ac{TLE}-dedicated low light sensitive cameras \cite{arnone2020} acting synergistically.

Because of a high sensitivity to \ac{UV} emissions in the atmosphere paired with high time resolutions, UHECR-dedicated observational experiments have proved to be able to greatly contribute to the study of \acp{TLE}. This is particularly true for their capability to record the dynamical evolution of ELVEs, even at their faintest threshold. Due to this, the \acs{TUS}~\cite{Klimov:2017lwx} and \acs{mini-EUSO}~\cite{Bacholle:2020emk} space missions, and the Auger Observatory on the ground have all made significant contributions to the study of \acp{TLE}. 

The Pierre Auger Observatory from its location in the Mendoza province of Argentina has a viewing footprint for ELVE observations of 3$\cdot$10$^6$ km$^2$, reaching areas above both the Pacific and Atlantic Oceans, as well as the C\'ordoba region, which is known for severe convective thunderstorms. Primarily designed for \ac{UHECR} observations, the Auger \ac{FD} turned out to be very sensitive to the \ac{UV} emission in ELVEs.
The first serendipitous observation of an ELVE candidate in Auger occurred in 2005 during the commissioning stage~\cite{Mussa:2012dq}. At the time, the criteria for rejection of close lightning were preventing the efficient detection of these phenomena. Nevertheless, further studies done in the following years lead to  the development of a simple selection algorithm and data-taking format dedicated to their observation, which was finally commissioned in 2013~\cite{Tonachini:2013lif}. The Pierre Auger Observatory reported observation of ELVEs from 2014 to 2016, recorded with unprecedented numbers at regional level (about 1,600) and time resolution (100\,ns) using the fluorescence detector~\cite{PierreAuger:2020lri}. It was found that within this 3-year sample, 72\% of the ELVEs correlate with the far-field radiation measurements of the World Wide Lightning Location Network. 
In 2017, the trigger was upgraded and the data taking format was further extended to detect light from the full region of maximum emission. From the new data it was found that the measured light profiles of 18\% of the ELVE events have more than one peak, compatible with intracloud activity~\cite{GRL:GRL53200}. Additionally, the fine time resolution of the \ac{FD} allowed for the first observations of triple ELVEs.
Starting 2021, the three \ac{HEAT} fluorescence telescopes, which overlook the array with angles between 30 and 60 degrees in elevation, started collecting data on ELVEs generated by lightning closer than 250\,km from the center of the array, which allows for more detailed studies of the region of maximum emission. To the best of our knowledge, the Pierre Auger Observatory is the only facility on Earth that both measures ELVEs year-round and has full coverage of the horizon.

\begin{figure}[!htb]
	\centering
	\includegraphics[width=.32\textwidth]{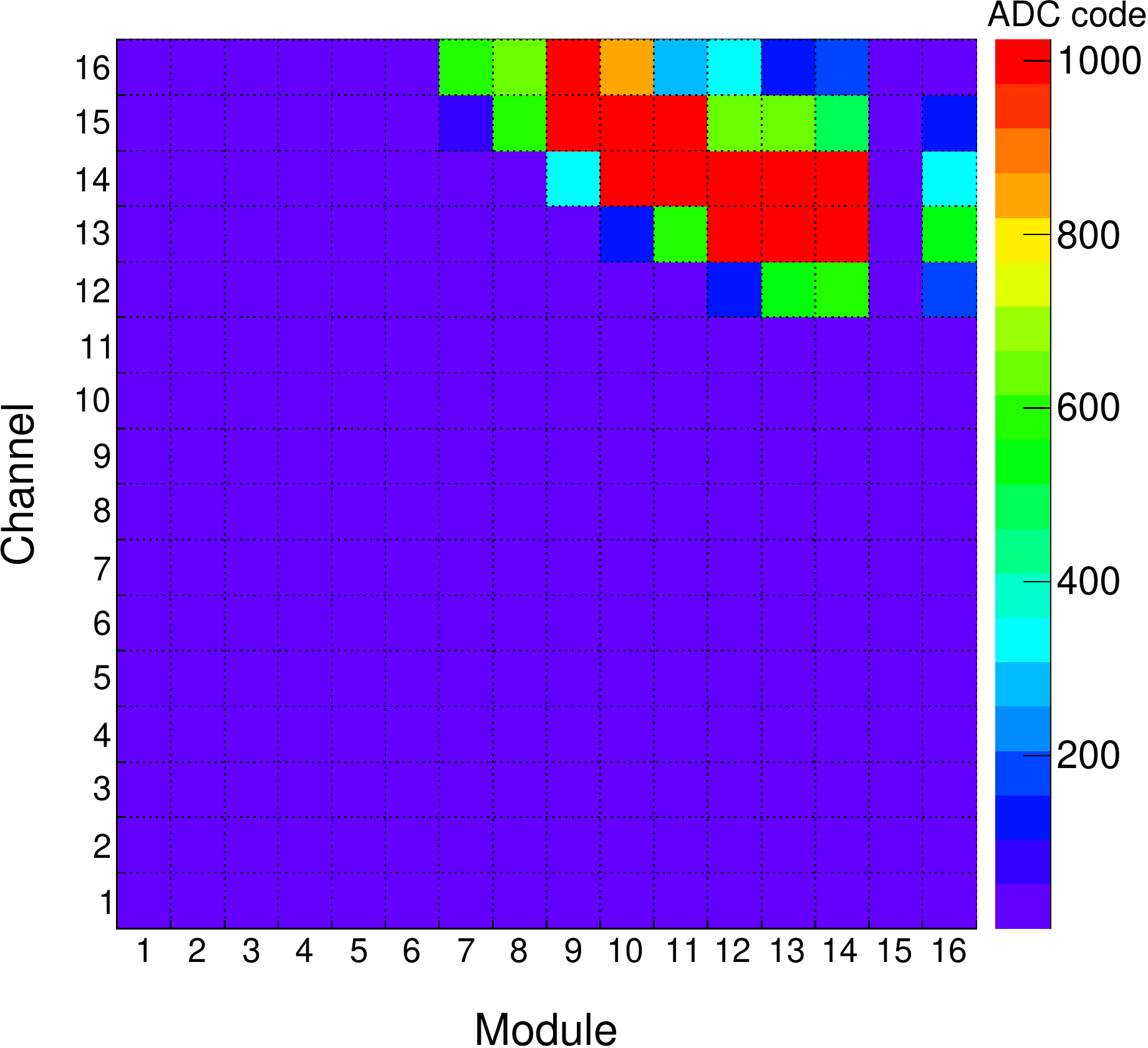}
	\includegraphics[width=.32\textwidth]{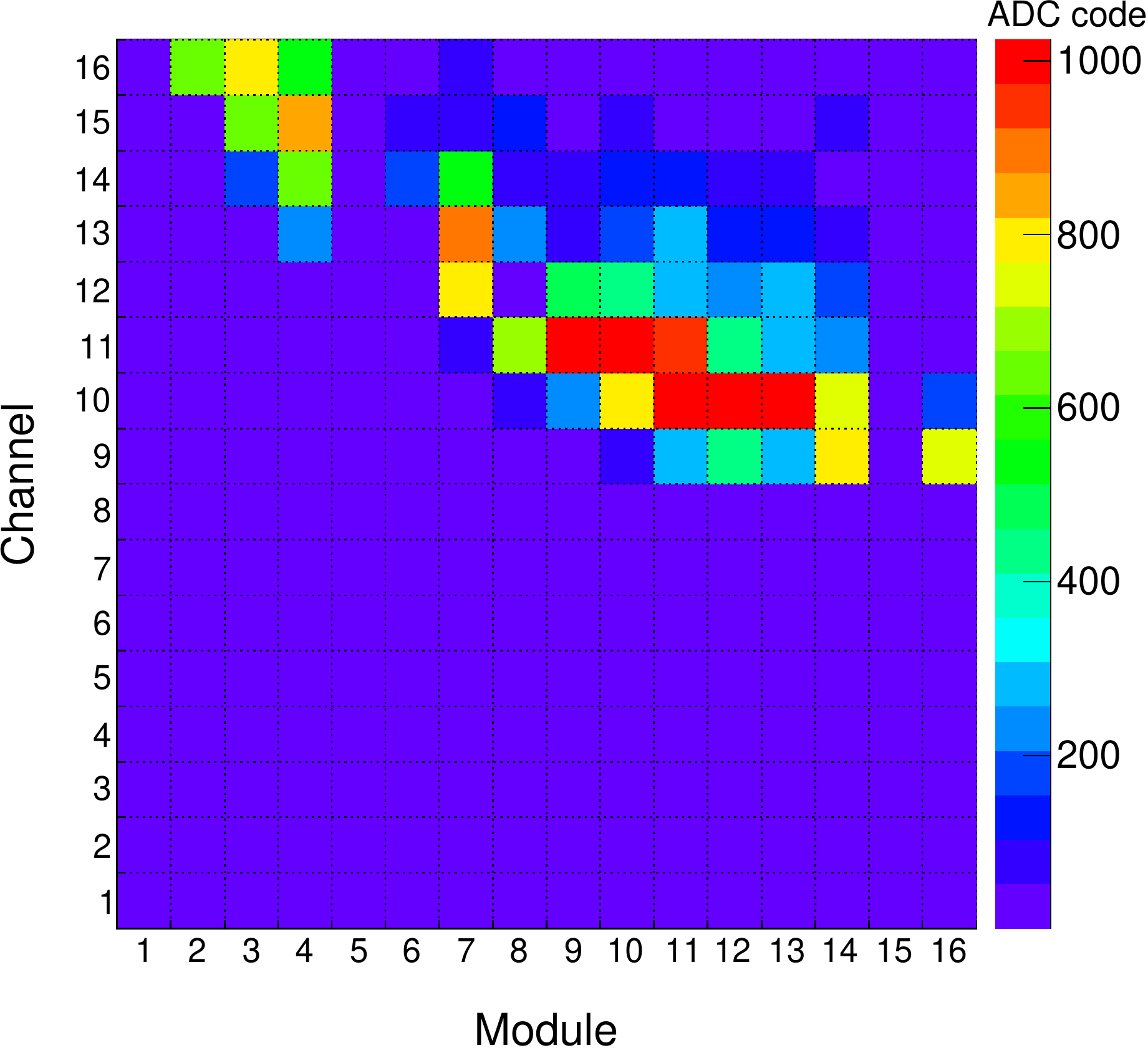}
	\includegraphics[width=.32\textwidth]{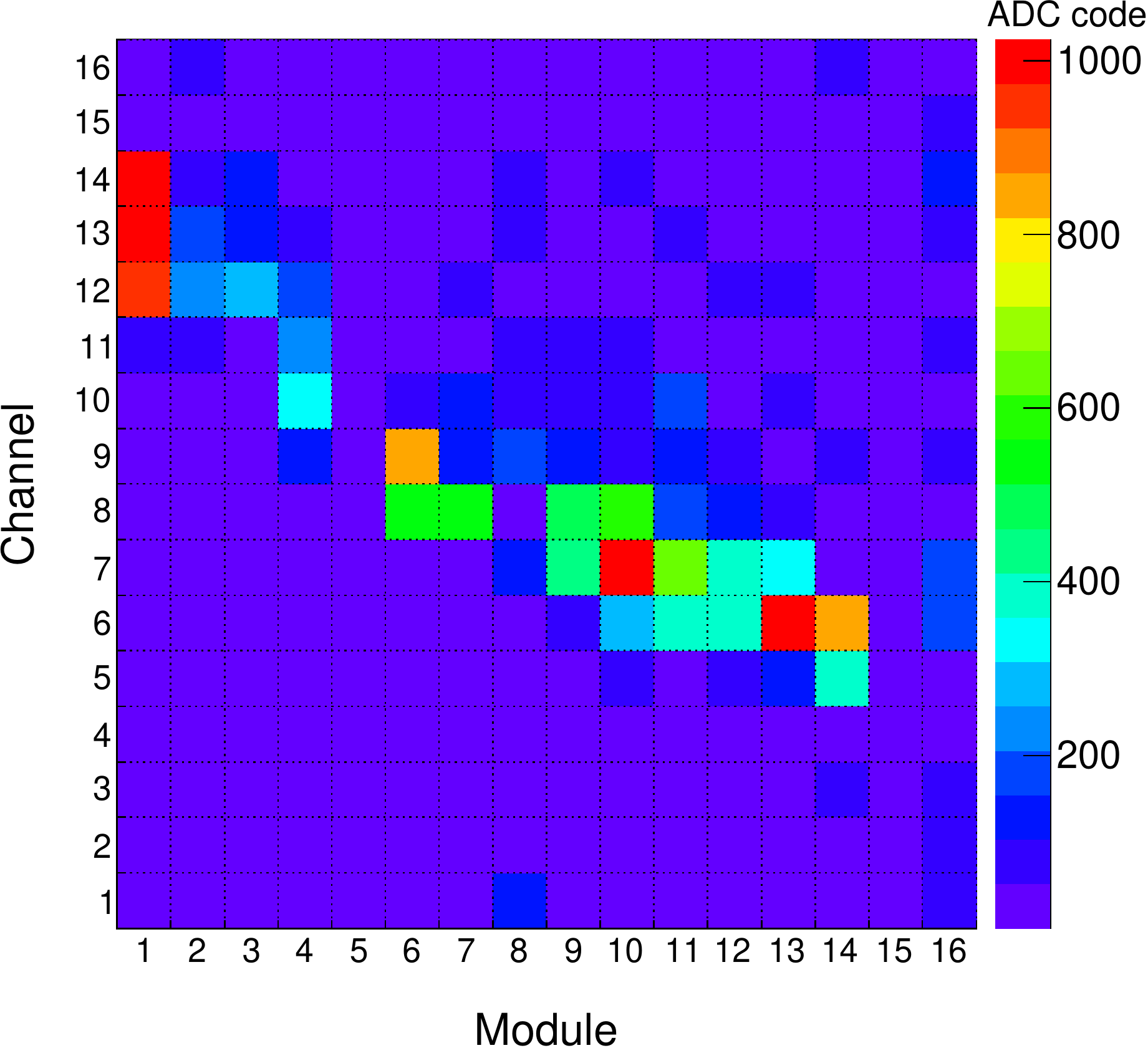}
	\caption{Snapshots of the focal plane show the arc-like shape and 	movement of an image of the ELVE registered on 23/08/17,
	in the detector \ac{FoV}. The snapshots were taken at 136\,$\mu$s, 168~$\mu$s and 200~$\mu$s from the beginning of the record. Colors denote the signal amplitude.}
	\label{fig_ELVE-map}
\end{figure}

UHECR experiments in space, such as the \acf{TUS} detector, are also able to record and classify various \ac{UV} transient events in the atmosphere. \ac{TUS} had several modes of operation with different temporal resolutions which allowed for the measurement of events in a wide range of time scales: from \acp{EAS} with durations of a hundred microseconds and time resolution of~0.8\,$\mu$s, up to maximum durations of 1.7\,s and time resolutions of 6.6\,ms. A total of 25 ELVEs were found in the \ac{TUS} data~\cite{klimov2019remote}, including ELVEs with $\sim$\,$4$ orders of magnitude in brightness less than those measured by previous space based experiments~\cite{Chen2008} (see Fig.~\ref{fig_ELVE-map}). In fact, the large aperture of the \ac{TUS} optical system allowed for the measurement of the faint emission from transient atmospheric events like ELVEs produced by lightning discharges with low peak current, pointing to a lower threshold in the lightning peak-current needed for ELVE production than previously thought~\cite{Barrington1999}.
\unskip
Interestingly, \ac{TUS} also made observations of so-called doublets, i.e., ELVEs with two rings~\cite{Newsome2010}. These phenomena in particular are easier to observed from space as nadir observations are more suitable for resolving the multi-ring structure of such ELVEs due to the simpler geometry and higher transparency of the upper atmosphere.

ELVEs have also been observed from space using the \acf{mini-EUSO} detector~\cite{Bacholle:2020emk}, a UV-telescope installed inside the \ac{ISS} in 2019 on the UV-transparent window of the Zvezda module, which is still taking data. \Ac{mini-EUSO} detects \ac{UV} emissions of cosmic, atmospheric and terrestrial origin on different time scales, starting from a few $\mu$s upwards. Due to its high spatial resolution ($\simeq$\,$4.7$\,km at the ionosphere altitude (90 km)), and sampling speed (2.5\,$\mu$s), \ac{mini-EUSO} is well suited for the observation of \ac{TLE} \ac{UV} emissions~\cite{Bacholle:2020emk, Marcelli:2021uX}. During the first year of data acquisition, \ac{mini-EUSO} detected 17 ELVEs, mainly in the equatorial zone, including three double-ringed ELVEs and one three-ringed ELVE. The analysis of the data acquired by the instrument makes it possible to reconstruct the expansion speed of single-ringed and multi-ringed ELVEs and thereby can help to shed light on the various phenomena involved in the multi-ring phenomena~\cite{Marcelli:2021uX}.

The sensitivity of \ac{UHECR} experiments may go beyond traditional \acp{TLE} to the detection and study of \emph{unusual} transient luminosity in the atmosphere, which emerge without any obvious connections to thunderstorm regions. These can occur with no powerful lightning near the events, nor at the conjugate region of the geomagnetic field~\cite{Charman1972, Elliot1972, Nemzek1989, Yair2005}. A dedicated analyses of the Vernov satellite~\cite{panasyuk2016relec} data was made to search for far-from-thunderstorm flashes~\cite{klimov2018uv}. A further six events with complicated temporal structure and not associated with lightning activity were found locally or at the geomagnetic conjugate point. The nature of such events is still unknown.

UHECR experiments may therefore contribute to key open questions in our understanding of \acp{TLE} and their impact of the atmosphere and climate. Of particularly importance is the ability of \ac{UHECR} experiments to measure the rate of occurrence of \acp{TLE} and constrain the chemical impact of such processes. This is becasue if \acp{TLE} are confirmed to perturb greenhouse gases, then they will have to be included in our picture of climate as is the case with other solar terrestrial processes. Other strengths of \ac{UHECR} experiments include a higher sensitivity of detection, which can increase our current estimates of \ac{TLE} occurrence rates by lowering the current and electric field thresholds currently known for their production, and an unprecedented ability of imaging the dynamical evolution of \acp{TLE} at high temporal resolution, which coupled with corollary measurements can disclose further details in their production mechanisms and relationship to the neutral atmosphere. For the ELVEs detection from ground, it is recommended that all future arrays include triggers that allow for the capture of a longer traces to allow for the study ELVEs. While this recommendation may not drive the design of the UHECR observatory itself, it is still worth pointing out as it represnts a small modification of the design which would lead to considerable increase to the range science goals which can be leveraged by the next generation experiments.

New UHECR detectors in space, such as K-EUSO~\cite{Klimov:2022jzk} or \ac{POEMMA} will have much lower threshold than \ac{TUS} and \ac{mini-EUSO}, and can measure even fainter ELVEs and other \acp{TLE} with a high temporal resolution. This will allow one to obtain fine profiles of the spatio-temporal dynamics of events and which will enable the study their formation mechanisms. For example, accurate measurements of the delay of the second ring an the ELVE with respect to the first one will allow for the estimation of the altitude of the \ac{EMP} source responsible.

\subsection{Terrestrial gamma-ray flashes}
\label{sec:tgf}

\Acp{TGF} are sub-millisecond bursts of gamma radiation up to several tens of MeV produced within thunderclouds and are associated with lightning activity. They are the manifestation of the most energetic natural particle acceleration processes on earth, and are at the core of a multidisciplinary field termed \emph{High-energy atmospheric Physics}~\cite{Dwyer2012}, which sits at the crossroads of atmospheric sciences, high energy physics and space science. First reported in 1994~\cite{Fishman1994}, \acp{TGF} have been routinely observed from space by spacecrafts dedicated to high-energy astrophysics~\cite{Smith2005, Marisaldi:2010zz, Briggs2010}. Since 2018 \acp{TGF} have also been observed by the \ac{ASIM} mission onboard the \ac{ISS} \cite{Neubert2019,Ostgaard2019}, the first mission specifically designed to observe \acp{TGF}, which provides simultaneous observations in gamma-rays and in optical bands. A general theoretical framework for the understanding of \acp{TGF} has been developed during the past three decades. In it, TGF are described as Bremsstrahlung emission from a large population of relativistic runaway electrons resulting from avalanche processes in the electric fields of either large thunderclouds~\cite{Gurevich1992,Dwyer2005} or at lightning leader tips~\cite{Celestin2011}, which has been possibly enhanced by the so-called \textit{relativistic feedback mechanism}~\cite{Dwyer2007}. Despite this working model, several knowledge gaps still need to be filled in order to advance the field beyond its current state, namely: 
\begin{itemize}[topsep=2pt]\setlength\itemsep{-0.2em}
    \item What is the exact relation between lightning leader, large-scale electric field, and \acp{TGF}?
    \item  What is the topology of \acp{TGF} (beaming angle, vertical tilt, fine time structure) and its variability?
    \item What is the relationship between \acp{TGF} and quasi-stationary gamma-ray emissions termed \emph{gamma-ray glows}?
    \item Do these high-energy atmospheric phenomena have any impact on atmospheric chemistry and dynamics?
\end{itemize}

Although large catalogs of \acp{TGF} counting thousands of events are now available from most of the TGF-detecting missions~\cite{Roberts2018, Lindanger2020, Maiorana2020, Smith2020}, major advancements in the field now come from simultaneous observations at different frequency bands, ranging from the radio to optical (see~\cite{Cummer2015, Ostgaard2021} for instance). A breakthrough in the field for example could come from observation of \acp{TGF} from space coupled with simultaneous high spatial and time resolution lightning measurements by ground-based interferometers (a goal so far eluded because of the sporadic nature of \acp{TGF} events and the limited range of lightning interferometers). Observing capabilities in the \ac{UV} spectrum provided by the \acp{FD} of current and future generation of UHECR observatories will provide a better understanding of the link between \acp{TGF} and ionospheric emission known as ELVEs, whcih were recently observed simultaneously for the first time by \ac{ASIM}~\cite{Neubert2020}. For this purpose, data from the \ac{mini-EUSO} experiment onboard the \ac{ISS} can already be exploited in association with \ac{ASIM} observations. 

No other space missions dedicated to \acp{TGF} are planned after \ac{ASIM}, with the exception of the two-CubeSat project, TRYAD~\cite{Briggs2019}, currently in construction phase. Observations from space in gamma-rays by a single instrument cannot be used to extract accurate \acp{TGF} source parameters by spectral analysis only~\cite{Lindanger2021}, even assuming a ten-fold increase in effective area for future instruments. Therefore, it is foreseen that advances in this field will require, in addition to a tight correlation with ground-based lightning instrumentation, the use of dedicated observing platforms such as aircraft~\cite{Smith2011,Bowers2018}, possibly flying at high altitude~\cite{Ostgaard2019b}, or balloons. THe synergy with the next generation space observatories for \acp{GRB} could also be enhanced, for example by including \acp{TGF} detection capabilities when designing the trigger logic for these missions.

First evidences of downward going \acp{TGF} in ultra-high energy cosmic ray observatories occurred in the early 2010’s, when some anomalous ring-shaped events were detected by the \ac{SD} of the Pierre Auger Observatory~\cite{Colalillo:2017lnj}. A major breakthrough in these searches was achieved a few years later by measurements made with the Telescope Array \ac{SD}. With the addition of a \ac{LMA} and a slow electric field antenna, the Telescope Array Collaboration succeeded in corroborating the correlation between the \ac{SD} events and lightning activity \cite{Abbasi:2019xan}. The observed bursts of gamma rays (which made of up to five individual pulses) were detected in the first 1-2\,ms of the downward negative breakdown prior to cloud-to-ground lightning strikes. The shower sources were found to be located at altitudes of a few kilometers above ground level by the \ac{LMA} detector. The measured events were found to have a an overall duration of several hundred microseconds and a footprint on the ground typically of 3--5\,km in diameter.

\subsection{Aurorae}

\label{sec:aurorae}

Aurorae are natural phenomena that appear in Earth's upper atmosphere at the altitudes of approximately 80-250\,km. They are characterized by the luminous photon emissions from atoms and molecules of the atmosphere which have ben excited by energetic charged particles that precipitate from Earth's magnetosphere~\cite{Vallance}. Aurorae are commonly observed by the ground-based optical equipment of different kinds. The spectral, spatial and temporal resolution of these observations depends on aims of the investigations. Most often, these data are used in studies of the dynamics of the magnetosphere-ionosphere system, where observations of brightest O\textsubscript{I} (557.7~nm) emission or even panchromatic emission with a relatively low temporal resolution ($>1$~s) are enough.

However, these ground-based observations require good weather (no clouds in the field of view of the instruments), and allow for obtaining information in only one local region of the sky (this problem is partially solved by combining data from cameras with different \acp{FoV} located close to each other). When observing from a satellite, the cloud cover is significantly below the glow region, which allows measurements regardless of weather conditions. Also, due to the precision of the orbit over the ground, it is possible to measure in the entire range of longitudes with one instrument. Observation from space also has its own problems however, as it is impossible to observe one geographical area or event for a long time, spatial resolution is usually worse due to the movement of the instrument, and data traffic limits mean observations either need to be rationed or subjected to heavy compression.

An interesting type of aurora, with a quasi-periodic intensity modulations of extended forms, known as \ac{PsA} were documented for the first time in 1968~\cite{Beach1968}, and up to now do not have a fully exhaustive explanation.
They occur predominantly in the midnight to morning \ac{MLT} sector following an auroral oval expansion and during the sub-storm recovery phase. They appear as irregular patches of luminosity with quasi-periodic (2--20\,s or longer) temporal fluctuations, which are often accompanied by fast complex motions of their bright part synchronized with their luminosity changes~\cite{Yamamoto1988}. In some cases, so-called ``internal modulation" is observed, which is characteristic of much faster pulsations in the luminosity ($\sim$\,3~Hz), enclosed in a single pulse of the main pulsation~\cite{Nishiyama2014}. The observations in specific aurorae lines (for example, the 391.4\,nm and 427.8\,nm lines of the \textit{first negative} system of $N_2^+$) are needed to register these fast pulsations.

\begin{figure}[!htb]
\centering
    \includegraphics{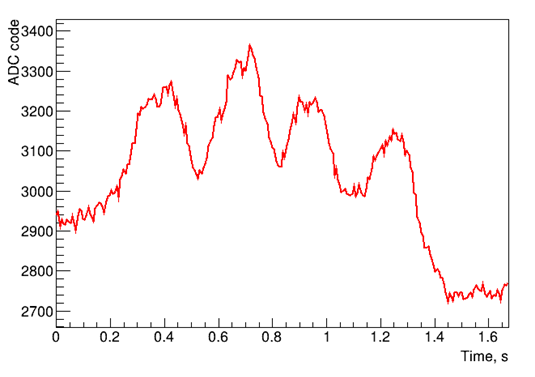}
\caption{Waveform from a single pixel in an event measured on November 10, 2017 at 13:31 UTC by \ac{TUS}.
    }
    \label{Aurora_waveforms}
\end{figure}

As already mentioned, space based \ac{UHECR} detectors are highly sensitive fluorescent telescopes looking downward to the Earth atmosphere~\cite{SSR, adams2015jem}. Thus, if a \ac{UHECR} space-telescope follows a polar orbit, it will fly above the regions of active emissions related to geomagnetic activity, i.e., aurora oval and can make observations of aurora. 
In the slow data acquisition operation mode of the \ac{TUS} detector (with a 6.6\,ms temporal resolution), about 2500\,events were measured at latitudes $>50^\circ$ in Northern hemisphere. Among them, 66\,events with interesting temporal structures were selected. These signals differ from clouds, cities and other well-known sources of light in the atmosphere and occur above both the land and ocean. The observed signals have a very diverse structures with characteristic frequencies of the order of 1--10\,Hz. The most frequently recorded pulsations lay in the 3--5\,Hz range, but there are also events with frequencies up to 20\,Hz. One example waveform is shown in \cref{Aurora_waveforms}. The luminescence regions are localized spatially with a characteristic size of about 10\,km. Several different pulsation regions with different temporal structures (waveforms) were observed simultaneously in the \ac{FoV} of the telescope.
An analysis of the geographical distribution and geomagnetic conditions indicates that these events were measured at the equatorial border of the aurora zone. 
Pulsating events locations obviously repeat shape of the aurora oval. The maximum portion of the pulsations is recorded in L-shells ranging from 4 to 6 and the frequency of events' occurrence correlates with geomagnetic activity.

The spatio-temporal structure of the events is similar to pulsating or flickering auroras observed earlier (for example, \cite{Sakanoi2005}) and have internal modulations. Due to high sensitivity of the telescope and near \ac{UV} spectrum of measurements (which corresponds to a $N_2^+$ first negative emission dominating deep in the atmosphere), measured events are related to a high-energy part ($\gtrsim$200~keV) of precipitating electrons caused by \textit{lower band chorus waves}~\cite{Miyoshi2020a}. 

However, the nature and mechanism of \ac{PsA} occurrence are not fully clear. To study and clarify the nature of this phenomenon, further experiments on high-sensitivity orbital detectors, as well as the comparison of data obtained on satellites with data from ground-based observatories, are needed. Moreover joint observations of atmospheric emission, magnetospheric electrons fluxes and electromagnetic waves onboard one satellite are needed. Despite the fact that future space-based UHECR observatories like K-EUSO and \ac{POEMMA} are not expected to orbit around the poles, it is important to recall the utility such \ac{UHECR} orbital experiments could have in this contest if they would monitor polar regions.

\subsection{Meteors}
\label{sec:meteors}

Meteors are generated by the interaction of a cosmic body with the Earth's atmosphere. The physical characteristics of the interacting body, as well as the entry angle, determine the magnitude and duration of these phenomena~\cite{Ceplecha1976, Gritsevich2011, Moreno2015, Moreno2017}. Estimates suggest that, on average, meteoroids cumulatively deposit 5 to 300\,t of extraterrestrial material every day, mostly into the Earth's atmosphere~\cite{Plane2012, Silber2018, Vinkovic2020}. Only a tiny fraction of this material is delivered to the Earth’s surface in a form of meteorite falls. Dust and small grains (up to 1\,cm), typically of cometary origin, are responsible for the so-called meteor showers that can be seen periodically when the Earth crosses near the orbit of a comet. Larger meteoroids generate brighter meteors, called fireballs or bolides. They are usually considered of sporadic origin and the search for a clear evidence of a correlation of this type of meteors with a common progenitor body is ongoing~\cite{Kornos2008, Trigo2015}.

The observations of meteors are valuable as they provide information about the physical properties of the body entering the atmosphere and, on a larger scale, serve as important input data for the situational awareness of nearby space~\cite{Lal2018}. The observations are also used to distinguish the meteoroids which fully ablate in the atmosphere from the less frequent events that survive all the way down to the ground and may be subsequently recovered in form of meteorites~\cite{Gritsevich2012, Sansom2019, Moreno2020,Boaca_2022}. Fireball observations can be also used to infer the individual trajectories of fragments resulting from atmospheric fragmentation. Together with modelling the dark flight, 
which constitutes the lower part of the trajectory following the termination of the luminous flight, this leads to a construction of a strewn field map showing where meteorites could be potentially recovered on the ground~\cite{Vinnikov2016, Moilanen2021}.


A computed meteor trajectory allows for the determination of the pre-impact orbit of the meteoroid, unveiling its origin in the Solar system \cite{Dmitriev2015, Jansen2019, Pena2021}. The derived orbits can be linked to a possible progenitor body and, in cases when fragments are recovered, with the physical and chemical characterization of the meteorite. Until now 38 meteorites have been recovered together with quality observations that have allowed for the reconstruction of the pre-atmospheric orbit of the meteoroid~\cite{Colas2020,Gardiol2021}.

Gathering sufficient statistics for meteoroid orbits enables more thorough investigations into the link between different meteorite classes and their origin in the Solar system. For these reasons, and with the overarching goal of tracing meteorite-producing events, many ground-based observational networks have been developed since the first double-station meteor photographic program initiated by Fred Whipple in the Smithsonian Astrophysical Observatory in 1936. 
Run by both, amateur and professional astronomers, these networks have a shared goal of continuously monitoring the night-sky and detecting meteor events. The scientific outcome for this kind of survey is twofold: First it provides a unique tool to discover new meteor showers by focusing on the faint but predominant component of the detected events, and secondly it allows for capturing the more rare occurrence of meteorite-dropping fireballs. In selected cases, the efforts are complemented by multi-instruments aircraft campaigns, e.g., to observe a predicted meteor shower outburst~\cite{Jenniskens2000, vaubaillon20152011}.

Orbital devices dedicated to meteor monitoring have advantages over the ground-based meteor observations. The performance of a space-based detection system is less dependent on weather or atmospheric conditions. It offers a wider spatial coverage and an unrestricted and extinction-free spectral domain. Also, the optimal orbit for achieving maximum detection rates can be calculated with the mass index of the meteoroid populations~\cite{Bouquet2014}. In this respect, a remarkable achievement is the observation of a meteorite-dropping fireball from both ground- and space-based instruments, together with the recovery of the meteorite residue on ground. This has already been accomplished a few times in very bright events detected from both the ground by fireball networks and from space by U.S. government orbital sensors, and in recent years also by the Geostationary Lightning Mapper on the GOES-16 satellite~\cite{Jenniskens2018}.

\begin{figure}[!htb]
\centering
    \includegraphics[width=.9\textwidth]{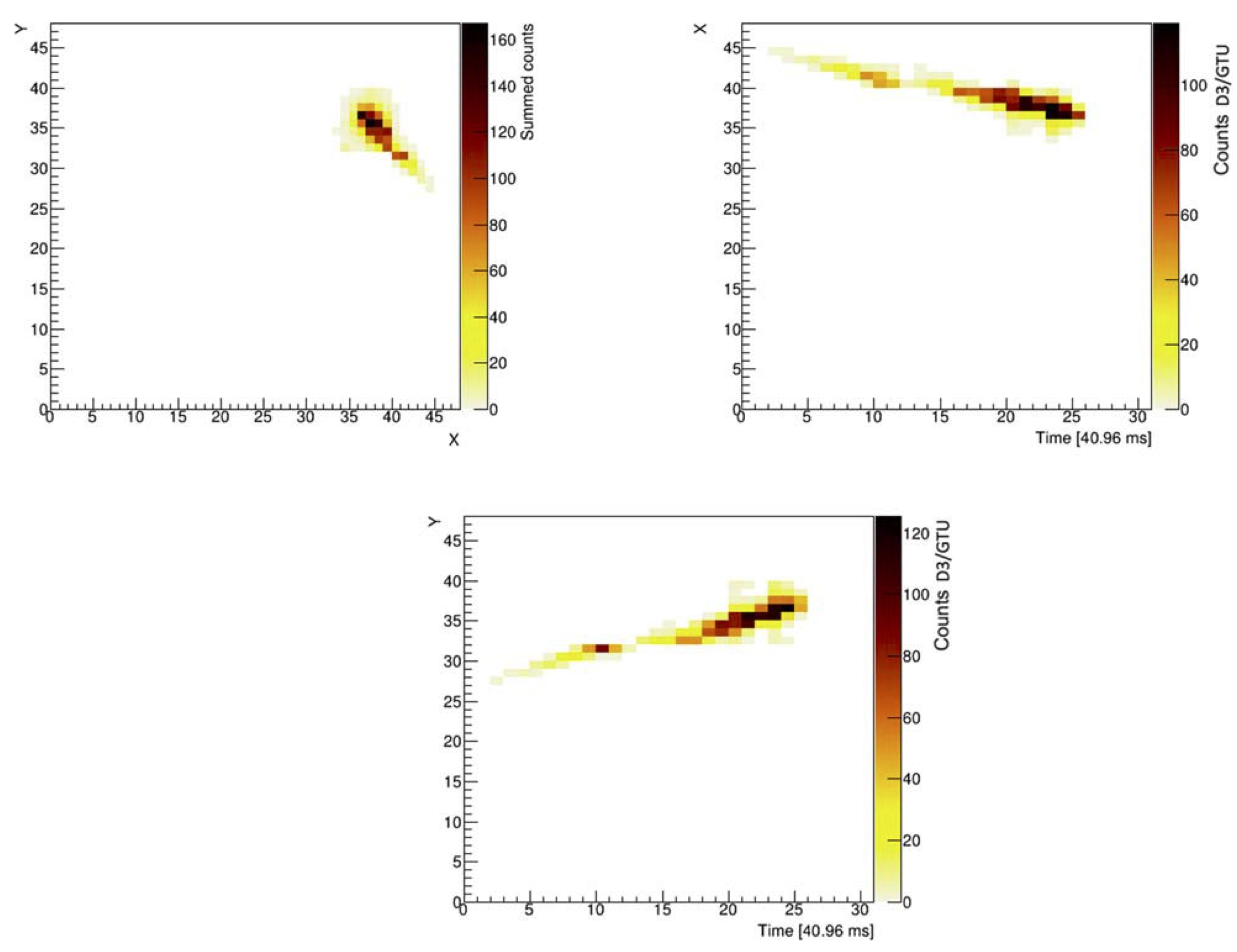}
\caption{A meteor track detected by \ac{mini-EUSO} projected on
the focal surface ($x$-$y$, left), and on the $x$-$t$ and $y$-$t$ profiles (center and right, respectively). Color denotes counts per GTU (1\,GTU = 2.5\,$\mu$s). Image taken from~\cite{Bacholle:2020emk}.
    }
    \label{fig:minieuso-meteor}
\end{figure}

A space based UHECR detector also has the potential to capture the passage of meteors in its field of view, as it looks to the Earth's atmosphere from above. This fact has been shown by the \ac{mini-EUSO} telescope which has observed thousands of meteors since the beginning of its operations in late 2019~\cite{Bacholle:2020emk, Barghini2020, Barghini2021}. An example meteor observation made by \ac{mini-EUSO} is shown in \cref{fig:minieuso-meteor}.

Systematic monitoring of meteors in the near \ac{UV} is almost unprecedented in meteor science. A space-based observation allows for capturing the emission lines of elements and compounds in this spectral range that otherwise are greatly attenuated below 300\,nm of wavelength by atmospheric ozone when observing from ground~\cite{Abdellaoui2017}. The spectral sensitivity of sensors deployed in meteor and fireball network stations is typically confined to a range above 300--400\,nm, and even observation surveys dedicated to meteor spectroscopy are limited to the visible range of 400--800\,nm of wavelength~\cite{Vojacek2015, Rudawska2016, Rudawska2020}. 

It is therefore evident that space-based observations of meteors are complementary to the observational efforts from ground which have been taking place continuously for almost a century. Even experiments that are not specifically dedicated to meteor science can contribute to advances in this field by exploiting their supplementary data and/or implementing dedicated triggers that can operate in parallel on the timescales of $10^{-1}-10^{-3}$ seconds per frame. Increasing the statistics of meteors observation is fundamental in modern planetary science, since a deeper understanding of the population of small bodies in the Solar System and its dynamic provides major insights into the formation and evolution mechanisms of planetary systems.


\subsection{Space debris remediation}
\label{sec:debris}

In the, so far, 60 year history of spaceflight, more than 30,000 rockets and satellites have been launched into space. 
As a result, the quantity of space debris has increased considerably, and particularly so in both the low and geostationary orbits. Added to the fragments produced gradually through normal space activities (disused satellites, rocket stages, parts of instruments, flecks of paint) there are also those which come in bursts due to the voluntary destruction of satellites (for instance by the USA in 1985, China in 2007, India in 2019, and Russia in 2021). 
Currently, it is estimated that at least 3,000\,t of non-operational debris remains in \ac{LEO} (300-600\,km). Overall estimations place the total number of objects in orbit around earth, mostly in \ac{LEO}, with $d<1$\,cm to be around 128 million, while objects in the $1 < d <10 $\,cm range are estimated to number around 900,000.
Given the high orbital speeds involved (about 7\,km/s), collisions with debris of once cm in size or greater can disable or completely destroy the objects involved, which produce additional fragments which in turn cause increased risks to spaceflight.  For instance, the first collision between the Iridium-33 and Kosmos-2251 satellites took place in 2009, leading to the destruction of both and the eventual creation of cloud of fragments at about 740 km of height. 
Even in the absence of destructive collisions, debris of a few millimeters in size cause the continuous degradation of solar panels.

Therefore, both satellites and the International Space Station are often
forced to correct their orbit to avoid potential collisions, which results in the consumption of extra propellant and in turn a reduction of their lifetime.
In the presence of the continuous launch of satellites, especially in \ac{LEO} (for example the Starlink project plans to launch 12,000 satellites with limited orbital correction capabilities), the risk of   \textit{Kessler syndrome}, a chain reaction in which the collision of space objects produces an exponential growth in debris eventually blocking space flight, increases. Given their design and sensitivity, \ac{UHECR} detectors in space would be capable of observing the reflected light from satellites and  space debris in the \ac{UV} band, allowing for the assessment of the space debris problem and may, as outlined below, potentially contribute to its solution. 

Reference \cite{debris2015} proposes a design for a staged implementation of an orbiting debris remediation system comprised of a super-wide \ac{FoV} telescope (JEM-EUSO or other space based \ac{UHECR} observatory) and a novel high-efficiency fibre-based laser system (CAN).
The, basic idea outlined is that the JEM-EUSO telescope could be used to detect detection high velocity fragmentation debris in orbit, which would then pass its location and trajectory info to a CAN system. Further tracking, characterisation and remediation are to be performed by a CAN laser system operating in tandem with the JEM-EUSO telescope. Assuming full scale versions of both instruments, the range of the detection/removal operation would be as large as 100\,km. A proof of concept of this technique is on-going on the \ac{ISS} with the \ac{mini-EUSO} telescope. Given the nadir-oriented observation geometry the experiment is restricted to the local twilight period of the orbit, taking place for about 5 minutes every 90 minutes \cite{debris2015}. A confirmation of  the potential of \ac{mini-EUSO} in this respect has been obtained through the \ac{mini-EUSO} Engineering Model (EM) on ground prior to the launch.
Additionally, already an orbiting rocket body that hosted a telecommunication satellite was detected by the \ac{mini-EUSO} EM, which was later identified as the ``Meteor 1-31 Rocket''~\cite{Bisconti2021}. 
This measurement could then be translated to an equivalent observation performed by a Mini-EUSO-like detector hosted on the ISS. In this case, such a detector (with a single pixel \ac{FoV} of $\sim$\,0.8$^\circ$ $\times$ 0.8$^\circ$) would observe the event with a speed of
$\sim$\,1.4 pixels/s, which would correspond to the observation of space debris with an apparent speed of $\sim$\,1\,km/s at a distance of 50\,km demonstrating the potential of the technique. 
The planned K-EUSO and POEMMA experiments could further prove this approach thanks to their much larger sensitivity and angular resolution.

%

\subsection{Relativistic dust grains}
\label{sec:dustgrains}

Back in 1972, based on a number of earlier works~\cite{Spitzer1949, 1954Tell....6..232A, 1956Tell....8..268H}, Hayakawa suggested that cosmic rays with energies as high as $10^{20}$~eV may consist of relativistic dust grains~\cite{Hayakawa1972}.  The idea was revisited in 1999 by Bingham and Tsytovich~\cite{1999APh....12...35B}. They argued that dust particles can be accelerated during the maximum luminosity stage of a supernova explosion to energies of the order of $10^{20}$~eV. It was concluded that dust particles with $\gamma\lesssim10^4\text{--}10^5$ would be able to reach Earth while interacting with solar radiation.  In early 2000s, Anchordoqui and his collaborators addressed the hypothesis of \acp{RDG} being responsible for a part of the highest-energy cosmic rays from another point of view by performing detailed simulations of \acp{EAS} produced by dust grains~\cite{PhysRevD.61.087302, 2001NuPhS..97..203A}.
One of the main conclusions of the studies was that the dependence of the longitudinal profile of \acp{RDG} on the Lorentz factor is rather weak, and while \ac{RDG} air showers must be regarded as highly speculative, they cannot be completely ruled out.

This hypothesis was criticized from the very beginning. In particular, Berezinsky and Prilutsky argued that \acp{RDG} with Lorentz factors $\gamma>30\text{--}50$ will be destroyed due to interaction with solar photons and other mechanisms~~\cite{1973Ap&SS..21..475B, 1977ICRC....2..358B}.  However, Elenskii and Suvorov immediately suggested a mechanism for how \acp{RDG} could survive transit to and through the solar system~\cite{1977Afz....13..731E}. They argued that dust grains of metallic nature with Lorentz factor $\gamma<360$ and initial radii $3\text{--}6\times10^{-6}$\,cm can traverse even cosmological distances. Another criticism came from John~Linsley~\cite{1980ApJ...235L.167L, 1981ICRC....2..141L} in early 1980s. Based on the superposition principle, Linsley argued that the atmospheric depth at which air showers initiated by dust grains would reach maximum development is much less than the depths observed experimentally. As a result, he concluded that few if any EASs observed by that time were due to \acp{RDG}.

In the latest study dedicated to the possible relation of \acp{RDG} to ultra-high energy cosmic rays, Hoang et~al.\ confirmed that dust grains can be accelerated to relativistic speeds by radiation pressure e.g., from active galactic nuclei, diffusive shocks, and other acceleration mechanisms~\cite{Hoang:2014bba}.  However, they found that Lorentz factor will be $<2$, which is much lower than the earlier estimates discussed above.  It was concluded that \acp{RDG} originating in other galaxies would be destroyed before reaching the Earth's atmosphere and is unlikely to account for \acp{UHECR}. However, dust grains of ideal strength with $\gamma<10\text{--}100$ arriving from distances with a gas column $\sim$\,$10^{20}$~cm$^{-2}$ in the Galaxy would survive both the interstellar medium and solar radiation to reach the Earth's atmosphere.

The idea that a part of \acp{UHECR} originate from relativistic dust grains remains speculative, but the parameter space of sizes and Lorentz factors of \acp{RDG} that can survive on their way to Earth is still non-empty. Taking into account the fact that statistics of events beyond the GZK cut-off are very limited, and only a handful of \acp{UHECR} with energies above 100~EeV have been registered~\cite{PierreAuger:2020qqz}, one cannot completely exclude the possibility that a small fraction of cosmic rays of the highest energies are produced by relativistic dust grains. In early 2000s it was proposed that orbital fluorescence telescopes aimed at observing \acp{UHECR} will be able provide an interesting opportunity for studying relativistic dust grains~~\cite{Khrenov:2000xt}. Interest in \acp{RDG} as a research subject with such detectors has been reignited after \ac{TUS}, the world's first orbital telescope aimed at studying \acp{UHECR} from a low-Earth orbit, registered an event that demonstrated the light curve and kinematics of the signal expected from an EAS, but was must brighter than can be produced by an ultra-high-energy nucleus~\cite{JCAP2020, Khrenov:2021df}. The \ac{mini-EUSO} telescope~\cite{Bacholle:2020emk} that is currently operating on the \ac{ISS}, as well as the future EUSO-SPB2~\cite{Adams:2017fjh}, K-EUSO~\cite{Fenu:2021wub} and \ac{POEMMA}~\cite{POEMMA:2020ykm} missions can extend the capabilities of the ground-based detectors and shed new light on this hypothesis.


\subsection{Clouds, dust, and climate}
\label{sec:clouds}

\begin{figure}[!htb]
\centering
    \includegraphics[scale=1.0]{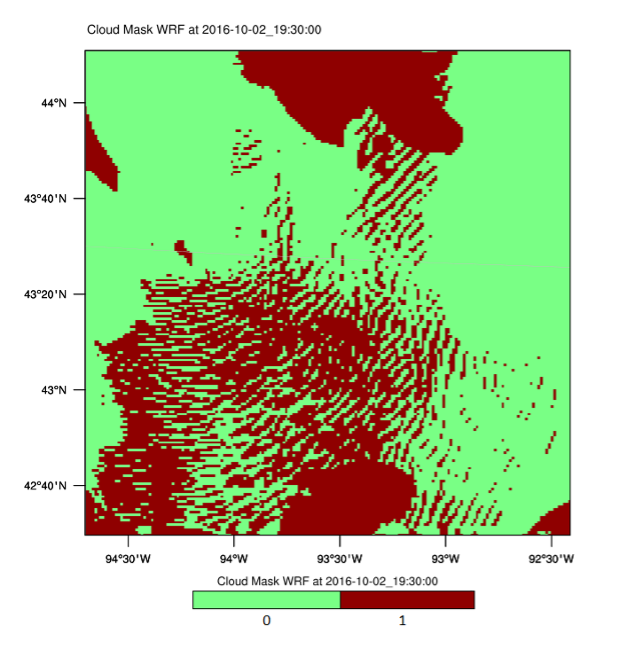}
\caption{Example of cloud mask as simulated by an \ac{WRF} model.
    }
    \label{fig:WRF}
\end{figure}

Clouds play a fundamental role in atmospheric physics and are involved both in weather forecasting and in climate change studies. In particular, they influence the hydrological cycle through precipitation and they interact with shortwave solar and longwave thermal radiation determining the variability of the energy balance of our planet~\cite{Liou}.
The forecasting of cloud localization and layer thickness is a difficult task due to a variety of quantities and processes. These include factors such as water vapor quantities, relative humidity, wind intensity, presence of cloud condensation nuclei, evaporation and condensation rates, heat fluxes and radiative budgets, all of which influence cloud formation and evolution~\cite{Liu_clouds}.
\ac{NWP} models solve the atmospheric primary equations on a three-dimensional grid and simulate different variables, such as temperature, pressure, relative humidity, for every grid point. Other variables, for example shortwave and longwave radiation, vapour, cloud water, rain water, ice, snow mixing ratio, cloud fraction are obtained on the same grid by applying parametrizations. 
Cloud masks (index of presence or absence of clouds) and \ac{CTH} can be computed with post-processing algorithms. \cref{fig:WRF} shows an example of a cloud mask computed using the outputs of the regional meteorological \ac{WRF} model~\cite{SkamarockWRF}.
 
The identification of the position, thickness and evolution of the cloud layer is a challenge for current global and regional models. With the aim of testing microphysical schemes and improving meteorological forecasts, model output like cloud fraction fields and cloud masks are regularly compared with the observations made from both the surface (i.e., lidar ceilometers) and space (satellites)~\cite{Anzalone}.
While high and thin clouds like cirrus are very important to calculating the planetary radiation balance with important implications in climate models, they are difficult to simulate with atmospheric models. Also, most satellites have difficulties in correctly identifying their presence and positions. In this contest, the \acs{UV} lasers (wavelength 355\,nm) which are expected to be employed along with space-based observatories~\cite{Toscano:2014bga} will be able to produce useful data needed to test the microphysical schemes in meteorological models. In fact they would be able to measure the \ac{CTH} which is a fundamental parameter in detection of high clouds (e.g., cirrus).

Mineral dust particles from major dust emitting regions in Africa and Asia also can have a global impact on the Earth's climate through both direct and indirect climate forcing, changing the chemical composition of the atmosphere through heterogeneous reactions, and on the biogeochemistry of the oceans through dust deposition~\cite{B805153D}. In particular, a number of laboratory studies have shown that mineral dust particles serve as potent heterogeneous ice nuclei, provided they can reach altitudes sufficiently high for ice super-saturation. A recent trajectory modelling study explores the availability of mineral dust ice nuclei for interactions with cirrus, mixed-phase and warm clouds. The results of the study suggest that the likelihood for the dust particles being lifted to altitudes where homogeneous ice nucleation can take place is small, whereas by far the largest fraction of cloud forming trajectories entered conditions of mixed-phase clouds~\cite{Wiacek2009}. However, only a few studies have so far made rigorous use of space-born satellite data to investigate the transport of desert dust to high altitudes and its potential interaction with cirrus or mixed-phase clouds. 
West Saharan dust could be measured by a space-based instrument like \ac{POEMMA}, providing measurement tracks which are approximately 200\,km apart. Over a time scale of two days mineral dust would typically move around 1500\,km westward, where it can be mapped again by \ac{POEMMA}. If in the meantime the dust interacted with clouds this interaction will leave a \emph{fingerprint} in the dust distribution. Given the high frequency of such events there should be ample opportunity to match the same dust-laden air masses and to record and analyze the fingerprints of the dust-cloud interactions.
Moreover, \ac{POEMMA} will allow for synergy with missions that belong to Morning or Afternoon Constellations.

\subsection{Bio-luminescence}
\label{sec:bioluminescence}

Since 1915, there have been 255 documented reports of {\it milky sea} (Great Britain Meteorological Office Marine Division, 1993) and even more events have been reported historically. The {\it milky sea} or {\it mareel} is a term used to describe conditions where large areas of the ocean surface (up to 16,000\,km$^2$) appear to glow during the night for periods of up to several days. The condition is poorly understood, but typically attributed to the bioluminescence of the luminous bacteria {\it Vibrio harveyi} in connection with the presence of colonies of the phytoplankton {\it Phaeocystis}. The bioluminescent bacteria have been shown in the laboratory to have an emission spectra which peaks at 490\,nm with a bandwidth of 140\,nm \cite{Hastings1991}. There has been a single report of satellite observations of this phenomenon, confirmed by a ship-based account \cite{Miller2005}. Space-based observatories for \acp{UHECR} could contribute to the search of these phenomena. As an example whilst the BG3 filter on the \ac{mini-EUSO} \acp{MAPMT} is optimised for the 300–400\,nm band, it extends up to 500 nm and therefor \ac{mini-EUSO} is able to detect $\sim$\,20\% of the bioluminescence spectrum. Taking this into account, the typical limiting source radiance of the bacteria should be $\sim$\,10$^{10}$ photons/cm$^2$/s. This number should be regarded as approximate as the true sensitivity also depends on the spatial extent of the signal on the focal plane and the background level, which is dependent on the atmospheric conditions at the time of observation. It is important to underline that this estimate gives an order of magnitude higher sensitivity than the value of 1.4 $\cdot$ 10$^{11}$ photons/cm$^2$/s reported in Ref.~\cite{Miller2005}, following a successful detection. Further detection of the {\it milky sea} events from space could deeply enhance the understanding of this elusive phenomena, as well as the distribution and transport of phytoplankton on a global scale. Experiments like K-\acs{EUSO} and \ac{POEMMA} with their much higher sensitivity could search for even fainter signals on the oceans.
\fakesection{Collaboration road-map}
\vspace{3cm}
{\noindent \LARGE \textbf{Chapter 8}}\\[.8cm]
\textbf{\noindent \huge Collaboration road-map:}\\[3mm]
\textbf{\LARGE  Organizing ourselves for the future}
\label{sec:NewChallenges}
\vspace{1cm}


When discussing the future of \ac{UHECR} science, this white paper has so far focused on the instrumentation, technologies, and analysis techniques that will be vital for continuation of progress in our field. However, focusing only on these concrete matters risks forgetting the most important aspect of \ac{UHECR} science infrastructure, the scientists themselves. Throughout the history of \ac{UHECR} physics, there has been a consistent trend of moving from isolated scientists toward larger and larger collaborations which should be expected to continue. This is not only because, science, like so many other aspects of society, benefits from a wide and open range of opinions and viewpoints, but also because the very nature of \ac{UHECR} phenomena requires large, coordinated, efforts over massive areas. Because of this inherent need for collaboration, it is clear that in order to continue to grow as a field, we must also continue to grow as a community. Therefore, it is critical to have a clear picture of what is important when organizing and building the next generations of \ac{UHECR} science. Though there are a great number of factors to consider, it is essential that we as a community make a firm commitment to increasing the diversity of scientists in our field, make real efforts to democratize access to our data through the tenants of Open Science, and take deliberate steps to meet our societal responsibility to minimize our carbon footprint.

\subsection[Commitment to diversity]{Commitment to diversity: Diversifying our perspectives}
\label{sec:Diversity}
Physics remains one of the least diverse fields in \ac{STEM}. In the most recent report from the American Institute of Physics, 19\% of physics PhDs awarded in the US in 2019 were to women, and among the physics PhDs awarded to US citizens 1\% of were awarded to African Americans and 4\% to Hispanic Americans~\cite{PhysPhDTrends2019}. 
A similar trend is seen at the undergraduate level, where 22\% of physics bachelor's degrees were awarded to women in 2018 while 4\% were awarded to African Americans and 9\% to Hispanic Americans in 2017-2018~\cite{PhysBachelors2018}.  
These numbers are in stark contrast to the 2017 college population where 14\% of students were African American, 19\% were Hispanic, 
and 54.9\% were women~\cite{StudentsCensus2018}. Diverse perspectives and backgrounds are important for carrying out research and increasing diversity and inclusion in the field is important to ensure scientific progress. This is in addition to an ethical and social justice motivation to creating more equitable opportunities and work places. 

Large scientific collaborations increasingly play a significant role in a scientist's professional career. Daily, even hourly, interactions with colleagues from around the world are not uncommon in today’s physics and astrophysics experiments. The climate and culture of collaborations matters and there is opportunity for collaborations to pursue inclusive and equitable practices.

\subsubsection*{Community of practice as a model}

\noindent Multi-messenger astronomy depends on the principle of collaboration to enable previously impossible discoveries. The \ac{MDN}~\cite{MDNWeb}, formed in 2018, takes this same principle and applies it to broadening participation in the field. Participating collaborations currently include the Dark Energy Spectroscopic Instrument, Fermi Gamma-ray Space Telescope, IceCube Neutrino Observatory, LISA, Vera C. Rubin Observatory, LIGO Scientific Collaboration, North American Nanohertz Observatory for Gravitational Waves, Pierre Auger Observatory, Neil Gehrels Swift Observatory, Very Energetic Radiation Imaging Telescope Array System, and Virgo. The \ac{MDN} is a community of practice, or a group of individuals who care about and carry out shared activities and resources on a subject. As such, the group operates around and promotes six elements to advance equity, diversity, and inclusion in multi-messenger collaborations: 
\begin{enumerate}[topsep=2pt]\setlength\itemsep{-0.2em}
    \item opportunity to go beyond individual accomplishments,
    \item structure through organizational principles and tools,
    \item training for members,
    \item support from each other and for current and future STEM professionals,
    \item presence at conferences, on websites, and on media outlets, and
    \item legitimacy in broadening participation efforts.
\end{enumerate}
These core elements underpin monthly meetings where support and knowledge are shared; the meetings motivate our participation in conferences and field-wide planning efforts (such as the Decadal Survey and Snowmass), and provide collaboration opportunities. 

\begin{figure}[!ht]
     \centering
     \includegraphics[width=\textwidth]{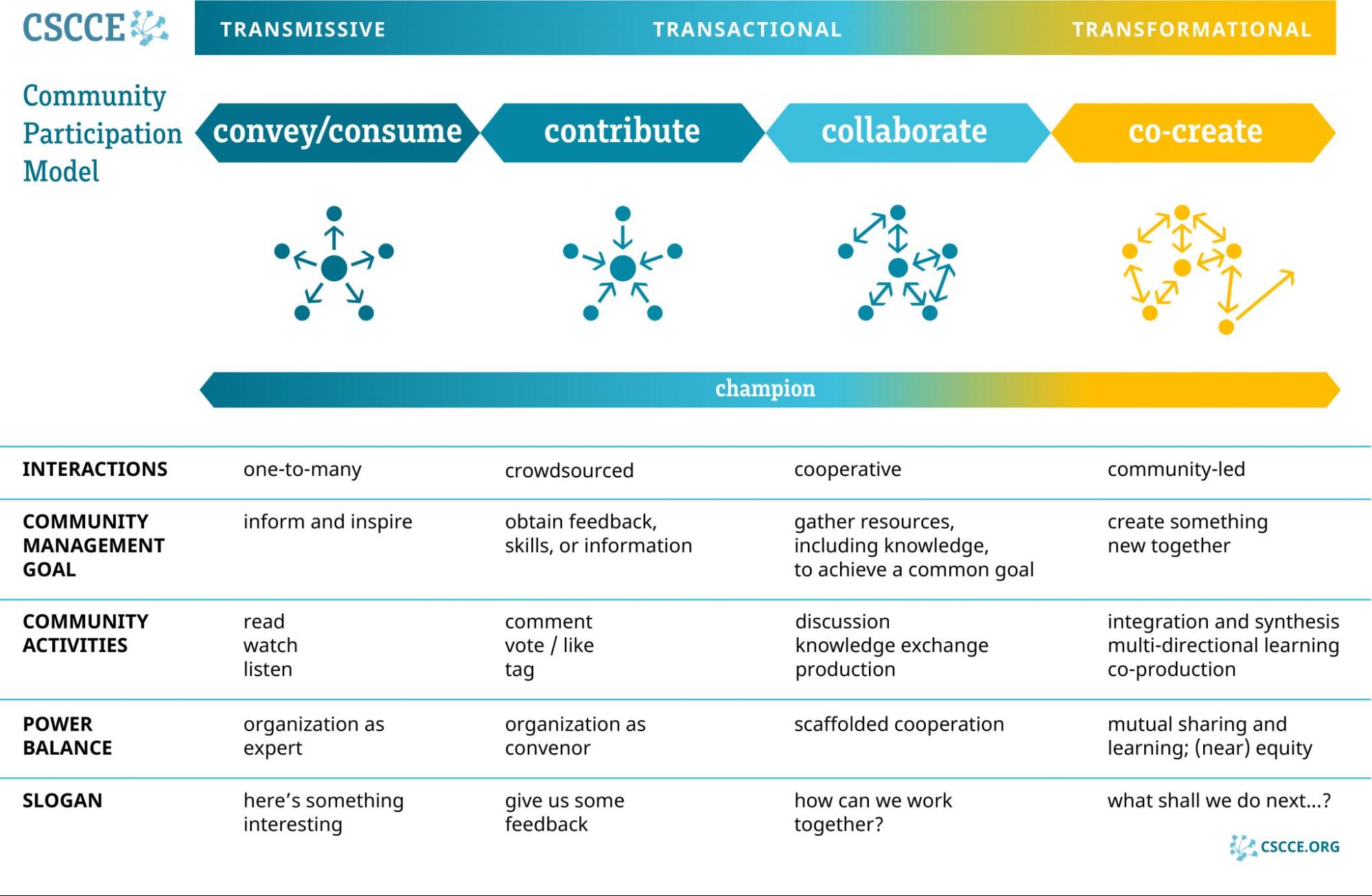}
     \caption{The Community Participation Model from the \ac{CSCCE} describes four participation modes: Convey/Consume, Contribute, Collaborate, and Co-Create. Each mode is described with participation characteristics. The \ac{MDN} most often spends time shifting between Collaborate and Co-Create.}
     \label{fig:CommPartModel}
 \end{figure}

The \emph{Community Participation Model} (see \cref{fig:CommPartModel}) was introduced to the \ac{MDN} in a 2019 community engagement workshop led by Lou Woodley, Director of the \ac{CSCCE} \cite{cscce}, and has been an especially helpful tool when considering the life-cycle of the \ac{MDN}. In this model, Woodley and Pratt~\cite{Woodley:2020cscce} posit that communities often start in a ``convey/consume'' phase of information transfer and move along a continuum towards a ``co-create'' phase where members develop something new collaboratively. phase where members develop something new collaboratively. Reflecting on the \ac{MDN} community of practice, it has occupied each participation phase and commonly advances and retreats between ``collaborate'' and ``co-create'' for which the goals and activities of these community phases are well-aligned with those of the \ac{MDN}.

\subsubsection*{Activities}

\noindent The \ac{MDN} holds monthly calls, often with guest speakers who talk about a range of topics. There is often time to share success, challenges, and opportunities during each meeting. Additionally, the group contributed to the Astro2020 Decadal Survey, has run a joint campaign for the International Day of Women and Girls in Science, maintains a website and hopes to grow a repository of resources, and is planning for upcoming activities. A community manager with dedicated time to work on \ac{MDN} helps sustain and drive efforts, sending out regular communications and scheduling guest speakers.

\subsubsection*{Impact}

\noindent Community connections are a primary strength of the \ac{MDN}. Collaborations are able to share experiences, describe lessons learned, present models of a variety of \ac{EDI} efforts, and exchange documentation and policies. The community offers a place to raise awareness of \ac{EDI} efforts within participating collaborations as well as others in the field at-large through invited speakers. Having a safe place to share knowledge and experiences around \ac{EDI} efforts is important and should be considered a vital part of increasing inclusion of science collaborations. Examples of discussions we have held within the \ac{MDN} include those on consensus-building when developing a code of conduct or conducting a climate survey, the pros and cons of using external ombuds, and how to create sustainable \ac{EDI} efforts.

\noindent There are also more tangible examples of the impact of the \ac{MDN}: 
\begin{itemize}[topsep=2pt]\setlength\itemsep{-0.2em}
\item The IceCube Impact Award inspired and modeled the VERITAS Outstanding Contribution Award.
\item The \textit{Fermi}-LAT mentoring program is a model for an IceCube mentoring program that is in the planning stages.
\item Examples from several participating collaborations provided a point of departure for a charter for the LISA \ac{EDI} effort.
\item The \ac{MDN} began with four collaborations and has since grown to include eleven collaborations, and two additional groups are in the process of joining. This is clear evidence of the impact of and need for communities of practice such as the \ac{MDN}. 
\end{itemize}

\subsection[Open Science]{Open Science: Democratization of access}
\label{sec:OpenScience}
Basic research in the fields of particle physics, astroparticle physics, nuclear physics, astrophysics, and astronomy is performed in large international collaborations, mostly with huge dedicated instruments which produce large amounts of valuable scientific data. To efficiently use the totality of information produced in these experiments to solve the many open questions about the universe, a broad, simple, and sustainable plan for open access to the valuable data from these publicly funded infrastructures needs to be developed and implemented. 

In general, there are currently several efforts underway to develop a (distributed) global data and analysis center. This is a difficult process as such a facility must deliver the following pillars of not only open data and open science, but also \acs{FAIR}~\cite{FAIR} (\acl{FAIR}) data management:
\begin{itemize}[topsep=2pt]\setlength\itemsep{-0.2em}
    \item Data availability: all participating researchers of the individual experiments or facilities need a fast and simple access to the relevant data;
    \item Data analysis: A fast access to the Big Data from measurements and simulations is needed;
    \item Simulations \& methods development: To prepare the analyses of the data the researchers need substantial computing power for the production of relevant simulations and the development of new methods, e.g., by deep machine learning;
    \item Education in data science: The handling of the center as well as the processing of the data needs specialized education in ``Big Data”;
    \item Open access: It is becoming more and more important to provide the scientific data not only to the internal research community, but also to the interested public: Public Data for Public Money! 
    \item Data archive: The valuable scientific data needs to be preserved for a later use as all possible future uses of the data can not be foreseen.
\end{itemize}
Whereas in both astronomy and particle physics data centers which fulfill a part of these requirements are already well established, in cosmic-ray physics only first attempts are presently under development. For example, \ac{KCDC} has made a public release of the scientific data (from the KASCADE-Grande experiment), and the Pierre Auger Observatory has published 10\% of their high-level data. 
In addition, some public IceCube or Auger data can be found in Astronomical Virtual Observatories and data repositories.

\subsubsection{Examples of open data in UHECR science}

Two examples of nascent open data initiatives are \ac{KCDC} and the Pierre Auger Open Data will be discussed in detail below. Generally, the main difference between them is that \ac{KCDC} has published the complete data set of the KASCADE-Grande experiment down to the raw data level (low-level data), whereas Auger has so far only made parts of the data set available and only in the form of reconstructed parameters (high-level data).
This highlights the two different concepts of an outreach driven project on the one hand (Auger) and a service for the entire community including the society on the other (\ac{KCDC}). 
Besides the scientific data, both approaches also provide analysis examples and tools for different target groups. This is important as open science will only work if the full data cycle including the workflows is made available.   
In any case, all efforts in this direction do not only provide a service to the society, but also both the publishing collaboration and the \ac{UHECR} community general benefit from it (for example by the acquisition of new students/collaboration members and an easier documentation of any analyses or workflow within the collaboration).

\paragraph*{The KASCADE cosmic-ray data center}

\ac{KCDC}, \url{https://kcdc.iap.kit.edu/}~\cite{Haungs:2018xpw}, is a web-based interface where, initially, the scientific data collected by and simulated for the completed air-shower experiment {KASCADE-Grande} was made available for the astroparticle community, as well as for the interested public. Over the past seven years, the collaboration has continuously extended the data shop with various releases which increased both the number of detector components from the KASCADE-Grande experiment with available data along with the corresponding simulations needed to interpret it. 

The aim of \ac{KCDC} was the installation and establishment of a public data center for high-energy astroparticle physics based on the data of the KASCADE-Grande experiment. The web portal as interface between the data archive, the data centre's software and the user is one of the most important parts of \ac{KCDC}. It provides the door to the open data publication, where the baseline concept follows the ‘Berlin Declaration on Open Data and Open Access’~\cite{Berlin}, which explicitly requests the use of web technologies and free, unlimited access for everyone. In addition, \ac{KCDC} provides the conceptual design, how the data can be treated and processed so that they are also usable outside the community of experts in the research field. 

With the latest releases, a new and independent data shop was added for a specific KASCADE-Grande event selection, which in turn created the technology for integrating further data shops as well as the data of other experiments, like the data of the air-shower experiment MAKET-ANI in Armenia. 
In addition, educational examples on how to use the data are available, more than 100 cosmic ray energy spectra from various experiments, and a public server with access to Jupyter notebooks covering various analyses.

For the future, \ac{KCDC} aims for an integration into a larger Science Data Platform. Doing so, \ac{KCDC} will benefit from the community's overarching synergistic development of a coherent data and metadata description. In addition, \ac{KCDC} can be the test base for a coherent concept for data storage and access, as well as for an eventual \ac{AAI} infrastructure developed with the goal of enabling a global multi-experimental and multi-messenger analysis platform.

\paragraph*{The Pierre Auger open data}

The Pierre Auger 2021 Open Data \url{https://opendata.auger.org/}~\cite{AugerOD} consist of a cosmic-ray dataset of 22731 showers measured with the surface detector array (SD events) and of 3156 hybrid events (i.e., showers that have been recorded simultaneously by the \ac{SD} and \ac{FD}). 
These data are available as pseudo-raw data in JSON format and as a summary CSV file containing the reconstructed parameters. 
The open data set also includes the counting rates of the surface detectors, recorded with scalers and averaged over every 15 minutes from 2005 to 2020, and atmospheric data acquired with weather stations. 
The collaboration provides the data via its own website.

All Auger Open Data have a unique DOI under zenodo that users are requested to cite in any applications or publications. The Auger Collaboration does not endorse any work, scientific or otherwise, produced using these data, even if available on, or linked from, this portal.

\subsubsection{The near future}
\label{sec:OpenScience10year}

Open data and open science have largely become a funding condition for large-scale facilities financed by tax payer's money.   
This is because open data and science are clearly drivers of innovation and not only for information technology, but also for the science itself.
Despite this, most of the original research data available in astroparticle physics has so far been primarily exploited by the researchers or research institutions who directly participated in its production. This stands in contrast to what is already standard in the astrophysics and astronomy community, where open data has been very successfully employed for some time. This is a pity as the current situation restricts the ability for outsiders to carry out a secondary exploitation of the data, and in particular for multi-messenger, i.e., multi-experimental analyses.

\begin{figure}[!htb]
\centering
\includegraphics[width=0.60\textwidth]{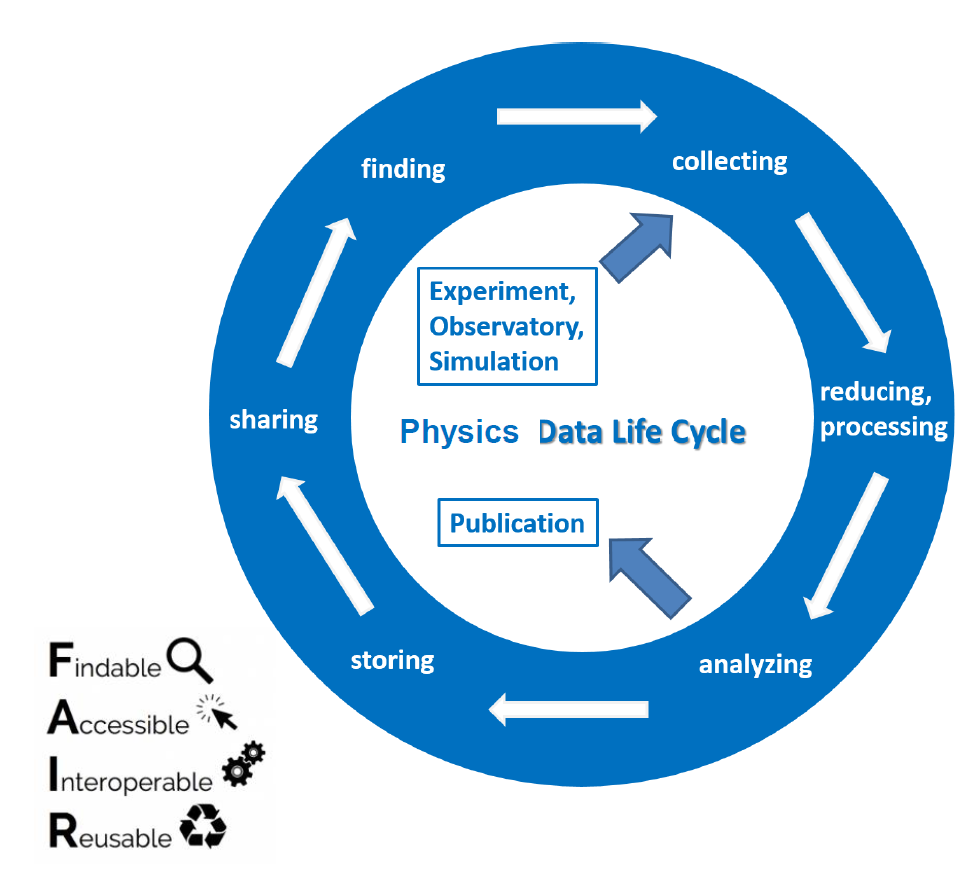}
\caption{General data life cycle scheme in physics. For a useful and efficient life cycle each step must be based on a \ac{FAIR} data and metadata treatment. For open science and open data, public access should be possible at each step of the data life cycle.} 
\label{dlcfig}
\end{figure}

Modern large-scale physics experiments generate a huge data stream, and the lifetime of their active operation can reach several decades.
Because of this, the amount of accumulated data can exceed one hundred petabytes and possibly even up to several Exabytes in the future.
In this context, it is clear that the issue of active and on-going management of the data, as well as the continued development of modern and sophisticated analyses methods throughout their life cycle, is a very important and highly topical issue. \cref{dlcfig} shows a typical data life cycle of a physics experiment, and for example closely follows the practical cycle for the Pierre Auger Observatory or the Telescope Array.
The concept of open data and open science requires collaborations to provide mechanisms, tools, and processes following the principals of a \ac{FAIR} data treatment over this entire cycle. 
Because this is not yet the standard in the \ac{UHECR} community, over the coming years it is crucial that the field pursues the adoption of \ac{FAIR} practices through a coherent approach as it is critical to full exploitation of the data as well as the ability for the community to efficiently pursue multi-messenger astroparticle physics. There are efforts underway to address this issue such as the above described expansion of \ac{KCDC}~\cite{Haungs:2018xpw}, the \acl{AMON}~\cite{AyalaSolares:2019iiy}, and the \ac{SCIMMA}~\cite{Allen:2018yvz} project, to name a few. For \ac{UHECR} science to progress, it is critical that these and other efforts are given wide support as the benefits of their formation widely outweigh the financial and research-hour costs of their development.

So far, when low-level data and their metadata have already been made openly available (like in \ac{KCDC}), their use has often been hampered by the highly variable definition of their metadata, missing interoperability, and also by sociological barriers to common projects between communities. 
To solve this problem, a common approach to access and a standardization of metadata definitions needs to implemented.
Furthermore, in order to exploit the full scientific potential of \ac{UHECR} research, cross-experiment, cross-project, and cross-community working groups are becoming increasingly necessary, for which the open exchange of data will be required. 
To achieve this however means the pursuit of not only open data, but open science as well, as without access to analysis methods and scientific know-how, the usefulness of open data is significantly diminished.

Under the catchword Citizen Science, activities are taking place that achieve a very high visibility in society and are also fun for the general public to participate in. In the field of cosmic ray research, however, such activities are rather scarce, which is a lost opportunity. 
There is broad public interest in new discoveries in astronomy and particle physics, in particular by a dedicated amateur community.
The younger generation’s increasing digital literacy, coupled with the ever more diverse nature of communication technology and social interactions, provide ample opportunity to engage citizens in novel ways and to serve their interest in astronomy while leveraging the power of their collective minds. 
Today, interested citizens can be easily invited to work on state-of-the-art research data, allowing them to share the research and discovery experience, and receive recognition for valuable contributions to science.
Furthermore, being an active part of an international scientific mission also helps to bridge differences in geography, culture, religion, ethnicity, and gender increasing the strengthening society.
In the astro- and astroparticle physics community, educational initiatives such as Zooniverse (incl. the Radio Galaxy Zoo), Muon Hunter, Einstein@home, or CREDO increasingly engage the public in a more active role.  
Such active participation cultivates the understanding of the scientific method and reasoning, and additionally increases the identification with the \ac{UHECR} field providing tangible benefits to the \ac{UHECR} community. 
Therefore, the \ac{UHECR} community should take advantage of the increasing digital literacy and diversifying communication of the public to actively engage them in citizen science projects whenever possible.  
We therefore need to create sufficient incentives and access to data infrastructures and methods to involve the public in ongoing research.

\subsubsection{Open science and next generation UHECR observatories}
\label{sec:OpenScience20year}

The need for Big and Exa-scale data management is primarily driven by the development of large-scale instrumentation as the scientific harvesting of their data requires high-performance systems for data ingestion, selection, transfer, and storage. Also due to the increasing complexity of analyses, research data management is also of central importance for all areas of future astroparticle physics facilities and can be decisive for the success of research projects. To manage these crucial aspects of future \ac{UHECR} projects and initiatives, it is vitally important that the \ac{FAIR} principles be implemented and tenants of open data be followed. Lastly, as base for effective open data and open science policy, current solutions for Exa-scale data management need to be developed and federated data storage infrastructures such as data lakes need to be built.

In order to fully deploy a successful open data policy, especially in regards to pursuing efficient multi-collaboration multi-messenger studies in UHECR science, the following is required:
\begin{itemize}[topsep=2pt]\setlength\itemsep{-0.2em}
    \item Federated data management solutions for high data rates, the reduction or compression of data and large publicly available data volumes, such as data lakes all need to be developed;
    \item Metadata systems and workflows that cover the entire life cycle of collected and generated data up to and including publication in accordance with \ac{FAIR} principles must be refined;
    \item \ac{FAIR} data management and open data needs to be promoted via the international collaborations, or via experiment overarching platforms (i.e., the CERN Open Data platform);
    \item Dedicated large-scale, federated analysis and data storage centers need to be established as infrastructure for multi-messenger astroparticle physics;
    \item The wide scale adoption and migration to the most modern computing, storage and data access concepts (data lakes) which will also open the possibility of developing specific analysis methods and corresponding simulations in one environment is required;
    \item A standardization of data formats and open storage following the \ac{FAIR} principals and thereby make it more accessible and attractive to a broader user community must be implemented.
\end{itemize}

As a further recommendation when designing a next-generation \ac{UHECR} observatory, the realisation of modern data management, including the public provision of data and an open science policy, must be considered from the outset. This should be organised via a separate working group within collaborations, comparable to a simulation or detector group, and is not possible without the provision of dedicated manpower on the order of 3 to 4 full time members. This of course is difficult to establish without dedicated support for such efforts from funding agencies as these efforts are both time consuming and often times have a low visibility.


\subsection[The low carbon future]{The low carbon future: Meeting our societal responsibilities}
\label{sec:CarbonFootprint}
Mankind is facing a worldwide, potentially existence-threatening anthropogenic climate crisis. Its consequences have already been experienced for decades in many endangered regions - yet the consequences are now also being observed in temperate climatic zones: droughts, floods, more frequently occurring local temperature records, increased forest fires. Worldwide temperatures have already risen by more than 1$^\circ$C on average compared to pre-industrial times, and even more so locally in many cases.  Apart from sea level rise as the most important, albeit abstract, threat of the past, the climate crisis has now arrived for most of humanity~\cite{ipcc}. The benchmark for how relevant this part might be is the "allowable" carbon footprint per person. The Paris Agreement~\cite{paris2015} offers a scientific estimate of the worldwide remaining \COT\ emission budget that limits global warming to a maximum of +1.5$^\circ$C with a probability of 50\%. This budget corresponds to a global residual emission of about 410\;Gt \COT\ as of 2022~\cite{sru2020}. Assuming global climate neutrality by 2050 and a disputable equal sharing amongst 7 billion people, this would allow each of us to emit a total of 60\,t \COT{} by 2050, or about 2\,t \COT\ per year if we start in 2022. The scientific community must face these realities and be proactive in responding to them in the design and implementation of projects, in travel, in data processing, and in the production of scientific results. The astroparticle physics community has also perceived this development~\cite{Grinberg:2022jah} and has begun to respond to it (see e.g.~\cite{Aujoux:2021kub}).

\subsubsection{Options for action}
In contrast to the current average emission per person per year in the US of about 15t \COT{}, the calculated per-scientist \COT\ emissions per year for e.g.,  \ac{MPIA}~\cite{jahnke2020} in 2018 amounts to 18\,t \COT\  emissions while the Australian astronomy community~\cite{Stevens:2019ntb} reports even beyond 40\,t \COT -- and these are both only work-related calculations which come on top of personal emissions. The key question now is how to reduce emissions or even prevent them from being generated in the first place.

\paragraph*{Green Computing}
Even though IT server farms are becoming increasingly efficient, the continued high demand for more computing power is currently more than offsetting the energy savings and resulting in a steady increase in energy demand. In particular, as should be clear from \cref{sec:10yrComputation} and \cref{sec:ComputationFutureInfrastructure} the use of machine learning and large scale computing within the astroparticle physics community is only set to grow, which poses a challenge that should be addressed early on.

The most obvious way to lower the \COT{} footprint of computing is to primarily employ renewable energy sources in powering the computational centers used in \ac{UHECR} analyses. This can be done by locating computational infrastructure in locations with ample wind, solar, nuclear or hydroelectric generation options. Care however should be taken that this effort expands the share of \COT{} neutral power being used by society rather primarily shifting the \COT{} footprint burden to other sectors. Beyond this step there is also a substantial need to employ so called Green Computing methods which can be defined as efficient computing that provides the identical results with less energy consumption and therefore less environmental impact. Based on this definition, three key points may be identified for the discussion. 
Firstly, data centers that house supercomputers typically require a large amount of energy for cooling systems and maintaining uninterruptible power supplies.  The Green IT Cube at FAIR/GSI is an example of an highly efficient design of the IT infrastructure. In short, the excess heat of the IT equipment is transferred to cooling water in a smart way. Since the thermal capacity of water exceeds that of air by a factor of 4000, the equivalent flow rates and temperature differences are correspondingly smaller~\cite{Pat1,Pat2}.

Furthermore, the computer architectures used as well as the implemented algorithms themselves have to be mentioned, where different implementations can differ greatly in energy consumption and also performance, sometimes even by several orders of magnitude. These two aspects are inherently interdependent and are therefore described jointly here.
Today's processor architectures offer an increasing number of vector instructions, although features differ between architectures. In addition, current computer architectures provide a deep number of vector registers. If only single precision or double precision is used for calculation, the performance of the processor suffers. Computer code that is properly vectorized works just as efficiently with \acp{GPU}. Therefore, it is necessary from the beginning that data structures and the algorithms themselves are designed and implemented appropriately. Subsequent vectorization of existing computer code typically requires refactoring of these data structures. This should therefore be avoided as far as possible. As existing examples show, porting to \ac{GPU} architecture can take place with great success. The optimisation by porting the program Open-CL lattice \ac{QCD} increased its run-time performance by a factor of 10 and shows good scalability on \ac{GPU} machines~\cite{Bach:2012iw}. 
Other examples such as the hadronic interaction generator \ac{UrQMD}, which was rewritten, are accelerated by a factor of 150 and more~\cite{Gerhard:2012uf}.

In general, it can be said that efficient computing, wherever possible and appropriate, should be based on massive parallel computing in the future, in financial terms and also in their energy efficiency, \acp{GPU} are vastly superior to \acp{CPU}. Due to the expected necessary computing power that future, more complex \ac{UHECR} experiments will require, these are almost impossible without the paradigm shift described here.

\paragraph*{Green experiments}
The remote and sparse nature of the arrays required to pursue \ac{UHECR} science (see \cref{sec:GoBigOrHome} and \cref{sec:FutureDetectors} for example) naturally results in our detectors largely being self-powered through solar and other renewable energy sources. This results in a largely carbon-neutral operation of these detectors.
However, when averaging the ecological footprint of design, construction and deployment of a large detector over its lifetime, the impact of choices at the initial stages of the experiment can be as large as that of computing and travel \cite{Aujoux:2021kub}. Therefore, ecological considerations must be taken into account from the start. 

In the process of selecting materials for construction, the ecological footprint of creating the raw material, as well as the possibility of re-using components are to be taken into account. Even though dismantling of the detector is still many decades in the future, its ecological effects, including possibilities of re-using elements and materials, should be considered from the start.

Even though renewable energy sources are used, a reduction of the energy consumption leads to a reduced material budget, for instance for solar panels and the required support structure, as well as a reduced requirement for energy storage for the same data taking efficiency. Furthermore, there is an ecological aspect in the trade-off between data taking efficiency, by requiring a minimal battery capacity, and enlarging the effective area of a detector, e.g., through the number of detector units, while still obtaining at the same statistical power.

Contrary to past and current practice, shipment of materials should be reduced by sourcing components locally and moving the production of the detector units as close to the site of deployment as possible while ensuring a minimal impact to the local environment. This will have an effect on funding options as the direct benefits for industries and laboratories in most of the participating countries will be in design and prototyping of the experiment rather than mass production, while local economies will further benefit from hosting observatories. This however can also have the effect of only shifting the \COT{} footprint further up the supply chain, which means \COT{} cost for component manufacturing should also be considered when sourcing parts. Regardless, in situations where international shipping is unavoidable, carbon neutral shipping options are becoming increasingly available and should be relied on as much as possible, even if there is a premium on their use.

Lastly, site-locations should be open for different collaborations, including those with other scientific goals. This includes the expanded use of \ac{UHECR} observatory data, as covered in \cref{sec:SuppSci}, but should go beyond this to supporting the co-hosting of entirely different experiments and observatories. This allows for the sharing of infrastructures thus reducing the overall emission of the scientific community. Good examples of such sharing of infrastructures exist already today, such as the study of marine mammals in the \ac{KM3NeT} area \cite{Guidi:2021dbl}, and such options should be included in the design of the infrastructures themselves. Additionally, tools, such as the \ac{ADEME} database \cite{ademe}, provide valuable information in reducing the carbon emission in all aspects of experimental planning and construction.

\paragraph*{Eco-friendly conferencing, meetings and travel}

The nature of conferences and collaboration meetings has altered significantly during the COVID-19 pandemic towards an almost fully online experience. This clearly provided a reduction in the ecological footprint of these events, as well as enhanced options for participation. A Nature poll \cite{Remmel:2021} shows that scientists in general appreciate these aspects of the virtual meetings. 
Drawbacks, such as the lack of networking possibilities, the different time zones in which the participants are located, and fatigue of online meetings mentioned in the same poll should also be taken seriously.

The challenge for current and future collaborations, is to balance in-person, hybrid and virtual meetings such that community building will take place while significantly reducing the environmental impact of travelling. In general, this affects the geographical locations and frequencies of in person meetings, as well as their duration, placing the burden of the overall reduction of emissions on a limited number of groups that could be compensated. The latter requires a different method of sharing costs between the different groups participating in a collaboration.

The connection between the scientific collaboration and the remote local communities that hosts the experiments benefits from visibility of the collaboration within the community. The option of having on-site scientific activities coincide with local events can be used to strengthen the bond between the community and the experiment. This is demonstrated by the Auger collaboration that holds a collaboration meeting in Malarg\"{u}e when collaborators are able to participate in the local holiday events \cite{PierreAuger:2021ddg}. In all cases, it is possible to further reduce the \COT{} footprint of travel and conferences by electing to pay for carbon offsetting for the flights and possibly including carbon offsetting costs for events directly in conference fees, but it is important to ensure this increased financial burden does not decrease accessibility to in-person meetings.

\subsubsection{Summary}
Reducing \COT{} emissions through green computing, green experiments and infrastructures, as well as eco-friendly conferences and travel, is not only an essential instrument for the sustainability of scientific practice, but also an essential message from the scientific community to society and policy makers and a wake-up call to act against climate change.  In any case, the political will in different parts of the world and the pressing necessity of transformation will demand action with vigour and we need be prepared for it. Indeed, the constraints of funding means that many of above strategies are already being partially followed and should be largely familiar to the community. However by giving climate impact more weight in \ac{UHECR} research decisions, we as a community can meet our obligation to become carbon neutral while also ensuring the money allocated to \ac{UHECR} research is leveraged to its maximum extent. It is however important to note that efforts such as carbon offsetting and sequestration represent new line items to the already tight budgets of \ac{UHECR} experiments. It is therefore hoped that the monetary resources needed to pursue such projects would be considered by government agencies when considering funding levels for \ac{UHECR} science.

\fakesection{Executive Summary}
\vspace{3cm}
{\noindent \LARGE \textbf{Chapter 9}}\\[.8cm]
\textbf{\noindent \huge Executive Summary:}\\[3mm]
\textbf{\LARGE The coming golden age of UHECR physics}
\label{sec:ExecutiveSummary}
\vspace{1cm}


\noindent \Acp{UHECR}, $E>100$\,PeV for the purpose of this white paper\footnote{While we do recognize the importance of cosmic-ray physics at lower energies and dedicated future projects, such as SWGO and others, this white paper was written to focus on the highest energies.}, sit in a unique position at the intersection of the Cosmic and Energy Frontiers. They can simultaneously inform our knowledge of the most extreme processes in the Universe and of particle physics well beyond the energies reachable by terrestrial accelerators.

\vspace{-4mm}
\subparagraph{\bf  Twenty years of UHECR discoveries} The past twenty years have been rich in fundamental advances in the field thanks to the Pierre Auger Observatory (Auger) in Argentina, \ac{TA} in the US, and the IceCube Neutrino Observatory (IceCube) in Antarctica, the first giant arrays of their kind. Far from the old and simplistic view of \acp{UHECR} dominated by protons at the highest energies, the experiments have uncovered a much more complex and nuanced picture originating mainly from the observation that the primary composition is a mixture of protons and heavier nuclei which changes significantly as a function of energy. At the Cosmic Frontier, the identification of the \ac{UHECR} sources is made more challenging by this as heavier (higher charged) primaries undergo larger deflections in galactic and extragalactic magnetic fields. Yet, the extragalactic origins of \acp{UHECR} beyond $8$\,EeV has been demonstrated through the observation of a large scale dipole in arrival direction. At the highest energies, there is also evidence for anisotropy at intermediate angular scales ($10$--$20^{\circ}$) with regional ``hot spots" both in the northern and southern hemispheres and growing signals of correlations with candidate source classes. At the Energy Frontier, particle physics measurements, such as cross sections at energies far beyond those available at terrestrial accelerators, can only be performed if the nature of the \ac{UHECR} beam at Earth is known.
Hence, measurements of nuclei-air cross sections have so far been with the tails of distributions in an energy range where there is wide agreement that protons are a substantial fraction the flux.

\vspace{-4mm}
\subparagraph{\bf Particle physics at the Cosmic Frontier} 

\begin{figure}[!ht]
     \centering
     \vspace{-1em}
     
     \includegraphics[width=0.75\textwidth]{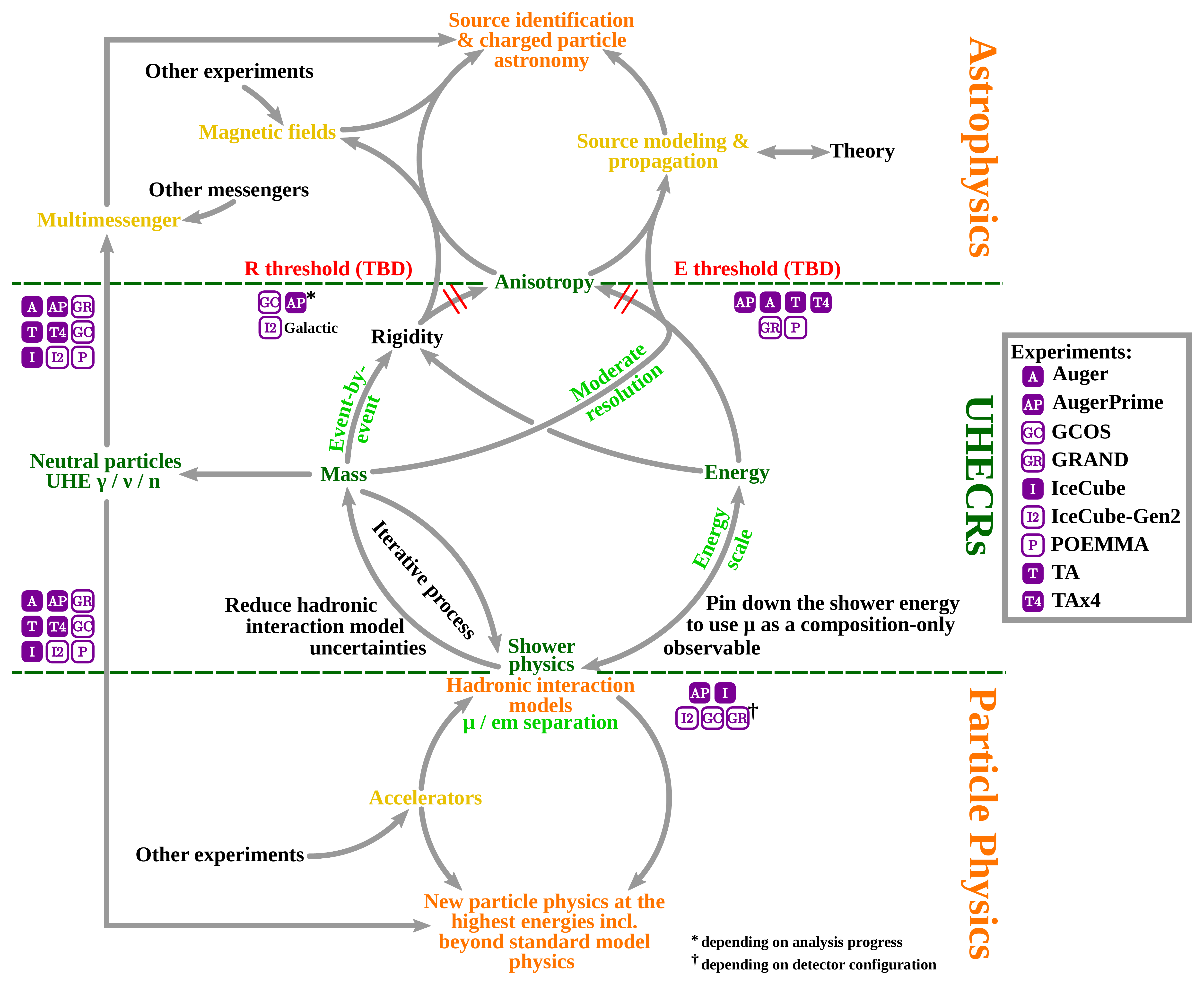}

     \caption{Diagram summarizing the strong connections of \acp{UHECR} with particle physics and astrophysics, the fundamental objectives of the field (in orange) for the next two decades, and the complementarity of current and next-generation experiments in addressing them.}
     \label{fig:the_diagram_execsummary}
\vspace{-4mm}
 \end{figure}

Hadronic interaction models, continuously informed by new accelerator data, play a key role in our understanding of the physics driving the production of \acp{EAS} induced by \acp{UHECR} in the atmosphere. Thanks to ever more precise measurements from \ac{UHECR} experiments, there are now strong indications that our understanding is incomplete. In particular, all of the hadronic models underestimate the number of muons produced in \acp{EAS}, hinting at new particle physics processes at the highest energies. Reducing the systematic uncertainties between models and incorporating the missing ingredients are major goals at the interface of the field of \acp{UHECR} and particle physics as shown in the summary diagram of \cref{fig:the_diagram_execsummary}.
%
The general strategy to solve the ``Muon Puzzle'' relies on the accurate determination of the energy scale combined with a precise set of measurements over a large parameter space, that can together disentangle the electromagnetic and muon components of \acp{EAS}. A muon-number resolution of $<$\,$15\%$ 
is within reach with upgraded detectors in the next decade using hybrid measurements. Achieving the prime goal of $<$\,$10\%$
will likely require a purposely-built next-generation observatory. Our ability to precisely determine the \ac{UHECR} mass composition hinges on our understanding of the physics driving the production of \acp{EAS}. Hence, solving the Muon Puzzle will allow for a better determination of the primary mass groups, possibly on an event-by-event basis. 

\vspace{-4mm}
\subparagraph{\bf A sensitive probe to BSM physics and dark matter} There is also the possibility that the Muon Puzzle does not originate from an incomplete understanding of the forward particle physics involved in shower physics. In this case, \ac{UHECR} measurements would provide a unique probe of new \ac{BSM} physics with a high potential for discovery. One main objective of the particle physics program is to discover the connection between \acf{DM} and the \acf{SM}. In addition to the searches for \acs{BSM} physics in \acs{EAS}, UHECR observatories offer a unique probe of the dark matter mass spectrum near the scale of \acfp{GUT}. The origin of \acf{SHDM} particles can be connected to inflationary cosmologies and their decay to instanton-induced processes, which would produce a cosmic flux of \ac{UHE} neutrinos and photons. While their non-observation sets restrictive constraints on the gauge couplings of the \acs{DM} models, the unambiguous detection of a single \acs{UHE} photon or neutrino would be a game changer in the quest to identify the \acs{DM} properties.  \acs{UHECR} experiments could be also sensitive to interactions induced by macroscopic \acs{DM} or nuclearites in the atmosphere, offering further windows to identify the nature of \ac{DM}.


\vspace{-4mm}
\subparagraph{\bf Astrophysics at the Energy Frontier} The ability to precisely measure both energy and mass composition on an event-by-event basis simultaneously is critical as together they would give access to each primary particle's rigidity as a new observable.
Given the natural relationship between rigidity and magnetic deflection, rigidity-based measurements will facilitate revealing the nature and origin(s) of \acp{UHECR} and enable charged-particle astronomy, the ability to study individual (classes of) sources with \acp{UHECR}.
At the highest energies, the classic approach of maximizing exposure and achieving good energy resolution and moderate mass discrimination may well be sufficient if the composition is pure or is bimodal comprising a mix of only protons and Fe nuclei, for example. We already know however that this is not the case at energies below the flux suppression. 
Thus, a purposely-built observatory combining excellent energy resolution and mass discrimination will be complementary to instruments with possibly larger exposure. It is also clear that both approaches will benefit from the reduction of systematic uncertainties between hadronic interaction models. \acp{UHECR} also have an important role to play in multi-messenger astrophysics, not only as cosmic messengers themselves but also as the source of \ac{UHE} photons and neutrinos.

\vspace{-4mm}
\subparagraph{\bf Upgrades of the current giant arrays}
To address the paradigm shift arising from the results of the current generation of experiments, three upgrades are either planned or already underway.
\TAxFour, a 4-fold expansion of \ac{TA}, will allow for Auger-like exposure in the northern hemisphere with the aim of identifying (classes of) \ac{UHECR} sources and further investigating potential differences between the northern and southern skies. AugerPrime, the upgrade of Auger, focuses on achieving mass-composition sensitivity for each \ac{EAS} measured by its upgraded surface detector through multi-hybrid observations. IceCube-Gen2, IceCube's planned upgrade, will include an expansion of the surface array to measure \acp{UHECR} with energies of up to a few EeV, providing a unique laboratory to study cosmic-ray physics, such as the insufficiently understood prompt particle-decays in \ac{EAS}.
It will also be used to study the transition from galactic to extragalactic sources, by combining the mass-sensitive observables of the surface and deep in-ice detectors.
The upgrades benefit from recent technological advances, including the resurgence of the radio technique as a competitive method 
and the development of machine learning as a powerful new analysis technique. Through extrapolation from the current state of analyses, the energy-dependent resolutions for mass observables in AugerPrime may reach as low as $20$\,\gcm{} for the atmospheric depth of the shower maximum, \xmax{}, and $10\%$ for the muon number at the highest energies ($E > 10$\,EeV). If these resolutions are achieved, AugerPrime 
should be able to distinguish between iron and proton on an event-by-event basis at $90\%$~C.L. and even separate iron from the CNO group at better than $50\%$~C.L.,
allowing for composition-enhanced anisotropy studies. One of its design goals is to identify the possible existence of a $10\%$ proton fraction at the highest energies. 

\begin{figure}[!htb]
     \centering
     \includegraphics[width=\textwidth]{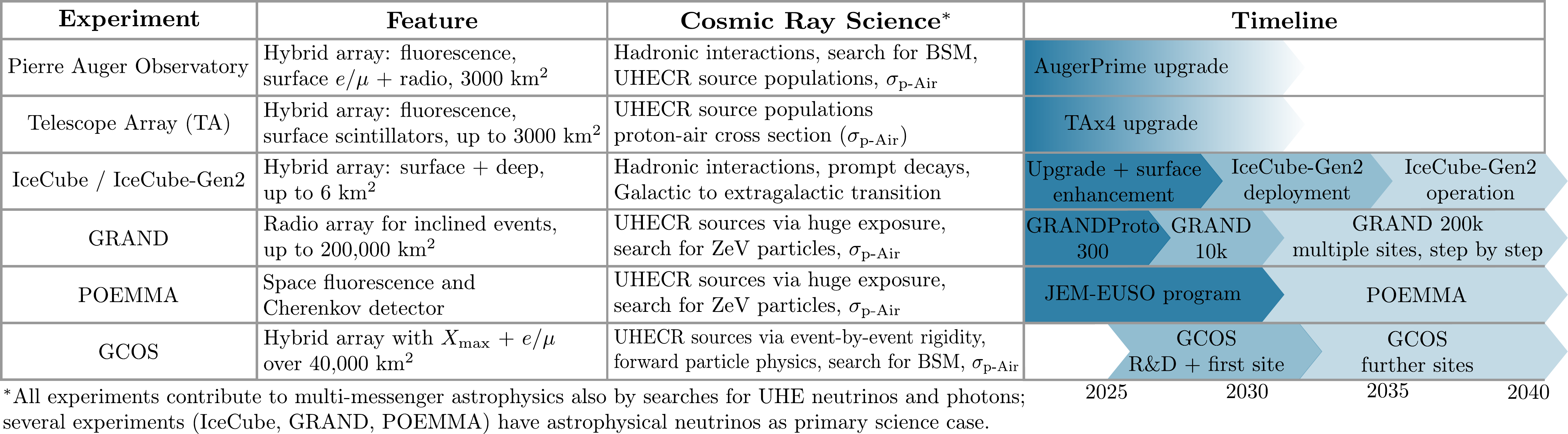}
     \vspace{-4mm}
     \caption{Upgraded and next-generation \ac{UHECR} instruments with their defining features, main scientific goals, and timeline.}
     \label{fig:the_table_execsummary}
    \vspace{-4mm}
 \end{figure}

\vspace{-4mm}
\subparagraph{\bf The exciting future ahead} Thanks to increasingly precise measurements, achieving the primary goals outlined at the top and bottom of \cref{fig:the_diagram_execsummary} are within reach in the next two decades. This will be done through complementary approaches taken by the upgraded and next-generation \ac{UHECR} detectors. The \acf{POEMMA} space observatory and the multi-site \acf{GRAND} ground observatory are two instruments that will measure both \ac{UHE} neutrinos and cosmic rays. Thanks to their large exposure, both \ac{POEMMA} and \ac{GRAND} will be able to search for \ac{UHECR} sources and ZeV particles beyond the flux suppression. The \acf{GCOS}, a $40,000$\,km$^{2}$ ground array likely split in at least two locations, one or more of them possibly co-located with a \ac{GRAND} site, will be the purposely-built precision multi-instrument ground array mentioned earlier. Its design will need to meet the goal of $<$\,$10\%$ muon-number resolution
to leverage our improved understanding of hadronic interactions. With these capabilities, \ac{GCOS} will be able to study particle and \ac{BSM} physics at the Energy Frontier while determining mass composition on an event-by-event basis to enable rigidity-based studies of \ac{UHECR} sources at the Cosmic Frontier. \cref{fig:the_table_execsummary} summarizes the features, complementary goals, and timeline of the upgraded and next-generation instruments.

\vspace{-4mm}
\subparagraph{\bf Interdisciplinary science and broader impact} The study of \acp{UHECR} leverages the atmosphere as a detector, providing many opportunities to study atmospheric science in particular. \ac{UHECR} detectors are extremely well suited for detecting transient events induced by the weather and even a variety of other exotic phenomena.
From a broader impact perspective, big science uses a lot of resources and the \ac{UHECR} community needs to be more aware of its societal and environmental impacts. For example, a community-wide effort to achieve carbon neutrality could not only help mitigate the effects of climate change, but also set a new standard to be followed outside of the scientific community. Likewise, a commitment to the principles of open science and open data can only benefit the \ac{UHECR} community by reducing the scientific gap between countries and increasing the potential for discoveries in the future. Most importantly, as we look two decades into the future, there has to be a strong renewed pledge for a diverse, equitable, and inclusive community -- ensuring equal opportunities for success and  transforming the workforce of our field.

\subparagraph{\bf Recommendations:} 
\vspace{-4mm}

\begin{itemize}
    \setlength\itemsep{0.2mm}
    \item Even in the most optimistic scenario, the first next-generation experiment will not be operational until around 2030. AugerPrime and \TAxFour should continue operation until at least 2032.
    \item IceCube and IceCube-Gen2 provide a unique laboratory to study particle physics in air showers. For this purpose, 
    the deep detector in the ice should be complemented by a hybrid surface array for sufficiently accurate measurements of the air showers.
    \item A robust effort in R\&D should continue in detector developments and cross-calibrations for all air-shower components, and also in computing techniques.
    This effort should include, whenever possible, optimized triggers for photons, neutrinos and transient events.
    \item \textcolor{red}{Closer collaboration between theorists and experimentalists is required. Clear pathways for theorists to propose analyses and receive feedback should be established, as testable predictions leading to specific measurement goals are needed to inform design choices. }
    \item To achieve the high precision \ac{UHECR} particle physics studies needed to provide strong constraints for leveraging by accelerator experiments at extreme energies, even finer grained calibration methods, of the absolute energy-scale for example, should be rigorously pursued.
    \item The next-generation experiments (\ac{GCOS}, \ac{GRAND}, and \ac{POEMMA}) will provide complementary information needed to meet the goals of the \ac{UHECR} community in the next two decades. They should proceed through their respective next stages of planning and prototyping.
    \item At least one next-generation experiment needs to be able to make high-precision measurements to explore new particle physics and measure particle rigidity on an event-by-event basis. 
    Of the planned next-generation experiments, \ac{GCOS} is the best positioned to meet this recommendation.
    \item As a complementary effort, experiments with sufficient exposure ($\gtrsim$\,$5\times10^5$\,km$^2$\,sr\,yr) are needed to search for 
    \ac{LIV}, \ac{SHDM}, and other \ac{BSM} physics at the Cosmic and Energy Frontiers, and to identify \ac{UHECR} sources at the highest energies.
    \item Full-sky coverage with low cross-hemisphere systematic uncertainties is critical for astrophysical studies. To this end, next generation experiments should be space-based or multi-site. Common sites between experiments are encouraged.
    \item Based on the productive results from inter-collaboration and inter-disciplinary work, we recommend the continued progress/formation of joint analyses between experiments and with other intersecting fields of research (e.g., magnetic fields).
    
    \item The \ac{UHECR} community should continue its efforts to advance diversity, equity, inclusion, and accessibility. It also needs to take steps to reduce its environmental impacts and improve open access to its data to reduce the scientific gap between countries.
    
\end{itemize}
\section*{Acknowledgements}
\addcontentsline{toc}{section}{Acknowledgements}

The authors would like to first thank all of those who participated in the UHECRs white-paper planning and review meetings, as well as all of the authors who contributed letters of interest to the original Snowmass call for submissions. We would like to especially acknowledge those who provided editorial feedback on the content and presentation of the materials of this white paper. In particular, we would like to thank Hernan Asorey, Peter L. Biermann, Johannes Bl\"{u}mer, Mauricio Bustamante, Carola Dobrigkeit, Bianca Keilhauer, Jim Matthews, Silvia Mollerach, Hiroyuki Sagawa, Max Stadelmaier, {\color{red} Fabian Sch\"ussler,} Franco Vazza, and Alan Watson
for their timely and critical feedback. We would like to also thank the IceCube, Pierre Auger, and Telescope Array collaborations for their work, commentary and vetting. Finally, we would like to particularly recognize the funding agencies, organizations, individuals, governments and, above all, tax payers who have financed the on-going study of \acp{UHECR}; thank you, this white-paper would have been impossible without your support.\bigskip

\noindent Additionally: 
$\bullet$ J.~Alvarez-Mu\~niz and E.~Zas would like to acknowledge funding from Xunta de Galicia (Centro singular deinvestigaci\'on de Galicia accreditation 2019-2022), from European Union ERDF, from the ''Mar\'\i a deMaeztu'' Units of Excellence program MDM-2016-0692, the Spanish Research State Agency and from Ministerio de Ciencia e Innovaci\'on PID2019-105544GB-I00 and RED2018-102661-T (RENATA).
$\bullet$ R.~Alves Batista acknowledges the support of the ``la Caixa'' Foundation (ID 100010434) and the European Union's Horizon~2020 research and innovation program under the Marie Skłodowska-Curie grant agreement No\,847648, fellowship code LCF/BQ/PI21/11830030.
$\bullet$ L.~A.~Anchordoqui is supported by the US National Science Foundation NSF Grant PHY-2112527.
{\color{red} $\bullet$ E.~Bechtol would like to acknowledge the Multimessenger Diversity Network and the support of the National Science Foundation.
$\bullet$ D.R. Bergman acknowledges support from the U.S National Science Foundation under the grants PHY-2012934 and PHY-2112904 and from the U.S. National Space and Aeronautics Administration under grant 80NSSC19K0485.}
$\bullet$ P.~B.~Denton acknowledges support from the US Department of Energy under Grant Contract DE-SC0012704.
$\bullet$ H.~Dujmovic and F.~Schroeder would like to acknowledge that this project has received funding from the European Research Council (ERC) under the European Union’s Horizon 2020 research and innovation programme (grant agreement No 802729). F.~Schroeder was also supported by grants NSF EPSCoR RII Track-2 FEC award \#2019597 and NSF CAREER award \#2046386 as was A.~Coleman.
{\color{red} $\bullet$ R.~Concei\c{c}\~ao and F.~Riehn would like to acknowledge the support of Funda\c{c}\~ao para a Ci\^encia e Tecnologia via DL57/2016/cP1330/cT0002 and CERN/FIS-PAR/0020/2021.}
$\bullet$ J.~Eser was supported by NASA grant 16-APRA16-0113.
$\bullet$ N.~Globus' research is supported by the Simons Foundation, The Chancellor Fellowship at UCSC and the Vera Rubin Presidential Chair.
$\bullet$ J.~Glombitza would like to acknowledge the support by the Ministry of Innovation, Science and Research of the State of North Rhine-Westphalia, and the Federal Ministry of Education and Research (BMBF)
$\bullet$ G.~Golup would like to acknowledge the support of CONICET (PIP 11220200100565CO) and ANPCyT(PICT 2018-03069).
$\bullet$ M.~Gritsevich acknowledges the Academy of Finland project nos. 325806 and 338042.
{\color{red} $\bullet$ P.~Klimov was supported by the Russian Science Foundation grant 22-62-00010.
$\bullet$ K.~Kotera would like to acknowledge the support of the APACHE grant (ANR-16-CE31-0001) of the French Agence Nationale de la Recherche.}
$\bullet$ J.~F.~Krizmanic acknowledges support by NASA grant 80NSSC19K0626 at the University of Maryland, Baltimore County under proposal 17-APRA17-0066 at NASA/GSFC and JPL and NASA grant 16-APROBES-0023.
{\color{red} $\bullet$ A.\,V.~Olinto was supported by NASA grant 80NSSC18K0246.}
$\bullet$ E.~Mayotte, S.~Mayotte, and F.~Sarazin would like to acknowledge the support of the NSF through grant \#2013146 and NASA through grant \#80NSSC19K0460.
{\color{red} $\bullet$ J. Madsen would like to acknowledge the support of the U.S. National Science Foundation (NSF) Office of Polar Programs, the NSF-Physics Division, and the NSF Division of Research on Learning.
$\bullet$ I.\,C.~Mari\c{s} would like to acknowledge the support of Fonds de la Recherche Scientifique (FNRS).
$\bullet$ J.\,N.~Matthews is grateful for the long standing support of the University of Utah Cosmic Ray group by US.
National Science Foundation under numerous grants.}
$\bullet$ M.~S.~Muzio would like to acknowledge support by the NSF MPS-Ascend Postdoctoral Award \#2138121.
$\bullet$ M.~Plum was supported by the grant NSF EPSCoR RII Track-2 FEC award 2019597.
{\color{red} $\bullet$ L.\,W.~Piotrowski acknowledges financing by the Polish National Agency for Academic Exchange within Polish Returns Programme no. PPN/PPO/2020/1/00024/U/00001 and National Science Centre, Poland grant no. 2021/01/1/ST2/00002”.}
$\bullet$ E.~Santos, J.~V\'{i}cha, and A.~Yushov would like to acknowledge the support of the Czech Science Foundation via 21-02226M and MSMT CR via CZ.02.1.01/0.0/0.0/16\_013/0001402, CZ.02.1.01/0.0/0.0/18\_046/0016010, CZ.02.1.01/0.0/0.0/17\_0\\49/0008422, LTT18004, LM2015038, and LM2018102.
$\bullet$ F.~Schl\"{u}ter is supported by the Helmholtz International Research School for Astroparticle Physics and Enabling Technologies (HIRSAP) (grant number HIRS-0009).
{\color{red} $\bullet$ D.~H.~Shoemaker thanks the NSF for support of the LIGO Lab, CA \#1764464.}
$\bullet$ D.~Soldin acknowledges support from the US NSF Grant PHY-1913607.
{\color{red} $\bullet$ Y.~Tsunesada is supported by Japanese Society for the Promotion of Science(JSPS) through Science Research (A)JP18H03705.}
$\bullet$ T.~Venters  would like to acknowledge the support of NASA through grant 17-APRA17-0066.
$\bullet$ The LAGO Collaboration would like to thank the Pierre Auger Collaboration for its continuous support.
$\bullet$ Mini-EUSO was supported by Italian Space Agency through the ASI INFN agreement n. 2020-26-HH.0 and contract n. 2016-1-U.0.
$\bullet$ The conceptual design of POEMMA was supported by NASA Probe Mission Concept Study grant NNX17AJ82G for the 2020 Decadal Survey Planning. Additional contributions to POEMMA were supported in part by NASA awards 16-APROBES16-0023, NNX17AJ82G, NNX13AH54G, 80NSSC18K0246, and 80NSSC18K0473.
$\bullet$ The co-authors at German institutions would like to generally acknowledge and thank the Bundesministerium f\"{u}r Bildung und Forschung (BMBF) and the Deutsche Forschungsgemeinschaft (DFG) for their support.\bigskip

{\color{red} Lastly, it is important to note that} the opinions and conclusions expressed herein are those of the authors and do not necessarily represent {\color{red} those of the above listed} funding agencies.

\bibliography{References}

\begin{thebibliography}{1000}
\expandafter\ifx\csname url\endcsname\relax
  \def\url#1{\texttt{#1}}\fi
\expandafter\ifx\csname urlprefix\endcsname\relax\def\urlprefix{URL }\fi
\expandafter\ifx\csname href\endcsname\relax
  \def\href#1#2{#2} \def\path#1{#1}\fi

\bibitem{Sarazin:2019fjz}
F.~Sarazin, et~al., {What is the nature and origin of the highest-energy
  particles in the universe?}, Bull. Am. Astron. Soc. 51~(3) (2019) 93.
\newblock \href {http://arxiv.org/abs/1903.04063} {\path{arXiv:1903.04063}}.

\bibitem{AlvesBatista:2019tlv}
R.~Alves~Batista, et~al., {Open Questions in Cosmic-Ray Research at Ultrahigh
  Energies}, Front. Astron. Space Sci. 6 (2019) 23.
\newblock \href {http://arxiv.org/abs/1903.06714} {\path{arXiv:1903.06714}},
  \href {http://dx.doi.org/10.3389/fspas.2019.00023}
  {\path{doi:10.3389/fspas.2019.00023}}.

\bibitem{Anchordoqui:2018qom}
L.~A. Anchordoqui, {Ultra-High-Energy Cosmic Rays}, Phys. Rept. 801 (2019)
  1--93.
\newblock \href {http://arxiv.org/abs/1807.09645} {\path{arXiv:1807.09645}},
  \href {http://dx.doi.org/10.1016/j.physrep.2019.01.002}
  {\path{doi:10.1016/j.physrep.2019.01.002}}.

\bibitem{halzen2003multi}
F.~Halzen, Multi-messenger astronomy: cosmic rays, gamma-rays, and neutrinos,
  in: Texas In Tuscany, World Scientific, 2003, pp. 117--131.

\bibitem{Barwick:1998xq}
S.~W. Barwick, {High-energy cosmic neutrinos}, Phys. Scripta T 85 (2000)
  106--116.
\newblock \href {http://arxiv.org/abs/astro-ph/9903467}
  {\path{arXiv:astro-ph/9903467}}, \href
  {http://dx.doi.org/10.1238/Physica.Topical.085a00106}
  {\path{doi:10.1238/Physica.Topical.085a00106}}.

\bibitem{Kotera:2010yn}
K.~Kotera, D.~Allard, A.~V. Olinto, {Cosmogenic Neutrinos: parameter space and
  detectabilty from PeV to ZeV}, JCAP 10 (2010) 013.
\newblock \href {http://arxiv.org/abs/1009.1382} {\path{arXiv:1009.1382}},
  \href {http://dx.doi.org/10.1088/1475-7516/2010/10/013}
  {\path{doi:10.1088/1475-7516/2010/10/013}}.

\bibitem{Ulrich:2010rg}
R.~Ulrich, R.~Engel, M.~Unger, {Hadronic Multiparticle Production at Ultra-High
  Energies and Extensive Air Showers}, Phys. Rev. D 83 (2011) 054026.
\newblock \href {http://arxiv.org/abs/1010.4310} {\path{arXiv:1010.4310}},
  \href {http://dx.doi.org/10.1103/PhysRevD.83.054026}
  {\path{doi:10.1103/PhysRevD.83.054026}}.

\bibitem{Albrecht:2021cxw}
J.~Albrecht, et~al., {The Muon Puzzle in cosmic-ray induced air showers and its
  connection to the Large Hadron Collider}, Astrophys. Space Sci. 367~(3)
  (2022) 27.
\newblock \href {http://arxiv.org/abs/2105.06148} {\path{arXiv:2105.06148}},
  \href {http://dx.doi.org/10.1007/s10509-022-04054-5}
  {\path{doi:10.1007/s10509-022-04054-5}}.

\bibitem{PierreAuger:2021tog}
P.~Abreu, et~al., {\bf Pierre Auger} Collaboration, {Testing effects of Lorentz
  invariance violation in the propagation of astroparticles with the Pierre
  Auger Observatory}\href {http://arxiv.org/abs/2112.06773}
  {\path{arXiv:2112.06773}}.

\bibitem{Saveliev:2011vw}
A.~Saveliev, L.~Maccione, G.~Sigl, {Lorentz Invariance Violation and Chemical
  Composition of Ultra High Energy Cosmic Rays}, JCAP 03 (2011) 046.
\newblock \href {http://arxiv.org/abs/1101.2903} {\path{arXiv:1101.2903}},
  \href {http://dx.doi.org/10.1088/1475-7516/2011/03/046}
  {\path{doi:10.1088/1475-7516/2011/03/046}}.

\bibitem{Diaz:2016dpk}
J.~S. Diaz, F.~R. Klinkhamer, M.~Risse, {Changes in extensive air showers from
  isotropic Lorentz violation in the photon sector}, Phys. Rev. D 94~(8) (2016)
  085025.
\newblock \href {http://arxiv.org/abs/1607.02099} {\path{arXiv:1607.02099}},
  \href {http://dx.doi.org/10.1103/PhysRevD.94.085025}
  {\path{doi:10.1103/PhysRevD.94.085025}}.

\bibitem{Anchordoqui:2017pmf}
L.~A. Anchordoqui, J.~F. Soriano, {New test of Lorentz symmetry using
  ultrahigh-energy cosmic rays}, Phys. Rev. D 97~(4) (2018) 043010.
\newblock \href {http://arxiv.org/abs/1710.00750} {\path{arXiv:1710.00750}},
  \href {http://dx.doi.org/10.1103/PhysRevD.97.043010}
  {\path{doi:10.1103/PhysRevD.97.043010}}.

\bibitem{Klinkhamer:2017puj}
F.~R. Klinkhamer, M.~Niechciol, M.~Risse, {Improved bound on isotropic Lorentz
  violation in the photon sector from extensive air showers}, Phys. Rev. D
  96~(11) (2017) 116011.
\newblock \href {http://arxiv.org/abs/1710.02507} {\path{arXiv:1710.02507}},
  \href {http://dx.doi.org/10.1103/PhysRevD.96.116011}
  {\path{doi:10.1103/PhysRevD.96.116011}}.

\bibitem{GuedesLang:2017sfl}
R.~Guedes~Lang, H.~Mart\'\i{}nez-Huerta, V.~de~Souza, {Limits on the Lorentz
  Invariance Violation from UHECR astrophysics}, Astrophys. J. 853~(1) (2018)
  23.
\newblock \href {http://arxiv.org/abs/1701.04865} {\path{arXiv:1701.04865}},
  \href {http://dx.doi.org/10.3847/1538-4357/aa9f2c}
  {\path{doi:10.3847/1538-4357/aa9f2c}}.

\bibitem{Torri:2020fao}
M.~D.~C. Torri, L.~Caccianiga, A.~di~Matteo, A.~Maino, L.~Miramonti,
  {Predictions of Ultra-High Energy Cosmic Ray Propagation in the Context of
  Homogeneously Modified Special Relativity}, Symmetry 12~(12) (2020) 1961.
\newblock \href {http://arxiv.org/abs/2110.09900} {\path{arXiv:2110.09900}},
  \href {http://dx.doi.org/10.3390/sym12121961}
  {\path{doi:10.3390/sym12121961}}.

\bibitem{Duenkel:2021gkq}
F.~Duenkel, M.~Niechciol, M.~Risse, {Photon decay in ultrahigh-energy air
  showers: Stringent bound on Lorentz violation}, Phys. Rev. D 104~(1) (2021)
  015010.
\newblock \href {http://arxiv.org/abs/2106.01012} {\path{arXiv:2106.01012}},
  \href {http://dx.doi.org/10.1103/PhysRevD.104.015010}
  {\path{doi:10.1103/PhysRevD.104.015010}}.

\bibitem{PierreAuger:2021mve}
P.~Abreu, et~al., {\bf Pierre Auger} Collaboration, {Constraining Lorentz
  Invariance Violation using the muon content of extensive air showers measured
  at the Pierre Auger Observatory}, PoS ICRC2021 (2021) 340.
\newblock \href {http://dx.doi.org/10.22323/1.395.0340}
  {\path{doi:10.22323/1.395.0340}}.

\bibitem{ThePierreAuger:2022}
A.~Aab, et~al., {\bf Pierre Auger} Collaboration, {Constraints on
  gravitationally-produced super-heavy dark matter particles in the early
  Universe using the Pierre Auger Observatory}, in preparation.

\bibitem{EAS-MSU:2019kmv}
H.~P. Dembinski, et~al., {\bf EAS-MSU, IceCube, KASCADE-Grande, NEVOD-DECOR,
  Pierre Auger, SUGAR, Telescope Array, Yakutsk EAS Array} Collaboration,
  {Report on Tests and Measurements of Hadronic Interaction Properties with Air
  Showers}, EPJ Web Conf. 210 (2019) 02004.
\newblock \href {http://arxiv.org/abs/1902.08124} {\path{arXiv:1902.08124}},
  \href {http://dx.doi.org/10.1051/epjconf/201921002004}
  {\path{doi:10.1051/epjconf/201921002004}}.

\bibitem{Cazon:2020zhx}
L.~Cazon, {\bf EAS-MSU, IceCube, KASCADE Grande, NEVOD-DECOR, Pierre Auger,
  SUGAR, Telescope Array, Yakutsk EAS Array} Collaboration, {Working Group
  Report on the Combined Analysis of Muon Density Measurements from Eight Air
  Shower Experiments}, PoS ICRC2019 (2020) 214.
\newblock \href {http://arxiv.org/abs/2001.07508} {\path{arXiv:2001.07508}},
  \href {http://dx.doi.org/10.22323/1.358.0214}
  {\path{doi:10.22323/1.358.0214}}.

\bibitem{Soldin:2021wyv}
D.~Soldin, {\bf EAS-MSU, IceCube, KASCADE-Grande, NEVOD-DECOR, Pierre Auger,
  SUGAR, Telescope Array, Yakutsk EAS Array} Collaboration, {Update on the
  Combined Analysis of Muon Measurements from Nine Air Shower Experiments}, PoS
  ICRC2021 (2021) 349.
\newblock \href {http://arxiv.org/abs/2108.08341} {\path{arXiv:2108.08341}},
  \href {http://dx.doi.org/10.22323/1.395.0349}
  {\path{doi:10.22323/1.395.0349}}.

\bibitem{PierreAuger:2016qzd}
A.~Aab, et~al., {\bf Pierre Auger} Collaboration, {The Pierre Auger Observatory
  Upgrade - Preliminary Design Report}\href {http://arxiv.org/abs/1604.03637}
  {\path{arXiv:1604.03637}}.

\bibitem{Kido:2019enj}
E.~Kido, {\bf Telsecope Array} Collaboration, {Status and prospects of the TAx4
  experiment}, EPJ Web Conf. 210 (2019) 06001.
\newblock \href {http://dx.doi.org/10.1051/epjconf/201921006001}
  {\path{doi:10.1051/epjconf/201921006001}}.

\bibitem{IceCube:2014gqr}
M.~G. Aartsen, et~al., {\bf IceCube} Collaboration, {IceCube-Gen2: A Vision for
  the Future of Neutrino Astronomy in Antarctica}\href
  {http://arxiv.org/abs/1412.5106} {\path{arXiv:1412.5106}}.

\bibitem{POEMMA:2020ykm}
A.~V. Olinto, et~al., {\bf POEMMA} Collaboration, {The POEMMA (Probe of Extreme
  Multi-Messenger Astrophysics) observatory}, JCAP 06 (2021) 007.
\newblock \href {http://arxiv.org/abs/2012.07945} {\path{arXiv:2012.07945}},
  \href {http://dx.doi.org/10.1088/1475-7516/2021/06/007}
  {\path{doi:10.1088/1475-7516/2021/06/007}}.

\bibitem{GRAND:2018iaj}
J.~\'Alvarez-Mu\~niz, et~al., {\bf GRAND} Collaboration, {The Giant Radio Array
  for Neutrino Detection (GRAND): Science and Design}, Sci. China Phys. Mech.
  Astron. 63~(1) (2020) 219501.
\newblock \href {http://arxiv.org/abs/1810.09994} {\path{arXiv:1810.09994}},
  \href {http://dx.doi.org/10.1007/s11433-018-9385-7}
  {\path{doi:10.1007/s11433-018-9385-7}}.

\bibitem{Horandel:2021prj}
J.~R. H\"orandel, {\bf GCOS} Collaboration, {GCOS -- The Global Cosmic Ray
  Observatory}, PoS ICRC2021 (2021) 027.
\newblock \href {http://arxiv.org/abs/2203.01127} {\path{arXiv:2203.01127}},
  \href {http://dx.doi.org/10.22323/1.395.0027}
  {\path{doi:10.22323/1.395.0027}}.

\bibitem{Bird:1994uy}
D.~J. Bird, et~al., {Detection of a cosmic ray with measured energy well beyond
  the expected spectral cutoff due to cosmic microwave radiation}, Astrophys.
  J. 441 (1995) 144--150.
\newblock \href {http://arxiv.org/abs/astro-ph/9410067}
  {\path{arXiv:astro-ph/9410067}}, \href {http://dx.doi.org/10.1086/175344}
  {\path{doi:10.1086/175344}}.

\bibitem{PierreAuger:2015eyc}
A.~Aab, et~al., {\bf Pierre Auger} Collaboration, {The Pierre Auger Cosmic Ray
  Observatory}, Nucl. Instrum. Meth. A 798 (2015) 172--213.
\newblock \href {http://arxiv.org/abs/1502.01323} {\path{arXiv:1502.01323}},
  \href {http://dx.doi.org/10.1016/j.nima.2015.06.058}
  {\path{doi:10.1016/j.nima.2015.06.058}}.

\bibitem{TelescopeArray:2008toq}
H.~Kawai, et~al., {\bf Telescope Array} Collaboration, {Telescope array
  experiment}, Nucl. Phys. B Proc. Suppl. 175-176 (2008) 221--226.
\newblock \href {http://dx.doi.org/10.1016/j.nuclphysbps.2007.11.002}
  {\path{doi:10.1016/j.nuclphysbps.2007.11.002}}.

\bibitem{IceCube:2016zyt}
M.~G. Aartsen, et~al., {\bf IceCube} Collaboration, {The IceCube Neutrino
  Observatory: Instrumentation and Online Systems}, JINST 12~(03) (2017)
  P03012.
\newblock \href {http://arxiv.org/abs/1612.05093} {\path{arXiv:1612.05093}},
  \href {http://dx.doi.org/10.1088/1748-0221/12/03/P03012}
  {\path{doi:10.1088/1748-0221/12/03/P03012}}.

\bibitem{Kampert:2012mx}
K.-H. Kampert, M.~Unger, {Measurements of the Cosmic Ray Composition with Air
  Shower Experiments}, Astropart. Phys. 35 (2012) 660--678.
\newblock \href {http://arxiv.org/abs/1201.0018} {\path{arXiv:1201.0018}},
  \href {http://dx.doi.org/10.1016/j.astropartphys.2012.02.004}
  {\path{doi:10.1016/j.astropartphys.2012.02.004}}.

\bibitem{PierreAuger:2020qqz}
A.~Aab, et~al., {\bf Pierre Auger} Collaboration, {Measurement of the
  cosmic-ray energy spectrum above $2.5{\times} 10^{18}$ eV using the Pierre
  Auger Observatory}, Phys. Rev. D 102~(6) (2020) 062005.
\newblock \href {http://arxiv.org/abs/2008.06486} {\path{arXiv:2008.06486}},
  \href {http://dx.doi.org/10.1103/PhysRevD.102.062005}
  {\path{doi:10.1103/PhysRevD.102.062005}}.

\bibitem{TelescopeArray:2021zox}
Y.~Tsunesada, {\bf Telescope Array, Pierre Auger} Collaboration, {Joint
  analysis of the energy spectrum of ultra-high-energy cosmic rays as measured
  at the Pierre Auger Observatory and the Telescope Array}, PoS ICRC2021 (2021)
  337.
\newblock \href {http://dx.doi.org/10.22323/1.395.0337}
  {\path{doi:10.22323/1.395.0337}}.

\bibitem{Greisen:1966jv}
K.~Greisen, {End to the cosmic ray spectrum?}, Phys. Rev. Lett. 16 (1966)
  748--750.
\newblock \href {http://dx.doi.org/10.1103/PhysRevLett.16.748}
  {\path{doi:10.1103/PhysRevLett.16.748}}.

\bibitem{Zatsepin:1966jv}
G.~T. Zatsepin, V.~A. Kuzmin, {Upper limit of the spectrum of cosmic rays},
  JETP Lett. 4 (1966) 78--80.

\bibitem{PierreAuger:2017pzq}
A.~Aab, et~al., {\bf Pierre Auger} Collaboration, {Observation of a Large-scale
  Anisotropy in the Arrival Directions of Cosmic Rays above $8 \times 10^{18}$
  eV}, Science 357~(6537) (2017) 1266--1270.
\newblock \href {http://arxiv.org/abs/1709.07321} {\path{arXiv:1709.07321}},
  \href {http://dx.doi.org/10.1126/science.aan4338}
  {\path{doi:10.1126/science.aan4338}}.

\bibitem{TelescopeArray:2014tsd}
R.~U. Abbasi, et~al., {\bf Telescope Array} Collaboration, {Indications of
  Intermediate-Scale Anisotropy of Cosmic Rays with Energy Greater Than 57 EeV
  in the Northern Sky Measured with the Surface Detector of the Telescope Array
  Experiment}, Astrophys. J. Lett. 790 (2014) L21.
\newblock \href {http://arxiv.org/abs/1404.5890} {\path{arXiv:1404.5890}},
  \href {http://dx.doi.org/10.1088/2041-8205/790/2/L21}
  {\path{doi:10.1088/2041-8205/790/2/L21}}.

\bibitem{PierreAuger:2018qvk}
A.~Aab, et~al., {\bf Pierre Auger} Collaboration, {An Indication of anisotropy
  in arrival directions of ultra-high-energy cosmic rays through comparison to
  the flux pattern of extragalactic gamma-ray sources}, Astrophys. J. Lett.
  853~(2) (2018) L29.
\newblock \href {http://arxiv.org/abs/1801.06160} {\path{arXiv:1801.06160}},
  \href {http://dx.doi.org/10.3847/2041-8213/aaa66d}
  {\path{doi:10.3847/2041-8213/aaa66d}}.

\bibitem{PierreAuger:2012egl}
P.~Abreu, et~al., {\bf Pierre Auger} Collaboration, {Measurement of the
  proton-air cross-section at $\sqrt{s}=57$ TeV with the Pierre Auger
  Observatory}, Phys. Rev. Lett. 109 (2012) 062002.
\newblock \href {http://arxiv.org/abs/1208.1520} {\path{arXiv:1208.1520}},
  \href {http://dx.doi.org/10.1103/PhysRevLett.109.062002}
  {\path{doi:10.1103/PhysRevLett.109.062002}}.

\bibitem{Ulrich:2015yoo}
R.~Ulrich, {\bf Pierre Auger} Collaboration, {Extension of the measurement of
  the proton-air cross section with the Pierre Auger Observatory}, PoS ICRC2015
  (2016) 401.
\newblock \href {http://dx.doi.org/10.22323/1.236.0401}
  {\path{doi:10.22323/1.236.0401}}.

\bibitem{Rautenberg:2021vvt}
J.~Rautenberg, {\bf Pierre Auger} Collaboration, {Limits on ultra-high energy
  photons with the Pierre Auger Observatory}, PoS ICRC2019 (2021) 398.
\newblock \href {http://dx.doi.org/10.22323/1.358.0398}
  {\path{doi:10.22323/1.358.0398}}.

\bibitem{PierreAuger:2014ucz}
A.~Aab, et~al., {\bf Pierre Auger} Collaboration, {Muons in Air Showers at the
  Pierre Auger Observatory: Mean Number in Highly Inclined Events}, Phys. Rev.
  D 91~(3) (2015) 032003, [Erratum: Phys.Rev.D 91, 059901 (2015)].
\newblock \href {http://arxiv.org/abs/1408.1421} {\path{arXiv:1408.1421}},
  \href {http://dx.doi.org/10.1103/PhysRevD.91.032003}
  {\path{doi:10.1103/PhysRevD.91.032003}}.

\bibitem{PierreAuger:2016nfk}
A.~Aab, et~al., {\bf Pierre Auger} Collaboration, {Testing Hadronic
  Interactions at Ultrahigh Energies with Air Showers Measured by the Pierre
  Auger Observatory}, Phys. Rev. Lett. 117~(19) (2016) 192001.
\newblock \href {http://arxiv.org/abs/1610.08509} {\path{arXiv:1610.08509}},
  \href {http://dx.doi.org/10.1103/PhysRevLett.117.192001}
  {\path{doi:10.1103/PhysRevLett.117.192001}}.

\bibitem{Pierog:2013ria}
T.~Pierog, I.~Karpenko, J.~M. Katzy, E.~Yatsenko, K.~Werner, {EPOS LHC: Test of
  collective hadronization with data measured at the CERN Large Hadron
  Collider}, Phys. Rev. C 92~(3) (2015) 034906.
\newblock \href {http://arxiv.org/abs/1306.0121} {\path{arXiv:1306.0121}},
  \href {http://dx.doi.org/10.1103/PhysRevC.92.034906}
  {\path{doi:10.1103/PhysRevC.92.034906}}.

\bibitem{Riehn:2019jet}
F.~Riehn, R.~Engel, A.~Fedynitch, T.~K. Gaisser, T.~Stanev, {Hadronic
  interaction model Sibyll 2.3d and extensive air showers}, Phys. Rev. D
  102~(6) (2020) 063002.
\newblock \href {http://arxiv.org/abs/1912.03300} {\path{arXiv:1912.03300}},
  \href {http://dx.doi.org/10.1103/PhysRevD.102.063002}
  {\path{doi:10.1103/PhysRevD.102.063002}}.

\bibitem{Bacholle:2020emk}
S.~Bacholle, et~al., {Mini-EUSO Mission to Study Earth UV Emissions on board
  the ISS}, Astrophys. J. Suppl. 253~(2) (2021) 36.
\newblock \href {http://arxiv.org/abs/2010.01937} {\path{arXiv:2010.01937}},
  \href {http://dx.doi.org/10.3847/1538-4365/abd93d}
  {\path{doi:10.3847/1538-4365/abd93d}}.

\bibitem{PierreAuger:2021hun}
P.~Abreu, et~al., {\bf Pierre Auger} Collaboration, {The energy spectrum of
  cosmic rays beyond the turn-down around $10^{17}$ eV as measured with the
  surface detector of the Pierre Auger Observatory}, Eur. Phys. J. C 81 (2021)
  966.
\newblock \href {http://arxiv.org/abs/2109.13400} {\path{arXiv:2109.13400}},
  \href {http://dx.doi.org/10.1140/epjc/s10052-021-09700-w}
  {\path{doi:10.1140/epjc/s10052-021-09700-w}}.

\bibitem{PierreAuger:2021tmd}
P.~Abreu, et~al., {\bf Pierre Auger} Collaboration, {Performance of the 433 m
  surface array of the Pierre Auger Observatory}, PoS ICRC2021 (2021) 224.
\newblock \href {http://dx.doi.org/10.22323/1.395.0224}
  {\path{doi:10.22323/1.395.0224}}.

\bibitem{augerWebsite}
\href{https://auger.org}{Figures for publications}, official website of the
  Pierre Auger Collaboration.
\newline\urlprefix\url{https://auger.org}

\bibitem{PierreAuger:2021rfz}
P.~Abreu, et~al., {\bf Pierre Auger} Collaboration, {The ultra-high-energy
  cosmic-ray sky above 32 EeV viewed from the Pierre Auger Observatory}, PoS
  ICRC2021 (2021) 307.
\newblock \href {http://dx.doi.org/10.22323/1.395.0307}
  {\path{doi:10.22323/1.395.0307}}.

\bibitem{PierreAuger:2021dqp}
R.~de~Almeida, et~al., {\bf Pierre Auger} Collaboration, {Large-scale and
  multipolar anisotropies of cosmic rays detected at the Pierre Auger
  Observatory with energies above 4 EeV}, PoS ICRC2021 (2021) 335.
\newblock \href {http://dx.doi.org/10.22323/1.395.0335}
  {\path{doi:10.22323/1.395.0335}}.

\bibitem{TelescopeArray:2021gxg}
A.~di~Matteo, et~al., {\bf Telescope Array, Pierre Auger} Collaboration, {UHECR
  arrival directions in the latest data from the original Auger and TA surface
  detectors and nearby galaxies}, PoS ICRC2021 (2021) 308.
\newblock \href {http://arxiv.org/abs/2111.12366} {\path{arXiv:2111.12366}},
  \href {http://dx.doi.org/10.22323/1.395.0308}
  {\path{doi:10.22323/1.395.0308}}.

\bibitem{Yushkov:2020nhr}
A.~Yushkov, {\bf Pierre Auger} Collaboration, {Mass Composition of Cosmic Rays
  with Energies above 10$^{17.2}$ eV from the Hybrid Data of the Pierre Auger
  Observatory}, PoS ICRC2019 (2020) 482.
\newblock \href {http://dx.doi.org/10.22323/1.358.0482}
  {\path{doi:10.22323/1.358.0482}}.

\bibitem{PierreAuger:2021jlg}
E.~Mayotte, et~al., {\bf Pierre Auger} Collaboration, {Indication of a
  mass-dependent anisotropy above $10^{18.7}\,$eV in the hybrid data of the
  Pierre Auger Observatory}, PoS ICRC2021 (2021) 321.
\newblock \href {http://dx.doi.org/10.22323/1.395.0321}
  {\path{doi:10.22323/1.395.0321}}.

\bibitem{PierreAuger:2021qsd}
A.~Aab, et~al., {\bf Pierre Auger} Collaboration, {Measurement of the
  Fluctuations in the Number of Muons in Extensive Air Showers with the Pierre
  Auger Observatory}, Phys. Rev. Lett. 126~(15) (2021) 152002.
\newblock \href {http://arxiv.org/abs/2102.07797} {\path{arXiv:2102.07797}},
  \href {http://dx.doi.org/10.1103/PhysRevLett.126.152002}
  {\path{doi:10.1103/PhysRevLett.126.152002}}.

\bibitem{PierreAuger:2021fkf}
A.~Aab, et~al., {\bf Pierre Auger} Collaboration, {Deep-learning based
  reconstruction of the shower maximum $X_{max}$ using the water-Cherenkov
  detectors of the Pierre Auger Observatory}, JINST 16~(07) (2021) P07019.
\newblock \href {http://arxiv.org/abs/2101.02946} {\path{arXiv:2101.02946}},
  \href {http://dx.doi.org/10.1088/1748-0221/16/07/P07019}
  {\path{doi:10.1088/1748-0221/16/07/P07019}}.

\bibitem{PierreAuger:2021nsq}
A.~Aab, et~al., {\bf Pierre Auger} Collaboration, {Extraction of the muon
  signals recorded with the surface detector of the Pierre Auger Observatory
  using recurrent neural networks}, JINST 16~(07) (2021) P07016.
\newblock \href {http://arxiv.org/abs/2103.11983} {\path{arXiv:2103.11983}},
  \href {http://dx.doi.org/10.1088/1748-0221/16/07/P07016}
  {\path{doi:10.1088/1748-0221/16/07/P07016}}.

\bibitem{PierreAuger:2019fdm}
A.~Aab, et~al., {\bf Pierre Auger} Collaboration, {Multi-Messenger Physics with
  the Pierre Auger Observatory}, Front. Astron. Space Sci. 6 (2019) 24.
\newblock \href {http://arxiv.org/abs/1904.11918} {\path{arXiv:1904.11918}},
  \href {http://dx.doi.org/10.3389/fspas.2019.00024}
  {\path{doi:10.3389/fspas.2019.00024}}.

\bibitem{Pedreira:2021gcl}
F.~Pedreira, {\bf Pierre Auger} Collaboration, {Bounds on diffuse and point
  source fluxes of ultra-high energy neutrinos with the Pierre Auger
  Observatory}, PoS ICRC2019 (2021) 979.
\newblock \href {http://dx.doi.org/10.22323/1.358.0979}
  {\path{doi:10.22323/1.358.0979}}.

\bibitem{PierreAuger:2019ens}
A.~Aab, et~al., {\bf Pierre Auger} Collaboration, {Probing the origin of
  ultra-high-energy cosmic rays with neutrinos in the EeV energy range using
  the Pierre Auger Observatory}, JCAP 10 (2019) 022.
\newblock \href {http://arxiv.org/abs/1906.07422} {\path{arXiv:1906.07422}},
  \href {http://dx.doi.org/10.1088/1475-7516/2019/10/022}
  {\path{doi:10.1088/1475-7516/2019/10/022}}.

\bibitem{PierreAuger:2021asv}
M.~Schimp, et~al., {\bf Pierre Auger} Collaboration, {Combined Search for UHE
  Neutrinos from Binary Black Hole Mergers with the Pierre Auger Observatory},
  PoS ICRC2021 (2021) 968.
\newblock \href {http://dx.doi.org/10.22323/1.395.0968}
  {\path{doi:10.22323/1.395.0968}}.

\bibitem{PierreAuger:2021oks}
P.~Abreu, et~al., {\bf Pierre Auger} Collaboration, {Follow-up Search for UHE
  Photons from Gravitational Wave Sources with the Pierre Auger Observatory},
  PoS ICRC2021 (2021) 973.
\newblock \href {http://dx.doi.org/10.22323/1.395.0973}
  {\path{doi:10.22323/1.395.0973}}.

\bibitem{PierreAuger:2020llu}
A.~Aab, et~al., {\bf Pierre Auger} Collaboration, {A Search for
  Ultra-high-energy Neutrinos from TXS 0506+056 Using the Pierre Auger
  Observatory}, Astrophys. J. 902~(2) (2020) 105.
\newblock \href {http://arxiv.org/abs/2010.10953} {\path{arXiv:2010.10953}},
  \href {http://dx.doi.org/10.3847/1538-4357/abb476}
  {\path{doi:10.3847/1538-4357/abb476}}.

\bibitem{LIGOScientific:2017ync}
B.~P. Abbott, et~al., {\bf LIGO Scientific, Virgo, Fermi GBM, INTEGRAL,
  IceCube, AstroSat Cadmium Zinc Telluride Imager Team, IPN, Insight-Hxmt,
  ANTARES, Swift, AGILE Team, 1M2H Team, Dark Energy Camera GW-EM, DES, DLT40,
  GRAWITA, Fermi-LAT, ATCA, ASKAP, Las Cumbres Observatory Group, OzGrav, DWF
  (Deeper Wider Faster Program), AST3, CAASTRO, VINROUGE, MASTER, J-GEM,
  GROWTH, JAGWAR, CaltechNRAO, TTU-NRAO, NuSTAR, Pan-STARRS, MAXI Team, TZAC
  Consortium, KU, Nordic Optical Telescope, ePESSTO, GROND, Texas Tech
  University, SALT Group, TOROS, BOOTES, MWA, CALET, IKI-GW Follow-up,
  H.E.S.S., LOFAR, LWA, HAWC, Pierre Auger, ALMA, Euro VLBI Team, Pi of Sky,
  Chandra Team at McGill University, DFN, ATLAS Telescopes, High Time
  Resolution Universe Survey, RIMAS, RATIR, SKA South Africa/MeerKAT}
  Collaboration, {Multi-messenger Observations of a Binary Neutron Star
  Merger}, Astrophys. J. Lett. 848~(2) (2017) L12.
\newblock \href {http://arxiv.org/abs/1710.05833} {\path{arXiv:1710.05833}},
  \href {http://dx.doi.org/10.3847/2041-8213/aa91c9}
  {\path{doi:10.3847/2041-8213/aa91c9}}.

\bibitem{ANTARES:2017bia}
A.~Albert, et~al., {\bf ANTARES, IceCube, Pierre Auger, LIGO Scientific, Virgo}
  Collaboration, {Search for High-energy Neutrinos from Binary Neutron Star
  Merger GW170817 with ANTARES, IceCube, and the Pierre Auger Observatory},
  Astrophys. J. Lett. 850~(2) (2017) L35.
\newblock \href {http://arxiv.org/abs/1710.05839} {\path{arXiv:1710.05839}},
  \href {http://dx.doi.org/10.3847/2041-8213/aa9aed}
  {\path{doi:10.3847/2041-8213/aa9aed}}.

\bibitem{PierreAuger:2021ibw}
V.~Novotn\'y, {\bf Pierre Auger} Collaboration, {Energy spectrum of cosmic rays
  measured using the Pierre Auger Observatory}, PoS ICRC2021 (2021) 324.
\newblock \href {http://dx.doi.org/10.22323/1.395.0324}
  {\path{doi:10.22323/1.395.0324}}.

\bibitem{PierreAuger:2020fbi}
A.~Aab, et~al., {\bf Pierre Auger} Collaboration, {Cosmic-ray anisotropies in
  right ascension measured by the Pierre Auger Observatory}, Astrophys. J. 891
  (2020) 142.
\newblock \href {http://arxiv.org/abs/2002.06172} {\path{arXiv:2002.06172}},
  \href {http://dx.doi.org/10.3847/1538-4357/ab7236}
  {\path{doi:10.3847/1538-4357/ab7236}}.

\bibitem{PierreAuger:2020lri}
A.~Aab, et~al., {\bf Pierre Auger} Collaboration, {A Three Year Sample of
  Almost 1600 Elves Recorded Above South America by the Pierre Auger Cosmic-Ray
  Observatory}, Earth Space Sci. 7~(4) (2020) e2019EA000582.
\newblock \href {http://dx.doi.org/10.1029/2019ea000582}
  {\path{doi:10.1029/2019ea000582}}.

\bibitem{PierreAuger:2021int}
P.~Abreu, et~al., {\bf Pierre Auger} Collaboration, {Downward Terrestrial
  Gamma-ray Flashes at the Pierre Auger Observatory?}, PoS ICRC2021 (2021) 395.
\newblock \href {http://dx.doi.org/10.22323/1.395.0395}
  {\path{doi:10.22323/1.395.0395}}.

\bibitem{Wiencke:2012dj}
L.~Wiencke, {\bf Pierre Auger} Collaboration, {The Pierre Auger Observatory and
  interdisciplinary science}, Eur. Phys. J. Plus 127 (2012) 98.
\newblock \href {http://dx.doi.org/10.1140/epjp/i2012-12098-6}
  {\path{doi:10.1140/epjp/i2012-12098-6}}.

\bibitem{PACollab-2017-ICRC-35}
T.~P.~A. collaboration, The pierre auger observatory: Contributions to the 35th
  international cosmic ray conference (icrc 2017), in: Proceedings of the 35th
  ICRC, Busan, 2017, [arXiv:1708.06592].

\bibitem{TelescopeArray:2018bya}
R.~U. Abbasi, et~al., {\bf Telescope Array} Collaboration, {The Cosmic-Ray
  Energy Spectrum between 2 PeV and 2 EeV Observed with the TALE detector in
  monocular mode}, Astrophys. J. 865~(1) (2018) 74.
\newblock \href {http://arxiv.org/abs/1803.01288} {\path{arXiv:1803.01288}},
  \href {http://dx.doi.org/10.3847/1538-4357/aada05}
  {\path{doi:10.3847/1538-4357/aada05}}.

\bibitem{Abbasi:2014sfa}
R.~U. Abbasi, et~al., {Study of Ultra-High Energy Cosmic Ray composition using
  Telescope Array\textquoteright{}s Middle Drum detector and surface array in
  hybrid mode}, Astropart. Phys. 64 (2015) 49--62.
\newblock \href {http://arxiv.org/abs/1408.1726} {\path{arXiv:1408.1726}},
  \href {http://dx.doi.org/10.1016/j.astropartphys.2014.11.004}
  {\path{doi:10.1016/j.astropartphys.2014.11.004}}.

\bibitem{TelescopeArray:2018xyi}
R.~U. Abbasi, et~al., {\bf Telescope Array} Collaboration, {Depth of Ultra High
  Energy Cosmic Ray Induced Air Shower Maxima Measured by the Telescope Array
  Black Rock and Long Ridge FADC Fluorescence Detectors and Surface Array in
  Hybrid Mode}, Astrophys. J. 858~(2) (2018) 76.
\newblock \href {http://arxiv.org/abs/1801.09784} {\path{arXiv:1801.09784}},
  \href {http://dx.doi.org/10.3847/1538-4357/aabad7}
  {\path{doi:10.3847/1538-4357/aabad7}}.

\bibitem{Lundquist:2017fjo}
P.~J. Lundquist, P.~Sokolsky, P.~Tinyakov, {\bf Telescope Array} Collaboration,
  {Evidence of Intermediate-Scale Energy Spectrum Anisotropy in the Northern
  Hemisphere from Telescope Array}, PoS ICRC2017 (2018) 513.
\newblock \href {http://dx.doi.org/10.22323/1.301.0513}
  {\path{doi:10.22323/1.301.0513}}.

\bibitem{Kawata-2019-ICRC-36-310}
K.~Kawata, et~al., Updated results on the uhecr hotspot observed by the
  telescope array experiment, in: Proceedings of the 36th ICRC, Madison, Vol.
  PoS(ICRC2019)310, 2019.

\bibitem{Ivanov:2017rwl}
D.~Ivanov, {\bf Telescope Array} Collaboration, {Declination Dependence of the
  Telescope Array Surface Detector Spectrum}, PoS ICRC2017 (2018) 496.
\newblock \href {http://dx.doi.org/10.22323/1.301.0496}
  {\path{doi:10.22323/1.301.0496}}.

\bibitem{Abbasi:2018ygn}
R.~U. Abbasi, et~al., {Evidence for Declination Dependence of the Ultrahigh
  Energy Cosmic Ray Spectrum in the Northern Hemisphere}\href
  {http://arxiv.org/abs/1801.07820} {\path{arXiv:1801.07820}}.

\bibitem{TelescopeArray:2018rtg}
R.~U. Abbasi, et~al., {\bf Telescope Array} Collaboration, {Evidence of
  Intermediate-Scale Energy Spectrum Anisotropy of Cosmic Rays
  E$\geq$10$^{19.2}$ eV with the Telescope Array Surface Detector}, Astrophys.
  J. 862~(2) (2018) 91.
\newblock \href {http://arxiv.org/abs/1802.05003} {\path{arXiv:1802.05003}},
  \href {http://dx.doi.org/10.3847/1538-4357/aac9c8}
  {\path{doi:10.3847/1538-4357/aac9c8}}.

\bibitem{IceCube:2019hmk}
M.~G. Aartsen, et~al., {\bf IceCube} Collaboration, {Cosmic ray spectrum and
  composition from PeV to EeV using 3 years of data from IceTop and IceCube},
  Phys. Rev. D 100~(8) (2019) 082002.
\newblock \href {http://arxiv.org/abs/1906.04317} {\path{arXiv:1906.04317}},
  \href {http://dx.doi.org/10.1103/PhysRevD.100.082002}
  {\path{doi:10.1103/PhysRevD.100.082002}}.

\bibitem{DeRidder:2017alk}
S.~De~Ridder, E.~Dvorak, T.~K. Gaisser, {\bf IceCube} Collaboration,
  {Sensitivity of IceCube Cosmic-Ray measurements to the hadronic interaction
  models}, PoS ICRC2017 (2018) 319.
\newblock \href {http://dx.doi.org/10.22323/1.301.0319}
  {\path{doi:10.22323/1.301.0319}}.

\bibitem{IceCube:2021ixw}
S.~Verpoest, et~al., {\bf IceCube} Collaboration, {Testing Hadronic Interaction
  Models with Cosmic Ray Measurements at the IceCube Neutrino Observatory}, PoS
  ICRC2021 (2021) 357.
\newblock \href {http://arxiv.org/abs/2107.09387} {\path{arXiv:2107.09387}},
  \href {http://dx.doi.org/10.22323/1.395.0357}
  {\path{doi:10.22323/1.395.0357}}.

\bibitem{IceCube:2012nn}
R.~Abbasi, et~al., {\bf IceCube} Collaboration, {IceTop: The surface component
  of IceCube}, Nucl. Instrum. Meth. A 700 (2013) 188--220.
\newblock \href {http://arxiv.org/abs/1207.6326} {\path{arXiv:1207.6326}},
  \href {http://dx.doi.org/10.1016/j.nima.2012.10.067}
  {\path{doi:10.1016/j.nima.2012.10.067}}.

\bibitem{Aartsen:2013wda}
M.~G. Aartsen, et~al., {\bf IceCube} Collaboration, {Measurement of the cosmic
  ray energy spectrum with IceTop-73}, Phys. Rev. D88~(4) (2013) 042004.
\newblock \href {http://arxiv.org/abs/1307.3795} {\path{arXiv:1307.3795}},
  \href {http://dx.doi.org/10.1103/PhysRevD.88.042004}
  {\path{doi:10.1103/PhysRevD.88.042004}}.

\bibitem{Aartsen_2020_IceTopLE}
M.~G. Aartsen, et~al., {\bf IceCube} Collaboration,
  \href{http://dx.doi.org/10.1103/PhysRevD.102.122001}{{Cosmic ray spectrum
  from 250 TeV to 10 PeV using IceTop}}, Physical Review D 102~(12) (2020)
  122001.
\newblock \href {http://arxiv.org/abs/2006.05215} {\path{arXiv:2006.05215}},
  \href {http://dx.doi.org/10.1103/physrevd.102.122001}
  {\path{doi:10.1103/physrevd.102.122001}}.
\newline\urlprefix\url{http://dx.doi.org/10.1103/PhysRevD.102.122001}

\bibitem{IceCube:2012vku}
M.~G. Aartsen, et~al., {\bf IceCube} Collaboration, {Search for Galactic PeV
  Gamma Rays with the IceCube Neutrino Observatory}, Phys. Rev. D 87~(6) (2013)
  062002.
\newblock \href {http://arxiv.org/abs/1210.7992} {\path{arXiv:1210.7992}},
  \href {http://dx.doi.org/10.1103/PhysRevD.87.062002}
  {\path{doi:10.1103/PhysRevD.87.062002}}.

\bibitem{IceCube:2019scr}
M.~G. Aartsen, et~al., {\bf IceCube} Collaboration, {Search for PeV Gamma-Ray
  Emission from the Southern Hemisphere with 5 Years of Data from the IceCube
  Observatory}, Astrophys. J. 891 (2019) 9.
\newblock \href {http://arxiv.org/abs/1908.09918} {\path{arXiv:1908.09918}},
  \href {http://dx.doi.org/10.3847/1538-4357/ab6d67}
  {\path{doi:10.3847/1538-4357/ab6d67}}.

\bibitem{Huang:2021hjc}
T.-Q. Huang, Z.~Li, {Neutrino Observations of LHAASO Sources: Present and
  Prospect}\href {http://arxiv.org/abs/2112.14062} {\path{arXiv:2112.14062}}.

\bibitem{LHAASO:2021cbz}
Z.~Cao, et~al., {\bf LHAASO*\textdagger{}, LHAASO} Collaboration,
  {Peta\textendash{}electron volt gamma-ray emission from the Crab Nebula},
  Science 373~(6553) (2021) 425--430.
\newblock \href {http://arxiv.org/abs/2111.06545} {\path{arXiv:2111.06545}},
  \href {http://dx.doi.org/10.1126/science.abg5137}
  {\path{doi:10.1126/science.abg5137}}.

\bibitem{Gonzalez:2019epd}
J.~G. Gonzalez, {\bf IceCube} Collaboration, {Muon Measurements with IceTop},
  EPJ Web Conf. 208 (2019) 03003.
\newblock \href
  {http://dx.doi.org/http://dx.doi.org/10.1051/epjconf/201920803003}
  {\path{doi:http://dx.doi.org/10.1051/epjconf/201920803003}}.

\bibitem{IceCube:2021tuv}
D.~Soldin, {\bf IceCube} Collaboration, {Density of GeV Muons Measured with
  IceTop}, PoS ICRC2021 (2021) 342.
\newblock \href {http://arxiv.org/abs/2107.09583} {\path{arXiv:2107.09583}},
  \href {http://dx.doi.org/10.22323/1.395.0342}
  {\path{doi:10.22323/1.395.0342}}.

\bibitem{IceCube:2022yap}
R.~Abbasi, et~al., {\bf IceCube} Collaboration, Density of gev muons in air
  showers measured with icetop, Submitted to Phys. Rev. D\href
  {http://arxiv.org/abs/2201.12635} {\path{arXiv:2201.12635}}.

\bibitem{Abbasi:2012kza}
R.~Abbasi, et~al., {\bf IceCube} Collaboration, {Lateral Distribution of Muons
  in IceCube Cosmic Ray Events}, Phys. Rev. D87~(1) (2013) 012005.
\newblock \href {http://arxiv.org/abs/1208.2979} {\path{arXiv:1208.2979}},
  \href {http://dx.doi.org/10.1103/PhysRevD.87.012005}
  {\path{doi:10.1103/PhysRevD.87.012005}}.

\bibitem{Soldin:2015iaa}
D.~Soldin, {\bf IceCube} Collaboration, {High $p_\mathrm{T}$ muons from cosmic
  ray air showers in IceCube}, PoS ICRC2015 (2016) 256.
\newblock \href {http://dx.doi.org/10.22323/1.236.0256}
  {\path{doi:10.22323/1.236.0256}}.

\bibitem{Soldin:2018vak}
D.~Soldin, {\bf IceCube} Collaboration, {Atmospheric Muons Measured with
  IceCube}, in: {20th International Symposium on Very High Energy Cosmic Ray
  Interactions (ISVHECRI 2018) Nagoya, Japan, May 21-25, 2018}, 2018.
\newblock \href {http://arxiv.org/abs/1811.03651} {\path{arXiv:1811.03651}}.

\bibitem{IceCube:2015wro}
M.~G. Aartsen, et~al., {\bf IceCube} Collaboration, {Characterization of the
  Atmospheric Muon Flux in IceCube}, Astropart. Phys. 78 (2016) 1--27.
\newblock \href {http://arxiv.org/abs/1506.07981} {\path{arXiv:1506.07981}},
  \href {http://dx.doi.org/10.1016/j.astropartphys.2016.01.006}
  {\path{doi:10.1016/j.astropartphys.2016.01.006}}.

\bibitem{Fuchs:2017nuo}
T.~Fuchs, {\bf IceCube} Collaboration, {Development of a Machine Learning Based
  Analysis Chain for the Measurement of Atmospheric Muon Spectra with IceCube},
  in: {25th European Cosmic Ray Symposium}, 2017.
\newblock \href {http://arxiv.org/abs/1701.04067} {\path{arXiv:1701.04067}}.

\bibitem{Desiati:2011hea}
P.~Desiati, T.~Kuwabara, T.~K. Gaisser, S.~Tilav, D.~Rocco, {\bf IceCube}
  Collaboration, {Seasonal Variations of High Energy Cosmic Ray Muons Observed
  by the IceCube Observatory as a Probe of Kaon/Pion Ratio}, in: {32nd
  International Cosmic Ray Conference}, Vol.~1, 2011, pp. 78--81.
\newblock \href {http://dx.doi.org/10.7529/ICRC2011/V01/0662}
  {\path{doi:10.7529/ICRC2011/V01/0662}}.

\bibitem{Tilav:2010hj}
S.~Tilav, P.~Desiati, T.~Kuwabara, D.~Rocco, F.~Rothmaier, M.~Simmons,
  H.~Wissing, {\bf IceCube} Collaboration, {Atmospheric Variations as Observed
  by IceCube}, 2010.
\newblock \href {http://arxiv.org/abs/1001.0776} {\path{arXiv:1001.0776}}.

\bibitem{Gaisser:2013lrk}
T.~Gaisser, {\bf IceCube} Collaboration, {Seasonal variation of atmospheric
  neutrinos in IceCube}, in: {33rd International Cosmic Ray Conference}, 2013,
  p. 0492.

\bibitem{Tilav:2019xmf}
S.~Tilav, T.~K. Gaisser, D.~Soldin, P.~Desiati, {\bf IceCube} Collaboration,
  {Seasonal variation of atmospheric muons in IceCube}, PoS ICRC2019 (2020)
  894.
\newblock \href {http://arxiv.org/abs/1909.01406} {\path{arXiv:1909.01406}},
  \href {http://dx.doi.org/10.22323/1.358.0894}
  {\path{doi:10.22323/1.358.0894}}.

\bibitem{IceCube:2016biq}
M.~G. Aartsen, et~al., {\bf IceCube} Collaboration, {Anisotropy in Cosmic-ray
  Arrival Directions in the Southern Hemisphere Based on six Years of Data From
  the IceCube Detector}, Astrophys. J. 826~(2) (2016) 220.
\newblock \href {http://arxiv.org/abs/1603.01227} {\path{arXiv:1603.01227}},
  \href {http://dx.doi.org/10.3847/0004-637X/826/2/220}
  {\path{doi:10.3847/0004-637X/826/2/220}}.

\bibitem{IceCube:2011fxx}
R.~Abbasi, et~al., {\bf IceCube} Collaboration, {Observation of Anisotropy in
  the Arrival Directions of Galactic Cosmic Rays at Multiple Angular Scales
  with IceCube}, Astrophys. J. 740 (2011) 16.
\newblock \href {http://arxiv.org/abs/1105.2326} {\path{arXiv:1105.2326}},
  \href {http://dx.doi.org/10.1088/0004-637X/740/1/16}
  {\path{doi:10.1088/0004-637X/740/1/16}}.

\bibitem{HAWC:2018wju}
A.~U. Abeysekara, et~al., {\bf HAWC, IceCube} Collaboration, {All-Sky
  Measurement of the Anisotropy of Cosmic Rays at 10 TeV and Mapping of the
  Local Interstellar Magnetic Field}, Astrophys. J. 871~(1) (2019) 96.
\newblock \href {http://arxiv.org/abs/1812.05682} {\path{arXiv:1812.05682}},
  \href {http://dx.doi.org/10.3847/1538-4357/aaf5cc}
  {\path{doi:10.3847/1538-4357/aaf5cc}}.

\bibitem{IceCube:2021epf}
R.~Abbasi, et~al., {\bf IceCube} Collaboration, {First air-shower measurements
  with the prototype station of the IceCube surface enhancement}, PoS ICRC2021
  (2021) 314.
\newblock \href {http://arxiv.org/abs/2107.08750} {\path{arXiv:2107.08750}},
  \href {http://dx.doi.org/10.22323/1.395.0314}
  {\path{doi:10.22323/1.395.0314}}.

\bibitem{PierreAuger:2021mmt}
E.~Guido, et~al., {\bf Pierre Auger} Collaboration, {Combined fit of the energy
  spectrum and mass composition across the ankle with the data measured at the
  Pierre Auger Observatory}, PoS ICRC2021 (2021) 311.
\newblock \href {http://dx.doi.org/10.22323/1.395.0311}
  {\path{doi:10.22323/1.395.0311}}.

\bibitem{Bergman:2021djm}
D.~Bergman, {\bf Telescope Array} Collaboration, {Telescope Array Combined Fit
  to Cosmic Ray Spectrum and Composition}, PoS ICRC2021 (2021) 338.
\newblock \href {http://dx.doi.org/10.22323/1.395.0338}
  {\path{doi:10.22323/1.395.0338}}.

\bibitem{Ivanov:2020rqn}
D.~Ivanov, {\bf Telescope Array} Collaboration, {Energy Spectrum Measured by
  the Telescope Array}, PoS ICRC2019 (2020) 298.
\newblock \href {http://dx.doi.org/10.22323/1.358.0298}
  {\path{doi:10.22323/1.358.0298}}.

\bibitem{Knurenko:2013dia}
S.~P. Knurenko, Z.~E. Petrov, R.~Sidorov, I.~Y. Sleptsov, S.~K. Starostin,
  G.~G. Struchkov, {Cosmic ray spectrum in the energy range 1.0E15-1.0E18 eV
  and the second knee according to the small Cherenkov setup at the Yakutsk EAS
  array}\href {http://arxiv.org/abs/1310.1978} {\path{arXiv:1310.1978}}.

\bibitem{KASCADEGrande:2017gtn}
C.~J. Arteaga-Vel\'azquez, et~al., {\bf KASCADE Grande} Collaboration,
  {Measurements of the muon content of EAS in KASCADE-Grande compared with
  SIBYLL 2.3 predictions}, PoS ICRC2017 (2018) 316.
\newblock \href {http://dx.doi.org/10.22323/1.301.0316}
  {\path{doi:10.22323/1.301.0316}}.

\bibitem{Budnev:2020oad}
N.~M. Budnev, et~al., {The primary cosmic-ray energy spectrum measured with the
  Tunka-133 array}, Astropart. Phys. 117 (2020) 102406.
\newblock \href {http://arxiv.org/abs/2104.03599} {\path{arXiv:2104.03599}},
  \href {http://dx.doi.org/10.1016/j.astropartphys.2019.102406}
  {\path{doi:10.1016/j.astropartphys.2019.102406}}.

\bibitem{Linsley:1963km}
J.~Linsley, {Evidence for a primary cosmic-ray particle with energy 10**20-eV},
  Phys. Rev. Lett. 10 (1963) 146--148.
\newblock \href {http://dx.doi.org/10.1103/PhysRevLett.10.146}
  {\path{doi:10.1103/PhysRevLett.10.146}}.

\bibitem{Chiba:1991nf}
N.~Chiba, et~al., {Akeno giant air shower array (AGASA) covering 100-km**2
  area}, Nucl. Instrum. Meth. A 311 (1992) 338--349.
\newblock \href {http://dx.doi.org/10.1016/0168-9002(92)90882-5}
  {\path{doi:10.1016/0168-9002(92)90882-5}}.

\bibitem{Nagano:2000ve}
M.~Nagano, A.~A. Watson, {Observations and implications of the ultrahigh-energy
  cosmic rays}, Rev. Mod. Phys. 72 (2000) 689--732.
\newblock \href {http://dx.doi.org/10.1103/RevModPhys.72.689}
  {\path{doi:10.1103/RevModPhys.72.689}}.

\bibitem{Greisen:1960wc}
K.~Greisen, {Cosmic ray showers}, Ann. Rev. Nucl. Part. Sci. 10 (1960) 63--108.
\newblock \href {http://dx.doi.org/10.1146/annurev.ns.10.120160.000431}
  {\path{doi:10.1146/annurev.ns.10.120160.000431}}.

\bibitem{delvaille1962spectrum}
J.~Delvaille, F.~Kendziorski, K.~Greisen, Spectrum and isotropy of eas, J.
  Phys. Soc. Japan 17~(Suppl A).

\bibitem{Suga1962}
K.~Suga, A.~Chudakov, Proc 5th Inter-American Seminar on Cosmic Rays, La Paz 2
  (1962) XLIX--1--5.

\bibitem{Baltrusaitis:1985mx}
R.~M. Baltrusaitis, et~al., {THE UTAH FLY'S EYE DETECTOR}, Nucl. Instrum. Meth.
  A 240 (1985) 410--428.
\newblock \href {http://dx.doi.org/10.1016/0168-9002(85)90658-8}
  {\path{doi:10.1016/0168-9002(85)90658-8}}.

\bibitem{Abu-Zayyad:2000vin}
T.~Abu-Zayyad, et~al., {The prototype high-resolution Fly's Eye cosmic ray
  detector}, Nucl. Instrum. Meth. A 450 (2000) 253--269.
\newblock \href {http://dx.doi.org/10.1016/S0168-9002(00)00307-7}
  {\path{doi:10.1016/S0168-9002(00)00307-7}}.

\bibitem{Takeda:2002at}
M.~Takeda, et~al., {Energy determination in the Akeno Giant Air Shower Array
  experiment}, Astropart. Phys. 19 (2003) 447--462.
\newblock \href {http://arxiv.org/abs/astro-ph/0209422}
  {\path{arXiv:astro-ph/0209422}}, \href
  {http://dx.doi.org/10.1016/S0927-6505(02)00243-8}
  {\path{doi:10.1016/S0927-6505(02)00243-8}}.

\bibitem{HiRes:2007lra}
R.~U. Abbasi, et~al., {\bf HiRes} Collaboration, {First observation of the
  Greisen-Zatsepin-Kuzmin suppression}, Phys. Rev. Lett. 100 (2008) 101101.
\newblock \href {http://arxiv.org/abs/astro-ph/0703099}
  {\path{arXiv:astro-ph/0703099}}, \href
  {http://dx.doi.org/10.1103/PhysRevLett.100.101101}
  {\path{doi:10.1103/PhysRevLett.100.101101}}.

\bibitem{Ivanov:2021mkn}
D.~Ivanov, D.~Bergman, G.~Furlich, R.~Gonzalez, G.~Thomson, Y.~Tsunesada, {\bf
  Telescope Array} Collaboration, {Recent measurement of the Telescope Array
  energy spectrum and observation of the shoulder feature in the Northern
  Hemisphere}, PoS ICRC2021 (2021) 341.
\newblock \href {http://dx.doi.org/10.22323/1.395.0341}
  {\path{doi:10.22323/1.395.0341}}.

\bibitem{Hersil:1961zz}
J.~Hersil, I.~Escobar, D.~Scott, G.~Clark, S.~Olbert, {Observations of
  Extensive Air Showers near the Maximum of Their Longitudinal Development},
  Phys. Rev. Lett. 6 (1961) 22--23.
\newblock \href {http://dx.doi.org/10.1103/PhysRevLett.6.22}
  {\path{doi:10.1103/PhysRevLett.6.22}}.

\bibitem{TelescopeArray:2012qqu}
T.~Abu-Zayyad, et~al., {\bf Telescope Array} Collaboration, {The Cosmic Ray
  Energy Spectrum Observed with the Surface Detector of the Telescope Array
  Experiment}, Astrophys. J. Lett. 768 (2013) L1.
\newblock \href {http://arxiv.org/abs/1205.5067} {\path{arXiv:1205.5067}},
  \href {http://dx.doi.org/10.1088/2041-8205/768/1/L1}
  {\path{doi:10.1088/2041-8205/768/1/L1}}.

\bibitem{Dawson:2020bkp}
B.~Dawson, {\bf Pierre Auger} Collaboration, {The Energy Scale of the Pierre
  Auger Observatory}, PoS ICRC2019 (2020) 231.
\newblock \href {http://dx.doi.org/10.22323/1.358.0231}
  {\path{doi:10.22323/1.358.0231}}.

\bibitem{Abu-Zayyad:2011ugz}
T.~Abu-Zayyad, M.~Allen, E.~Barcikowski, {TA Energy Scale: Methods and
  Photometry}, in: {32nd International Cosmic Ray Conference}, Vol.~2, 2011, p.
  250.
\newblock \href {http://dx.doi.org/10.7529/ICRC2011/V02/1270}
  {\path{doi:10.7529/ICRC2011/V02/1270}}.

\bibitem{PierreAuger:2008rol}
J.~Abraham, et~al., {\bf Pierre Auger} Collaboration, {Observation of the
  suppression of the flux of cosmic rays above $4\times 10^{19}$eV}, Phys. Rev.
  Lett. 101 (2008) 061101.
\newblock \href {http://arxiv.org/abs/0806.4302} {\path{arXiv:0806.4302}},
  \href {http://dx.doi.org/10.1103/PhysRevLett.101.061101}
  {\path{doi:10.1103/PhysRevLett.101.061101}}.

\bibitem{PierreAuger:2020kuy}
A.~Aab, et~al., {\bf Pierre Auger} Collaboration, {Features of the Energy
  Spectrum of Cosmic Rays above 2.5\texttimes{}10$^{18}$ eV Using the Pierre
  Auger Observatory}, Phys. Rev. Lett. 125~(12) (2020) 121106.
\newblock \href {http://arxiv.org/abs/2008.06488} {\path{arXiv:2008.06488}},
  \href {http://dx.doi.org/10.1103/PhysRevLett.125.121106}
  {\path{doi:10.1103/PhysRevLett.125.121106}}.

\bibitem{PierreAuger:2013dyy}
B.~R. Dawson, I.~C. Maris, M.~Roth, F.~Salamida, T.~Abu-Zayyad, D.~Ikeda,
  D.~Ivanov, Y.~Tsunesada, M.~I. Pravdin, A.~V. Sabourov, {\bf Pierre Auger,
  Yakutsk, Telescope Array} Collaboration, {The energy spectrum of cosmic rays
  at the highest energies}, EPJ Web Conf. 53 (2013) 01005.
\newblock \href {http://arxiv.org/abs/1306.6138} {\path{arXiv:1306.6138}},
  \href {http://dx.doi.org/10.1051/epjconf/20135301005}
  {\path{doi:10.1051/epjconf/20135301005}}.

\bibitem{Verzi:2017hro}
V.~Verzi, D.~Ivanov, Y.~Tsunesada, {Measurement of Energy Spectrum of
  Ultra-High Energy Cosmic Rays}, PTEP 2017~(12) (2017) 12A103.
\newblock \href {http://arxiv.org/abs/1705.09111} {\path{arXiv:1705.09111}},
  \href {http://dx.doi.org/10.1093/ptep/ptx082}
  {\path{doi:10.1093/ptep/ptx082}}.

\bibitem{Ivanov:2017juh}
D.~Ivanov, {\bf Telescope-Array, Pierre Auger} Collaboration, {Report of the
  Telescope Array - Pierre Auger Observatory Working Group on Energy Spectrum},
  PoS ICRC2017 (2018) 498.
\newblock \href {http://dx.doi.org/10.22323/1.301.0498}
  {\path{doi:10.22323/1.301.0498}}.

\bibitem{AbuZayyad:2018aua}
T.~AbuZayyad, et~al., {The Energy Spectrum of Cosmic Rays at the Highest
  Energies}, JPS Conf. Proc. 19 (2018) 011003.
\newblock \href {http://dx.doi.org/10.7566/JPSCP.19.011003}
  {\path{doi:10.7566/JPSCP.19.011003}}.

\bibitem{PierreAuger:2019vpk}
T.~AbuZayyad, et~al., {\bf Pierre Auger, Telescope Array} Collaboration,
  {Auger-TA energy spectrum working group report}, EPJ Web Conf. 210 (2019)
  01002.
\newblock \href {http://dx.doi.org/10.1051/epjconf/201921001002}
  {\path{doi:10.1051/epjconf/201921001002}}.

\bibitem{Deligny:2020gzq}
O.~Deligny, {\bf Pierre Auger, Telescope Array} Collaboration, {The energy
  spectrum of ultra-high energy cosmic rays measured at the Pierre Auger
  Observatory and at the Telescope Array}, PoS ICRC2019 (2020) 234.
\newblock \href {http://arxiv.org/abs/2001.08811} {\path{arXiv:2001.08811}},
  \href {http://dx.doi.org/10.22323/1.358.0234}
  {\path{doi:10.22323/1.358.0234}}.

\bibitem{AIRFLY:2012msg}
M.~Ave, et~al., {\bf AIRFLY} Collaboration, {Precise measurement of the
  absolute fluorescence yield of the 337 nm band in atmospheric gases},
  Astropart. Phys. 42 (2013) 90--102.
\newblock \href {http://arxiv.org/abs/1210.6734} {\path{arXiv:1210.6734}},
  \href {http://dx.doi.org/10.1016/j.astropartphys.2012.12.006}
  {\path{doi:10.1016/j.astropartphys.2012.12.006}}.

\bibitem{AIRFLY:2007msg}
M.~Ave, et~al., {\bf AIRFLY} Collaboration, {Measurement of the pressure
  dependence of air fluorescence emission induced by electrons}, Astropart.
  Phys. 28 (2007) 41--57.
\newblock \href {http://arxiv.org/abs/0703132} {\path{arXiv:0703132}}, \href
  {http://dx.doi.org/10.1016/j.astropartphys.2007.04.006}
  {\path{doi:10.1016/j.astropartphys.2007.04.006}}.

\bibitem{Kakimoto:1995pr}
F.~Kakimoto, E.~C. Loh, M.~Nagano, H.~Okuno, M.~Teshima, S.~Ueno, {A
  Measurement of the air fluorescence yield}, Nucl. Instrum. Meth. A 372 (1996)
  527--533.
\newblock \href {http://dx.doi.org/10.1016/0168-9002(95)01423-3}
  {\path{doi:10.1016/0168-9002(95)01423-3}}.

\bibitem{Abbasi:2007am}
R.~Abbasi, et~al., {Air fluorescence measurements in the spectral range 300-420
  nm using a 28.5-GeV electron beam}, Astropart. Phys. 29 (2008) 77--86.
\newblock \href {http://arxiv.org/abs/0708.3116} {\path{arXiv:0708.3116}},
  \href {http://dx.doi.org/10.1016/j.astropartphys.2007.11.010}
  {\path{doi:10.1016/j.astropartphys.2007.11.010}}.

\bibitem{PierreAuger:2019dhr}
A.~Aab, et~al., {\bf Pierre Auger} Collaboration, {Data-driven estimation of
  the invisible energy of cosmic ray showers with the Pierre Auger
  Observatory}, Phys. Rev. D 100~(8) (2019) 082003.
\newblock \href {http://arxiv.org/abs/1901.08040} {\path{arXiv:1901.08040}},
  \href {http://dx.doi.org/10.1103/PhysRevD.100.082003}
  {\path{doi:10.1103/PhysRevD.100.082003}}.

\bibitem{Tsunesada:2011mp}
Y.~Tsunesada, {\bf Telescope Array} Collaboration, {Highlights from Telescope
  Array}, in: {32nd International Cosmic Ray Conference}, Vol.~c, 2011, p.~67.
\newblock \href {http://arxiv.org/abs/1111.2507} {\path{arXiv:1111.2507}},
  \href {http://dx.doi.org/10.7529/ICRC2011/V12/H05}
  {\path{doi:10.7529/ICRC2011/V12/H05}}.

\bibitem{Epimakhov:2013cnw}
S.~Epimakhov, et~al., {Elemental Composition of Cosmic Rays above the Knee from
  X$_{max}$ measurements of the Tunka Array}, in: {33rd International Cosmic
  Ray Conference}, 2013, p. 0326.

\bibitem{TelescopeArray:2020bfv}
R.~U. Abbasi, et~al., {\bf Telescope Array} Collaboration, {The Cosmic-Ray
  Composition between 2 PeV and 2 EeV Observed with the TALE Detector in
  Monocular Mode}, Astrophys. J. 909~(2) (2021) 178.
\newblock \href {http://arxiv.org/abs/2012.10372} {\path{arXiv:2012.10372}},
  \href {http://dx.doi.org/10.3847/1538-4357/abdd30}
  {\path{doi:10.3847/1538-4357/abdd30}}.

\bibitem{Blasi:2013rva}
P.~Blasi, {The Origin of Galactic Cosmic Rays}, Astron. Astrophys. Rev. 21
  (2013) 70.
\newblock \href {http://arxiv.org/abs/1311.7346} {\path{arXiv:1311.7346}},
  \href {http://dx.doi.org/10.1007/s00159-013-0070-7}
  {\path{doi:10.1007/s00159-013-0070-7}}.

\bibitem{Hillas:2005cs}
A.~M. Hillas, {Can diffusive shock acceleration in supernova remnants account
  for high-energy galactic cosmic rays?}, J. Phys. G 31 (2005) R95--R131.
\newblock \href {http://dx.doi.org/10.1088/0954-3899/31/5/R02}
  {\path{doi:10.1088/0954-3899/31/5/R02}}.

\bibitem{Aloisio:2013hya}
R.~Aloisio, V.~Berezinsky, P.~Blasi, {Ultra high energy cosmic rays:
  implications of Auger data for source spectra and chemical composition}, JCAP
  10 (2014) 020.
\newblock \href {http://arxiv.org/abs/1312.7459} {\path{arXiv:1312.7459}},
  \href {http://dx.doi.org/10.1088/1475-7516/2014/10/020}
  {\path{doi:10.1088/1475-7516/2014/10/020}}.

\bibitem{Abbasi:2020chd}
R.~U. Abbasi, et~al., {Measurement of the proton-air cross section with
  Telescope Array\textquoteright{}s Black Rock Mesa and Long Ridge fluorescence
  detectors, and surface array in hybrid mode}, Phys. Rev. D 102~(6) (2020)
  062004.
\newblock \href {http://arxiv.org/abs/2006.05012} {\path{arXiv:2006.05012}},
  \href {http://dx.doi.org/10.1103/PhysRevD.102.062004}
  {\path{doi:10.1103/PhysRevD.102.062004}}.

\bibitem{TelescopeArray:2018eph}
R.~U. Abbasi, et~al., {\bf Telescope Array} Collaboration, {Study of muons from
  ultrahigh energy cosmic ray air showers measured with the Telescope Array
  experiment}, Phys. Rev. D 98~(2) (2018) 022002.
\newblock \href {http://arxiv.org/abs/1804.03877} {\path{arXiv:1804.03877}},
  \href {http://dx.doi.org/10.1103/PhysRevD.98.022002}
  {\path{doi:10.1103/PhysRevD.98.022002}}.

\bibitem{Farrar:2019cid}
G.~R. Farrar, {Particle Physics at Ultrahigh Energies}, in: {18th International
  Symposium on Very High Energy Cosmic Ray Interactions}, 2019.
\newblock \href {http://arxiv.org/abs/1902.11271} {\path{arXiv:1902.11271}}.

\bibitem{Anchordoqui:2016oxy}
L.~A. Anchordoqui, H.~Goldberg, T.~J. Weiler, {Strange fireball as an
  explanation of the muon excess in Auger data}, Phys. Rev. D 95~(6) (2017)
  063005.
\newblock \href {http://arxiv.org/abs/1612.07328} {\path{arXiv:1612.07328}},
  \href {http://dx.doi.org/10.1103/PhysRevD.95.063005}
  {\path{doi:10.1103/PhysRevD.95.063005}}.

\bibitem{Anchordoqui:2019laz}
L.~A. Anchordoqui, C.~Garc\'\i{}a~Canal, S.~J. Sciutto, J.~F. Soriano, {Through
  the looking-glass with ALICE into the quark-gluon plasma: A new test for
  hadronic interaction models used in air shower simulations}, Phys. Lett. B
  810 (2020) 135837.
\newblock \href {http://arxiv.org/abs/1907.09816} {\path{arXiv:1907.09816}},
  \href {http://dx.doi.org/10.1016/j.physletb.2020.135837}
  {\path{doi:10.1016/j.physletb.2020.135837}}.

\bibitem{Alcantara:2019sco}
E.~Alcantara, L.~A. Anchordoqui, J.~F. Soriano, {Hunting for superheavy dark
  matter with the highest-energy cosmic rays}, Phys. Rev. D 99~(10) (2019)
  103016.
\newblock \href {http://arxiv.org/abs/1903.05429} {\path{arXiv:1903.05429}},
  \href {http://dx.doi.org/10.1103/PhysRevD.99.103016}
  {\path{doi:10.1103/PhysRevD.99.103016}}.

\bibitem{Supanitsky:2019ayx}
A.~D. Supanitsky, G.~Medina-Tanco, {Ultra high energy cosmic rays from
  super-heavy dark matter in the context of large exposure observatories}, JCAP
  11 (2019) 036.
\newblock \href {http://arxiv.org/abs/1909.09191} {\path{arXiv:1909.09191}},
  \href {http://dx.doi.org/10.1088/1475-7516/2019/11/036}
  {\path{doi:10.1088/1475-7516/2019/11/036}}.

\bibitem{Aloisio:2015lva}
R.~Aloisio, S.~Matarrese, A.~V. Olinto, {Super Heavy Dark Matter in light of
  BICEP2, Planck and Ultra High Energy Cosmic Rays Observations}, JCAP 08
  (2015) 024.
\newblock \href {http://arxiv.org/abs/1504.01319} {\path{arXiv:1504.01319}},
  \href {http://dx.doi.org/10.1088/1475-7516/2015/08/024}
  {\path{doi:10.1088/1475-7516/2015/08/024}}.

\bibitem{Ishiwata:2019aet}
K.~Ishiwata, O.~Macias, S.~Ando, M.~Arimoto, {Probing heavy dark matter decays
  with multi-messenger astrophysical data}, JCAP 01 (2020) 003.
\newblock \href {http://arxiv.org/abs/1907.11671} {\path{arXiv:1907.11671}},
  \href {http://dx.doi.org/10.1088/1475-7516/2020/01/003}
  {\path{doi:10.1088/1475-7516/2020/01/003}}.

\bibitem{Garzelli:2016xmx}
M.~V. Garzelli, S.~Moch, O.~Zenaiev, A.~Cooper-Sarkar, A.~Geiser, K.~Lipka,
  R.~Placakyte, G.~Sigl, {\bf PROSA} Collaboration, {Prompt neutrino fluxes in
  the atmosphere with PROSA parton distribution functions}, JHEP 05 (2017) 004.
\newblock \href {http://arxiv.org/abs/1611.03815} {\path{arXiv:1611.03815}},
  \href {http://dx.doi.org/10.1007/JHEP05(2017)004}
  {\path{doi:10.1007/JHEP05(2017)004}}.

\bibitem{Zenaiev:2019ktw}
O.~Zenaiev, M.~V. Garzelli, K.~Lipka, S.~O. Moch, A.~Cooper-Sarkar, F.~Olness,
  A.~Geiser, G.~Sigl, {\bf PROSA} Collaboration, {Improved constraints on
  parton distributions using LHCb, ALICE and HERA heavy-flavour measurements
  and implications for the predictions for prompt atmospheric-neutrino fluxes},
  JHEP 04 (2020) 118.
\newblock \href {http://arxiv.org/abs/1911.13164} {\path{arXiv:1911.13164}},
  \href {http://dx.doi.org/10.1007/JHEP04(2020)118}
  {\path{doi:10.1007/JHEP04(2020)118}}.

\bibitem{Sommers:2008ji}
P.~Sommers, S.~Westerhoff, {Cosmic Ray Astronomy}, New J. Phys. 11 (2009)
  055004.
\newblock \href {http://arxiv.org/abs/0802.1267} {\path{arXiv:0802.1267}},
  \href {http://dx.doi.org/10.1088/1367-2630/11/5/055004}
  {\path{doi:10.1088/1367-2630/11/5/055004}}.

\bibitem{Erdmann:2016vle}
M.~Erdmann, G.~M\"uller, M.~Urban, M.~Wirtz, {The Nuclear Window to the
  Extragalactic Universe}, Astropart. Phys. 85 (2016) 54--64.
\newblock \href {http://arxiv.org/abs/1607.01645} {\path{arXiv:1607.01645}},
  \href {http://dx.doi.org/10.1016/j.astropartphys.2016.10.002}
  {\path{doi:10.1016/j.astropartphys.2016.10.002}}.

\bibitem{PierreAuger:2016use}
A.~Aab, et~al., {\bf Pierre Auger} Collaboration, {Combined fit of spectrum and
  composition data as measured by the Pierre Auger Observatory}, JCAP 04 (2017)
  038, [Erratum: JCAP 03, E02 (2018)].
\newblock \href {http://arxiv.org/abs/1612.07155} {\path{arXiv:1612.07155}},
  \href {http://dx.doi.org/10.1088/1475-7516/2017/04/038}
  {\path{doi:10.1088/1475-7516/2017/04/038}}.

\bibitem{Unger:2015laa}
M.~Unger, G.~R. Farrar, L.~A. Anchordoqui, {Origin of the ankle in the
  ultrahigh energy cosmic ray spectrum, and of the extragalactic protons below
  it}, Phys. Rev. D 92~(12) (2015) 123001.
\newblock \href {http://arxiv.org/abs/1505.02153} {\path{arXiv:1505.02153}},
  \href {http://dx.doi.org/10.1103/PhysRevD.92.123001}
  {\path{doi:10.1103/PhysRevD.92.123001}}.

\bibitem{Heinze:2019jou}
J.~Heinze, A.~Fedynitch, D.~Boncioli, W.~Winter, {A new view on Auger data and
  cosmogenic neutrinos in light of different nuclear disintegration and
  air-shower models}, Astrophys. J. 873~(1) (2019) 88.
\newblock \href {http://arxiv.org/abs/1901.03338} {\path{arXiv:1901.03338}},
  \href {http://dx.doi.org/10.3847/1538-4357/ab05ce}
  {\path{doi:10.3847/1538-4357/ab05ce}}.

\bibitem{AlvesBatista:2019rhs}
R.~Alves~Batista, D.~Boncioli, A.~di~Matteo, A.~van Vliet, {Secondary neutrino
  and gamma-ray fluxes from SimProp and CRPropa}, JCAP 05 (2019) 006.
\newblock \href {http://arxiv.org/abs/1901.01244} {\path{arXiv:1901.01244}},
  \href {http://dx.doi.org/10.1088/1475-7516/2019/05/006}
  {\path{doi:10.1088/1475-7516/2019/05/006}}.

\bibitem{Fang:2016hop}
K.~Fang, K.~Kotera, M.~C. Miller, K.~Murase, F.~Oikonomou, {Identifying
  Ultrahigh-Energy Cosmic-Ray Accelerators with Future Ultrahigh-Energy
  Neutrino Detectors}, JCAP 12 (2016) 017.
\newblock \href {http://arxiv.org/abs/1609.08027} {\path{arXiv:1609.08027}},
  \href {http://dx.doi.org/10.1088/1475-7516/2016/12/017}
  {\path{doi:10.1088/1475-7516/2016/12/017}}.

\bibitem{Globus:2007bi}
N.~Globus, D.~Allard, E.~Parizot, {Propagation of high-energy cosmic rays in
  extragalactic turbulent magnetic fields: resulting energy spectrum and
  composition}, Astron. Astrophys. 479 (2008) 97.
\newblock \href {http://arxiv.org/abs/0709.1541} {\path{arXiv:0709.1541}},
  \href {http://dx.doi.org/10.1051/0004-6361:20078653}
  {\path{doi:10.1051/0004-6361:20078653}}.

\bibitem{Ding:2021emg}
C.~Ding, N.~Globus, G.~R. Farrar, {The Imprint of Large Scale Structure on the
  Ultra-High-Energy Cosmic Ray Sky}, Astrophys. J. Lett. 913~(1) (2021) L13.
\newblock \href {http://arxiv.org/abs/2101.04564} {\path{arXiv:2101.04564}},
  \href {http://dx.doi.org/10.3847/2041-8213/abf11e}
  {\path{doi:10.3847/2041-8213/abf11e}}.

\bibitem{Olinto:2019euf}
A.~V. Olinto, et~al., {The POEMMA (Probe of Extreme
  Multi-MessengerAstrophysics) mission}, PoS ICRC2019 (2020) 378.
\newblock \href {http://arxiv.org/abs/1909.09466} {\path{arXiv:1909.09466}},
  \href {http://dx.doi.org/10.22323/1.358.0378}
  {\path{doi:10.22323/1.358.0378}}.

\bibitem{Anchordoqui:2019omw}
L.~A. Anchordoqui, et~al., {Performance and science reach of the Probe of
  Extreme Multimessenger Astrophysics for ultrahigh-energy particles}, Phys.
  Rev. D 101~(2) (2020) 023012.
\newblock \href {http://arxiv.org/abs/1907.03694} {\path{arXiv:1907.03694}},
  \href {http://dx.doi.org/10.1103/PhysRevD.101.023012}
  {\path{doi:10.1103/PhysRevD.101.023012}}.

\bibitem{Tameda:2019wmj}
Y.~Tameda, T.~Tomida, M.~Yamamoto, H.~Iwakura, D.~Ikeda, K.~Yamazaki, {Air
  shower observation by a simple structured Fresnel lens telescope with a
  single pixel for the next generation of ultra-high-energy cosmic ray
  observatories}, PTEP 2019~(4) (2019) 043F01.
\newblock \href {http://arxiv.org/abs/1903.01626} {\path{arXiv:1903.01626}},
  \href {http://dx.doi.org/10.1093/ptep/ptz025}
  {\path{doi:10.1093/ptep/ptz025}}.

\bibitem{Fujii:2015dra}
T.~Fujii, et~al., {Detection of ultra-high energy cosmic ray showers with a
  single-pixel fluorescence telescope}, Astropart. Phys. 74 (2016) 64--72.
\newblock \href {http://arxiv.org/abs/1504.00692} {\path{arXiv:1504.00692}},
  \href {http://dx.doi.org/10.1016/j.astropartphys.2015.10.006}
  {\path{doi:10.1016/j.astropartphys.2015.10.006}}.

\bibitem{Malacari:2019uqw}
M.~Malacari, et~al., {The First Full-Scale Prototypes of the Fluorescence
  detector Array of Single-pixel Telescopes}, Astropart. Phys. 119 (2020)
  102430.
\newblock \href {http://arxiv.org/abs/1911.05285} {\path{arXiv:1911.05285}},
  \href {http://dx.doi.org/10.1016/j.astropartphys.2020.102430}
  {\path{doi:10.1016/j.astropartphys.2020.102430}}.

\bibitem{PierreAuger:2014zay}
A.~Aab, et~al., {\bf Pierre Auger} Collaboration, {Muons in Air Showers at the
  Pierre Auger Observatory: Measurement of Atmospheric Production Depth}, Phys.
  Rev. D 90~(1) (2014) 012012, [Addendum: Phys.Rev.D 90, 039904 (2014),
  Erratum: Phys.Rev.D 92, 019903 (2015)].
\newblock \href {http://arxiv.org/abs/1407.5919} {\path{arXiv:1407.5919}},
  \href {http://dx.doi.org/10.1103/PhysRevD.90.012012}
  {\path{doi:10.1103/PhysRevD.90.012012}}.

\bibitem{PierreAuger:2017tlx}
A.~Aab, et~al., {\bf Pierre Auger} Collaboration, {Inferences on mass
  composition and tests of hadronic interactions from 0.3 to 100 EeV using the
  water-Cherenkov detectors of the Pierre Auger Observatory}, Phys. Rev. D
  96~(12) (2017) 122003.
\newblock \href {http://arxiv.org/abs/1710.07249} {\path{arXiv:1710.07249}},
  \href {http://dx.doi.org/10.1103/PhysRevD.96.122003}
  {\path{doi:10.1103/PhysRevD.96.122003}}.

\bibitem{PierreAuger:2020gxz}
A.~Aab, et~al., {\bf Pierre Auger} Collaboration, {Direct measurement of the
  muonic content of extensive air showers between $\mathbf { 2\times 10^{17}}$
  and $\mathbf {2\times 10^{18}}~$eV at the Pierre Auger Observatory}, Eur.
  Phys. J. C 80~(8) (2020) 751.
\newblock \href {http://dx.doi.org/10.1140/epjc/s10052-020-8055-y}
  {\path{doi:10.1140/epjc/s10052-020-8055-y}}.

\bibitem{Taboada:2020spx}
A.~Taboada, {\bf Pierre Auger} Collaboration, {Analysis of Data from Surface
  Detector Stations of the AugerPrime Upgrade}, PoS ICRC2019 (2020) 434.
\newblock \href {http://dx.doi.org/10.22323/1.358.0434}
  {\path{doi:10.22323/1.358.0434}}.

\bibitem{Pont:2021pwd}
B.~Pont, {\bf Pierre Auger} Collaboration, {A Large Radio Detector at the
  Pierre Auger Observatory - Measuring the Properties of Cosmic Rays up to the
  Highest Energies}, PoS ICRC2019 (2021) 395.
\newblock \href {http://dx.doi.org/10.22323/1.358.0395}
  {\path{doi:10.22323/1.358.0395}}.

\bibitem{gcos_uhecr2018}
R.~Engel, et~al.,
  \href{https://indico.in2p3.fr/event/17063/contributions/66403/}{{Towards a
  Global Cosmic Ray Observatory (GCOS) - requirements for a future
  observatory}}, uHECR (2018).
\newline\urlprefix\url{https://indico.in2p3.fr/event/17063/contributions/66403/}

\bibitem{TAIGA:2021fgx}
N.~M. Budnev, et~al., {\bf TAIGA} Collaboration, {TAIGA - an advanced hybrid
  detector complex for astroparticle physics, cosmic ray physics and gamma-ray
  astronomy}, PoS ICRC2021 (2021) 731.
\newblock \href {http://dx.doi.org/10.22323/1.395.0731}
  {\path{doi:10.22323/1.395.0731}}.

\bibitem{Knurenko:2019oil}
S.~Knurenko, I.~Petrov, {Mass composition of cosmic rays above 0.1 EeV by the
  Yakutsk array data}, Adv. Space Res. 64~(12) (2019) 2570--2577.
\newblock \href {http://arxiv.org/abs/1908.01508} {\path{arXiv:1908.01508}},
  \href {http://dx.doi.org/10.1016/j.asr.2019.07.019}
  {\path{doi:10.1016/j.asr.2019.07.019}}.

\bibitem{PierreAuger:2021rio}
B.~Pont, et~al., {\bf Pierre Auger} Collaboration, {The depth of the shower
  maximum of air showers measured with AERA}, PoS ICRC2021 (2021) 387.
\newblock \href {http://dx.doi.org/10.22323/1.395.0387}
  {\path{doi:10.22323/1.395.0387}}.

\bibitem{Corstanje:2021kik}
A.~Corstanje, et~al., {Depth of shower maximum and mass composition of cosmic
  rays from 50 PeV to 2 EeV measured with the LOFAR radio telescope}, Phys.
  Rev. D 103~(10) (2021) 102006.
\newblock \href {http://arxiv.org/abs/2103.12549} {\path{arXiv:2103.12549}},
  \href {http://dx.doi.org/10.1103/PhysRevD.103.102006}
  {\path{doi:10.1103/PhysRevD.103.102006}}.

\bibitem{Bezyazeekov:2018yjw}
P.~A. Bezyazeekov, et~al., {Reconstruction of cosmic ray air showers with
  Tunka-Rex data using template fitting of radio pulses}, Phys. Rev. D 97~(12)
  (2018) 122004.
\newblock \href {http://arxiv.org/abs/1803.06862} {\path{arXiv:1803.06862}},
  \href {http://dx.doi.org/10.1103/PhysRevD.97.122004}
  {\path{doi:10.1103/PhysRevD.97.122004}}.

\bibitem{Petrov:2020edv}
I.~Petrov, S.~Knurenko, {Results of Ultra-high Energy Cosmic Ray Study by Radio
  Technique at Yakutsk Array}, PoS ICRC2019 (2020) 385.
\newblock \href {http://dx.doi.org/10.22323/1.358.0385}
  {\path{doi:10.22323/1.358.0385}}.

\bibitem{ARGO-YBJ:2015isx}
B.~Bartoli, et~al., {\bf ARGO-YBJ, LHAASO} Collaboration, {Knee of the cosmic
  hydrogen and helium spectrum below 1 PeV measured by ARGO-YBJ and a Cherenkov
  telescope of LHAASO}, Phys. Rev. D 92~(9) (2015) 092005.
\newblock \href {http://arxiv.org/abs/1502.03164} {\path{arXiv:1502.03164}},
  \href {http://dx.doi.org/10.1103/PhysRevD.92.092005}
  {\path{doi:10.1103/PhysRevD.92.092005}}.

\bibitem{KASCADEGrande:2011kpw}
W.~D. Apel, et~al., {\bf KASCADE Grande} Collaboration, {Kneelike structure in
  the spectrum of the heavy component of cosmic rays observed with
  KASCADE-Grande}, Phys. Rev. Lett. 107 (2011) 171104.
\newblock \href {http://arxiv.org/abs/1107.5885} {\path{arXiv:1107.5885}},
  \href {http://dx.doi.org/10.1103/PhysRevLett.107.171104}
  {\path{doi:10.1103/PhysRevLett.107.171104}}.

\bibitem{Apel:2013ura}
W.~D. Apel, et~al., {Ankle-like Feature in the Energy Spectrum of Light
  Elements of Cosmic Rays Observed with KASCADE-Grande}, Phys. Rev. D 87 (2013)
  081101.
\newblock \href {http://arxiv.org/abs/1304.7114} {\path{arXiv:1304.7114}},
  \href {http://dx.doi.org/10.1103/PhysRevD.87.081101}
  {\path{doi:10.1103/PhysRevD.87.081101}}.

\bibitem{Bellido:2017cgf}
J.~Bellido, {\bf Pierre Auger} Collaboration, {Depth of maximum of air-shower
  profiles at the Pierre Auger Observatory: Measurements above $10^{17.2}$ eV
  and Composition Implications}, PoS ICRC2017 (2018) 506.
\newblock \href {http://dx.doi.org/10.22323/1.301.0506}
  {\path{doi:10.22323/1.301.0506}}.

\bibitem{PierreAuger:2010ymv}
J.~Abraham, et~al., {\bf Pierre Auger} Collaboration, {Measurement of the Depth
  of Maximum of Extensive Air Showers above $10^{18}$ eV}, Phys. Rev. Lett. 104
  (2010) 091101.
\newblock \href {http://arxiv.org/abs/1002.0699} {\path{arXiv:1002.0699}},
  \href {http://dx.doi.org/10.1103/PhysRevLett.104.091101}
  {\path{doi:10.1103/PhysRevLett.104.091101}}.

\bibitem{PierreAuger:2014sui}
A.~Aab, et~al., {\bf Pierre Auger} Collaboration, {Depth of Maximum of
  Air-Shower Profiles at the Pierre Auger Observatory: Measurements at Energies
  above $10^{17.8}$ eV}, Phys. Rev. D 90~(12) (2014) 122005.
\newblock \href {http://arxiv.org/abs/1409.4809} {\path{arXiv:1409.4809}},
  \href {http://dx.doi.org/10.1103/PhysRevD.90.122005}
  {\path{doi:10.1103/PhysRevD.90.122005}}.

\bibitem{Hanlon:2018dhz}
W.~Hanlon, J.~Bellido, J.~Belz, S.~Blaess, V.~de~Souza, D.~Ikeda, P.~Sokolsky,
  Y.~Tsunesada, M.~Unger, A.~Yushkov, {Report of the Working Group on the Mass
  Composition of Ultrahigh Energy Cosmic Rays}, JPS Conf. Proc. 19 (2018)
  011013.
\newblock \href {http://dx.doi.org/10.7566/JPSCP.19.011013}
  {\path{doi:10.7566/JPSCP.19.011013}}.

\bibitem{deSouza:2017wgx}
V.~de~Souza, {\bf Pierre Auger, Telescope Array} Collaboration, {Testing the
  agreement between the $X_\mathrm{max}$ distributions measured by the Pierre
  Auger and Telescope Array Observatories}, PoS ICRC2017 (2018) 522.
\newblock \href {http://dx.doi.org/10.22323/1.301.0522}
  {\path{doi:10.22323/1.301.0522}}.

\bibitem{Yushkov:2019hoh}
A.~Yushkov, J.~Bellido, J.~Belz, V.~de~Souza, W.~Hanlon, D.~Ikeda, P.~Sokolsky,
  Y.~Tsunesada, M.~Unger, {\bf Pierre Auger, Telescope Array} Collaboration,
  {Depth of maximum of air-shower profiles: testing the compatibility of
  measurements performed at the Pierre Auger Observatory and the Telescope
  Array experiment}, EPJ Web Conf. 210 (2019) 01009.
\newblock \href {http://arxiv.org/abs/1905.06245} {\path{arXiv:1905.06245}},
  \href {http://dx.doi.org/10.1051/epjconf/201921001009}
  {\path{doi:10.1051/epjconf/201921001009}}.

\bibitem{ToderoPeixoto:2020rta}
C.~J. Todero~Peixoto, {\bf Pierre Auger} Collaboration, {Estimating the Depth
  of Shower Maximum using the Surface Detectors of the Pierre Auger
  Observatory}, PoS ICRC2019 (2020) 440.
\newblock \href {http://dx.doi.org/10.22323/1.358.0440}
  {\path{doi:10.22323/1.358.0440}}.

\bibitem{PierreAuger:2016qzj}
A.~Aab, et~al., {\bf Pierre Auger} Collaboration, {Evidence for a mixed mass
  composition at the \textquoteleft{}ankle\textquoteright{} in the cosmic-ray
  spectrum}, Phys. Lett. B 762 (2016) 288--295.
\newblock \href {http://arxiv.org/abs/1609.08567} {\path{arXiv:1609.08567}},
  \href {http://dx.doi.org/10.1016/j.physletb.2016.09.039}
  {\path{doi:10.1016/j.physletb.2016.09.039}}.

\bibitem{PierreAuger:2014gko}
A.~Aab, et~al., {\bf Pierre Auger} Collaboration, {Depth of maximum of
  air-shower profiles at the Pierre Auger Observatory. II. Composition
  implications}, Phys. Rev. D 90~(12) (2014) 122006.
\newblock \href {http://arxiv.org/abs/1409.5083} {\path{arXiv:1409.5083}},
  \href {http://dx.doi.org/10.1103/PhysRevD.90.122006}
  {\path{doi:10.1103/PhysRevD.90.122006}}.

\bibitem{Hanlon:2019onl}
W.~Hanlon, {\bf Telescope Array} Collaboration, {Telescope Array 10 Year
  Composition}, PoS ICRC2019 (2021) 280.
\newblock \href {http://arxiv.org/abs/1908.01356} {\path{arXiv:1908.01356}},
  \href {http://dx.doi.org/10.22323/1.358.0280}
  {\path{doi:10.22323/1.358.0280}}.

\bibitem{Sokolsky:2021xto}
P.~Sokolsky, R.~D'Avignon, {The Unreasonable Effectiveness of the
  Air-Fluorescence Technique in Determining the EAS Shower Maximum}\href
  {http://arxiv.org/abs/2110.09588} {\path{arXiv:2110.09588}}.

\bibitem{Watson:2021rfb}
A.~A. Watson, {Further evidence for an increase of the mean mass of the
  highest-energy cosmic-rays with energy}, JHEAp 33 (2022) 130.
\newblock \href {http://arxiv.org/abs/2112.06525} {\path{arXiv:2112.06525}},
  \href {http://dx.doi.org/10.1016/j.jheap.2021.11.001}
  {\path{doi:10.1016/j.jheap.2021.11.001}}.

\bibitem{PierreAuger:2021xnt}
J.~Glombitza, et~al., {\bf Pierre Auger} Collaboration, {Event-by-event
  reconstruction of the shower maximum $X_{\mathrm{max}}$ with the Surface
  Detector of the Pierre Auger Observatory using deep learning}, PoS ICRC2021
  (2021) 359.
\newblock \href {http://dx.doi.org/10.22323/1.395.0359}
  {\path{doi:10.22323/1.395.0359}}.

\bibitem{Engel:2011zzb}
R.~Engel, D.~Heck, T.~Pierog, {Extensive air showers and hadronic interactions
  at high energy}, Ann. Rev. Nucl. Part. Sci. 61 (2011) 467--489.
\newblock \href {http://dx.doi.org/10.1146/annurev.nucl.012809.104544}
  {\path{doi:10.1146/annurev.nucl.012809.104544}}.

\bibitem{Abbasi:2018bzb}
R.~Abbasi, G.~Thomson, {$\langle X_{\rm{max}} \rangle$ Uncertainty from
  Extrapolation of Cosmic Ray Air Shower Parameters}, JPS Conf. Proc. 19 (2018)
  011015.
\newblock \href {http://dx.doi.org/10.7566/JPSCP.19.011015}
  {\path{doi:10.7566/JPSCP.19.011015}}.

\bibitem{LoI_EFmeetsCF21}
L.~A. Anchordoqui, et~al., {Synergy of astro-particle physics and collider
  physics}, sNOWMASS21-CF7 LoI (2020) (2020).

\bibitem{LoI_WHISP21}
D.~Soldin, et~al., {Origin of the Muon Excess in Cosmic Ray Air Showers},
  sNOWMASS21-CF7 LoI (2020) (2020).

\bibitem{TelescopeArray:2018bep}
R.~U. Abbasi, et~al., {\bf Telescope Array} Collaboration, {Mass composition of
  ultrahigh-energy cosmic rays with the Telescope Array Surface Detector data},
  Phys. Rev. D 99~(2) (2019) 022002.
\newblock \href {http://arxiv.org/abs/1808.03680} {\path{arXiv:1808.03680}},
  \href {http://dx.doi.org/10.1103/PhysRevD.99.022002}
  {\path{doi:10.1103/PhysRevD.99.022002}}.

\bibitem{PierreAuger:2021oxo}
T.~Bister, et~al., {\bf Pierre Auger} Collaboration, {A combined fit of energy
  spectrum, shower depth distribution and arrival directions to constrain
  astrophysical models of UHECR sources}, PoS ICRC2021 (2021) 368.
\newblock \href {http://dx.doi.org/10.22323/1.395.0368}
  {\path{doi:10.22323/1.395.0368}}.

\bibitem{Porcelli:2015jli}
A.~Porcelli, et~al., {Measurements of the first two moments of the depth of
  shower maximum over nearly three decades of energy, combining data from}, PoS
  ICRC2015 (2016) 420.
\newblock \href {http://dx.doi.org/10.22323/1.236.0420}
  {\path{doi:10.22323/1.236.0420}}.

\bibitem{PierreAuger:2016tar}
A.~Aab, et~al., {\bf Pierre Auger} Collaboration, {Azimuthal Asymmetry in the
  Risetime of the Surface Detector Signals of the Pierre Auger Observatory},
  Phys. Rev. D 93~(7) (2016) 072006.
\newblock \href {http://arxiv.org/abs/1604.00978} {\path{arXiv:1604.00978}},
  \href {http://dx.doi.org/10.1103/PhysRevD.93.072006}
  {\path{doi:10.1103/PhysRevD.93.072006}}.

\bibitem{PierreAuger:2021xah}
J.~Vicha, et~al., {\bf Pierre Auger} Collaboration, {Adjustments to Model
  Predictions of Depth of Shower Maximum and Signals at Ground Level using
  Hybrid Events of the Pierre Auger Observatory}, PoS ICRC2021 (2021) 310.
\newblock \href {http://dx.doi.org/10.22323/1.395.0310}
  {\path{doi:10.22323/1.395.0310}}.

\bibitem{IceCube:2020wum}
R.~Abbasi, et~al., {\bf IceCube} Collaboration, {The IceCube high-energy
  starting event sample: Description and flux characterization with 7.5 years
  of data}, Phys. Rev. D 104 (2021) 022002.
\newblock \href {http://arxiv.org/abs/2011.03545} {\path{arXiv:2011.03545}},
  \href {http://dx.doi.org/10.1103/PhysRevD.104.022002}
  {\path{doi:10.1103/PhysRevD.104.022002}}.

\bibitem{PierreAuger:2018zqu}
A.~Aab, et~al., {\bf Pierre Auger} Collaboration, {Large-scale cosmic-ray
  anisotropies above 4 EeV measured by the Pierre Auger Observatory},
  Astrophys. J. 868~(1) (2018) 4.
\newblock \href {http://arxiv.org/abs/1808.03579} {\path{arXiv:1808.03579}},
  \href {http://dx.doi.org/10.3847/1538-4357/aae689}
  {\path{doi:10.3847/1538-4357/aae689}}.

\bibitem{TelescopeArray:2020cbq}
R.~U. Abbasi, et~al., {\bf Telescope Array} Collaboration, {Search for
  Large-scale Anisotropy on Arrival Directions of Ultra-high-energy Cosmic Rays
  Observed with the Telescope Array Experiment}, Astrophys. J. Lett. 898~(2)
  (2020) L28.
\newblock \href {http://arxiv.org/abs/2007.00023} {\path{arXiv:2007.00023}},
  \href {http://dx.doi.org/10.3847/2041-8213/aba0bc}
  {\path{doi:10.3847/2041-8213/aba0bc}}.

\bibitem{TelescopeArray:2021ygq}
P.~Tinyakov, et~al., {\bf Telescope Array, Pierre Auger} Collaboration, {The
  UHECR dipole and quadrupole in the latest data from the original Auger and TA
  surface detectors}, PoS ICRC2021 (2021) 375.
\newblock \href {http://dx.doi.org/10.22323/1.395.0375}
  {\path{doi:10.22323/1.395.0375}}.

\bibitem{Allard:2021ioh}
D.~Allard, J.~Aublin, B.~Baret, E.~Parizot, {What can be learnt from UHECR
  anisotropies observations? Paper I : large-scale anisotropies and composition
  features}\href {http://arxiv.org/abs/2110.10761} {\path{arXiv:2110.10761}}.

\bibitem{PierreAuger:2014yba}
A.~Aab, et~al., {\bf Pierre Auger} Collaboration, {Searches for Anisotropies in
  the Arrival Directions of the Highest Energy Cosmic Rays Detected by the
  Pierre Auger Observatory}, Astrophys. J. 804~(1) (2015) 15.
\newblock \href {http://arxiv.org/abs/1411.6111} {\path{arXiv:1411.6111}},
  \href {http://dx.doi.org/10.1088/0004-637X/804/1/15}
  {\path{doi:10.1088/0004-637X/804/1/15}}.

\bibitem{Kim:2021Aj}
J.~Kim, D.~Ivanov, K.~Kawata, H.~Sagawa, G.~Thomson, {Hotspot Update, and a new
  Excess of Events on the Sky Seen by the Telescope Array Experiment}, PoS
  ICRC2021 (2021) 328.
\newblock \href {http://dx.doi.org/10.22323/1.395.0328}
  {\path{doi:10.22323/1.395.0328}}.

\bibitem{TelescopeArray:2021dfb}
R.~U. Abbasi, et~al., {\bf Telescope Array} Collaboration, {Indications of a
  Cosmic Ray Source in the Perseus-Pisces Supercluster}\href
  {http://arxiv.org/abs/2110.14827} {\path{arXiv:2110.14827}}.

\bibitem{IceCube:2015xib}
M.~G. Aartsen, et~al., {\bf IceCube, Pierre Auger, Telescope Array}
  Collaboration, {The IceCube Neutrino Observatory, the Pierre Auger
  Observatory and the Telescope Array: Joint Contribution to the 34th
  International Cosmic Ray Conference (ICRC 2015)}\href
  {http://arxiv.org/abs/1511.02109} {\path{arXiv:1511.02109}}.

\bibitem{IceCube:2015afa}
M.~G. Aartsen, et~al., {\bf IceCube, Pierre Auger, Telescope Array}
  Collaboration, {Search for correlations between the arrival directions of
  IceCube neutrino events and ultrahigh-energy cosmic rays detected by the
  Pierre Auger Observatory and the Telescope Array}, JCAP 01 (2016) 037.
\newblock \href {http://arxiv.org/abs/1511.09408} {\path{arXiv:1511.09408}},
  \href {http://dx.doi.org/10.1088/1475-7516/2016/01/037}
  {\path{doi:10.1088/1475-7516/2016/01/037}}.

\bibitem{IceCube:2017qeh}
M.~G. Aartsen, et~al., {\bf IceCube} Collaboration, {The IceCube Neutrino
  Observatory - Contributions to ICRC 2017 Part I: Searches for the Sources of
  Astrophysical Neutrinos}\href {http://arxiv.org/abs/1710.01179}
  {\path{arXiv:1710.01179}}.

\bibitem{ANTARES:2019ufk}
J.~Aublin, et~al., {\bf ANTARES, IceCube, Pierre Auger, Telescope Array}
  Collaboration, {Search for a correlation between the UHECRs measured by the
  Pierre Auger Observatory and the Telescope Array and the neutrino candidate
  events from IceCube and ANTARES}, EPJ Web Conf. 210 (2019) 03003.
\newblock \href {http://arxiv.org/abs/1905.03997} {\path{arXiv:1905.03997}},
  \href {http://dx.doi.org/10.1051/epjconf/201921003003}
  {\path{doi:10.1051/epjconf/201921003003}}.

\bibitem{Barbano:2020scg}
A.~Barbano, {\bf IceCube, Pierre Auger, Telescope Array, ANTARES}
  Collaboration, {Search for correlations of high-energy neutrinos and
  ultra-high energy cosmic rays}, PoS ICRC2019 (2020) 842.
\newblock \href {http://arxiv.org/abs/2001.09057} {\path{arXiv:2001.09057}},
  \href {http://dx.doi.org/10.22323/1.358.0842}
  {\path{doi:10.22323/1.358.0842}}.

\bibitem{PierreAuger:2012dqz}
P.~Abreu, et~al., {\bf Pierre Auger} Collaboration, {A Search for Point Sources
  of EeV Neutrons}, Astrophys. J. 760 (2012) 148.
\newblock \href {http://arxiv.org/abs/1211.4901} {\path{arXiv:1211.4901}},
  \href {http://dx.doi.org/10.1088/0004-637X/760/2/148}
  {\path{doi:10.1088/0004-637X/760/2/148}}.

\bibitem{PierreAuger:2014tey}
A.~Aab, et~al., {\bf Pierre Auger} Collaboration, {A Targeted Search for Point
  Sources of EeV Neutrons}, Astrophys. J. Lett. 789 (2014) L34.
\newblock \href {http://arxiv.org/abs/1406.4038} {\path{arXiv:1406.4038}},
  \href {http://dx.doi.org/10.1088/2041-8205/789/2/L34}
  {\path{doi:10.1088/2041-8205/789/2/L34}}.

\bibitem{LIGOScientific:2017vwq}
B.~P. Abbott, et~al., {\bf LIGO Scientific, Virgo} Collaboration, {GW170817:
  Observation of Gravitational Waves from a Binary Neutron Star Inspiral},
  Phys. Rev. Lett. 119~(16) (2017) 161101.
\newblock \href {http://arxiv.org/abs/1710.05832} {\path{arXiv:1710.05832}},
  \href {http://dx.doi.org/10.1103/PhysRevLett.119.161101}
  {\path{doi:10.1103/PhysRevLett.119.161101}}.

\bibitem{LIGOScientific:2017adf}
B.~P. Abbott, et~al., {\bf LIGO Scientific, Virgo, 1M2H, Dark Energy Camera
  GW-E, DES, DLT40, Las Cumbres Observatory, VINROUGE, MASTER} Collaboration,
  {A gravitational-wave standard siren measurement of the Hubble constant},
  Nature 551~(7678) (2017) 85--88.
\newblock \href {http://arxiv.org/abs/1710.05835} {\path{arXiv:1710.05835}},
  \href {http://dx.doi.org/10.1038/nature24471}
  {\path{doi:10.1038/nature24471}}.

\bibitem{Drout:2017ijr}
M.~R. Drout, et~al., {Light Curves of the Neutron Star Merger GW170817/SSS17a:
  Implications for R-Process Nucleosynthesis}, Science 358 (2017) 1570--1574.
\newblock \href {http://arxiv.org/abs/1710.05443} {\path{arXiv:1710.05443}},
  \href {http://dx.doi.org/10.1126/science.aaq0049}
  {\path{doi:10.1126/science.aaq0049}}.

\bibitem{Metzger:2017wot}
B.~D. Metzger, {Welcome to the Multi-Messenger Era! Lessons from a Neutron Star
  Merger and the Landscape Ahead}\href {http://arxiv.org/abs/1710.05931}
  {\path{arXiv:1710.05931}}.

\bibitem{AlvesBatista:2018zui}
R.~Alves~Batista, R.~M. de~Almeida, B.~Lago, K.~Kotera, {Cosmogenic photon and
  neutrino fluxes in the Auger era}, JCAP 01 (2019) 002.
\newblock \href {http://arxiv.org/abs/1806.10879} {\path{arXiv:1806.10879}},
  \href {http://dx.doi.org/10.1088/1475-7516/2019/01/002}
  {\path{doi:10.1088/1475-7516/2019/01/002}}.

\bibitem{AlvesBatista:2016vpy}
R.~Alves~Batista, A.~Dundovic, M.~Erdmann, K.-H. Kampert, D.~Kuempel,
  G.~M\"uller, G.~Sigl, A.~van Vliet, D.~Walz, T.~Winchen, {CRPropa 3 - a
  Public Astrophysical Simulation Framework for Propagating Extraterrestrial
  Ultra-High Energy Particles}, JCAP 05 (2016) 038.
\newblock \href {http://arxiv.org/abs/1603.07142} {\path{arXiv:1603.07142}},
  \href {http://dx.doi.org/10.1088/1475-7516/2016/05/038}
  {\path{doi:10.1088/1475-7516/2016/05/038}}.

\bibitem{IceCube:2013low}
M.~G. Aartsen, et~al., {\bf IceCube} Collaboration, {Evidence for High-Energy
  Extraterrestrial Neutrinos at the IceCube Detector}, Science 342 (2013)
  1242856.
\newblock \href {http://arxiv.org/abs/1311.5238} {\path{arXiv:1311.5238}},
  \href {http://dx.doi.org/10.1126/science.1242856}
  {\path{doi:10.1126/science.1242856}}.

\bibitem{IceCube:2014stg}
M.~G. Aartsen, et~al., {\bf IceCube} Collaboration, {Observation of High-Energy
  Astrophysical Neutrinos in Three Years of IceCube Data}, Phys. Rev. Lett. 113
  (2014) 101101.
\newblock \href {http://arxiv.org/abs/1405.5303} {\path{arXiv:1405.5303}},
  \href {http://dx.doi.org/10.1103/PhysRevLett.113.101101}
  {\path{doi:10.1103/PhysRevLett.113.101101}}.

\bibitem{IceCube:2015qii}
M.~G. Aartsen, et~al., {\bf IceCube} Collaboration, {Evidence for Astrophysical
  Muon Neutrinos from the Northern Sky with IceCube}, Phys. Rev. Lett. 115~(8)
  (2015) 081102.
\newblock \href {http://arxiv.org/abs/1507.04005} {\path{arXiv:1507.04005}},
  \href {http://dx.doi.org/10.1103/PhysRevLett.115.081102}
  {\path{doi:10.1103/PhysRevLett.115.081102}}.

\bibitem{Kowalski:2021oda}
M.~Kowalski, {Resolving the high-energy neutrino sky at 3$\sigma$}, Nature
  Astron. 5~(8) (2021) 732--734.
\newblock \href {http://dx.doi.org/10.1038/s41550-021-01431-y}
  {\path{doi:10.1038/s41550-021-01431-y}}.

\bibitem{IceCube:2018dnn}
M.~G. Aartsen, et~al., {\bf IceCube, Fermi-LAT, MAGIC, AGILE, ASAS-SN, HAWC,
  H.E.S.S., INTEGRAL, Kanata, Kiso, Kapteyn, Liverpool Telescope, Subaru, Swift
  NuSTAR, VERITAS, VLA/17B-403} Collaboration, {Multimessenger observations of
  a flaring blazar coincident with high-energy neutrino IceCube-170922A},
  Science 361~(6398) (2018) eaat1378.
\newblock \href {http://arxiv.org/abs/1807.08816} {\path{arXiv:1807.08816}},
  \href {http://dx.doi.org/10.1126/science.aat1378}
  {\path{doi:10.1126/science.aat1378}}.

\bibitem{IceCube:2018cha}
M.~G. Aartsen, et~al., {\bf IceCube} Collaboration, {Neutrino emission from the
  direction of the blazar TXS 0506+056 prior to the IceCube-170922A alert},
  Science 361~(6398) (2018) 147--151.
\newblock \href {http://arxiv.org/abs/1807.08794} {\path{arXiv:1807.08794}},
  \href {http://dx.doi.org/10.1126/science.aat2890}
  {\path{doi:10.1126/science.aat2890}}.

\bibitem{Ahlers:2018fkn}
M.~Ahlers, F.~Halzen, {Opening a New Window onto the Universe with IceCube},
  Prog. Part. Nucl. Phys. 102 (2018) 73--88.
\newblock \href {http://arxiv.org/abs/1805.11112} {\path{arXiv:1805.11112}},
  \href {http://dx.doi.org/10.1016/j.ppnp.2018.05.001}
  {\path{doi:10.1016/j.ppnp.2018.05.001}}.

\bibitem{Waxman:1998yy}
E.~Waxman, J.~N. Bahcall, {High-energy neutrinos from astrophysical sources: An
  Upper bound}, Phys. Rev. D 59 (1999) 023002.
\newblock \href {http://arxiv.org/abs/hep-ph/9807282}
  {\path{arXiv:hep-ph/9807282}}, \href
  {http://dx.doi.org/10.1103/PhysRevD.59.023002}
  {\path{doi:10.1103/PhysRevD.59.023002}}.

\bibitem{Mannheim:1998wp}
K.~Mannheim, R.~J. Protheroe, J.~P. Rachen, {On the cosmic ray bound for models
  of extragalactic neutrino production}, Phys. Rev. D 63 (2001) 023003.
\newblock \href {http://arxiv.org/abs/astro-ph/9812398}
  {\path{arXiv:astro-ph/9812398}}, \href
  {http://dx.doi.org/10.1103/PhysRevD.63.023003}
  {\path{doi:10.1103/PhysRevD.63.023003}}.

\bibitem{Ave:2000nd}
M.~Ave, J.~A. Hinton, R.~A. Vazquez, A.~A. Watson, E.~Zas, {New constraints
  from Haverah Park data on the photon and iron fluxes of UHE cosmic rays},
  Phys. Rev. Lett. 85 (2000) 2244--2247.
\newblock \href {http://arxiv.org/abs/astro-ph/0007386}
  {\path{arXiv:astro-ph/0007386}}, \href
  {http://dx.doi.org/10.1103/PhysRevLett.85.2244}
  {\path{doi:10.1103/PhysRevLett.85.2244}}.

\bibitem{PierreAuger:2007hjd}
J.~Abraham, et~al., {\bf Pierre Auger} Collaboration, {Upper limit on the
  cosmic-ray photon flux above 10$^{19}$ eV using the surface detector of the
  Pierre Auger Observatory}, Astropart. Phys. 29 (2008) 243--256.
\newblock \href {http://arxiv.org/abs/0712.1147} {\path{arXiv:0712.1147}},
  \href {http://dx.doi.org/10.1016/j.astropartphys.2008.01.003}
  {\path{doi:10.1016/j.astropartphys.2008.01.003}}.

\bibitem{TelescopeArray:2018rbt}
R.~U. Abbasi, et~al., {\bf Telescope Array} Collaboration, {Constraints on the
  diffuse photon flux with energies above $10^{18}$ eV using the surface
  detector of the Telescope Array experiment}, Astropart. Phys. 110 (2019)
  8--14.
\newblock \href {http://arxiv.org/abs/1811.03920} {\path{arXiv:1811.03920}},
  \href {http://dx.doi.org/10.1016/j.astropartphys.2019.03.003}
  {\path{doi:10.1016/j.astropartphys.2019.03.003}}.

\bibitem{PierreAuger:2021mjh}
P.~Abreu, et~al., {\bf Pierre Auger} Collaboration, {A search for
  ultra-high-energy photons at the Pierre Auger Observatory exploiting
  air-shower universality}, PoS ICRC2021 (2021) 373.
\newblock \href {http://dx.doi.org/10.22323/1.395.0373}
  {\path{doi:10.22323/1.395.0373}}.

\bibitem{PierreAuger:2007vvh}
J.~Abraham, et~al., {\bf Pierre Auger} Collaboration, {Upper limit on the
  diffuse flux of UHE tau neutrinos from the Pierre Auger Observatory}, Phys.
  Rev. Lett. 100 (2008) 211101.
\newblock \href {http://arxiv.org/abs/0712.1909} {\path{arXiv:0712.1909}},
  \href {http://dx.doi.org/10.1103/PhysRevLett.100.211101}
  {\path{doi:10.1103/PhysRevLett.100.211101}}.

\bibitem{PierreAuger:2015ihf}
A.~Aab, et~al., {\bf Pierre Auger} Collaboration, {Improved limit to the
  diffuse flux of ultrahigh energy neutrinos from the Pierre Auger
  Observatory}, Phys. Rev. D 91~(9) (2015) 092008.
\newblock \href {http://arxiv.org/abs/1504.05397} {\path{arXiv:1504.05397}},
  \href {http://dx.doi.org/10.1103/PhysRevD.91.092008}
  {\path{doi:10.1103/PhysRevD.91.092008}}.

\bibitem{Bhattacharjee:1991zm}
P.~Bhattacharjee, C.~T. Hill, D.~N. Schramm, {Grand unified theories,
  topological defects and ultrahigh-energy cosmic rays}, Phys. Rev. Lett. 69
  (1992) 567--570.
\newblock \href {http://dx.doi.org/10.1103/PhysRevLett.69.567}
  {\path{doi:10.1103/PhysRevLett.69.567}}.

\bibitem{Gelmini:2005wu}
G.~Gelmini, O.~E. Kalashev, D.~V. Semikoz, {GZK photons as ultra high energy
  cosmic rays}, J. Exp. Theor. Phys. 106 (2008) 1061--1082.
\newblock \href {http://arxiv.org/abs/astro-ph/0506128}
  {\path{arXiv:astro-ph/0506128}}, \href
  {http://dx.doi.org/10.1134/S106377610806006X}
  {\path{doi:10.1134/S106377610806006X}}.

\bibitem{Kampert:2011hkm}
K.-H. Kampert, et~al., {Ultra-High Energy Photon and Neutrino Fluxes in
  Realistic Astrophysical Scenarios}, Proc. 32nd International Cosmic Ray
  Conference (Beijing, China)\href
  {http://dx.doi.org/10.7529/ICRC2011/V02/1087}
  {\path{doi:10.7529/ICRC2011/V02/1087}}.

\bibitem{Muzio:2019leu}
M.~S. Muzio, M.~Unger, G.~R. Farrar, {Progress towards characterizing ultrahigh
  energy cosmic ray sources}, Phys. Rev. D 100~(10) (2019) 103008.
\newblock \href {http://arxiv.org/abs/1906.06233} {\path{arXiv:1906.06233}},
  \href {http://dx.doi.org/10.1103/PhysRevD.100.103008}
  {\path{doi:10.1103/PhysRevD.100.103008}}.

\bibitem{Bobrikova:2021kuj}
A.~Bobrikova, M.~Niechciol, M.~Risse, P.~Ruehl, {Predicting the UHE photon flux
  from GZK-interactions of hadronic cosmic rays using CRPropa 3}, Proc. 37th
  International Cosmic Ray Conference (Berlin, Germany) PoS (ICRC2021) (2021)
  449.
\newblock \href {http://dx.doi.org/10.22323/1.395.0449}
  {\path{doi:10.22323/1.395.0449}}.

\bibitem{vanVliet:2019nse}
A.~van Vliet, R.~Alves~Batista, J.~R. H\"orandel, {Determining the fraction of
  cosmic-ray protons at ultrahigh energies with cosmogenic neutrinos}, Phys.
  Rev. D 100~(2) (2019) 021302.
\newblock \href {http://arxiv.org/abs/1901.01899} {\path{arXiv:1901.01899}},
  \href {http://dx.doi.org/10.1103/PhysRevD.100.021302}
  {\path{doi:10.1103/PhysRevD.100.021302}}.

\bibitem{PierreAuger:2019azx}
A.~Aab, et~al., {\bf Pierre Auger} Collaboration, {Limits on point-like sources
  of ultra-high-energy neutrinos with the Pierre Auger Observatory}, JCAP 11
  (2019) 004.
\newblock \href {http://arxiv.org/abs/1906.07419} {\path{arXiv:1906.07419}},
  \href {http://dx.doi.org/10.1088/1475-7516/2019/11/004}
  {\path{doi:10.1088/1475-7516/2019/11/004}}.

\bibitem{IceCube:2018fhm}
M.~G. Aartsen, et~al., {\bf IceCube} Collaboration, {Differential limit on the
  extremely-high-energy cosmic neutrino flux in the presence of astrophysical
  background from nine years of IceCube data}, Phys. Rev. D 98~(6) (2018)
  062003.
\newblock \href {http://arxiv.org/abs/1807.01820} {\path{arXiv:1807.01820}},
  \href {http://dx.doi.org/10.1103/PhysRevD.98.062003}
  {\path{doi:10.1103/PhysRevD.98.062003}}.

\bibitem{Heinze:2015hhp}
J.~Heinze, D.~Boncioli, M.~Bustamante, W.~Winter, {Cosmogenic Neutrinos
  Challenge the Cosmic Ray Proton Dip Model}, Astrophys. J. 825~(2) (2016) 122.
\newblock \href {http://arxiv.org/abs/1512.05988} {\path{arXiv:1512.05988}},
  \href {http://dx.doi.org/10.3847/0004-637X/825/2/122}
  {\path{doi:10.3847/0004-637X/825/2/122}}.

\bibitem{Muzio:2021zud}
M.~S. Muzio, G.~R. Farrar, M.~Unger, {Ultrahigh energy cosmic rays and high
  energy astrophysical neutrinos}\href {http://arxiv.org/abs/2108.05512}
  {\path{arXiv:2108.05512}}.

\bibitem{Rodrigues:2020pli}
X.~Rodrigues, J.~Heinze, A.~Palladino, A.~van Vliet, W.~Winter, {Active
  Galactic Nuclei Jets as the Origin of Ultrahigh-Energy Cosmic Rays and
  Perspectives for the Detection of Astrophysical Source Neutrinos at EeV
  Energies}, Phys. Rev. Lett. 126~(19) (2021) 191101.
\newblock \href {http://arxiv.org/abs/2003.08392} {\path{arXiv:2003.08392}},
  \href {http://dx.doi.org/10.1103/PhysRevLett.126.191101}
  {\path{doi:10.1103/PhysRevLett.126.191101}}.

\bibitem{IceCube:2020acn}
M.~G. Aartsen, et~al., {\bf IceCube} Collaboration, {Characteristics of the
  diffuse astrophysical electron and tau neutrino flux with six years of
  IceCube high energy cascade data}, Phys. Rev. Lett. 125~(12) (2020) 121104.
\newblock \href {http://arxiv.org/abs/2001.09520} {\path{arXiv:2001.09520}},
  \href {http://dx.doi.org/10.1103/PhysRevLett.125.121104}
  {\path{doi:10.1103/PhysRevLett.125.121104}}.

\bibitem{Stettner:2019tok}
J.~Stettner, {\bf IceCube} Collaboration, {Measurement of the Diffuse
  Astrophysical Muon-Neutrino Spectrum with Ten Years of IceCube Data}, PoS
  ICRC2019 (2020) 1017.
\newblock \href {http://arxiv.org/abs/1908.09551} {\path{arXiv:1908.09551}},
  \href {http://dx.doi.org/10.22323/1.358.1017}
  {\path{doi:10.22323/1.358.1017}}.

\bibitem{IceCube:2021rpz}
M.~G. Aartsen, et~al., {\bf IceCube} Collaboration, {Detection of a particle
  shower at the Glashow resonance with IceCube}, Nature 591~(7849) (2021)
  220--224, [Erratum: Nature 592, E11 (2021)].
\newblock \href {http://arxiv.org/abs/2110.15051} {\path{arXiv:2110.15051}},
  \href {http://dx.doi.org/10.1038/s41586-021-03256-1}
  {\path{doi:10.1038/s41586-021-03256-1}}.

\bibitem{ANITA:2019wyx}
P.~W. Gorham, et~al., {\bf ANITA} Collaboration, {Constraints on the
  ultrahigh-energy cosmic neutrino flux from the fourth flight of ANITA}, Phys.
  Rev. D 99~(12) (2019) 122001.
\newblock \href {http://arxiv.org/abs/1902.04005} {\path{arXiv:1902.04005}},
  \href {http://dx.doi.org/10.1103/PhysRevD.99.122001}
  {\path{doi:10.1103/PhysRevD.99.122001}}.

\bibitem{Kaeaepae-icrc21}
{A. K\"a\"ap\"a, K.-H. Kampert, and E. Mayotte}, {The effects of the GMF on the
  transition from Galactic to extragalactic cosmic rays}, PoS ICRC2021 (2021)
  004.
\newblock \href {http://dx.doi.org/10.22323/1.395.0004}
  {\path{doi:10.22323/1.395.0004}}.

\bibitem{ANTARES:2022pdr}
A.~Albert, et~al., {\bf ANTARES, IceCube, Auger, Telescope Array}
  Collaboration, {Search for Spatial Correlations of Neutrinos with
  Ultra-High-Energy Cosmic Rays}\href {http://arxiv.org/abs/2201.07313}
  {\path{arXiv:2201.07313}}.

\bibitem{Schumacher:2019qdx}
L.~{Schumacher}, {Search for correlations of high-energy neutrinos and
  ultrahigh- energy cosmic rays}, in: European Physical Journal Web of
  Conferences, Vol. 207 of European Physical Journal Web of Conferences, 2019,
  p. 02010.
\newblock \href {http://arxiv.org/abs/1905.10111} {\path{arXiv:1905.10111}},
  \href {http://dx.doi.org/10.1051/epjconf/201920702010}
  {\path{doi:10.1051/epjconf/201920702010}}.

\bibitem{Berezinsky:1969erk}
V.~S. Berezinsky, G.~T. Zatsepin, {Cosmic rays at ultrahigh-energies
  (neutrino?)}, Phys. Lett. B 28 (1969) 423--424.
\newblock \href {http://dx.doi.org/10.1016/0370-2693(69)90341-4}
  {\path{doi:10.1016/0370-2693(69)90341-4}}.

\bibitem{Zas:2005zz}
E.~Zas, {Neutrino detection with inclined air showers}, New J. Phys. 7 (2005)
  130.
\newblock \href {http://arxiv.org/abs/astro-ph/0504610}
  {\path{arXiv:astro-ph/0504610}}, \href
  {http://dx.doi.org/10.1088/1367-2630/7/1/130}
  {\path{doi:10.1088/1367-2630/7/1/130}}.

\bibitem{Learned:1994wg}
J.~G. Learned, S.~Pakvasa, {Detecting tau-neutrino oscillations at PeV
  energies}, Astropart. Phys. 3 (1995) 267--274.
\newblock \href {http://arxiv.org/abs/hep-ph/9405296}
  {\path{arXiv:hep-ph/9405296}}, \href
  {http://dx.doi.org/10.1016/0927-6505(94)00043-3}
  {\path{doi:10.1016/0927-6505(94)00043-3}}.

\bibitem{Athar:2000yw}
H.~Athar, M.~Jezabek, O.~Yasuda, {Effects of neutrino mixing on high-energy
  cosmic neutrino flux}, Phys. Rev. D 62 (2000) 103007.
\newblock \href {http://arxiv.org/abs/hep-ph/0005104}
  {\path{arXiv:hep-ph/0005104}}, \href
  {http://dx.doi.org/10.1103/PhysRevD.62.103007}
  {\path{doi:10.1103/PhysRevD.62.103007}}.

\bibitem{Fargion:2000iz}
D.~Fargion, {Discovering Ultra High Energy Neutrinos by Horizontal and Upward
  tau Air-Showers: Evidences in Terrestrial Gamma Flashes?}, Astrophys. J. 570
  (2002) 909--925.
\newblock \href {http://arxiv.org/abs/astro-ph/0002453}
  {\path{arXiv:astro-ph/0002453}}, \href {http://dx.doi.org/10.1086/339772}
  {\path{doi:10.1086/339772}}.

\bibitem{Bertou:2001vm}
X.~Bertou, P.~Billoir, O.~Deligny, C.~Lachaud, A.~Letessier-Selvon, {Tau
  neutrinos in the Auger Observatory: A New window to UHECR sources},
  Astropart. Phys. 17 (2002) 183--193.
\newblock \href {http://arxiv.org/abs/astro-ph/0104452}
  {\path{arXiv:astro-ph/0104452}}, \href
  {http://dx.doi.org/10.1016/S0927-6505(01)00147-5}
  {\path{doi:10.1016/S0927-6505(01)00147-5}}.

\bibitem{PierreAuger:2016kuz}
A.~Aab, et~al., {\bf Pierre Auger} Collaboration, {Search for photons with
  energies above 10$^{18}$ eV using the hybrid detector of the Pierre Auger
  Observatory}, JCAP 04 (2017) 009, [Erratum: JCAP 09, E02 (2020)].
\newblock \href {http://arxiv.org/abs/1612.01517} {\path{arXiv:1612.01517}},
  \href {http://dx.doi.org/10.1088/1475-7516/2017/04/009}
  {\path{doi:10.1088/1475-7516/2017/04/009}}.

\bibitem{PierreAuger:2016ppv}
A.~Aab, et~al., {\bf Pierre Auger} Collaboration, {A targeted search for point
  sources of EeV photons with the Pierre Auger Observatory}, Astrophys. J.
  Lett. 837~(2) (2017) L25.
\newblock \href {http://arxiv.org/abs/1612.04155} {\path{arXiv:1612.04155}},
  \href {http://dx.doi.org/10.3847/2041-8213/aa61a5}
  {\path{doi:10.3847/2041-8213/aa61a5}}.

\bibitem{Kalashev:2016cre}
O.~K. Kalashev, M.~Y. Kuznetsov, {Constraining heavy decaying dark matter with
  the high energy gamma-ray limits}, Phys. Rev. D 94~(6) (2016) 063535.
\newblock \href {http://arxiv.org/abs/1606.07354} {\path{arXiv:1606.07354}},
  \href {http://dx.doi.org/10.1103/PhysRevD.94.063535}
  {\path{doi:10.1103/PhysRevD.94.063535}}.

\bibitem{Anchordoqui:2021crl}
L.~A. Anchordoqui, et~al., {Hunting super-heavy dark matter with ultra-high
  energy photons}, Astropart. Phys. 132 (2021) 102614.
\newblock \href {http://arxiv.org/abs/2105.12895} {\path{arXiv:2105.12895}},
  \href {http://dx.doi.org/10.1016/j.astropartphys.2021.102614}
  {\path{doi:10.1016/j.astropartphys.2021.102614}}.

\bibitem{TelescopeArray:2013yze}
T.~Abu-Zayyad, et~al., {\bf Telescope Array} Collaboration, {Upper limit on the
  flux of photons with energies above $10^{19}$ eV using the Telescope Array
  surface detector}, Phys. Rev. D 88~(11) (2013) 112005.
\newblock \href {http://arxiv.org/abs/1304.5614} {\path{arXiv:1304.5614}},
  \href {http://dx.doi.org/10.1103/PhysRevD.88.112005}
  {\path{doi:10.1103/PhysRevD.88.112005}}.

\bibitem{TelescopeArray:2020hey}
R.~U. Abbasi, et~al., {\bf Telescope Array} Collaboration, {Search for point
  sources of ultra-high-energy photons with the Telescope Array surface
  detector}, Mon. Not. Roy. Astron. Soc. 492~(3) (2020) 3984--3993.
\newblock \href {http://dx.doi.org/10.1093/mnras/stz3618}
  {\path{doi:10.1093/mnras/stz3618}}.

\bibitem{Kalashev:2020hqc}
O.~Kalashev, M.~Kuznetsov, Y.~Zhezher, {Constraining superheavy decaying dark
  matter with directional ultra-high energy gamma-ray limits}, JCAP 11 (2021)
  016.
\newblock \href {http://arxiv.org/abs/2005.04085} {\path{arXiv:2005.04085}},
  \href {http://dx.doi.org/10.1088/1475-7516/2021/11/016}
  {\path{doi:10.1088/1475-7516/2021/11/016}}.

\bibitem{Maccione:2010sv}
L.~Maccione, S.~Liberati, G.~Sigl, {Ultra high energy photons as probes of
  Lorentz symmetry violations in stringy space-time foam models}, Phys. Rev.
  Lett. 105 (2010) 021101.
\newblock \href {http://arxiv.org/abs/1003.5468} {\path{arXiv:1003.5468}},
  \href {http://dx.doi.org/10.1103/PhysRevLett.105.021101}
  {\path{doi:10.1103/PhysRevLett.105.021101}}.

\bibitem{Galaverni:2007tq}
M.~Galaverni, G.~Sigl, {Lorentz Violation in the Photon Sector and Ultra-High
  Energy Cosmic Rays}, Phys. Rev. Lett. 100 (2008) 021102.
\newblock \href {http://arxiv.org/abs/0708.1737} {\path{arXiv:0708.1737}},
  \href {http://dx.doi.org/10.1103/PhysRevLett.100.021102}
  {\path{doi:10.1103/PhysRevLett.100.021102}}.

\bibitem{Galaverni:2008yj}
M.~Galaverni, G.~Sigl, {Lorentz Violation and Ultrahigh-Energy Photons}, Phys.
  Rev. D 78 (2008) 063003.
\newblock \href {http://arxiv.org/abs/0807.1210} {\path{arXiv:0807.1210}},
  \href {http://dx.doi.org/10.1103/PhysRevD.78.063003}
  {\path{doi:10.1103/PhysRevD.78.063003}}.

\bibitem{Stecker:2017gdy}
F.~W. Stecker, {Testing Lorentz Symmetry using High Energy Astrophysics
  Observations}, Symmetry 9~(10) (2017) 201.
\newblock \href {http://arxiv.org/abs/1708.05672} {\path{arXiv:1708.05672}},
  \href {http://dx.doi.org/10.3390/sym9100201} {\path{doi:10.3390/sym9100201}}.

\bibitem{PierreAuger:2016efk}
A.~Aab, et~al., {\bf Pierre Auger} Collaboration, {Ultrahigh-Energy Neutrino
  Follow-Up of Gravitational Wave Events GW150914 and GW151226 with the Pierre
  Auger Observatory}, Phys. Rev. D 94~(12) (2016) 122007.
\newblock \href {http://arxiv.org/abs/1608.07378} {\path{arXiv:1608.07378}},
  \href {http://dx.doi.org/10.1103/PhysRevD.94.122007}
  {\path{doi:10.1103/PhysRevD.94.122007}}.

\bibitem{Schimp:2020gxx}
M.~Schimp, {\bf Pierre Auger} Collaboration, {Follow-up searches for ultra-high
  energy neutrinos from transient astrophysical sources with the Pierre Auger
  Observatory}, PoS ICRC2019 (2020) 415.
\newblock \href {http://dx.doi.org/10.22323/1.358.0415}
  {\path{doi:10.22323/1.358.0415}}.

\bibitem{AyalaSolares:2019iiy}
H.~A. Ayala~Solares, et~al., {The Astrophysical Multimessenger Observatory
  Network (AMON): Performance and science program}, Astropart. Phys. 114 (2020)
  68--76.
\newblock \href {http://arxiv.org/abs/1903.08714} {\path{arXiv:1903.08714}},
  \href {http://dx.doi.org/10.1016/j.astropartphys.2019.06.007}
  {\path{doi:10.1016/j.astropartphys.2019.06.007}}.

\bibitem{Ahn:2009wx}
E.-J. Ahn, R.~Engel, T.~K. Gaisser, P.~Lipari, T.~Stanev, {Cosmic ray
  interaction event generator SIBYLL 2.1}, Phys. Rev. D 80 (2009) 094003.
\newblock \href {http://arxiv.org/abs/0906.4113} {\path{arXiv:0906.4113}},
  \href {http://dx.doi.org/10.1103/PhysRevD.80.094003}
  {\path{doi:10.1103/PhysRevD.80.094003}}.

\bibitem{Riehn:2015oba}
F.~Riehn, et~al., {A new version of the event generator Sibyll}, PoS ICRC2015
  (2016) 558.
\newblock \href {http://arxiv.org/abs/1510.00568} {\path{arXiv:1510.00568}},
  \href {http://dx.doi.org/10.22323/1.236.0558}
  {\path{doi:10.22323/1.236.0558}}.

\bibitem{Riehn:2017mfm}
F.~Riehn, et~al., {The hadronic interaction model SIBYLL 2.3c and Feynman
  scaling}, PoS ICRC2017 (2018) 301.
\newblock \href {http://arxiv.org/abs/1709.07227} {\path{arXiv:1709.07227}},
  \href {http://dx.doi.org/10.22323/1.301.0301}
  {\path{doi:10.22323/1.301.0301}}.

\bibitem{Ostapchenko:2005nj}
S.~Ostapchenko, {Nonlinear screening effects in high energy hadronic
  interactions}, Phys. Rev. D 74~(1) (2006) 014026.
\newblock \href {http://arxiv.org/abs/hep-ph/0505259}
  {\path{arXiv:hep-ph/0505259}}, \href
  {http://dx.doi.org/10.1103/PhysRevD.74.014026}
  {\path{doi:10.1103/PhysRevD.74.014026}}.

\bibitem{Ostapchenko:2006vr}
S.~Ostapchenko, {On the re-summation of enhanced Pomeron diagrams}, Phys. Lett.
  B 636 (2006) 40--45.
\newblock \href {http://arxiv.org/abs/hep-ph/0602139}
  {\path{arXiv:hep-ph/0602139}}, \href
  {http://dx.doi.org/10.1016/j.physletb.2006.03.026}
  {\path{doi:10.1016/j.physletb.2006.03.026}}.

\bibitem{Ostapchenko:2010vb}
S.~Ostapchenko, {Monte Carlo treatment of hadronic interactions in enhanced
  Pomeron scheme: I. QGSJET-II model}, Phys. Rev. D 83 (2011) 014018.
\newblock \href {http://arxiv.org/abs/1010.1869} {\path{arXiv:1010.1869}},
  \href {http://dx.doi.org/10.1103/PhysRevD.83.014018}
  {\path{doi:10.1103/PhysRevD.83.014018}}.

\bibitem{Ostapchenko:2019few}
S.~Ostapchenko, {QGSJET-III model: physics and preliminary results}, EPJ Web
  Conf. 208 (2019) 11001.
\newblock \href {http://dx.doi.org/10.1051/epjconf/201920811001}
  {\path{doi:10.1051/epjconf/201920811001}}.

\bibitem{Werner:2005jf}
K.~Werner, F.-M. Liu, T.~Pierog, {Parton ladder splitting and the rapidity
  dependence of transverse momentum spectra in deuteron gold collisions at
  RHIC}, Phys. Rev. C74 (2006) 044902.
\newblock \href {http://arxiv.org/abs/hep-ph/0506232}
  {\path{arXiv:hep-ph/0506232}}, \href
  {http://dx.doi.org/10.1103/PhysRevC.74.044902}
  {\path{doi:10.1103/PhysRevC.74.044902}}.

\bibitem{Pierog:2009zt}
T.~Pierog, K.~Werner, {EPOS Model and Ultra High Energy Cosmic Rays}, Nucl.
  Phys. B Proc. Suppl. 196 (2009) 102--105.
\newblock \href {http://arxiv.org/abs/0905.1198} {\path{arXiv:0905.1198}},
  \href {http://dx.doi.org/10.1016/j.nuclphysbps.2009.09.017}
  {\path{doi:10.1016/j.nuclphysbps.2009.09.017}}.

\bibitem{Pierog:2017awp}
T.~Pierog, {Air Shower Simulation with a New Generation of post-LHC Hadronic
  Interaction Models in CORSIKA}, PoS ICRC2017 (2018) 1100.
\newblock \href {http://dx.doi.org/10.22323/1.301.1100}
  {\path{doi:10.22323/1.301.1100}}.

\bibitem{Ranft:1994fd}
J.~Ranft, {The Dual parton model at cosmic ray energies}, Phys. Rev. D 51
  (1995) 64--84.
\newblock \href {http://dx.doi.org/10.1103/PhysRevD.51.64}
  {\path{doi:10.1103/PhysRevD.51.64}}.

\bibitem{Ranft:1999fy}
J.~Ranft, {New features in DPMJET version II.5}\href
  {http://arxiv.org/abs/hep-ph/9911213} {\path{arXiv:hep-ph/9911213}}.

\bibitem{Ranft:2002rj}
J.~Ranft, R.~Engel, S.~Roesler, {DPMJET-III, learning from RHIC data for cosmic
  ray particle production}, Nucl. Phys. B Proc. Suppl. 122 (2003) 392--395.
\newblock \href {http://dx.doi.org/10.1016/S0920-5632(03)80426-7}
  {\path{doi:10.1016/S0920-5632(03)80426-7}}.

\bibitem{Roesler:2000he}
S.~Roesler, R.~Engel, J.~Ranft, {The Monte Carlo event generator
  DPMJET-III}\href {http://arxiv.org/abs/hep-ph/0012252}
  {\path{arXiv:hep-ph/0012252}}.

\bibitem{Fedynitch:2015kcn}
A.~Fedynitch, {Cascade equations and hadronic interactions at very high
  energies}, Ph.D. thesis, KIT, Karlsruhe, Dept. Phys. (11 2015).
\newblock \href {http://dx.doi.org/10.5445/IR/1000055433}
  {\path{doi:10.5445/IR/1000055433}}.

\bibitem{Abbasi:2015fdr}
R.~U. Abbasi, et~al., {\bf Telescope Array} Collaboration, {Measurement of the
  proton-air cross section with Telescope Array\textquoteright{}s Middle Drum
  detector and surface array in hybrid mode}, Phys. Rev. D 92~(3) (2015)
  032007.
\newblock \href {http://arxiv.org/abs/1505.01860} {\path{arXiv:1505.01860}},
  \href {http://dx.doi.org/10.1103/PhysRevD.92.032007}
  {\path{doi:10.1103/PhysRevD.92.032007}}.

\bibitem{HiRes:1999ioa}
T.~Abu-Zayyad, et~al., {\bf HiRes, MIA} Collaboration, {Evidence for Changing
  of Cosmic Ray Composition between 10**17-eV and 10**18-eV from Multicomponent
  Measurements}, Phys. Rev. Lett. 84 (2000) 4276--4279.
\newblock \href {http://arxiv.org/abs/astro-ph/9911144}
  {\path{arXiv:astro-ph/9911144}}, \href
  {http://dx.doi.org/10.1103/PhysRevLett.84.4276}
  {\path{doi:10.1103/PhysRevLett.84.4276}}.

\bibitem{Cazon:2019mtd}
L.~Cazon, {Probing High-Energy Hadronic Interactions with Extensive Air
  Showers}, PoS ICRC2019 (2020) 005.
\newblock \href {http://arxiv.org/abs/1909.02962} {\path{arXiv:1909.02962}},
  \href {http://dx.doi.org/10.22323/1.358.0005}
  {\path{doi:10.22323/1.358.0005}}.

\bibitem{Apel:2017thr}
W.~D. Apel, et~al., {\bf KASCADE-Grande} Collaboration, {Probing the evolution
  of the EAS muon content in the atmosphere with KASCADE-Grande}, Astropart.
  Phys. 95 (2017) 25--43.
\newblock \href {http://arxiv.org/abs/1801.05513} {\path{arXiv:1801.05513}},
  \href
  {http://dx.doi.org/http://dx.doi.org/10.1016/j.astropartphys.2017.07.001}
  {\path{doi:http://dx.doi.org/10.1016/j.astropartphys.2017.07.001}}.

\bibitem{Bogdanov:2018sfw}
A.~G. Bogdanov, et~al., {\bf NEVOD-DECOR} Collaboration, {Investigation of very
  high energy cosmic rays by means of inclined muon bundles}, Astropart. Phys.
  98 (2018) 13--20.
\newblock \href
  {http://dx.doi.org/http://dx.doi.org/10.1016/j.astropartphys.2018.01.003}
  {\path{doi:http://dx.doi.org/10.1016/j.astropartphys.2018.01.003}}.

\bibitem{Glushkov}
A.~Glushkov, M.~Pravdin, A.~Sabourov, {\bf Yakutsk} Collaboration, Priv. Comm.

\bibitem{Fomin:2016kul}
Y.~A. Fomin, et~al., {\bf EAS-MSU} Collaboration, {No muon excess in extensive
  air showers at 100\textendash{}500 PeV primary energy: EAS\textendash{}MSU
  results}, Astropart. Phys. 92 (2017) 1--6.
\newblock \href {http://arxiv.org/abs/1609.05764} {\path{arXiv:1609.05764}},
  \href
  {http://dx.doi.org/http://dx.doi.org/10.1016/j.astropartphys.2017.04.001}
  {\path{doi:http://dx.doi.org/10.1016/j.astropartphys.2017.04.001}}.

\bibitem{Bellido:2018toz}
J.~A. Bellido, et~al., {\bf SUGAR} Collaboration, {Muon content of extensive
  air showers: comparison of the energy spectra obtained by the Sydney
  University Giant Air-shower Recorder and by the Pierre Auger Observatory},
  Phys. Rev. D 98~(2) (2018) 023014.
\newblock \href {http://arxiv.org/abs/1803.08662} {\path{arXiv:1803.08662}},
  \href {http://dx.doi.org/http://dx.doi.org/10.1103/PhysRevD.98.023014}
  {\path{doi:http://dx.doi.org/10.1103/PhysRevD.98.023014}}.

\bibitem{Gesualdi:2021yay}
F.~Gesualdi, et~al., {On the muon scale of air showers and its application to
  the AGASA data}, PoS ICRC2021 (2021) 473.
\newblock \href {http://arxiv.org/abs/2108.04824} {\path{arXiv:2108.04824}},
  \href {http://dx.doi.org/10.22323/1.395.0473}
  {\path{doi:10.22323/1.395.0473}}.

\bibitem{Dembinski:2017kpa}
H.~P. Dembinski, {Computing mean logarithmic mass from muon counts in air
  shower experiments}, Astropart. Phys. 102 (2018) 89--94.
\newblock \href {http://arxiv.org/abs/1711.05737} {\path{arXiv:1711.05737}},
  \href {http://dx.doi.org/10.1016/j.astropartphys.2018.05.008}
  {\path{doi:10.1016/j.astropartphys.2018.05.008}}.

\bibitem{Dembinski:2015xtn}
H.~P. Dembinski, J.~Gonzalez, {\bf IceCube} Collaboration, {Surface muons in
  IceTop}, PoS ICRC2015 (2016) 267.
\newblock \href {http://dx.doi.org/10.22323/1.236.0267}
  {\path{doi:10.22323/1.236.0267}}.

\bibitem{KASCADE-Grande:2017wfe}
W.~D. Apel, et~al., {\bf KASCADE-Grande} Collaboration, {Probing the evolution
  of the EAS muon content in the atmosphere with KASCADE-Grande}, Astropart.
  Phys. 95 (2017) 25--43.
\newblock \href {http://arxiv.org/abs/1801.05513} {\path{arXiv:1801.05513}},
  \href {http://dx.doi.org/10.1016/j.astropartphys.2017.07.001}
  {\path{doi:10.1016/j.astropartphys.2017.07.001}}.

\bibitem{Evans:2008zzb}
{LHC Machine}, JINST 3 (2008) S08001.
\newblock \href {http://dx.doi.org/10.1088/1748-0221/3/08/S08001}
  {\path{doi:10.1088/1748-0221/3/08/S08001}}.

\bibitem{Citron:2018lsq}
Z.~Citron, et~al., {Report from Working Group 5}: {Future physics opportunities
  for high-density QCD at the LHC with heavy-ion and proton beams}, CERN Yellow
  Rep. Monogr. 7 (2019) 1159--1410.
\newblock \href {http://arxiv.org/abs/1812.06772} {\path{arXiv:1812.06772}},
  \href {http://dx.doi.org/10.23731/CYRM-2019-007.1159}
  {\path{doi:10.23731/CYRM-2019-007.1159}}.

\bibitem{Bruce:2021hii}
R.~Bruce, et~al., {Performance and luminosity models for heavy-ion operation at
  the CERN Large Hadron Collider}, Eur. Phys. J. Plus 136~(7) (2021) 745.
\newblock \href {http://arxiv.org/abs/2107.09560} {\path{arXiv:2107.09560}},
  \href {http://dx.doi.org/10.1140/epjp/s13360-021-01685-5}
  {\path{doi:10.1140/epjp/s13360-021-01685-5}}.

\bibitem{French-Soviet:1975yfh}
P.~Beilliere, et~al., {\bf French-Soviet, CERN-Soviet} Collaboration, {Neutral
  K, Lambda and anti-Lambda Production in K- p and K+ p Interactions at
  32-GeV/c}, Nucl. Phys. B 90 (1975) 20--34.
\newblock \href {http://dx.doi.org/10.1016/0550-3213(75)90631-8}
  {\path{doi:10.1016/0550-3213(75)90631-8}}.

\bibitem{Brick:1981sy}
D.~Brick, et~al., {The Effective Energy Dependence of the Charged Particle's
  Multiplicity in $P / \pi^+ / K^+$ Interactions on Protons at 147-{GeV}/c},
  Phys. Lett. B 103 (1981) 241--246.
\newblock \href {http://dx.doi.org/10.1016/0370-2693(81)90750-4}
  {\path{doi:10.1016/0370-2693(81)90750-4}}.

\bibitem{DeWolf:1986mb}
E.~A. De~Wolf, et~al., {$K^+$ Fragmentation and Prompt Kaon Production in $K^+
  P$ Collisions at 70-{GeV}/c}, Z. Phys. C 31 (1986) 13--19.
\newblock \href {http://dx.doi.org/10.1007/BF01559587}
  {\path{doi:10.1007/BF01559587}}.

\bibitem{EHSNA22:1990otw}
I.~V. Ajinenko, et~al., {\bf EHS/NA22} Collaboration, {Neutral kaon production
  in K+ p and pi+ p interactions at 250-GeV/c}, Z. Phys. C 46 (1990) 525--536.
\newblock \href {http://dx.doi.org/10.1007/BF01560253}
  {\path{doi:10.1007/BF01560253}}.

\bibitem{Prado:2017hub}
R.~R. Prado, {\bf NA61/SHINE} Collaboration, {Measurements of Hadron Production
  in Pion-Carbon Interactions with NA61/SHINE at the CERN SPS}, PoS ICRC2017
  (2018) 315.
\newblock \href {http://arxiv.org/abs/1707.07902} {\path{arXiv:1707.07902}},
  \href {http://dx.doi.org/10.22323/1.301.0315}
  {\path{doi:10.22323/1.301.0315}}.

\bibitem{CDF:1993wpv}
F.~Abe, et~al., {\bf CDF} Collaboration, {Measurement of the $\bar{p}p$ Total
  Cross-Section at $\sqrt{s} = 546$ GeV and 1800 GeV}, Phys. Rev. D 50 (1994)
  5550--5561.
\newblock \href {http://dx.doi.org/10.1103/PhysRevD.50.5550}
  {\path{doi:10.1103/PhysRevD.50.5550}}.

\bibitem{E710:1991bcl}
N.~A. Amos, et~al., {\bf E710} Collaboration, {Measurement of $\rho$, the Ratio
  of the Real to Imaginary Part of the $\bar{p} p$ Forward Elastic Scattering
  Amplitude, at $\sqrt{s}$ = 1.8-TeV}, Phys. Rev. Lett. 68 (1992) 2433--2436.
\newblock \href {http://dx.doi.org/10.1103/PhysRevLett.68.2433}
  {\path{doi:10.1103/PhysRevLett.68.2433}}.

\bibitem{CMS:2015nfb}
V.~Khachatryan, et~al., {\bf CMS} Collaboration, {Measurement of the inelastic
  cross section in proton\textendash{}lead collisions at $\sqrt {s_{NN}}=$ 5.02
  TeV}, Phys. Lett. B 759 (2016) 641--662.
\newblock \href {http://arxiv.org/abs/1509.03893} {\path{arXiv:1509.03893}},
  \href {http://dx.doi.org/10.1016/j.physletb.2016.06.027}
  {\path{doi:10.1016/j.physletb.2016.06.027}}.

\bibitem{ALICE:2017pcy}
S.~Acharya, et~al., {\bf ALICE} Collaboration, {Charged-particle multiplicity
  distributions over a wide pseudorapidity range in proton-proton collisions at
  $\sqrt{s}=$ 0.9, 7, and 8 TeV}, Eur. Phys. J. C 77~(12) (2017) 852.
\newblock \href {http://arxiv.org/abs/1708.01435} {\path{arXiv:1708.01435}},
  \href {http://dx.doi.org/10.1140/epjc/s10052-017-5412-6}
  {\path{doi:10.1140/epjc/s10052-017-5412-6}}.

\bibitem{LHCb:2021abm}
R.~Aaij, et~al., {\bf LHCb} Collaboration, {Measurement of prompt
  charged-particle production in proton-proton collisions at a centre-of-mass
  energy of 13 TeV}\href {http://arxiv.org/abs/2107.10090}
  {\path{arXiv:2107.10090}}.

\bibitem{LHCb:2021vww}
R.~Aaij, et~al., {\bf LHCb} Collaboration, {Measurement of the nuclear
  modification factor and prompt charged particle production in $p\mathrm{Pb}$
  and $pp$ collisions at
  $\sqrt{s_{\scriptscriptstyle\mathrm{NN}}}=5\,\mathrm{TeV}$}\href
  {http://arxiv.org/abs/2108.13115} {\path{arXiv:2108.13115}}.

\bibitem{CMS:2014kix}
S.~Chatrchyan, et~al., {\bf CMS, TOTEM} Collaboration, {Measurement of
  pseudorapidity distributions of charged particles in proton-proton collisions
  at $\sqrt{s}$ = 8 TeV by the CMS and TOTEM experiments}, Eur. Phys. J. C
  74~(10) (2014) 3053.
\newblock \href {http://arxiv.org/abs/1405.0722} {\path{arXiv:1405.0722}},
  \href {http://dx.doi.org/10.1140/epjc/s10052-014-3053-6}
  {\path{doi:10.1140/epjc/s10052-014-3053-6}}.

\bibitem{LHCb:2012gpm}
R.~Aaij, et~al., {\bf LHCb} Collaboration, {Measurement of the forward energy
  flow in $pp$ collisions at $\sqrt{s}=7$ TeV}, Eur. Phys. J. C 73 (2013) 2421.
\newblock \href {http://arxiv.org/abs/1212.4755} {\path{arXiv:1212.4755}},
  \href {http://dx.doi.org/10.1140/epjc/s10052-013-2421-y}
  {\path{doi:10.1140/epjc/s10052-013-2421-y}}.

\bibitem{CMS:2011xjg}
S.~Chatrchyan, et~al., {\bf CMS} Collaboration, {Measurement of energy flow at
  large pseudorapidities in $pp$ collisions at $\sqrt{s} = 0.9$ and 7 TeV},
  JHEP 11 (2011) 148, [Erratum: JHEP 02, 055 (2012)].
\newblock \href {http://arxiv.org/abs/1110.0211} {\path{arXiv:1110.0211}},
  \href {http://dx.doi.org/10.1007/JHEP11(2011)148}
  {\path{doi:10.1007/JHEP11(2011)148}}.

\bibitem{CMS:2017dou}
A.~M. Sirunyan, et~al., {\bf CMS} Collaboration, {Measurement of the inclusive
  energy spectrum in the very forward direction in proton-proton collisions at
  $ \sqrt{s}=13 $ TeV}, JHEP 08 (2017) 046.
\newblock \href {http://arxiv.org/abs/1701.08695} {\path{arXiv:1701.08695}},
  \href {http://dx.doi.org/10.1007/JHEP08(2017)046}
  {\path{doi:10.1007/JHEP08(2017)046}}.

\bibitem{CMS:2018lqt}
A.~M. Sirunyan, et~al., {\bf CMS} Collaboration, {Measurement of the energy
  density as a function of pseudorapidity in proton-proton collisions at
  $\sqrt{s} =$ 13 TeV}, Eur. Phys. J. C 79~(5) (2019) 391.
\newblock \href {http://arxiv.org/abs/1812.04095} {\path{arXiv:1812.04095}},
  \href {http://dx.doi.org/10.1140/epjc/s10052-019-6861-x}
  {\path{doi:10.1140/epjc/s10052-019-6861-x}}.

\bibitem{CMS:2019kap}
A.~M. Sirunyan, et~al., {\bf CMS} Collaboration, {Measurement of the average
  very forward energy as a function of the track multiplicity at central
  pseudorapidities in proton-proton collisions at $\sqrt{s}=13\,\mathrm {TeV}
  $}, Eur. Phys. J. C 79~(11) (2019) 893.
\newblock \href {http://arxiv.org/abs/1908.01750} {\path{arXiv:1908.01750}},
  \href {http://dx.doi.org/10.1140/epjc/s10052-019-7402-3}
  {\path{doi:10.1140/epjc/s10052-019-7402-3}}.

\bibitem{LHCf:2020hjf}
O.~Adriani, et~al., {\bf LHCf} Collaboration, {Measurement of energy flow,
  cross section and average inelasticity of forward neutrons produced in $
  \sqrt{s} $ = 13 TeV proton-proton collisions with the LHCf Arm2 detector},
  JHEP 07 (2020) 016.
\newblock \href {http://arxiv.org/abs/2003.02192} {\path{arXiv:2003.02192}},
  \href {http://dx.doi.org/10.1007/JHEP07(2020)016}
  {\path{doi:10.1007/JHEP07(2020)016}}.

\bibitem{LHCf:2011hln}
O.~Adriani, et~al., {\bf LHCf} Collaboration, {Measurement of zero degree
  single photon energy spectra for $\sqrt{s} =$ 7 TeV proton-proton collisions
  at LHC}, Phys. Lett. B 703 (2011) 128--134.
\newblock \href {http://arxiv.org/abs/1104.5294} {\path{arXiv:1104.5294}},
  \href {http://dx.doi.org/10.1016/j.physletb.2011.07.077}
  {\path{doi:10.1016/j.physletb.2011.07.077}}.

\bibitem{LHCf:2012stt}
O.~Adriani, et~al., {\bf LHCf} Collaboration, {Measurement of zero degree
  inclusive photon energy spectra for $\sqrt{s}=$ 900 GeV proton-proton
  collisions at LHC}, Phys. Lett. B 715 (2012) 298--303.
\newblock \href {http://arxiv.org/abs/1207.7183} {\path{arXiv:1207.7183}},
  \href {http://dx.doi.org/10.1016/j.physletb.2012.07.065}
  {\path{doi:10.1016/j.physletb.2012.07.065}}.

\bibitem{LHCf:2012mtr}
O.~Adriani, et~al., {\bf LHCf} Collaboration, {Measurement of forward neutral
  pion transverse momentum spectra for $\sqrt{s}$ = 7TeV proton-proton
  collisions at LHC}, Phys. Rev. D 86 (2012) 092001.
\newblock \href {http://arxiv.org/abs/1205.4578} {\path{arXiv:1205.4578}},
  \href {http://dx.doi.org/10.1103/PhysRevD.86.092001}
  {\path{doi:10.1103/PhysRevD.86.092001}}.

\bibitem{LHCf:2014gqm}
O.~Adriani, et~al., {\bf LHCf} Collaboration, {Transverse-momentum distribution
  and nuclear modification factor for neutral pions in the forward-rapidity
  region in proton-lead collisions at $\sqrt{s_{NN}} = 5.02$ TeV}, Phys. Rev. C
  89~(6) (2014) 065209.
\newblock \href {http://arxiv.org/abs/1403.7845} {\path{arXiv:1403.7845}},
  \href {http://dx.doi.org/10.1103/PhysRevC.89.065209}
  {\path{doi:10.1103/PhysRevC.89.065209}}.

\bibitem{LHCf:2015nel}
O.~Adriani, et~al., {\bf LHCf} Collaboration, {Measurement of very forward
  neutron energy spectra for 7 TeV proton\textendash{}proton collisions at the
  Large Hadron Collider}, Phys. Lett. B 750 (2015) 360--366.
\newblock \href {http://arxiv.org/abs/1503.03505} {\path{arXiv:1503.03505}},
  \href {http://dx.doi.org/10.1016/j.physletb.2015.09.041}
  {\path{doi:10.1016/j.physletb.2015.09.041}}.

\bibitem{LHCf:2015rcj}
O.~Adriani, et~al., {\bf LHCf} Collaboration, {Measurements of longitudinal and
  transverse momentum distributions for neutral pions in the forward-rapidity
  region with the LHCf detector}, Phys. Rev. D 94~(3) (2016) 032007.
\newblock \href {http://arxiv.org/abs/1507.08764} {\path{arXiv:1507.08764}},
  \href {http://dx.doi.org/10.1103/PhysRevD.94.032007}
  {\path{doi:10.1103/PhysRevD.94.032007}}.

\bibitem{LHCf:2018gbv}
O.~Adriani, et~al., {\bf LHCf} Collaboration, {Measurement of inclusive forward
  neutron production cross section in proton-proton collisions at $ \sqrt{s}=13
  $ TeV with the LHCf Arm2 detector}, JHEP 11 (2018) 073.
\newblock \href {http://arxiv.org/abs/1808.09877} {\path{arXiv:1808.09877}},
  \href {http://dx.doi.org/10.1007/JHEP11(2018)073}
  {\path{doi:10.1007/JHEP11(2018)073}}.

\bibitem{ALICE:2016fzo}
J.~Adam, et~al., {\bf ALICE} Collaboration, {Enhanced production of
  multi-strange hadrons in high-multiplicity proton-proton collisions}, Nature
  Phys. 13 (2017) 535--539.
\newblock \href {http://arxiv.org/abs/1606.07424} {\path{arXiv:1606.07424}},
  \href {http://dx.doi.org/10.1038/nphys4111} {\path{doi:10.1038/nphys4111}}.

\bibitem{Vasileiou:2020rov}
M.~Vasileiou, {\bf ALICE} Collaboration, {Strangeness production with ALICE at
  the LHC}, Phys. Scripta 95~(6) (2020) 064007.
\newblock \href {http://dx.doi.org/10.1088/1402-4896/ab85fc}
  {\path{doi:10.1088/1402-4896/ab85fc}}.

\bibitem{Baur:2019cpv}
S.~Baur, et~al., {Core-corona effect in hadron collisions and muon production
  in air showers}\href {http://arxiv.org/abs/1902.09265}
  {\path{arXiv:1902.09265}}.

\bibitem{Prado:2018wsv}
R.~R. Prado, {\bf NA61/SHINE} Collaboration, {Recent results from the cosmic
  ray program of the NA61/SHINE experiment}, EPJ Web Conf. 208 (2019) 05006.
\newblock \href {http://arxiv.org/abs/1810.00642} {\path{arXiv:1810.00642}},
  \href {http://dx.doi.org/10.1051/epjconf/201920805006}
  {\path{doi:10.1051/epjconf/201920805006}}.

\bibitem{Unger:2019nus}
M.~Unger, {\bf NA61/SHINE} Collaboration, {New Results from the Cosmic-Ray
  Program of the NA61/SHINE facility at the CERN SPS}, PoS ICRC2019 (2020) 446.
\newblock \href {http://arxiv.org/abs/1909.07136} {\path{arXiv:1909.07136}},
  \href {http://dx.doi.org/10.22323/1.358.0446}
  {\path{doi:10.22323/1.358.0446}}.

\bibitem{Meurer:2005dt}
C.~Meurer, et~al., {Muon production in extensive air showers and its relation
  to hadronic interactions}, Czech. J. Phys. 56 (2006) A211.
\newblock \href {http://arxiv.org/abs/astro-ph/0512536}
  {\path{arXiv:astro-ph/0512536}}, \href
  {http://dx.doi.org/10.1007/s10582-006-0156-9}
  {\path{doi:10.1007/s10582-006-0156-9}}.

\bibitem{NA61:2014lfx}
N.~Abgrall, et~al., {\bf NA61} Collaboration, {NA61/SHINE facility at the CERN
  SPS: beams and detector system}, JINST 9 (2014) P06005.
\newblock \href {http://arxiv.org/abs/1401.4699} {\path{arXiv:1401.4699}},
  \href {http://dx.doi.org/10.1088/1748-0221/9/06/P06005}
  {\path{doi:10.1088/1748-0221/9/06/P06005}}.

\bibitem{NA61SHINE:2017vqs}
A.~Aduszkiewicz, et~al., {\bf NA61/SHINE} Collaboration, {Measurement of meson
  resonance production in $\pi ^-+$ C interactions at SPS energies}, Eur. Phys.
  J. C 77~(9) (2017) 626.
\newblock \href {http://arxiv.org/abs/1705.08206} {\path{arXiv:1705.08206}},
  \href {http://dx.doi.org/10.1140/epjc/s10052-017-5184-z}
  {\path{doi:10.1140/epjc/s10052-017-5184-z}}.

\bibitem{LHCb:2014vhh}
R.~Aaij, et~al., {\bf LHCb} Collaboration, {Precision luminosity measurements
  at LHCb}, JINST 9~(12) (2014) P12005.
\newblock \href {http://arxiv.org/abs/1410.0149} {\path{arXiv:1410.0149}},
  \href {http://dx.doi.org/10.1088/1748-0221/9/12/P12005}
  {\path{doi:10.1088/1748-0221/9/12/P12005}}.

\bibitem{LHCb:2018jry}
R.~Aaij, et~al., {\bf LHCb} Collaboration, {First Measurement of Charm
  Production in its Fixed-Target Configuration at the LHC}, Phys. Rev. Lett.
  122~(13) (2019) 132002.
\newblock \href {http://arxiv.org/abs/1810.07907} {\path{arXiv:1810.07907}},
  \href {http://dx.doi.org/10.1103/PhysRevLett.122.132002}
  {\path{doi:10.1103/PhysRevLett.122.132002}}.

\bibitem{LHCb:2018ygc}
R.~Aaij, et~al., {\bf LHCb} Collaboration, {Measurement of Antiproton
  Production in ${\rm p He}$ Collisions at $\sqrt{s_{NN}}=110$ GeV}, Phys. Rev.
  Lett. 121~(22) (2018) 222001.
\newblock \href {http://arxiv.org/abs/1808.06127} {\path{arXiv:1808.06127}},
  \href {http://dx.doi.org/10.1103/PhysRevLett.121.222001}
  {\path{doi:10.1103/PhysRevLett.121.222001}}.

\bibitem{Farrar:2013sfa}
G.~R. Farrar, J.~D. Allen, {A new physical phenomenon in ultra-high energy
  collisions}, EPJ Web Conf. 53 (2013) 07007.
\newblock \href {http://arxiv.org/abs/1307.2322} {\path{arXiv:1307.2322}},
  \href {http://dx.doi.org/10.1051/epjconf/20135307007}
  {\path{doi:10.1051/epjconf/20135307007}}.

\bibitem{Planck:2018nkj}
N.~Aghanim, et~al., {\bf Planck} Collaboration, {Planck 2018 results. I.
  Overview and the cosmological legacy of Planck}, Astron. Astrophys. 641
  (2020) A1.
\newblock \href {http://arxiv.org/abs/1807.06205} {\path{arXiv:1807.06205}},
  \href {http://dx.doi.org/10.1051/0004-6361/201833880}
  {\path{doi:10.1051/0004-6361/201833880}}.

\bibitem{Jungman:1995df}
G.~Jungman, M.~Kamionkowski, K.~Griest, {Supersymmetric dark matter}, Phys.
  Rept. 267 (1996) 195--373.
\newblock \href {http://arxiv.org/abs/hep-ph/9506380}
  {\path{arXiv:hep-ph/9506380}}, \href
  {http://dx.doi.org/10.1016/0370-1573(95)00058-5}
  {\path{doi:10.1016/0370-1573(95)00058-5}}.

\bibitem{MarrodanUndagoitia:2015veg}
T.~Marrod\'an~Undagoitia, L.~Rauch, {Dark matter direct-detection experiments},
  J. Phys. G 43~(1) (2016) 013001.
\newblock \href {http://arxiv.org/abs/1509.08767} {\path{arXiv:1509.08767}},
  \href {http://dx.doi.org/10.1088/0954-3899/43/1/013001}
  {\path{doi:10.1088/0954-3899/43/1/013001}}.

\bibitem{Gaskins:2016cha}
J.~M. Gaskins, {A review of indirect searches for particle dark matter},
  Contemp. Phys. 57~(4) (2016) 496--525.
\newblock \href {http://arxiv.org/abs/1604.00014} {\path{arXiv:1604.00014}},
  \href {http://dx.doi.org/10.1080/00107514.2016.1175160}
  {\path{doi:10.1080/00107514.2016.1175160}}.

\bibitem{Rappoccio:2018qxp}
S.~Rappoccio, {The experimental status of direct searches for exotic physics
  beyond the standard model at the Large Hadron Collider}, Rev. Phys. 4 (2019)
  100027.
\newblock \href {http://arxiv.org/abs/1810.10579} {\path{arXiv:1810.10579}},
  \href {http://dx.doi.org/10.1016/j.revip.2018.100027}
  {\path{doi:10.1016/j.revip.2018.100027}}.

\bibitem{Ellis:1983ew}
J.~R. Ellis, J.~S. Hagelin, D.~V. Nanopoulos, K.~A. Olive, M.~Srednicki,
  {Supersymmetric Relics from the Big Bang}, Nucl. Phys. B 238 (1984) 453--476.
\newblock \href {http://dx.doi.org/10.1016/0550-3213(84)90461-9}
  {\path{doi:10.1016/0550-3213(84)90461-9}}.

\bibitem{Nanopoulos:1982bv}
D.~V. Nanopoulos, K.~A. Olive, M.~Srednicki, K.~Tamvakis, {Primordial Inflation
  in Simple Supergravity}, Phys. Lett. B 123 (1983) 41--44.
\newblock \href {http://dx.doi.org/10.1016/0370-2693(83)90954-1}
  {\path{doi:10.1016/0370-2693(83)90954-1}}.

\bibitem{Khlopov:1984pf}
M.~Y. Khlopov, A.~D. Linde, {Is It Easy to Save the Gravitino?}, Phys. Lett. B
  138 (1984) 265--268.
\newblock \href {http://dx.doi.org/10.1016/0370-2693(84)91656-3}
  {\path{doi:10.1016/0370-2693(84)91656-3}}.

\bibitem{Olive:1984bi}
K.~A. Olive, D.~N. Schramm, M.~Srednicki, {Gravitinos as the Cold Dark Matter
  in an $\Omega$ = 1 Universe}, Nucl. Phys. B 255 (1985) 495--504.
\newblock \href {http://dx.doi.org/10.1016/0550-3213(85)90149-X}
  {\path{doi:10.1016/0550-3213(85)90149-X}}.

\bibitem{Ellis:1990iu}
J.~R. Ellis, J.~L. Lopez, D.~V. Nanopoulos, {Confinement of fractional charges
  yields integer charged relics in string models}, Phys. Lett. B 247 (1990)
  257--264.
\newblock \href {http://dx.doi.org/10.1016/0370-2693(90)90893-B}
  {\path{doi:10.1016/0370-2693(90)90893-B}}.

\bibitem{Ellis:1990nb}
J.~R. Ellis, G.~Gelmini, J.~L. Lopez, D.~V. Nanopoulos, S.~Sarkar,
  {Astrophysical constraints on massive unstable neutral relic particles},
  Nucl. Phys. B 373 (1992) 399--437.
\newblock \href {http://dx.doi.org/10.1016/0550-3213(92)90438-H}
  {\path{doi:10.1016/0550-3213(92)90438-H}}.

\bibitem{Berezinsky:1997hy}
V.~Berezinsky, M.~Kachelriess, A.~Vilenkin, {Ultrahigh-energy cosmic rays
  without GZK cutoff}, Phys. Rev. Lett. 79 (1997) 4302--4305.
\newblock \href {http://arxiv.org/abs/astro-ph/9708217}
  {\path{arXiv:astro-ph/9708217}}, \href
  {http://dx.doi.org/10.1103/PhysRevLett.79.4302}
  {\path{doi:10.1103/PhysRevLett.79.4302}}.

\bibitem{Chung:1998zb}
D.~J. Chung, E.~W. Kolb, A.~Riotto, {Superheavy dark matter}, Phys. Rev. D 59
  (1998) 023501.
\newblock \href {http://arxiv.org/abs/hep-ph/9802238}
  {\path{arXiv:hep-ph/9802238}}, \href
  {http://dx.doi.org/10.1103/PhysRevD.59.023501}
  {\path{doi:10.1103/PhysRevD.59.023501}}.

\bibitem{Kuzmin:1997jua}
V.~Kuzmin, V.~Rubakov, {Ultrahigh-energy cosmic rays: A Window to
  postinflationary reheating epoch of the universe?}, Phys. Atom. Nucl. 61
  (1998) 1028.
\newblock \href {http://arxiv.org/abs/astro-ph/9709187}
  {\path{arXiv:astro-ph/9709187}}.

\bibitem{Birkel:1998nx}
M.~Birkel, S.~Sarkar, {Extremely high-energy cosmic rays from relic particle
  decays}, Astropart. Phys. 9 (1998) 297--309.
\newblock \href {http://arxiv.org/abs/hep-ph/9804285}
  {\path{arXiv:hep-ph/9804285}}, \href
  {http://dx.doi.org/10.1016/S0927-6505(98)00028-0}
  {\path{doi:10.1016/S0927-6505(98)00028-0}}.

\bibitem{Berezinsky:1998ft}
V.~Berezinsky, P.~Blasi, A.~Vilenkin, {Ultrahigh-energy gamma-rays as signature
  of topological defects}, Phys. Rev. D 58 (1998) 103515.
\newblock \href {http://arxiv.org/abs/astro-ph/9803271}
  {\path{arXiv:astro-ph/9803271}}, \href
  {http://dx.doi.org/10.1103/PhysRevD.58.103515}
  {\path{doi:10.1103/PhysRevD.58.103515}}.

\bibitem{Garny:2015sjg}
M.~Garny, M.~Sandora, M.~S. Sloth, {Planckian Interacting Massive Particles as
  Dark Matter}, Phys. Rev. Lett. 116~(10) (2016) 101302.
\newblock \href {http://arxiv.org/abs/1511.03278} {\path{arXiv:1511.03278}},
  \href {http://dx.doi.org/10.1103/PhysRevLett.116.101302}
  {\path{doi:10.1103/PhysRevLett.116.101302}}.

\bibitem{Ellis:2015jpg}
J.~Ellis, M.~A.~G. Garcia, D.~V. Nanopoulos, K.~A. Olive, M.~Peloso,
  {Post-Inflationary Gravitino Production Revisited}, JCAP 03 (2016) 008.
\newblock \href {http://arxiv.org/abs/1512.05701} {\path{arXiv:1512.05701}},
  \href {http://dx.doi.org/10.1088/1475-7516/2016/03/008}
  {\path{doi:10.1088/1475-7516/2016/03/008}}.

\bibitem{Dudas:2017rpa}
E.~Dudas, Y.~Mambrini, K.~Olive, {Case for an EeV Gravitino}, Phys. Rev. Lett.
  119~(5) (2017) 051801.
\newblock \href {http://arxiv.org/abs/1704.03008} {\path{arXiv:1704.03008}},
  \href {http://dx.doi.org/10.1103/PhysRevLett.119.051801}
  {\path{doi:10.1103/PhysRevLett.119.051801}}.

\bibitem{Kaneta:2019zgw}
K.~Kaneta, Y.~Mambrini, K.~A. Olive, {Radiative production of nonthermal dark
  matter}, Phys. Rev. D 99~(6) (2019) 063508.
\newblock \href {http://arxiv.org/abs/1901.04449} {\path{arXiv:1901.04449}},
  \href {http://dx.doi.org/10.1103/PhysRevD.99.063508}
  {\path{doi:10.1103/PhysRevD.99.063508}}.

\bibitem{Mambrini:2021zpp}
Y.~Mambrini, K.~A. Olive, {Gravitational Production of Dark Matter during
  Reheating}, Phys. Rev. D 103~(11) (2021) 115009.
\newblock \href {http://arxiv.org/abs/2102.06214} {\path{arXiv:2102.06214}},
  \href {http://dx.doi.org/10.1103/PhysRevD.103.115009}
  {\path{doi:10.1103/PhysRevD.103.115009}}.

\bibitem{Buttazzo:2013uya}
D.~Buttazzo, G.~Degrassi, P.~P. Giardino, G.~F. Giudice, F.~Sala, A.~Salvio,
  A.~Strumia, {Investigating the near-criticality of the Higgs boson}, JHEP 12
  (2013) 089.
\newblock \href {http://arxiv.org/abs/1307.3536} {\path{arXiv:1307.3536}},
  \href {http://dx.doi.org/10.1007/JHEP12(2013)089}
  {\path{doi:10.1007/JHEP12(2013)089}}.

\bibitem{Alekhin:2012py}
S.~Alekhin, A.~Djouadi, S.~Moch, {The top quark and Higgs boson masses and the
  stability of the electroweak vacuum}, Phys. Lett. B 716 (2012) 214--219.
\newblock \href {http://arxiv.org/abs/1207.0980} {\path{arXiv:1207.0980}},
  \href {http://dx.doi.org/10.1016/j.physletb.2012.08.024}
  {\path{doi:10.1016/j.physletb.2012.08.024}}.

\bibitem{Bednyakov:2015sca}
A.~V. Bednyakov, B.~A. Kniehl, A.~F. Pikelner, O.~L. Veretin, {Stability of the
  Electroweak Vacuum: Gauge Independence and Advanced Precision}, Phys. Rev.
  Lett. 115~(20) (2015) 201802.
\newblock \href {http://arxiv.org/abs/1507.08833} {\path{arXiv:1507.08833}},
  \href {http://dx.doi.org/10.1103/PhysRevLett.115.201802}
  {\path{doi:10.1103/PhysRevLett.115.201802}}.

\bibitem{Markkanen:2018pdo}
T.~Markkanen, A.~Rajantie, S.~Stopyra, {Cosmological Aspects of Higgs Vacuum
  Metastability}, Front. Astron. Space Sci. 5 (2018) 40.
\newblock \href {http://arxiv.org/abs/1809.06923} {\path{arXiv:1809.06923}},
  \href {http://dx.doi.org/10.3389/fspas.2018.00040}
  {\path{doi:10.3389/fspas.2018.00040}}.

\bibitem{Berat:2022iea}
C.~B\'erat, C.~Bleve, O.~Deligny, F.~Montanet, P.~Savina, Z.~Torr\`es, {Diffuse
  flux of ultra-high energy photons from cosmic-ray interactions in the disk of
  the Galaxy and implications for the search for decaying super-heavy dark
  matter}\href {http://arxiv.org/abs/2203.08751} {\path{arXiv:2203.08751}}.

\bibitem{ParticleDataGroup:2020ssz}
P.~A. Zyla, et~al., {\bf Particle Data Group} Collaboration, {Review of
  Particle Physics}, PTEP 2020~(8) (2020) 083C01.
\newblock \href {http://dx.doi.org/10.1093/ptep/ptaa104}
  {\path{doi:10.1093/ptep/ptaa104}}.

\bibitem{Cazon:2020jla}
L.~Cazon, R.~Concei\c{c}\~ao, M.~A. Martins, F.~Riehn, {Constraining the energy
  spectrum of neutral pions in ultra-high-energy proton-air interactions},
  Phys. Rev. D 103~(2) (2021) 022001.
\newblock \href {http://arxiv.org/abs/2006.11303} {\path{arXiv:2006.11303}},
  \href {http://dx.doi.org/10.1103/PhysRevD.103.022001}
  {\path{doi:10.1103/PhysRevD.103.022001}}.

\bibitem{Cazon:2018gww}
L.~Cazon, R.~Concei\c{c}\~ao, F.~Riehn, {Probing the energy spectrum of hadrons
  in proton air interactions at ultrahigh energies through the fluctuations of
  the muon content of extensive air showers}, Phys. Lett. B 784 (2018) 68--76.
\newblock \href {http://arxiv.org/abs/1803.05699} {\path{arXiv:1803.05699}},
  \href {http://dx.doi.org/10.1016/j.physletb.2018.07.026}
  {\path{doi:10.1016/j.physletb.2018.07.026}}.

\bibitem{PierreAuger:2021ece}
P.~Abreu, et~al., {\bf Pierre Auger} Collaboration, {Expected performance of
  the AugerPrime Radio Detector}, PoS ICRC2021 (2021) 262.
\newblock \href {http://dx.doi.org/10.22323/1.395.0262}
  {\path{doi:10.22323/1.395.0262}}.

\bibitem{Decoene:2019sgx}
V.~Decoene, {\bf GRAND} Collaboration, {GRANDProto300 experiment: a pathfinder
  with rich astroparticle and radio-astronomy science case}, PoS ICRC2019
  (2020) 233.
\newblock \href {http://arxiv.org/abs/1909.04893} {\path{arXiv:1909.04893}},
  \href {http://dx.doi.org/10.22323/1.358.0233}
  {\path{doi:10.22323/1.358.0233}}.

\bibitem{Glaser:2017fgg}
C.~Glaser, {\bf Pierre Auger} Collaboration, {A novel method for the absolute
  energy calibration of large-scale cosmic-ray detectors using radio emission
  of extensive air showers}, in: {52nd Rencontres de Moriond on Very High
  Energy Phenomena in the Universe}, 2017, pp. 55--64.
\newblock \href {http://arxiv.org/abs/1706.01451} {\path{arXiv:1706.01451}}.

\bibitem{Goosthesis}
I.~Goos, {{Study of the $X_{\rm max}$ - $N_\mu$ anticorrelation to infer
  physical properties of high-energy hadronic interactions}}, Ph.D. thesis,
  KIT, Karlsruhe, Dept. Phys. (02 2022).

\bibitem{Ostapchenko:2016bir}
S.~Ostapchenko, M.~Bleicher, {Constraining pion interactions at very high
  energies by cosmic ray data}, Phys. Rev. D 93~(5) (2016) 051501.
\newblock \href {http://arxiv.org/abs/1601.06567} {\path{arXiv:1601.06567}},
  \href {http://dx.doi.org/10.1103/PhysRevD.93.051501}
  {\path{doi:10.1103/PhysRevD.93.051501}}.

\bibitem{Pierog:2015ifw}
T.~Pierog, B.~Guiot, K.~Werner, {Air Shower Development, pion interactions and
  modified EPOS Model}, PoS ICRC2015 (2016) 337.
\newblock \href {http://dx.doi.org/10.22323/1.236.0337}
  {\path{doi:10.22323/1.236.0337}}.

\bibitem{Ostapchenko:2019ubd}
S.~Ostapchenko, {High energy cosmic ray interactions and UHECR composition
  problem}, EPJ Web Conf. 210 (2019) 02001.
\newblock \href {http://dx.doi.org/10.1051/epjconf/201921002001}
  {\path{doi:10.1051/epjconf/201921002001}}.

\bibitem{Haungs:2019ylq}
A.~Haungs, {\bf IceCube} Collaboration, {A Scintillator and Radio Enhancement
  of the IceCube Surface Detector Array}, EPJ Web Conf. 210 (2019) 06009.
\newblock \href {http://arxiv.org/abs/1903.04117} {\path{arXiv:1903.04117}},
  \href {http://dx.doi.org/10.1051/epjconf/201921006009}
  {\path{doi:10.1051/epjconf/201921006009}}.

\bibitem{Espadanal:2016jse}
J.~Espadanal, L.~Cazon, R.~Concei\c{c}\~ao, {Sensitivity of EAS measurements to
  the energy spectrum of muons}, Astropart. Phys. 86 (2017) 32--40.
\newblock \href {http://arxiv.org/abs/1607.06760} {\path{arXiv:1607.06760}},
  \href {http://dx.doi.org/10.1016/j.astropartphys.2016.11.003}
  {\path{doi:10.1016/j.astropartphys.2016.11.003}}.

\bibitem{FCC:2018bvk}
A.~Abada, et~al., {\bf FCC} Collaboration, {HE-LHC: The High-Energy Large
  Hadron Collider}: {Future Circular Collider Conceptual Design Report Volume
  4}, Eur. Phys. J. ST 228~(5) (2019) 1109--1382.
\newblock \href {http://dx.doi.org/10.1140/epjst/e2019-900088-6}
  {\path{doi:10.1140/epjst/e2019-900088-6}}.

\bibitem{LHCf:2021qrz}
H.~Menjo, et~al., {\bf LHCf, RHICf} Collaboration, {Status and Prospects of the
  LHCf and RHICf experiments}, PoS ICRC2021 (2021) 301.
\newblock \href {http://dx.doi.org/10.22323/1.395.0301}
  {\path{doi:10.22323/1.395.0301}}.

\bibitem{Barschel:2020drr}
C.~Barschel, et~al., {LHC fixed target experiments : Report from the LHC Fixed
  Target Working Group of the CERN Physics Beyond Colliders Forum}, Vol. 4/2020
  of CERN Yellow Reports: Monographs, CERN, Geneva, 2020.
\newblock \href {http://dx.doi.org/10.23731/CYRM-2020-004}
  {\path{doi:10.23731/CYRM-2020-004}}.

\bibitem{Feng:2017uoz}
J.~L. Feng, I.~Galon, F.~Kling, S.~Trojanowski, {ForwArd Search ExpeRiment at
  the LHC}, Phys. Rev. D 97~(3) (2018) 035001.
\newblock \href {http://arxiv.org/abs/1708.09389} {\path{arXiv:1708.09389}},
  \href {http://dx.doi.org/10.1103/PhysRevD.97.035001}
  {\path{doi:10.1103/PhysRevD.97.035001}}.

\bibitem{FASER:2018ceo}
A.~Ariga, et~al., {\bf FASER} Collaboration, {Letter of Intent for FASER:
  ForwArd Search ExpeRiment at the LHC}\href {http://arxiv.org/abs/1811.10243}
  {\path{arXiv:1811.10243}}.

\bibitem{FASER:2018bac}
A.~Ariga, et~al., {\bf FASER} Collaboration, {Technical Proposal for FASER:
  ForwArd Search ExpeRiment at the LHC}\href {http://arxiv.org/abs/1812.09139}
  {\path{arXiv:1812.09139}}.

\bibitem{FASER:2018eoc}
A.~Ariga, et~al., {\bf FASER} Collaboration, {FASER\textquoteright{}s physics
  reach for long-lived particles}, Phys. Rev. D 99~(9) (2019) 095011.
\newblock \href {http://arxiv.org/abs/1811.12522} {\path{arXiv:1811.12522}},
  \href {http://dx.doi.org/10.1103/PhysRevD.99.095011}
  {\path{doi:10.1103/PhysRevD.99.095011}}.

\bibitem{FASER:2020gpr}
H.~Abreu, et~al., {\bf FASER} Collaboration, {Technical Proposal: FASERnu}\href
  {http://arxiv.org/abs/2001.03073} {\path{arXiv:2001.03073}}.

\bibitem{Abreu:2019yak}
H.~Abreu, et~al., {\bf FASER} Collaboration, {Detecting and Studying
  High-Energy Collider Neutrinos with FASER at the LHC}, Eur. Phys. J. C 80~(1)
  (2020) 61.
\newblock \href {http://arxiv.org/abs/1908.02310} {\path{arXiv:1908.02310}},
  \href {http://dx.doi.org/10.1140/epjc/s10052-020-7631-5}
  {\path{doi:10.1140/epjc/s10052-020-7631-5}}.

\bibitem{FASER:2021mtu}
H.~Abreu, et~al., {\bf FASER} Collaboration, {First neutrino interaction
  candidates at the LHC}, Phys. Rev. D 104~(9) (2021) L091101.
\newblock \href {http://arxiv.org/abs/2105.06197} {\path{arXiv:2105.06197}},
  \href {http://dx.doi.org/10.1103/PhysRevD.104.L091101}
  {\path{doi:10.1103/PhysRevD.104.L091101}}.

\bibitem{Anchordoqui:2022fpn}
L.~A. Anchordoqui, C.~G. Canal, F.~Kling, S.~J. Sciutto, J.~F. Soriano, {An
  explanation of the muon puzzle of ultrahigh-energy cosmic rays and the role
  of the Forward Physics Facility for model improvement}, JHEAp 34 (2022)
  19--32.
\newblock \href {http://arxiv.org/abs/2202.03095} {\path{arXiv:2202.03095}},
  \href {http://dx.doi.org/10.1016/j.jheap.2022.03.004}
  {\path{doi:10.1016/j.jheap.2022.03.004}}.

\bibitem{Zyla:2020zbs}
P.~Zyla, et~al., {\bf Particle Data Group} Collaboration, {Review of Particle
  Physics}, PTEP 2020~(8) (2020) 083C01.
\newblock \href {http://dx.doi.org/10.1093/ptep/ptaa104}
  {\path{doi:10.1093/ptep/ptaa104}}.

\bibitem{Starkman:2020sbz}
N.~Starkman, J.~Sidhu, H.~Winch, G.~Starkman, {Straight lightning as a
  signature of macroscopic dark matter}, Phys. Rev. D 103~(6) (2021) 063024.
\newblock \href {http://arxiv.org/abs/2006.16272} {\path{arXiv:2006.16272}},
  \href {http://dx.doi.org/10.1103/PhysRevD.103.063024}
  {\path{doi:10.1103/PhysRevD.103.063024}}.

\bibitem{Paul:2021bhh}
T.~C. Paul, S.~T. Reese, L.~A. Anchordoqui, A.~V. Olinto, {EUSO-SPB2
  sensitivity to macroscopic dark matter}, PoS ICRC2021 (2021) 519.
\newblock \href {http://arxiv.org/abs/2104.01152} {\path{arXiv:2104.01152}},
  \href {http://dx.doi.org/10.22323/1.395.0519}
  {\path{doi:10.22323/1.395.0519}}.

\bibitem{Mini-EUSO:nuclearites}
L.~W. Piotrowski, et~al., {Towards observations of nuclearites in Mini-EUSO},
  PoS 395.
\newblock \href {http://dx.doi.org/10.22323/1.395.0503}
  {\path{doi:10.22323/1.395.0503}}.

\bibitem{Price:1988ge}
P.~B. Price, {Limits on Contribution of Cosmic Nuclearites to Galactic Dark
  Matter}, Physical Review D 38 (1988) 3813--3814.
\newblock \href {http://dx.doi.org/10.1103/PhysRevD.38.3813}
  {\path{doi:10.1103/PhysRevD.38.3813}}.

\bibitem{MACRO:1999ijn}
M.~Ambrosio, et~al., {\bf MACRO} Collaboration, {Nuclearite search with the
  macro detector at Gran Sasso}, Eur. Phys. J. C 13 (2000) 453--458.
\newblock \href {http://arxiv.org/abs/hep-ex/9904031}
  {\path{arXiv:hep-ex/9904031}}, \href
  {http://dx.doi.org/10.1007/s100520050708} {\path{doi:10.1007/s100520050708}}.

\bibitem{Piotrowski:2019}
L.~W. Piotrowski, et~al.,
  \href{https://link.aps.org/doi/10.1103/PhysRevLett.125.091101}{Limits on the
  flux of nuclearites and other heavy compact objects from the pi of the sky
  project}, Phys. Rev. Lett. 125 (2020) 091101.
\newblock \href {http://dx.doi.org/10.1103/PhysRevLett.125.091101}
  {\path{doi:10.1103/PhysRevLett.125.091101}}.
\newline\urlprefix\url{https://link.aps.org/doi/10.1103/PhysRevLett.125.091101}

\bibitem{Sidhu:2019fgg}
J.~S. Sidhu, G.~Starkman, {Macroscopic Dark Matter Constraints from Bolide
  Camera Networks}, Phys. Rev. D 100~(12) (2019) 123008.
\newblock \href {http://arxiv.org/abs/1908.00557} {\path{arXiv:1908.00557}},
  \href {http://dx.doi.org/10.1103/PhysRevD.100.123008}
  {\path{doi:10.1103/PhysRevD.100.123008}}.

\bibitem{Abe:2021ugf}
S.~Abe, et~al., {DIMS Experiment for Dark Matter and Interstellar Meteoroid
  Study}, PoS ICRC2021 (2021) 554.
\newblock \href {http://dx.doi.org/10.22323/1.395.0554}
  {\path{doi:10.22323/1.395.0554}}.

\bibitem{DIMS:2021ipc}
D.~Barghini, et~al., {\bf DIMS} Collaboration, {Characterization of the DIMS
  system based on astronomical meteor techniques for macroscopic dark matter
  search}, PoS ICRC2021 (2021) 500.
\newblock \href {http://dx.doi.org/10.22323/1.395.0500}
  {\path{doi:10.22323/1.395.0500}}.

\bibitem{Cooray:2021dvp}
V.~Cooray, G.~Cooray, M.~Rubinstein, F.~Rachidi, {Could macroscopic dark matter
  (macros) give rise to mini-lightning flashes out of a blue sky without
  clouds?}\href {http://arxiv.org/abs/2107.05338} {\path{arXiv:2107.05338}},
  \href {http://dx.doi.org/10.3390/atmos12091230}
  {\path{doi:10.3390/atmos12091230}}.

\bibitem{SinghSidhu:2018oqs}
J.~Singh~Sidhu, R.~M. Abraham, C.~Covault, G.~Starkman, {Macro detection using
  fluorescence detectors}, JCAP 02 (2019) 037.
\newblock \href {http://arxiv.org/abs/1808.06978} {\path{arXiv:1808.06978}},
  \href {http://dx.doi.org/10.1088/1475-7516/2019/02/037}
  {\path{doi:10.1088/1475-7516/2019/02/037}}.

\bibitem{Anchordoqui:2021xhu}
L.~A. Anchordoqui, et~al., {Prospects for macroscopic dark matter detection at
  space-based and suborbital experiments}, EPL 135~(5) (2021) 51001.
\newblock \href {http://arxiv.org/abs/2104.05131} {\path{arXiv:2104.05131}},
  \href {http://dx.doi.org/10.1209/0295-5075/ac115f}
  {\path{doi:10.1209/0295-5075/ac115f}}.

\bibitem{DeRujula:1984axn}
A.~De~Rujula, S.~L. Glashow, {Nuclearites: A Novel Form of Cosmic Radiation},
  Nature 312 (1984) 734--737.
\newblock \href {http://dx.doi.org/10.1038/312734a0}
  {\path{doi:10.1038/312734a0}}.

\bibitem{Bai:2018dxf}
Y.~Bai, A.~J. Long, S.~Lu, {Dark Quark Nuggets}, Phys. Rev. D 99~(5) (2019)
  055047.
\newblock \href {http://arxiv.org/abs/1810.04360} {\path{arXiv:1810.04360}},
  \href {http://dx.doi.org/10.1103/PhysRevD.99.055047}
  {\path{doi:10.1103/PhysRevD.99.055047}}.

\bibitem{Strong:2007nh}
A.~W. Strong, I.~V. Moskalenko, V.~S. Ptuskin, {Cosmic-ray propagation and
  interactions in the Galaxy}, Ann. Rev. Nucl. Part. Sci. 57 (2007) 285--327.
\newblock \href {http://arxiv.org/abs/astro-ph/0701517}
  {\path{arXiv:astro-ph/0701517}}, \href
  {http://dx.doi.org/10.1146/annurev.nucl.57.090506.123011}
  {\path{doi:10.1146/annurev.nucl.57.090506.123011}}.

\bibitem{Fermi-LAT:2013iui}
M.~Ackermann, et~al., {\bf Fermi-LAT} Collaboration, {Detection of the
  Characteristic Pion-Decay Signature in Supernova Remnants}, Science 339
  (2013) 807.
\newblock \href {http://arxiv.org/abs/1302.3307} {\path{arXiv:1302.3307}},
  \href {http://dx.doi.org/10.1126/science.1231160}
  {\path{doi:10.1126/science.1231160}}.

\bibitem{Amato:2021yyv}
E.~Amato, S.~Casanova, {On particle acceleration and transport in plasmas in
  the Galaxy: theory and observations}, J. Plasma Phys. 87~(1) (2021)
  845870101.
\newblock \href {http://arxiv.org/abs/2104.12428} {\path{arXiv:2104.12428}},
  \href {http://dx.doi.org/10.1017/S0022377821000064}
  {\path{doi:10.1017/S0022377821000064}}.

\bibitem{Haungs:2015ema}
A.~Haungs, {Cosmic Rays from the Knee to the Ankle}, Phys. Procedia 61 (2015)
  425--434.
\newblock \href {http://arxiv.org/abs/1504.01859} {\path{arXiv:1504.01859}},
  \href {http://dx.doi.org/10.1016/j.phpro.2014.12.094}
  {\path{doi:10.1016/j.phpro.2014.12.094}}.

\bibitem{Globus:2015xga}
N.~Globus, D.~Allard, E.~Parizot, {A complete model of the cosmic ray spectrum
  and composition across the Galactic to extragalactic transition}, Phys. Rev.
  D 92~(2) (2015) 021302.
\newblock \href {http://arxiv.org/abs/1505.01377} {\path{arXiv:1505.01377}},
  \href {http://dx.doi.org/10.1103/PhysRevD.92.021302}
  {\path{doi:10.1103/PhysRevD.92.021302}}.

\bibitem{2019EPJWC.21004003K}
M.~{Kachelrie{\ss}}, {Transition from Galactic to Extragalactic Cosmic Rays},
  in: European Physical Journal Web of Conferences, Vol. 210 of European
  Physical Journal Web of Conferences, 2019, p. 04003.
\newblock \href {http://dx.doi.org/10.1051/epjconf/201921004003}
  {\path{doi:10.1051/epjconf/201921004003}}.

\bibitem{Kaapa:2021Fk}
{K\"{a}\"{a}p\"{a}, Alex and Kampert, Karl-Heinz and Mayotte, Eric}, {The
  effects of the GMF on the transition from Galactic to extragalactic cosmic
  rays}, PoS ICRC2021 (2021) 004.
\newblock \href {http://dx.doi.org/10.22323/1.395.0004}
  {\path{doi:10.22323/1.395.0004}}.

\bibitem{Mollerach:2018lkt}
S.~Mollerach, E.~Roulet, {A scenario for the Galactic cosmic rays between the
  knee and the second-knee}, JCAP 03 (2019) 017.
\newblock \href {http://arxiv.org/abs/1812.04026} {\path{arXiv:1812.04026}},
  \href {http://dx.doi.org/10.1088/1475-7516/2019/03/017}
  {\path{doi:10.1088/1475-7516/2019/03/017}}.

\bibitem{Candia:2003dk}
J.~Candia, S.~Mollerach, E.~Roulet, {Cosmic ray spectrum and anisotropies from
  the knee to the second knee}, JCAP 05 (2003) 003.
\newblock \href {http://arxiv.org/abs/astro-ph/0302082}
  {\path{arXiv:astro-ph/0302082}}, \href
  {http://dx.doi.org/10.1088/1475-7516/2003/05/003}
  {\path{doi:10.1088/1475-7516/2003/05/003}}.

\bibitem{Giacinti:2011ww}
G.~Giacinti, M.~Kachelriess, D.~V. Semikoz, G.~Sigl, {Cosmic Ray Anisotropy as
  Signature for the Transition from Galactic to Extragalactic Cosmic Rays},
  JCAP 07 (2012) 031.
\newblock \href {http://arxiv.org/abs/1112.5599} {\path{arXiv:1112.5599}},
  \href {http://dx.doi.org/10.1088/1475-7516/2012/07/031}
  {\path{doi:10.1088/1475-7516/2012/07/031}}.

\bibitem{Katz:2008xx}
B.~Katz, R.~Budnik, E.~Waxman, {The energy production rate \& the generation
  spectrum of UHECRs}, JCAP 03 (2009) 020.
\newblock \href {http://arxiv.org/abs/0811.3759} {\path{arXiv:0811.3759}},
  \href {http://dx.doi.org/10.1088/1475-7516/2009/03/020}
  {\path{doi:10.1088/1475-7516/2009/03/020}}.

\bibitem{Allard:2008gj}
D.~Allard, N.~G. Busca, G.~Decerprit, A.~V. Olinto, E.~Parizot, {Implications
  of the cosmic ray spectrum for the mass composition at the highest energies},
  JCAP 10 (2008) 033.
\newblock \href {http://arxiv.org/abs/0805.4779} {\path{arXiv:0805.4779}},
  \href {http://dx.doi.org/10.1088/1475-7516/2008/10/033}
  {\path{doi:10.1088/1475-7516/2008/10/033}}.

\bibitem{Aloisio:2009sj}
R.~Aloisio, V.~Berezinsky, A.~Gazizov, {Ultra High Energy Cosmic Rays: The
  disappointing model}, Astropart. Phys. 34 (2011) 620--626.
\newblock \href {http://arxiv.org/abs/0907.5194} {\path{arXiv:0907.5194}},
  \href {http://dx.doi.org/10.1016/j.astropartphys.2010.12.008}
  {\path{doi:10.1016/j.astropartphys.2010.12.008}}.

\bibitem{Pruet:2002hi}
J.~Pruet, S.~Guiles, G.~M. Fuller, {Light element synthesis in high entropy
  relativistic flows associated with gamma-ray bursts}, Astrophys. J. 580
  (2002) 368--373.
\newblock \href {http://arxiv.org/abs/astro-ph/0205056}
  {\path{arXiv:astro-ph/0205056}}, \href {http://dx.doi.org/10.1086/342838}
  {\path{doi:10.1086/342838}}.

\bibitem{Lemoine:2002vg}
M.~Lemoine, {Nucleosynthesis in gamma-ray bursts outflows}, Astron. Astrophys.
  390 (2002) L31.
\newblock \href {http://arxiv.org/abs/astro-ph/0205093}
  {\path{arXiv:astro-ph/0205093}}, \href
  {http://dx.doi.org/10.1051/0004-6361:20020939}
  {\path{doi:10.1051/0004-6361:20020939}}.

\bibitem{Wang:2007xj}
X.-Y. Wang, S.~Razzaque, P.~Meszaros, {On the Origin and Survival of UHE
  Cosmic-Ray Nuclei in GRBs and Hypernovae}, Astrophys. J. 677 (2008) 432--440.
\newblock \href {http://arxiv.org/abs/0711.2065} {\path{arXiv:0711.2065}},
  \href {http://dx.doi.org/10.1086/529018} {\path{doi:10.1086/529018}}.

\bibitem{Murase:2008mr}
K.~Murase, K.~Ioka, S.~Nagataki, T.~Nakamura, {High-energy cosmic-ray nuclei
  from high- and low-luminosity gamma-ray bursts and implications for
  multi-messenger astronomy}, Phys. Rev. D 78 (2008) 023005.
\newblock \href {http://arxiv.org/abs/0801.2861} {\path{arXiv:0801.2861}},
  \href {http://dx.doi.org/10.1103/PhysRevD.78.023005}
  {\path{doi:10.1103/PhysRevD.78.023005}}.

\bibitem{Murase:2011cy}
K.~Murase, C.~D. Dermer, H.~Takami, G.~Migliori, {Blazars as Ultra-High-Energy
  Cosmic-Ray Sources: Implications for TeV Gamma-Ray Observations}, Astrophys.
  J. 749 (2012) 63.
\newblock \href {http://arxiv.org/abs/1107.5576} {\path{arXiv:1107.5576}},
  \href {http://dx.doi.org/10.1088/0004-637X/749/1/63}
  {\path{doi:10.1088/0004-637X/749/1/63}}.

\bibitem{Horiuchi:2012by}
S.~Horiuchi, K.~Murase, K.~Ioka, P.~Meszaros, {The survival of nuclei in jets
  associated with core-collapse supernovae and gamma-ray bursts}, Astrophys. J.
  753 (2012) 69.
\newblock \href {http://arxiv.org/abs/1203.0296} {\path{arXiv:1203.0296}},
  \href {http://dx.doi.org/10.1088/0004-637X/753/1/69}
  {\path{doi:10.1088/0004-637X/753/1/69}}.

\bibitem{Fang:2012rx}
K.~Fang, K.~Kotera, A.~V. Olinto, {Newly-born pulsars as sources of ultrahigh
  energy cosmic rays}, Astrophys. J. 750 (2012) 118.
\newblock \href {http://arxiv.org/abs/1201.5197} {\path{arXiv:1201.5197}},
  \href {http://dx.doi.org/10.1088/0004-637X/750/2/118}
  {\path{doi:10.1088/0004-637X/750/2/118}}.

\bibitem{Kotera:2015pya}
K.~Kotera, E.~Amato, P.~Blasi, {The fate of ultrahigh energy nuclei in the
  immediate environment of young fast-rotating pulsars}, JCAP 08 (2015) 026.
\newblock \href {http://arxiv.org/abs/1503.07907} {\path{arXiv:1503.07907}},
  \href {http://dx.doi.org/10.1088/1475-7516/2015/08/026}
  {\path{doi:10.1088/1475-7516/2015/08/026}}.

\bibitem{Globus:2015bko}
N.~Globus, D.~Allard, R.~Mochkovitch, E.~Parizot, {UHECR acceleration at GRB
  internal shocks}, PoS ICRC2015 (2016) 516.
\newblock \href {http://dx.doi.org/10.22323/1.236.0516}
  {\path{doi:10.22323/1.236.0516}}.

\bibitem{Guepin:2017abw}
C.~Gu\'epin, K.~Kotera, E.~Barausse, K.~Fang, K.~Murase, {Ultra-High Energy
  Cosmic Rays and Neutrinos from Tidal Disruptions by Massive Black Holes},
  Astron. Astrophys. 616 (2018) A179, [Erratum: Astron.Astrophys. 636, C3
  (2020)].
\newblock \href {http://arxiv.org/abs/1711.11274} {\path{arXiv:1711.11274}},
  \href {http://dx.doi.org/10.1051/0004-6361/201732392}
  {\path{doi:10.1051/0004-6361/201732392}}.

\bibitem{Biehl:2017hnb}
D.~Biehl, D.~Boncioli, C.~Lunardini, W.~Winter, {Tidally disrupted stars as a
  possible origin of both cosmic rays and neutrinos at the highest energies},
  Sci. Rep. 8~(1) (2018) 10828.
\newblock \href {http://arxiv.org/abs/1711.03555} {\path{arXiv:1711.03555}},
  \href {http://dx.doi.org/10.1038/s41598-018-29022-4}
  {\path{doi:10.1038/s41598-018-29022-4}}.

\bibitem{Bhattacharya:2021cjc}
M.~Bhattacharya, S.~Horiuchi, K.~Murase, {On the synthesis of heavy nuclei in
  protomagnetar outflows and implications for ultra-high energy cosmic
  rays}\href {http://arxiv.org/abs/2111.05863} {\path{arXiv:2111.05863}}.

\bibitem{Ekanger:2022tia}
N.~Ekanger, M.~Bhattacharya, S.~Horiuchi, {Systematic exploration of heavy
  element nucleosynthesis in protomagnetar outflows}\href
  {http://arxiv.org/abs/2201.03576} {\path{arXiv:2201.03576}}.

\bibitem{Blaksley:2013eho}
C.~Blaksley, E.~Parizot, G.~Decerprit, D.~Allard, {Ultra-high-energy cosmic ray
  source statistics in the GZK energy range}, Astron. Astrophys. 552 (2013)
  A125.
\newblock \href {http://dx.doi.org/10.1051/0004-6361/201220178}
  {\path{doi:10.1051/0004-6361/201220178}}.

\bibitem{dOrfeuil:2014qgw}
B.~R. d'Orfeuil, D.~Allard, C.~Lachaud, E.~Parizot, C.~Blaksley, S.~Nagataki,
  {Anisotropy expectations for ultra-high-energy cosmic rays with future high
  statistics experiments}, Astron. Astrophys. 567 (2014) A81.
\newblock \href {http://arxiv.org/abs/1401.1119} {\path{arXiv:1401.1119}},
  \href {http://dx.doi.org/10.1051/0004-6361/201423462}
  {\path{doi:10.1051/0004-6361/201423462}}.

\bibitem{Oikonomou:2014zva}
F.~Oikonomou, K.~Kotera, F.~B. Abdalla, {Simulations for a next-generation
  UHECR observatory}, JCAP 01 (2015) 030.
\newblock \href {http://arxiv.org/abs/1409.1925} {\path{arXiv:1409.1925}},
  \href {http://dx.doi.org/10.1088/1475-7516/2015/01/030}
  {\path{doi:10.1088/1475-7516/2015/01/030}}.

\bibitem{Kashti:2008bw}
T.~Kashti, E.~Waxman, {Searching for a Correlation Between Cosmic-Ray Sources
  Above 10\textasciicircum{}{19} eV and Large-Scale Structure}, JCAP 05 (2008)
  006.
\newblock \href {http://arxiv.org/abs/0801.4516} {\path{arXiv:0801.4516}},
  \href {http://dx.doi.org/10.1088/1475-7516/2008/05/006}
  {\path{doi:10.1088/1475-7516/2008/05/006}}.

\bibitem{Takami:2011nn}
H.~Takami, K.~Murase, {The Role of Structured Magnetic Fields on Constraining
  Properties of Transient Sources of Ultra-high-energy Cosmic Rays}, Astrophys.
  J. 748 (2012) 9.
\newblock \href {http://arxiv.org/abs/1110.3245} {\path{arXiv:1110.3245}},
  \href {http://dx.doi.org/10.1088/0004-637X/748/1/9}
  {\path{doi:10.1088/0004-637X/748/1/9}}.

\bibitem{PierreAuger:2013waq}
P.~Abreu, et~al., {\bf Pierre Auger} Collaboration, {Bounds on the density of
  sources of ultra-high energy cosmic rays from the Pierre Auger Observatory},
  JCAP 05 (2013) 009.
\newblock \href {http://arxiv.org/abs/1305.1576} {\path{arXiv:1305.1576}},
  \href {http://dx.doi.org/10.1088/1475-7516/2013/05/009}
  {\path{doi:10.1088/1475-7516/2013/05/009}}.

\bibitem{Murase:2008sa}
K.~Murase, H.~Takami, {Implications of Ultra-High-Energy Cosmic Rays for
  Transient Sources in the Auger Era}, Astrophys. J. Lett. 690 (2009) L14--L17.
\newblock \href {http://arxiv.org/abs/0810.1813} {\path{arXiv:0810.1813}},
  \href {http://dx.doi.org/10.1088/0004-637X/690/1/L14}
  {\path{doi:10.1088/0004-637X/690/1/L14}}.

\bibitem{Waxman:2008bj}
E.~Waxman, A.~Loeb, {Constraints on the Local Sources of Ultra High-Energy
  Cosmic Rays}, JCAP 08 (2009) 026.
\newblock \href {http://arxiv.org/abs/0809.3788} {\path{arXiv:0809.3788}},
  \href {http://dx.doi.org/10.1088/1475-7516/2009/08/026}
  {\path{doi:10.1088/1475-7516/2009/08/026}}.

\bibitem{Kotera:2008ae}
K.~Kotera, M.~Lemoine, {The optical depth of the Universe for ultra-high energy
  cosmic ray scattering in the magnetized large scale structure}, Phys. Rev. D
  77 (2008) 123003.
\newblock \href {http://arxiv.org/abs/0801.1450} {\path{arXiv:0801.1450}},
  \href {http://dx.doi.org/10.1103/PhysRevD.77.123003}
  {\path{doi:10.1103/PhysRevD.77.123003}}.

\bibitem{2011A&A...528A.109K}
S.~{Kalli}, M.~{Lemoine}, K.~{Kotera}, {Distortion of the ultrahigh energy
  cosmic ray flux from rare transient sources in inhomogeneous extragalactic
  magnetic fields}, ArXiv e-prints\href {http://arxiv.org/abs/1101.3801}
  {\path{arXiv:1101.3801}}.

\bibitem{PierreAuger:2012gro}
P.~Abreu, et~al., {\bf Pierre Auger} Collaboration, {Constraints on the origin
  of cosmic rays above $10^{18}$ eV from large scale anisotropy searches in
  data of the Pierre Auger Observatory}, Astrophys. J. Lett. 762 (2012) L13.
\newblock \href {http://arxiv.org/abs/1212.3083} {\path{arXiv:1212.3083}},
  \href {http://dx.doi.org/10.1088/2041-8205/762/1/L13}
  {\path{doi:10.1088/2041-8205/762/1/L13}}.

\bibitem{Fang:2016ewe}
K.~Fang, K.~Kotera, {The Highest-Energy Cosmic Rays Cannot be Dominantly
  Protons from Steady Sources}, Astrophys. J. Lett. 832~(1) (2016) L17.
\newblock \href {http://arxiv.org/abs/1610.08055} {\path{arXiv:1610.08055}},
  \href {http://dx.doi.org/10.3847/2041-8205/832/1/L17}
  {\path{doi:10.3847/2041-8205/832/1/L17}}.

\bibitem{Palladino:2019hsk}
A.~Palladino, A.~van Vliet, W.~Winter, A.~Franckowiak, {Can astrophysical
  neutrinos trace the origin of the detected ultra-high energy cosmic rays?},
  Mon. Not. Roy. Astron. Soc. 494~(3) (2020) 4255--4265.
\newblock \href {http://arxiv.org/abs/1911.05756} {\path{arXiv:1911.05756}},
  \href {http://dx.doi.org/10.1093/mnras/staa1003}
  {\path{doi:10.1093/mnras/staa1003}}.

\bibitem{GKO22}
C.~Guepin, K.~Kotera, F.~Oikonomou, High-energy neutrino transients: the future
  of multi-messenger astronomy, journal publication in preparation.

\bibitem{Murase:2009ah}
K.~Murase, {Ultrahigh-Energy Photons as a Probe of Nearby Transient
  Ultrahigh-Energy Cosmic-Ray Sources and Possible Lorentz-Invariance
  Violation}, Phys. Rev. Lett. 103 (2009) 081102.
\newblock \href {http://arxiv.org/abs/0904.2087} {\path{arXiv:0904.2087}},
  \href {http://dx.doi.org/10.1103/PhysRevLett.103.081102}
  {\path{doi:10.1103/PhysRevLett.103.081102}}.

\bibitem{Matthews:2020lig}
J.~Matthews, A.~Bell, K.~Blundell, {Particle acceleration in astrophysical
  jets}, New Astron. Rev. 89 (2020) 101543.
\newblock \href {http://arxiv.org/abs/2003.06587} {\path{arXiv:2003.06587}},
  \href {http://dx.doi.org/10.1016/j.newar.2020.101543}
  {\path{doi:10.1016/j.newar.2020.101543}}.

\bibitem{Achterberg:2001rx}
A.~Achterberg, Y.~A. Gallant, J.~G. Kirk, A.~W. Guthmann, {Particle
  acceleration by ultrarelativistic shocks: Theory and simulations}, Mon. Not.
  Roy. Astron. Soc. 328 (2001) 393.
\newblock \href {http://arxiv.org/abs/astro-ph/0107530}
  {\path{arXiv:astro-ph/0107530}}, \href
  {http://dx.doi.org/10.1046/j.1365-8711.2001.04851.x}
  {\path{doi:10.1046/j.1365-8711.2001.04851.x}}.

\bibitem{Bell:2019nnf}
A.~Bell, J.~Matthews, K.~Blundell, A.~Araudo, {Cosmic Ray Acceleration in
  Hydromagnetic Flux Tubes}, Mon. Not. Roy. Astron. Soc. 487~(4) (2019)
  4571--4579.
\newblock \href {http://arxiv.org/abs/1906.02508} {\path{arXiv:1906.02508}},
  \href {http://dx.doi.org/10.1093/mnras/stz1604}
  {\path{doi:10.1093/mnras/stz1604}}.

\bibitem{Hillas:1984ijl}
A.~M. Hillas, {The Origin of Ultrahigh-Energy Cosmic Rays}, Ann. Rev. Astron.
  Astrophys. 22 (1984) 425--444.
\newblock \href {http://dx.doi.org/10.1146/annurev.aa.22.090184.002233}
  {\path{doi:10.1146/annurev.aa.22.090184.002233}}.

\bibitem{Murase:2018utn}
K.~Murase, M.~Fukugita, {Energetics of High-Energy Cosmic Radiations}, Phys.
  Rev. D 99~(6) (2019) 063012.
\newblock \href {http://arxiv.org/abs/1806.04194} {\path{arXiv:1806.04194}},
  \href {http://dx.doi.org/10.1103/PhysRevD.99.063012}
  {\path{doi:10.1103/PhysRevD.99.063012}}.

\bibitem{Jiang:2020arb}
Y.~Jiang, B.~T. Zhang, K.~Murase, {Energetics of ultrahigh-energy cosmic-ray
  nuclei}, Phys. Rev. D 104~(4) (2021) 043017.
\newblock \href {http://arxiv.org/abs/2012.03122} {\path{arXiv:2012.03122}},
  \href {http://dx.doi.org/10.1103/PhysRevD.104.043017}
  {\path{doi:10.1103/PhysRevD.104.043017}}.

\bibitem{Waxman:1995vg}
E.~Waxman, {Cosmological gamma-ray bursts and the highest energy cosmic rays},
  Phys. Rev. Lett. 75 (1995) 386--389.
\newblock \href {http://arxiv.org/abs/astro-ph/9505082}
  {\path{arXiv:astro-ph/9505082}}, \href
  {http://dx.doi.org/10.1103/PhysRevLett.75.386}
  {\path{doi:10.1103/PhysRevLett.75.386}}.

\bibitem{Vietri:1995hs}
M.~Vietri, {On the acceleration of ultrahigh-energy cosmic rays in gamma-ray
  bursts}, Astrophys. J. 453 (1995) 883--889.
\newblock \href {http://arxiv.org/abs/astro-ph/9506081}
  {\path{arXiv:astro-ph/9506081}}, \href {http://dx.doi.org/10.1086/176448}
  {\path{doi:10.1086/176448}}.

\bibitem{Murase:2006mm}
K.~Murase, K.~Ioka, S.~Nagataki, T.~Nakamura, {High Energy Neutrinos and
  Cosmic-Rays from Low-Luminosity Gamma-Ray Bursts?}, Astrophys. J. Lett. 651
  (2006) L5--L8.
\newblock \href {http://arxiv.org/abs/astro-ph/0607104}
  {\path{arXiv:astro-ph/0607104}}, \href {http://dx.doi.org/10.1086/509323}
  {\path{doi:10.1086/509323}}.

\bibitem{Zhang:2017moz}
B.~T. Zhang, K.~Murase, S.~S. Kimura, S.~Horiuchi, P.~M\'esz\'aros,
  {Low-luminosity gamma-ray bursts as the sources of ultrahigh-energy cosmic
  ray nuclei}, Phys. Rev. D 97~(8) (2018) 083010.
\newblock \href {http://arxiv.org/abs/1712.09984} {\path{arXiv:1712.09984}},
  \href {http://dx.doi.org/10.1103/PhysRevD.97.083010}
  {\path{doi:10.1103/PhysRevD.97.083010}}.

\bibitem{Zhang:2018agl}
B.~T. Zhang, K.~Murase, {Ultrahigh-energy cosmic-ray nuclei and neutrinos from
  engine-driven supernovae}, Phys. Rev. D 100~(10) (2019) 103004.
\newblock \href {http://arxiv.org/abs/1812.10289} {\path{arXiv:1812.10289}},
  \href {http://dx.doi.org/10.1103/PhysRevD.100.103004}
  {\path{doi:10.1103/PhysRevD.100.103004}}.

\bibitem{Globus:2014fka}
N.~Globus, D.~Allard, R.~Mochkovitch, E.~Parizot, {UHECR acceleration at GRB
  internal shocks}, Mon. Not. Roy. Astron. Soc. 451~(1) (2015) 751--790.
\newblock \href {http://arxiv.org/abs/1409.1271} {\path{arXiv:1409.1271}},
  \href {http://dx.doi.org/10.1093/mnras/stv893}
  {\path{doi:10.1093/mnras/stv893}}.

\bibitem{Baerwald:2014zga}
P.~Baerwald, M.~Bustamante, W.~Winter, {Are gamma-ray bursts the sources of
  ultra-high energy cosmic rays?}, Astropart. Phys. 62 (2015) 66--91.
\newblock \href {http://arxiv.org/abs/1401.1820} {\path{arXiv:1401.1820}},
  \href {http://dx.doi.org/10.1016/j.astropartphys.2014.07.007}
  {\path{doi:10.1016/j.astropartphys.2014.07.007}}.

\bibitem{Boncioli:2018lrv}
D.~Boncioli, D.~Biehl, W.~Winter, {On the common origin of cosmic rays across
  the ankle and diffuse neutrinos at the highest energies from low-luminosity
  Gamma-Ray Bursts}, Astrophys. J. 872~(1) (2019) 110.
\newblock \href {http://arxiv.org/abs/1808.07481} {\path{arXiv:1808.07481}},
  \href {http://dx.doi.org/10.3847/1538-4357/aafda7}
  {\path{doi:10.3847/1538-4357/aafda7}}.

\bibitem{Bustamante:2014oka}
M.~Bustamante, P.~Baerwald, K.~Murase, W.~Winter, {Neutrino and cosmic-ray
  emission from multiple internal shocks in gamma-ray bursts}, Nature Commun. 6
  (2015) 6783.
\newblock \href {http://arxiv.org/abs/1409.2874} {\path{arXiv:1409.2874}},
  \href {http://dx.doi.org/10.1038/ncomms7783} {\path{doi:10.1038/ncomms7783}}.

\bibitem{Caprioli:2015zka}
D.~Caprioli, {''Espresso'' Acceleration of Ultra-high-energy Cosmic Rays},
  Astrophys. J. Lett. 811~(2) (2015) L38.
\newblock \href {http://arxiv.org/abs/1505.06739} {\path{arXiv:1505.06739}},
  \href {http://dx.doi.org/10.1088/2041-8205/811/2/L38}
  {\path{doi:10.1088/2041-8205/811/2/L38}}.

\bibitem{Kimura:2017ubz}
S.~S. Kimura, K.~Murase, B.~T. Zhang, {Ultrahigh-energy Cosmic-ray Nuclei from
  Black Hole Jets: Recycling Galactic Cosmic Rays through Shear Acceleration},
  Phys. Rev. D 97~(2) (2018) 023026.
\newblock \href {http://arxiv.org/abs/1705.05027} {\path{arXiv:1705.05027}},
  \href {http://dx.doi.org/10.1103/PhysRevD.97.023026}
  {\path{doi:10.1103/PhysRevD.97.023026}}.

\bibitem{Farrar:2008ex}
G.~R. Farrar, A.~Gruzinov, {Giant AGN Flares and Cosmic Ray Bursts}, Astrophys.
  J. 693 (2009) 329--332.
\newblock \href {http://arxiv.org/abs/0802.1074} {\path{arXiv:0802.1074}},
  \href {http://dx.doi.org/10.1088/0004-637X/693/1/329}
  {\path{doi:10.1088/0004-637X/693/1/329}}.

\bibitem{Farrar:2014yla}
G.~R. Farrar, T.~Piran, {Tidal disruption jets as the source of Ultra-High
  Energy Cosmic Rays}\href {http://arxiv.org/abs/1411.0704}
  {\path{arXiv:1411.0704}}.

\bibitem{Zhang:2017hom}
B.~T. Zhang, K.~Murase, F.~Oikonomou, Z.~Li, {High-energy cosmic ray nuclei
  from tidal disruption events: Origin, survival, and implications}, Phys. Rev.
  D 96~(6) (2017) 063007, [Addendum: Phys.Rev.D 96, 069902 (2017)].
\newblock \href {http://arxiv.org/abs/1706.00391} {\path{arXiv:1706.00391}},
  \href {http://dx.doi.org/10.1103/PhysRevD.96.063007}
  {\path{doi:10.1103/PhysRevD.96.063007}}.

\bibitem{Stein:2020xhk}
R.~Stein, et~al., {A tidal disruption event coincident with a high-energy
  neutrino}, Nature Astron. 5~(5) (2021) 510--518.
\newblock \href {http://arxiv.org/abs/2005.05340} {\path{arXiv:2005.05340}},
  \href {http://dx.doi.org/10.1038/s41550-020-01295-8}
  {\path{doi:10.1038/s41550-020-01295-8}}.

\bibitem{Anchordoqui:1999cu}
L.~A. Anchordoqui, G.~E. Romero, J.~A. Combi, {Heavy nuclei at the end of the
  cosmic ray spectrum?}, Phys. Rev. D 60 (1999) 103001.
\newblock \href {http://arxiv.org/abs/astro-ph/9903145}
  {\path{arXiv:astro-ph/9903145}}, \href
  {http://dx.doi.org/10.1103/PhysRevD.60.103001}
  {\path{doi:10.1103/PhysRevD.60.103001}}.

\bibitem{Romero:2018mnb}
G.~E. Romero, A.~L. M\"uller, M.~Roth, {Particle acceleration in the superwinds
  of starburst galaxies}, Astron. Astrophys. 616 (2018) A57.
\newblock \href {http://arxiv.org/abs/1801.06483} {\path{arXiv:1801.06483}},
  \href {http://dx.doi.org/10.1051/0004-6361/201832666}
  {\path{doi:10.1051/0004-6361/201832666}}.

\bibitem{Anchordoqui:2018vji}
L.~A. Anchordoqui, {Acceleration of ultrahigh-energy cosmic rays in starburst
  superwinds}, Phys. Rev. D 97~(6) (2018) 063010.
\newblock \href {http://arxiv.org/abs/1801.07170} {\path{arXiv:1801.07170}},
  \href {http://dx.doi.org/10.1103/PhysRevD.97.063010}
  {\path{doi:10.1103/PhysRevD.97.063010}}.

\bibitem{Anchordoqui:2020otc}
L.~A. Anchordoqui, D.~F. Torres, {Exploring the superwind mechanism for
  generating ultrahigh-energy cosmic rays using large-scale modeling of
  starbursts}, Phys. Rev. D 102~(2) (2020) 023034.
\newblock \href {http://arxiv.org/abs/2004.09378} {\path{arXiv:2004.09378}},
  \href {http://dx.doi.org/10.1103/PhysRevD.102.023034}
  {\path{doi:10.1103/PhysRevD.102.023034}}.

\bibitem{Anchordoqui:2019ncn}
L.~A. Anchordoqui, C.~Mechmann, J.~F. Soriano, {Toward a robust inference
  method for the likelihood of low-luminosity gamma-ray bursts to be
  progenitors of ultrahigh-energy cosmic rays correlating with starburst
  galaxies}, JHEAp 25 (2020) 23--28.
\newblock \href {http://arxiv.org/abs/1910.07311} {\path{arXiv:1910.07311}},
  \href {http://dx.doi.org/10.1016/j.jheap.2020.01.001}
  {\path{doi:10.1016/j.jheap.2020.01.001}}.

\bibitem{Kang:1996rp}
H.~Kang, J.~P. Rachen, P.~L. Biermann, {Contributions to the cosmic ray flux
  above the ankle: Clusters of galaxies}, Mon. Not. Roy. Astron. Soc. 286
  (1997) 257.
\newblock \href {http://arxiv.org/abs/astro-ph/9608071}
  {\path{arXiv:astro-ph/9608071}}, \href
  {http://dx.doi.org/10.1093/mnras/286.2.257}
  {\path{doi:10.1093/mnras/286.2.257}}.

\bibitem{Murase:2008yt}
K.~Murase, S.~Inoue, S.~Nagataki, {Cosmic Rays Above the Second Knee from
  Clusters of Galaxies and Associated High-Energy Neutrino Emission},
  Astrophys. J. Lett. 689 (2008) L105.
\newblock \href {http://arxiv.org/abs/0805.0104} {\path{arXiv:0805.0104}},
  \href {http://dx.doi.org/10.1086/595882} {\path{doi:10.1086/595882}}.

\bibitem{Kotera:2009ms}
K.~Kotera, D.~Allard, K.~Murase, J.~Aoi, Y.~Dubois, T.~Pierog, S.~Nagataki,
  {Propagation of ultrahigh energy nuclei in clusters of galaxies: resulting
  composition and secondary emissions}, Astrophys. J. 707 (2009) 370--386.
\newblock \href {http://arxiv.org/abs/0907.2433} {\path{arXiv:0907.2433}},
  \href {http://dx.doi.org/10.1088/0004-637X/707/1/370}
  {\path{doi:10.1088/0004-637X/707/1/370}}.

\bibitem{Fang:2017zjf}
K.~Fang, K.~Murase, {Linking High-Energy Cosmic Particles by Black Hole Jets
  Embedded in Large-Scale Structures}, Nature Phys. 14~(4) (2018) 396--398.
\newblock \href {http://arxiv.org/abs/1704.00015} {\path{arXiv:1704.00015}},
  \href {http://dx.doi.org/10.1038/s41567-017-0025-4}
  {\path{doi:10.1038/s41567-017-0025-4}}.

\bibitem{Cooray:2016jrk}
A.~Cooray, {Extragalactic Background Light: Measurements and Applications}\href
  {http://arxiv.org/abs/1602.03512} {\path{arXiv:1602.03512}}.

\bibitem{Allard:2011aa}
D.~Allard, {Extragalactic propagation of ultrahigh energy cosmic-rays},
  Astropart. Phys. 39-40 (2012) 33--43.
\newblock \href {http://arxiv.org/abs/1111.3290} {\path{arXiv:1111.3290}},
  \href {http://dx.doi.org/10.1016/j.astropartphys.2011.10.011}
  {\path{doi:10.1016/j.astropartphys.2011.10.011}}.

\bibitem{PhysRev.180.1264}
F.~W. Stecker,
  \href{https://link.aps.org/doi/10.1103/PhysRev.180.1264}{Photodisintegration
  of ultrahigh-energy cosmic rays by the universal radiation field}, Phys. Rev.
  180 (1969) 1264--1266.
\newblock \href {http://dx.doi.org/10.1103/PhysRev.180.1264}
  {\path{doi:10.1103/PhysRev.180.1264}}.
\newline\urlprefix\url{https://link.aps.org/doi/10.1103/PhysRev.180.1264}

\bibitem{Puget:1976nz}
J.~L. Puget, F.~W. Stecker, J.~H. Bredekamp, {Photonuclear Interactions of
  Ultrahigh-Energy Cosmic Rays and their Astrophysical Consequences},
  Astrophys. J. 205 (1976) 638--654.
\newblock \href {http://dx.doi.org/10.1086/154321} {\path{doi:10.1086/154321}}.

\bibitem{Epele:1998ia}
L.~N. Epele, E.~Roulet, {On the propagation of the highest energy cosmic ray
  nuclei}, JHEP 10 (1998) 009.
\newblock \href {http://arxiv.org/abs/astro-ph/9808104}
  {\path{arXiv:astro-ph/9808104}}, \href
  {http://dx.doi.org/10.1088/1126-6708/1998/10/009}
  {\path{doi:10.1088/1126-6708/1998/10/009}}.

\bibitem{Stecker:1998ib}
F.~W. Stecker, M.~H. Salamon, {Photodisintegration of ultrahigh-energy cosmic
  rays: A New determination}, Astrophys. J. 512 (1999) 521--526.
\newblock \href {http://arxiv.org/abs/astro-ph/9808110}
  {\path{arXiv:astro-ph/9808110}}, \href {http://dx.doi.org/10.1086/306816}
  {\path{doi:10.1086/306816}}.

\bibitem{Huchra:2011ii}
J.~P. Huchra, et~al., {The 2MASS Redshift Survey - Description and Data
  Release}, Astrophys. J. Suppl. 199 (2012) 26.
\newblock \href {http://arxiv.org/abs/1108.0669} {\path{arXiv:1108.0669}},
  \href {http://dx.doi.org/10.1088/0067-0049/199/2/26}
  {\path{doi:10.1088/0067-0049/199/2/26}}.

\bibitem{Mucke:1999yb}
A.~Mucke, R.~Engel, J.~P. Rachen, R.~J. Protheroe, T.~Stanev, {SOPHIA: Monte
  Carlo simulations of photohadronic processes in astrophysics}, Comput. Phys.
  Commun. 124 (2000) 290--314.
\newblock \href {http://arxiv.org/abs/astro-ph/9903478}
  {\path{arXiv:astro-ph/9903478}}, \href
  {http://dx.doi.org/10.1016/S0010-4655(99)00446-4}
  {\path{doi:10.1016/S0010-4655(99)00446-4}}.

\bibitem{Jaffe:2019iuk}
T.~R. Jaffe, {Practical Modeling of Large-Scale Galactic Magnetic Fields:
  Status and Prospects}, Galaxies 7~(2) (2019) 52.
\newblock \href {http://arxiv.org/abs/1904.12689} {\path{arXiv:1904.12689}},
  \href {http://dx.doi.org/10.3390/galaxies7020052}
  {\path{doi:10.3390/galaxies7020052}}.

\bibitem{Harari:2010wq}
D.~Harari, S.~Mollerach, E.~Roulet, {Effects of the galactic magnetic field
  upon large scale anisotropies of extragalactic Cosmic Rays}, JCAP 11 (2010)
  033.
\newblock \href {http://arxiv.org/abs/1009.5891} {\path{arXiv:1009.5891}},
  \href {http://dx.doi.org/10.1088/1475-7516/2010/11/033}
  {\path{doi:10.1088/1475-7516/2010/11/033}}.

\bibitem{Globus:2017fym}
N.~Globus, T.~Piran, {The extragalactic ultra-high energy cosmic-ray dipole},
  Astrophys. J. Lett. 850~(2) (2017) L25.
\newblock \href {http://arxiv.org/abs/1709.10110} {\path{arXiv:1709.10110}},
  \href {http://dx.doi.org/10.3847/2041-8213/aa991b}
  {\path{doi:10.3847/2041-8213/aa991b}}.

\bibitem{diMatteo:2017dtg}
A.~di~Matteo, P.~Tinyakov, {How isotropic can the UHECR flux be?}, Mon. Not.
  Roy. Astron. Soc. 476~(1) (2018) 715--723.
\newblock \href {http://arxiv.org/abs/1706.02534} {\path{arXiv:1706.02534}},
  \href {http://dx.doi.org/10.1093/mnras/sty277/4835522}
  {\path{doi:10.1093/mnras/sty277/4835522}}.

\bibitem{Wittkowski:2017nfd}
D.~Wittkowski, K.-H. Kampert, {On the anisotropy in the arrival directions of
  ultra-high-energy cosmic rays}, Astrophys. J. Lett. 854~(1) (2018) L3.
\newblock \href {http://arxiv.org/abs/1710.05617} {\path{arXiv:1710.05617}},
  \href {http://dx.doi.org/10.3847/2041-8213/aaa2f9}
  {\path{doi:10.3847/2041-8213/aaa2f9}}.

\bibitem{Globus:2018svy}
N.~Globus, T.~Piran, Y.~Hoffman, E.~Carlesi, D.~Pomar\`ede, {Cosmic-Ray
  Anisotropy from Large Scale Structure and the effect of magnetic horizons},
  Mon. Not. Roy. Astron. Soc. 484~(3) (2019) 4167--4173.
\newblock \href {http://arxiv.org/abs/1808.02048} {\path{arXiv:1808.02048}},
  \href {http://dx.doi.org/10.1093/mnras/stz164}
  {\path{doi:10.1093/mnras/stz164}}.

\bibitem{Mollerach:2019wne}
S.~Mollerach, E.~Roulet, {Ultrahigh energy cosmic rays from a nearby
  extragalactic source in the diffusive regime}, Phys. Rev. D 99~(10) (2019)
  103010.
\newblock \href {http://arxiv.org/abs/1903.05722} {\path{arXiv:1903.05722}},
  \href {http://dx.doi.org/10.1103/PhysRevD.99.103010}
  {\path{doi:10.1103/PhysRevD.99.103010}}.

\bibitem{Eichmann:2020adn}
B.~Eichmann, T.~Winchen, {Galactic Magnetic Field Bias on Inferences from UHECR
  Data}, JCAP 04 (2020) 047.
\newblock \href {http://arxiv.org/abs/2001.01530} {\path{arXiv:2001.01530}},
  \href {http://dx.doi.org/10.1088/1475-7516/2020/04/047}
  {\path{doi:10.1088/1475-7516/2020/04/047}}.

\bibitem{Mollerach:2021ifa}
S.~Mollerach, E.~Roulet, {Anisotropies of ultrahigh-energy cosmic rays in a
  scenario with nearby sources}\href {http://arxiv.org/abs/2111.00560}
  {\path{arXiv:2111.00560}}.

\bibitem{PierreAuger:2011mup}
P.~Abreu, et~al., {\bf Pierre Auger} Collaboration, {Search for signatures of
  magnetically-induced alignment in the arrival directions measured by the
  Pierre Auger Observatory}, Astropart. Phys. 35 (2012) 354--361.
\newblock \href {http://arxiv.org/abs/1111.2472} {\path{arXiv:1111.2472}},
  \href {http://dx.doi.org/10.1016/j.astropartphys.2011.10.004}
  {\path{doi:10.1016/j.astropartphys.2011.10.004}}.

\bibitem{PierreAuger:2020zii}
A.~Aab, et~al., {\bf Pierre Auger} Collaboration, {Search for
  magnetically-induced signatures in the arrival directions of
  ultra-high-energy cosmic rays measured at the Pierre Auger Observatory}, JCAP
  06 (2020) 017.
\newblock \href {http://arxiv.org/abs/2004.10591} {\path{arXiv:2004.10591}},
  \href {http://dx.doi.org/10.1088/1475-7516/2020/06/017}
  {\path{doi:10.1088/1475-7516/2020/06/017}}.

\bibitem{Harari:2000az}
D.~Harari, S.~Mollerach, E.~Roulet, {Signatures of galactic magnetic lensing
  upon ultrahigh-energy cosmic rays}, JHEP 02 (2000) 035.
\newblock \href {http://arxiv.org/abs/astro-ph/0001084}
  {\path{arXiv:astro-ph/0001084}}, \href
  {http://dx.doi.org/10.1088/1126-6708/2000/02/035}
  {\path{doi:10.1088/1126-6708/2000/02/035}}.

\bibitem{Harari:2002dy}
D.~Harari, S.~Mollerach, E.~Roulet, F.~Sanchez, {Lensing of ultrahigh-energy
  cosmic rays in turbulent magnetic fields}, JHEP 03 (2002) 045.
\newblock \href {http://arxiv.org/abs/astro-ph/0202362}
  {\path{arXiv:astro-ph/0202362}}, \href
  {http://dx.doi.org/10.1088/1126-6708/2002/03/045}
  {\path{doi:10.1088/1126-6708/2002/03/045}}.

\bibitem{Dolag:2008py}
K.~Dolag, M.~Kachelrie\ss{}, D.~V. Semikoz, {UHECR observations and lensing in
  the magnetic field of the Virgo cluster}, JCAP 01 (2009) 033.
\newblock \href {http://arxiv.org/abs/0809.5055} {\path{arXiv:0809.5055}},
  \href {http://dx.doi.org/10.1088/1475-7516/2009/01/033}
  {\path{doi:10.1088/1475-7516/2009/01/033}}.

\bibitem{Giacinti:2009fy}
G.~Giacinti, X.~Derkx, D.~V. Semikoz, {Search for single sources of ultra high
  energy cosmic rays on the sky}, JCAP 03 (2010) 022.
\newblock \href {http://arxiv.org/abs/0907.1035} {\path{arXiv:0907.1035}},
  \href {http://dx.doi.org/10.1088/1475-7516/2010/03/022}
  {\path{doi:10.1088/1475-7516/2010/03/022}}.

\bibitem{Takami:2009qz}
H.~Takami, K.~Sato, {Does Galactic Magnetic Field Disturb the Correlation of
  the Highest Energy Cosmic Rays with their Sources?}, Astrophys. J. 724 (2010)
  1456--1472.
\newblock \href {http://arxiv.org/abs/0909.1532} {\path{arXiv:0909.1532}},
  \href {http://dx.doi.org/10.1088/0004-637X/724/2/1456}
  {\path{doi:10.1088/0004-637X/724/2/1456}}.

\bibitem{Keivani:2014kua}
A.~Keivani, G.~R. Farrar, M.~Sutherland, {Magnetic Deflections of Ultra-High
  Energy Cosmic Rays from Centaurus A}, Astropart. Phys. 61 (2014) 47--55.
\newblock \href {http://arxiv.org/abs/1406.5249} {\path{arXiv:1406.5249}},
  \href {http://dx.doi.org/10.1016/j.astropartphys.2014.07.001}
  {\path{doi:10.1016/j.astropartphys.2014.07.001}}.

\bibitem{Smida:2015kga}
R.~Smida, R.~Engel, {The ultra-high energy cosmic rays image of Virgo A}, PoS
  ICRC2015 (2016) 470.
\newblock \href {http://arxiv.org/abs/1509.09033} {\path{arXiv:1509.09033}},
  \href {http://dx.doi.org/10.22323/1.236.0470}
  {\path{doi:10.22323/1.236.0470}}.

\bibitem{AlvesBatista:2017vob}
R.~Alves~Batista, M.-S. Shin, J.~Devriendt, D.~Semikoz, G.~Sigl, {Implications
  of strong intergalactic magnetic fields for ultrahigh-energy cosmic-ray
  astronomy}, Phys. Rev. D 96~(2) (2017) 023010.
\newblock \href {http://arxiv.org/abs/1704.05869} {\path{arXiv:1704.05869}},
  \href {http://dx.doi.org/10.1103/PhysRevD.96.023010}
  {\path{doi:10.1103/PhysRevD.96.023010}}.

\bibitem{Farrar:2017lhm}
G.~R. Farrar, M.~S. Sutherland, {Deflections of UHECRs in the Galactic magnetic
  field}, JCAP 05 (2019) 004.
\newblock \href {http://arxiv.org/abs/1711.02730} {\path{arXiv:1711.02730}},
  \href {http://dx.doi.org/10.1088/1475-7516/2019/05/004}
  {\path{doi:10.1088/1475-7516/2019/05/004}}.

\bibitem{TelescopeArray:2021cvw}
K.~Kawata, et~al., {\bf Telescope Array} Collaboration, {Effects of Galactic
  magnetic field on the UHECR anisotropy studies}, PoS ICRC2021 (2021) 358.
\newblock \href {http://dx.doi.org/10.22323/1.395.0358}
  {\path{doi:10.22323/1.395.0358}}.

\bibitem{deOliveira:2021ckh}
C.~a. de~Oliveira, V.~de~Souza, {Magnetically Induced Anisotropies in the
  Arrival Directions of Ultra-high-energy Cosmic Rays from Nearby Radio
  Galaxies}\href {http://arxiv.org/abs/2112.02415} {\path{arXiv:2112.02415}}.

\bibitem{Carpio:2015ewa}
J.~A. Carpio, A.~M. Gago, {Impact of Galactic magnetic field modeling on
  searches of point sources via ultrahigh energy cosmic ray-neutrino
  correlations}, Phys. Rev. D 93~(2) (2016) 023004.
\newblock \href {http://arxiv.org/abs/1507.02781} {\path{arXiv:1507.02781}},
  \href {http://dx.doi.org/10.1103/PhysRevD.93.023004}
  {\path{doi:10.1103/PhysRevD.93.023004}}.

\bibitem{Resconi:2016ggj}
E.~Resconi, S.~Coenders, P.~Padovani, P.~Giommi, L.~Caccianiga, {Connecting
  blazars with ultrahigh-energy cosmic rays and astrophysical neutrinos}, Mon.
  Not. Roy. Astron. Soc. 468~(1) (2017) 597--606.
\newblock \href {http://arxiv.org/abs/1611.06022} {\path{arXiv:1611.06022}},
  \href {http://dx.doi.org/10.1093/mnras/stx498}
  {\path{doi:10.1093/mnras/stx498}}.

\bibitem{Carpio:2016wec}
J.~A. Carpio, A.~M. Gago, {Roadmap for searching cosmic rays correlated with
  the extraterrestrial neutrinos seen at IceCube}, Phys. Rev. D 95~(12) (2017)
  123009.
\newblock \href {http://arxiv.org/abs/1608.05099} {\path{arXiv:1608.05099}},
  \href {http://dx.doi.org/10.1103/PhysRevD.95.123009}
  {\path{doi:10.1103/PhysRevD.95.123009}}.

\bibitem{Golup:2009cv}
G.~Golup, D.~Harari, S.~Mollerach, E.~Roulet, {Source position reconstruction
  and constraints on the galactic magnetic field from ultra-high energy cosmic
  rays}, Astropart. Phys. 32 (2009) 269--277.
\newblock \href {http://arxiv.org/abs/0902.1742} {\path{arXiv:0902.1742}},
  \href {http://dx.doi.org/10.1016/j.astropartphys.2009.09.003}
  {\path{doi:10.1016/j.astropartphys.2009.09.003}}.

\bibitem{PierreAuger:2014tos}
A.~Aab, et~al., {\bf Pierre Auger} Collaboration, {Search for patterns by
  combining cosmic-ray energy and arrival directions at the Pierre Auger
  Observatory}, Eur. Phys. J. C 75~(6) (2015) 269.
\newblock \href {http://arxiv.org/abs/1410.0515} {\path{arXiv:1410.0515}},
  \href {http://dx.doi.org/10.1140/epjc/s10052-015-3471-0}
  {\path{doi:10.1140/epjc/s10052-015-3471-0}}.

\bibitem{TelescopeArray:2020acv}
R.~U. Abbasi, et~al., {\bf Telescope Array} Collaboration, {Evidence for a
  Supergalactic Structure of Magnetic Deflection Multiplets of Ultra-High
  Energy Cosmic Rays}, Astrophys. J. 899~(1) (2020) 86.
\newblock \href {http://arxiv.org/abs/2005.07312} {\path{arXiv:2005.07312}},
  \href {http://dx.doi.org/10.3847/1538-4357/aba26c}
  {\path{doi:10.3847/1538-4357/aba26c}}.

\bibitem{Unger:2019xct}
M.~Unger, G.~Farrar, {Progress in the Global Modeling of the Galactic Magnetic
  Field}, EPJ Web Conf. 210 (2019) 04005.
\newblock \href {http://arxiv.org/abs/1901.04720} {\path{arXiv:1901.04720}},
  \href {http://dx.doi.org/10.1051/epjconf/201921004005}
  {\path{doi:10.1051/epjconf/201921004005}}.

\bibitem{Mbarek:2021bay}
R.~Mbarek, D.~Caprioli, {Espresso and Stochastic Acceleration of
  Ultra-high-energy Cosmic Rays in Relativistic Jets}, Astrophys. J. 921~(1)
  (2021) 85.
\newblock \href {http://arxiv.org/abs/2105.05262} {\path{arXiv:2105.05262}},
  \href {http://dx.doi.org/10.3847/1538-4357/ac1da8}
  {\path{doi:10.3847/1538-4357/ac1da8}}.

\bibitem{Miralda-Escude:1996twc}
J.~Miralda-Escude, E.~Waxman, {Signatures of the origin of high-energy cosmic
  rays in cosmological gamma-ray bursts}, Astrophys. J. Lett. 462 (1996)
  L59--L62.
\newblock \href {http://arxiv.org/abs/astro-ph/9601012}
  {\path{arXiv:astro-ph/9601012}}, \href {http://dx.doi.org/10.1086/310042}
  {\path{doi:10.1086/310042}}.

\bibitem{Stadelmaier:2021ajo}
M.~Stadelmaier, M.~Roth, D.~Schmidt, D.~Veberic, {A complete model of the
  signal in surface detector arrays and its application for the reconstruction
  of mass-sensitive observables}, PoS ICRC2021 (2021) 432.
\newblock \href {http://dx.doi.org/10.22323/1.395.0432}
  {\path{doi:10.22323/1.395.0432}}.

\bibitem{2018arXiv180104341S}
T.~{Steininger}, et~al., {Inferring Galactic magnetic field model parameters
  using IMAGINE - An Interstellar MAGnetic field INference Engine}, arXiv
  e-prints (2018) arXiv:1801.04341\href {http://arxiv.org/abs/1801.04341}
  {\path{arXiv:1801.04341}}.

\bibitem{Erdmann:2018cvz}
M.~Erdmann, L.~Geiger, D.~Schmidt, M.~Urban, M.~Wirtz, {Origins of
  Extragalactic Cosmic Ray Nuclei by Contracting Alignment Patterns induced in
  the Galactic Magnetic Field}, Astropart. Phys. 108 (2019) 74--83.
\newblock \href {http://arxiv.org/abs/1807.08734} {\path{arXiv:1807.08734}},
  \href {http://dx.doi.org/10.1016/j.astropartphys.2018.11.004}
  {\path{doi:10.1016/j.astropartphys.2018.11.004}}.

\bibitem{Bister:2020rfv}
T.~Bister, M.~Erdmann, J.~Glombitza, N.~Langner, J.~Schulte, M.~Wirtz,
  {Identification of patterns in cosmic-ray arrival directions using dynamic
  graph convolutional neural networks}, Astropart. Phys. 126 (2021) 102527.
\newblock \href {http://arxiv.org/abs/2003.13038} {\path{arXiv:2003.13038}},
  \href {http://dx.doi.org/10.1016/j.astropartphys.2020.102527}
  {\path{doi:10.1016/j.astropartphys.2020.102527}}.

\bibitem{Wirtz:2021ifo}
M.~Wirtz, T.~Bister, M.~Erdmann, {Towards extracting cosmic magnetic field
  structures from cosmic-ray arrival directions}, Eur. Phys. J. C 81~(9) (2021)
  794.
\newblock \href {http://arxiv.org/abs/2101.02890} {\path{arXiv:2101.02890}},
  \href {http://dx.doi.org/10.1140/epjc/s10052-021-09575-x}
  {\path{doi:10.1140/epjc/s10052-021-09575-x}}.

\bibitem{Bray:2018ipq}
J.~D. Bray, A.~M.~M. Scaife, {An upper limit on the strength of the
  extragalactic magnetic field from ultra-high-energy cosmic-ray anisotropy},
  Astrophys. J. 861~(1) (2018) 3.
\newblock \href {http://arxiv.org/abs/1805.07995} {\path{arXiv:1805.07995}},
  \href {http://dx.doi.org/10.3847/1538-4357/aac777}
  {\path{doi:10.3847/1538-4357/aac777}}.

\bibitem{VanVliet:2021sbc}
A.~Van~Vliet, A.~Palladino, A.~Taylor, W.~Winter, {Extragalactic magnetic field
  constraints from ultra-high-energy cosmic rays from local galaxies}\href
  {http://arxiv.org/abs/2104.05732} {\path{arXiv:2104.05732}}.

\bibitem{Neronov:2021xua}
A.~Neronov, D.~Semikoz, O.~Kalashev, {Limit on intergalactic magnetic field
  from ultra-high-energy cosmic ray hotspot in Perseus-Pisces region}\href
  {http://arxiv.org/abs/2112.08202} {\path{arXiv:2112.08202}}.

\bibitem{McDonald:2001vt}
J.~McDonald, {Thermally generated gauge singlet scalars as selfinteracting dark
  matter}, Phys. Rev. Lett. 88 (2002) 091304.
\newblock \href {http://arxiv.org/abs/hep-ph/0106249}
  {\path{arXiv:hep-ph/0106249}}, \href
  {http://dx.doi.org/10.1103/PhysRevLett.88.091304}
  {\path{doi:10.1103/PhysRevLett.88.091304}}.

\bibitem{Hall:2009bx}
L.~J. Hall, K.~Jedamzik, J.~March-Russell, S.~M. West, {Freeze-In Production of
  FIMP Dark Matter}, JHEP 03 (2010) 080.
\newblock \href {http://arxiv.org/abs/0911.1120} {\path{arXiv:0911.1120}},
  \href {http://dx.doi.org/10.1007/JHEP03(2010)080}
  {\path{doi:10.1007/JHEP03(2010)080}}.

\bibitem{Bernal:2017kxu}
N.~Bernal, M.~Heikinheimo, T.~Tenkanen, K.~Tuominen, V.~Vaskonen, {The Dawn of
  FIMP Dark Matter: A Review of Models and Constraints}, Int. J. Mod. Phys. A
  32~(27) (2017) 1730023.
\newblock \href {http://arxiv.org/abs/1706.07442} {\path{arXiv:1706.07442}},
  \href {http://dx.doi.org/10.1142/S0217751X1730023X}
  {\path{doi:10.1142/S0217751X1730023X}}.

\bibitem{Enqvist:2014zqa}
K.~Enqvist, S.~Nurmi, T.~Tenkanen, K.~Tuominen, {Standard Model with a real
  singlet scalar and inflation}, JCAP 08 (2014) 035.
\newblock \href {http://arxiv.org/abs/1407.0659} {\path{arXiv:1407.0659}},
  \href {http://dx.doi.org/10.1088/1475-7516/2014/08/035}
  {\path{doi:10.1088/1475-7516/2014/08/035}}.

\bibitem{SHDM}
P.~Abreu, et~al., {\bf Pierre Auger} Collaboration, Limits on dark-sector gauge
  coupling from non-observation of instanton-induced decay of super-heavy
  particles in the data of the {Pierre Auger Observatory}, in preparation.

\bibitem{Markkanen:2018bfx}
T.~Markkanen, S.~Nurmi, A.~Rajantie, S.~Stopyra, {The 1-loop effective
  potential for the Standard Model in curved spacetime}, JHEP 06 (2018) 040.
\newblock \href {http://arxiv.org/abs/1804.02020} {\path{arXiv:1804.02020}},
  \href {http://dx.doi.org/10.1007/JHEP06(2018)040}
  {\path{doi:10.1007/JHEP06(2018)040}}.

\bibitem{Guth:1980zm}
A.~H. Guth, {The Inflationary Universe: A Possible Solution to the Horizon and
  Flatness Problems}, Phys. Rev. D 23 (1981) 347--356.
\newblock \href {http://dx.doi.org/10.1103/PhysRevD.23.347}
  {\path{doi:10.1103/PhysRevD.23.347}}.

\bibitem{Linde:1981mu}
A.~D. Linde, {A New Inflationary Universe Scenario: A Possible Solution of the
  Horizon, Flatness, Homogeneity, Isotropy and Primordial Monopole Problems},
  Phys. Lett. B 108 (1982) 389--393.
\newblock \href {http://dx.doi.org/10.1016/0370-2693(82)91219-9}
  {\path{doi:10.1016/0370-2693(82)91219-9}}.

\bibitem{Guth:2013epa}
A.~H. Guth, {Quantum Fluctuations in Cosmology and How They Lead to a
  Multiverse}, in: {25th Solvay Conference on Physics}: {The Theory of the
  Quantum World}, 2013.
\newblock \href {http://arxiv.org/abs/1312.7340} {\path{arXiv:1312.7340}}.

\bibitem{Drees:2017iod}
M.~Drees, F.~Hajkarim, {Dark Matter Production in an Early Matter Dominated
  Era}, JCAP 02 (2018) 057.
\newblock \href {http://arxiv.org/abs/1711.05007} {\path{arXiv:1711.05007}},
  \href {http://dx.doi.org/10.1088/1475-7516/2018/02/057}
  {\path{doi:10.1088/1475-7516/2018/02/057}}.

\bibitem{Kamionkowski:1996zd}
M.~Kamionkowski, A.~Kosowsky, A.~Stebbins, {A Probe of primordial gravity waves
  and vorticity}, Phys. Rev. Lett. 78 (1997) 2058--2061.
\newblock \href {http://arxiv.org/abs/astro-ph/9609132}
  {\path{arXiv:astro-ph/9609132}}, \href
  {http://dx.doi.org/10.1103/PhysRevLett.78.2058}
  {\path{doi:10.1103/PhysRevLett.78.2058}}.

\bibitem{Zaldarriaga:1996xe}
M.~Zaldarriaga, U.~Seljak, {An all sky analysis of polarization in the
  microwave background}, Phys. Rev. D 55 (1997) 1830--1840.
\newblock \href {http://arxiv.org/abs/astro-ph/9609170}
  {\path{arXiv:astro-ph/9609170}}, \href
  {http://dx.doi.org/10.1103/PhysRevD.55.1830}
  {\path{doi:10.1103/PhysRevD.55.1830}}.

\bibitem{Kamionkowski:2015yta}
M.~Kamionkowski, E.~D. Kovetz, {The Quest for B Modes from Inflationary
  Gravitational Waves}, Ann. Rev. Astron. Astrophys. 54 (2016) 227--269.
\newblock \href {http://arxiv.org/abs/1510.06042} {\path{arXiv:1510.06042}},
  \href {http://dx.doi.org/10.1146/annurev-astro-081915-023433}
  {\path{doi:10.1146/annurev-astro-081915-023433}}.

\bibitem{CMB-S4:2016ple}
K.~N. Abazajian, et~al., {\bf CMB-S4} Collaboration, {CMB-S4 Science Book,
  First Edition}\href {http://arxiv.org/abs/1610.02743}
  {\path{arXiv:1610.02743}}.

\bibitem{Marcowith:2020vho}
A.~Marcowith, G.~Ferrand, M.~Grech, Z.~Meliani, I.~Plotnikov, R.~Walder,
  {Multi-scale simulations of particle acceleration in astrophysical systems},
  Liv. Rev. Comput. Astrophys. 6 (2020) 1.
\newblock \href {http://arxiv.org/abs/2002.09411} {\path{arXiv:2002.09411}},
  \href {http://dx.doi.org/10.1007/s41115-020-0007-6}
  {\path{doi:10.1007/s41115-020-0007-6}}.

\bibitem{Fermi:1949ee}
E.~Fermi, {On the Origin of the Cosmic Radiation}, Phys. Rev. 75 (1949)
  1169--1174.
\newblock \href {http://dx.doi.org/10.1103/PhysRev.75.1169}
  {\path{doi:10.1103/PhysRev.75.1169}}.

\bibitem{Parker:1958zza}
E.~N. Parker, {Origin and Dynamics of Cosmic Rays}, Phys. Rev. 109 (1958)
  1328--1344.
\newblock \href {http://dx.doi.org/10.1103/PhysRev.109.1328}
  {\path{doi:10.1103/PhysRev.109.1328}}.

\bibitem{1963ApJ...137..135W}
D.~G. {Wentzel}, {Fermi Acceleration of Charged Particles.}, \apj 137 (1963)
  135.
\newblock \href {http://dx.doi.org/10.1086/147490} {\path{doi:10.1086/147490}}.

\bibitem{Marcowith:2016vzl}
A.~Marcowith, et~al., {The microphysics of collisionless shock waves}, Rept.
  Prog. Phys. 79 (2016) 046901.
\newblock \href {http://arxiv.org/abs/1604.00318} {\path{arXiv:1604.00318}},
  \href {http://dx.doi.org/10.1088/0034-4885/79/4/046901}
  {\path{doi:10.1088/0034-4885/79/4/046901}}.

\bibitem{1977ICRC...11..132A}
W.~I. {Axford}, E.~{Leer}, G.~{Skadron}, {The Acceleration of Cosmic Rays by
  Shock Waves}, in: International Cosmic Ray Conference, Vol.~11 of
  International Cosmic Ray Conference, 1977, p. 132.

\bibitem{1977SPhD...22..327K}
G.~F. {Krymskii}, {A regular mechanism for the acceleration of charged
  particles on the front of a shock wave}, Soviet Physics Doklady 22 (1977)
  327.

\bibitem{Bell:1978zc}
A.~R. Bell, {The Acceleration of cosmic rays in shock fronts. I}, Mon. Not.
  Roy. Astron. Soc. 182 (1978) 147--156.

\bibitem{Blandford:1978ky}
R.~D. Blandford, J.~P. Ostriker, {Particle Acceleration by Astrophysical
  Shocks}, Astrophys. J. Lett. 221 (1978) L29--L32.
\newblock \href {http://dx.doi.org/10.1086/182658} {\path{doi:10.1086/182658}}.

\bibitem{1981ApJ...248..344D}
L.~O. {Drury}, J.~H. {Voelk}, {Hydromagnetic shock structure in the presence of
  cosmic rays}, \apj 248 (1981) 344--351.
\newblock \href {http://dx.doi.org/10.1086/159159} {\path{doi:10.1086/159159}}.

\bibitem{Drury:1983zz}
L.~O. Drury, {An introduction to the theory of diffusive shock acceleration of
  energetic particles in tenuous plasmas}, Rept. Prog. Phys. 46 (1983)
  973--1027.
\newblock \href {http://dx.doi.org/10.1088/0034-4885/46/8/002}
  {\path{doi:10.1088/0034-4885/46/8/002}}.

\bibitem{1986MNRAS.223..353D}
L.~O. {Drury}, S.~A.~E.~G. {Falle}, {On the Stability of Shocks Modified by
  Particle Acceleration}, \mnras 223 (1986) 353.
\newblock \href {http://dx.doi.org/10.1093/mnras/223.2.353}
  {\path{doi:10.1093/mnras/223.2.353}}.

\bibitem{2000MNRAS.314...65L}
S.~G. {Lucek}, A.~R. {Bell}, {Non-linear amplification of a magnetic field
  driven by cosmic ray streaming}, \mnras 314~(1) (2000) 65--74.
\newblock \href {http://dx.doi.org/10.1046/j.1365-8711.2000.03363.x}
  {\path{doi:10.1046/j.1365-8711.2000.03363.x}}.

\bibitem{2004MNRAS.353..550B}
A.~R. {Bell}, {Turbulent amplification of magnetic field and diffusive shock
  acceleration of cosmic rays}, \mnras 353~(2) (2004) 550--558.
\newblock \href {http://dx.doi.org/10.1111/j.1365-2966.2004.08097.x}
  {\path{doi:10.1111/j.1365-2966.2004.08097.x}}.

\bibitem{Zirakashvili:2008bg}
V.~N. Zirakashvili, V.~S. Ptuskin, {Diffusive Shock Acceleration with Magnetic
  Amplification by Non-resonant Streaming Instability in SNRs}, Astrophys. J.
  678 (2008) 939.
\newblock \href {http://arxiv.org/abs/0801.4488} {\path{arXiv:0801.4488}},
  \href {http://dx.doi.org/10.1086/529580} {\path{doi:10.1086/529580}}.

\bibitem{Reville:2008bp}
B.~Reville, S.~O'Sullivan, P.~Duffy, J.~G. Kirk, {The transport of cosmic rays
  in self-excited magnetic turbulence}, Mon. Not. Roy. Astron. Soc. 386 (2008)
  509.
\newblock \href {http://arxiv.org/abs/0802.0109} {\path{arXiv:0802.0109}},
  \href {http://dx.doi.org/10.1111/j.1365-2966.2008.13059.x}
  {\path{doi:10.1111/j.1365-2966.2008.13059.x}}.

\bibitem{Caprioli:2008st}
D.~Caprioli, P.~Blasi, E.~Amato, M.~Vietri, {Dynamical Feedback of
  Self-generated Magnetic Fields in Cosmic Rays Modified Shocks}, Mon. Not.
  Roy. Astron. Soc. 395 (2009) 895--906.
\newblock \href {http://arxiv.org/abs/0807.4261} {\path{arXiv:0807.4261}},
  \href {http://dx.doi.org/10.1111/j.1365-2966.2009.14570.x}
  {\path{doi:10.1111/j.1365-2966.2009.14570.x}}.

\bibitem{Matthews:2017apu}
J.~H. Matthews, A.~R. Bell, K.~M. Blundell, A.~T. Araudo, {Amplification of
  perpendicular and parallel magnetic fields by cosmic ray currents}, Mon. Not.
  Roy. Astron. Soc. 469~(2) (2017) 1849--1860.
\newblock \href {http://arxiv.org/abs/1704.02985} {\path{arXiv:1704.02985}},
  \href {http://dx.doi.org/10.1093/mnras/stx905}
  {\path{doi:10.1093/mnras/stx905}}.

\bibitem{Bell:2019uia}
A.~R. Bell, J.~H. Matthews, K.~M. Blundell, {Cosmic ray acceleration by shocks:
  spectral steepening due to turbulent magnetic field amplification}, Mon. Not.
  Roy. Astron. Soc. 488~(2) (2019) 2466--2472.
\newblock \href {http://arxiv.org/abs/1906.12240} {\path{arXiv:1906.12240}},
  \href {http://dx.doi.org/10.1093/mnras/stz1805}
  {\path{doi:10.1093/mnras/stz1805}}.

\bibitem{Malkov:2019pgq}
M.~Malkov, F.~Aharonian, {Cosmic Ray Spectrum Steepening in Supernova Remnants
  -- I. Loss-Free Self-Similar Solution}, Astrophys. J. 881 (2019) 2.
\newblock \href {http://arxiv.org/abs/1901.01284} {\path{arXiv:1901.01284}},
  \href {http://dx.doi.org/10.3847/1538-4357/ab2c01}
  {\path{doi:10.3847/1538-4357/ab2c01}}.

\bibitem{Ptuskin:2010zn}
V.~Ptuskin, V.~Zirakashvili, E.-S. Seo, {Spectrum of Galactic Cosmic Rays
  Accelerated in Supernova Remnants}, Astrophys. J. 718 (2010) 31--36.
\newblock \href {http://arxiv.org/abs/1006.0034} {\path{arXiv:1006.0034}},
  \href {http://dx.doi.org/10.1088/0004-637X/718/1/31}
  {\path{doi:10.1088/0004-637X/718/1/31}}.

\bibitem{Lemoine:2009vr}
M.~Lemoine, G.~Pelletier, {On electromagnetic instabilities at
  ultra-relativistic shock waves}, Mon. Not. Roy. Astron. Soc. 402 (2010) 321.
\newblock \href {http://arxiv.org/abs/0904.2657} {\path{arXiv:0904.2657}},
  \href {http://dx.doi.org/10.1111/j.1365-2966.2009.15869.x}
  {\path{doi:10.1111/j.1365-2966.2009.15869.x}}.

\bibitem{Reville:2014mta}
B.~Reville, A.~R. Bell, {On the maximum energy of shock-accelerated cosmic rays
  at ultra-relativistic shocks}, Mon. Not. Roy. Astron. Soc. 439~(2) (2014)
  2050--2059.
\newblock \href {http://arxiv.org/abs/1401.2803} {\path{arXiv:1401.2803}},
  \href {http://dx.doi.org/10.1093/mnras/stu088}
  {\path{doi:10.1093/mnras/stu088}}.

\bibitem{Bell:2017zzx}
A.~R. Bell, A.~T. Araudo, J.~H. Matthews, K.~M. Blundell, {Cosmic Ray
  Acceleration by Relativistic Shocks: Limits and Estimates}, Mon. Not. Roy.
  Astron. Soc. 473~(2) (2018) 2364--2371.
\newblock \href {http://arxiv.org/abs/1709.07793} {\path{arXiv:1709.07793}},
  \href {http://dx.doi.org/10.1093/mnras/stx2485}
  {\path{doi:10.1093/mnras/stx2485}}.

\bibitem{Mertens:2016rhi}
F.~Mertens, A.~P. Lobanov, R.~C. Walker, P.~E. Hardee, {Kinematics of the jet
  in M 87 on scales of 100\textendash{}1000 Schwarzschild radii}, Astron.
  Astrophys. 595 (2016) A54.
\newblock \href {http://arxiv.org/abs/1608.05063} {\path{arXiv:1608.05063}},
  \href {http://dx.doi.org/10.1051/0004-6361/201628829}
  {\path{doi:10.1051/0004-6361/201628829}}.

\bibitem{1832RSPT..122..125F}
M.~{Faraday}, {Experimental Researches in Electricity}, Philosophical
  Transactions of the Royal Society of London Series I 122 (1832) 125--162.

\bibitem{Gunn:1969ej}
J.~E. Gunn, J.~P. Ostriker, {Acceleration of high-energy cosmic rays by
  pulsars}, Phys. Rev. Lett. 22 (1969) 728--731.
\newblock \href {http://dx.doi.org/10.1103/PhysRevLett.22.728}
  {\path{doi:10.1103/PhysRevLett.22.728}}.

\bibitem{Blasi:2000xm}
P.~Blasi, R.~I. Epstein, A.~V. Olinto, {Ultrahigh-energy cosmic rays from young
  neutron star winds}, Astrophys. J. Lett. 533 (2000) L123.
\newblock \href {http://arxiv.org/abs/astro-ph/9912240}
  {\path{arXiv:astro-ph/9912240}}, \href {http://dx.doi.org/10.1086/312626}
  {\path{doi:10.1086/312626}}.

\bibitem{Arons:2002yj}
J.~Arons, {Magnetars in the metagalaxy: an origin for ultrahigh-energy cosmic
  rays in the nearby universe}, Astrophys. J. 589 (2003) 871--892.
\newblock \href {http://arxiv.org/abs/astro-ph/0208444}
  {\path{arXiv:astro-ph/0208444}}, \href {http://dx.doi.org/10.1086/374776}
  {\path{doi:10.1086/374776}}.

\bibitem{Fang:2013cba}
K.~Fang, K.~Kotera, A.~V. Olinto, {Ultrahigh Energy Cosmic Ray Nuclei from
  Extragalactic Pulsars and the effect of their Galactic counterparts}, JCAP 03
  (2013) 010.
\newblock \href {http://arxiv.org/abs/1302.4482} {\path{arXiv:1302.4482}},
  \href {http://dx.doi.org/10.1088/1475-7516/2013/03/010}
  {\path{doi:10.1088/1475-7516/2013/03/010}}.

\bibitem{Boldt:1999ge}
E.~Boldt, P.~Ghosh, {Cosmic rays from remnants of quasars?}, Mon. Not. Roy.
  Astron. Soc. 307 (1999) 491--494.
\newblock \href {http://arxiv.org/abs/astro-ph/9902342}
  {\path{arXiv:astro-ph/9902342}}, \href
  {http://dx.doi.org/10.1046/j.1365-8711.1999.02600.x}
  {\path{doi:10.1046/j.1365-8711.1999.02600.x}}.

\bibitem{Boldt:2000dx}
E.~Boldt, M.~Loewenstein, {Cosmic ray generation by quasar remnants:
  Constraints and implications}, Mon. Not. Roy. Astron. Soc. 316 (2000) L29.
\newblock \href {http://arxiv.org/abs/astro-ph/0006221}
  {\path{arXiv:astro-ph/0006221}}, \href
  {http://dx.doi.org/10.1046/j.1365-8711.2000.03768.x}
  {\path{doi:10.1046/j.1365-8711.2000.03768.x}}.

\bibitem{Neronov:2009zz}
A.~Y. Neronov, D.~V. Semikoz, I.~I. Tkachev, {Ultra-High Energy Cosmic Ray
  production in the polar cap regions of black hole magnetospheres}, New J.
  Phys. 11 (2009) 065015.
\newblock \href {http://arxiv.org/abs/0712.1737} {\path{arXiv:0712.1737}},
  \href {http://dx.doi.org/10.1088/1367-2630/11/6/065015}
  {\path{doi:10.1088/1367-2630/11/6/065015}}.

\bibitem{Cerutti:2016ttn}
B.~Cerutti, A.~Beloborodov, {Electrodynamics of pulsar magnetospheres}, Space
  Sci. Rev. 207~(1-4) (2017) 111--136.
\newblock \href {http://arxiv.org/abs/1611.04331} {\path{arXiv:1611.04331}},
  \href {http://dx.doi.org/10.1007/s11214-016-0315-7}
  {\path{doi:10.1007/s11214-016-0315-7}}.

\bibitem{Kotera:2011cp}
K.~Kotera, A.~V. Olinto, {The Astrophysics of Ultrahigh Energy Cosmic Rays},
  Ann. Rev. Astron. Astrophys. 49 (2011) 119--153.
\newblock \href {http://arxiv.org/abs/1101.4256} {\path{arXiv:1101.4256}},
  \href {http://dx.doi.org/10.1146/annurev-astro-081710-102620}
  {\path{doi:10.1146/annurev-astro-081710-102620}}.

\bibitem{Lander:2019guk}
S.~K. Lander, D.~I. Jones, {Magnetar birth: rotation rates and
  gravitational-wave emission}, Mon. Not. Roy. Astron. Soc. 494~(4) (2020)
  4838--4847.
\newblock \href {http://arxiv.org/abs/1910.14336} {\path{arXiv:1910.14336}},
  \href {http://dx.doi.org/10.1093/mnras/staa966}
  {\path{doi:10.1093/mnras/staa966}}.

\bibitem{Blandford:2017chu}
R.~Blandford, Y.~Yuan, M.~Hoshino, L.~Sironi, {Magnetoluminescence}, Space Sci.
  Rev. 207~(1-4) (2017) 291--317.
\newblock \href {http://arxiv.org/abs/1705.02021} {\path{arXiv:1705.02021}},
  \href {http://dx.doi.org/10.1007/s11214-017-0376-2}
  {\path{doi:10.1007/s11214-017-0376-2}}.

\bibitem{Kagan:2014hea}
D.~Kagan, L.~Sironi, B.~Cerutti, D.~Giannios, {Relativistic magnetic
  reconnection in pair plasmas and its astrophysical applications}, Space Sci.
  Rev. 191~(1-4) (2015) 545--573.
\newblock \href {http://arxiv.org/abs/1412.2451} {\path{arXiv:1412.2451}},
  \href {http://dx.doi.org/10.1007/s11214-014-0132-9}
  {\path{doi:10.1007/s11214-014-0132-9}}.

\bibitem{1958IAUS....6..123S}
P.~A. {Sweet}, {The Neutral Point Theory of Solar Flares}, in: B.~{Lehnert}
  (Ed.), Electromagnetic Phenomena in Cosmical Physics, Vol.~6, 1958, p. 123.

\bibitem{Parker:1957abc}
E.~N. Parker,
  \href{https://agupubs.onlinelibrary.wiley.com/doi/abs/10.1029/JZ062i004p00509}{Sweet's
  mechanism for merging magnetic fields in conducting fluids}, Journal of
  Geophysical Research (1896-1977) 62~(4) (1957) 509--520.
\newblock \href
  {http://arxiv.org/abs/https://agupubs.onlinelibrary.wiley.com/doi/pdf/10.1029/JZ062i004p00509}
  {\path{arXiv:https://agupubs.onlinelibrary.wiley.com/doi/pdf/10.1029/JZ062i004p00509}},
  \href {http://dx.doi.org/https://doi.org/10.1029/JZ062i004p00509}
  {\path{doi:https://doi.org/10.1029/JZ062i004p00509}}.
\newline\urlprefix\url{https://agupubs.onlinelibrary.wiley.com/doi/abs/10.1029/JZ062i004p00509}

\bibitem{Melzani:2014jsa}
M.~Melzani, R.~Walder, D.~Folini, C.~Winisdoerffer, J.~M. Favre, {Relativistic
  magnetic reconnection in collisionless ion-electron plasmas explored with
  particle-in-cell simulations}, Astron. Astrophys. 570 (2014) A111.
\newblock \href {http://arxiv.org/abs/1404.7366} {\path{arXiv:1404.7366}},
  \href {http://dx.doi.org/10.1051/0004-6361/201424083}
  {\path{doi:10.1051/0004-6361/201424083}}.

\bibitem{Lazarian:1998wd}
A.~Lazarian, E.~T. Vishniac, {Reconnection in a weakly stochastic field},
  Astrophys. J. 517 (1999) 700--718.
\newblock \href {http://arxiv.org/abs/astro-ph/9811037}
  {\path{arXiv:astro-ph/9811037}}, \href {http://dx.doi.org/10.1086/307233}
  {\path{doi:10.1086/307233}}.

\bibitem{Loureiro:2007gv}
N.~F. Loureiro, A.~A. Schekochihin, S.~C. Cowley, {Instability of current
  sheets and formation of plasmoid chains}, Phys. Plasmas 14 (2007) 100703.
\newblock \href {http://arxiv.org/abs/astro-ph/0703631}
  {\path{arXiv:astro-ph/0703631}}, \href {http://dx.doi.org/10.1063/1.2783986}
  {\path{doi:10.1063/1.2783986}}.

\bibitem{2010PhRvL.105w5002U}
D.~A. {Uzdensky}, N.~F. {Loureiro}, A.~A. {Schekochihin}, {Fast Magnetic
  Reconnection in the Plasmoid-Dominated Regime}, \prl 105~(23) (2010) 235002.
\newblock \href {http://arxiv.org/abs/1008.3330} {\path{arXiv:1008.3330}},
  \href {http://dx.doi.org/10.1103/PhysRevLett.105.235002}
  {\path{doi:10.1103/PhysRevLett.105.235002}}.

\bibitem{DelZanna:2016uiq}
L.~Del~Zanna, E.~Papini, S.~Landi, M.~Bugli, N.~Bucciantini, {Fast reconnection
  in relativistic plasmas: the magnetohydrodynamics tearing instability
  revisited}, Mon. Not. Roy. Astron. Soc. 460~(4) (2016) 3753--3765.
\newblock \href {http://arxiv.org/abs/1605.06331} {\path{arXiv:1605.06331}},
  \href {http://dx.doi.org/10.1093/mnras/stw1242}
  {\path{doi:10.1093/mnras/stw1242}}.

\bibitem{Lyubarsky:2005zt}
Y.~E. Lyubarsky, {On the relativistic magnetic reconnection}, Mon. Not. Roy.
  Astron. Soc. 358 (2005) 113--119.
\newblock \href {http://arxiv.org/abs/astro-ph/0501392}
  {\path{arXiv:astro-ph/0501392}}, \href
  {http://dx.doi.org/10.1111/j.1365-2966.2005.08767.x}
  {\path{doi:10.1111/j.1365-2966.2005.08767.x}}.

\bibitem{Sironi:2014jfa}
L.~Sironi, A.~Spitkovsky, {Relativistic Reconnection: an Efficient Source of
  Non-Thermal Particles}, Astrophys. J. Lett. 783 (2014) L21.
\newblock \href {http://arxiv.org/abs/1401.5471} {\path{arXiv:1401.5471}},
  \href {http://dx.doi.org/10.1088/2041-8205/783/1/L21}
  {\path{doi:10.1088/2041-8205/783/1/L21}}.

\bibitem{Werner:2014spa}
G.~R. Werner, D.~A. Uzdensky, B.~Cerutti, K.~Nalewajko, M.~C. Begelman, {The
  extent of power-law energy spectra in collisionless relativistic magnetic
  reconnection in pair plasmas}, Astrophys. J. Lett. 816~(1) (2016) L8.
\newblock \href {http://arxiv.org/abs/1409.8262} {\path{arXiv:1409.8262}},
  \href {http://dx.doi.org/10.3847/2041-8205/816/1/L8}
  {\path{doi:10.3847/2041-8205/816/1/L8}}.

\bibitem{Guo:2014via}
F.~Guo, H.~Li, W.~Daughton, Y.-H. Liu, {Formation of Hard Power-laws in the
  Energetic Particle Spectra Resulting from Relativistic Magnetic
  Reconnection}, Phys. Rev. Lett. 113 (2014) 155005.
\newblock \href {http://arxiv.org/abs/1405.4040} {\path{arXiv:1405.4040}},
  \href {http://dx.doi.org/10.1103/PhysRevLett.113.155005}
  {\path{doi:10.1103/PhysRevLett.113.155005}}.

\bibitem{1996ApJ...462..997L}
Y.~E. {Litvinenko}, {Particle Acceleration in Reconnecting Current Sheets with
  a Nonzero Magnetic Field}, \apj 462 (1996) 997.
\newblock \href {http://dx.doi.org/10.1086/177213} {\path{doi:10.1086/177213}}.

\bibitem{Kirk:2004ap}
J.~G. Kirk, {Particle acceleration in relativistic current sheets}, Phys. Rev.
  Lett. 92 (2004) 181101.
\newblock \href {http://arxiv.org/abs/astro-ph/0403516}
  {\path{arXiv:astro-ph/0403516}}, \href
  {http://dx.doi.org/10.1103/PhysRevLett.92.181101}
  {\path{doi:10.1103/PhysRevLett.92.181101}}.

\bibitem{deGouveiaDalPino:2003mu}
E.~M. de~Gouveia Dal~Pino, A.~Lazarian, {Production of the large scale
  superluminal ejections of the microquasar GRS 1915+105 by violent magnetic
  reconnection}\href {http://arxiv.org/abs/astro-ph/0307054}
  {\path{arXiv:astro-ph/0307054}}.

\bibitem{2012MNRAS.422.2474D}
L.~O. {Drury}, {First-order Fermi acceleration driven by magnetic
  reconnection}, \mnras 422~(3) (2012) 2474--2476.
\newblock \href {http://arxiv.org/abs/1201.6612} {\path{arXiv:1201.6612}},
  \href {http://dx.doi.org/10.1111/j.1365-2966.2012.20804.x}
  {\path{doi:10.1111/j.1365-2966.2012.20804.x}}.

\bibitem{deGouveiaDalPino:2013qng}
E.~M. de~Gouveia Dal~Pino, G.~Kowal, {Particle Acceleration by Magnetic
  Reconnection}\href {http://arxiv.org/abs/1302.4374} {\path{arXiv:1302.4374}},
  \href {http://dx.doi.org/10.1007/978-3-662-44625-6_13}
  {\path{doi:10.1007/978-3-662-44625-6_13}}.

\bibitem{Guo:2015ydj}
F.~Guo, X.~Li, H.~Li, W.~Daughton, B.~Zhang, N.~Lloyd-Ronning, Y.-H. Liu,
  H.~Zhang, W.~Deng, {Efficient Production of High-energy Nonthermal Particles
  During Magnetic Reconnection in a Magnetically Dominated
  Ion\textendash{}electron Plasma}, Astrophys. J. Lett. 818~(1) (2016) L9.
\newblock \href {http://arxiv.org/abs/1511.01434} {\path{arXiv:1511.01434}},
  \href {http://dx.doi.org/10.3847/2041-8205/818/1/L9}
  {\path{doi:10.3847/2041-8205/818/1/L9}}.

\bibitem{broadbent}
A.~Broadbent, C.~G.~T. Haslam, J.~L. Osborne, {A Detailed Model of the
  Synchrotron Radiation in the Galactic Disk}, Proc.\ 21st ICRC 3 (1990) 229.

\bibitem{Jansson:2012pc}
R.~Jansson, G.~R. Farrar, {A New Model of the Galactic Magnetic Field},
  Astrophys. J. 757 (2012) 14.
\newblock \href {http://arxiv.org/abs/1204.3662} {\path{arXiv:1204.3662}},
  \href {http://dx.doi.org/10.1088/0004-637X/757/1/14}
  {\path{doi:10.1088/0004-637X/757/1/14}}.

\bibitem{Beck:2013bxa}
R.~Beck, R.~Wielebinski, {Magnetic Fields in the Milky Way and in Galaxies},
  Springer Netherlands, 2013.
\newblock \href {http://arxiv.org/abs/1302.5663} {\path{arXiv:1302.5663}},
  \href {http://dx.doi.org/10.1007/978-94-007-5612-0_13}
  {\path{doi:10.1007/978-94-007-5612-0_13}}.

\bibitem{Farrar:2014hma}
G.~R. Farrar, {The Galactic Magnetic Field and Ultrahigh-Energy Cosmic Ray
  Deflections}, Comptes Rendus Physique 15 (2014) 339--348.
\newblock \href {http://arxiv.org/abs/1405.3680} {\path{arXiv:1405.3680}},
  \href {http://dx.doi.org/10.1016/j.crhy.2014.04.002}
  {\path{doi:10.1016/j.crhy.2014.04.002}}.

\bibitem{2015A&ARv..24....4B}
R.~{Beck}, {Magnetic fields in spiral galaxies}, A\&A~Rev. 24 (2015) 4.
\newblock \href {http://arxiv.org/abs/1509.04522} {\path{arXiv:1509.04522}},
  \href {http://dx.doi.org/10.1007/s00159-015-0084-4}
  {\path{doi:10.1007/s00159-015-0084-4}}.

\bibitem{haverkorn:2014}
M.~{Haverkorn}, {Magnetic Fields in the Milky Way}, in: A.~{Lazarian}, E.~M.
  {de Gouveia Dal Pino}, C.~{Melioli} (Eds.), Magnetic Fields in Diffuse Media,
  Vol. 407 of Astrophysics and Space Science Library, 2015, p. 483.
\newblock \href {http://arxiv.org/abs/1406.0283} {\path{arXiv:1406.0283}},
  \href {http://dx.doi.org/10.1007/978-3-662-44625-6\_17}
  {\path{doi:10.1007/978-3-662-44625-6\_17}}.

\bibitem{Unger:2017kfh}
M.~Unger, G.~R. Farrar, Uncertainties in the magnetic field of the milky way,
  2017.
\newblock \href {http://arxiv.org/abs/1707.02339} {\path{arXiv:1707.02339}}.

\bibitem{Jaffe:2013yi}
T.~R. Jaffe, et~al., {Comparing Polarised Synchrotron and Thermal Dust Emission
  in the Galactic Plane}, Mon. Not. Roy. Astron. Soc. 431 (2013) 683.
\newblock \href {http://arxiv.org/abs/1302.0143} {\path{arXiv:1302.0143}},
  \href {http://dx.doi.org/10.1093/mnras/stt200}
  {\path{doi:10.1093/mnras/stt200}}.

\bibitem{Orlando:2013ysa}
E.~Orlando, A.~Strong, {Galactic synchrotron emission with cosmic ray
  propagation models}, Mon. Not. Roy. Astron. Soc. 436 (2013) 2127.
\newblock \href {http://arxiv.org/abs/1309.2947} {\path{arXiv:1309.2947}},
  \href {http://dx.doi.org/10.1093/mnras/stt1718}
  {\path{doi:10.1093/mnras/stt1718}}.

\bibitem{Brown:2007qv}
J.~C. Brown, et~al., {Rotation Measures of Extragalactic Sources Behind the
  Southern Galactic Plane: New Insights into the Large-Scale Magnetic Field of
  the Inner Milky Way}, Astrophys. J. 663 (2007) 258--266.
\newblock \href {http://arxiv.org/abs/0704.0458} {\path{arXiv:0704.0458}},
  \href {http://dx.doi.org/10.1086/518499} {\path{doi:10.1086/518499}}.

\bibitem{Braun:2015zta}
R.~Braun, T.~Bourke, J.~A. Green, E.~Keane, J.~Wagg, {Advancing Astrophysics
  with the Square Kilometre Array}, PoS AASKA14 (2015) 174.
\newblock \href {http://dx.doi.org/10.22323/1.215.0174}
  {\path{doi:10.22323/1.215.0174}}.

\bibitem{SKAMagnetismScienceWorkingGroup:2020xim}
G.~Heald, et~al., {\bf SKA Magnetism Science Working Group} Collaboration,
  {Magnetism Science with the Square Kilometre Array}, Galaxies 8~(3) (2020)
  53.
\newblock \href {http://arxiv.org/abs/2006.03172} {\path{arXiv:2006.03172}},
  \href {http://dx.doi.org/10.3390/galaxies8030053}
  {\path{doi:10.3390/galaxies8030053}}.

\bibitem{Smits:2011zh}
R.~Smits, S.~J. Tingay, N.~Wex, M.~Kramer, B.~Stappers, {Prospects for accurate
  distance measurements of pulsars with the SKA: enabling fundamental physics},
  Astron. Astrophys. 528 (2011) A108.
\newblock \href {http://arxiv.org/abs/1101.5971} {\path{arXiv:1101.5971}},
  \href {http://dx.doi.org/10.1051/0004-6361/201016141}
  {\path{doi:10.1051/0004-6361/201016141}}.

\bibitem{ASKAP:2007rlq}
S.~Johnston, et~al., {\bf ASKAP} Collaboration, {Science With The Australian
  Square Kilometre Array Pathfinder}, PoS MRU (2007) 006.
\newblock \href {http://arxiv.org/abs/0711.2103} {\path{arXiv:0711.2103}},
  \href {http://dx.doi.org/10.1071/AS07033} {\path{doi:10.1071/AS07033}}.

\bibitem{2010AAS...21547013G}
B.~M. {Gaensler}, et~al., {\bf POSSUM} Collaboration, {Survey Science with
  ASKAP: Polarization Sky Survey of the Universe's Magnetism (POSSUM)}, Vol.
  215 of American Astronomical Society Meeting Abstracts, 2010, p. 470.13.

\bibitem{vanHaarlem:2013dsa}
M.~P. van Haarlem, et~al., {LOFAR: The LOw-Frequency ARray}, Astron. Astrophys.
  556 (2013) A2.
\newblock \href {http://arxiv.org/abs/1305.3550} {\path{arXiv:1305.3550}},
  \href {http://dx.doi.org/10.1051/0004-6361/201220873}
  {\path{doi:10.1051/0004-6361/201220873}}.

\bibitem{Jarvis:2017aml}
M.~J. Jarvis, et~al., {The MeerKAT International GHz Tiered Extragalactic
  Exploration (MIGHTEE) Survey}, PoS MeerKAT2016 (2018) 006.
\newblock \href {http://arxiv.org/abs/1709.01901} {\path{arXiv:1709.01901}},
  \href {http://dx.doi.org/10.22323/1.277.0006}
  {\path{doi:10.22323/1.277.0006}}.

\bibitem{2020A&A...639A.111S}
Y.~{Stein}, et~al., {CHANG-ES. XXI. Transport processes and the X-shaped
  magnetic field of NGC 4217: off-center superbubble structure}, \aap 639
  (2020) A111.
\newblock \href {http://arxiv.org/abs/2007.03002} {\path{arXiv:2007.03002}},
  \href {http://dx.doi.org/10.1051/0004-6361/202037675}
  {\path{doi:10.1051/0004-6361/202037675}}.

\bibitem{2020A&A...639A.112K}
M.~{Krause}, et~al., {CHANG-ES. XXII. Coherent magnetic fields in the halos of
  spiral galaxies}, \aap 639 (2020) A112.
\newblock \href {http://arxiv.org/abs/2004.14383} {\path{arXiv:2004.14383}},
  \href {http://dx.doi.org/10.1051/0004-6361/202037780}
  {\path{doi:10.1051/0004-6361/202037780}}.

\bibitem{Heald:2021wnt}
G.~H. Heald, et~al., {CHANG-ES XXIII: influence of a galactic wind in
  NGC~5775}, Mon. Not. Roy. Astron. Soc. 509~(1) (2021) 658--684.
\newblock \href {http://arxiv.org/abs/2109.12267} {\path{arXiv:2109.12267}},
  \href {http://dx.doi.org/10.1093/mnras/stab2804}
  {\path{doi:10.1093/mnras/stab2804}}.

\bibitem{2018arXiv181005652T}
K.~{Tassis}, et~al., {PASIPHAE: A high-Galactic-latitude, high-accuracy
  optopolarimetric survey}, arXiv e-prints (2018) arXiv:1810.05652\href
  {http://arxiv.org/abs/1810.05652} {\path{arXiv:1810.05652}}.

\bibitem{Boulanger:2018zrk}
F.~Boulanger, et~al., {IMAGINE: A comprehensive view of the interstellar
  medium, Galactic magnetic fields and cosmic rays}, JCAP 08 (2018) 049.
\newblock \href {http://arxiv.org/abs/1805.02496} {\path{arXiv:1805.02496}},
  \href {http://dx.doi.org/10.1088/1475-7516/2018/08/049}
  {\path{doi:10.1088/1475-7516/2018/08/049}}.

\bibitem{Keane:2014vja}
E.~F. Keane, et~al., {A Cosmic Census of Radio Pulsars with the SKA}, PoS
  AASKA14 (2015) 040.
\newblock \href {http://arxiv.org/abs/1501.00056} {\path{arXiv:1501.00056}},
  \href {http://dx.doi.org/10.22323/1.215.0040}
  {\path{doi:10.22323/1.215.0040}}.

\bibitem{Kulsrud:2007an}
R.~M. Kulsrud, E.~G. Zweibel, {The Origin of Astrophysical Magnetic Fields},
  Rept. Prog. Phys. 71 (2008) 0046091.
\newblock \href {http://arxiv.org/abs/0707.2783} {\path{arXiv:0707.2783}},
  \href {http://dx.doi.org/10.1088/0034-4885/71/4/046901}
  {\path{doi:10.1088/0034-4885/71/4/046901}}.

\bibitem{Vachaspati:2020blt}
T.~Vachaspati, {Progress on cosmological magnetic fields}, Rept. Prog. Phys.
  84~(7) (2021) 074901.
\newblock \href {http://arxiv.org/abs/2010.10525} {\path{arXiv:2010.10525}},
  \href {http://dx.doi.org/10.1088/1361-6633/ac03a9}
  {\path{doi:10.1088/1361-6633/ac03a9}}.

\bibitem{Barrow:1997mj}
J.~D. Barrow, P.~G. Ferreira, J.~Silk, {Constraints on a primordial magnetic
  field}, Phys. Rev. Lett. 78 (1997) 3610--3613.
\newblock \href {http://arxiv.org/abs/astro-ph/9701063}
  {\path{arXiv:astro-ph/9701063}}, \href
  {http://dx.doi.org/10.1103/PhysRevLett.78.3610}
  {\path{doi:10.1103/PhysRevLett.78.3610}}.

\bibitem{Jedamzik:1999bm}
K.~Jedamzik, V.~Katalinic, A.~V. Olinto, {A Limit on primordial small scale
  magnetic fields from CMB distortions}, Phys. Rev. Lett. 85 (2000) 700--703.
\newblock \href {http://arxiv.org/abs/astro-ph/9911100}
  {\path{arXiv:astro-ph/9911100}}, \href
  {http://dx.doi.org/10.1103/PhysRevLett.85.700}
  {\path{doi:10.1103/PhysRevLett.85.700}}.

\bibitem{Planck:2013pxb}
P.~A.~R. Ade, et~al., {\bf Planck} Collaboration, {Planck 2013 results. XVI.
  Cosmological parameters}, Astron. Astrophys. 571 (2014) A16.
\newblock \href {http://arxiv.org/abs/1303.5076} {\path{arXiv:1303.5076}},
  \href {http://dx.doi.org/10.1051/0004-6361/201321591}
  {\path{doi:10.1051/0004-6361/201321591}}.

\bibitem{Jedamzik:2018itu}
K.~Jedamzik, A.~Saveliev, {Stringent Limit on Primordial Magnetic Fields from
  the Cosmic Microwave Background Radiation}, Phys. Rev. Lett. 123~(2) (2019)
  021301.
\newblock \href {http://arxiv.org/abs/1804.06115} {\path{arXiv:1804.06115}},
  \href {http://dx.doi.org/10.1103/PhysRevLett.123.021301}
  {\path{doi:10.1103/PhysRevLett.123.021301}}.

\bibitem{Akahori:2014qja}
T.~Akahori, K.~Kumazaki, K.~Takahashi, D.~Ryu, {Exploring the Intergalactic
  Magnetic Field by Means of Faraday Tomography}, Publ. Astron. Soc. Jap.
  66~(3) (2014) 65.
\newblock \href {http://arxiv.org/abs/1403.0325} {\path{arXiv:1403.0325}},
  \href {http://dx.doi.org/10.1093/pasj/psu033}
  {\path{doi:10.1093/pasj/psu033}}.

\bibitem{Ravi:2016kfj}
V.~Ravi, et~al., {The magnetic field and turbulence of the cosmic web measured
  using a brilliant fast radio burst}, Science 354 (2016) 1249.
\newblock \href {http://arxiv.org/abs/1611.05758} {\path{arXiv:1611.05758}},
  \href {http://dx.doi.org/10.1126/science.aaf6807}
  {\path{doi:10.1126/science.aaf6807}}.

\bibitem{OSullivan:2018shr}
S.~P. O'Sullivan, et~al., {The intergalactic magnetic field probed by a giant
  radio galaxy}, Astron. Astrophys. 622 (2019) A16.
\newblock \href {http://arxiv.org/abs/1811.07934} {\path{arXiv:1811.07934}},
  \href {http://dx.doi.org/10.1051/0004-6361/201833832}
  {\path{doi:10.1051/0004-6361/201833832}}.

\bibitem{OSullivan:2018adp}
S.~P. O'Sullivan, et~al., {Untangling Cosmic Magnetic Fields: Faraday
  Tomography at Metre Wavelengths with LOFAR}, Galaxies 6~(4) (2018) 126.
\newblock \href {http://arxiv.org/abs/1811.12732} {\path{arXiv:1811.12732}},
  \href {http://dx.doi.org/10.3390/galaxies6040126}
  {\path{doi:10.3390/galaxies6040126}}.

\bibitem{Vernstrom:2019gjr}
T.~Vernstrom, B.~Gaensler, L.~Rudnick, H.~Andernach, {Differences in Faraday
  Rotation Between Adjacent Extragalactic Radio Sources as a Probe of Cosmic
  Magnetic Fields}, Astrophys. J. 878~(2) (2019) 92.
\newblock \href {http://arxiv.org/abs/1905.02410} {\path{arXiv:1905.02410}},
  \href {http://dx.doi.org/10.3847/1538-4357/ab1f83}
  {\path{doi:10.3847/1538-4357/ab1f83}}.

\bibitem{Amaral:2021mly}
A.~D. Amaral, T.~Vernstrom, B.~M. Gaensler, {Constraints on Large-Scale
  Magnetic Fields in the Intergalactic Medium Using Cross-Correlation Methods},
  Mon. Not. Roy. Astron. Soc. 503~(2) (2021) 2913--2926.
\newblock \href {http://arxiv.org/abs/2102.11312} {\path{arXiv:2102.11312}},
  \href {http://dx.doi.org/10.1093/mnras/stab564}
  {\path{doi:10.1093/mnras/stab564}}.

\bibitem{Vernstrom:2017jvh}
T.~Vernstrom, et~al., {Low Frequency Radio Constraints on the Synchrotron
  Cosmic Web}, Mon. Not. Roy. Astron. Soc. 467~(4) (2017) 4914--4936.
\newblock \href {http://arxiv.org/abs/1702.05069} {\path{arXiv:1702.05069}},
  \href {http://dx.doi.org/10.1093/mnras/stx424}
  {\path{doi:10.1093/mnras/stx424}}.

\bibitem{2019Sci...364..981G}
F.~{Govoni}, et~al., {A radio ridge connecting two galaxy clusters in a
  filament of the cosmic web}, Science 364~(6444) (2019) 981--984.
\newblock \href {http://arxiv.org/abs/1906.07584} {\path{arXiv:1906.07584}},
  \href {http://dx.doi.org/10.1126/science.aat7500}
  {\path{doi:10.1126/science.aat7500}}.

\bibitem{Locatelli:2021byc}
N.~Locatelli, F.~Vazza, A.~Bonafede, S.~Banfi, G.~Bernardi, C.~Gheller,
  A.~Botteon, T.~Shimwell, {New constraints on the magnetic field in filaments
  of the cosmic web}, Astron. Astrophys. 652 (2021) A80.
\newblock \href {http://arxiv.org/abs/2101.06051} {\path{arXiv:2101.06051}},
  \href {http://dx.doi.org/10.1051/0004-6361/202140526}
  {\path{doi:10.1051/0004-6361/202140526}}.

\bibitem{Vernstrom:2021hru}
T.~Vernstrom, G.~Heald, F.~Vazza, T.~J. Galvin, J.~West, N.~Locatelli,
  N.~Fornengo, E.~Pinetti, {Discovery of magnetic fields along stacked cosmic
  filaments as revealed by radio and X-ray emission}, Mon. Not. Roy. Astron.
  Soc. 505~(3) (2021) 4178--4196.
\newblock \href {http://arxiv.org/abs/2101.09331} {\path{arXiv:2101.09331}},
  \href {http://dx.doi.org/10.1093/mnras/stab1301}
  {\path{doi:10.1093/mnras/stab1301}}.

\bibitem{Hodgson:2021rzd}
T.~Hodgson, M.~Johnston-Hollitt, B.~McKinley, N.~Hurley-Walker, {On Detecting
  the Synchrotron Cosmic Web Twice}\href {http://arxiv.org/abs/2112.01754}
  {\path{arXiv:2112.01754}}.

\bibitem{Aharonian:1993vz}
F.~A. Aharonian, P.~S. Coppi, H.~J. Voelk, {Very high-energy gamma-rays from
  AGN: Cascading on the cosmic background radiation fields and the formation of
  pair halos}, Astrophys. J. Lett. 423 (1994) L5--L8.
\newblock \href {http://arxiv.org/abs/astro-ph/9312045}
  {\path{arXiv:astro-ph/9312045}}, \href {http://dx.doi.org/10.1086/187222}
  {\path{doi:10.1086/187222}}.

\bibitem{Plaga:1995ins}
R.~Plaga, {Detecting intergalactic magnetic fields using time delays in pulses
  of \ensuremath{\gamma}-rays}, Nature 374~(6521) (1995) 430--432.
\newblock \href {http://dx.doi.org/10.1038/374430a0}
  {\path{doi:10.1038/374430a0}}.

\bibitem{Ryu:2011hu}
D.~Ryu, D.~R.~G. Schleicher, R.~A. Treumann, C.~G. Tsagas, L.~M. Widrow,
  {Magnetic fields in the Large-Scale Structure of the Universe}, Space Sci.
  Rev. 166 (2012) 1--35.
\newblock \href {http://arxiv.org/abs/1109.4055} {\path{arXiv:1109.4055}},
  \href {http://dx.doi.org/10.1007/s11214-011-9839-z}
  {\path{doi:10.1007/s11214-011-9839-z}}.

\bibitem{OSullivan:2020pll}
S.~P. O'Sullivan, et~al., {New constraints on the magnetization of the cosmic
  web using LOFAR Faraday rotation observations}, Mon. Not. Roy. Astron. Soc.
  495~(3) (2020) 2607--2619.
\newblock \href {http://arxiv.org/abs/2002.06924} {\path{arXiv:2002.06924}},
  \href {http://dx.doi.org/10.1093/mnras/staa1395}
  {\path{doi:10.1093/mnras/staa1395}}.

\bibitem{Gheller:2019wlf}
C.~Gheller, F.~Vazza, {A survey of the thermal and non-thermal properties of
  cosmic filaments}, Mon. Not. Roy. Astron. Soc. 486~(1) (2019) 981--1002.
\newblock \href {http://arxiv.org/abs/1903.08401} {\path{arXiv:1903.08401}},
  \href {http://dx.doi.org/10.1093/mnras/stz843}
  {\path{doi:10.1093/mnras/stz843}}.

\bibitem{Vazza:2021vwy}
F.~Vazza, et~al., {Magnetogenesis and the Cosmic Web: A Joint Challenge for
  Radio Observations and Numerical Simulations}, Galaxies 9~(4) (2021) 109.
\newblock \href {http://arxiv.org/abs/2111.09129} {\path{arXiv:2111.09129}},
  \href {http://dx.doi.org/10.3390/galaxies9040109}
  {\path{doi:10.3390/galaxies9040109}}.

\bibitem{Vazza:2017qge}
F.~Vazza, M.~Br\"uggen, C.~Gheller, S.~Hackstein, D.~Wittor, P.~M. Hinz,
  {Simulations of extragalactic magnetic fields and of their observables},
  Class. Quant. Grav. 34~(23) (2017) 234001.
\newblock \href {http://arxiv.org/abs/1711.02669} {\path{arXiv:1711.02669}},
  \href {http://dx.doi.org/10.1088/1361-6382/aa8e60}
  {\path{doi:10.1088/1361-6382/aa8e60}}.

\bibitem{Neronov:2010gir}
A.~Neronov, I.~Vovk, {Evidence for strong extragalactic magnetic fields from
  Fermi observations of TeV blazars}, Science 328 (2010) 73--75.
\newblock \href {http://arxiv.org/abs/1006.3504} {\path{arXiv:1006.3504}},
  \href {http://dx.doi.org/10.1126/science.1184192}
  {\path{doi:10.1126/science.1184192}}.

\bibitem{Tavecchio:2010ja}
F.~Tavecchio, G.~Ghisellini, G.~Bonnoli, L.~Foschini, {Extreme TeV blazars and
  the intergalactic magnetic field}, Mon. Not. Roy. Astron. Soc. 414 (2011)
  3566.
\newblock \href {http://arxiv.org/abs/1009.1048} {\path{arXiv:1009.1048}},
  \href {http://dx.doi.org/10.1111/j.1365-2966.2011.18657.x}
  {\path{doi:10.1111/j.1365-2966.2011.18657.x}}.

\bibitem{Dermer:2010mm}
C.~D. Dermer, M.~Cavadini, S.~Razzaque, J.~D. Finke, J.~Chiang, B.~Lott, {Time
  Delay of Cascade Radiation for TeV Blazars and the Measurement of the
  Intergalactic Magnetic Field}, Astrophys. J. Lett. 733 (2011) L21.
\newblock \href {http://arxiv.org/abs/1011.6660} {\path{arXiv:1011.6660}},
  \href {http://dx.doi.org/10.1088/2041-8205/733/2/L21}
  {\path{doi:10.1088/2041-8205/733/2/L21}}.

\bibitem{Finke:2015ona}
J.~D. Finke, L.~C. Reyes, M.~Georganopoulos, K.~Reynolds, M.~Ajello, S.~J.
  Fegan, K.~McCann, {Constraints on the Intergalactic Magnetic Field with
  Gamma-Ray Observations of Blazars}, Astrophys. J. 814~(1) (2015) 20.
\newblock \href {http://arxiv.org/abs/1510.02485} {\path{arXiv:1510.02485}},
  \href {http://dx.doi.org/10.1088/0004-637X/814/1/20}
  {\path{doi:10.1088/0004-637X/814/1/20}}.

\bibitem{Veres:2017aou}
P.~Veres, C.~D. Dermer, K.~S. Dhuga, {Properties of the Intergalactic Magnetic
  Field Constrained by Gamma-ray Observations of Gamma-Ray Bursts}, Astrophys.
  J. 847~(1) (2017) 39.
\newblock \href {http://arxiv.org/abs/1705.08531} {\path{arXiv:1705.08531}},
  \href {http://dx.doi.org/10.3847/1538-4357/aa87b1}
  {\path{doi:10.3847/1538-4357/aa87b1}}.

\bibitem{Fermi-LAT:2018jdy}
M.~Ackermann, et~al., {\bf Fermi-LAT} Collaboration, {The Search for Spatial
  Extension in High-latitude Sources Detected by the $Fermi$ Large Area
  Telescope}, Astrophys. J. Suppl. 237~(2) (2018) 32.
\newblock \href {http://arxiv.org/abs/1804.08035} {\path{arXiv:1804.08035}},
  \href {http://dx.doi.org/10.3847/1538-4365/aacdf7}
  {\path{doi:10.3847/1538-4365/aacdf7}}.

\bibitem{AlvesBatista:2020oio}
R.~Alves~Batista, A.~Saveliev, {Multimessenger Constraints on Intergalactic
  Magnetic Fields from the Flare of TXS 0506+056}, Astrophys. J. Lett. 902~(1)
  (2020) L11.
\newblock \href {http://arxiv.org/abs/2009.12161} {\path{arXiv:2009.12161}},
  \href {http://dx.doi.org/10.3847/2041-8213/abb816}
  {\path{doi:10.3847/2041-8213/abb816}}.

\bibitem{Garaldi:2020xos}
E.~Garaldi, R.~Pakmor, V.~Springel, {Magnetogenesis around the first galaxies:
  the impact of different field seeding processes on galaxy formation}, Mon.
  Not. Roy. Astron. Soc. 502~(4) (2021) 5726--5744.
\newblock \href {http://arxiv.org/abs/2010.09729} {\path{arXiv:2010.09729}},
  \href {http://dx.doi.org/10.1093/mnras/stab086}
  {\path{doi:10.1093/mnras/stab086}}.

\bibitem{Martin-Alvarez:2020enk}
S.~Martin-Alvarez, H.~Katz, D.~Sijacki, J.~Devriendt, A.~Slyz, {Unravelling the
  origin of magnetic fields in galaxies}, Mon. Not. Roy. Astron. Soc. 504~(2)
  (2021) 2517.
\newblock \href {http://arxiv.org/abs/2011.11648} {\path{arXiv:2011.11648}},
  \href {http://dx.doi.org/10.1093/mnras/stab968}
  {\path{doi:10.1093/mnras/stab968}}.

\bibitem{Katz:2021iou}
H.~Katz, et~al., {Introducing SPHINX-MHD: the impact of primordial magnetic
  fields on the first galaxies, reionization, and the global 21-cm signal},
  Mon. Not. Roy. Astron. Soc. 507~(1) (2021) 1254--1282.
\newblock \href {http://arxiv.org/abs/2101.11624} {\path{arXiv:2101.11624}},
  \href {http://dx.doi.org/10.1093/mnras/stab2148}
  {\path{doi:10.1093/mnras/stab2148}}.

\bibitem{Mtchedlidze:2021bfy}
S.~Mtchedlidze, P.~Dom\'\i{}nguez-Fern\'andez, X.~Du, A.~Brandenburg,
  T.~Kahniashvili, S.~O'Sullivan, W.~Schmidt, M.~Br\"uggen, {Evolution of
  primordial magnetic fields during large-scale structure formation}\href
  {http://arxiv.org/abs/2109.13520} {\path{arXiv:2109.13520}}.

\bibitem{Batista:2021rgm}
R.~Alves~Batista, A.~Saveliev, {The Gamma-ray Window to Intergalactic
  Magnetism}, Universe 7~(7) (2021) 223.
\newblock \href {http://arxiv.org/abs/2105.12020} {\path{arXiv:2105.12020}},
  \href {http://dx.doi.org/10.3390/universe7070223}
  {\path{doi:10.3390/universe7070223}}.

\bibitem{Sigl:2003ay}
G.~Sigl, F.~Miniati, T.~A. Ensslin, {Ultrahigh-energy cosmic rays in a
  structured and magnetized universe}, Phys. Rev. D 68 (2003) 043002.
\newblock \href {http://arxiv.org/abs/astro-ph/0302388}
  {\path{arXiv:astro-ph/0302388}}, \href
  {http://dx.doi.org/10.1103/PhysRevD.68.043002}
  {\path{doi:10.1103/PhysRevD.68.043002}}.

\bibitem{Dolag:2004kp}
K.~Dolag, D.~Grasso, V.~Springel, I.~Tkachev, {Constrained simulations of the
  magnetic field in the local Universe and the propagation of UHECRs}, JCAP 01
  (2005) 009.
\newblock \href {http://arxiv.org/abs/astro-ph/0410419}
  {\path{arXiv:astro-ph/0410419}}, \href
  {http://dx.doi.org/10.1088/1475-7516/2005/01/009}
  {\path{doi:10.1088/1475-7516/2005/01/009}}.

\bibitem{Hackstein:2017pex}
S.~Hackstein, F.~Vazza, M.~Br\"uggen, J.~G. Sorce, S.~Gottl\"ober, {Simulations
  of ultra-high Energy Cosmic Rays in the local Universe and the origin of
  Cosmic Magnetic Fields}, Mon. Not. Roy. Astron. Soc. 475~(2) (2018)
  2519--2529.
\newblock \href {http://arxiv.org/abs/1710.01353} {\path{arXiv:1710.01353}},
  \href {http://dx.doi.org/10.1093/mnras/stx3354}
  {\path{doi:10.1093/mnras/stx3354}}.

\bibitem{Forero-Romero:2008svv}
J.~E. Forero-Romero, Y.~Hoffman, S.~Gottloeber, A.~Klypin, G.~Yepes, {A
  Dynamical Classification of the Cosmic Web}, Mon. Not. Roy. Astron. Soc. 396
  (2009) 1815--1824.
\newblock \href {http://arxiv.org/abs/0809.4135} {\path{arXiv:0809.4135}},
  \href {http://dx.doi.org/10.1111/j.1365-2966.2009.14885.x}
  {\path{doi:10.1111/j.1365-2966.2009.14885.x}}.

\bibitem{Kahniashvili:2005yp}
T.~Kahniashvili, T.~Vachaspati, {On the detection of magnetic helicity}, Phys.
  Rev. D 73 (2006) 063507.
\newblock \href {http://arxiv.org/abs/astro-ph/0511373}
  {\path{arXiv:astro-ph/0511373}}, \href
  {http://dx.doi.org/10.1103/PhysRevD.73.063507}
  {\path{doi:10.1103/PhysRevD.73.063507}}.

\bibitem{AlvesBatista:2018owq}
R.~Alves~Batista, A.~Saveliev, {On the Measurement of the Helicity of
  Intergalactic Magnetic Fields Using Ultra-High-Energy Cosmic Rays}, JCAP 03
  (2019) 011.
\newblock \href {http://arxiv.org/abs/1808.04182} {\path{arXiv:1808.04182}},
  \href {http://dx.doi.org/10.1088/1475-7516/2019/03/011}
  {\path{doi:10.1088/1475-7516/2019/03/011}}.

\bibitem{Sigl:2004yk}
G.~Sigl, F.~Miniati, T.~A. Ensslin, {Ultrahigh energy cosmic ray probes of
  large scale structure and magnetic fields}, Phys. Rev. D 70 (2004) 043007.
\newblock \href {http://arxiv.org/abs/astro-ph/0401084}
  {\path{arXiv:astro-ph/0401084}}, \href
  {http://dx.doi.org/10.1103/PhysRevD.70.043007}
  {\path{doi:10.1103/PhysRevD.70.043007}}.

\bibitem{Armengaud:2004yt}
E.~Armengaud, G.~Sigl, F.~Miniati, {Ultrahigh energy nuclei propagation in a
  structured, magnetized Universe}, Phys. Rev. D 72 (2005) 043009.
\newblock \href {http://arxiv.org/abs/astro-ph/0412525}
  {\path{arXiv:astro-ph/0412525}}, \href
  {http://dx.doi.org/10.1103/PhysRevD.72.043009}
  {\path{doi:10.1103/PhysRevD.72.043009}}.

\bibitem{Das:2008vb}
S.~Das, H.~Kang, D.~Ryu, J.~Cho, {Propagation of UHE Protons through Magnetized
  Cosmic Web}, Astrophys. J. 682 (2008) 29.
\newblock \href {http://arxiv.org/abs/0801.0371} {\path{arXiv:0801.0371}},
  \href {http://dx.doi.org/10.1086/588278} {\path{doi:10.1086/588278}}.

\bibitem{Garcia:2021cgu}
A.~Arámburo-García, K.~Bondarenko, A.~Boyarsky, D.~Nelson, A.~Pillepich,
  A.~Sokolenko, {Ultrahigh energy cosmic ray deflection by the intergalactic
  magnetic field}, Phys. Rev. D 104~(8) (2021) 083017.
\newblock \href {http://arxiv.org/abs/2101.07207} {\path{arXiv:2101.07207}},
  \href {http://dx.doi.org/10.1103/PhysRevD.104.083017}
  {\path{doi:10.1103/PhysRevD.104.083017}}.

\bibitem{Fang:2016amf}
K.~Fang, A.~V. Olinto, {High-energy neutrinos from sources in clusters of
  galaxies}, Astrophys. J. 828~(1) (2016) 37.
\newblock \href {http://arxiv.org/abs/1607.00380} {\path{arXiv:1607.00380}},
  \href {http://dx.doi.org/10.3847/0004-637X/828/1/37}
  {\path{doi:10.3847/0004-637X/828/1/37}}.

\bibitem{Hussain:2021kye}
S.~Hussain, R.~A. Alves~Batista, E.~d. de~Gouveia Dal~Pino, K.~Dolag,
  {High-Energy Neutrino Production in Clusters of Galaxies}, PoS ICRC2021
  (2021) 1212.
\newblock \href {http://arxiv.org/abs/2110.13958} {\path{arXiv:2110.13958}},
  \href {http://dx.doi.org/10.22323/1.395.1212}
  {\path{doi:10.22323/1.395.1212}}.

\bibitem{Kotera:2007ca}
K.~Kotera, M.~Lemoine, {Inhomogeneous extragalactic magnetic fields and the
  second knee in the cosmic ray spectrum}, Phys. Rev. D 77 (2008) 023005.
\newblock \href {http://arxiv.org/abs/0706.1891} {\path{arXiv:0706.1891}},
  \href {http://dx.doi.org/10.1103/PhysRevD.77.023005}
  {\path{doi:10.1103/PhysRevD.77.023005}}.

\bibitem{Mollerach:2013dza}
S.~Mollerach, E.~Roulet, {Magnetic diffusion effects on the ultra-high energy
  cosmic ray spectrum and composition}, JCAP 10 (2013) 013.
\newblock \href {http://arxiv.org/abs/1305.6519} {\path{arXiv:1305.6519}},
  \href {http://dx.doi.org/10.1088/1475-7516/2013/10/013}
  {\path{doi:10.1088/1475-7516/2013/10/013}}.

\bibitem{AlvesBatista:2014aky}
R.~Alves~Batista, G.~Sigl, {Diffusion of cosmic rays at EeV energies in
  inhomogeneous extragalactic magnetic fields}, JCAP 11 (2014) 031.
\newblock \href {http://arxiv.org/abs/1407.6150} {\path{arXiv:1407.6150}},
  \href {http://dx.doi.org/10.1088/1475-7516/2014/11/031}
  {\path{doi:10.1088/1475-7516/2014/11/031}}.

\bibitem{AlvesBatista:2018kup}
R.~Alves~Batista, E.~M. de~Gouveia Dal~Pino, K.~Dolag, S.~Hussain, {Cosmic-ray
  propagation in the turbulent intergalactic medium}, in: {30th General
  Assembly of the International Astronomical Union}, 2018.
\newblock \href {http://arxiv.org/abs/1811.03062} {\path{arXiv:1811.03062}}.

\bibitem{Lacy:2019rfe}
M.~Lacy, et~al., {The Karl G. Jansky Very Large Array Sky Survey (VLASS).
  Science case and survey design}, Publ. Astron. Soc. Pac. 132~(1009) (2020)
  035001.
\newblock \href {http://arxiv.org/abs/1907.01981} {\path{arXiv:1907.01981}},
  \href {http://dx.doi.org/10.1088/1538-3873/ab63eb}
  {\path{doi:10.1088/1538-3873/ab63eb}}.

\bibitem{Akahori:2016ami}
T.~Akahori, D.~Ryu, B.~M. Gaensler, {Fast Radio Bursts as Probes of Magnetic
  Fields in the Intergalactic Medium}, Astrophys. J. 824~(2) (2016) 105.
\newblock \href {http://arxiv.org/abs/1602.03235} {\path{arXiv:1602.03235}},
  \href {http://dx.doi.org/10.3847/0004-637X/824/2/105}
  {\path{doi:10.3847/0004-637X/824/2/105}}.

\bibitem{Hackstein:2019abb}
S.~Hackstein, otherss, {Fast Radio Burst dispersion measures and rotation
  measures and the origin of intergalactic magnetic fields}, Mon. Not. Roy.
  Astron. Soc. 488~(3) (2019) 4220--4238.
\newblock \href {http://arxiv.org/abs/1907.09650} {\path{arXiv:1907.09650}},
  \href {http://dx.doi.org/10.1093/mnras/stz2033}
  {\path{doi:10.1093/mnras/stz2033}}.

\bibitem{CTAConsortium:2017dvg}
B.~S. Acharya, et~al., {Science with the Cherenkov Telescope Array}, WSP, 2018.
\newblock \href {http://arxiv.org/abs/1709.07997} {\path{arXiv:1709.07997}},
  \href {http://dx.doi.org/10.1142/10986} {\path{doi:10.1142/10986}}.

\bibitem{CTA:2020hii}
H.~Abdalla, et~al., {\bf CTA} Collaboration, {Sensitivity of the Cherenkov
  Telescope Array for probing cosmology and fundamental physics with gamma-ray
  propagation}, JCAP 02 (2021) 048.
\newblock \href {http://arxiv.org/abs/2010.01349} {\path{arXiv:2010.01349}},
  \href {http://dx.doi.org/10.1088/1475-7516/2021/02/048}
  {\path{doi:10.1088/1475-7516/2021/02/048}}.

\bibitem{AlvesBatista:2021gzc}
R.~Alves~Batista, et~al., {EuCAPT White Paper: Opportunities and Challenges for
  Theoretical Astroparticle Physics in the Next Decade} (2021).
\newblock \href {http://arxiv.org/abs/2110.10074} {\path{arXiv:2110.10074}}.

\bibitem{Broderick:2011av}
A.~E. Broderick, P.~Chang, C.~Pfrommer, {The Cosmological Impact of Luminous
  TeV Blazars I: Implications of Plasma Instabilities for the Intergalactic
  Magnetic Field and Extragalactic Gamma-Ray Background}, Astrophys. J. 752
  (2012) 22.
\newblock \href {http://arxiv.org/abs/1106.5494} {\path{arXiv:1106.5494}},
  \href {http://dx.doi.org/10.1088/0004-637X/752/1/22}
  {\path{doi:10.1088/0004-637X/752/1/22}}.

\bibitem{2012ApJ...758..102S}
R.~{Schlickeiser}, D.~{Ibscher}, M.~{Supsar}, {Plasma Effects on Fast Pair
  Beams in Cosmic Voids}, ApJ 758~(2) (2012) 102.
\newblock \href {http://dx.doi.org/10.1088/0004-637X/758/2/102}
  {\path{doi:10.1088/0004-637X/758/2/102}}.

\bibitem{Broderick:2018nqf}
A.~E. Broderick, P.~Tiede, P.~Chang, A.~Lamberts, C.~Pfrommer, E.~Puchwein,
  M.~Shalaby, M.~Werhahn, {Missing Gamma-ray Halos and the Need for New Physics
  in the Gamma-ray Sky}, Astrophys. J. 868~(2) (2018) 87.
\newblock \href {http://arxiv.org/abs/1808.02959} {\path{arXiv:1808.02959}},
  \href {http://dx.doi.org/10.3847/1538-4357/aae5f2}
  {\path{doi:10.3847/1538-4357/aae5f2}}.

\bibitem{Yan:2018pca}
D.~Yan, J.~Zhou, P.~Zhang, Q.~Zhu, J.~Wang, {Impact of Plasma Instability on
  Constraint of the Intergalactic Magnetic Field}, Astrophys. J. 870~(1) (2019)
  17.
\newblock \href {http://arxiv.org/abs/1810.07013} {\path{arXiv:1810.07013}},
  \href {http://dx.doi.org/10.3847/1538-4357/aaef7d}
  {\path{doi:10.3847/1538-4357/aaef7d}}.

\bibitem{AlvesBatista:2019ipr}
R.~Alves~Batista, A.~Saveliev, E.~M. de~Gouveia Dal~Pino, {The Impact of Plasma
  Instabilities on the Spectra of TeV Blazars}, Mon. Not. Roy. Astron. Soc.
  489~(3) (2019) 3836--3849.
\newblock \href {http://arxiv.org/abs/1904.13345} {\path{arXiv:1904.13345}},
  \href {http://dx.doi.org/10.1093/mnras/stz2389}
  {\path{doi:10.1093/mnras/stz2389}}.

\bibitem{AMEGO:2019gny}
R.~Caputo, et~al., {\bf AMEGO} Collaboration, {All-sky Medium Energy Gamma-ray
  Observatory: Exploring the Extreme Multimessenger Universe}\href
  {http://arxiv.org/abs/1907.07558} {\path{arXiv:1907.07558}}.

\bibitem{Schael:2019lvx}
S.~Schael, et~al., {AMS-100: The next generation magnetic spectrometer in space
  \textendash{} An international science platform for physics and astrophysics
  at Lagrange point 2}, Nucl. Instrum. Meth. A 944 (2019) 162561.
\newblock \href {http://arxiv.org/abs/1907.04168} {\path{arXiv:1907.04168}},
  \href {http://dx.doi.org/10.1016/j.nima.2019.162561}
  {\path{doi:10.1016/j.nima.2019.162561}}.

\bibitem{Topchiev:2021uer}
N.~P. Topchiev, et~al., {Gamma- and Cosmic-Ray Observations with the GAMMA-400
  Gamma-Ray Telescope}\href {http://arxiv.org/abs/2108.12609}
  {\path{arXiv:2108.12609}}.

\bibitem{Castellina:2019irv}
A.~Castellina, {\bf Pierre Auger} Collaboration, {AugerPrime: the Pierre Auger
  Observatory Upgrade}, EPJ Web Conf. 210 (2019) 06002.
\newblock \href {http://arxiv.org/abs/1905.04472} {\path{arXiv:1905.04472}},
  \href {http://dx.doi.org/10.1051/epjconf/201921006002}
  {\path{doi:10.1051/epjconf/201921006002}}.

\bibitem{PierreAuger:2021ccl}
G.~Cataldi, et~al., {\bf Pierre Auger} Collaboration, {The upgrade of the
  Pierre Auger Observatory with the Scintillator Surface Detector}, PoS
  ICRC2021 (2021) 251.
\newblock \href {http://dx.doi.org/10.22323/1.395.0251}
  {\path{doi:10.22323/1.395.0251}}.

\bibitem{PierreAuger:2021nnx}
P.~Abreu, et~al., {\bf Pierre Auger} Collaboration, {AugerPrime Upgraded
  Electronics}, PoS ICRC2021 (2021) 230.
\newblock \href {http://dx.doi.org/10.22323/1.395.0230}
  {\path{doi:10.22323/1.395.0230}}.

\bibitem{PierreAuger:2021bwp}
P.~Abreu, et~al., {\bf Pierre Auger} Collaboration, {First results from the
  AugerPrime Radio Detector}, PoS ICRC2021 (2021) 270.
\newblock \href {http://dx.doi.org/10.22323/1.395.0270}
  {\path{doi:10.22323/1.395.0270}}.

\bibitem{PierreAuger:2021nob}
P.~Abreu, et~al., {\bf Pierre Auger} Collaboration, {Status and performance of
  the underground muon detector of the Pierre Auger Observatory}, PoS ICRC2021
  (2021) 233.
\newblock \href {http://dx.doi.org/10.22323/1.395.0233}
  {\path{doi:10.22323/1.395.0233}}.

\bibitem{Allen:2013hfa}
J.~Allen, G.~Farrar, {Testing models of new physics with UHE air shower
  observations}, in: {33rd International Cosmic Ray Conference}, 2013, p. 1182.
\newblock \href {http://arxiv.org/abs/1307.7131} {\path{arXiv:1307.7131}}.

\bibitem{FengSnowmass}
J.~L. Feng, et~al., {The Forward Physics Facility at the High-Luminosity LHC},
  in: {2022 Snowmass Summer Study}, 2022.
\newblock \href {http://arxiv.org/abs/2203.05090} {\path{arXiv:2203.05090}}.

\bibitem{IceCube:2021htd}
R.~Abbasi, et~al., {\bf IceCube} Collaboration, {Hybrid cosmic ray measurements
  using the IceAct telescopes in coincidence with the IceCube and IceTop
  detectors}, PoS ICRC2021 (2021) 276.
\newblock \href {http://arxiv.org/abs/2108.05572} {\path{arXiv:2108.05572}},
  \href {http://dx.doi.org/10.22323/1.395.0276}
  {\path{doi:10.22323/1.395.0276}}.

\bibitem{Ishihara:2019aao}
A.~Ishihara, {\bf IceCube} Collaboration, {The IceCube Upgrade - Design and
  Science Goals}, PoS ICRC2019 (2021) 1031.
\newblock \href {http://arxiv.org/abs/1908.09441} {\path{arXiv:1908.09441}},
  \href {http://dx.doi.org/10.22323/1.358.1031}
  {\path{doi:10.22323/1.358.1031}}.

\bibitem{IceCube-Gen2:2021jce}
R.~Abbasi, et~al., {\bf IceCube-Gen2} Collaboration, {Simulation study for the
  future IceCube-Gen2 surface array}, PoS ICRC2021 (2021) 411.
\newblock \href {http://arxiv.org/abs/2108.04307} {\path{arXiv:2108.04307}},
  \href {http://dx.doi.org/10.22323/1.395.0411}
  {\path{doi:10.22323/1.395.0411}}.

\bibitem{Dembinski:2021szp}
H.~Dembinski, et~al., {The Muon Puzzle in air showers and its connection to the
  LHC}, PoS ICRC2021 (2021) 037.
\newblock \href {http://dx.doi.org/10.22323/1.395.0037}
  {\path{doi:10.22323/1.395.0037}}.

\bibitem{Anchordoqui:2021ghd}
L.~A. Anchordoqui, et~al., {The Forward Physics Facility: Sites, Experiments,
  and Physics Potential}\href {http://arxiv.org/abs/2109.10905}
  {\path{arXiv:2109.10905}}.

\bibitem{Fedynitch:2018cbl}
A.~Fedynitch, F.~Riehn, R.~Engel, T.~K. Gaisser, T.~Stanev, {Hadronic
  interaction model sibyll 2.3c and inclusive lepton fluxes}, Phys. Rev. D
  100~(10) (2019) 103018.
\newblock \href {http://arxiv.org/abs/1806.04140} {\path{arXiv:1806.04140}},
  \href {http://dx.doi.org/10.1103/PhysRevD.100.103018}
  {\path{doi:10.1103/PhysRevD.100.103018}}.

\bibitem{Schroder:2019rxp}
F.~G. Schr\"oder, et~al., {High-Energy Galactic Cosmic Rays (Astro2020 Science
  White Paper)}, Bull. Am. Astron. Soc. 51 (2019) 131.
\newblock \href {http://arxiv.org/abs/1903.07713} {\path{arXiv:1903.07713}}.

\bibitem{Baack:2020jwb}
D.~Baack, W.~Rhode, {GPU based photon propagation for CORSIKA 8}, J. Phys.
  Conf. Ser. 1690~(1) (2020) 012073.
\newblock \href {http://dx.doi.org/10.1088/1742-6596/1690/1/012073}
  {\path{doi:10.1088/1742-6596/1690/1/012073}}.

\bibitem{c8}
A.~A. Alves, M.~Reininghaus, A.~Schmidt, R.~Prechelt, R.~Ulrich, {\bf CORSIKA}
  Collaboration, {CORSIKA 8 - A novel high-performance computing tool for
  particle cascade Monte Carlo simulations}, EPJ Web Conf. 251 (2021) 03038.
\newblock \href {http://dx.doi.org/10.1051/epjconf/202125103038}
  {\path{doi:10.1051/epjconf/202125103038}}.

\bibitem{Huege:2016jvc}
T.~Huege, et~al., {Ultimate precision in cosmic-ray radio detection
  \textemdash{} the SKA}, EPJ Web Conf. 135 (2017) 02003.
\newblock \href {http://arxiv.org/abs/1608.08869} {\path{arXiv:1608.08869}},
  \href {http://dx.doi.org/10.1051/epjconf/201713502003}
  {\path{doi:10.1051/epjconf/201713502003}}.

\bibitem{dlbook}
I.~Goodfellow, Y.~Bengio, A.~Courville, Deep Learning, MIT Press, 2016,
  \url{http://www.deeplearningbook.org}.

\bibitem{lecun_deep_2015}
Y.~LeCun, Y.~Bengio, G.~Hinton, Deep learning, Nature 521~(7553) (2015)
  436--444.
\newblock \href {http://dx.doi.org/10.1038/nature14539}
  {\path{doi:10.1038/nature14539}}.

\bibitem{ml_physics_review}
{G. Carleo et al.}, Machine learning and the physical sciences, Reviews of
  Modern Physics 91~(4).
\newblock \href {http://dx.doi.org/10.1103/revmodphys.91.045002}
  {\path{doi:10.1103/revmodphys.91.045002}}.

\bibitem{dlfpr}
M.~Erdmann, J.~Glombitza, G.~Kasieczka, U.~Klemradt, Deep Learning for Physics
  Research, {WORLD SCIENTIFIC}, 2022.
\newblock \href {http://dx.doi.org/10.1142/12294} {\path{doi:10.1142/12294}}.

\bibitem{iact_2019}
{I. Shilon et al.}, {Application of deep learning methods to analysis of
  imaging atmospheric Cherenkov telescopes data}, Astroparticle Physics 105
  (2019) 44–53.
\newblock \href {http://dx.doi.org/10.1016/j.astropartphys.2018.10.003}
  {\path{doi:10.1016/j.astropartphys.2018.10.003}}.

\bibitem{icecube_cnn}
{R. Abbasi et al.}, {\bf IceCube} Collaboration, {A convolutional neural
  network based cascade reconstruction for the {IceCube} Neutrino Observatory},
  Journal of Instrumentation 16~(07) (2021) P07041.
\newblock \href {http://dx.doi.org/10.1088/1748-0221/16/07/p07041}
  {\path{doi:10.1088/1748-0221/16/07/p07041}}.

\bibitem{ligo_2018}
D.~George, E.~Huerta, {Deep Learning for real-time gravitational wave detection
  and parameter estimation: Results with Advanced LIGO data}, Physics Letters B
  778 (2018) 64–70.
\newblock \href {http://dx.doi.org/10.1016/j.physletb.2017.12.053}
  {\path{doi:10.1016/j.physletb.2017.12.053}}.

\bibitem{Carrillo-Perez2021}
F.~Carrillo-Perez, L.~J. Herrera, J.~M. Carceller, A.~Guill{\'e}n, Deep
  learning to classify ultra-high-energy cosmic rays by means of pmt signals,
  Neural Computing and Applications 33~(15) (2021) 9153--9169.
\newblock \href {http://dx.doi.org/10.1007/s00521-020-05679-9}
  {\path{doi:10.1007/s00521-020-05679-9}}.

\bibitem{Erdmann:2019nie}
M.~Erdmann, F.~Schl\"uter, R.~Smida, {Classification and Recovery of Radio
  Signals from Cosmic Ray Induced Air Showers with Deep Learning}, JINST
  14~(04) (2019) P04005.
\newblock \href {http://arxiv.org/abs/1901.04079} {\path{arXiv:1901.04079}},
  \href {http://dx.doi.org/10.1088/1748-0221/14/04/P04005}
  {\path{doi:10.1088/1748-0221/14/04/P04005}}.

\bibitem{Erdmann:2017str}
M.~Erdmann, J.~Glombitza, D.~Walz, {A deep learning-based reconstruction of
  cosmic ray-induced air showers}, Astropart. Phys. 97 (2018) 46--53.
\newblock \href {http://arxiv.org/abs/1708.00647} {\path{arXiv:1708.00647}},
  \href {http://dx.doi.org/10.1016/j.astropartphys.2017.10.006}
  {\path{doi:10.1016/j.astropartphys.2017.10.006}}.

\bibitem{IceCube:2021ldz}
R.~Abbasi, et~al., {\bf IceCube} Collaboration, {Study of mass composition of
  cosmic rays with IceTop and IceCube}, PoS ICRC2021 (2021) 323.
\newblock \href {http://arxiv.org/abs/2107.09626} {\path{arXiv:2107.09626}},
  \href {http://dx.doi.org/10.22323/1.395.0323}
  {\path{doi:10.22323/1.395.0323}}.

\bibitem{Glombitza:2020yhw}
J.~Glombitza, {\bf Pierre Auger} Collaboration, {Air-Shower Reconstruction at
  the Pierre Auger Observatory based on Deep Learning}, PoS ICRC2019 (2020)
  270.
\newblock \href {http://dx.doi.org/10.22323/1.358.0270}
  {\path{doi:10.22323/1.358.0270}}.

\bibitem{carceller}
F.~Carrillo-Perez, L.~J. Herrera, J.~M. Carceller, A.~Guill{\'e}n, Improving
  classification of ultra-high energy cosmic rays using spacial locality by
  means of a convolutional dnn, in: I.~Rojas, G.~Joya, A.~Catala (Eds.),
  Advances in Computational Intelligence, Springer International Publishing,
  Cham, 2019, pp. 222--232.

\bibitem{Kalashev:2020fzg}
O.~Kalashev, {\bf Telescope Array} Collaboration, {Using Deep Learning in
  Ultra-High Energy Cosmic Ray Experiments}, J. Phys. Conf. Ser. 1525~(1)
  (2020) 012001.
\newblock \href {http://dx.doi.org/10.1088/1742-6596/1525/1/012001}
  {\path{doi:10.1088/1742-6596/1525/1/012001}}.

\bibitem{ta_ml}
D.~Ivanov, O.~E. Kalashev, M.~Y. Kuznetsov, G.~I. Rubtsov, T.~Sako,
  Y.~Tsunesada, Y.~V. Zhezher, Using deep learning to enhance event geometry
  reconstruction for the telescope array surface detector, Machine Learning:
  Science and Technology 2~(1) (2020) 015006.
\newblock \href {http://dx.doi.org/10.1088/2632-2153/abae74}
  {\path{doi:10.1088/2632-2153/abae74}}.

\bibitem{Rehman:2021nw}
A.~Rehman, A.~Coleman, F.~G. Schr\"oder, D.~Kostunin, {Classification and
  Denoising of Cosmic-Ray Radio Signals using Deep Learning}, PoS ICRC2021
  (2021) 417.
\newblock \href {http://dx.doi.org/10.22323/1.395.0417}
  {\path{doi:10.22323/1.395.0417}}.

\bibitem{unfolding}
M.~Erdmann, K.~Hafner, J.~Schulte, M.~Straub, {Autoencoder-extended Conditional
  Invertible Neural Networks for Unfolding Signal Traces} ACAT 2021, to appear
  in the ACAT proceedings.

\bibitem{Zehrer_2021}
L.~Zehrer,
  \href{http://repozitorij.ung.si/IzpisGradiva.php?lang=slv&id=6815}{Application
  of machine learning techniques for cosmic ray event classification and
  implementation of a real-time ultra-high energy photon search with the
  surface detector of the pierre auger observatory: dissertation}, Ph.D.
  thesis, Univerza v Novi Gorici, Fakulteta za podiplomski študij (2021).
\newline\urlprefix\url{http://repozitorij.ung.si/IzpisGradiva.php?lang=slv&id=6815}

\bibitem{Bister:2021arb}
T.~Bister, M.~Erdmann, U.~K\"othe, J.~Schulte, {Inference of cosmic-ray source
  properties by conditional invertible neural networks}, Eur. Phys. J. C 82~(2)
  (2022) 171.
\newblock \href {http://arxiv.org/abs/2110.09493} {\path{arXiv:2110.09493}},
  \href {http://dx.doi.org/10.1140/epjc/s10052-022-10138-x}
  {\path{doi:10.1140/epjc/s10052-022-10138-x}}.

\bibitem{Kalashev:2019skq}
O.~Kalashev, M.~Pshirkov, M.~Zotov, {Identifying nearby sources of
  ultra-high-energy cosmic rays with deep learning}, JCAP 11 (2020) 005.
\newblock \href {http://arxiv.org/abs/1912.00625} {\path{arXiv:1912.00625}},
  \href {http://dx.doi.org/10.1088/1475-7516/2020/11/005}
  {\path{doi:10.1088/1475-7516/2020/11/005}}.

\bibitem{Erdmann2018}
M.~Erdmann, L.~Geiger, J.~Glombitza, D.~Schmidt, Generating and refining
  particle detector simulations using the wasserstein distance in adversarial
  networks, Computing and Software for Big Science 2~(1) (2018) 4.
\newblock \href {http://dx.doi.org/10.1007/s41781-018-0008-x}
  {\path{doi:10.1007/s41781-018-0008-x}}.

\bibitem{TelescopeArray:2021auw}
R.~Abbasi, et~al., {\bf Telescope Array} Collaboration, {Cosmic ray energy
  spectrum in the 2nd knee region measured by the TALE-SD array}, PoS ICRC2021
  (2021) 362.
\newblock \href {http://dx.doi.org/10.22323/1.395.0362}
  {\path{doi:10.22323/1.395.0362}}.

\bibitem{Auger:2017nlp}
S.~Quinn, S.~Colognes, B.~Courty, B.~Genolini, L.~Guglielmi, P.~Lebrun,
  M.~Marton, E.~Rauly, T.~Trung, O.~Wolf, {\bf Auger, Telescope Array}
  Collaboration, {Auger at the Telescope Array: toward a direct
  cross-calibration of surface-detector stations}, PoS ICRC2017 (2018) 395.
\newblock \href {http://dx.doi.org/10.22323/1.301.0395}
  {\path{doi:10.22323/1.301.0395}}.

\bibitem{PierreAuger:2019mun}
F.~Sarazin, et~al., {\bf Pierre Auger, Telescope Array} Collaboration,
  {Overview of the Auger{\@}TA project and preliminary results from Phase I},
  EPJ Web Conf. 210 (2019) 05002.
\newblock \href {http://dx.doi.org/10.1051/epjconf/201921005002}
  {\path{doi:10.1051/epjconf/201921005002}}.

\bibitem{PierreAuger:2019vqa}
C.~Covault, et~al., {\bf Pierre Auger, Telescope Array} Collaboration, {The
  Auger{\@}TA Project: Phase II Progress and Plans}, EPJ Web Conf. 210 (2019)
  05004.
\newblock \href {http://dx.doi.org/10.1051/epjconf/201921005004}
  {\path{doi:10.1051/epjconf/201921005004}}.

\bibitem{Mulrey:2020oqe}
K.~Mulrey, et~al., {On the cosmic-ray energy scale of the LOFAR radio
  telescope}, JCAP 11 (2020) 017.
\newblock \href {http://arxiv.org/abs/2005.13441} {\path{arXiv:2005.13441}},
  \href {http://dx.doi.org/10.1088/1475-7516/2020/11/017}
  {\path{doi:10.1088/1475-7516/2020/11/017}}.

\bibitem{Mulrey:2021fqw}
K.~Mulrey, {Cross-calibrating the energy scales of cosmic-ray experiments using
  a portable radio array}, PoS ICRC2021 (2021) 414.
\newblock \href {http://dx.doi.org/10.22323/1.395.0414}
  {\path{doi:10.22323/1.395.0414}}.

\bibitem{TelescopeArray:2012uws}
T.~Abu-Zayyad, et~al., {\bf Telescope Array} Collaboration, {The surface
  detector array of the Telescope Array experiment}, Nucl. Instrum. Meth. A 689
  (2013) 87--97.
\newblock \href {http://arxiv.org/abs/1201.4964} {\path{arXiv:1201.4964}},
  \href {http://dx.doi.org/10.1016/j.nima.2012.05.079}
  {\path{doi:10.1016/j.nima.2012.05.079}}.

\bibitem{PierreAuger:2016vya}
A.~Aab, et~al., {\bf Pierre Auger} Collaboration, {Measurement of the Radiation
  Energy in the Radio Signal of Extensive Air Showers as a Universal Estimator
  of Cosmic-Ray Energy}, Phys. Rev. Lett. 116~(24) (2016) 241101.
\newblock \href {http://arxiv.org/abs/1605.02564} {\path{arXiv:1605.02564}},
  \href {http://dx.doi.org/10.1103/PhysRevLett.116.241101}
  {\path{doi:10.1103/PhysRevLett.116.241101}}.

\bibitem{PierreAuger:2015hbf}
A.~Aab, et~al., {\bf Pierre Auger} Collaboration, {Energy Estimation of Cosmic
  Rays with the Engineering Radio Array of the Pierre Auger Observatory}, Phys.
  Rev. D 93~(12) (2016) 122005.
\newblock \href {http://arxiv.org/abs/1508.04267} {\path{arXiv:1508.04267}},
  \href {http://dx.doi.org/10.1103/PhysRevD.93.122005}
  {\path{doi:10.1103/PhysRevD.93.122005}}.

\bibitem{Verzi:2013ajy}
V.~Verzi, {\bf Pierre Auger} Collaboration, {The Energy Scale of the Pierre
  Auger Observatory}, in: {33rd International Cosmic Ray Conference}, 2013, p.
  0928.

\bibitem{PierreAuger:2021cwj}
P.~Abreu, et~al., {\bf Pierre Auger} Collaboration, {The XY Scanner - A
  Versatile Method of the Absolute End-to-End Calibration of Fluorescence
  Detectors}, PoS ICRC2021 (2021) 220.
\newblock \href {http://dx.doi.org/10.22323/1.395.0220}
  {\path{doi:10.22323/1.395.0220}}.

\bibitem{PierreAuger:2017xgp}
A.~Aab, et~al., {\bf Pierre Auger} Collaboration, {Calibration of the
  logarithmic-periodic dipole antenna (LPDA) radio stations at the Pierre Auger
  Observatory using an octocopter}, JINST 12~(10) (2017) T10005.
\newblock \href {http://arxiv.org/abs/1702.01392} {\path{arXiv:1702.01392}},
  \href {http://dx.doi.org/10.1088/1748-0221/12/10/T10005}
  {\path{doi:10.1088/1748-0221/12/10/T10005}}.

\bibitem{Mulrey:2019vtz}
K.~Mulrey, et~al., {Calibration of the LOFAR low-band antennas using the Galaxy
  and a model of the signal chain}, Astropart. Phys. 111 (2019) 1--11.
\newblock \href {http://arxiv.org/abs/1903.05988} {\path{arXiv:1903.05988}},
  \href {http://dx.doi.org/10.1016/j.astropartphys.2019.03.004}
  {\path{doi:10.1016/j.astropartphys.2019.03.004}}.

\bibitem{PierreAuger:2012ker}
P.~Abreu, et~al., {\bf Pierre Auger} Collaboration, {Antennas for the Detection
  of Radio Emission Pulses from Cosmic-Ray}, JINST 7 (2012) P10011.
\newblock \href {http://arxiv.org/abs/1209.3840} {\path{arXiv:1209.3840}},
  \href {http://dx.doi.org/10.1088/1748-0221/7/10/P10011}
  {\path{doi:10.1088/1748-0221/7/10/P10011}}.

\bibitem{Klages:2007zza}
H.~O. Klages, {\bf Pierre Auger} Collaboration, {HEAT \textendash{} Enhancement
  Telescopes for the Pierre Auger Southern Observatory}, in: {30th
  International Cosmic Ray Conference}, Vol.~5, 2007, pp. 849--852.

\bibitem{Thomson:2011gke}
G.~B. Thomson, P.~Sokolsky, C.~C.~H. Jui, {The Telescope Array Low Energy
  Extension (TALE)}, in: {32nd International Cosmic Ray Conference}, Vol.~3,
  2011, p. 338.
\newblock \href {http://dx.doi.org/10.7529/ICRC2011/V03/1307}
  {\path{doi:10.7529/ICRC2011/V03/1307}}.

\bibitem{PierreAuger:2016crs}
A.~Aab, et~al., {\bf Pierre Auger} Collaboration, {Prototype muon detectors for
  the AMIGA component of the Pierre Auger Observatory}, JINST 11~(02) (2016)
  P02012.
\newblock \href {http://arxiv.org/abs/1605.01625} {\path{arXiv:1605.01625}},
  \href {http://dx.doi.org/10.1088/1748-0221/11/02/P02012}
  {\path{doi:10.1088/1748-0221/11/02/P02012}}.

\bibitem{Ogio:2020fyf}
S.~Ogio, {\bf Pierre Auger} Collaboration, {Telescope Array Low energy
  Extension(TALE) Hybrid}, PoS ICRC2019 (2020) 375.
\newblock \href {http://dx.doi.org/10.22323/1.358.0375}
  {\path{doi:10.22323/1.358.0375}}.

\bibitem{Bergman:2019pnx}
D.~R. Bergman, Y.~Tsunesada, J.~F. Krizmanic, Y.~Omura, {NICHE: Air-Cherenkov
  observation at the TA site}, EPJ Web Conf. 210 (2019) 05001.
\newblock \href {http://dx.doi.org/10.1051/epjconf/201921005001}
  {\path{doi:10.1051/epjconf/201921005001}}.

\bibitem{Holt:2019tja}
E.~M. Holt, {\bf Pierre Auger} Collaboration, {Estimating the mass of cosmic
  rays by combining radio and muon measurements}, EPJ Web Conf. 216 (2019)
  02002.
\newblock \href {http://dx.doi.org/10.1051/epjconf/201921602002}
  {\path{doi:10.1051/epjconf/201921602002}}.

\bibitem{Zhezher:2021qke}
Y.~Zhezher, {\bf Telescope Array} Collaboration, {Cosmic-ray mass composition
  with the TA SD 12-year data}, PoS ICRC2021 (2021) 300.
\newblock \href {http://dx.doi.org/10.22323/1.395.0300}
  {\path{doi:10.22323/1.395.0300}}.

\bibitem{TelescopeArray:2021gov}
R.~Abbasi, et~al., {\bf Telescope Array} Collaboration, {Mass composition of
  Telescope Array's surface detectors events using deep learning}, PoS ICRC2021
  (2021) 384.
\newblock \href {http://dx.doi.org/10.22323/1.395.0384}
  {\path{doi:10.22323/1.395.0384}}.

\bibitem{Andringa:2011zz}
S.~Andringa, R.~Conceicao, M.~Pimenta, {Mass composition and cross-section from
  the shape of cosmic ray shower longitudinal profiles}, Astropart. Phys. 34
  (2011) 360--367.
\newblock \href {http://dx.doi.org/10.1016/j.astropartphys.2010.10.002}
  {\path{doi:10.1016/j.astropartphys.2010.10.002}}.

\bibitem{Zhezher:2021qug}
Y.~Zhezher, {\bf Telescope Array} Collaboration, {Mass composition anisotropy
  with the Telescope Array Surface Detector data}, PoS ICRC2021 (2021) 299.
\newblock \href {http://dx.doi.org/10.22323/1.395.0299}
  {\path{doi:10.22323/1.395.0299}}.

\bibitem{Younk:2012mp}
P.~Younk, M.~Risse, {Sensitivity of the correlation between the depth of shower
  maximum and the muon shower size to the cosmic ray composition}, Astropart.
  Phys. 35 (2012) 807--812.
\newblock \href {http://arxiv.org/abs/1203.3732} {\path{arXiv:1203.3732}},
  \href {http://dx.doi.org/10.1016/j.astropartphys.2012.03.001}
  {\path{doi:10.1016/j.astropartphys.2012.03.001}}.

\bibitem{Yushkov:2016xiz}
A.~Yushkov, M.~Risse, M.~Werner, J.~Krieg, {Determination of the
  proton-to-helium ratio in cosmic rays at ultra-high energies from the tail of
  the $X_{max}$ distribution}, Astropart. Phys. 85 (2016) 29--34.
\newblock \href {http://arxiv.org/abs/1609.08586} {\path{arXiv:1609.08586}},
  \href {http://dx.doi.org/10.1016/j.astropartphys.2016.09.007}
  {\path{doi:10.1016/j.astropartphys.2016.09.007}}.

\bibitem{Karpikov:2018itx}
I.~I. Karpikov, G.~I. Rubtsov, Y.~V. Zhezher, {Lower limit on the
  ultrahigh-energy proton-to-helium ratio from the measurements of the tail of
  the $X_{max}$ distribution}, Phys. Rev. D 98~(10) (2018) 103002.
\newblock \href {http://arxiv.org/abs/1805.04080} {\path{arXiv:1805.04080}},
  \href {http://dx.doi.org/10.1103/PhysRevD.98.103002}
  {\path{doi:10.1103/PhysRevD.98.103002}}.

\bibitem{PierreAuger:2013xim}
P.~Abreu, et~al., {\bf Pierre Auger} Collaboration, {Interpretation of the
  Depths of Maximum of Extensive Air Showers Measured by the Pierre Auger
  Observatory}, JCAP 02 (2013) 026.
\newblock \href {http://arxiv.org/abs/1301.6637} {\path{arXiv:1301.6637}},
  \href {http://dx.doi.org/10.1088/1475-7516/2013/02/026}
  {\path{doi:10.1088/1475-7516/2013/02/026}}.

\bibitem{Blaess:2018msv}
S.~Blaess, J.~A. Bellido, B.~R. Dawson, {Extracting a less model dependent
  cosmic ray composition from $X_\mathrm{max}$ distributions}\href
  {http://arxiv.org/abs/1803.02520} {\path{arXiv:1803.02520}}.

\bibitem{Tkachenko:2021bja}
O.~Tkachenko, R.~Engel, R.~Ulrich, M.~Unger, {Study on the Combined Estimate of
  the Cosmic-Ray Composition and Particle Cross Sections at Ultrahigh
  Energies}, PoS ICRC2021 (2021) 438.
\newblock \href {http://dx.doi.org/10.22323/1.395.0438}
  {\path{doi:10.22323/1.395.0438}}.

\bibitem{Vicha:2020sfv}
J.~Vicha, A.~Yushkov, D.~Nosek, P.~Travnicek, E.~Santos, {Testing Hadronic
  Interactions Using Hybrid Observables}, PoS ICRC2019 (2020) 452.
\newblock \href {http://dx.doi.org/10.22323/1.358.0452}
  {\path{doi:10.22323/1.358.0452}}.

\bibitem{Kachelriess:2005qm}
M.~Kachelriess, P.~D. Serpico, M.~Teshima, {The Galactic magnetic field as
  spectrograph for ultrahigh energy cosmic rays}, Astropart. Phys. 26 (2006)
  378--386.
\newblock \href {http://arxiv.org/abs/astro-ph/0510444}
  {\path{arXiv:astro-ph/0510444}}, \href
  {http://dx.doi.org/10.1016/j.astropartphys.2006.08.004}
  {\path{doi:10.1016/j.astropartphys.2006.08.004}}.

\bibitem{Anchordoqui:2017abg}
L.~A. Anchordoqui, V.~Barger, T.~J. Weiler, {Cosmic Mass Spectrometer}, JHEAp
  17 (2018) 38--49.
\newblock \href {http://arxiv.org/abs/1707.05408} {\path{arXiv:1707.05408}},
  \href {http://dx.doi.org/10.1016/j.jheap.2017.12.001}
  {\path{doi:10.1016/j.jheap.2017.12.001}}.

\bibitem{Sjostrand:2006za}
T.~Sjostrand, S.~Mrenna, P.~Z. Skands, {PYTHIA 6.4 Physics and Manual}, JHEP 05
  (2006) 026.
\newblock \href {http://arxiv.org/abs/hep-ph/0603175}
  {\path{arXiv:hep-ph/0603175}}, \href
  {http://dx.doi.org/10.1088/1126-6708/2006/05/026}
  {\path{doi:10.1088/1126-6708/2006/05/026}}.

\bibitem{Sjostrand:2007gs}
T.~Sjostrand, S.~Mrenna, P.~Z. Skands, {A Brief Introduction to PYTHIA 8.1},
  Comput. Phys. Commun. 178 (2008) 852--867.
\newblock \href {http://arxiv.org/abs/0710.3820} {\path{arXiv:0710.3820}},
  \href {http://dx.doi.org/10.1016/j.cpc.2008.01.036}
  {\path{doi:10.1016/j.cpc.2008.01.036}}.

\bibitem{Sjostrand:2014zea}
T.~Sj\"ostrand, et~al., {An introduction to PYTHIA 8.2}, Comput. Phys. Commun.
  191 (2015) 159--177.
\newblock \href {http://arxiv.org/abs/1410.3012} {\path{arXiv:1410.3012}},
  \href {http://dx.doi.org/10.1016/j.cpc.2015.01.024}
  {\path{doi:10.1016/j.cpc.2015.01.024}}.

\bibitem{Sjostrand:2019zhc}
T.~Sj{\"o}strand, {The PYTHIA Event Generator: Past, Present and Future},
  Comput. Phys. Commun. 246 (2020) 106910.
\newblock \href {http://arxiv.org/abs/1907.09874} {\path{arXiv:1907.09874}},
  \href {http://dx.doi.org/10.1016/j.cpc.2019.106910}
  {\path{doi:10.1016/j.cpc.2019.106910}}.

\bibitem{Sjostrand:2021dal}
T.~Sj\"ostrand, M.~Utheim, {Hadron interactions for arbitrary energies and
  species, with applications to cosmic rays}, Eur. Phys. J. C 82~(1) (2022) 21.
\newblock \href {http://arxiv.org/abs/2108.03481} {\path{arXiv:2108.03481}},
  \href {http://dx.doi.org/10.1140/epjc/s10052-021-09953-5}
  {\path{doi:10.1140/epjc/s10052-021-09953-5}}.

\bibitem{GeneratorsWhitepaper}
S.~Alioli, et~al., {Event Generators for High-Energy Physics Experiments},
  White paper for Snowmass 2021 (in preparation).

\bibitem{UHEnuWhitepaper}
M.~Ackermann, et~al., {High-Energy and Ultra-High-Energy Neutrinos}, in: {2022
  Snowmass Summer Study}, 2022.
\newblock \href {http://arxiv.org/abs/2203.08096} {\path{arXiv:2203.08096}}.

\bibitem{MMWhitepaper}
K.~Engel, et~al., {Advancing the Landscape of Multimessenger Science in the
  Next Decade}, White Paper for Snowmass 2021 (in preparation).

\bibitem{Albrow:2018kxz}
M.~Albrow, {A Very Forward Hadron Spectrometer for the LHC and Cosmic Ray
  Physics}, PoS EDSU2018 (2018) 048.
\newblock \href {http://arxiv.org/abs/1811.02047} {\path{arXiv:1811.02047}},
  \href {http://dx.doi.org/10.22323/1.335.0048}
  {\path{doi:10.22323/1.335.0048}}.

\bibitem{GEANT4:2002zbu}
S.~Agostinelli, et~al., {\bf GEANT4} Collaboration, {GEANT4--a simulation
  toolkit}, Nucl. Instrum. Meth. A 506 (2003) 250--303.
\newblock \href {http://dx.doi.org/10.1016/S0168-9002(03)01368-8}
  {\path{doi:10.1016/S0168-9002(03)01368-8}}.

\bibitem{Allison:2006ve}
J.~Allison, et~al., {Geant4 developments and applications}, IEEE Trans. Nucl.
  Sci. 53 (2006) 270.
\newblock \href {http://dx.doi.org/10.1109/TNS.2006.869826}
  {\path{doi:10.1109/TNS.2006.869826}}.

\bibitem{Jiang:2021cit}
W.~Jiang, et~al., {\bf DAMPE} Collaboration, {Simulation of the DAMPE
  detector}, PoS ICRC2021 (2021) 082.
\newblock \href {http://dx.doi.org/10.22323/1.395.0082}
  {\path{doi:10.22323/1.395.0082}}.

\bibitem{TelescopeArray:2021dri}
R.~U. Abbasi, et~al., {\bf Telescope Array} Collaboration, {Surface detectors
  of the TAx4 experiment}, Nucl. Instrum. Meth. A 1019 (2021) 165726.
\newblock \href {http://arxiv.org/abs/2103.01086} {\path{arXiv:2103.01086}},
  \href {http://dx.doi.org/10.1016/j.nima.2021.165726}
  {\path{doi:10.1016/j.nima.2021.165726}}.

\bibitem{Bridgeman:2017rcv}
A.~Bridgeman, {\bf Pierre Auger} Collaboration, {Shower universality
  reconstruction of data from the Pierre Auger Observatory and validations with
  hadronic interaction models}: {Shower universality with Auger data} (2017)
  81--88\href {http://dx.doi.org/10.22323/1.301.0323}
  {\path{doi:10.22323/1.301.0323}}.

\bibitem{Berezinsky:2016jys}
V.~Berezinsky, A.~Gazizov, O.~Kalashev, {Cascade photons as test of protons in
  UHECR}, Astropart. Phys. 84 (2016) 52--61.
\newblock \href {http://arxiv.org/abs/1606.09293} {\path{arXiv:1606.09293}},
  \href {http://dx.doi.org/10.1016/j.astropartphys.2016.08.007}
  {\path{doi:10.1016/j.astropartphys.2016.08.007}}.

\bibitem{Supanitsky:2016gke}
A.~D. Supanitsky, {Implications of gamma-ray observations on proton models of
  ultrahigh energy cosmic rays}, Phys. Rev. D 94~(6) (2016) 063002.
\newblock \href {http://arxiv.org/abs/1607.00290} {\path{arXiv:1607.00290}},
  \href {http://dx.doi.org/10.1103/PhysRevD.94.063002}
  {\path{doi:10.1103/PhysRevD.94.063002}}.

\bibitem{Hussain:2013ui}
S.~Hussain, {\bf IceCube} Collaboration, {Measurements of the cosmic ray
  spectrum and average mass with IceCube}, Adv. Space Res. 53 (2014)
  1470--1475.
\newblock \href {http://arxiv.org/abs/1301.6619} {\path{arXiv:1301.6619}},
  \href {http://dx.doi.org/10.1016/j.asr.2013.06.023}
  {\path{doi:10.1016/j.asr.2013.06.023}}.

\bibitem{Radel:2017ule}
L.~R\"adel, {Measurement of High-Energy Muon Neutrinos with the IceCube
  Neutrino Observatory}, Ph.D. thesis, RWTH Aachen U. (2017).
\newblock \href {http://dx.doi.org/10.18154/RWTH-2017-10054}
  {\path{doi:10.18154/RWTH-2017-10054}}.

\bibitem{showerSchematics_photo1}
{Pierre Auger Observatory} (2015).
\newblock
  \href{https://www.flickr.com/photos/134252569@N07/20092256145/in/album-72157654097333143/}{[link]}.
\newline\urlprefix\url{https://www.flickr.com/photos/134252569@N07/20092256145/in/album-72157654097333143/}

\bibitem{showerSchematics_photo2}
{H.E.S.S.} (2012).
\newblock
  \href{https://www.mpi-hd.mpg.de/hfm/HESS/pages/press/2012/HESS_II_first_light/images/Image_73.png}{[link]}.
\newline\urlprefix\url{https://www.mpi-hd.mpg.de/hfm/HESS/pages/press/2012/HESS_II_first_light/images/Image_73.png}

\bibitem{showerSchematics_photo3}
{CMS} (2016).
\newblock
  \href{https://twiki.cern.ch/twiki/pub/CMSPublic/DisplacedMuonsRun2/cosmic-event-display_276872_11_258060_RhoPhi.png}{[link]}.
\newline\urlprefix\url{https://twiki.cern.ch/twiki/pub/CMSPublic/DisplacedMuonsRun2/cosmic-event-display_276872_11_258060_RhoPhi.png}

\bibitem{PierreAuger:2004naf}
J.~Abraham, et~al., {\bf Pierre Auger} Collaboration, {Properties and
  performance of the prototype instrument for the Pierre Auger Observatory},
  Nucl. Instrum. Meth. A 523 (2004) 50--95.
\newblock \href {http://dx.doi.org/10.1016/j.nima.2003.12.012}
  {\path{doi:10.1016/j.nima.2003.12.012}}.

\bibitem{Schroder:2019suq}
F.~G. Schr\"oder, {\bf IceCube} Collaboration, {Science Case of a Scintillator
  and Radio Surface Array at IceCube}, PoS ICRC2019 (2020) 418.
\newblock \href {http://arxiv.org/abs/1908.11469} {\path{arXiv:1908.11469}},
  \href {http://dx.doi.org/10.22323/1.358.0418}
  {\path{doi:10.22323/1.358.0418}}.

\bibitem{Letessier-Selvon:2014sga}
A.~Letessier-Selvon, P.~Billoir, M.~Blanco, I.~C. Mari\c{s}, M.~Settimo,
  {Layered water Cherenkov detector for the study of ultra high energy cosmic
  rays}, Nucl. Instrum. Meth. A 767 (2014) 41--49.
\newblock \href {http://arxiv.org/abs/1405.5699} {\path{arXiv:1405.5699}},
  \href {http://dx.doi.org/10.1016/j.nima.2014.08.029}
  {\path{doi:10.1016/j.nima.2014.08.029}}.

\bibitem{FAST:2021bkj}
T.~Fujii, et~al., {\bf FAST} Collaboration, {Latest results of
  ultra-high-energy cosmic ray measurements with prototypes of the Fluorescence
  detector Array of Single-pixel Telescopes (FAST)}, PoS ICRC2021 (2021) 402.
\newblock \href {http://arxiv.org/abs/2107.02949} {\path{arXiv:2107.02949}},
  \href {http://dx.doi.org/10.22323/1.395.0402}
  {\path{doi:10.22323/1.395.0402}}.

\bibitem{FAST:2017vfb}
D.~Mandat, et~al., {\bf FAST} Collaboration, {The prototype opto-mechanical
  system for the Fluorescence detector Array of Single-pixel Telescopes}, JINST
  12~(07) (2017) T07001.
\newblock \href {http://dx.doi.org/10.1088/1748-0221/12/07/T07001}
  {\path{doi:10.1088/1748-0221/12/07/T07001}}.

\bibitem{FAST:2020blj}
L.~Chytka, et~al., {\bf FAST} Collaboration, {An automated all-sky atmospheric
  monitoring camera for a next-generation ultrahigh-energy cosmic-ray
  observatory}, JINST 15~(10) (2020) T10009.
\newblock \href {http://dx.doi.org/10.1088/1748-0221/15/10/T10009}
  {\path{doi:10.1088/1748-0221/15/10/T10009}}.

\bibitem{Justin:2021phd}
J.~Albury, {Extending the Energy Range of Ultra-High Energy Cosmic Ray
  Fluorescence Detectors}, Ph.D. thesis (2021) University of Adelaide.

\bibitem{Heck:1998vt}
D.~Heck, J.~Knapp, J.~N. Capdevielle, G.~Schatz, T.~Thouw, {CORSIKA: A Monte
  Carlo code to simulate extensive air showers}\href
  {http://dx.doi.org/10.5445/IR/270043064} {\path{doi:10.5445/IR/270043064}}.

\bibitem{Hillas:1982wz}
A.~M. Hillas, {THE SENSITIVITY OF CHERENKOV RADIATION PULSES TO THE
  LONGITUDINAL DEVELOPMENT OF COSMIC RAY SHOWERS}, J. Phys. G 8 (1982)
  1475--1492.
\newblock \href {http://dx.doi.org/10.1088/0305-4616/8/10/017}
  {\path{doi:10.1088/0305-4616/8/10/017}}.

\bibitem{Patterson:1983qj}
J.~R. Patterson, A.~M. Hillas, {THE RELATION OF THE LATERAL DISTRIBUTION OF
  CHERENKOV LIGHT FROM COSMIC RAY SHOWERS TO THE DISTANCE OF MAXIMUM
  DEVELOPMENT}, J. Phys. G 9 (1983) 1433--1452.
\newblock \href {http://dx.doi.org/10.1088/0305-4616/9/11/015}
  {\path{doi:10.1088/0305-4616/9/11/015}}.

\bibitem{Karle:1995dk}
A.~Karle, {Design and performance of the angle integrating Cherenkov array
  AIROBICC}, Astropart. Phys. 3 (1995) 321--347.
\newblock \href {http://dx.doi.org/10.1016/0927-6505(95)00009-6}
  {\path{doi:10.1016/0927-6505(95)00009-6}}.

\bibitem{Fowler:2000si}
J.~W. Fowler, L.~F. Fortson, C.~C.~H. Jui, D.~B. Kieda, R.~A. Ong, C.~L. Pryke,
  P.~Sommers, {A Measurement of the cosmic ray spectrum and composition at the
  knee}, Astropart. Phys. 15 (2001) 49--64.
\newblock \href {http://arxiv.org/abs/astro-ph/0003190}
  {\path{arXiv:astro-ph/0003190}}, \href
  {http://dx.doi.org/10.1016/S0927-6505(00)00139-0}
  {\path{doi:10.1016/S0927-6505(00)00139-0}}.

\bibitem{Ivanov:2014tra}
A.~A. Ivanov, et~al., {Wide field of view Cherenkov telescope to detect cosmic
  rays in coincidence with surface detectors of the extensive air shower
  array}, Nucl. Instrum. Meth. A 772 (2015) 34--42.
\newblock \href {http://arxiv.org/abs/1404.6595} {\path{arXiv:1404.6595}},
  \href {http://dx.doi.org/10.1016/j.nima.2014.10.029}
  {\path{doi:10.1016/j.nima.2014.10.029}}.

\bibitem{Prosin:2021nad}
V.~V. Prosin, et~al., {Depth of the Maximum of Extensive Air Showers (EASes)
  and the Mean Mass Composition of Primary Cosmic Rays in the
  10$^{15}$\textendash{}10$^{18}$ eV Range of Energies, According to Data from
  the TUNKA-133 and TAIGA-HiSCORE Arrays for Detecting EAS Cherenkov Light in
  the Tunkinsk Valley}, Bull. Russ. Acad. Sci. Phys. 85~(4) (2021) 395--397.
\newblock \href {http://dx.doi.org/10.3103/S1062873821040298}
  {\path{doi:10.3103/S1062873821040298}}.

\bibitem{Omura:2021nkh}
Y.~Omura, R.~Tsuda, Y.~Tsunesada, D.~R. Bergman, J.~F. Krizmanic, {Energy
  spectrum and the shower maxima of cosmic rays above the knee region measured
  with the NICHE detectors at the TA site}, PoS ICRC2021 (2021) 329.
\newblock \href {http://dx.doi.org/10.22323/1.395.0329}
  {\path{doi:10.22323/1.395.0329}}.

\bibitem{1985ICRC....3..445H}
A.~M. {Hillas}, {Cerenkov Light Images of EAS Produced by Primary Gamma Rays
  and by Nuclei}, in: 19th International Cosmic Ray Conference (ICRC19), Volume
  3, Vol.~3 of International Cosmic Ray Conference, 1985, p. 445.

\bibitem{Forster:2014rra}
A.~F\"orster, {\bf H.E.S.S.} Collaboration, {Gamma-ray astronomy with H.E.S.S},
  Nucl. Instrum. Meth. A 766 (2014) 69--72.
\newblock \href {http://dx.doi.org/10.1016/j.nima.2014.05.038}
  {\path{doi:10.1016/j.nima.2014.05.038}}.

\bibitem{Gueta:2021vrf}
O.~Gueta, {\bf CTA Consortium, CTA Observatory} Collaboration, {The Cherenkov
  Telescope Array: layout, design and performance}, PoS ICRC2021 (2021) 885.
\newblock \href {http://arxiv.org/abs/2108.04512} {\path{arXiv:2108.04512}},
  \href {http://dx.doi.org/10.22323/1.395.0885}
  {\path{doi:10.22323/1.395.0885}}.

\bibitem{Jankowsky:2020baz}
D.~Jankowsky, {Measurement of the Cosmic Ray Proton Spectrum with H.E.S.S. and
  Characterization of the TARGET ASICs for the CTA}, Ph.D. thesis, Erlangen -
  Nuremberg U. (2020).

\bibitem{VERITAS:2018gjd}
A.~Archer, et~al., {\bf VERITAS} Collaboration, {Measurement of the Iron
  Spectrum in Cosmic Rays by VERITAS}, Phys. Rev. D 98~(2) (2018) 022009.
\newblock \href {http://arxiv.org/abs/1807.08010} {\path{arXiv:1807.08010}},
  \href {http://dx.doi.org/10.1103/PhysRevD.98.022009}
  {\path{doi:10.1103/PhysRevD.98.022009}}.

\bibitem{larissa_paul_2022_6354743}
L.~Paul, et~al., {\bf IceCube} Collaboration, Air shower reconstruction using a
  graph neural network for the iceact telescopes (Feb. 2022).
\newblock \href {http://dx.doi.org/10.5281/zenodo.6354743}
  {\path{doi:10.5281/zenodo.6354743}}.

\bibitem{Bergman:2020pjh}
D.~Bergman, J.~F. Krizmanic, K.~Nakai, Y.~Omura, Y.~Tsunesada, {\bf Telescope
  Array} Collaboration, {First Results from NICHE and the NICHE-TALE Hybrid
  Detector}, PoS ICRC2019 (2020) 189.
\newblock \href {http://dx.doi.org/10.22323/1.358.0189}
  {\path{doi:10.22323/1.358.0189}}.

\bibitem{Novotny:2021lfu}
V.~Novotny, {\bf Pierre Auger} Collaboration, {Measurement of the spectrum of
  cosmic rays above $10^{16.5}$ eV with Cherenkov-dominated events at the
  Pierre Auger Observatory}, PoS ICRC2019 (2021) 374.
\newblock \href {http://dx.doi.org/10.22323/1.358.0374}
  {\path{doi:10.22323/1.358.0374}}.

\bibitem{Huege:2016veh}
T.~Huege, {Radio detection of cosmic ray air showers in the digital era}, Phys.
  Rept. 620 (2016) 1--52.
\newblock \href {http://arxiv.org/abs/1601.07426} {\path{arXiv:1601.07426}},
  \href {http://dx.doi.org/10.1016/j.physrep.2016.02.001}
  {\path{doi:10.1016/j.physrep.2016.02.001}}.

\bibitem{Schroder:2016hrv}
F.~G. Schr\"oder, {Radio detection of Cosmic-Ray Air Showers and High-Energy
  Neutrinos}, Prog. Part. Nucl. Phys. 93 (2017) 1--68.
\newblock \href {http://arxiv.org/abs/1607.08781} {\path{arXiv:1607.08781}},
  \href {http://dx.doi.org/10.1016/j.ppnp.2016.12.002}
  {\path{doi:10.1016/j.ppnp.2016.12.002}}.

\bibitem{Huege:2013vt}
T.~Huege, M.~Ludwig, C.~W. James, {Simulating radio emission from air showers
  with CoREAS}, AIP Conf. Proc. 1535~(1) (2013) 128.
\newblock \href {http://arxiv.org/abs/1301.2132} {\path{arXiv:1301.2132}},
  \href {http://dx.doi.org/10.1063/1.4807534} {\path{doi:10.1063/1.4807534}}.

\bibitem{Alvarez-Muniz:2010wjm}
J.~Alvarez-Muniz, A.~Romero-Wolf, E.~Zas, {Cherenkov radio pulses from
  electromagnetic showers in the time-domain}, Phys. Rev. D 81 (2010) 123009.
\newblock \href {http://arxiv.org/abs/1002.3873} {\path{arXiv:1002.3873}},
  \href {http://dx.doi.org/10.1103/PhysRevD.81.123009}
  {\path{doi:10.1103/PhysRevD.81.123009}}.

\bibitem{Gottowik:2019yih}
M.~Gottowik, C.~Glaser, T.~Huege, J.~Rautenberg, {Systematic uncertainty of
  first-principle calculations of the radiation energy emitted by extensive air
  showers}, EPJ Web Conf. 216 (2019) 03008.
\newblock \href {http://dx.doi.org/10.1051/epjconf/201921603008}
  {\path{doi:10.1051/epjconf/201921603008}}.

\bibitem{T-510:2015pyu}
K.~Belov, et~al., {\bf T-510} Collaboration, {Accelerator measurements of
  magnetically-induced radio emission from particle cascades with applications
  to cosmic-ray air showers}, Phys. Rev. Lett. 116~(14) (2016) 141103.
\newblock \href {http://arxiv.org/abs/1507.07296} {\path{arXiv:1507.07296}},
  \href {http://dx.doi.org/10.1103/PhysRevLett.116.141103}
  {\path{doi:10.1103/PhysRevLett.116.141103}}.

\bibitem{Scholten:2016gmj}
O.~Scholten, et~al., {Measurement of the circular polarization in radio
  emission from extensive air showers confirms emission mechanisms}, Phys. Rev.
  D 94~(10) (2016) 103010.
\newblock \href {http://arxiv.org/abs/1611.00758} {\path{arXiv:1611.00758}},
  \href {http://dx.doi.org/10.1103/PhysRevD.94.103010}
  {\path{doi:10.1103/PhysRevD.94.103010}}.

\bibitem{Karastathis:2021akf}
N.~Karastathis, R.~Prechelt, T.~Huege, J.~Ammerman-Yebra, {\bf CORSIKA 8}
  Collaboration, {Simulations of radio emission from air showers with CORSIKA
  8}, PoS ICRC2021 (2021) 427.
\newblock \href {http://dx.doi.org/10.22323/1.395.0427}
  {\path{doi:10.22323/1.395.0427}}.

\bibitem{Huege:2015jga}
T.~Huege, et~al., {High-precision measurements of extensive air showers with
  the SKA}, PoS ICRC2015 (2016) 309.
\newblock \href {http://arxiv.org/abs/1508.03465} {\path{arXiv:1508.03465}},
  \href {http://dx.doi.org/10.22323/1.236.0309}
  {\path{doi:10.22323/1.236.0309}}.

\bibitem{Scholten:2007ky}
O.~Scholten, K.~Werner, F.~Rusydi, {A Macroscopic Description of Coherent
  Geo-Magnetic Radiation from Cosmic Ray Air Showers}, Astropart. Phys. 29
  (2008) 94--103.
\newblock \href {http://arxiv.org/abs/0709.2872} {\path{arXiv:0709.2872}},
  \href {http://dx.doi.org/10.1016/j.astropartphys.2007.11.012}
  {\path{doi:10.1016/j.astropartphys.2007.11.012}}.

\bibitem{Scholten:2017tcr}
O.~Scholten, T.~N.~G. Trinh, K.~D. de~Vries, B.~M. Hare, {Analytic calculation
  of radio emission from parametrized extensive air showers: A tool to extract
  shower parameters}, Phys. Rev. D 97~(2) (2018) 023005.
\newblock \href {http://arxiv.org/abs/1711.10164} {\path{arXiv:1711.10164}},
  \href {http://dx.doi.org/10.1103/PhysRevD.97.023005}
  {\path{doi:10.1103/PhysRevD.97.023005}}.

\bibitem{Butler:2019kde}
D.~Butler, T.~Huege, R.~Engel, O.~Scholten, {Universality and template
  synthesis of cosmic ray air shower radio emission}, PoS ICRC2019 (2020) 295.
\newblock \href {http://arxiv.org/abs/1908.09543} {\path{arXiv:1908.09543}},
  \href {http://dx.doi.org/10.22323/1.358.0295}
  {\path{doi:10.22323/1.358.0295}}.

\bibitem{Chiche:2021iin}
S.~Chiche, O.~Martineau-Huynh, K.~Kotera, M.~Tueros, K.~D.~de Vries,
  {Radio-Morphing: a fast, efficient and accurate tool to compute the radio
  signals from air-showers}, PoS ICRC2021 (2021) 194.
\newblock \href {http://dx.doi.org/10.22323/1.395.0194}
  {\path{doi:10.22323/1.395.0194}}.

\bibitem{PierreAuger:2018pmw}
A.~Aab, et~al., {\bf Pierre Auger} Collaboration, {Observation of inclined EeV
  air showers with the radio detector of the Pierre Auger Observatory}, JCAP 10
  (2018) 026.
\newblock \href {http://arxiv.org/abs/1806.05386} {\path{arXiv:1806.05386}},
  \href {http://dx.doi.org/10.1088/1475-7516/2018/10/026}
  {\path{doi:10.1088/1475-7516/2018/10/026}}.

\bibitem{Alvarez-Muniz:2018bhp}
J.~\'Alvarez-Mu\~niz, et~al., {\bf GRAND} Collaboration, {The Giant Radio Array
  for Neutrino Detection (GRAND): Science and Design}, Sci. China Phys. Mech.
  Astron. 63~(1) (2020) 219501.
\newblock \href {http://arxiv.org/abs/1810.09994} {\path{arXiv:1810.09994}},
  \href {http://dx.doi.org/10.1007/s11433-018-9385-7}
  {\path{doi:10.1007/s11433-018-9385-7}}.

\bibitem{Schluter:2020tdz}
F.~Schl\"uter, M.~Gottowik, T.~Huege, J.~Rautenberg, {Refractive displacement
  of the radio-emission footprint of inclined air showers simulated with
  CoREAS}, Eur. Phys. J. C 80~(7) (2020) 643.
\newblock \href {http://arxiv.org/abs/2005.06775} {\path{arXiv:2005.06775}},
  \href {http://dx.doi.org/10.1140/epjc/s10052-020-8216-z}
  {\path{doi:10.1140/epjc/s10052-020-8216-z}}.

\bibitem{DeKockere:2021qka}
S.~De~Kockere, K.~de~Vries, N.~van Eijndhoven, {Simulation of the propagation
  of CR air shower cores in ice}, PoS ICRC2021 (2021) 1032.
\newblock \href {http://dx.doi.org/10.22323/1.395.1032}
  {\path{doi:10.22323/1.395.1032}}.

\bibitem{Kostunin:2015taa}
D.~Kostunin, P.~A. Bezyazeekov, R.~Hiller, F.~G. Schr\"oder, V.~Lenok,
  E.~Levinson, {Reconstruction of air-shower parameters for large-scale radio
  detectors using the lateral distribution}, Astropart. Phys. 74 (2016) 79--86.
\newblock \href {http://arxiv.org/abs/1504.05083} {\path{arXiv:1504.05083}},
  \href {http://dx.doi.org/10.1016/j.astropartphys.2015.10.004}
  {\path{doi:10.1016/j.astropartphys.2015.10.004}}.

\bibitem{LOPES:2014bps}
W.~D. Apel, et~al., {\bf LOPES} Collaboration, {Reconstruction of the energy
  and depth of maximum of cosmic-ray air-showers from LOPES radio
  measurements}, Phys. Rev. D 90~(6) (2014) 062001.
\newblock \href {http://arxiv.org/abs/1408.2346} {\path{arXiv:1408.2346}},
  \href {http://dx.doi.org/10.1103/PhysRevD.90.062001}
  {\path{doi:10.1103/PhysRevD.90.062001}}.

\bibitem{Buitink:2014eqa}
S.~Buitink, et~al., {Method for high precision reconstruction of air shower
  $X_{max}$ using two-dimensional radio intensity profiles}, Phys. Rev. D
  90~(8) (2014) 082003.
\newblock \href {http://arxiv.org/abs/1408.7001} {\path{arXiv:1408.7001}},
  \href {http://dx.doi.org/10.1103/PhysRevD.90.082003}
  {\path{doi:10.1103/PhysRevD.90.082003}}.

\bibitem{Glaser:2016tng}
C.~Glaser, M.~Erdmann, J.~R. H\"orandel, T.~Huege, J.~Schulz, {Simulation of
  the Radiation Energy Release in Air Showers}, EPJ Web Conf. 135 (2017) 01016.
\newblock \href {http://arxiv.org/abs/1609.05743} {\path{arXiv:1609.05743}},
  \href {http://dx.doi.org/10.1051/epjconf/201713501016}
  {\path{doi:10.1051/epjconf/201713501016}}.

\bibitem{Nelles:2014xaa}
A.~Nelles, S.~Buitink, H.~Falcke, J.~H\"orandel, T.~Huege, P.~Schellart, {A
  parameterization for the radio emission of air showers as predicted by CoREAS
  simulations and applied to LOFAR measurements}, Astropart. Phys. 60 (2015)
  13--24.
\newblock \href {http://arxiv.org/abs/1402.2872} {\path{arXiv:1402.2872}},
  \href {http://dx.doi.org/10.1016/j.astropartphys.2014.05.001}
  {\path{doi:10.1016/j.astropartphys.2014.05.001}}.

\bibitem{Tunka-Rex:2016gcn}
R.~Hiller, et~al., {\bf Tunka-Rex} Collaboration, {Tunka-Rex: energy
  reconstruction with a single antenna station}, EPJ Web Conf. 135 (2017)
  01004.
\newblock \href {http://arxiv.org/abs/1611.09614} {\path{arXiv:1611.09614}},
  \href {http://dx.doi.org/10.1051/epjconf/201713501004}
  {\path{doi:10.1051/epjconf/201713501004}}.

\bibitem{Schoorlemmer:2015ujs}
H.~Schoorlemmer, et~al., {Energy and Flux Measurements of Ultra-High Energy
  Cosmic Rays Observed During the First ANITA Flight}, PoS ICRC2015 (2016) 272.
\newblock \href {http://dx.doi.org/10.22323/1.236.0272}
  {\path{doi:10.22323/1.236.0272}}.

\bibitem{Welling:2019scz}
C.~Welling, C.~Glaser, A.~Nelles, {Reconstructing the cosmic-ray energy from
  the radio signal measured in one single station}, JCAP 10 (2019) 075.
\newblock \href {http://arxiv.org/abs/1905.11185} {\path{arXiv:1905.11185}},
  \href {http://dx.doi.org/10.1088/1475-7516/2019/10/075}
  {\path{doi:10.1088/1475-7516/2019/10/075}}.

\bibitem{Schroder:2018dvb}
F.~G. Schr\"oder, {\bf IceCube-Gen2} Collaboration, {Physics Potential of a
  Radio Surface Array at the South Pole}, EPJ Web Conf. 216 (2019) 01007.
\newblock \href {http://arxiv.org/abs/1811.00599} {\path{arXiv:1811.00599}},
  \href {http://dx.doi.org/10.1051/epjconf/201921601007}
  {\path{doi:10.1051/epjconf/201921601007}}.

\bibitem{Buitink:2021pkz}
S.~Buitink, et~al., {Performance of SKA as an air shower observatory}, PoS
  ICRC2021 (2021) 415.
\newblock \href {http://dx.doi.org/10.22323/1.395.0415}
  {\path{doi:10.22323/1.395.0415}}.

\bibitem{HiRes-MIA:2000ook}
T.~Abu-Zayyad, et~al., {\bf HiRes-MIA} Collaboration, {Measurement of the
  cosmic ray energy spectrum and composition from 10**17-eV to 10**18.3-eV
  using a hybrid fluorescence technique}, Astrophys. J. 557 (2001) 686--699.
\newblock \href {http://arxiv.org/abs/astro-ph/0010652}
  {\path{arXiv:astro-ph/0010652}}, \href {http://dx.doi.org/10.1086/322240}
  {\path{doi:10.1086/322240}}.

\bibitem{ref:radio_xmax_Allan1971}
H.~R. Allan, {Radio emission from extensive air showers}, Progress in
  Elementary Particle and Cosmic Ray Physics 10 (1971) 169--302.

\bibitem{ref:radio_xmax_AllanICRC71}
H.~R. Allan, {The lateral distribution of the radio emission and its dependence
  on the longitudinal structure of the air shower}, Proceedings of the 12th
  International Conference on Cosmic Rays, Tasmania, Australia 3 (1971) 1108.

\bibitem{LOPES:2012xou}
W.~D. Apel, et~al., {\bf LOPES} Collaboration, {Experimental evidence for the
  sensitivity of the air-shower radio signal to the longitudinal shower
  development}, Phys. Rev. D 85 (2012) 071101.
\newblock \href {http://arxiv.org/abs/1203.3971} {\path{arXiv:1203.3971}},
  \href {http://dx.doi.org/10.1103/PhysRevD.85.071101}
  {\path{doi:10.1103/PhysRevD.85.071101}}.

\bibitem{Palmieri:2013kvf}
N.~Palmieri, et~al., {Reconstructing energy and X$_{max}$ of cosmic ray air
  showers using the radio lateral distribution measured with LOPES}, AIP Conf.
  Proc. 1535~(1) (2013) 89--93.
\newblock \href {http://arxiv.org/abs/1308.0053} {\path{arXiv:1308.0053}},
  \href {http://dx.doi.org/10.1063/1.4807527} {\path{doi:10.1063/1.4807527}}.

\bibitem{Glaser:2018byo}
C.~Glaser, S.~de~Jong, M.~Erdmann, J.~R. H\"orandel, {An analytic description
  of the radio emission of air showers based on its emission mechanisms},
  Astropart. Phys. 104 (2019) 64--77.
\newblock \href {http://arxiv.org/abs/1806.03620} {\path{arXiv:1806.03620}},
  \href {http://dx.doi.org/10.1016/j.astropartphys.2018.08.004}
  {\path{doi:10.1016/j.astropartphys.2018.08.004}}.

\bibitem{Tunka-Rex:2015zsa}
P.~A. Bezyazeekov, et~al., {\bf Tunka-Rex} Collaboration, {Radio measurements
  of the energy and the depth of the shower maximum of cosmic-ray air showers
  by Tunka-Rex}, JCAP 01 (2016) 052.
\newblock \href {http://arxiv.org/abs/1509.05652} {\path{arXiv:1509.05652}},
  \href {http://dx.doi.org/10.1088/1475-7516/2016/01/052}
  {\path{doi:10.1088/1475-7516/2016/01/052}}.

\bibitem{Jansen:2016sjo}
S.~Jansen, {Radio for the masses}: {Cosmic ray mass composition measurements in
  the radio frequency domain}, Ph.D. thesis, Nijmegen U. (2016).

\bibitem{Canfora:2021xkh}
F.~Canfora, {Cosmic-Ray Composition Measurements Using Radio Signals}, Ph.D.
  thesis (2021).

\bibitem{Apel:2014usa}
W.~D. Apel, et~al., {The wavefront of the radio signal emitted by cosmic ray
  air showers}, JCAP 09 (2014) 025.
\newblock \href {http://arxiv.org/abs/1404.3283} {\path{arXiv:1404.3283}},
  \href {http://dx.doi.org/10.1088/1475-7516/2014/09/025}
  {\path{doi:10.1088/1475-7516/2014/09/025}}.

\bibitem{Mitra:2020mza}
P.~Mitra, et~al., {Reconstructing air shower parameters with LOFAR using event
  specific GDAS atmosphere}, Astropart. Phys. 123 (2020) 102470.
\newblock \href {http://arxiv.org/abs/2006.02228} {\path{arXiv:2006.02228}},
  \href {http://dx.doi.org/10.1016/j.astropartphys.2020.102470}
  {\path{doi:10.1016/j.astropartphys.2020.102470}}.

\bibitem{ANITA:2008mzi}
P.~W. Gorham, et~al., {\bf ANITA} Collaboration, {The Antarctic Impulsive
  Transient Antenna Ultra-high Energy Neutrino Detector Design, Performance,
  and Sensitivity for 2006-2007 Balloon Flight}, Astropart. Phys. 32 (2009)
  10--41.
\newblock \href {http://arxiv.org/abs/0812.1920} {\path{arXiv:0812.1920}},
  \href {http://dx.doi.org/10.1016/j.astropartphys.2009.05.003}
  {\path{doi:10.1016/j.astropartphys.2009.05.003}}.

\bibitem{LOPES:2021ipp}
W.~D. Apel, et~al., {\bf LOPES} Collaboration, {Final results of the LOPES
  radio interferometer for cosmic-ray air showers}, Eur. Phys. J. C 81~(2)
  (2021) 176.
\newblock \href {http://arxiv.org/abs/2102.03928} {\path{arXiv:2102.03928}},
  \href {http://dx.doi.org/10.1140/epjc/s10052-021-08912-4}
  {\path{doi:10.1140/epjc/s10052-021-08912-4}}.

\bibitem{Schroder:2010sa}
F.~G. Schroder, et~al., {New method for the time calibration of an
  interferometric radio antenna array}, Nucl. Instrum. Meth. A 615 (2010)
  277--284.
\newblock \href {http://arxiv.org/abs/1002.3775} {\path{arXiv:1002.3775}},
  \href {http://dx.doi.org/10.1016/j.nima.2010.01.072}
  {\path{doi:10.1016/j.nima.2010.01.072}}.

\bibitem{Schoorlemmer:2020low}
H.~Schoorlemmer, W.~R. Carvalho, {Radio interferometry applied to the
  observation of cosmic-ray induced extensive air showers}, Eur. Phys. J. C
  81~(12) (2021) 1120.
\newblock \href {http://arxiv.org/abs/2006.10348} {\path{arXiv:2006.10348}},
  \href {http://dx.doi.org/10.1140/epjc/s10052-021-09925-9}
  {\path{doi:10.1140/epjc/s10052-021-09925-9}}.

\bibitem{Schluter:2021egm}
F.~Schl\"uter, T.~Huege, {Expected performance of air-shower measurements with
  the radio-interferometric technique}, JINST 16~(07) (2021) P07048.
\newblock \href {http://arxiv.org/abs/2102.13577} {\path{arXiv:2102.13577}},
  \href {http://dx.doi.org/10.1088/1748-0221/16/07/P07048}
  {\path{doi:10.1088/1748-0221/16/07/P07048}}.

\bibitem{Plant:2021iqx}
K.~Plant, A.~Romero-Wolf, W.~Carvalho, K.~Belov, G.~Hallinan, {Updates from the
  OVRO-LWA: Commissioning a Full-Duty-Cycle Radio-Only Cosmic Ray Detector},
  PoS ICRC2021 (2021) 204.
\newblock \href {http://dx.doi.org/10.22323/1.395.0204}
  {\path{doi:10.22323/1.395.0204}}.

\bibitem{9076025}
J.~E. Gilligan, E.~M. Konitzer, E.~Siman-Tov, J.~W. Zobel, E.~J. Adles, White
  rabbit time and frequency transfer over wireless millimeter-wave carriers,
  IEEE Transactions on Ultrasonics, Ferroelectrics, and Frequency Control
  67~(9) (2020) 1946--1952.
\newblock \href {http://dx.doi.org/10.1109/TUFFC.2020.2989667}
  {\path{doi:10.1109/TUFFC.2020.2989667}}.

\bibitem{Allison:2018ynt}
P.~Allison, et~al., {Design and performance of an interferometric trigger array
  for radio detection of high-energy neutrinos}, Nucl. Instrum. Meth. A 930
  (2019) 112--125.
\newblock \href {http://arxiv.org/abs/1809.04573} {\path{arXiv:1809.04573}},
  \href {http://dx.doi.org/10.1016/j.nima.2019.01.067}
  {\path{doi:10.1016/j.nima.2019.01.067}}.

\bibitem{RNO-G:2020rmc}
J.~A. Aguilar, et~al., {\bf RNO-G} Collaboration, {Design and Sensitivity of
  the Radio Neutrino Observatory in Greenland (RNO-G)}, JINST 16~(03) (2021)
  P03025.
\newblock \href {http://arxiv.org/abs/2010.12279} {\path{arXiv:2010.12279}},
  \href {http://dx.doi.org/10.1088/1748-0221/16/03/P03025}
  {\path{doi:10.1088/1748-0221/16/03/P03025}}.

\bibitem{Escudie:2019tlt}
A.~Escudie, D.~Charrier, R.~Dallier, D.~Garc\'\i{}a-Fern\'andez, A.~Lecacheux,
  L.~Martin, B.~Revenu, {Radio detection of atmospheric air showers of
  particles}\href {http://arxiv.org/abs/1903.02889} {\path{arXiv:1903.02889}}.

\bibitem{Hughes:2020ghq}
K.~Hughes, et~al., {Towards Interferometric Triggering on Air Showers Induced
  by Tau Neutrino Interactions}, PoS ICRC2019 (2020) 917.
\newblock \href {http://dx.doi.org/10.22323/1.358.0917}
  {\path{doi:10.22323/1.358.0917}}.

\bibitem{Barwick:2016mxm}
S.~W. Barwick, et~al., {Radio detection of air showers with the ARIANNA
  experiment on the Ross Ice Shelf}, Astropart. Phys. 90 (2017) 50--68.
\newblock \href {http://arxiv.org/abs/1612.04473} {\path{arXiv:1612.04473}},
  \href {http://dx.doi.org/10.1016/j.astropartphys.2017.02.003}
  {\path{doi:10.1016/j.astropartphys.2017.02.003}}.

\bibitem{ANITA:2010ect}
S.~Hoover, et~al., {\bf ANITA} Collaboration, {Observation of Ultra-high-energy
  Cosmic Rays with the ANITA Balloon-borne Radio Interferometer}, Phys. Rev.
  Lett. 105 (2010) 151101.
\newblock \href {http://arxiv.org/abs/1005.0035} {\path{arXiv:1005.0035}},
  \href {http://dx.doi.org/10.1103/PhysRevLett.105.151101}
  {\path{doi:10.1103/PhysRevLett.105.151101}}.

\bibitem{ANITA:2010hzc}
P.~W. Gorham, et~al., {\bf ANITA} Collaboration, {Observational Constraints on
  the Ultra-high Energy Cosmic Neutrino Flux from the Second Flight of the
  ANITA Experiment}, Phys. Rev. D 82 (2010) 022004, [Erratum: Phys.Rev.D 85,
  049901 (2012)].
\newblock \href {http://arxiv.org/abs/1003.2961} {\path{arXiv:1003.2961}},
  \href {http://dx.doi.org/10.1103/PhysRevD.82.022004}
  {\path{doi:10.1103/PhysRevD.82.022004}}.

\bibitem{ANITA:2018vwl}
P.~W. Gorham, et~al., {\bf ANITA} Collaboration, {Constraints on the diffuse
  high-energy neutrino flux from the third flight of ANITA}, Phys. Rev. D
  98~(2) (2018) 022001.
\newblock \href {http://arxiv.org/abs/1803.02719} {\path{arXiv:1803.02719}},
  \href {http://dx.doi.org/10.1103/PhysRevD.98.022001}
  {\path{doi:10.1103/PhysRevD.98.022001}}.

\bibitem{PUEO:2020bnn}
Q.~Abarr, et~al., {\bf PUEO} Collaboration, {The Payload for Ultrahigh Energy
  Observations (PUEO): a white paper}, JINST 16~(08) (2021) P08035.
\newblock \href {http://arxiv.org/abs/2010.02892} {\path{arXiv:2010.02892}},
  \href {http://dx.doi.org/10.1088/1748-0221/16/08/P08035}
  {\path{doi:10.1088/1748-0221/16/08/P08035}}.

\bibitem{IceCube-Gen2:2020qha}
M.~G. Aartsen, et~al., {\bf IceCube-Gen2} Collaboration, {IceCube-Gen2: the
  window to the extreme Universe}, J. Phys. G 48~(6) (2021) 060501.
\newblock \href {http://arxiv.org/abs/2008.04323} {\path{arXiv:2008.04323}},
  \href {http://dx.doi.org/10.1088/1361-6471/abbd48}
  {\path{doi:10.1088/1361-6471/abbd48}}.

\bibitem{Huege:2019ufo}
T.~Huege, C.~B. Welling, {\bf Pierre Auger} Collaboration, {Reconstruction of
  air-shower measurements with AERA in the presence of pulsed radio-frequency
  interference}, EPJ Web Conf. 216 (2019) 03007.
\newblock \href {http://arxiv.org/abs/1906.05148} {\path{arXiv:1906.05148}},
  \href {http://dx.doi.org/10.1051/epjconf/201921603007}
  {\path{doi:10.1051/epjconf/201921603007}}.

\bibitem{Schmidt:2011vue}
A.~Schmidt, {Realization of a Self-Triggered Detector for the Radio Emission of
  Cosmic Rays}, Ph.D. thesis, KIT, Karlsruhe (2012).
\newblock \href {http://dx.doi.org/10.5445/IR/1000030957}
  {\path{doi:10.5445/IR/1000030957}}.

\bibitem{PierreAuger:2012gwg}
P.~Abreu, et~al., {\bf Pierre Auger} Collaboration, {Results of a
  Self-Triggered Prototype System for Radio-Detection of Extensive Air Showers
  at the Pierre Auger Observatory}, JINST 7 (2012) P11023.
\newblock \href {http://arxiv.org/abs/1211.0572} {\path{arXiv:1211.0572}},
  \href {http://dx.doi.org/10.1088/1748-0221/7/11/P11023}
  {\path{doi:10.1088/1748-0221/7/11/P11023}}.

\bibitem{Charrier:2018fle}
D.~Charrier, et~al., {Autonomous radio detection of air showers with the
  TREND50 antenna array}, Astropart. Phys. 110 (2019) 15--29.
\newblock \href {http://arxiv.org/abs/1810.03070} {\path{arXiv:1810.03070}},
  \href {http://dx.doi.org/10.1016/j.astropartphys.2019.03.002}
  {\path{doi:10.1016/j.astropartphys.2019.03.002}}.

\bibitem{Ardouin:2010gz}
D.~Ardouin, et~al., {First detection of extensive air showers by the TREND
  self-triggering radio experiment}, Astropart. Phys. 34 (2011) 717--731.
\newblock \href {http://arxiv.org/abs/1007.4359} {\path{arXiv:1007.4359}},
  \href {http://dx.doi.org/10.1016/j.astropartphys.2011.01.002}
  {\path{doi:10.1016/j.astropartphys.2011.01.002}}.

\bibitem{Zhang:2021Tk}
Y.~Zhang, {Self-trigger radio prototype array for GRAND}, PoS ICRC2021 (2021)
  1035.
\newblock \href {http://dx.doi.org/10.22323/1.395.1035}
  {\path{doi:10.22323/1.395.1035}}.

\bibitem{Holt:2019fnj}
E.~M. Holt, F.~G. Schr\"oder, A.~Haungs, {Enhancing the cosmic-ray mass
  sensitivity of air-shower arrays by combining radio and muon detectors}, Eur.
  Phys. J. C 79~(5) (2019) 371.
\newblock \href {http://arxiv.org/abs/1905.01409} {\path{arXiv:1905.01409}},
  \href {http://dx.doi.org/10.1140/epjc/s10052-019-6859-4}
  {\path{doi:10.1140/epjc/s10052-019-6859-4}}.

\bibitem{Shipilov:2018wph}
D.~Shipilov, et~al., {Signal recognition and background suppression by matched
  filters and neural networks for Tunka-Rex}, EPJ Web Conf. 216 (2019) 02003.
\newblock \href {http://arxiv.org/abs/1812.03347} {\path{arXiv:1812.03347}},
  \href {http://dx.doi.org/10.1051/epjconf/201921602003}
  {\path{doi:10.1051/epjconf/201921602003}}.

\bibitem{Rehman:2021oby}
A.~Rehman, A.~Coleman, F.~G. Schr\"oder, D.~Kostunin, {Classification and
  Denoising of Cosmic-Ray Radio Signals using Deep Learning}, PoS ICRC2021
  (2021) 417.
\newblock \href {http://dx.doi.org/10.22323/1.395.0417}
  {\path{doi:10.22323/1.395.0417}}.

\bibitem{Sadovnichy2011}
V.~A. Sadovnichy, et~al., Investigations of the space environment aboard the
  universitetsky-tat'yana and universitetsky-tat'yana-2 microsatellites, Solar
  System Research 45~(1) (2011) 3--29.
\newblock \href {http://dx.doi.org/10.1134/S0038094611010096}
  {\path{doi:10.1134/S0038094611010096}}.

\bibitem{Klimov:2017lwx}
P.~A. Klimov, et~al., {The TUS detector of extreme energy cosmic rays on board
  the Lomonosov satellite}, Space Sci. Rev. 212~(3-4) (2017) 1687--1703.
\newblock \href {http://arxiv.org/abs/1706.04976} {\path{arXiv:1706.04976}},
  \href {http://dx.doi.org/10.1007/s11214-017-0403-3}
  {\path{doi:10.1007/s11214-017-0403-3}}.

\bibitem{Adams:2015pec}
J.~H. Adams, {\bf JEM-EUSO} Collaboration, {The EUSO-Balloon pathfinder},
  Exper. Astron. 40~(1) (2015) 281--299.
\newblock \href {http://dx.doi.org/10.1007/s10686-015-9467-9}
  {\path{doi:10.1007/s10686-015-9467-9}}.

\bibitem{Abdellaoui:2019qmg}
G.~Abdellaoui, et~al., {Ultra-violet imaging of the night-time earth by
  EUSO-Balloon towards space-based ultra-high energy cosmic ray observations},
  Astropart. Phys. 111 (2019) 54--71.
\newblock \href {http://dx.doi.org/10.1016/j.astropartphys.2018.10.008}
  {\path{doi:10.1016/j.astropartphys.2018.10.008}}.

\bibitem{Adams:2022oko}
J.~H. Adams, et~al., {A Review of the EUSO-Balloon Pathfinder for the JEM-EUSO
  Program}, Space Sci. Rev. 218~(1) (2022) 3.
\newblock \href {http://dx.doi.org/10.1007/s11214-022-00870-x}
  {\path{doi:10.1007/s11214-022-00870-x}}.

\bibitem{Bacholle:2017dye}
S.~Bacholle, {\bf JEM-EUSO} Collaboration, {The EUSO-SPB instrument}, PoS
  ICRC2017 (2018) 384.
\newblock \href {http://dx.doi.org/10.22323/1.301.0384}
  {\path{doi:10.22323/1.301.0384}}.

\bibitem{Kungel:2021anx}
V.~Kungel, et~al., {EUSO-SPB2 Telescope Optics and Testing}, PoS ICRC2021
  (2021) 412.
\newblock \href {http://dx.doi.org/10.22323/1.395.0412}
  {\path{doi:10.22323/1.395.0412}}.

\bibitem{Abdellaoui:2018rkw}
G.~Abdellaoui, et~al., {EUSO-TA \textendash{} First results from a ground-based
  EUSO telescope}, Astropart. Phys. 102 (2018) 98--111.
\newblock \href {http://dx.doi.org/10.1016/j.astropartphys.2018.05.007}
  {\path{doi:10.1016/j.astropartphys.2018.05.007}}.

\bibitem{Adams:2017fjh}
J.~H. Adams, et~al., {White paper on EUSO-SPB2}\href
  {http://arxiv.org/abs/1703.04513} {\path{arXiv:1703.04513}}.

\bibitem{Blin:2018tjp}
S.~Blin, et~al., {\bf JEM-EUSO} Collaboration, {SPACIROC3: 100 MHz photon
  counting ASIC for EUSO-SPB}, Nucl. Instrum. Meth. A 912 (2018) 363--367.
\newblock \href {http://dx.doi.org/10.1016/j.nima.2017.12.060}
  {\path{doi:10.1016/j.nima.2017.12.060}}.

\bibitem{Casolino:2021pss}
M.~Casolino, G.~Cambie', L.~Marcelli, E.~Reali, {SiPM development for
  space-borne and ground detectors: From Lazio-Sirad and Mini-EUSO to Lanfos},
  Nucl. Instrum. Meth. A 986 (2021) 164649.
\newblock \href {http://dx.doi.org/10.1016/j.nima.2020.164649}
  {\path{doi:10.1016/j.nima.2020.164649}}.

\bibitem{Otte:2019lbq}
A.~N. Otte, et~al., {\bf JEM-EUSO, POEMMA} Collaboration, {Development of a
  Cherenkov Telescope for the Detection of Ultra-High Energy Neutrinos with
  EUSO-SPB2 and POEMMA}, PoS ICRC2019~(977) (2021) 977.
\newblock \href {http://arxiv.org/abs/1907.08728} {\path{arXiv:1907.08728}},
  \href {http://dx.doi.org/10.22323/1.358.0977}
  {\path{doi:10.22323/1.358.0977}}.

\bibitem{cvmfs}
{CernVM-File System}, \url{https://cernvm.cern.ch/fs/}.

\bibitem{cloud_1_2020}
I.~Sfiligoi, et~al., Running a pre-exascale, geographically distributed,
  multi-cloud scientific simulation, High Performance Computing (2020)
  {23–40, }\href {http://dx.doi.org/10.1007/978-3-030-50743-5_2}
  {\path{doi:10.1007/978-3-030-50743-5_2}}.

\bibitem{cloud_2_2020}
I.~Sfiligoi, et~al., Demonstrating a pre-exascale, cost-effective multi-cloud
  environment for scientific computing, {Practice and Experience in Advanced
  Research Computing, }\href {http://dx.doi.org/10.1145/3311790.3396625}
  {\path{doi:10.1145/3311790.3396625}}.

\bibitem{cloud_2021}
I.~Sfiligoi, et~al., Managing cloud networking costs for data-intensive
  applications by provisioning dedicated network links, Practice and Experience
  in Advanced Research Computing\href
  {http://dx.doi.org/10.1145/3437359.3465563}
  {\path{doi:10.1145/3437359.3465563}}.

\bibitem{Arguelles:2019phs}
C.~A. Arg\"uelles, B.~J.~P. Jones, {Neutrino Oscillations in a Quantum
  Processor}, Phys. Rev. Research. 1 (2019) 033176.
\newblock \href {http://arxiv.org/abs/1904.10559} {\path{arXiv:1904.10559}},
  \href {http://dx.doi.org/10.1103/PhysRevResearch.1.033176}
  {\path{doi:10.1103/PhysRevResearch.1.033176}}.

\bibitem{Bauer:2019qxa}
C.~W. Bauer, W.~A. de~Jong, B.~Nachman, D.~Provasoli, {Quantum Algorithm for
  High Energy Physics Simulations}, Phys. Rev. Lett. 126~(6) (2021) 062001.
\newblock \href {http://arxiv.org/abs/1904.03196} {\path{arXiv:1904.03196}},
  \href {http://dx.doi.org/10.1103/PhysRevLett.126.062001}
  {\path{doi:10.1103/PhysRevLett.126.062001}}.

\bibitem{Wei:2019rqy}
A.~Y. Wei, P.~Naik, A.~W. Harrow, J.~Thaler, {Quantum Algorithms for Jet
  Clustering}, Phys. Rev. D 101~(9) (2020) 094015.
\newblock \href {http://arxiv.org/abs/1908.08949} {\path{arXiv:1908.08949}},
  \href {http://dx.doi.org/10.1103/PhysRevD.101.094015}
  {\path{doi:10.1103/PhysRevD.101.094015}}.

\bibitem{Benato:2021olt}
L.~Benato, et~al., {Shared Data and Algorithms for Deep Learning in Fundamental
  Physics }\href {http://arxiv.org/abs/2107.00656} {\path{arXiv:2107.00656}}.

\bibitem{geometric_dl}
M.~M. Bronstein, J.~Bruna, Y.~LeCun, A.~Szlam, P.~Vandergheynst, Geometric deep
  learning: Going beyond euclidean data, IEEE Signal Processing Magazine 34~(4)
  (2017) 18–42.
\newblock \href {http://dx.doi.org/10.1109/msp.2017.2693418}
  {\path{doi:10.1109/msp.2017.2693418}}.

\bibitem{vaswani2017attention}
A.~Vaswani, N.~Shazeer, N.~Parmar, J.~Uszkoreit, L.~Jones, A.~N. Gomez,
  L.~Kaiser, I.~Polosukhin, Attention is all you need (2017).
\newblock \href {http://arxiv.org/abs/1706.03762} {\path{arXiv:1706.03762}}.

\bibitem{wiencke2019extreme}
L.~Wiencke, A.~Olinto, The extreme universe space observatory on a
  super-pressure balloon ii mission (2019).
\newblock \href {http://arxiv.org/abs/1909.12835} {\path{arXiv:1909.12835}}.

\bibitem{Filippatos:2021noz}
G.~Filippatos, M.~Battisti, M.~E. Bertaina, F.~Bisconti, J.~Eser, G.~Osteria,
  F.~Sarazin, L.~Wiencke, C.~Heaton, {\bf JEM-EUSO} Collaboration, {Expected
  Performance of the EUSO-SPB2 Fluorescence Telescope}, PoS ICRC2021 (2021)
  405.
\newblock \href {http://dx.doi.org/10.22323/1.395.0405}
  {\path{doi:10.22323/1.395.0405}}.

\bibitem{goodfellow2014generative}
I.~J. Goodfellow, J.~Pouget-Abadie, M.~Mirza, B.~Xu, D.~Warde-Farley, S.~Ozair,
  A.~Courville, Y.~Bengio, Generative adversarial networks (2014).
\newblock \href {http://arxiv.org/abs/1406.2661} {\path{arXiv:1406.2661}}.

\bibitem{Paganini:2017hrr}
M.~Paganini, L.~de~Oliveira, B.~Nachman, {Accelerating Science with Generative
  Adversarial Networks: An Application to 3D Particle Showers in Multilayer
  Calorimeters}, Phys. Rev. Lett. 120~(4) (2018) 042003.
\newblock \href {http://arxiv.org/abs/1705.02355} {\path{arXiv:1705.02355}},
  \href {http://dx.doi.org/10.1103/PhysRevLett.120.042003}
  {\path{doi:10.1103/PhysRevLett.120.042003}}.

\bibitem{iris-hep}
Institute for research and innovation in software for high energy physics
  (iris-hep), \url{https://iris-hep.org}.

\bibitem{collin2018pathintegrals}
G.~H. Collin, Using path integrals for the propagation of light in a scattering
  dominated medium (2018).
\newblock \href {http://arxiv.org/abs/1811.04156} {\path{arXiv:1811.04156}}.

\bibitem{quantum_ml_2014}
M.~Schuld, I.~Sinayskiy, F.~Petruccione,
  \href{http://dx.doi.org/10.1080/00107514.2014.964942}{An introduction to
  quantum machine learning}, Contemporary Physics 56~(2) (2014) 172–185.
\newblock \href {http://dx.doi.org/10.1080/00107514.2014.964942}
  {\path{doi:10.1080/00107514.2014.964942}}.
\newline\urlprefix\url{http://dx.doi.org/10.1080/00107514.2014.964942}

\bibitem{quantum_ml_2017}
J.~Biamonte, P.~Wittek, N.~Pancotti, P.~Rebentrost, N.~Wiebe, S.~Lloyd, Quantum
  machine learning, Nature 549~(7671) (2017) 195–202.
\newblock \href {http://dx.doi.org/10.1038/nature23474}
  {\path{doi:10.1038/nature23474}}.

\bibitem{Reno:2019jtr}
M.~H. Reno, J.~F. Krizmanic, T.~M. Venters, {Cosmic tau neutrino detection via
  Cherenkov signals from air showers from Earth-emerging taus}, Phys. Rev. D
  100~(6) (2019) 063010.
\newblock \href {http://arxiv.org/abs/1902.11287} {\path{arXiv:1902.11287}},
  \href {http://dx.doi.org/10.1103/PhysRevD.100.063010}
  {\path{doi:10.1103/PhysRevD.100.063010}}.

\bibitem{Venters:2019xwi}
T.~M. Venters, M.~H. Reno, J.~F. Krizmanic, L.~A. Anchordoqui, C.~Gu\'epin,
  A.~V. Olinto, {POEMMA's Target of Opportunity Sensitivity to Cosmic Neutrino
  Transient Sources}, Phys. Rev. D 102 (2020) 123013.
\newblock \href {http://arxiv.org/abs/1906.07209} {\path{arXiv:1906.07209}},
  \href {http://dx.doi.org/10.1103/PhysRevD.102.123013}
  {\path{doi:10.1103/PhysRevD.102.123013}}.

\bibitem{Krizmanic:2013tea}
J.~Krizmanic, D.~Bergman, P.~Sokolsky, {The modeling of the nuclear composition
  measurement performance of the Non-Imaging CHErenkov Array (NICHE)}, in:
  {33rd International Cosmic Ray Conference}, 2013, p. 0366.
\newblock \href {http://arxiv.org/abs/1307.3918} {\path{arXiv:1307.3918}}.

\bibitem{PierreAuger:2019phh}
{The Pierre Auger Observatory: Contributions to the 36th International Cosmic
  Ray Conference (ICRC 2019)}: {Madison, Wisconsin, USA, July 24- August 1,
  2019}.
\newblock \href {http://arxiv.org/abs/1909.09073} {\path{arXiv:1909.09073}}.

\bibitem{Guepin:2021ljb}
C.~Gu\'epin, et~al., {Indirect dark matter searches at ultrahigh energy
  neutrino detectors}, Phys. Rev. D 104~(8) (2021) 083002.
\newblock \href {http://arxiv.org/abs/2106.04446} {\path{arXiv:2106.04446}},
  \href {http://dx.doi.org/10.1103/PhysRevD.104.083002}
  {\path{doi:10.1103/PhysRevD.104.083002}}.

\bibitem{Cummings:2021bhg}
A.~Cummings, R.~Aloisio, J.~Eser, J.~Krizmanic, {Modeling the optical Cherenkov
  signals by cosmic ray extensive air showers directly observed from suborbital
  and orbital altitudes}, Phys. Rev. D 104~(6) (2021) 063029.
\newblock \href {http://arxiv.org/abs/2105.03255} {\path{arXiv:2105.03255}},
  \href {http://dx.doi.org/10.1103/PhysRevD.104.063029}
  {\path{doi:10.1103/PhysRevD.104.063029}}.

\bibitem{Fenu:2021wub}
F.~Fenu, et~al., {\bf JEM-EUSO} Collaboration, {Expected performance of the
  K-EUSO space-based observatory}, PoS ICRC2021 (2021) 409.
\newblock \href {http://arxiv.org/abs/2112.11302} {\path{arXiv:2112.11302}},
  \href {http://dx.doi.org/10.22323/1.395.0409}
  {\path{doi:10.22323/1.395.0409}}.

\bibitem{Kotera:2021hbp}
K.~Kotera, {\bf GRAND} Collaboration, {The Giant Radio Array for Neutrino
  Detection (GRAND) project}, PoS ICRC2021 (2021) 1181.
\newblock \href {http://arxiv.org/abs/2108.00032} {\path{arXiv:2108.00032}},
  \href {http://dx.doi.org/10.22323/1.395.1181}
  {\path{doi:10.22323/1.395.1181}}.

\bibitem{Ardouin:2009zp}
D.~Ardouin, et~al., {Geomagnetic origin of the radio emission from cosmic ray
  induced air showers observed by CODALEMA}, Astropart. Phys. 31 (2009)
  192--200.
\newblock \href {http://arxiv.org/abs/0901.4502} {\path{arXiv:0901.4502}},
  \href {http://dx.doi.org/10.1016/j.astropartphys.2009.01.001}
  {\path{doi:10.1016/j.astropartphys.2009.01.001}}.

\bibitem{Charrier:2012zz}
D.~Charrier, {\bf CODALEMA} Collaboration, {Antenna development for
  astroparticle and radioastronomy experiments}, Nucl. Instrum. Meth. A 662
  (2012) S142--S145.
\newblock \href {http://dx.doi.org/10.1016/j.nima.2010.10.141}
  {\path{doi:10.1016/j.nima.2010.10.141}}.

\bibitem{Nelles:2014dja}
A.~Nelles, et~al., {Measuring a Cherenkov ring in the radio emission from air
  showers at 110\textendash{}190 MHz with LOFAR}, Astropart. Phys. 65 (2015)
  11--21.
\newblock \href {http://arxiv.org/abs/1411.6865} {\path{arXiv:1411.6865}},
  \href {http://dx.doi.org/10.1016/j.astropartphys.2014.11.006}
  {\path{doi:10.1016/j.astropartphys.2014.11.006}}.

\bibitem{Corstanje:2014waa}
A.~Corstanje, et~al., {The shape of the radio wavefront of extensive air
  showers as measured with LOFAR}, Astropart. Phys. 61 (2015) 22--31.
\newblock \href {http://arxiv.org/abs/1404.3907} {\path{arXiv:1404.3907}},
  \href {http://dx.doi.org/10.1016/j.astropartphys.2014.06.001}
  {\path{doi:10.1016/j.astropartphys.2014.06.001}}.

\bibitem{Buitink:2016nkf}
S.~Buitink, et~al., {A large light-mass component of cosmic rays at
  10\textasciicircum{}{17} - 10\textasciicircum{}{17.5} eV from radio
  observations}, Nature 531 (2016) 70.
\newblock \href {http://arxiv.org/abs/1603.01594} {\path{arXiv:1603.01594}},
  \href {http://dx.doi.org/10.1038/nature16976}
  {\path{doi:10.1038/nature16976}}.

\bibitem{Prosin:2016jev}
V.~V. Prosin, et~al., {Results and perspectives of cosmic ray mass composition
  studies with EAS arrays in the Tunka Valley}, J. Phys. Conf. Ser. 718~(5)
  (2016) 052031.
\newblock \href {http://dx.doi.org/10.1088/1742-6596/718/5/052031}
  {\path{doi:10.1088/1742-6596/718/5/052031}}.

\bibitem{Martineau-Huynh:2021oqi}
O.~Martineau-Huynh, {The path towards the Giant Radio Array for Neutrino
  Detection}, \url{https://tel.archives-ouvertes.fr/tel-03332202} (2021).

\bibitem{Feng:2001ue}
J.~L. Feng, P.~Fisher, F.~Wilczek, T.~M. Yu, {Observability of earth skimming
  ultrahigh-energy neutrinos}, Phys. Rev. Lett. 88 (2002) 161102.
\newblock \href {http://arxiv.org/abs/hep-ph/0105067}
  {\path{arXiv:hep-ph/0105067}}, \href
  {http://dx.doi.org/10.1103/PhysRevLett.88.161102}
  {\path{doi:10.1103/PhysRevLett.88.161102}}.

\bibitem{TelescopeArray:2015dcv}
R.~U. Abbasi, et~al., {\bf Telescope Array} Collaboration, {The energy spectrum
  of cosmic rays above 10$^{17.2}$ eV measured by the fluorescence detectors of
  the Telescope Array experiment in seven years}, Astropart. Phys. 80 (2016)
  131--140.
\newblock \href {http://arxiv.org/abs/1511.07510} {\path{arXiv:1511.07510}},
  \href {http://dx.doi.org/10.1016/j.astropartphys.2016.04.002}
  {\path{doi:10.1016/j.astropartphys.2016.04.002}}.

\bibitem{Billoir:2007kb}
P.~Billoir, O.~Deligny, {Estimates of multipolar coefficients to search for
  cosmic ray anisotropies with non-uniform or partial sky coverage}, JCAP 02
  (2008) 009.
\newblock \href {http://arxiv.org/abs/0710.2290} {\path{arXiv:0710.2290}},
  \href {http://dx.doi.org/10.1088/1475-7516/2008/02/009}
  {\path{doi:10.1088/1475-7516/2008/02/009}}.

\bibitem{Denton:2015bga}
P.~B. Denton, T.~J. Weiler, {Sensitivity of full-sky experiments to large scale
  cosmic ray anisotropies}, JHEAp 8 (2015) 1--9.
\newblock \href {http://arxiv.org/abs/1505.03922} {\path{arXiv:1505.03922}},
  \href {http://dx.doi.org/10.1016/j.jheap.2015.06.002}
  {\path{doi:10.1016/j.jheap.2015.06.002}}.

\bibitem{Deligny:2017wbx}
O.~Deligny, K.~Kawata, P.~Tinyakov, {Measurement of anisotropy and the search
  for ultra high energy cosmic ray sources}, PTEP 2017~(12) (2017) 12A104.
\newblock \href {http://arxiv.org/abs/1702.07209} {\path{arXiv:1702.07209}},
  \href {http://dx.doi.org/10.1093/ptep/ptx043}
  {\path{doi:10.1093/ptep/ptx043}}.

\bibitem{PierreAuger:2016gkp}
A.~Aab, et~al., {\bf Pierre Auger} Collaboration, {Multi-resolution anisotropy
  studies of ultrahigh-energy cosmic rays detected at the Pierre Auger
  Observatory}, JCAP 06 (2017) 026.
\newblock \href {http://arxiv.org/abs/1611.06812} {\path{arXiv:1611.06812}},
  \href {http://dx.doi.org/10.1088/1475-7516/2017/06/026}
  {\path{doi:10.1088/1475-7516/2017/06/026}}.

\bibitem{Arsene:2021inm}
N.~Arsene, {Mass Composition of UHECRs from $X_{\rm max}$ Distributions
  Recorded by the Pierre Auger and Telescope Array Observatories}\href
  {http://arxiv.org/abs/2109.03626} {\path{arXiv:2109.03626}}, \href
  {http://dx.doi.org/10.3390/universe7090321}
  {\path{doi:10.3390/universe7090321}}.

\bibitem{Decoene:2021ncf}
V.~Decoene, O.~Martineau-Huynh, M.~Tueros, S.~Chiche, {A reconstruction
  procedure for very inclined extensive air showers based on radio signals},
  PoS ICRC2021 (2021) 211.
\newblock \href {http://arxiv.org/abs/2107.03206} {\path{arXiv:2107.03206}},
  \href {http://dx.doi.org/10.22323/1.395.0211}
  {\path{doi:10.22323/1.395.0211}}.

\bibitem{Decoene:2021atj}
V.~Decoene, O.~Martineau-Huynh, M.~Tueros, {Radio wavefront of very inclined
  extensive air-showers observed with extended and sparse radio arrays}\href
  {http://arxiv.org/abs/2112.07542} {\path{arXiv:2112.07542}}.

\bibitem{Moller:2018isk}
K.~M\o{}ller, P.~B. Denton, I.~Tamborra, {Cosmogenic Neutrinos Through the
  GRAND Lens Unveil the Nature of Cosmic Accelerators}, JCAP 05 (2019) 047.
\newblock \href {http://arxiv.org/abs/1809.04866} {\path{arXiv:1809.04866}},
  \href {http://dx.doi.org/10.1088/1475-7516/2019/05/047}
  {\path{doi:10.1088/1475-7516/2019/05/047}}.

\bibitem{Lipari:2017qpu}
P.~Lipari, {Cosmic Rays, Gamma rays, Neutrinos and Gravitational Waves}, Nuovo
  Cim. C 40~(3) (2017) 144.
\newblock \href {http://arxiv.org/abs/1707.02732} {\path{arXiv:1707.02732}},
  \href {http://dx.doi.org/10.1393/ncc/i2017-17144-0}
  {\path{doi:10.1393/ncc/i2017-17144-0}}.

\bibitem{Branchesi:2016vef}
M.~Branchesi, {Multi-messenger astronomy: gravitational waves, neutrinos,
  photons, and cosmic rays}, J. Phys. Conf. Ser. 718~(2) (2016) 022004.
\newblock \href {http://dx.doi.org/10.1088/1742-6596/718/2/022004}
  {\path{doi:10.1088/1742-6596/718/2/022004}}.

\bibitem{Gergely:2007ny}
L.~A. Gergely, P.~L. Biermann, {The Spin-Flip Phenomenon in Supermassive Black
  hole binary mergers}, Astrophys. J. 697 (2009) 1621--1633.
\newblock \href {http://arxiv.org/abs/0704.1968} {\path{arXiv:0704.1968}},
  \href {http://dx.doi.org/10.1088/0004-637X/697/2/1621}
  {\path{doi:10.1088/0004-637X/697/2/1621}}.

\bibitem{Gergely:2008dw}
L.~A. Gergely, P.~L. Biermann, {Supermassive binary black hole mergers}, J.
  Phys. Conf. Ser. 122 (2008) 012040.
\newblock \href {http://arxiv.org/abs/0805.4582} {\path{arXiv:0805.4582}},
  \href {http://dx.doi.org/10.1088/1742-6596/122/1/012040}
  {\path{doi:10.1088/1742-6596/122/1/012040}}.

\bibitem{Gergely:2010xr}
L.~A. Gergely, P.~L. Biermann, L.~I. Caramete, {Supermassive black hole
  spin-flip during the inspiral}, Class. Quant. Grav. 27 (2010) 194009.
\newblock \href {http://arxiv.org/abs/1005.2287} {\path{arXiv:1005.2287}},
  \href {http://dx.doi.org/10.1088/0264-9381/27/19/194009}
  {\path{doi:10.1088/0264-9381/27/19/194009}}.

\bibitem{Tapai:2013jza}
M.~T\'apai, et~al., {Supermassive black hole mergers as dual sources for
  electromagnetic flares in the jet emission and gravitational waves}, Astron.
  Nachr. 334 (2013) 1032.
\newblock \href {http://arxiv.org/abs/1309.1831} {\path{arXiv:1309.1831}},
  \href {http://dx.doi.org/10.1002/asna.201211988}
  {\path{doi:10.1002/asna.201211988}}.

\bibitem{Klinkhamer:2008ss}
F.~R. Klinkhamer, M.~Risse, {Addendum: Ultrahigh-energy cosmic-ray bounds on
  nonbirefringent modified-Maxwell theory}, Phys. Rev. D 77 (2008) 117901.
\newblock \href {http://arxiv.org/abs/0806.4351} {\path{arXiv:0806.4351}},
  \href {http://dx.doi.org/10.1103/PhysRevD.77.117901}
  {\path{doi:10.1103/PhysRevD.77.117901}}.

\bibitem{Aloisio:2000cm}
R.~Aloisio, P.~Blasi, P.~L. Ghia, A.~F. Grillo, {Probing the structure of
  space-time with cosmic rays}, Phys. Rev. D 62 (2000) 053010.
\newblock \href {http://arxiv.org/abs/astro-ph/0001258}
  {\path{arXiv:astro-ph/0001258}}, \href
  {http://dx.doi.org/10.1103/PhysRevD.62.053010}
  {\path{doi:10.1103/PhysRevD.62.053010}}.

\bibitem{Cowsik:2012qm}
R.~Cowsik, T.~Madziwa-Nussinov, S.~Nussinov, U.~Sarkar, {Testing Violations of
  Lorentz Invariance with Cosmic-Rays}, Phys. Rev. D 86 (2012) 045024.
\newblock \href {http://arxiv.org/abs/1206.0713} {\path{arXiv:1206.0713}},
  \href {http://dx.doi.org/10.1103/PhysRevD.86.045024}
  {\path{doi:10.1103/PhysRevD.86.045024}}.

\bibitem{Martinez-Huerta:2017ulw}
H.~Mart\'\i{}nez-Huerta, A.~P\'erez-Lorenzana, {Restrictions from Lorentz
  invariance violation on cosmic ray propagation}, Phys. Rev. D 95~(6) (2017)
  063001.
\newblock \href {http://arxiv.org/abs/1610.00047} {\path{arXiv:1610.00047}},
  \href {http://dx.doi.org/10.1103/PhysRevD.95.063001}
  {\path{doi:10.1103/PhysRevD.95.063001}}.

\bibitem{Trimarelli:2021Ul}
C.~Trimarelli, {\bf Pierre Auger} Collaboration, {Constraining Lorentz
  Invariance Violation using the muon content of extensive air showers measured
  at the Pierre Auger Observatory}, PoS ICRC2021 (2021) 340.
\newblock \href {http://dx.doi.org/10.22323/1.395.0340}
  {\path{doi:10.22323/1.395.0340}}.

\bibitem{LHAASO:2019qwt}
C.~Zhen, et~al., {\bf LHAASO} Collaboration, {Introduction to Large High
  Altitude Air Shower Observatory (LHAASO)}, Chin. Astron. Astrophys. 43 (2019)
  457--478.
\newblock \href {http://dx.doi.org/10.1016/j.chinastron.2019.11.001}
  {\path{doi:10.1016/j.chinastron.2019.11.001}}.

\bibitem{Hinton:2021rvp}
J.~Hinton, {\bf SWGO} Collaboration, {The Southern Wide-field Gamma-ray
  Observatory: Status and Prospects}, PoS ICRC2021 (2021) 023.
\newblock \href {http://arxiv.org/abs/2111.13158} {\path{arXiv:2111.13158}}.

\bibitem{Apel:2010zz}
W.~D. Apel, et~al., {The KASCADE-Grande experiment}, Nucl. Instrum. Meth. A 620
  (2010) 202--216.
\newblock \href {http://dx.doi.org/10.1016/j.nima.2010.03.147}
  {\path{doi:10.1016/j.nima.2010.03.147}}.

\bibitem{SWGO:LoI191}
A.~Albert, {Constraining the Local Positron Contribution from TeVHalos with the
  Southern Wide-field Gamma-ray Observatory (SWGO)}, Snowmass Letter of Intent
  CF1\_CF7~(191).

\bibitem{SWGO:LoI226}
K.~L. Engel, {Cosmic Rays in the TeV to PeV Energy Range}, Snowmass Letter of
  Intent CF7\_CF0~(226).

\bibitem{CREDO:2020pzy}
P.~Homola, et~al., {\bf CREDO} Collaboration, {Cosmic Ray Extremely Distributed
  Observatory}, Symmetry 12~(11) (2020) 1835.
\newblock \href {http://arxiv.org/abs/2010.08351} {\path{arXiv:2010.08351}},
  \href {http://dx.doi.org/10.3390/sym12111835}
  {\path{doi:10.3390/sym12111835}}.

\bibitem{CREDO:2018ghi}
N.~Dhital, et~al., {\bf CREDO} Collaboration, {Cosmic ray ensembles as
  signatures of ultra-high energy photons interacting with the solar magnetic
  field}, JCAP 03 (2022) 038.
\newblock \href {http://arxiv.org/abs/1811.10334} {\path{arXiv:1811.10334}},
  \href {http://dx.doi.org/10.1088/1475-7516/2022/03/038}
  {\path{doi:10.1088/1475-7516/2022/03/038}}.

\bibitem{CREDO:2022qpa}
R.~Clay, et~al., {\bf CREDO} Collaboration, {A Search for Cosmic Ray Bursts at
  0.1 PeV with a Small Air Shower Array}, Symmetry 14~(3) (2022) 501.
\newblock \href {http://dx.doi.org/10.3390/sym14030501}
  {\path{doi:10.3390/sym14030501}}.

\bibitem{Allard:2008zz}
D.~Allard, et~al., {Use of water-Cherenkov detectors to detect gamma ray bursts
  at the Large Aperture GRB Observatory (LAGO)}, Nucl. Instrum. Meth. A 595
  (2008) 70--72.
\newblock \href {http://dx.doi.org/10.1016/j.nima.2008.07.041}
  {\path{doi:10.1016/j.nima.2008.07.041}}.

\bibitem{Sidelnik:2015yky}
I.~Sidelnik, {\bf LAGO} Collaboration, {The Sites of the Latin American Giant
  Observatory}, PoS ICRC2015 (2016) 665.
\newblock \href {http://dx.doi.org/10.22323/1.236.0665}
  {\path{doi:10.22323/1.236.0665}}.

\bibitem{Suarez-Duran:2015pas}
M.~Su\'arez-Dur\'an, et~al., {\bf LAGO} Collaboration, {The LAGO Space Weather
  Program: Directional Geomagnetic Effects, Background Fluence Calculations and
  Multi-Spectral Data Analysis}, PoS ICRC2015 (2016) 142.
\newblock \href {http://dx.doi.org/10.22323/1.236.0142}
  {\path{doi:10.22323/1.236.0142}}.

\bibitem{Dasso:2015lfq}
S.~Dasso, et~al., {\bf LAGO} Collaboration, A project to install
  water-cherenkov detectors in the antarctic peninsula as part of the lago
  detection network, PoS ICRC2015 (2016) 105.
\newblock \href {http://dx.doi.org/10.22323/1.236.0105}
  {\path{doi:10.22323/1.236.0105}}.

\bibitem{Dasso2019}
V.~Lanabere, S.~Dasso, A.~Gulisano, V.~López, A.~Niemelä-Celeda, Space
  weather service activities and initiatives at lamp (argentinean space weather
  laboratory group), Advances in Space Research 65~(9) (2020) 2223--2234.
\newblock \href {http://dx.doi.org/10.1016/j.asr.2019.08.016}
  {\path{doi:10.1016/j.asr.2019.08.016}}.

\bibitem{Asorey:2016pru}
H.~Asorey, S.~Dasso, S.~Dasso, {\bf LAGO} Collaboration, {LAGO: the Latin
  American Giant Observatory}, PoS ICRC2015 (2016) 247.
\newblock \href {http://dx.doi.org/10.22323/1.236.0247}
  {\path{doi:10.22323/1.236.0247}}.

\bibitem{Klimov:2022jzk}
P.~Klimov, et~al., {Status of the K-EUSO Orbital Detector of Ultra-high Energy
  Cosmic Rays}, Universe 8 (2022) 88.
\newblock \href {http://arxiv.org/abs/2201.12766} {\path{arXiv:2201.12766}}.

\bibitem{Taylor:2011ta}
A.~M. Taylor, M.~Ahlers, F.~A. Aharonian, {The need for a local source of UHE
  CR nuclei}, Phys. Rev. D 84 (2011) 105007.
\newblock \href {http://arxiv.org/abs/1107.2055} {\path{arXiv:1107.2055}},
  \href {http://dx.doi.org/10.1103/PhysRevD.84.105007}
  {\path{doi:10.1103/PhysRevD.84.105007}}.

\bibitem{Gonzalez:2021ajv}
J.~M. Gonz\'alez, S.~Mollerach, E.~Roulet, {Magnetic diffusion and interaction
  effects on ultrahigh energy cosmic rays: Protons and nuclei}, Phys. Rev. D
  104~(6) (2021) 063005.
\newblock \href {http://arxiv.org/abs/2105.08138} {\path{arXiv:2105.08138}},
  \href {http://dx.doi.org/10.1103/PhysRevD.104.063005}
  {\path{doi:10.1103/PhysRevD.104.063005}}.

\bibitem{Hillas:1985is}
A.~Hillas, {The Origin of Ultrahigh-Energy Cosmic Rays}, Ann. Rev. Astron.
  Astrophys. 22 (1984) 425--444.
\newblock \href {http://dx.doi.org/10.1146/annurev.aa.22.090184.002233}
  {\path{doi:10.1146/annurev.aa.22.090184.002233}}.

\bibitem{Scully:2008jp}
S.~T. Scully, F.~W. Stecker, {Lorentz Invariance Violation and the Observed
  Spectrum of Ultrahigh Energy Cosmic Rays}, Astropart. Phys. 31 (2009)
  220--225.
\newblock \href {http://arxiv.org/abs/0811.2230} {\path{arXiv:0811.2230}},
  \href {http://dx.doi.org/10.1016/j.astropartphys.2009.01.002}
  {\path{doi:10.1016/j.astropartphys.2009.01.002}}.

\bibitem{Colladay:1998fq}
D.~Colladay, V.~A. Kostelecky, {Lorentz violating extension of the standard
  model}, Phys. Rev. D 58 (1998) 116002.
\newblock \href {http://arxiv.org/abs/hep-ph/9809521}
  {\path{arXiv:hep-ph/9809521}}, \href
  {http://dx.doi.org/10.1103/PhysRevD.58.116002}
  {\path{doi:10.1103/PhysRevD.58.116002}}.

\bibitem{AmelinoCamelia:2000zs}
G.~Amelino-Camelia, T.~Piran, {Planck scale deformation of Lorentz symmetry as
  a solution to the UHECR and the TeV gamma paradoxes}, Phys. Rev. D 64 (2001)
  036005.
\newblock \href {http://arxiv.org/abs/astro-ph/0008107}
  {\path{arXiv:astro-ph/0008107}}, \href
  {http://dx.doi.org/10.1103/PhysRevD.64.036005}
  {\path{doi:10.1103/PhysRevD.64.036005}}.

\bibitem{Torri:2019gud}
M.~D.~C. Torri, V.~Antonelli, L.~Miramonti, {Homogeneously Modified Special
  relativity (HMSR)}: {A new possible way to introduce an isotropic Lorentz
  invariance violation in particle standard model}, Eur. Phys. J. C 79~(9)
  (2019) 808.
\newblock \href {http://arxiv.org/abs/1906.05595} {\path{arXiv:1906.05595}},
  \href {http://dx.doi.org/10.1140/epjc/s10052-019-7301-7}
  {\path{doi:10.1140/epjc/s10052-019-7301-7}}.

\bibitem{private:AlanColeman}
A.~Coleman, F.~Schr{\"o}der, private communication, 2022-18-01.

\bibitem{Waxman:1996zn}
E.~Waxman, J.~Miralda-Escude, {Images of bursting sources of high-energy cosmic
  rays. 1. Effects of magnetic fields}, Astrophys. J. Lett. 472 (1996)
  L89--L92.
\newblock \href {http://arxiv.org/abs/astro-ph/9607059}
  {\path{arXiv:astro-ph/9607059}}, \href {http://dx.doi.org/10.1086/310367}
  {\path{doi:10.1086/310367}}.

\bibitem{2011AsBio..11..551D}
L.~R. {Dartnell}, {Ionizing Radiation and Life}, Astrobiology 11~(6) (2011)
  551--582.
\newblock \href {http://dx.doi.org/10.1089/ast.2010.0528}
  {\path{doi:10.1089/ast.2010.0528}}.

\bibitem{Griessmeier:2015zna}
J.~M. Grie\ss{}meier, F.~Tabataba-Vakili, A.~Stadelmann, J.~L. Grenfell,
  D.~Atri, {Galactic cosmic rays on extrasolar Earth-like planets I. Cosmic ray
  flux}, Astron. Astrophys. 581 (2015) A44.
\newblock \href {http://arxiv.org/abs/1509.00735} {\path{arXiv:1509.00735}},
  \href {http://dx.doi.org/10.1051/0004-6361/201425451}
  {\path{doi:10.1051/0004-6361/201425451}}.

\bibitem{griessmeier2016}
J.-M. Grie{\ss}meier, F.~Tabataba-Vakili, A.~Stadelmann, J.~Grenfell, D.~Atri,
  Galactic cosmic rays on extrasolar earth-like planets-ii. atmospheric
  implications, Astronomy \& Astrophysics 587 (2016) A159.

\bibitem{ferrari2009}
F.~Ferrari, E.~Szuszkiewicz, Cosmic rays: a review for astrobiologists,
  Astrobiology 9~(4) (2009) 413--436.

\bibitem{Globus:2020gud}
N.~Globus, R.~D. Blandford, {The Chiral Puzzle of Life}, Astrophys. J. Lett.
  895~(1) (2020) L11.
\newblock \href {http://arxiv.org/abs/2002.12138} {\path{arXiv:2002.12138}},
  \href {http://dx.doi.org/10.3847/2041-8213/ab8dc6}
  {\path{doi:10.3847/2041-8213/ab8dc6}}.

\bibitem{kiefer1996}
J.~Kiefer, K.~Schenk-Meuser, M.~Kost, Radiation biology, in: Biological and
  medical research in space, Springer, 1996, pp. 300--367.

\bibitem{ellis1995}
J.~Ellis, D.~N. Schramm, Could a nearby supernova explosion have caused a mass
  extinction?, Proceedings of the National Academy of Sciences 92~(1) (1995)
  235--238.

\bibitem{dar1998}
A.~Dar, A.~Laor, N.~J. Shaviv, Life extinctions by cosmic ray jets, Physical
  review letters 80~(26) (1998) 5813.

\bibitem{Thomas:2016fcp}
B.~C. Thomas, E.~E. Engler, M.~Kachelrie\ss{}, A.~L. Melott, A.~C. Overholt,
  D.~V. Semikoz, {Terrestrial Effects Of Nearby Supernovae In The Early
  Pleistocene}, Astrophys. J. Lett. 826 (2016) L3.
\newblock \href {http://arxiv.org/abs/1605.04926} {\path{arXiv:1605.04926}},
  \href {http://dx.doi.org/10.3847/2041-8205/826/1/L3}
  {\path{doi:10.3847/2041-8205/826/1/L3}}.

\bibitem{melott2017}
A.~Melott, B.~Thomas, M.~Kachelriess, D.~Semikoz, A.~Overholt, A supernova at
  50 pc: effects on the earth's atmosphere and biota, The Astrophysical Journal
  840~(2) (2017) 105.

\bibitem{globus2021}
N.~Globus, A.~Fedynitch, R.~D. Blandford, Polarized radiation and the emergence
  of biological homochirality on earth and beyond, The Astrophysical Journal
  910~(2) (2021) 85.

\bibitem{Lipari:1993hd}
P.~Lipari, {Lepton spectra in the earth's atmosphere}, Astropart. Phys. 1
  (1993) 195--227.
\newblock \href {http://dx.doi.org/10.1016/0927-6505(93)90022-6}
  {\path{doi:10.1016/0927-6505(93)90022-6}}.

\bibitem{limaye2021venus}
S.~S. Limaye, et~al., Venus, an astrobiology target, Astrobiology 21~(10)
  (2021) 1163--1185.

\bibitem{takahashi2019}
J.-i. Takahashi, K.~Kobayashi, Origin of terrestrial bioorganic homochirality
  and symmetry breaking in the universe, Symmetry 11~(7) (2019) 919.

\bibitem{avnir2021}
D.~Avnir, Critical review of chirality indicators of extraterrestrial life, New
  Astronomy Reviews 92 (2021) 101596.

\bibitem{marinho2014}
F.~Marinho, L.~Paulucci, D.~Galante, Propagation and energy deposition of
  cosmic rays’ muons on terrestrial environments, International Journal of
  Astrobiology 13~(4) (2014) 319--323.

\bibitem{zimmer2000Europa}
C.~Zimmer, K.~K. Khurana, M.~G. Kivelson, Subsurface oceans on europa and
  callisto: Constraints from galileo magnetometer observations, Icarus 147~(2)
  (2000) 329--347.

\bibitem{thomas2016enceladus}
P.~Thomas, R.~Tajeddine, M.~Tiscareno, J.~Burns, J.~Joseph, T.~Loredo,
  P.~Helfenstein, C.~Porco, Enceladus’s measured physical libration requires
  a global subsurface ocean, Icarus 264 (2016) 37--47.

\bibitem{orosei2018radar}
R.~Orosei, S.~Lauro, E.~Pettinelli, A.~Cicchetti, M.~Coradini, B.~Cosciotti,
  F.~Di~Paolo, E.~Flamini, E.~Mattei, M.~Pajola, et~al., Radar evidence of
  subglacial liquid water on mars, Science 361~(6401) (2018) 490--493.

\bibitem{damer2020hot}
B.~Damer, D.~Deamer, The hot spring hypothesis for an origin of life,
  Astrobiology 20~(4) (2020) 429--452.

\bibitem{teece2020biomolecules}
B.~L. Teece, S.~C. George, T.~Djokic, K.~A. Campbell, S.~W. Ruff, M.~J.
  Van~Kranendonk, Biomolecules from fossilized hot spring sinters: implications
  for the search for life on mars, Astrobiology 20~(4) (2020) 537--551.

\bibitem{checinska2019habitability}
A.~Checinska~Sielaff, S.~A. Smith, Habitability of mars: How welcoming are the
  surface and subsurface to life on the red planet?, Geosciences 9~(9) (2019)
  361.

\bibitem{Rodger1999}
C.~J. Rodger, Red sprites, upward lightning, and vlf perturbations, Reviews of
  Geophysics 37~(3) (1999) 317--336.

\bibitem{Pasko2011}
V.~P. {Pasko}, Y.~{Yair}, C.-L. {Kuo}, {Lightning Related Transient Luminous
  Events at High Altitude in the Earth's Atmosphere: Phenomenology, Mechanisms
  and Effects}, Space Science Reviews 168 (2012) 475--516.
\newblock \href {http://dx.doi.org/10.1007/s11214-011-9813-9}
  {\path{doi:10.1007/s11214-011-9813-9}}.

\bibitem{Vazquez2021}
F.~J. {Gordillo-V{\'a}zquez}, F.~J. {P{\'e}rez-Invern{\'o}n}, {A review of the
  impact of transient luminous events on the atmospheric chemistry: Past,
  present, and future}, Atmospheric Research 252 (2021) 105432.
\newblock \href {http://dx.doi.org/10.1016/j.atmosres.2020.105432}
  {\path{doi:10.1016/j.atmosres.2020.105432}}.

\bibitem{soler2021}
S.~{Soler}, et~al., {Global Frequency and Geographical Distribution of
  Nighttime Streamer Corona Discharges (BLUEs) in Thunderclouds}, Geophysical
  Research Letters 48~(18) (2021) e94657.
\newblock \href {http://dx.doi.org/10.1029/2021GL094657}
  {\path{doi:10.1029/2021GL094657}}.

\bibitem{Chen2008}
A.~B. {Chen}, et~al., {Global distributions and occurrence rates of transient
  luminous events}, Journal of Geophysical Research (Space Physics) 113 (2008)
  A08306.
\newblock \href {http://dx.doi.org/10.1029/2008JA013101}
  {\path{doi:10.1029/2008JA013101}}.

\bibitem{GRL:GRL9448}
H.~Fukunishi, et~al., \href{http://dx.doi.org/10.1029/96GL01979}{Elves:
  Lightning-induced transient luminous events in the lower ionosphere},
  Geophysical Research Letters 23~(16) (1996) 2157--2160.
\newblock \href {http://dx.doi.org/10.1029/96GL01979}
  {\path{doi:10.1029/96GL01979}}.
\newline\urlprefix\url{http://dx.doi.org/10.1029/96GL01979}

\bibitem{GRL:GRL53200}
R.~A. Marshall, C.~L. da~Silva, V.~P. Pasko, Elve doublets and compact
  intracloud discharges, Geophysical Research Letters 42~(14) (2015)
  6112--6119, 2015GL064862.
\newblock \href {http://dx.doi.org/10.1002/2015GL064862}
  {\path{doi:10.1002/2015GL064862}}.

\bibitem{pasko1998}
V.~P. {Pasko}, U.~S. {Inan}, T.~F. {Bell}, {Spatial structure of sprites},
  Geophysical Research Letters 25~(12) (1998) 2123--2126.
\newblock \href {http://dx.doi.org/10.1029/98GL01242}
  {\path{doi:10.1029/98GL01242}}.

\bibitem{cummer2006}
S.~A. {Cummer}, N.~{Jaugey}, J.~{Li}, W.~A. {Lyons}, T.~E. {Nelson}, E.~A.
  {Gerken}, {Submillisecond imaging of sprite development and structure},
  Geophysical Research Letters 33~(4) (2006) L04104.
\newblock \href {http://dx.doi.org/10.1029/2005GL024969}
  {\path{doi:10.1029/2005GL024969}}.

\bibitem{Wescott1998}
E.~M. {Wescott}, et~al., {Blue Jets: their relationship to lightning and very
  large hailfall, and their physical mechanisms for their production}, Journal
  of Atmospheric and Solar-Terrestrial Physics 60 (1998) 713--724.
\newblock \href {http://dx.doi.org/10.1016/S1364-6826(98)00018-2}
  {\path{doi:10.1016/S1364-6826(98)00018-2}}.

\bibitem{Su2003}
H.~T. {Su}, et~al., {Gigantic jets between a thundercloud and the ionosphere},
  Nature 423 (2003) 974--976.

\bibitem{Neubert2019}
T.~Neubert, et~al., {The ASIM Mission on the International Space Station},
  {SPACE SCIENCE REVIEWS} {215}~({2}).
\newblock \href {http://dx.doi.org/{10.1007/s11214-019-0592-z}}
  {\path{doi:{10.1007/s11214-019-0592-z}}}.

\bibitem{arnone2020}
E.~{Arnone}, et~al., {Climatology of Transient Luminous Events and Lightning
  Observed Above Europe and the Mediterranean Sea}, Surveys in Geophysics
  41~(2) (2019) 167--199.
\newblock \href {http://dx.doi.org/10.1007/s10712-019-09573-5}
  {\path{doi:10.1007/s10712-019-09573-5}}.

\bibitem{Mussa:2012dq}
R.~Mussa, G.~Ciaccio, {\bf Pierre Auger} Collaboration, {Observation of ELVES
  at the Pierre Auger Observatory}, Eur. Phys. J. Plus 127 (2012) 94.
\newblock \href {http://dx.doi.org/10.1140/epjp/i2012-12094-x}
  {\path{doi:10.1140/epjp/i2012-12094-x}}.

\bibitem{Tonachini:2013lif}
A.~S. Tonachini, {Observation of Elves at the Pierre Auger Observatory}, in:
  {33rd International Cosmic Ray Conference}, 2013, p. 0676.

\bibitem{klimov2019remote}
P.~Klimov, et~al., Remote sensing of the atmosphere by the ultraviolet detector
  tus onboard the lomonosov satellite, Remote Sensing 11~(20) (2019) 2449.

\bibitem{Barrington1999}
C.~P. Barrington-Leigh, U.~S. Inan, Elves triggered by positive and negative
  lightning discharges, Geophysical Research Letters 26~(6) (1999) 683--686.
\newblock \href {http://dx.doi.org/10.1029/1999GL900059}
  {\path{doi:10.1029/1999GL900059}}.

\bibitem{Newsome2010}
R.~T. Newsome, U.~S. Inan, Free-running ground-based photometric array imaging
  of transient luminous events, Journal of Geophysical Research: Space Physics
  115~(A7).
\newblock \href {http://dx.doi.org/10.1029/2009JA014834}
  {\path{doi:10.1029/2009JA014834}}.

\bibitem{Marcelli:2021uX}
L.~Marcelli, J.~Collaboration, {Observation of ELVES with Mini-EUSO telescope
  on board the International Space Station}, PoS ICRC2021 (2021) 367.
\newblock \href {http://dx.doi.org/10.22323/1.395.0367}
  {\path{doi:10.22323/1.395.0367}}.

\bibitem{Charman1972}
W.~N. {Charman}, J.~V. {Jelley}, {A search for pulses of fluorescence produced
  by supernovae in the upper atmosphere}, Journal of Physics A Mathematical
  General 5 (1972) 773--780.
\newblock \href {http://dx.doi.org/10.1088/0305-4470/5/5/019}
  {\path{doi:10.1088/0305-4470/5/5/019}}.

\bibitem{Elliot1972}
J.~L. {Elliot}, {Atmospheric Fluorescence as a Ground-Based Method of Detecting
  Cosmic X-Rays}, SAO Special Report 341.

\bibitem{Nemzek1989}
R.~J. {Nemzek}, J.~R. {Winckler}, {Observation and interpretation of fast
  sub-visual light pulses from the night sky}, Geophys.Res.Lett. 16 (1989)
  1015--1018.
\newblock \href {http://dx.doi.org/10.1029/GL016i009p01015}
  {\path{doi:10.1029/GL016i009p01015}}.

\bibitem{Yair2005}
Y.~{Yair}, et~al., {Space shuttle observation of an unusual transient
  atmospheric emission}, Geophys.Res.Lett. 32 (2005) L02801.
\newblock \href {http://dx.doi.org/10.1029/2004GL021551}
  {\path{doi:10.1029/2004GL021551}}.

\bibitem{panasyuk2016relec}
M.~Panasyuk, et~al., Relec mission: Relativistic electron precipitation and tle
  study on-board small spacecraft, Advances in Space Research 57~(3) (2016)
  835--849.

\bibitem{klimov2018uv}
P.~A. Klimov, et~al., Uv transient atmospheric events observed far from
  thunderstorms by the vernov satellite, IEEE Geoscience and Remote Sensing
  Letters 15~(8) (2018) 1139--1143.

\bibitem{Dwyer2012}
J.~R. {Dwyer}, D.~M. {Smith}, S.~A. {Cummer}, {High-Energy Atmospheric Physics:
  Terrestrial Gamma-Ray Flashes and Related Phenomena}, Space Science Reviews
  173~(1-4) (2012) 133--196.
\newblock \href {http://dx.doi.org/10.1007/s11214-012-9894-0}
  {\path{doi:10.1007/s11214-012-9894-0}}.

\bibitem{Fishman1994}
G.~J. {Fishman}, et~al., {Discovery of Intense Gamma-Ray Flashes of Atmospheric
  Origin}, Science 264~(5163) (1994) 1313--1316.
\newblock \href {http://dx.doi.org/10.1126/science.264.5163.1313}
  {\path{doi:10.1126/science.264.5163.1313}}.

\bibitem{Smith2005}
D.~M. Smith, et~al., {Terrestrial gamma-ray flashes observed up to 20 MeV},
  Science 307 (2005) 1085--1088.

\bibitem{Marisaldi:2010zz}
M.~Marisaldi, P.~Cattaneo, {\bf AGILE} Collaboration, {Detection of Terrestrial
  Gamma-Ray Flashes up to 40-MeV by the AGILE Satellite}, J. Geophys. Res.
  115~(A3) (2010) A00E13.
\newblock \href {http://dx.doi.org/10.1029/2009JA014502}
  {\path{doi:10.1029/2009JA014502}}.

\bibitem{Briggs2010}
M.~S. {Briggs}, et~al., {First results on terrestrial gamma ray flashes from
  the Fermi Gamma-ray Burst Monitor}, Journal of Geophysical Research (Space
  Physics) 115~(A7) (2010) A07323.
\newblock \href {http://dx.doi.org/10.1029/2009JA015242}
  {\path{doi:10.1029/2009JA015242}}.

\bibitem{Ostgaard2019}
N.~{{\O}stgaard}, et~al., {First 10 Months of TGF Observations by ASIM},
  Journal of Geophysical Research (Atmospheres) 124~(24) (2019) 14,024--14,036.
\newblock \href {http://dx.doi.org/10.1029/2019JD031214}
  {\path{doi:10.1029/2019JD031214}}.

\bibitem{Gurevich1992}
A.~V. {Gurevich}, G.~M. {Milikh}, R.~{Roussel-Dupre}, {Runaway electron
  mechanism of air breakdown and preconditioning during a thunderstorm},
  Physics Letters A 165~(5-6) (1992) 463--468.
\newblock \href {http://dx.doi.org/10.1016/0375-9601(92)90348-P}
  {\path{doi:10.1016/0375-9601(92)90348-P}}.

\bibitem{Dwyer2005}
J.~R. {Dwyer}, D.~M. {Smith}, {A comparison between Monte Carlo simulations of
  runaway breakdown and terrestrial gamma-ray flash observations}, GRL 32~(22)
  (2005) L22804.
\newblock \href {http://dx.doi.org/10.1029/2005GL023848}
  {\path{doi:10.1029/2005GL023848}}.

\bibitem{Celestin2011}
S.~{Celestin}, V.~P. {Pasko}, {Energy and fluxes of thermal runaway electrons
  produced by exponential growth of streamers during the stepping of lightning
  leaders and in transient luminous events}, Journal of Geophysical Research
  (Space Physics) 116~(A3) (2011) A03315.
\newblock \href {http://dx.doi.org/10.1029/2010JA016260}
  {\path{doi:10.1029/2010JA016260}}.

\bibitem{Dwyer2007}
J.~R. {Dwyer}, {Relativistic breakdown in planetary atmospheres}, Physics of
  Plasmas 14~(4) (2007) 042901--042901.
\newblock \href {http://dx.doi.org/10.1063/1.2709652}
  {\path{doi:10.1063/1.2709652}}.

\bibitem{Roberts2018}
O.~J. {Roberts}, et~al., {The First Fermi-GBM Terrestrial Gamma Ray Flash
  Catalog}, Journal of Geophysical Research (Space Physics) 123~(5) (2018)
  4381--4401.
\newblock \href {http://dx.doi.org/10.1029/2017JA024837}
  {\path{doi:10.1029/2017JA024837}}.

\bibitem{Lindanger2020}
A.~{Lindanger}, et~al., {The 3rd AGILE Terrestrial Gamma Ray Flash Catalog.
  Part I: Association to Lightning Sferics}, Journal of Geophysical Research
  (Atmospheres) 125~(11) (2020) e31985.
\newblock \href {http://dx.doi.org/10.1029/2019JD031985}
  {\path{doi:10.1029/2019JD031985}}.

\bibitem{Maiorana2020}
C.~{Maiorana}, et~al., {The 3rd AGILE Terrestrial Gamma-ray Flashes Catalog.
  Part II: Optimized Selection Criteria and Characteristics of the New Sample},
  Journal of Geophysical Research (Atmospheres) 125~(11) (2020) e31986.
\newblock \href {http://dx.doi.org/10.1029/2019JD031986}
  {\path{doi:10.1029/2019JD031986}}.

\bibitem{Smith2020}
D.~M. {Smith}, et~al., {Special Classes of Terrestrial Gamma Ray Flashes From
  RHESSI}, Journal of Geophysical Research (Atmospheres) 125~(20) (2020)
  e33043.
\newblock \href {http://dx.doi.org/10.1029/2020JD033043}
  {\path{doi:10.1029/2020JD033043}}.

\bibitem{Cummer2015}
S.~A. {Cummer}, et~al., {Lightning leader altitude progression in terrestrial
  gamma-ray flashes}, GRL 42~(18) (2015) 7792--7798.
\newblock \href {http://dx.doi.org/10.1002/2015GL065228}
  {\path{doi:10.1002/2015GL065228}}.

\bibitem{Ostgaard2021}
N.~{{\O}stgaard}, et~al., {Simultaneous Observations of EIP, TGF, Elve, and
  Optical Lightning}, Journal of Geophysical Research (Atmospheres) 126~(11)
  (2021) e33921.
\newblock \href {http://dx.doi.org/10.1029/2020JD033921}
  {\path{doi:10.1029/2020JD033921}}.

\bibitem{Neubert2020}
T.~{Neubert}, et~al., {A terrestrial gamma-ray flash and ionospheric
  ultraviolet emissions powered by lightning}, Science 367~(6474) (2020)
  183--186.
\newblock \href {http://dx.doi.org/10.1126/science.aax3872}
  {\path{doi:10.1126/science.aax3872}}.

\bibitem{Briggs2019}
M.~S. {Briggs}, et~al., {Development and Design of the Terrestrial RaYs
  Analysis and Detection (TRYAD) Science Instrument}, in: AGU Fall Meeting
  Abstracts, Vol. 2019, 2019, pp. AE33A--3125.

\bibitem{Lindanger2021}
A.~Lindanger, et~al., Spectral analysis of individual terrestrial gamma-ray
  flashes detected by asim, Journal of Geophysical Research: Atmospheres
  n/a~(n/a) (2021) e2021JD035347, e2021JD035347 2021JD035347.
\newblock \href {http://dx.doi.org/https://doi.org/10.1029/2021JD035347}
  {\path{doi:https://doi.org/10.1029/2021JD035347}}.

\bibitem{Smith2011}
D.~M. Smith, et~al., A terrestrial gamma ray flash observed from an aircraft,
  Journal of Geophysical Research: Atmospheres 116~(D20).
\newblock \href {http://dx.doi.org/https://doi.org/10.1029/2011JD016252}
  {\path{doi:https://doi.org/10.1029/2011JD016252}}.

\bibitem{Bowers2018}
G.~S. Bowers, et~al., A terrestrial gamma-ray flash inside the eyewall of
  hurricane patricia, Journal of Geophysical Research: Atmospheres 123~(10)
  (2018) 4977--4987.
\newblock \href {http://dx.doi.org/https://doi.org/10.1029/2017JD027771}
  {\path{doi:https://doi.org/10.1029/2017JD027771}}.

\bibitem{Ostgaard2019b}
N.~{{\O}stgaard}, et~al., {Gamma Ray Glow Observations at 20-km Altitude},
  Journal of Geophysical Research (Atmospheres) 124~(13) (2019) 7236--7254.
\newblock \href {http://arxiv.org/abs/2107.03181} {\path{arXiv:2107.03181}},
  \href {http://dx.doi.org/10.1029/2019JD030312}
  {\path{doi:10.1029/2019JD030312}}.

\bibitem{Colalillo:2017lnj}
R.~Colalillo, {\bf Pierre Auger} Collaboration, {Peculiar lightning-related
  events observed by the surface detector of the Pierre Auger Observatory}
  (2017) 138--145\href {http://dx.doi.org/10.22323/1.301.0314}
  {\path{doi:10.22323/1.301.0314}}.

\bibitem{Abbasi:2019xan}
R.~Abbasi, et~al., {Ground-Based Observations of Terrestrial Gamma Ray Flashes
  Associated with Downward-Directed Lightning Leaders}, EPJ Web Conf. 197
  (2019) 03002.
\newblock \href {http://dx.doi.org/10.1051/epjconf/201919703002}
  {\path{doi:10.1051/epjconf/201919703002}}.

\bibitem{Vallance}
V.~Jones, Aurora, Vol.~9 of Geophysics and Astrophysics Monographs, Springer,
  Dordrecht, 1974.

\bibitem{Beach1968}
R.~{Beach}, et~al., {Flickering, a 10-cps fluctuation within bright auroras},
  Planetary and Space Science 16~(12).
\newblock \href {http://dx.doi.org/10.1016/0032-0633(68)90064-0}
  {\path{doi:10.1016/0032-0633(68)90064-0}}.

\bibitem{Yamamoto1988}
T.~Yamamoto, On the temporal fluctuations of pulsating auroral luminosity,
  Journal of Geophysical Research: Space Physics 93~(A2) (1988) 897--911.
\newblock \href {http://dx.doi.org/10.1029/JA093iA02p00897}
  {\path{doi:10.1029/JA093iA02p00897}}.

\bibitem{Nishiyama2014}
T.~{Nishiyama}, et~al., {Multiscale temporal variations of pulsating auroras:
  On-off pulsation and a few Hz modulation}, Journal of Geophysical Research
  (Space Physics) 119~(5) (2014) 3514--3527.
\newblock \href {http://dx.doi.org/10.1002/2014JA019818}
  {\path{doi:10.1002/2014JA019818}}.

\bibitem{SSR}
P.~A. Klimov, et~al., The {TUS} detector of extreme energy cosmic rays on board
  the {L}omonosov satellite, Space Science Reviews (2017) 1--17\href
  {http://dx.doi.org/10.1007/s11214-017-0403-3}
  {\path{doi:10.1007/s11214-017-0403-3}}.

\bibitem{adams2015jem}
J.~Adams, S.~Ahmad, J.-N. Albert, D.~Allard, L.~Anchordoqui, V.~Andreev,
  A.~Anzalone, Y.~Arai, K.~Asano, M.~A. Pernas, et~al., The {JEM-EUSO}
  instrument, Experimental Astronomy 40~(1) (2015) 19--44.

\bibitem{Sakanoi2005}
K.~Sakanoi, H.~Fukunishi, Y.~Kasahara, A possible generation mechanism of
  temporal and spatial structures of flickering aurora, Journal of Geophysical
  Research 110.
\newblock \href {http://dx.doi.org/10.1029/2004JA010549}
  {\path{doi:10.1029/2004JA010549}}.

\bibitem{Miyoshi2020a}
Y.~{Miyoshi}, et~al., {Relativistic Electron Microbursts as High-Energy Tail of
  Pulsating Aurora Electrons}, Geophysical Research Letters 47~(21) (2020)
  e90360.
\newblock \href {http://dx.doi.org/10.1029/2020GL090360}
  {\path{doi:10.1029/2020GL090360}}.

\bibitem{Ceplecha1976}
Z.~{Ceplecha}, R.~E. {McCrosky}, {Fireball end heights: A diagnostic for the
  structure of meteoric material}, Journal of Geophysical Research 81~(B35)
  (1976) 6257--6275.
\newblock \href {http://dx.doi.org/10.1029/JB081i035p06257}
  {\path{doi:10.1029/JB081i035p06257}}.

\bibitem{Gritsevich2011}
M.~{Gritsevich}, D.~{Koschny}, {Constraining the luminous efficiency of
  meteors}, Icarus 212~(2) (2011) 877--884.
\newblock \href {http://dx.doi.org/10.1016/j.icarus.2011.01.033}
  {\path{doi:10.1016/j.icarus.2011.01.033}}.

\bibitem{Moreno2015}
M.~{Moreno-Ib{\'a}{\~n}ez}, M.~{Gritsevich}, J.~M. {Trigo-Rodr{\'\i}guez}, {New
  methodology to determine the terminal height of a fireball}, Icarus 250
  (2015) 544--552.
\newblock \href {http://arxiv.org/abs/1502.01898} {\path{arXiv:1502.01898}},
  \href {http://dx.doi.org/10.1016/j.icarus.2014.12.027}
  {\path{doi:10.1016/j.icarus.2014.12.027}}.

\bibitem{Moreno2017}
M.~{Moreno-Ib{\'a}{\~n}ez}, M.~{Gritsevich}, J.~M. {Trigo-Rodr{\'\i}guez},
  {Measuring the Terminal Heights of Bolides to Understand the Atmospheric
  Flight of Large Asteroidal Fragments}, in: J.~M. {Trigo-Rodr{\'\i}guez},
  M.~{Gritsevich}, H.~{Palme} (Eds.), Assessment and Mitigation of Asteroid
  Impact Hazards: Proceedings of the 2015 Barcelona Asteroid Day, Vol.~46 of
  Astrophysics and Space Science Proceedings, 2017, p. 129.
\newblock \href {http://dx.doi.org/10.1007/978-3-319-46179-3\_7}
  {\path{doi:10.1007/978-3-319-46179-3\_7}}.

\bibitem{Plane2012}
J.~M.~C. {Plane}, {Cosmic dust in the earth's atmosphere}, Chemical Society
  Reviews 41 (2012) 6507--6518.
\newblock \href {http://dx.doi.org/10.1039/c2cs35132c}
  {\path{doi:10.1039/c2cs35132c}}.

\bibitem{Silber2018}
E.~A. {Silber}, et~al., {Physics of meteor generated shock waves in the Earth's
  atmosphere - A review}, Advances in Space Research 62~(3) (2018) 489--532.
\newblock \href {http://arxiv.org/abs/1805.07842} {\path{arXiv:1805.07842}},
  \href {http://dx.doi.org/10.1016/j.asr.2018.05.010}
  {\path{doi:10.1016/j.asr.2018.05.010}}.

\bibitem{Vinkovic2020}
D.~Vinkovi{\'c}, M.~Gritsevich, The challenges in hypervelocity microphysics
  research on meteoroid impacts into the atmosphere, Journal of the
  Geographical Institute {"}Jovan Cvijic{"} SASA 70~(1) (2020) 45--55.
\newblock \href {http://dx.doi.org/10.2298/IJGI2001045V}
  {\path{doi:10.2298/IJGI2001045V}}.

\bibitem{Kornos2008}
L.~{Korno{\v{s}}}, J.~{T{\'o}th}, P.~{Vere{\v{s}}}, {Orbital Evolution of
  P{\v{r}}{\'\i}bram and Neuschwanstein}, Earth Moon and Planets 102~(1-4)
  (2008) 59--65.
\newblock \href {http://arxiv.org/abs/1104.3115} {\path{arXiv:1104.3115}},
  \href {http://dx.doi.org/10.1007/s11038-007-9213-z}
  {\path{doi:10.1007/s11038-007-9213-z}}.

\bibitem{Trigo2015}
J.~M. {Trigo-Rodr{\'\i}guez}, et~al., {Orbit and dynamic origin of the recently
  recovered Annama's H5 chondrite}, Monthly Notices of the Royal Astronomical
  Society 449~(2) (2015) 2119--2127.
\newblock \href {http://arxiv.org/abs/1507.04342} {\path{arXiv:1507.04342}},
  \href {http://dx.doi.org/10.1093/mnras/stv378}
  {\path{doi:10.1093/mnras/stv378}}.

\bibitem{Lal2018}
B.~Lal, A.~Balakrishnan, B.~M. Caldwell, R.~S. Buenconsejo, S.~A. Carioscia,
  Global trends in space situational awareness (ssa) and space traffic
  management (stm), Tech. rep. (2018).

\bibitem{Gritsevich2012}
M.~I. {Gritsevich}, V.~P. {Stulov}, L.~I. {Turchak}, {Consequences of
  collisions of natural cosmic bodies with the Earth's atmosphere and surface},
  Cosmic Research 50~(1) (2012) 56--64.
\newblock \href {http://dx.doi.org/10.1134/S0010952512010017}
  {\path{doi:10.1134/S0010952512010017}}.

\bibitem{Sansom2019}
E.~K. {Sansom}, et~al., {Determining Fireball Fates Using the
  {\ensuremath{\alpha}}-{\ensuremath{\beta}} Criterion}, The Astrophysical
  Journal 885~(2) (2019) 115.
\newblock \href {http://arxiv.org/abs/1909.11494} {\path{arXiv:1909.11494}},
  \href {http://dx.doi.org/10.3847/1538-4357/ab4516}
  {\path{doi:10.3847/1538-4357/ab4516}}.

\bibitem{Moreno2020}
M.~{Moreno-Ib{\'a}{\~n}ez}, et~al., {Physically based alternative to the PE
  criterion for meteoroids}, Monthly Notices of the Royal Astronomical Society
  494~(1) (2020) 316--324.
\newblock \href {http://arxiv.org/abs/2002.12842} {\path{arXiv:2002.12842}},
  \href {http://dx.doi.org/10.1093/mnras/staa646}
  {\path{doi:10.1093/mnras/staa646}}.

\bibitem{Boaca_2022}
I.~Boaca, M.~Gritsevich, M.~Birlan, A.~Nedelcu, T.~Boaca, F.~Colas,
  A.~Malgoyre, B.~Zanda, P.~Vernazza, Characterization of the fireballs
  detected by all-sky cameras in romania, The Astrophysical Journal 936~(2)
  (2022) 150.
\newblock \href {http://dx.doi.org/10.3847/1538-4357/ac8542}
  {\path{doi:10.3847/1538-4357/ac8542}}.

\bibitem{Vinnikov2016}
V.~V. {Vinnikov}, M.~I. {Gritsevich}, L.~I. {Turchak}, {Mathematical model for
  estimation of meteoroid dark flight trajectory}, in: Application of
  Mathematics in Technical and Natural Sciences: 8th International Conference
  for Promoting the Application of Mathematics in Technical and Natural
  Sciences - AMiTaNS'16, Vol. 1773 of American Institute of Physics Conference
  Series, 2016, p. 110016.
\newblock \href {http://dx.doi.org/10.1063/1.4965020}
  {\path{doi:10.1063/1.4965020}}.

\bibitem{Moilanen2021}
J.~{Moilanen}, M.~{Gritsevich}, E.~{Lyytinen}, {Determination of strewn fields
  for meteorite falls}, Monthly Notices of the Royal Astronomical Society
  503~(3) (2021) 3337--3350.
\newblock \href {http://dx.doi.org/10.1093/mnras/stab586}
  {\path{doi:10.1093/mnras/stab586}}.

\bibitem{Dmitriev2015}
V.~{Dmitriev}, V.~{Lupovka}, M.~{Gritsevich}, {Orbit determination based on
  meteor observations using numerical integration of equations of motion},
  Planetary and Space Science 117 (2015) 223--235.
\newblock \href {http://dx.doi.org/10.1016/j.pss.2015.06.015}
  {\path{doi:10.1016/j.pss.2015.06.015}}.

\bibitem{Jansen2019}
T.~{Jansen-Sturgeon}, E.~K. {Sansom}, P.~A. {Bland}, {Comparing analytical and
  numerical approaches to meteoroid orbit determination using Hayabusa
  telemetry}, Meteoritics and Planetary Science 54~(9) (2019) 2149--2162.
\newblock \href {http://arxiv.org/abs/1808.05768} {\path{arXiv:1808.05768}},
  \href {http://dx.doi.org/10.1111/maps.13376} {\path{doi:10.1111/maps.13376}}.

\bibitem{Pena2021}
E.~{Pe{\~n}a-Asensio}, et~al., {Accurate 3D fireball trajectory and orbit
  calculation using the 3D-FIRETOC automatic Python code}, Monthly Notices of
  the Royal Astronomical Society 504~(4) (2021) 4829--4840.
\newblock \href {http://arxiv.org/abs/2103.13758} {\path{arXiv:2103.13758}},
  \href {http://dx.doi.org/10.1093/mnras/stab999}
  {\path{doi:10.1093/mnras/stab999}}.

\bibitem{Colas2020}
F.~{Colas}, et~al., {FRIPON: a worldwide network to track incoming meteoroids},
  Astronomy and Astrophysics 644 (2020) A53.
\newblock \href {http://arxiv.org/abs/2012.00616} {\path{arXiv:2012.00616}},
  \href {http://dx.doi.org/10.1051/0004-6361/202038649}
  {\path{doi:10.1051/0004-6361/202038649}}.

\bibitem{Gardiol2021}
D.~{Gardiol}, et~al., {Cavezzo, the first Italian meteorite recovered by the
  PRISMA fireball network. Orbit, trajectory, and strewn-field}, Monthly
  Notices of the Royal Astronomical Society 501~(1) (2021) 1215--1227.
\newblock \href {http://dx.doi.org/10.1093/mnras/staa3646}
  {\path{doi:10.1093/mnras/staa3646}}.

\bibitem{Jenniskens2000}
P.~{Jenniskens}, E.~{Lyytinen}, {Possible Ursid Outburst on December 22, 2000},
  WGN, Journal of the International Meteor Organization 28~(6) (2000) 221--226.

\bibitem{vaubaillon20152011}
J.~Vaubaillon, P.~Koten, A.~Margonis, J.~Toth, R.~Rudawska, M.~Gritsevich,
  J.~Zender, J.~McAuliffe, P.-D. Pautet, P.~Jenniskens, et~al., The 2011
  draconids: the first european airborne meteor observation campaign, Earth,
  Moon, and Planets 114~(3) (2015) 137--157.
\newblock \href {http://dx.doi.org/10.1007/s11038-014-9455-5}
  {\path{doi:10.1007/s11038-014-9455-5}}.

\bibitem{Bouquet2014}
A.~{Bouquet}, et~al., {Simulation of the capabilities of an orbiter for
  monitoring the entry of interplanetary matter into the terrestrial
  atmosphere}, Planetary and Space Science 103 (2014) 238--249.
\newblock \href {http://dx.doi.org/10.1016/j.pss.2014.09.001}
  {\path{doi:10.1016/j.pss.2014.09.001}}.

\bibitem{Jenniskens2018}
P.~{Jenniskens}, et~al., {Detection of meteoroid impacts by the Geostationary
  Lightning Mapper on the GOES-16 satellite}, Meteoritics and Planetary Science
  53~(12) (2018) 2445--2469.
\newblock \href {http://dx.doi.org/10.1111/maps.13137}
  {\path{doi:10.1111/maps.13137}}.

\bibitem{Barghini2020}
D.~{Barghini}, et~al., {Meteor detection from space with Mini-EUSO telescope},
  in: European Planetary Science Congress, 2020, pp. EPSC2020--800.

\bibitem{Barghini2021}
D.~{Barghini}, et~al., {Analysis of meteors observed in the UV by the Mini-EUSO
  telescope onboard the International Space Station}, in: European Planetary
  Science Congress, 2021, pp. EPSC2021--243.

\bibitem{Abdellaoui2017}
G.~{Abdellaoui}, et~al., {Meteor studies in the framework of the JEM-EUSO
  program}, Planetary and Space Science 143 (2017) 245--255.
\newblock \href {http://dx.doi.org/10.1016/j.pss.2016.12.001}
  {\path{doi:10.1016/j.pss.2016.12.001}}.

\bibitem{Vojacek2015}
V.~{Voj{\'a}{\v{c}}ek}, J.~{Borovi{\v{c}}ka}, P.~{Koten}, P.~{Spurn{\'y}},
  R.~{{\v{S}}tork}, {Catalogue of representative meteor spectra}, Astronomy and
  Astrophysics 580 (2015) A67.
\newblock \href {http://dx.doi.org/10.1051/0004-6361/201425047}
  {\path{doi:10.1051/0004-6361/201425047}}.

\bibitem{Rudawska2016}
R.~{Rudawska}, J.~{T{\'o}th}, D.~{Kalman{\v{c}}ok}, P.~{Zigo},
  P.~{Matlovi{\v{c}}}, {Meteor spectra from AMOS video system}, Planetary and
  Space Science 123 (2016) 25--32.
\newblock \href {http://dx.doi.org/10.1016/j.pss.2015.11.018}
  {\path{doi:10.1016/j.pss.2015.11.018}}.

\bibitem{Rudawska2020}
R.~{Rudawska}, et~al., {A spectroscopy pipeline for the Canary island long
  baseline observatory meteor detection system}, Planetary and Space Science
  180 (2020) 104773.
\newblock \href {http://dx.doi.org/10.1016/j.pss.2019.104773}
  {\path{doi:10.1016/j.pss.2019.104773}}.

\bibitem{debris2015}
T.~Ebisuzaki, et~al.,
  \href{https://www.sciencedirect.com/science/article/pii/S0094576515000867}{Demonstration
  designs for the remediation of space debris from the international space
  station}, Acta Astronautica 112 (2015) 102--113.
\newblock \href
  {http://dx.doi.org/https://doi.org/10.1016/j.actaastro.2015.03.004}
  {\path{doi:https://doi.org/10.1016/j.actaastro.2015.03.004}}.
\newline\urlprefix\url{https://www.sciencedirect.com/science/article/pii/S0094576515000867}

\bibitem{Bisconti2021}
F.~e.~a. Bisconti, Pre-flight qualification tests of the mini-euso telescope
  engineering model, Experimental Astronomy\href
  {http://dx.doi.org/https://doi.org/10.1007/s10686-021-09805-w}
  {\path{doi:https://doi.org/10.1007/s10686-021-09805-w}}.

\bibitem{Spitzer1949}
L.~Spitzer, \href{https://link.aps.org/doi/10.1103/PhysRev.76.583}{On the
  origin of heavy cosmic-ray particles}, Phys. Rev. 76 (1949) 583--583.
\newblock \href {http://dx.doi.org/10.1103/PhysRev.76.583}
  {\path{doi:10.1103/PhysRev.76.583}}.
\newline\urlprefix\url{https://link.aps.org/doi/10.1103/PhysRev.76.583}

\bibitem{1954Tell....6..232A}
H.~{Alfv{\'e}n}, On the origin of cosmic radiation, Tellus 6~(3) (1954)
  232--253.
\newblock \href {http://dx.doi.org/10.3402/tellusa.v6i3.8739}
  {\path{doi:10.3402/tellusa.v6i3.8739}}.

\bibitem{1956Tell....8..268H}
N.~{Herlofson}, Accelerated dust grains and the highest cosmic ray energies,
  Tellus 8~(2) (1956) 268--273.
\newblock \href {http://dx.doi.org/10.3402/tellusa.v8i2.8954}
  {\path{doi:10.3402/tellusa.v8i2.8954}}.

\bibitem{Hayakawa1972}
S.~Hayakawa, \href{https://doi.org/10.1007/BF00642736}{Dust grain origin of
  cosmic ray air showers}, Astrophysics and Space Science 16~(2) (1972)
  238--240.
\newblock \href {http://dx.doi.org/10.1007/BF00642736}
  {\path{doi:10.1007/BF00642736}}.
\newline\urlprefix\url{https://doi.org/10.1007/BF00642736}

\bibitem{1999APh....12...35B}
R.~{Bingham}, V.~N. {Tsytovich}, Comments on relativistic dust particles
  forming the highest energy cosmic rays, Astroparticle Physics 12~(1-2) (1999)
  35--44.
\newblock \href {http://dx.doi.org/10.1016/S0927-6505(99)00015-8}
  {\path{doi:10.1016/S0927-6505(99)00015-8}}.

\bibitem{PhysRevD.61.087302}
L.~A. Anchordoqui,
  \href{https://link.aps.org/doi/10.1103/PhysRevD.61.087302}{Cosmic dust grains
  strike again}, Phys. Rev. D 61 (2000) 087302.
\newblock \href {http://dx.doi.org/10.1103/PhysRevD.61.087302}
  {\path{doi:10.1103/PhysRevD.61.087302}}.
\newline\urlprefix\url{https://link.aps.org/doi/10.1103/PhysRevD.61.087302}

\bibitem{2001NuPhS..97..203A}
L.~A. {Anchordoqui}, et~al., A pot of gold at the end of the cosmic
  ``raynbow''?, Nuclear Physics B Proceedings Supplements 97~(1-3) (2001)
  203--206.
\newblock \href {http://arxiv.org/abs/astro-ph/0006071}
  {\path{arXiv:astro-ph/0006071}}, \href
  {http://dx.doi.org/10.1016/S0920-5632(01)01264-6}
  {\path{doi:10.1016/S0920-5632(01)01264-6}}.

\bibitem{1973Ap&SS..21..475B}
V.~S. {Berezinsky}, O.~F. {Prilutsky}, On the hypothesis of dust grain origin
  of cosmic rays air showers, Astrophysics and Space Science 21~(2) (1973)
  475--476.
\newblock \href {http://dx.doi.org/10.1007/BF00643110}
  {\path{doi:10.1007/BF00643110}}.

\bibitem{1977ICRC....2..358B}
V.~S. {Berezinsky}, O.~F. {Prilutsky}, Mechanisms of interstellar dust particle
  destruction, in: International Cosmic Ray Conference, Vol.~2 of International
  Cosmic Ray Conference, 1977, p. 358.

\bibitem{1977Afz....13..731E}
I.~S. {Elenskii}, A.~L. {Suvorov}, Mechanism of ``surviving'' of cosmic
  relativistic specks of dust., Astrofizika 13 (1977) 731--735.

\bibitem{1980ApJ...235L.167L}
J.~{Linsley}, {Are extensive air showers produced by relativistic dust grains},
  Astrophysical Journal 235 (1980) L167--L169.
\newblock \href {http://dx.doi.org/10.1086/183183} {\path{doi:10.1086/183183}}.

\bibitem{1981ICRC....2..141L}
J.~{Linsley}, Large air showers and the dust grain hypothesis, in:
  International Cosmic Ray Conference, Vol.~2 of International Cosmic Ray
  Conference, 1981, p. 141.

\bibitem{Hoang:2014bba}
T.~Hoang, A.~Lazarian, R.~Schlickeiser, {On origin and destruction of
  relativistic dust and its implication for ultrahigh energy cosmic rays},
  Astrophys. J. 806~(2) (2015) 255.
\newblock \href {http://arxiv.org/abs/1412.0578} {\path{arXiv:1412.0578}},
  \href {http://dx.doi.org/10.1088/0004-637X/806/2/255}
  {\path{doi:10.1088/0004-637X/806/2/255}}.

\bibitem{Khrenov:2000xt}
B.~A. Khrenov, et~al., {Space program KOSMOTEPETL (projects KLYPVE and TUS) for
  the study of extremely high energy cosmic rays}, AIP Conf. Proc. 566~(1)
  (2001) 57--75.
\newblock \href {http://dx.doi.org/10.1063/1.1378622}
  {\path{doi:10.1063/1.1378622}}.

\bibitem{JCAP2020}
B.~A. {Khrenov}, et~al., An extensive-air-shower-like event registered with the
  {TUS} orbital detector, Journal of Cosmology and Astroparticle Physics
  2020~(3) (2020) 033.
\newblock \href {http://arxiv.org/abs/1907.06028} {\path{arXiv:1907.06028}},
  \href {http://dx.doi.org/10.1088/1475-7516/2020/03/033}
  {\path{doi:10.1088/1475-7516/2020/03/033}}.

\bibitem{Khrenov:2021df}
B.~Khrenov, et~al., {Relativistic dust grains: a new subject of research with
  orbital fluorescence detectors}, in: Proceedings of 37th International Cosmic
  Ray Conference {\textemdash} PoS(ICRC2021), Vol. 395, 2021, p. 315.
\newblock \href {http://arxiv.org/abs/2108.07021} {\path{arXiv:2108.07021}},
  \href {http://dx.doi.org/10.22323/1.395.0315}
  {\path{doi:10.22323/1.395.0315}}.

\bibitem{Liou}
K.~N. Liou, Radiation and cloud processes in the atmosphere. Theory,
  observation, and modeling, Oxford University Press, 1992.

\bibitem{Liu_clouds}
Z.~Ma, et~al., Application and evaluation of an explicit prognostic cloud-cover
  scheme in grapes global forecast system, Journal of Advances in Modeling
  Earth Systems 10 (2018) 652--667.
\newblock \href {http://dx.doi.org/10.1002/2017MS001234}
  {\path{doi:10.1002/2017MS001234}}.

\bibitem{SkamarockWRF}
W.~Skamarock, et~al., A description of the advanced research wrf model version
  4, Mesoscale and Microscale Meteorological Division NCAR, Boulder, Colorado,
  USA, Tech rep\href {http://dx.doi.org/10.5065/1dfh-6p97}
  {\path{doi:10.5065/1dfh-6p97}}.

\bibitem{Anzalone}
A.~Anzalone, et~al., Methods to retrieve the cloud-top height in the frame of
  the jem-euso mission, IEEE Transaction on Geoscience and remote sensing
  57~(1) (2018) 304--318.
\newblock \href {http://dx.doi.org/10.1109/TGRS.2018.2854296}
  {\path{doi:10.1109/TGRS.2018.2854296}}.

\bibitem{Toscano:2014bga}
S.~Toscano, A.~Neronov, M.~D. Rodr\'\i{}guez~Fr\'\i{}as, S.~Wada, {\bf
  JEM-EUSO} Collaboration, {The Atmospheric Monitoring system of the JEM-EUSO
  telescope}, Exper. Astron. 40~(1) (2015) 45--60.
\newblock \href {http://arxiv.org/abs/1402.6097} {\path{arXiv:1402.6097}},
  \href {http://dx.doi.org/10.1007/s10686-014-9378-1}
  {\path{doi:10.1007/s10686-014-9378-1}}.

\bibitem{B805153D}
C.~D. Hatch, V.~H. Grassian, \href{http://dx.doi.org/10.1039/B805153D}{10th
  anniversary review: Applications of analytical techniques in laboratory
  studies of the chemical and climatic impacts of mineral dust aerosol in the
  earth{'}s atmosphere}, J. Environ. Monit. 10 (2008) 919--934.
\newblock \href {http://dx.doi.org/10.1039/B805153D}
  {\path{doi:10.1039/B805153D}}.
\newline\urlprefix\url{http://dx.doi.org/10.1039/B805153D}

\bibitem{Wiacek2009}
A.~Wiacek, T.~Peter, On the availability of uncoated mineral dust ice nuclei in
  cold cloud regions, Geophysical Research Letters - GEOPHYS RES LETT 36.
\newblock \href {http://dx.doi.org/10.1029/2009GL039429}
  {\path{doi:10.1029/2009GL039429}}.

\bibitem{Hastings1991}
J.~Hastings, J.~Morin, C.~Prosser, Neural and integrative animal physiology,
  Wiley-Interscience, New York (1991) 131--170.

\bibitem{Miller2005}
S.~D. Miller, S.~H.~D. Haddock, C.~D. Elvidge, T.~F. Lee,
  \href{https://www.pnas.org/content/102/40/14181}{Detection of a
  bioluminescent milky sea from space}, Proceedings of the National Academy of
  Sciences 102~(40) (2005) 14181--14184.
\newblock \href
  {http://arxiv.org/abs/https://www.pnas.org/content/102/40/14181.full.pdf}
  {\path{arXiv:https://www.pnas.org/content/102/40/14181.full.pdf}}, \href
  {http://dx.doi.org/10.1073/pnas.0507253102}
  {\path{doi:10.1073/pnas.0507253102}}.
\newline\urlprefix\url{https://www.pnas.org/content/102/40/14181}

\bibitem{PhysPhDTrends2019}
P.~J. Mulvey, S.~Nicholson, J.~Pold,
  \href{https://www.aip.org/statistics/reports/trends-physics-phds-171819}{Trends
  in physics phds}, American Institute of Physics (2021).
\newline\urlprefix\url{https://www.aip.org/statistics/reports/trends-physics-phds-171819}

\bibitem{PhysBachelors2018}
P.~J. Mulvey, S.~Nicholson,
  \href{https://www.aip.org/statistics/reports/physics-bachelors-degrees-2018}{{Physics
  Bachelor's Degrees: 2018}}, American Institute of Physics (2020).
\newline\urlprefix\url{https://www.aip.org/statistics/reports/physics-bachelors-degrees-2018}

\bibitem{StudentsCensus2018}
\href{https://www.census.gov/newsroom/press-releases/2018/school-enrollment.html}{{More
  Than 76 Million Students Enrolled in U.S. Schools, Census Bureau Reports}},
  Census Bureau Reports, United States Census Bureau (2018).
\newline\urlprefix\url{https://www.census.gov/newsroom/press-releases/2018/school-enrollment.html}

\bibitem{MDNWeb}
\href{https://astromdn.github.io/}{The multimessenger diversity network},
  Github.
\newline\urlprefix\url{https://astromdn.github.io/}

\bibitem{cscce}
{Center for Scientific Collaboration and Community Engagement},
  \url{https://www.cscce.org/}.

\bibitem{Woodley:2020cscce}
L.~Woodley, K.~Pratt, \href{https://doi.org/10.5281/zenodo.3997802}{The cscce
  community participation model}, Center for Scientific Collaboration and
  Community Engagement (2020).
\newline\urlprefix\url{https://doi.org/10.5281/zenodo.3997802}

\bibitem{FAIR}
{\relax FAIR Data Principles},
  https://www.force11.org/group/fairgroup/fairprinciples (2020).

\bibitem{Haungs:2018xpw}
A.~Haungs, et~al., {The KASCADE Cosmic-ray Data Centre KCDC: Granting Open
  Access to Astroparticle Physics Research Data}, Eur. Phys. J. C 78~(9) (2018)
  741.
\newblock \href {http://arxiv.org/abs/1806.05493} {\path{arXiv:1806.05493}},
  \href {http://dx.doi.org/10.1140/epjc/s10052-018-6221-2}
  {\path{doi:10.1140/epjc/s10052-018-6221-2}}.

\bibitem{Berlin}
{\relax Berlin Declaration on Open Access to Knowledge in the Sciences and
  Humanities}, {https://openaccess.mpg.de/Berlin-Declaration} (2015).

\bibitem{AugerOD}
{\relax Pierre Auger Observatory}, Auger open data, https://opendata.auger.org/
  (2021).

\bibitem{Allen:2018yvz}
G.~Allen, et~al., {Multi-Messenger Astrophysics: Harnessing the Data
  Revolution}, 2018.
\newblock \href {http://arxiv.org/abs/1807.04780} {\path{arXiv:1807.04780}}.

\bibitem{ipcc}
{IPCC}, {Summary for Policymakers}, in: V.~Masson-Delmotte, et~al. (Eds.),
  Climate Change 2021: The Physical Science Basis. Contribution of Working
  Group I to the Sixth Assessment Report of the Intergovernmental Panel on
  Climate Change, Cambridge University Press, in press.

\bibitem{paris2015}
{United Nations}, {Paris Agreement},
  \url{https://treaties.un.org/pages/ViewDetails.aspx?chapter=27&clang=_en&mtdsg_no=XXVII-7-d&src=TREATY},
  treaty No. XXVII-7-d (12 2015).

\bibitem{sru2020}
{German Advisory Council on the Environment}, {Using the CO$_\mathrm{2}$ budget
  to meet the Paris climate targets, Environmental Report 2020, Chapter 2}
  (2020).

\bibitem{Grinberg:2022jah}
V.~Grinberg, K.~Jahnke, V.~Lindenstruth, C.~Markou, S.~Funk, U.~Katz, M.~Roth,
  {Sustainability in Astroparticle Physics}, PoS ICRC2021 (2022) 1401.
\newblock \href {http://dx.doi.org/10.22323/1.395.1401}
  {\path{doi:10.22323/1.395.1401}}.

\bibitem{Aujoux:2021kub}
C.~Aujoux, K.~Kotera, O.~Blanchard, {\bf GRAND} Collaboration, {Estimating the
  carbon footprint of the GRAND Project, a multi-decade astrophysics
  experiment}, Astropart. Phys. 131 (2021) 102587.
\newblock \href {http://arxiv.org/abs/2101.02049} {\path{arXiv:2101.02049}},
  \href {http://dx.doi.org/10.1016/j.astropartphys.2021.102587}
  {\path{doi:10.1016/j.astropartphys.2021.102587}}.

\bibitem{jahnke2020}
K.~{Jahnke}, C.~{Fendt}, M.~{Fouesneau}, I.~{Georgiev}, T.~{Herbst},
  M.~{Kaasinen}, D.~{Kossakowski}, J.~{Rybizki}, M.~{Schlecker}, G.~{Seidel},
  T.~{Henning}, L.~{Kreidberg}, H.-W. {Rix}, {An astronomical institute's
  perspective on meeting the challenges of the climate crisis}, Nature
  Astronomy 4 (2020) 812--815.
\newblock \href {http://arxiv.org/abs/2009.11307} {\path{arXiv:2009.11307}},
  \href {http://dx.doi.org/10.1038/s41550-020-1202-4}
  {\path{doi:10.1038/s41550-020-1202-4}}.

\bibitem{Stevens:2019ntb}
A.~R.~H. Stevens, S.~Bellstedt, P.~J. Elahi, M.~T. Murphy, {The imperative to
  reduce carbon emissions in astronomy}, Nature Astron. 4~(9) (2020) 843--851.
\newblock \href {http://arxiv.org/abs/1912.05834} {\path{arXiv:1912.05834}},
  \href {http://dx.doi.org/10.1038/s41550-020-1169-1}
  {\path{doi:10.1038/s41550-020-1169-1}}.

\bibitem{Pat1}
V.~Lindenstruth, H.~St{\"o}cker, Building for a computer centre with devices
  for efficient cooling, {US Patent 2011/0220324 A1}.

\bibitem{Pat2}
V.~Lindenstruth, H.~St{\"o}cker, Methods and apparatus for temperature control
  of computer racks and computer data centres, {US Patent 2017/0254551 A1}.

\bibitem{Bach:2012iw}
M.~Bach, V.~Lindenstruth, O.~Philipsen, C.~Pinke, {Lattice QCD based on
  OpenCL}, Comput. Phys. Commun. 184 (2013) 2042--2052.
\newblock \href {http://arxiv.org/abs/1209.5942} {\path{arXiv:1209.5942}},
  \href {http://dx.doi.org/10.1016/j.cpc.2013.03.020}
  {\path{doi:10.1016/j.cpc.2013.03.020}}.

\bibitem{Gerhard:2012uf}
J.~Gerhard, V.~Lindenstruth, M.~Bleicher, {Relativistic Hydrodynamics on
  Graphic Cards}, Comput. Phys. Commun. 184 (2013) 311--319.
\newblock \href {http://arxiv.org/abs/1206.0919} {\path{arXiv:1206.0919}},
  \href {http://dx.doi.org/10.1016/j.cpc.2012.09.013}
  {\path{doi:10.1016/j.cpc.2012.09.013}}.

\bibitem{Guidi:2021dbl}
C.~Guidi, M.~Bou-Cabo, G.~Lara, {\bf KM3NeT} Collaboration, {Passive acoustic
  monitoring of cetaceans with KM3NeT acoustic receivers}, JINST 16~(10) (2021)
  C10004.
\newblock \href {http://dx.doi.org/10.1088/1748-0221/16/10/C10004}
  {\path{doi:10.1088/1748-0221/16/10/C10004}}.

\bibitem{ademe}
\href{https://bilans-ges.ademe.fr/en/accueil}{Ademe database}.
\newline\urlprefix\url{https://bilans-ges.ademe.fr/en/accueil}

\bibitem{Remmel:2021}
A.~Remmel, {Scientists want virtual meetings to stay after the COVID pandemic},
  Nature 591 (2021) 185.

\bibitem{PierreAuger:2021ddg}
K.~S. Caballero~Mora, et~al., {\bf Pierre Auger} Collaboration, {Outreach
  activities at the Pierre Auger Observatory}, PoS ICRC2021 (2021) 1374.
\newblock \href {http://dx.doi.org/10.22323/1.395.1374}
  {\path{doi:10.22323/1.395.1374}}.

\end{thebibliography}

\end{document}